\documentclass[8pt,DIV=27,a4paper]{scrartcl}

\usepackage{amssymb}
\usepackage{amsmath}
\usepackage{array}
\usepackage{algorithm}
\usepackage{algpseudocode}
\usepackage{graphicx}
\usepackage{cite}
\usepackage{morefloats}
\usepackage{wrapfig}
\usepackage[numbers]{natbib}
\newcommand{\pderfoper}[1]{{\frac{\partial}{\partial #1}}}
\newcommand{\pdernoper}[2]{{\frac{\partial^{#2}}{\partial #1^{#2}}}}
\newcommand{\pderf}[2]{{\frac{\partial #1}{\partial #2}}}
\newcommand{\pdern}[3]{{\frac{\partial^{#3} #1}{\partial #2^{#3}}}}

\newcommand{\Dx}[0]{{\Delta x}}
\newcommand{\Dt}[0]{{\Delta t}}
\newcommand{\Dtn}[1]{{{\Delta t}^{#1}}}
\newcommand{\intrefdom}[1]{{\int\limits_{0}^{1} d{#1}}}
\newcommand{\uphi}{\smash[b]{\underline{\varphi}}}
\newcommand{\ophi}{\smash[b]{\overline{\varphi}}}
\newcommand{\uvecphi}{\smash[b]{\underline{\boldsymbol{\varphi}}}}
\newcommand{\ovecphi}{\smash[b]{\overline{\boldsymbol{\varphi}}}}

\graphicspath{{./figs/}}

\title{Theory and internal structure of ADER-DG method for partial differential equations}
\author{Popov I.S.\thanks{Department of Theoretical Physics, Dostoevsky Omsk State University, Omsk, Russia {\em diphosgen@mail.ru, popovis@omsu.ru}}}
\date{}

\begin{document}
\maketitle

\begin{abstract}
\noindent
Highly accurate stability boundary values for the ADER-DG method with the LST-DG predictor are obtained for arbitrary degrees $N$ of basis polynomials. It is found that, in the linear case, instability is violated precisely when one of the matrix eigenvalues reaches $\lambda = -1$, regardless of the phase $\theta$ of the numerical solution. A rigorous mathematical framework for the stability of the ADER-DG method is developed. The stability condition is significantly simplified, reducing it to the problem of calculating the roots of polynomials in the Courant number $\mathrm{CFL}$, from the initially complex problem of finding the maximum spectral radius of a matrix dependent on the continuous phase $\theta\in[0, 2\pi]$. The maximum of the Courant numbers $\mathrm{CFL}_{\rm max}(N)$ for the ADER-DG method are calculated. These results are new and very convenient for practical use. A comparison of the obtained results with existing results reveals differences that may be significant for the selection of calculation parameters using the ADER-DG method, especially for high degrees $N$. It is shown that widely used  existing estimates $\mathrm{CFL}_{\rm max}(N) \propto 1/(2N+1)$ are overestimated. An interesting qualitative asymptotic $\mathrm{CFL}_{\rm max}(N) \propto 1/(N+1)^{2}$ is obtained. A rigorous direct proof of the approximation of the ADER-DG method is presented. Approximation orders $p = N+1$ for arbitrary degrees $N$ are rigorously derived. A set of numerical experiments is carried out to apply the ADER-DG method to solving both a linear advection equation and an Euler system of equations. The results obtained in these calculations confirm the theoretical results well. In particular, an excess of the Courant number over the $\mathrm{CFL}_{\rm max}(N)$ by even 1\% in the linear case immediately leads to significant instability of the numerical solution. The obtained quantitative estimates of the boundary Courant number in the nonlinear case are somewhat underestimated -- by no more than 5\%, which is due to the diffusivity and stability of the approximate Riemann solver. Empirical convergence orders are obtained, which are in good agreement with the theoretical results $p = N+1$ for the orders of approximation.
\end{abstract}

\noindent
\noindent
\begin{minipage}[t]{0.35\textwidth}
\textbf{Highlights:}
\begin{itemize}
\setlength{\itemsep}{1pt}
\setlength{\parskip}{0pt}
\item A new deep stability theory for the ADER-DG method with an LST-DG predictor of arbitrarily high order is constructed.
\item Simple polynomial equations for the maximum Courant numbers for the ADER-DG method of arbitrarily high order are derived.
\item The maximum Courant numbers for the ADER-DG method are calculated for degrees $1$--$12$ of basis polynomials.
\item The stability limits of the ADER-DG method of arbitrarily high order are shown upon reaching the maximum Courant number for linear and nonlinear systems of equations.
\item A direct approximation theory for the ADER-DG method with an LST-DG predictor of arbitrarily high order is constructed.
\end{itemize}
\end{minipage}
\hfill
\begin{minipage}[t]{0.59\textwidth}
\centering
\small\textbf{Graphical Abstract}
\includegraphics[width=\textwidth]{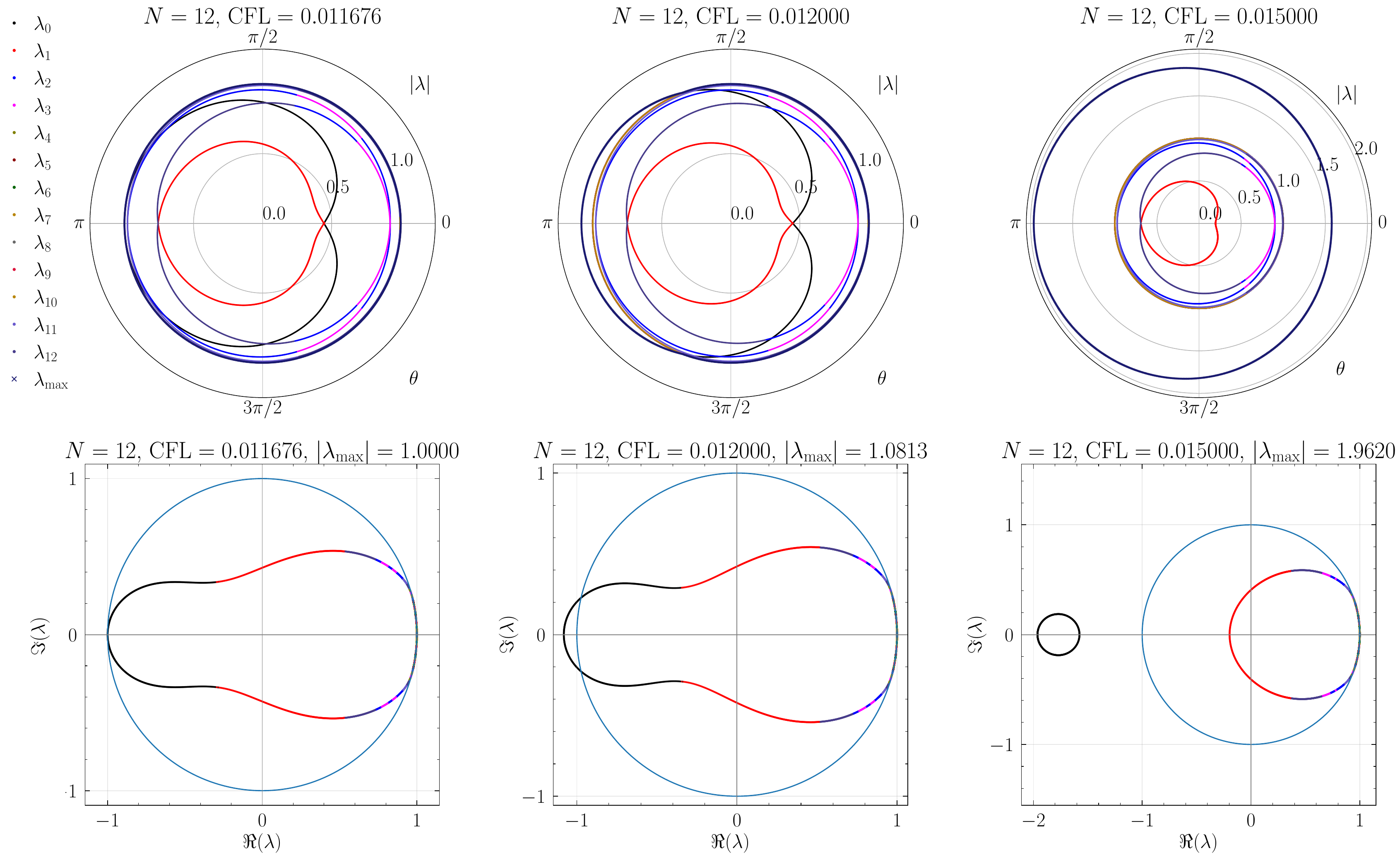}
\vfill
\end{minipage}


\section{Introduction}
\label{intro}

Finite-element ADER-DG numerical methods have an arbitrarily high order and are among the most accurate numerical methods for solving systems of partial differential equations (PDE systems)~\cite{ader_rev_2024}. Titarev and Toro~\cite{ader_init_1, ader_init_2} developed the ADER paradigm for use in finite-volume methods, and the main innovation was the use of a solution to the generalized Riemann problem based on calculating the evolution of a local space-time solution on the interfaces of finite-volume cells using the Cauchy-Kovalevskaya procedure, based on expanding the solution into high-order Taylor series in coordinates and time (see also~\cite{ader_init_3}). The ADER paradigm was significantly developed in subsequent works~\cite{ader_init_4, ader_init_5, ader_init_6} by Titarev and Toro. Dumbser \textit{et al}~\cite{ader_stiff_1, ader_stiff_2} modified the ADER paradigm for finite-volume methods based on the use of a discrete space-time solution obtained in a separate stage of the local space-time DG-predictor (LST-DG predictor), and showed higher solution accuracy than the original version, especially for stiff problems. Finite-element ADER-DG methods are fundamentally linear methods for solving PDE systems, so they are subject to the well-known Godunov theorem and such methods require limiters to preserve the monotonicity of the numerical solution. Currently, finite-element ADER-DG methods are based on the development of the Multi-dimensional Optimal Order Detection (MOOD) paradigm~\cite{mood_par_1, mood_par_2, mood_par_3, mood_par_4}, which resulted in the development~\cite{ader_dg_dev_1, ader_dg_dev_2} of finite-element ADER-DG methods with \textit{a posteriori} correction of the solution in subcells by a finite-volume limiter, which can be the finite-volume ADER-WENO method~\cite{ader_weno_lstdg_ideal, ader_weno_lstdg_diss}. This approach made it possible to preserve the monotonicity of the numerical solution and the subgrid resolution characteristic of DG methods~\cite{ader_dg_dev_1}. However, the further development of purely finite-volume methods of the ADER family, carried out in the works~\cite{ader_weno_new_1, ader_weno_new_2, ader_weno_new_3, ader_weno_new_4, ader_weno_new_5, ader_weno_new_6, ader_weno_new_7}, is not directly related to the ADER-DG method with a posteriori correction of the solution.

The current modern state and formulations of finite-element ADER-DG numerical methods with the LST-DG predictor are presented in~\cite{ader_rev_2024, ader_dg_simple_mod_2025, ader_dg_axioms, exahype, fron_phys, ader_dg_simple_mod_2016, ader_dg_mod_1, ader_dg_mod_2, ader_dg_mod_3}. It is necessary to note the historical aspect that the DG methods were created in the work~\cite{lasl_rep_dg_1973} for numerical solution of the neutron transport equations. Cockburn, Shu \textit{et al} in a series of works~\cite{Cockburn_base_1, Cockburn_base_2, Cockburn_base_3, Cockburn_base_4, Cockburn_base_5} created an accurate and thoroughly developed mathematical basis of DG methods, which stimulated their further development and use for solving a wide class of problems. The ADER-DG numerical methods with the LST-DG predictor are widely used in solving problems of classical ideal and dissipative hydrodynamics and magnetohydrodynamics~\cite{ader_dg_ideal_flows, ader_dg_diss_flows, ader_dg_semiexpl, PNPM_DG_2010}, relativistic hydrodynamics and magnetohydrodynamics~\cite{ader_dg_grmhd, ader_dg_PNPM, PNPM_DG_2008, PNPM_DG_2009}, which are characterized by essentially stiff terms. In recent years, significant interest~\cite{ader_dg_gr_prd, ader_dg_gr_z4_2024, ader_dg_gr_z4_2026} has been associated with the development and use of numerical ADER-DG methods for solving Einstein equations of general relativity in the formulations of first-order hyperbolic systems~\cite{rel_hyd_book_2013, rel_hyd_book_2007, rel_hyd_book_2008, rel_hyd_book_2009, rel_hyd_book_2010, rel_hyd_book_2012, rel_hyd_book_2021}. The ADER-DG numerical methods are also used to solve the equations of motion of deformable media~\cite{ader_dg_axioms, fron_phys, ader_dg_hyperelastic}, seismic wave propagation~\cite{ader_dg_axioms, exahype, ader_dg_seiemic, ader_dg_seiemic_underwater}, the equations of shallow water theory~\cite{ADER_GRP_LST_DG_1, ader_dg_wb_shwater_2018, ader_dg_wb_shwater_2022}, simulating blood flow~\cite{ader_eno_fv_blood_2022}, simulating compressible barotropic two‐fluid flows~\cite{ader_dg_barotropic} and nonlinear dispersive systems~\cite{ader_dg_dispersive}. A special place is occupied by problems associated with the study of the reacting flows, which are characterized by abnormally high stiffness, for which the finite element ADER-DG methods and finite volume ADER-WENO methods were developed and studied in the works~\cite{ader_stiff_1, ader_stiff_2, ader_stiff_3, ader_stiff_4}. Application of finite element ADER-DG method for solving a wide range of other problems are also presented in~\cite{ader_dg_axioms, exahype}.

Finite-element ADER-DG numerical method with the LST-DG predictor are developed for structured Cartesian meshes~\cite{ader_dg_dev_1, ader_dg_dev_2, ader_dg_grmhd, ader_dg_gr_prd, ader_dg_gr_z4_2024}, using the adaptive mesh refinement paradigm~\cite{ader_weno_lstdg_ideal, ader_weno_lstdg_diss, ader_dg_ideal_flows, ader_dg_diss_flows, ader_dg_PNPM}, unstructured triangular and tetrahedral meshes~\cite{ader_dg_simple_mod_2025, ader_dg_simple_mod_2016, PNPM_DG_2009, PNPM_DG_2010, ader_dg_axioms, ader_dg_wb_shwater_2018} and moving meshes~\cite{ader_dg_ale} (see also~\cite{ader_rev_2024}). A detailed discussion of the features of the mathematical and algorithmic formulations of numerical methods and their software implementation is presented in works~\cite{ader_dg_axioms, exahype, fron_phys, ader_dg_hpc_impl_1, ader_dg_hpc_impl_2, ader_dg_hpc_impl_3, ader_dg_hpc_impl_4, ader_eff_blas}. The implementation of the ADER-DG numerical method on a GPU is presented in~\cite{ader_dg_impl_gpu}. Dumbser \textit{et al} developed well-balanced ADER-DG method for the Euler equations in general relativity problems~\cite{ader_dg_gr_z4_2024} and for the multilayer shallow water equations~\cite{ader_dg_wb_shwater_2022}. Zhao \textit{et al} developed path-conservative discontinuous Galerkin (DG) method to solve shallow water equations~\cite{ader_dg_wb_pc_shwater_2024}. In~\cite{dg_entropy} Gaburro \textit{et al} developed entropy preserving ADER-DG schemes (see also~\cite{dg_entropy_add}). Based on the ADER paradigm, a SPH numerical method was created~\cite{ader_weno_sph}, which demonstrated a very high quality of modeling hydrodynamic and magnetohydrodynamic flows with discontinuities, compared to classical SPH methods.

The numerical methods of the ADER family are significantly interconnected with the deferred correction (DeC) numerical methods, which showed Han Veiga \textit{et al}~\cite{dec_vs_ader_2021}. DeC methods have a long history and their application goes back to~\cite{dec_src_1968}, and are effectively used to solve partial differential equations~\cite{dec_abgrall_2017, dec_abgrall_2019} and ordinary differential systems of equations~\cite{dec_dutt_2000, dec_dutt_2000, dec_minion_2003, dec_shu_2008}. DeC methods, like the methods of the ADER family, allow one to obtain an arbitrarily high order and are characterized by a high accuracy of the numerical solution. The current state of research on DeC methods~\cite{dec_vs_ader_2021} shows that they compete with methods of the ADER family. The relationship between the ADER paradigm and the DeC paradigm are discussed in~\cite{dec_vs_ader_2021, dec_vs_ader_2023}.

An important feature of the ADER-DG method, like other DG methods, is a very strict limitation on the time step and on the permissible maximum Courant number $\mathrm{CFL}_{\rm max}$, especially for high degrees $N$ of the basis polynomials used in the DG representation of the solution (in particular, Huynh developed~\cite{fr_init_2007, fr_init_2009, fr_init_rev} DG methods with flux reconstruction to weaken this limitation; see also the discussion in~\cite{fr_1, fr_2, fr_3, fr_4, fr_5}). ADER-DG methods are characterized by very small values of the Courant number $\mathrm{CFL}_{\rm max}$, which is used in particular for subcell limiting of the solution with finite-volume limiters, which do not have such a strict limitation on the time step and can reliably compute solutions even for smaller coordinate grid steps~\cite{ader_dg_dev_1, ader_dg_dev_2, ader_dg_ideal_flows, ader_dg_diss_flows, ader_dg_PNPM}.

There are several different estimates of the maximum Courant number $\mathrm{CFL}_{\rm max}(N)$ for the ADER-DG method with the LST-DG predictor. Many works (for example~\cite{ader_dg_dev_1, ader_dg_dev_2, ader_weno_lstdg_ideal, ader_dg_grmhd, ader_stiff_3, ader_stiff_4}) employ the widely used estimate $\mathrm{CFL}_{\rm max}(N) \propto 1/(2N+1)$, which, in particular, forms the basis for a certain rule for choosing the number of subcells $N_{s} = 2N+1$ on each cell for a posteriori subgrid constraint of a solution by a finite-volume limiter~\cite{ader_dg_dev_1, ader_dg_dev_2, ader_weno_lstdg_ideal}, for which the Courant number is usually expected to be approximately one. In~\cite{PNPM_DG_2008} the estimates of the maximum Courant number $\mathrm{CFL}_{\rm max}(N)$ for the ADER-DG method with the LST-DG predictor were first calculated for degrees $N = 1, \ldots, 4$ of basis polynomials. In a subsequent work~\cite{ader_dg_PNPM} estimates of the maximum Courant number $\mathrm{CFL}_{\rm max}(N)$ for the ADER-DG method with the LST-DG predictor were calculated for basis polynomials for degrees $N = 1, \ldots, 6$ of basis polynomials, which were in good agreement with the previous estimates~\cite{PNPM_DG_2008} in the intersection region $N \leqslant 4$. A new study~\cite{ader_dg_stab} has recently been published in which estimates of the maximum Courant number have been calculated for degrees $N = 1, \ldots, 9$ of basis polynomials, and here differences with existing results~\cite{PNPM_DG_2008, ader_dg_PNPM} have begun to arise, which are described in detail in~\cite{ader_dg_stab}. It should be noted that estimates~\cite{ader_dg_stab, ader_dg_PNPM, PNPM_DG_2008} for large degrees $N \gg 1$ of the basis polynomials differ significantly from the widely used estimate $\mathrm{CFL}_{\rm max}(N) \propto 1/(2N+1)$, which is significantly, sometimes by more than twofold, overestimated. Therefore, using estimates $\mathrm{CFL}_{\rm max}(N) \propto 1/(2N+1)$ requires introducing an additional empirical factor $\mathrm{C}(N) \leqslant 1$, with $\mathrm{C}(N) \ll 1$ for large degrees $N$ of the basis polynomials. The calculations in~\cite{ader_dg_stab, ader_dg_PNPM, PNPM_DG_2008} were based on the Von Neumann stability criterion (spectral stability criterion) and were technically implemented by calculating the maximum spectral radius of the amplification matrix (the matrix of the evolution operator for a single time step $\Dt$) for a continuous set of phase $\theta\in[0, 2\pi]$. Clearly, such calculations require an estimate of the robustness of the eigenvalues and an extremely detailed discretization of the continuous set of phase $\theta\in[0, 2\pi]$. Moreover, the evolution operator matrix itself for a single time step, when directly calculated, even in the one-dimensional case, requires multiplying $(N+1)\times(N+1)^{2}$ and $(N+1)^{2}\times(N+1)$ matrices to obtain a $(N+1)\times(N+1)$ matrix, which also affects the final calculation error for large degrees $N$ of the basis polynomials, when using standard machine arithmetic to represent real numbers with single- or double-precision floating-point numbers.

In this work, an attempt is made to construct the consistent and rigorous stability theory for a classical ADER-DG numerical method with the LST-DG predictor based on the Von Neumann stability criterion. This theory is completely free from the problem of enumerating a continuous set of phase $\theta\in[0, 2\pi]$. The theory is built on only one empirically substantiated assumption: the numerical method becomes unstable when one of the eigenvalues $\lambda_{k}$ of the matrix of the evolution operator for a single time step $\Dt$ reaches $\lambda = -1$, regardless of the phase for which this occurs. The resulting equations, the solution of which determines the maximum Courant numbers, are simple finite polynomial equations. These results are new and very convenient for practical use: a detailed Table~\ref{tab:cfls_max_data} of the maximum Courant number $\mathrm{CFL}_{\rm max}(N)$ for degrees $N = 1, \ldots, 12$ of basis polynomials is presented; closed-form polynomial equations for the maximum values of the Courant number $\mathrm{CFL}_{\rm max}(N)$ for an arbitrary degree $N$ of basis polynomials are derived. A rigorous, consistent, and constructive approximation theory for the ADER-DG numerical method with the LST-DG predictor is also constructed, based on direct derivation rather than on integral and variational estimates. The results of numerical experiments are presented, which are in good agreement with the developed theory and demonstrate the limitations associated with incorrect estimates of the maximum Courant number $\mathrm{CFL}_{\rm max}(N)$. The numerical experiments are carried out for a linear advection equations and a nonlinear system of equations --- the Euler system of equations.

The article is divided into four main Sections. Section~\ref{sec:descr} ``General description of the numerical method'' presents a detailed description of the ADER-DG numerical method with the LST-DG predictor for the linear advection equation, for which a direct rigorous approximation theory and stability theory are constructed. This Section introduces all the basic concepts and notations, as well as the main properties of the ADER-DG numerical method with the LST-DG predictor, which are used in the subsequent Sections for research. Section~\ref{sec:approx_anal} ``Approximation analysis'' presents the direct rigorous approximation theory of the ADER-DG numerical method with the LST-DG predictor. Subsection~\ref{sec:approx_anal:local_sol} ``Discrete space-time solution'' is devoted to the derivation of the order and approximation conditions for the local discrete spatiotemporal solution obtained at the LST-DG predictor stage. Subsection~\ref{sec:approx_anal:final_sol} ``Corrected solution at time steps'' is devoted to the derivation of the approximation order for the adjusted solution determined at time steps. This Section also introduces and rigorously proves properties of the matrices that arise during a detailed analysis of the numerical method and are used in the next, main Section of the article. Section~\ref{sec:stab_anal} ``Stability analysis'' is the key Section of this article and is divided into four Subsections. Subsection~\ref{sec:stab_anal:qual_cons} ``Qualitative considerations'' is devoted to the derivation and justification of a new empirical estimate for the maximum Courant number $\mathrm{CFL}_{\rm max}(N)$ of the ADER-DG numerical method with the LST-DG predictor, which has an asymptotic behavior $\mathrm{CFL}_{\rm max}(N) \propto 1/(N+1)^{2}$ for sufficiently large degrees $N \gg 1$ of the basis polynomials. Subsection~\ref{sec:stab_anal:statment_quant_cons} ``Statement of the problem of quantitative considerations'' is devoted to a detailed derivation of the matrices required for a detailed stability analysis of the ADER-DG numerical method with the LST-DG predictor, as well as the derivation and proof of the relations required for further research. The constant-preserving property of the ADER-DG numerical method with the LST-DG predictor is also rigorously proven. Subsection~\ref{sec:stab_anal:empir_cons} ``Empirical considerations'' is devoted to the justification of the empirical assumption that stability is violated when one of the (or several) eigenvalues $\lambda_{k}$ of the matrix of the evolution operator for a single time step $\Dt$ reaches the value of $\lambda = -1$. Subsection~\ref{sec:stab_anal:quant_cons} ``Rigorous quantitative considerations'' is devoted to a detailed derivation of equations whose solutions determine the maximum Courant numbers and are simple finite polynomial equations. This Subsection presents the main results, including the obtained values of the maximum Courant number $\mathrm{CFL}_{\rm max}(N)$ for degrees $N = 1, \ldots, 12$ of the basis polynomials and polynomial equations for the values of the maximum Courant number $\mathrm{CFL}_{\rm max}(N)$ for an arbitrary degree $N$. Section~\ref{sec:app} ``Applications'' contains problem statements and results of numerical experiments that confirm previously developed theoretical results. Subsection~\ref{sec:app:lin_adv} ``Linear advection equation'' is devoted to numerical experiments for a linear advection equation. Subsection~\ref{sec:app:euler} ``Nonlinear Euler equations'' is devoted to numerical experiments for the system of Euler equations, which was chosen as a good example of a nonlinear problem for which ADER-DG numerical method with the LST-DG predictor are widely applicable.


\section{General description of the numerical method}
\label{sec:descr}

This Section presents a detailed description of the ADER-DG numerical method with the LST-DG predictor for the linear advection equation, for which a direct rigorous approximation theory and stability theory are constructed. The advection equation is chosen in the following form:
\begin{equation}\label{eq:adv_eq_src}
\pderf{u}{t} + a\pderf{u}{x} = 0,\quad
(x, t)\in\Omega\times\mathcal{T},\quad
\Omega\subset\mathcal{R},\quad
\mathcal{T} = \left[t_{0}, t_{f}\right]\subset\mathcal{R},
\end{equation}
where $a\in\mathcal{R}$ is the advection velocity, $u: \Omega\times\mathcal{T}\rightarrow\mathcal{R}$ is the desired function. This equation has a well-known exact analytical solution $u(x, t) = u_{0}(x - a(t - t_{0}))$, where $u_{0}(x)$ is the solution at time $t_{0}$, which is a simple shift of the solution function in the direction of advection at velocity $a$.

It should be noted that numerical DG methods, which include the ADER-DG numerical method, are linear numerical methods of arbitrarily high order, and therefore they are very sensitive to the smoothness of the solution to the problem, even in the case of a linear advection equation (\ref{eq:adv_eq_src}). Therefore, in the subsequent presentation of the results of this work, it is assumed that the smoothness of the initial condition $u_{0}(x)$, determined at time $t_{0}$, is sufficient to achieve the expected orders of approximation and convergence of the ADER-DG numerical method. Condition $u_{0}(x) \in \mathcal{C}_{N}(\Omega)$, where $N$ is the degree of the polynomials used in the DG representation of the solution, is usually sufficient for this. Otherwise, a decrease in the order of the ADER-DG numerical method will be observed.

The coordinate domain $\Omega$ is discretized by a one-dimensional mesh represented by non-overlapping coordinate discretization domains $\Omega_{i}$, with nodes $x_{i}$ and a coordinate discretization step $\Dx = x_{i+1} - x_{i}$, which is assumed to be constant for simplicity, but without loss of generality. The time domain $\mathcal{T}$ is also discretized by a one-dimensional mesh with time steps $t^{n}$ and non-overlapping time discretization domains $\mathcal{T}^{n}$, with a discretization time step $\Dtn{n} = t^{n+1} - t^{n}$. The resulting coordinate-time mesh $\{\Xi_{i}^{n}\}$ is defined as follows:
\begin{equation}\label{eq:discr_domains}
\begin{split}
\Omega_{i} = \left[x_{i}, x_{i+1}\right],\ 
\Omega = \bigcup\limits_{i} \Omega_{i},\ 
\mathcal{T}^{n} = \left[t^{n}, t^{n+1}\right],\ 
\mathcal{T} = \bigcup\limits_{n} \mathcal{T}^{n},\ 
\Xi_{i}^{n} = \mathcal{T}^{n} \times \Omega_{i},\ 
\Omega\times\mathcal{T} = \bigcup\limits_{n, i} \Xi_{i}^{n},
\end{split}
\end{equation}
where $\Xi_{i}^{n}$ is the coordinate-time discretization domain.

The ADER-DG numerical method with the LST-DG predictor is constructed on a reference coordinate-time element $\omega^{2} = [0, 1]^{2}$ using the following reversible mapping:
\begin{equation}\label{eq:mapping}
t(\tau) = t^{n} + \tau\Dtn{n},\quad
x(\xi) = x_{i} + \xi\Dx,
\end{equation}
where $\tau\in\omega$ and $\xi\in\omega$ are the rescaled coordinate $x$ and time $t$. The advection equation (\ref{eq:adv_eq_src}) in the reference coordinate-time element $\omega^{2}$ is presented in the following form:
\begin{equation}\label{eq:adv_eq_mapped}
\pderf{u}{\tau} + \frac{a\Dtn{n}}{\Dx}\pderf{u}{\xi} = 0.
\end{equation}
The ADER-DG numerical method is formulated using the weak form of the equation (\ref{eq:adv_eq_mapped}):
\begin{equation}\label{eq:adv_eq_weak}
\intrefdom{\tau}\intrefdom{\xi}\,\varphi_{k}(\xi)\left[\pderf{u}{\tau} + \frac{a\Dtn{n}}{\Dx}\pderf{u}{\xi}\right] = 0,
\end{equation}
for which a finite set $\{\varphi_{k}\}$ of basis functions $\varphi_{k}: \omega\rightarrow\mathcal{R}$ is introduced.

In this work, a nodal polynomial basis $\{\varphi_{k}(\xi)\}$ is chosen, consisting of Lagrange interpolation polynomials $\varphi_{k}\in\mathcal{P}_{N}(\omega)$ with nodal points $\{\xi_{p}\}$ at the roots of shifted Legendre polynomials $\tilde{L}_{N+1}: \omega\rightarrow\mathcal{R}$. The basis functions $\{\varphi_{k}(\xi)\}$ are represented in the following form:
\begin{equation}\label{eq:phi_def}
\varphi_{p}(\xi) = \sum\limits_{k = 0}^{N} \varphi_{p, k} \xi^{k},\quad
\varphi_{p}(\xi) = \prod\limits_{k = 0}^{N}\frac{\xi - \xi_{k}}{\xi_{p} - \xi_{k}},\quad
\tilde{L}_{N+1}(\xi_{p}) = 0,\quad
\varphi_{p}(\xi_{q}) = \delta_{pq},
\end{equation}
where the last expression is used to calculate the polynomial coefficients $\{\varphi_{p, k}\}$, which is technically equivalent to calculating the inverse matrix of the Vandermonde matrix $\{\xi_{q}^{p}\}_{p,q}$. It is well known that Legendre polynomials $\tilde{L}_{N+1}$ do not have multiple roots $\xi_{p}$, therefore the corresponding Vandermonde matrix is non-singular, and hence the interpolation polynomials that make up the basis $\{\varphi_{k}\}$ are unique. The chosen functional basis is orthogonal in the sense of the $\mathcal{L}_{2}(\omega)$-scalar product.

The choice of this functional basis $\{\varphi_{k}\}$ (\ref{eq:phi_def}) allows for the convenient use of the Gauss-Legendre (GL) quadrature rule, in the following form:
\begin{equation}\label{eq:gl_rule}
\intrefdom{\xi}\,f(\xi) \approx \sum\limits_{p = 0}^{N} w_{p} f(\xi_{p}),\quad
\sum\limits_{p = 0}^{N} w_{p} = 1,\quad
w_{p} = \intrefdom{\xi}\,\varphi_{p}(\xi) = \intrefdom{\xi}\,\varphi_{p}^{2}(\xi) > 0,
\end{equation}
where $\{w_{p}\}$ is the weight of the GL quadrature rule. The GL quadrature rule is satisfied exactly for all functions $\mathcal{P}_{2N+1}(\omega)$. The chosen functional basis $\{\varphi_{k}\}$ (\ref{eq:phi_def}) allows for the use of the point-wise evaluation for the expansion coefficients of functions $f: \omega\rightarrow\mathcal{R}$ in this basis:
\begin{equation}\label{eq:pw_eval}
f(\xi) = \sum\limits_{p = 0}^{N} f_{p}\varphi_{p}(\xi) \mapsto \sum\limits_{p = 0}^{N} f(\xi_{p})\varphi_{p}(\xi),
\end{equation}
which is justified by the following chain of transformations:
\begin{equation}
f_{p} = \frac{1}{w_{p}} \intrefdom{\xi}\,f(\xi)\varphi_{p}(\xi) \approx
\frac{1}{w_{p}} \sum\limits_{q = 0}^{N} w_{q} f(\xi_{q})\varphi_{p}(\xi_{q}) =
\frac{1}{w_{p}} \sum\limits_{q = 0}^{N} w_{q} f(\xi_{q})\delta_{pq} = f(\xi_{p}),
\end{equation}
where it is used the orthogonality properties of functions $\{\varphi_{k}\}$, the nodal property of the interpolation polynomial (\ref{eq:phi_def}), and the GL quadrature rule (\ref{eq:gl_rule}). It should be noted that the representation (\ref{eq:pw_eval}) is exact for all functions $\mathcal{P}_{N}(\omega)$.

By introducing the DG representations of the solution at the current $t^{n}$ and next $t^{n+1}$ time steps in the following form:
\begin{equation}\label{eq:u_n_def}
u^{n}(\xi) = \sum\limits_{k = 0}^{N} \hat{u}^{n}_{k} \varphi_{k}(\xi),\quad
u^{n+1}(\xi) = \sum\limits_{k = 0}^{N} \hat{u}^{n+1}_{k} \varphi_{k}(\xi),
\end{equation}
and performing integration by parts in the flux term, the following form of the completely single-step discrete finite element scheme of the ADER-DG numerical method is obtained:
\begin{equation}\label{eq:ader_dg_corr_temp_1}
\begin{split}
&\sum\limits_{l = 0}^{N} \intrefdom{\xi}\,\varphi_{k}(\xi)\varphi_{l}(\xi)\left[\hat{u}^{n+1}_{l} - \hat{u}^{n}_{l}\right] -
\frac{a\Dtn{n}}{\Dx}\intrefdom{\tau}\intrefdom{\xi}\,\varphi_{k}'(\xi) q(\tau, \xi)\\ & +
\frac{\Dtn{n}}{\Dx}\intrefdom{\tau}\,\Big[
  \varphi_{k}(1) F_{\rm RP}\left(q(\tau, 1),\, q^{+}(\tau, 0)\right) -
  \varphi_{k}(0) F_{\rm RP}\left(q^{-}(\tau, 1),\, q(\tau, 0)\right)
\Big] = 0,
\end{split}
\end{equation}
where $q: \omega^{2}\rightarrow\mathcal{R}$ is the local discrete space-time solution used in ADER methods to calculate local and flux terms, and is determined for each coordinate-time element $\Xi_{i}^{n}$ separately, $F_{\rm RP}$ is the solver of the classical Riemann problem, which, in the case of the advection equation (\ref{eq:adv_eq_src}), is expressed in exact closed finite form:
\begin{equation}\label{eq:f_rp}
F_{\rm RP}\left(u_{L},\, u_{R}\right) = \frac{a+|a|}{2} u_{L} + \frac{a-|a|}{2} u_{R} = a \Big[\theta(a)u_{L} + \theta(-a)u_{R}\Big],
\end{equation}
where $\theta$ is the Heaviside $\theta$-function.

The local discrete space-time solution $q$ is a numerical solution of the equation (\ref{eq:adv_eq_mapped}) in coordinate-time discretization domain $\Xi_{i}^{n}$:
\begin{equation}\label{eq:lstdg_pred_eq_src}
\pderf{q}{\tau} + \frac{a\Dtn{n}}{\Dx}\pderf{q}{\xi} = 0,\quad q(0, \xi) = u^{n}(\xi),
\end{equation}
and is chosen in the form of an expansion over a functional basis $\{\Theta_{\mathrm{p}}(\tau, \xi)\, |\, \Theta_{\mathrm{p}}(\tau, \xi)\in\mathcal{P}_{N}(\omega)\times\mathcal{P}_{N}(\omega)\}$, chosen in the form of a tensor product of the functional basis $\{\varphi_{k}\}$:
\begin{equation}\label{eq:local_sol_def}
q(\tau, \xi) = \sum\limits_{\mathrm{p}} \hat{q}_{\mathrm{p}} \Theta_{\mathrm{p}}(\tau, \xi)
             = \sum\limits_{p_{0} = 0}^{N}\sum\limits_{p_{1} = 0}^{N} \hat{q}_{p_{0}p_{1}} \varphi_{p_{0}}(\tau)\varphi_{p_{1}}(\xi),
\end{equation}
where $\mathrm{p}$ is a multi-index for which projections $(p_{0}, p_{1})$ are defined:
\begin{equation}\label{eq:local_sol_idx}
\mathrm{p} = (p_{0},\, p_{1}),\quad p_{0} = \pi_{0}(\mathrm{p}) = \mathrm{idx}(\mathrm{p}),\quad p_{1} = \pi_{1}(\mathrm{p}),\quad
\mathrm{idx}(\mathrm{p}) = p_{0}\cdot(N+1) + p_{1},\quad 0 \leqslant p_{0},\, p_{1} \leqslant N.
\end{equation}
Choosing basis functions $\{\Theta_{\mathrm{p}}(\tau, \xi)\}$ in the form of a tensor product will allow us to conveniently represent the matrices of the linear transformations as Kronecker products of the matrices of the original basis $\{\varphi_{k}\}$.

The equation (\ref{eq:lstdg_pred_eq_src}) for obtaining the local discrete space-time solution $q$ is considered in weak form:
\begin{equation}\label{eq:lstdg_pred_eq_weak}
\intrefdom{\tau}\intrefdom{\xi}\,\Theta_{\mathrm{p}}(\tau, \xi)\left[\pderf{q}{\tau} + \frac{a\Dtn{n}}{\Dx}\pderf{q}{\xi}\right] = 0,\quad q(0, \xi) = u^{n}(\xi),
\end{equation}
which is the LST-DG predictor. It should be noted that the initial conditions in the problem (\ref{eq:lstdg_pred_eq_weak}) are also considered in weak form, so a discontinuity is allowed for the solution $q(\tau, \xi)$ at the initial time $\tau = 0$. Substituting the chosen solution representation (\ref{eq:local_sol_def}) into the equation (\ref{eq:lstdg_pred_eq_weak}) and calculating the terms included in the equation (\ref{eq:lstdg_pred_eq_weak}) leads to the following expression:
\begin{equation}\label{eq:lstdg_pred_eq_tmp_1}
\begin{split}
\sum\limits_{q_{0} = 0}^{N}\sum\limits_{q_{1} = 0}^{N} &\left[\vphantom{\intrefdom{\tau}}
  \varphi_{p_{0}}(1)\varphi_{q_{0}}(1) -
  \intrefdom{\tau}\,\varphi_{p_{0}}'(\tau)\varphi_{q_{0}}(\tau)
\right]\intrefdom{\xi}\,\varphi_{p_{1}}(\xi)\varphi_{q_{1}}(\xi)\cdot\hat{q}_{q_{0}q_{1}}\\
&+\frac{a\Dtn{n}}{\Dx}\,\sum\limits_{q_{0} = 0}^{N}\sum\limits_{q_{1} = 0}^{N}
  \intrefdom{\tau}\,\varphi_{p_{0}}(\tau)\varphi_{q_{0}}(\tau)\,\intrefdom{\xi}\,\varphi_{p_{1}}(\xi)\varphi_{q_{1}}'(\xi)\cdot\hat{q}_{q_{0}q_{1}} \\ & =
\sum\limits_{q_{0} = 0}^{N}\sum\limits_{q_{1} = 0}^{N} \varphi_{p_{0}}(0)\varphi_{q_{0}}(0)
\intrefdom{\xi}\,\varphi_{p_{1}}(\xi)\varphi_{q_{1}}(\xi)\cdot\hat{u}^{n}_{q_{1}},
\end{split}
\end{equation}
where integration by parts over time was performed in the time term to include the initial condition in weak form in the system of equations. To simplify the notation, the following matrix and vector elements are introduced:
\begin{equation}\label{eq:matrices_def}
\begin{split}
&\varrho_{pq} = \intrefdom{\xi}\, \varphi_{p}(\xi)\varphi_{q}'(\xi),\quad
m_{pq} = \intrefdom{\xi}\, \varphi_{p}(\xi)\varphi_{q}(\xi) = w_{p}\delta_{pq},\\
&\kappa_{pq} = \varphi_{p}(1)\varphi_{q}(1) - \intrefdom{\xi}\, \varphi_{p}'(\xi)\varphi_{q}(\xi)
             = \varphi_{p}(0)\varphi_{q}(0) + \intrefdom{\xi}\, \varphi_{p}(\xi)\varphi_{q}'(\xi),
\end{split}
\end{equation}
from which the following matrices and vectors are assembled:
\begin{equation}
\begin{split}
&\uphi_{p} = \varphi_{p}(0),\quad \ophi_{p} = \varphi_{p}(1),\quad
\uvecphi = \{\uphi_{p}\}_{p},\quad \ovecphi = \{\ophi_{p}\}_{p},\\
&\varrho = \{\varrho_{pq}\}_{p,q},\quad m = \{m_{pq}\}_{p,q} = \mathrm{diag}(w_{0}, \ldots, w_{N}),\\
&\kappa = \{\kappa_{pq}\}_{p,q} = \ovecphi\otimes\ovecphi^{T} - \varrho^{T} = \uvecphi\otimes\uvecphi^{T} + \varrho,
\end{split}
\end{equation}
where the $\mathcal{L}_{2}(\omega)$-orthogonality of the functional basis $\{\varphi_{k}\}$ is taken into account through the diagonality of the mass matrix $m$; matrix $\varrho$ is the stiffness matrix of the functional basis $\{\varphi_{k}\}$. The following two properties of matrix $\kappa$ can be verified by direct substitution and were derived in~\cite{ader_improving_2024, ader_dg_ivp_ode_sinum}:
\begin{equation}\label{eq:gl_rule:int_prop}
\begin{split}
\sum_{q = 0}^{N} \kappa_{pq} = \varphi_{p}(0),\qquad
\sum_{p = 0}^{N} \kappa_{pq} = \varphi_{q}(1),
\end{split}
\end{equation}
which can be rewritten in matrix form: $\kappa \mathbf{1} = \uvecphi$, $\kappa^{-1} \uvecphi = \mathbf{1}$, $\mathbf{1}^{T} \kappa = \ovecphi^{T}$, $\ovecphi^{T} \kappa^{-1} = \mathbf{1}^{T}$.

The matrices included in the system of algebraic equations (\ref{eq:lstdg_pred_eq_tmp_1}) of the LST-DG predictor are conveniently expressed through the matrices and vectors introduced above as follows:
\begin{equation}
\begin{split}
&\mathrm{K}^{(\tau)}_{\mathrm{p}\mathrm{q}} = \mathrm{K}^{(\tau)}_{p_{0}p_{1},q_{0}q_{1}} = \kappa_{p_{0}q_{0}}m_{p_{1}q_{1}} = \kappa_{p_{0}q_{0}}w_{p_{1}}\delta_{p_{1}q_{1}},\\
&\mathrm{K}^{(\xi)}_{\mathrm{p}\mathrm{q}} = \mathrm{K}^{(\xi)}_{p_{0}p_{1},q_{0}q_{1}} = m_{p_{0}q_{0}}\varrho_{p_{1}q_{1}} = w_{p_{0}}\delta_{p_{0}q_{0}}\varrho_{p_{1}q_{1}},\\
&\mathrm{F}^{(0)}_{\mathrm{p},k} = \mathrm{F}^{(0)}_{p_{0}p_{1},k} = \uphi_{p}m_{p_{1}q_{1}} = \uphi_{p}w_{p_{1}}\delta_{p_{1}q_{1}},
\end{split}
\end{equation}
where $\mathrm{p}$ and $\mathrm{q}$ are multi-indices (\ref{eq:local_sol_idx}). The resulting matrices, as well as the vectors of the coefficients of the DG representations (\ref{eq:u_n_def}) and the coefficients of the local discrete space-times solution (\ref{eq:local_sol_def}), can be expressed through the Kronecker products as follows:
\begin{equation}\label{eq:matrices_K_def}
\begin{split}
&\mathrm{K}^{(\tau)} = \left\{\mathrm{K}^{(\tau)}_{\mathrm{p}\mathrm{q}}\right\}_{\mathrm{p},\mathrm{q}} = \kappa \otimes m,\quad
\mathrm{K}^{(\xi)} = \left\{\mathrm{K}^{(\xi)}_{\mathrm{p}\mathrm{q}}\right\}_{\mathrm{p},\mathrm{q}} = m \otimes \varrho,\\
&\mathrm{F}^{(0)} = \left\{\mathrm{F}^{(0)}_{\mathrm{p},k}\right\}_{\mathrm{p},k} = \uvecphi \otimes m,\quad\,
\hat{\mathbf{q}} = \left\{\hat{q}_{\mathrm{p}}\right\}_{\mathrm{p}},\quad \hat{\mathbf{u}}^{n} = \left\{\hat{u}^{n}_{k}\right\}_{k},
\end{split}
\end{equation}
where $\otimes$ is the symbol of the Kronecker product.

\section{Approximation analysis}
\label{sec:approx_anal}

This Section is devoted to a rigorous direct analysis of the approximation of the ADER-DG numerical method with the LST-DG predictor. Subsection~\ref{sec:approx_anal:local_sol} ``Discrete space-time solution'' presents a rigorous direct analysis of the approximation of a local discrete space-time solution $q(\tau, \xi)$ (\ref{eq:local_sol_def}). Subsection~\ref{sec:approx_anal:final_sol} ``Corrected solution at time steps'' presents a rigorous direct analysis of the approximation of the numerical solution $u^{n+1}(\xi)$ at time steps obtained at the corrector stage (\ref{eq:ader_dg_corr}).

\subsection{Discrete space-time solution}
\label{sec:approx_anal:local_sol}

As a result of introducing matrices and vectors (\ref{eq:matrices_K_def}), the system of equations (\ref{eq:lstdg_pred_eq_tmp_1}) takes the classical form of a system of linear algebraic equations:
\begin{equation}\label{eq:lstdg_pred_eq_tmp_2}
\left(\mathrm{K}^{(\tau)} + \frac{a\Dtn{n}}{\Dx}\,\mathrm{K}^{(\xi)}\right)\hat{\mathbf{q}} = \mathrm{F}^{(0)}\mathbf{u}^{n},
\end{equation}
which leads to the following form:
\begin{equation}\label{eq:lstdg_pred_eq_tmp_3}
\left[I \otimes I + \frac{a\Dtn{n}}{\Dx}\,\left(\mathrm{K}^{(\tau)}\right)^{-1}\mathrm{K}^{(\xi)}\right]\hat{\mathbf{q}} =
\left[\left(\mathrm{K}^{(\tau)}\right)^{-1}\mathrm{F}^{(0)}\right]\hat{\mathbf{u}}^{n}.
\end{equation}
The matrices arising in this form of the system of equations (\ref{eq:lstdg_pred_eq_tmp_3}) are calculated as follows:
\begin{equation}
\begin{split}
&\left(\mathrm{K}^{(\tau)}\right)^{-1}\mathrm{F}^{(0)} = 
  \left[\kappa \otimes m\right]^{-1} \left[\uvecphi \otimes m\right] =
  \left[\kappa^{-1} \uvecphi\right] \otimes \left[m^{-1} m\right] = \mathbf{1} \otimes I,\\
&\left(\mathrm{K}^{(\tau)}\right)^{-1}\mathrm{K}^{(\xi)} = 
  \left[\kappa \otimes m\right]^{-1} \left[m \otimes \varrho\right] =
  \left[\kappa^{-1} m\right] \otimes \left[m^{-1} \varrho\right] = \chi \otimes \mathrm{D},
\end{split}
\end{equation}
where the notations for the two matrices $\chi$ and $\mathrm{D}$ are introduced:
\begin{equation}\label{eq:chi_and_tilde_ae_def}
\chi = \kappa^{-1} m,\quad \mathrm{D} = m^{-1} \varrho.
\end{equation}
This leads to the following explicit expression for the solution of the LST-DG predictor system of equations:
\begin{equation}\label{eq:lstdg_pred_eq_sol_temp_1}
\hat{\mathbf{q}} = \left(I \otimes I + \frac{a\Dtn{n}}{\Dx} \cdot \chi \otimes \mathrm{D}\right)^{-1}\Big(\mathbf{1} \otimes \hat{\mathbf{u}}^{n}\Big),
\end{equation}
that is then used to obtain the numerical solution (\ref{eq:u_n_def}) obtained by an ADER-DG numerical method with LST-DG predictor in a closed form convenient for analyzing the approximation of the method.

The resulting vector $\hat{\mathbf{q}}$ of expansion coefficients of the local discrete space-time solution $q$ (\ref{eq:local_sol_def}) in the original multi-index notation (\ref{eq:local_sol_idx}) is written in the following form:
\begin{equation}\label{eq:lstdg_pred_eq_sol_temp_1_ind_form}
\hat{q}_{\mathrm{p}} = 
  \sum\limits_{\mathrm{q}}
  \left[\left(I \otimes I + \frac{a\Dtn{n}}{\Dx} \cdot \chi \otimes \mathrm{D}\right)^{-1}\right]_{\mathrm{p}\mathrm{q}}\cdot\hat{u}^{n}_{\pi_{1}(\mathrm{q})} =
  \sum\limits_{k = 0}^{N} \sigma_{\mathrm{p}, k} \hat{u}^{n}_{k},
\end{equation}
where the following matrix $\sigma = \{\sigma_{\mathrm{p}, k}\}_{\mathrm{p}, k}$ is introduced:
\begin{equation}\label{eq:sigma_matrix_def}
\sigma_{\mathrm{p}, k} \equiv \sigma_{p_{0}p_{1}, k} = 
  \sum\limits_{l = 0}^{N} \left[\left(I \otimes I + \frac{a\Dtn{n}}{\Dx} \cdot \chi \otimes \mathrm{D}\right)^{-1}\right]_{p_{0}p_{1},lk},
\end{equation}
whose elements can be calculated in explicit closed form.

The explicit calculation of $\sigma_{\mathrm{p}, k}$ (\ref{eq:chi_and_tilde_ae_def}) is based on the properties of matrix $\mathrm{D}$, for which the meaning of $D_{pq}$ is clarified. Despite the fact that this property seems obvious (see, for example, Section~8.1 in the book~\cite{Funaro_book_1992} or Section~1.9 in the Zanotti's lectures on the ADER-DG method~\cite{Zanotti_lects_2016}), a rigorous proof was given by Jackson in~\cite{Jackson_2017}, as it was hidden by other properties of the ADER-DG method. In the notation of this paper, this property is derived similarly as follows. A direct calculation using the GL quadrature rule (\ref{eq:gl_rule}) shows
\begin{equation}
D_{pq} =
  \left[m^{-1} \varrho\right]_{pq} = \frac{\varrho_{pq}}{w_{p}} =
  \frac{1}{w_{p}} \intrefdom{\xi}\, \varphi_{p}(\xi)\varphi_{q}'(\xi) = \varphi_{q}'(\tau_{p}),
\end{equation}
that is exactly satisfied because the quadrature rule (\ref{eq:gl_rule}) exactly integrates polynomials in $\mathcal{P}_{2N+1}(\omega)$, and $(\varphi_{p}\cdot\varphi_{q}')\in\mathcal{P}_{2N-1}(\omega)\subset\mathcal{P}_{2N+1}(\omega)$. The arbitrary function $f\in\mathcal{P}_{N}(\omega)$ is considered, for which $f'\in\mathcal{P}_{N-1}(\omega)$ follow, which allowed the use of the following representation:
\begin{equation}\label{eq:tilde_ae_deriv_prop}
\begin{split}
&f(\tau) = \sum\limits_{p = 0}^{N} f^{(0)}_{p}\varphi_{p}(\tau),\quad f^{(0)}_{p} = f(\tau_{p}),\\
&f'(\tau) = \sum\limits_{p = 0}^{N} f^{(1)}_{p}\varphi_{p}(\tau),\quad
f^{(1)}_{p} = f'(\tau_{p}) = \sum\limits_{q = 0}^{N} f_{q}\varphi_{q}'(\tau_{p}) = \sum\limits_{q = 0}^{N} D_{pq} f_{q},
\end{split}
\end{equation}
which are exact, since the functional basis $\{\varphi_{p}\}$ (\ref{eq:phi_def}) allows for an exact representation of any function in $\mathcal{P}_{N}(\omega)$. The last expression shows that the matrix $\mathrm{D}$ (\ref{eq:chi_and_tilde_ae_def}) is a matrix of linear transformation of the vector of the expansion coefficients $\{f^{(0)}_{p}\}_{p}$ of the function $f$ into the vector of the expansion coefficients $\{f^{(1)}_{p}\}_{p}$ of the derivative $f'$ of the function $f$. For any function $f\in\mathcal{P}_{N}(\omega)$, the property $f^{(N+1)} \equiv 0$ is known, so $\mathrm{D}^{N+1} = 0$ holds --- the matrix $\mathrm{D}$ (\ref{eq:chi_and_tilde_ae_def}) is nilpotent with degree of nilpotency $N+1$.

Using the property of the nilpotency of matrix $\mathrm{D}$, the inverse matrix in the solution of the LST-DG predictor system of equations (\ref{eq:lstdg_pred_eq_sol_temp_1}) (and (\ref{eq:lstdg_pred_eq_sol_temp_1_ind_form})) can be exactly represented in the form of a binomial theorem expansion:
\begin{equation}\label{eq:local_sol_inv_matrix}
\left[I \otimes I + \frac{a\Dtn{n}}{\Dx} \cdot \chi \otimes \mathrm{D}\right]^{-1} =
  \sum\limits_{s = 0}^{N} (-1)^{s} \left(\frac{a\Dtn{n}}{\Dx}\right)^{s} \left(\chi^{s} \otimes \mathrm{D}^{s}\right),
\end{equation}
where only the finite part of the sum, containing $N+1$ terms, remains. Using expression (\ref{eq:local_sol_inv_matrix}), the expression (\ref{eq:sigma_matrix_def}) for the elements $\{\sigma_{\mathrm{p}, k}\}$ of the matrix $\sigma$ is rewritten as follows:
\begin{equation}\label{eq:sigma_matrix_temp_1}
\begin{split}
\sigma_{p_{0}p_{1}, k} & = 
\sum\limits_{l = 0}^{N} \left[\sum\limits_{s = 0}^{N} (-1)^{s} \left(\frac{a\Dtn{n}}{\Dx}\right)^{s}\left(\chi^{s} \otimes \mathrm{D}^{s}\right)\right]_{p_{0}p_{1},lk}\\ & = 
\sum\limits_{l = 0}^{N} \sum\limits_{s = 0}^{N} (-1)^{s} \left(\frac{a\Dtn{n}}{\Dx}\right)^{s} \left(\left[\chi^{s}\right]_{p_{0}l} \left[\mathrm{D}^{s}\right]_{p_{1}k}\right) = 
\sum\limits_{s = 0}^{N} (-1)^{s} \left(\frac{a\Dtn{n}}{\Dx}\right)^{s} \left(\sum\limits_{l = 0}^{N}\left[\chi^{s}\right]_{p_{0}l}\right) \left[\mathrm{D}^{s}\right]_{p_{1}k}.
\end{split}
\end{equation}
An explicit calculation of this sums in (\ref{eq:local_sol_inv_matrix}) and (\ref{eq:sigma_matrix_temp_1}) is carried out using the property of matrix elements $\{\chi_{pq}\}$ proved by Han Veiga \textit{et al} in~\cite{ader_improving_2024} (which is also interpreted as the simplifying condition $C(N)$~\cite{Butcher_book_2016, Hairer_book_1, Hairer_book_2, Dekker_Verwer_1984} in the case of using the ADER-DG method with LST-DG-predictor as Runge-Kutta method to solve the initial value problem for an ODE system~\cite{ader_improving_2024, ader_dg_ivp_ode_sinum} and~\cite{ader_dg_ivp_ode, ader_dg_ivp_dae}):
\begin{equation}\label{eq:c_prop_ode}
\sum\limits_{q = 0}^{N} \chi_{pq} \tau_{q}^{r} = \frac{\tau_{p}^{r+1}}{r+1},\quad 0 \leqslant p \leqslant N,\quad 0 \leqslant r \leqslant N-1.
\end{equation}
In the notation of this work, the property (\ref{eq:c_prop_ode}) is derived similarly as follows:
\begin{equation}
\begin{split}
\sum\limits_{q = 0}^{N} \frac{\kappa_{pq}\tau_{q}^{r+1}}{r+1} & = \sum\limits_{q = 0}^{N} \left[
	\smash[b]{\underline{\varphi}}_{p}\smash[b]{\underline{\varphi}}_{q} + \int\limits_{0}^{1} \varphi_{p}(\tau)\frac{d\varphi_{q}(\tau)}{d\tau}d\tau
\right] \frac{\tau_{q}^{r+1}}{r+1} =
\sum\limits_{q = 0}^{N} \frac{\tau_{q}^{r+1}}{r+1}\smash[b]{\underline{\varphi}}_{p}\smash[b]{\underline{\varphi}}_{q} +
\sum\limits_{q = 0}^{N} \frac{\tau_{q}^{r+1}}{r+1}\int\limits_{0}^{1} \varphi_{p}(\tau)\frac{d\varphi_{q}(\tau)}{d\tau}d\tau\\ & =
\frac{\smash[b]{\underline{\varphi}}_{p}}{r+1} \sum\limits_{q = 0}^{N} \tau_{q}^{r+1}\smash[b]{\underline{\varphi}}_{q} +
\int\limits_{0}^{1} \frac{\varphi_{p}(\tau)}{r+1} \frac{d}{d\tau}\left[\sum\limits_{q = 0}^{N}\tau_{q}^{r+1}\varphi_{q}(\tau)\right]d\tau =
\frac{\smash[b]{\underline{\varphi}}_{p}}{r+1}\cdot(0)^{r+1} +
\int\limits_{0}^{1} \frac{\varphi_{p}(\tau)}{r+1} \frac{d\left(\tau^{r+1}\right)}{d\tau}d\tau = \int\limits_{0}^{1} \tau^{r}\varphi_{p}(\tau) d\tau = w_{p}\tau_{p}^{r}.
\end{split}
\end{equation}
which yields the original expression (\ref{eq:c_prop_ode}) after using the definition (\ref{eq:chi_and_tilde_ae_def}) of the matrix $\chi$.

The calculation of the sum in the expression (\ref{eq:sigma_matrix_temp_1}) is based on the following expression:
\begin{equation}\label{eq:chi_int_prop}
\begin{split}
\sum\limits_{l = 0}^{N}\left[\chi^{s}\right]_{p_{0}l} & =
\sum\limits_{l = 0}^{N} \sum\limits_{i_{1} = 0}^{N}\ldots\sum\limits_{i_{s} = 0}^{N} \chi_{p_{0}i_{1}}\ldots\chi_{i_{s}l} = 
\sum\limits_{i_{1} = 0}^{N}\ldots\sum\limits_{i_{s} = 0}^{N} \chi_{p_{0}i_{1}}\ldots\chi_{i_{s-1}i_{s}}\left[\sum\limits_{l = 0}^{N}\chi_{i_{s}l}\right]\\ & =
\sum\limits_{i_{1} = 0}^{N}\ldots\sum\limits_{i_{s} = 0}^{N} \chi_{p_{0}i_{1}}\ldots\chi_{i_{s-1}i_{s}}\tau_{i_{s}} =
\sum\limits_{i_{1} = 0}^{N}\ldots\hspace{-1.5mm}\sum\limits_{i_{s-1} = 0}^{N} \chi_{p_{0}i_{1}}\ldots\chi_{i_{s-2}i_{s-1}}\left[
  \sum\limits_{i_{s} = 0}^{N}\chi_{i_{s-1}i_{s}}\tau_{i_{s}}
\right]\\ & = \sum\limits_{i_{1} = 0}^{N}\ldots\sum\limits_{i_{s-1} = 0}^{N} \chi_{p_{0}i_{1}}\ldots\chi_{i_{s-2}i_{s-1}}\frac{\tau_{i_{s-1}}^{2}}{2} =
\ldots = \sum\limits_{i_{1} = 0}^{N} \chi_{p_{0}i_{1}}\frac{\tau_{p_{0}}^{s-1}}{(s-1)!} = \frac{\tau_{p_{0}}^{s}}{s!},
\end{split}
\end{equation}
which is strictly satisfied for $0 \leqslant s \leqslant N$, leading to the expression for the elements $\{\sigma_{\mathrm{p}, k}\}$ of the matrix $\sigma$ (\ref{eq:sigma_matrix_temp_1}):
\begin{equation}
\sigma_{p_{0}p_{1}, k} =
\sum\limits_{s = 0}^{N} (-1)^{s} \left(\frac{a\Dtn{n}}{\Dx}\right)^{s} \frac{\tau_{p_{0}}^{s}}{s!} \left[\mathrm{D}^{s}\right]_{p_{1}k}.
\end{equation}
To obtain a local discrete space-time solution $q$ (\ref{eq:local_sol_def}), the property of matrix $\mathrm{D}$ (\ref{eq:tilde_ae_deriv_prop}) as a differentiation matrix of a function represented by a representation in the form of an expansion in basis functions $\{\varphi_{p}\}$ is used:
\begin{equation}
\frac{d^{s}u^{n}(\xi)}{d\xi^{s}} = \sum\limits_{k = 0}^{N} \hat{u}^{n}_{k} \varphi_{k}^{(s)}(\xi) = \sum\limits_{k = 0}^{N} \hat{u}^{s, n}_{k} \varphi_{k}(\xi),\quad
\hat{u}^{s, n}_{k} = \sum\limits_{l = 0}^{N} [\mathrm{D}^{s}]_{kl} \hat{u}^{n}_{l},
\end{equation}
where the set $\{\hat{u}^{s, n}_{k}\}$ corresponds to the representation of derivatives of $u^{n}(\xi)$ in the form of an expansion in basis functions $\{\varphi_{p}\}$, which, when using expression (\ref{eq:lstdg_pred_eq_sol_temp_1_ind_form}), leads to the following expression for the expansion coefficients $\hat{q}_{p_{0}p_{1}}$ of a local discrete space-time solution $q$:
\begin{equation}\label{eq:local_sol_coeffs}
\hat{q}_{p_{0}p_{1}} = 
  \sum\limits_{k = 0}^{N} \sigma_{p_{0}p_{1}, k} \hat{u}^{n}_{k} =
  \sum\limits_{s = 0}^{N} (-1)^{s} \left(\frac{a\Dtn{n}}{\Dx}\right)^{s} \frac{\tau_{p_{0}}^{s}}{s!} \, \hat{u}^{s, n}_{p_{1}}.
\end{equation}
The coordinate and time dependence of the local discrete space-time solution $q(\tau, \xi)$ (\ref{eq:local_sol_def}) itself is expressed as follows:
\begin{equation}\label{eq:lstdg_pred_eq_sol_temp_2}
\begin{split}
q(\tau, \xi) & = \sum\limits_{p_{0} = 0}^{N}\sum\limits_{p_{1} = 0}^{N}\sum\limits_{s = 0}^{N}
                 (-1)^{s} \left(\frac{a\Dtn{n}}{\Dx}\right)^{s} \frac{\tau_{p_{0}}^{s}}{s!} \, \hat{u}^{s, n}_{k} \varphi_{p_{0}}(\tau)\varphi_{p_{1}}(\xi)\\
             & = \sum\limits_{s = 0}^{N} \frac{(-1)^{s}}{s!} \left(\frac{a\Dtn{n}}{\Dx}\right)^{s}
                 \left(\sum\limits_{p_{0} = 0}^{N} \tau_{p_{0}}^{s}\varphi_{p_{0}}(\tau)\right)
                 \left(\sum\limits_{p_{1} = 0}^{N} \hat{u}^{s, n}_{k}\varphi_{p_{1}}(\xi)\right)\\
             & = \sum\limits_{s = 0}^{N} \frac{(-1)^{s}}{s!} \left(\frac{a\Dtn{n}}{\Dx}\right)^{s} \tau^{s}\,\frac{d^{s}u^{n}(\xi)}{d\xi^{s}}
               = \left\{ \frac{d^{s}u^{n}(\xi)}{\left(\Dx\right)^{s} d\xi^{s}} = \frac{d^{s}}{dx^{s}}u^{n}(x, t^{n}) \right\}\\
             & = \sum\limits_{s = 0}^{N} \frac{(-1)^{s}}{s!} \left(\frac{a\Dtn{n}}{\Dx}\right)^{s} \left(\frac{t(\tau) - t^{n}}{\Dtn{n}}\right)^{s}
                 \left(\Dx\right)^{s}\left.\left[\frac{d^{s}}{dx^{s}}u^{n}(x, t^{n})\right]\right|_{x(\xi)}\\
             & = \sum\limits_{s = 0}^{N} \frac{(-1)^{s}}{s!} \left[a\left(t(\tau) - t^{n}\right)\right]^{s}
                 \left.\left[\frac{d^{s}}{dx^{s}}u^{n}(x, t^{n})\right]\right|_{x(\xi)}.
\end{split}
\end{equation}
Due to $u^{n}(x, t^{n})\in\mathcal{P}_{N}(\omega)$ and $q(\tau, \xi)\in\mathcal{P}_{N}(\omega)\times\mathcal{P}_{N}(\omega)$, it is clear that the resulting sum (\ref{eq:lstdg_pred_eq_sol_temp_2}) is an exact expression for the Taylor series expansion of the following function:
\begin{equation}
q(\tau, \xi) = u^{n}\big(x(\xi) - a\left(t(\tau) - t^{n}\right), t^{n}\big),
\end{equation}
which is an exact solution to the original advection equation (\ref{eq:adv_eq_src}) (or (\ref{eq:adv_eq_mapped})) in the case where the initial condition is exactly represented by a polynomial representation in basis functions $\{\varphi_{p}\}$.

In the general case, when function $u(x(\xi), t^{n})$ is not exactly representable by a polynomial representation ($u(x(\xi), t^{n})|_{\Omega_{i}}\not\in\mathcal{P}_{N}(\omega)$), but are sufficiently smooth $u(x(\xi), t^{n})|_{\Omega_{i}}\in\mathcal{C}_{N+1}(\omega)$, the well-known estimates hold:
\begin{equation}\label{eq:errs_ests}
\begin{split}
&u^{n}(\xi) = \left.u(x(\xi), t^{n})\right|_{\Omega_{i}} + R^{(0)}_{u}(\xi),\quad
\frac{d^{s}}{dx^{s}}u^{n}(x, t^{n}) = \left.\pdernoper{x}{s}u(x(\xi), t^{n})\right|_{\Omega_{i}} + R^{(s)}_{u}(\xi),\\
&R^{(0)}_{u}(\xi) = \left[\frac{u^{(N+1)}(\xi_{0})}{(N+1)!}\hat{\omega}(\xi)\right]\Dx^{N+1} + o\left(\Dx^{N+1}\right) = O\left(\Dx^{N+1}\right),\\
&R^{(s)}_{u}(\xi) = \left[\frac{u^{(N+1)}(\xi_{s}(\xi))}{(N+1)!}\frac{d^{s}\hat{\omega}(\xi)}{d\xi^{s}}\right]\Dx^{N+1-s} + o\left(\Dx^{N+1-s}\right) = O\left(\Dx^{N+1-s}\right),
\end{split}
\end{equation}
where $R^{(0)}_{u}(\xi)$ and $R^{(s)}_{u}(\xi)$ are remainder terms, $\xi_{0}\in[0, 1]$ is a constant intermediate point (point from the mean value theorem), $\{\xi_{s}(\xi)\in\omega\}$ is a set of the intermediate points, $\hat{\omega}:\, \omega \rightarrow \mathcal{R}$ is a polynomial with leading coefficient~$1$ and roots at the roots $\{\xi_{p}\}$ of the shifted Legendre polynomial $\tilde{P}_{N+1}(\xi)$. This leads to the following estimate for the approximation of the local discrete space-time solution $q(\tau, \xi)$:
\begin{equation}\label{eq:local_sol_app_order}
\begin{split}
q(\tau, \xi) & = \sum\limits_{s = 0}^{N} \frac{(-1)^{s}}{s!} \left(t - t^{n}\right)^{s}
                 \left.\left[\pdernoper{x}{s}\left(a^{s} u(x, t^{n})\right)\right]\right|_{x(\xi)}
               + \sum\limits_{s = 0}^{N} \left(t - t^{n}\right)^{s} \cdot O\left(\Dx^{N+1-s}\right)\\
             & = \left\{ \pdernoper{x}{s}(a^{s} u) = (-1)^{s}\pdern{u}{t}{s} \right\}
               = \sum\limits_{s = 0}^{N} \frac{(t - t^{n})^{s}}{s!} \left.\left[\pdernoper{t}{s}u(x(\xi), t)\right]\right|_{t^{n}} + O\left(\Dx^{N+1}\right)\\
             & = u\big(x(\xi) - a\left(t(\tau) - t^{n}\right)\big) + O\left(\Dx^{N+1}\right),
\end{split}
\end{equation}
where constraints $|t - t^{n}| \leqslant \Dtn{n}$ and $\Dtn{n} \propto \Dx$ are assumed. The obtained approximation estimate showed that the local discrete space-time solution $q(\tau, \xi)$ (\ref{eq:local_sol_def}) has an approximation order $p_{\rm L} = N+1$, however, of course, this estimate takes place only locally within element $\Xi_{i}^{n}$ (\ref{eq:discr_domains}) only in the domain of influence $\{\smash[b]{\tilde{\Xi}}_{i}^{n}\subseteq\Xi_{i}^{n}\, |\, x\in\Omega_{i},\, t - t_{n}\leqslant \theta(a)a(x - x_{i}) - \theta(-a)a(x_{i+1} - x)\}$ of the discretization domain $\Omega_{i}$.

\subsection{Corrected solution at time steps}
\label{sec:approx_anal:final_sol}

Substitution of the diagonal matrix $m$ (\ref{eq:matrices_def}) for the basis functions $\{\varphi\}$ (\ref{eq:phi_def}) into the expression for the completely one-step ADER-DG scheme (\ref{eq:ader_dg_corr_temp_1}) (which, with respect to the LST-DG predictor step, is a corrector step) leads to the following expression:
\begin{equation}\label{eq:ader_dg_corr_temp_2}
\begin{split}
\left(\hat{u}^{n+1}_{k} - \hat{u}^{n}_{k}\right) -
\frac{1}{w_{k}}\frac{a\Dtn{n}}{\Dx}\intrefdom{\tau}\intrefdom{\xi}\,\varphi_{k}'(\xi) q(\tau, \xi) +
\frac{1}{w_{k}}\frac{\Dtn{n}}{\Dx}\intrefdom{\tau}\,\Big[ &
  \ophi_{k} F_{\rm RP}\left(q(\tau, 1),\, q^{+}(\tau, 0)\right)\\ - &
  \uphi_{k} F_{\rm RP}\left(q^{-}(\tau, 1),\, q(\tau, 0)\right)
\Big] = 0.
\end{split}
\end{equation}

The local term of the resulting expression (\ref{eq:ader_dg_corr_temp_2}), after detailed calculation and use of notations (\ref{eq:matrices_def}) and (\ref{eq:chi_and_tilde_ae_def}), is reduced to the form
\begin{equation}\label{eq:ader_dg_corr_local_term_temp_1}
\begin{split}
\frac{1}{w_{k}}\frac{a\Dtn{n}}{\Dx}&\intrefdom{\tau}\intrefdom{\xi}\,\varphi_{k}'(\xi) q(\tau, \xi)
    = \frac{1}{w_{k}}\frac{a\Dtn{n}}{\Dx}\sum\limits_{p_{0} = 0}^{N}\sum\limits_{p_{1} = 0}^{N}
      \left[\intrefdom{\tau}\, \varphi_{p_{0}}(\tau)\right] \left[\intrefdom{\xi}\, \varphi_{k}'(\xi)\varphi_{p_{1}}(\xi)\right]\,\hat{q}_{p_{0}p_{1}}\\
  & = \frac{1}{w_{k}}\frac{a\Dtn{n}}{\Dx}\sum\limits_{p_{0} = 0}^{N}\sum\limits_{p_{1} = 0}^{N} w_{p_{0}} \varrho_{p_{1}k} \,\hat{q}_{p_{0}p_{1}}
    = \frac{1}{w_{k}}\sum\limits_{p_{0} = 0}^{N}\sum\limits_{p_{1} = 0}^{N} w_{p_{0}} \varrho_{p_{1}k}
      \sum\limits_{s = 0}^{N} (-1)^{s} \left(\frac{a\Dtn{n}}{\Dx}\right)^{s+1} \frac{\tau_{p_{0}}^{s}}{s!} \, \hat{u}^{s, n}_{k}\\
  & = \frac{1}{w_{k}}\sum\limits_{s = 0}^{N} \frac{(-1)^{s}}{s!}\left(\frac{a\Dtn{n}}{\Dx}\right)^{s+1}
      \left[\sum\limits_{p_{0} = 0}^{N}w_{p_{0}}\tau_{p_{0}}^{s}\right] \left[\sum\limits_{p_{1} = 0}^{N} \varrho_{p_{1}k} \hat{u}^{s, n}_{k}\right].
\end{split}
\end{equation}
The first sum included in the expression (\ref{eq:ader_dg_corr_local_term_temp_1}) is explicitly calculated using the GL quadrature rule (\ref{eq:gl_rule}) as follows (which is also interpreted as the simplifying condition $B(2N+2)$~\cite{Butcher_book_2016, Hairer_book_1, Hairer_book_2, Dekker_Verwer_1984} in the case of using the ADER-DG method with LST-DG-predictor as Runge-Kutta method to solve the initial value problem for an ODE system~\cite{ader_improving_2024, ader_dg_ivp_ode_sinum}):
\begin{equation}\label{eq:b_prop_ode}
\sum\limits_{p_{0} = 0}^{N}w_{p_{0}}\tau_{p_{0}}^{s} = \frac{1}{s+1},\quad 0 \leqslant s \leqslant 2N+1,
\end{equation}
however, in expression (\ref{eq:ader_dg_corr_local_term_temp_1}), only cases $0 \leqslant s \leqslant N+1$ are necessary. The second sum in the expression (\ref{eq:ader_dg_corr_local_term_temp_1}) is also explicitly calculated using the GL quadrature rule (\ref{eq:gl_rule}) as follows:
\begin{equation}\label{eq:ader_dg_corr_local_term_temp_2}
\begin{split}
\sum\limits_{p_{1} = 0}^{N} \varrho_{p_{1}k} \hat{u}^{s, n}_{k}
  & = \sum\limits_{p_{1} = 0}^{N} w_{p_{1}} \varphi_{k}'(\xi_{p_{1}}) \hat{u}^{s, n}_{k}
    = \intrefdom{\xi}\,\varphi_{k}'(\xi)\,\frac{d^{s}u^{n}(\xi)}{d\xi^{s}}
    = \left.\left[\varphi_{k}(\xi)\,\frac{d^{s}u^{n}(\xi)}{d\xi^{s}}\right]\right|_{0}^{1}\\
  & - \intrefdom{\xi}\,\varphi_{k}(\xi)\,\frac{d^{s+1}u^{n}(\xi)}{d\xi^{s+1}}
    = \ophi_{k}\,\frac{d^{s}u^{n}}{d\xi^{s}}(1) - \uphi_{k}\,\frac{d^{s}u^{n}}{d\xi^{s}}(0) - w_{k}\,\hat{u}^{s+1, n}_{k},
\end{split}
\end{equation}
where integration by parts is performed and it is used that all integrands are contained in subspaces of $\mathcal{P}_{2N+1}(\omega)$. Substitution of the obtained results into the expression (\ref{eq:ader_dg_corr_local_term_temp_1}) leads to the following expression for the local term of ADER-DG method (\ref{eq:ader_dg_corr_temp_2}):
\begin{equation}\label{eq:ader_dg_corr_local_term}
\begin{split}
\frac{1}{w_{k}}\frac{a\Dtn{n}}{\Dx}&\intrefdom{\tau}\intrefdom{\xi}\,\varphi_{k}'(\xi) q(\tau, \xi)
= \sum\limits_{s = 0}^{N} \frac{(-1)^{s}}{(s+1)!}\left(\frac{a\Dtn{n}}{\Dx}\right)^{s+1}
  \Bigg[\frac{\ophi_{k}}{w_{k}}\,\frac{d^{s}u^{n}}{d\xi^{s}}(1) - \frac{\uphi_{k}}{w_{k}}\,\frac{d^{s}u^{n}}{d\xi^{s}}(0) - \hat{u}^{s+1, n}_{k}\Bigg].
\end{split}
\end{equation}
The last term in the expression (\ref{eq:ader_dg_corr_local_term}) is summed up explicitly:
\begin{equation}
\begin{split}
\sum\limits_{k = 0}^{N}\sum\limits_{s = 0}^{N}\frac{(-1)^{s+1}}{(s+1)!}&\left(\frac{a\Dtn{n}}{\Dx}\right)^{s+1}\hat{u}^{s+1, n}_{k}\varphi_{k}(\xi(x))
  = \sum\limits_{s = 0}^{N-1}\frac{(-1)^{s+1}}{(s+1)!}\left(\frac{a\Dtn{n}}{\Dx}\right)^{s+1}\left(\Dx\right)^{s}\frac{d^{s}}{dx^{s}}u^{n}(\xi(x), t^{n})\\
& = \sum\limits_{s = 0}^{N}\frac{(-1)^{s}}{s!}\left(a\Dtn{n}\right)^{s}\frac{d^{s}}{dx^{s}}u^{n}(\xi(x), t^{n}) - u^{n}(\xi(x))
  = u^{n}(\xi(x - a\Dtn{n}), t^{n}) - u^{n}(\xi(x)),
\end{split}
\end{equation}
which is why integration by parts was required when calculating expression (\ref{eq:ader_dg_corr_local_term_temp_2}), and the nilpotency property of matrix $\mathrm{D}$ ($\mathrm{D}^{N+1} \equiv 0$) was used (which is equivalent to relation $u^{(N+1)} \equiv 0$ for any function $u\in\mathcal{P}_{N}(\omega)$). The first two terms in the expression (\ref{eq:ader_dg_corr_local_term}) take into account the values of the local discrete space-time solution $q$ at the boundary points of discretization domain $\Omega_{i}$, which is evident from the factors $\ophi_{k}/w_{k}$ and $\uphi_{k}/w_{k}$ included in these terms. To substantiate this conclusion, the delta-function $\delta$ in polynomial space $\mathcal{P}_{N}(\omega)$ is considered, represented as an expansion in basis functions $\{\varphi_{p}\}$ (\ref{eq:phi_def}):
\begin{equation}\label{eq:delta_func_varphi_coeffs}
\delta_{\mathcal{P}_{N}}(\tau) = \sum\limits_{k = 0}^{N} \iota_{k}\varphi_{k}(\tau),\quad
\iota_{k} = \frac{1}{w_{k}} \intrefdom{\tau}\, \delta(\tau)\varphi_{k}(\tau) = \frac{\uphi_{k}}{w_{k}},
\end{equation}
where $\{\iota_{k}\}$ is the set of the expansion coefficients. From the resulting representation, it is clear that $\delta_{\mathcal{P}_{N}}$ accurately represents the sifting (filtering) property of the delta-function $\delta$ when acting on any function $f\in\mathcal{P}_{N}(\omega)$:
\begin{equation}\label{eq:delta_func_varphi_action}
\begin{split}
f(\tau) = \sum\limits_{k = 0}^{N} f(\tau_{k})\varphi_{k}(\tau),\ \Rightarrow\ 
\intrefdom{\tau}\, f(\tau)\delta_{\mathcal{P}_{N}}(\tau)
  = \sum\limits_{k = 0}^{N}\sum\limits_{l = 0}^{N} f(\tau_{k}) \frac{\uphi_{l}}{w_{l}} \intrefdom{\tau}\, \varphi_{k}(\tau)\varphi_{l}(\tau)
  = \sum\limits_{k = 0}^{N} f(\tau_{k})\uphi_{k} = f(0).
\end{split}
\end{equation}
It should be noted that despite the ``flux form'' of the boundary terms in the local term (\ref{eq:ader_dg_corr_local_term}), these are local contributions to the final numerical solution $\{u_{k}^{n+1}\}$ (\ref{eq:u_n_def}) --- only the local discrete space-time solution $q$ at this coordinate-time discretization domain $\Xi_{i}^{n}$ is used there.

The flux term in the expression (\ref{eq:ader_dg_corr_temp_2}) contains a local discrete space-time solution $q$ from two nearest coordinate-time discretization domains $\Xi_{i-1}^{n}$ and $\Xi_{i+1}^{n}$ (this is the case in the general form of recording the formulation of the numerical method, and in each specific case it contains a local discrete space-time solution from only one nearest cell --- left or right, depending on the sign of the advection velocity $a$ (\ref{eq:adv_eq_src}), in accordance with the upwind property, which is determined by the $\theta$-functions in the exact Riemann solver (\ref{eq:f_rp})) --- $q^{-}$ and $q^{+}$, which are selected in the following form:
\begin{equation}
\begin{split}
&q^{+}(\tau, \xi) = \sum\limits_{p_{0} = 0}^{N}\sum\limits_{p_{1} = 0}^{N} \hat{q}^{+}_{p_{0}p_{1}} \varphi_{p_{0}}(\tau)\varphi_{p_{1}}(\xi),\quad
\hat{q}^{+}_{p_{0}p_{1}} = 
  \sum\limits_{s = 0}^{N} (-1)^{s} \left(\frac{a\Dtn{n}}{\Dx}\right)^{s} \frac{\tau_{p_{0}}^{s}}{s!} \,
  \sum\limits_{l = 0}^{N} [\mathrm{D}^{s}]_{p_{1}l} \hat{u}^{+,n}_{l},\\
&q^{-}(\tau, \xi) = \sum\limits_{p_{0} = 0}^{N}\sum\limits_{p_{1} = 0}^{N} \hat{q}^{-}_{p_{0}p_{1}} \varphi_{p_{0}}(\tau)\varphi_{p_{1}}(\xi),\quad
\hat{q}^{-}_{p_{0}p_{1}} = 
  \sum\limits_{s = 0}^{N} (-1)^{s} \left(\frac{a\Dtn{n}}{\Dx}\right)^{s} \frac{\tau_{p_{0}}^{s}}{s!} \,
  \sum\limits_{l = 0}^{N} [\mathrm{D}^{s}]_{p_{1}l} \hat{u}^{-,n}_{l},\\
\end{split}
\end{equation}
where upper index ``$+$'' corresponds to the right cell, upper index ``$-$'' corresponds to the left cell, the last expressions for the expansion coefficients $\{\hat{q}^{+}_{p_{0}p_{1}}\}$ and $\{\hat{q}^{-}_{p_{0}p_{1}}\}$ correspond to formula (\ref{eq:local_sol_coeffs}). After substituting the Riemann solver (\ref{eq:f_rp}) expression into the flux term of the expression (\ref{eq:ader_dg_corr_temp_2}) and rearranging the components of the formula, the following expression is obtained:
\begin{equation}
\begin{split}
&\frac{1}{w_{k}}\frac{a\Dtn{n}}{\Dx}\intrefdom{\tau}\,\Big[\ophi_{k} F_{\rm RP}\left(q(\tau, 1),\, q^{+}(\tau, 0)\right) -
                                                           \uphi_{k} F_{\rm RP}\left(q^{-}(\tau, 1),\, q(\tau, 0)\right)\Big]\\ & =
\frac{1}{w_{k}}\frac{a\Dtn{n}}{\Dx}\intrefdom{\tau}\,\Bigg\{\ophi_{k}\Big[\theta(a)q(\tau, 1) + \theta(-a)q^{+}(\tau, 0)\Big] -
                                                            \uphi_{k}\Big[\theta(a)q^{-}(\tau, 1) + \theta(-a)q(\tau, 0)\Big]\Bigg\}\\ & =
\frac{1}{w_{k}}\frac{a\Dtn{n}}{\Dx}\sum\limits_{p = 0}^{N} w_{p}\Bigg\{\theta( a)\Big[\ophi_{k}q(\tau_{p}, 1) - \uphi_{k}q^{-}(\tau_{p}, 1)\Big] +
                                                                       \theta(-a)\Big[\ophi_{k}q^{+}(\tau_{p}, 0) - \uphi_{k}q(\tau_{p}, 0)\Big]\Bigg\},
\end{split}
\end{equation}
where the terms are explicitly separated in accordance with the sign of the advection velocity $a$. A detailed calculation and transformation of the first term, nonzero under condition $a > 0$, leads to the following expression:
\begin{equation}
\begin{split}
&\frac{1}{w_{k}}\frac{a\Dtn{n}}{\Dx}\sum\limits_{p = 0}^{N} w_{p} \Big[\ophi_{k}q(\tau_{p}, 1) - \uphi_{k}q^{-}(\tau_{p}, 1)\Big]\\ & =
\frac{1}{w_{k}} \sum\limits_{p = 0}^{N} w_{p} \sum\limits_{p_{0} = 0}^{N}\sum\limits_{p_{1} = 0}^{N}\sum\limits_{s = 0}^{N}
  (-1)^{s} \varphi_{p_{0}}(\tau_{p}) \ophi_{p_{1}} \left(\frac{a\Dtn{n}}{\Dx}\right)^{s+1} \frac{\tau_{p_{0}}^{s}}{s!}
  \sum\limits_{l = 0}^{N} [\mathrm{D}^{s}]_{p_{1}l} \Big[\ophi_{k}\hat{u}^{n}_{l} - \uphi_{k}\hat{u}^{-,n}_{l}\Big]\\ & =
\frac{1}{w_{k}} \sum\limits_{s = 0}^{N} (-1)^{s} \left(\frac{a\Dtn{n}}{\Dx}\right)^{s+1}
  \left[\sum\limits_{p_{0} = 0}^{N} w_{p_{0}} \frac{\tau_{p_{0}}^{s}}{s!}\right]
  \left[\sum\limits_{p_{1} = 0}^{N} \ophi_{p_{1}} \sum\limits_{l = 0}^{N} [\mathrm{D}^{s}]_{p_{1}l}
        \Big(\ophi_{k}\hat{u}^{n}_{l} - \uphi_{k}\hat{u}^{-,n}_{l}\Big)\right]\\ & =
\frac{1}{w_{k}} \sum\limits_{s = 0}^{N} \frac{(-1)^{s}}{(s+1)!} \left(\frac{a\Dtn{n}}{\Dx}\right)^{s+1}
  \left[\sum\limits_{p_{1} = 0}^{N} \ophi_{p_{1}} \left(\ophi_{k}\frac{d^{s}u^{n}}{d\xi^{s}}(\xi_{p_{1}}) -
                                                        \uphi_{k}\frac{d^{s}u^{-,n}}{d\xi^{s}}(\xi_{p_{1}})\right)\right]\\ & =
  \frac{1}{w_{k}} \sum\limits_{s = 0}^{N} \frac{(-1)^{s}}{(s+1)!} \left(\frac{a\Dtn{n}}{\Dx}\right)^{s+1}
  \left[\ophi_{k}\frac{d^{s}u^{n}}{d\xi^{s}}(1) - \uphi_{k}\frac{d^{s}u^{-,n}}{d\xi^{s}}(1)\right],
\end{split}
\end{equation}
where expressions (\ref{eq:b_prop_ode}) and (\ref{eq:tilde_ae_deriv_prop}) are used. A similar calculation and transformation of the second term, different from zero under condition $a < 0$, leads to the following expression:
\begin{equation}
\begin{split}
&\frac{1}{w_{k}}\frac{a\Dtn{n}}{\Dx}\sum\limits_{p = 0}^{N} w_{p} \Big[\ophi_{k}q^{+}(\tau_{p}, 0) - \uphi_{k}q(\tau_{p}, 0)\Big]\\ & =
\frac{1}{w_{k}} \sum\limits_{p = 0}^{N} w_{p} \sum\limits_{p_{0} = 0}^{N}\sum\limits_{p_{1} = 0}^{N}\sum\limits_{s = 0}^{N}
  (-1)^{s} \varphi_{p_{0}}(\tau_{p}) \uphi_{p_{1}} \left(\frac{a\Dtn{n}}{\Dx}\right)^{s+1} \frac{\tau_{p_{0}}^{s}}{s!}
  \sum\limits_{l = 0}^{N} [\mathrm{D}^{s}]_{p_{1}l} \Big[\ophi_{k}\hat{u}^{+,n}_{l} - \uphi_{k}\hat{u}^{n}_{l}\Big]\\ & =
\frac{1}{w_{k}} \sum\limits_{s = 0}^{N} (-1)^{s} \left(\frac{a\Dtn{n}}{\Dx}\right)^{s+1}
  \left[\sum\limits_{p_{0} = 0}^{N} w_{p_{0}} \frac{\tau_{p_{0}}^{s}}{s!}\right]
  \left[\sum\limits_{p_{1} = 0}^{N} \uphi_{p_{1}} \sum\limits_{l = 0}^{N} [\mathrm{D}^{s}]_{p_{1}l}
        \Big(\ophi_{k}\hat{u}^{+,n}_{l} - \uphi_{k}\hat{u}^{n}_{l}\Big)\right]\\ & =
\frac{1}{w_{k}} \sum\limits_{s = 0}^{N} \frac{(-1)^{s}}{(s+1)!} \left(\frac{a\Dtn{n}}{\Dx}\right)^{s+1}
  \left[\sum\limits_{p_{1} = 0}^{N} \uphi_{p_{1}} \left(\ophi_{k}\frac{d^{s}u^{+,n}}{d\xi^{s}}(\xi_{p_{1}}) -
                                                        \uphi_{k}\frac{d^{s}u^{n}}{d\xi^{s}}(\xi_{p_{1}})\right)\right]\\ & =
  \frac{1}{w_{k}} \sum\limits_{s = 0}^{N} \frac{(-1)^{s}}{(s+1)!} \left(\frac{a\Dtn{n}}{\Dx}\right)^{s+1}
  \left[\ophi_{k}\frac{d^{s}u^{+,n}}{d\xi^{s}}(0) - \uphi_{k}\frac{d^{s}u^{n}}{d\xi^{s}}(0)\right].
\end{split}
\end{equation}
The final expression for the flux term of the expression (\ref{eq:ader_dg_corr_temp_2}) takes the following form:
\begin{equation}\label{eq:ader_dg_corr_flux_term}
\begin{split}
&\frac{1}{w_{k}}\frac{a\Dtn{n}}{\Dx}\intrefdom{\tau}\,\Big[\ophi_{k} F_{\rm RP}\left(q(\tau, 1),\, q^{+}(\tau, 0)\right) -
                                                           \uphi_{k} F_{\rm RP}\left(q^{-}(\tau, 1),\, q(\tau, 0)\right)\Big]\\ & =
\frac{1}{w_{k}} \sum\limits_{s = 0}^{N} \frac{(-1)^{s}}{(s+1)!} \left(\frac{a\Dtn{n}}{\Dx}\right)^{s+1}
\Bigg\{
\theta( a)\left[\ophi_{k}\frac{d^{s}u^{n}}{d\xi^{s}}(1) - \uphi_{k}\frac{d^{s}u^{-,n}}{d\xi^{s}}(1)\right] +
\theta(-a)\left[\ophi_{k}\frac{d^{s}u^{+,n}}{d\xi^{s}}(0) - \uphi_{k}\frac{d^{s}u^{n}}{d\xi^{s}}(0)\right]
\Bigg\}.
\end{split}
\end{equation}

Substituting the resulting expressions for the local term (\ref{eq:ader_dg_corr_local_term}) and the flux term (\ref{eq:ader_dg_corr_flux_term}) into the original expression (\ref{eq:ader_dg_corr_temp_2}) for the ADER-DG numerical method leads to the following expression:
\begin{equation}
\begin{split}
&\hat{u}^{n+1}_{k} - \hat{u}^{n}_{k}
    = \sum\limits_{s = 0}^{N} \frac{(-1)^{s}}{(s+1)!}\left(\frac{a\Dtn{n}}{\Dx}\right)^{s+1}
      \Bigg\{
        \left[\frac{\ophi_{k}}{w_{k}}\,\frac{d^{s}u^{n}}{d\xi^{s}}(1) - \frac{\uphi_{k}}{w_{k}}\,\frac{d^{s}u^{n}}{d\xi^{s}}(0) - \hat{u}^{s+1, n}_{k}\right]\\ & -
        \frac{\theta( a)}{w_{k}}\left[\ophi_{k}\frac{d^{s}u^{n}}{d\xi^{s}}(1) - \uphi_{k}\frac{d^{s}u^{-,n}}{d\xi^{s}}(1)\right] -
        \frac{\theta(-a)}{w_{k}}\left[\ophi_{k}\frac{d^{s}u^{+,n}}{d\xi^{s}}(0) - \uphi_{k}\frac{d^{s}u^{n}}{d\xi^{s}}(0)\right]
      \Bigg\}\\ 
  & = \sum\limits_{s = 1}^{N} \frac{(-1)^{s}}{s!}\left(\frac{a\Dtn{n}}{\Dx}\right)^{s} \hat{u}^{s, n}_{k}
    + \sum\limits_{s = 0}^{N} \frac{(-1)^{s}}{(s+1)!}\left(\frac{a\Dtn{n}}{\Dx}\right)^{s+1}\\ & \cdot \Bigg[
        \theta( a)\frac{\uphi_{k}}{w_{k}} \left(\frac{d^{s}u^{-,n}}{d\xi^{s}}(1) - \frac{d^{s}u^{n}}{d\xi^{s}}(0)\right) +
        \theta(-a)\frac{\ophi_{k}}{w_{k}} \left(\frac{d^{s}u^{n}}{d\xi^{s}}(1) - \frac{d^{s}u^{+,n}}{d\xi^{s}}(0)\right)
      \Bigg],
\end{split}
\end{equation}
where in the first expression, a final zero term $\hat{u}^{N+1, n}_{k} = 0$ is added to ensure homogeneity of summation by index $s$. The last expression is written in the following equivalent form:
\begin{equation}\label{eq:ader_dg_corr}
\begin{split}
\hat{u}^{n+1}_{k} - \hat{u}^{n}_{k} = \sum\limits_{s = 1}^{N} \frac{(-1)^{s}}{s!}\left(\frac{a\Dtn{n}}{\Dx}\right)^{s} \hat{u}^{s, n}_{k}
   & + \theta( a)\frac{\uphi_{k}}{w_{k}} \sum\limits_{s = 0}^{N} \frac{(-1)^{s}}{(s+1)!}
       \left(\frac{a\Dtn{n}}{\Dx}\right)^{s+1}\left(\frac{d^{s}u^{-,n}}{d\xi^{s}}(1) - \frac{d^{s}u^{n}}{d\xi^{s}}(0)\right)\\
   & + \theta(-a)\frac{\ophi_{k}}{w_{k}} \sum\limits_{s = 0}^{N} \frac{(-1)^{s}}{(s+1)!}
       \left(\frac{a\Dtn{n}}{\Dx}\right)^{s+1}\left(\frac{d^{s}u^{n}}{d\xi^{s}}(1) - \frac{d^{s}u^{+,n}}{d\xi^{s}}(0)\right),
\end{split}
\end{equation}
which is convenient for approximation analysis, in particular, for calculating the order of approximation of the ADER-DG numerical method, and also for a qualitative (but not strictly quantitative) analysis of the stability of the ADER-DG numerical method (in Section~\ref{sec:stab_anal}).

First, an analysis of the numerical method approximation and the calculation of the approximation order are presented. For this purpose, the following estimates for approximating the interface values of the solution $u^{n}$ (\ref{eq:u_n_def}) are used, which are well known from knowledge of polynomial approximation and are similar to estimates (\ref{eq:errs_ests}):
\begin{equation}\label{eq:errs_ests_temp}
\begin{split}
&\frac{d^{s}u^{-,n}}{d\xi^{s}}(1) = 
  \left(\Dx\right)^{s} \left[\pdern{u}{x}{s}(x_{i}^{-}, t^{n}) + O\left(\Dx^{N+1-s}\right)\right] =
  (-1)^{s} \frac{\left(\Dx\right)^{s}}{a^{s}} \pdern{u}{t}{s}(x_{i}^{-}, t^{n}) + O\left(\Dx^{N+1}\right),\\
&\frac{d^{s}u^{n}}{d\xi^{s}}(0) = 
  \left(\Dx\right)^{s} \left[\pdern{u}{x}{s}(x_{i}^{+}, t^{n}) + O\left(\Dx^{N+1-s}\right)\right] =
  (-1)^{s} \frac{\left(\Dx\right)^{s}}{a^{s}} \pdern{u}{t}{s}(x_{i}^{+}, t^{n}) + O\left(\Dx^{N+1}\right),\\
&\frac{d^{s}u^{n}}{d\xi^{s}}(1) = 
    \left(\Dx\right)^{s} \left[\pdern{u}{x}{s}(x_{i+1}^{-}, t^{n}) + O\left(\Dx^{N+1-s}\right)\right] =
    (-1)^{s} \frac{\left(\Dx\right)^{s}}{a^{s}} \pdern{u}{t}{s}(x_{i}^{+}, t^{n}) + O\left(\Dx^{N+1}\right),\\
&\frac{d^{s}u^{+,n}}{d\xi^{s}}(0) = 
  \left(\Dx\right)^{s} \left[\pdern{u}{x}{s}(x_{i+1}^{+}, t^{n}) + O\left(\Dx^{N+1-s}\right)\right] =
  (-1)^{s} \frac{\left(\Dx\right)^{s}}{a^{s}} \pdern{u}{t}{s}(x_{i}^{-}, t^{n}) + O\left(\Dx^{N+1}\right).
\end{split}
\end{equation}
To obtain the result in closed form, an additional function $v: \Xi_{i}^{n} \rightarrow \mathcal{R}$ is introduced, with two constraints (boundary conditions) at time steps $t^{n}$ and $t^{n+1}$:
\begin{equation}\label{eq:add_function_v}
\begin{split}
&u(x, t) = \pderfoper{t} v(x, t),\ 
v(x, t) = \int\limits_{t^{n}}^{t} d\tilde{t}\, u\left(x, \tilde{t}\right),\ 
v(x, t^{n}) = 0,\ 
v(x, t^{n+1}) = \int\limits_{t^{n}}^{t^{n+1}} d\tilde{t}\, u\left(x, \tilde{t}\right),\ 
\pderfoper{t}\left[\pderf{v}{t} + a\pderf{v}{x}\right] = 0,
\end{split}
\end{equation}
which is necessary to form an expansion of the solution $u^{n+1}$ at time step $t^{n+1}$ equivalent to the finite part of the first terms of the Taylor series of solution.
It should also be noted that $v\in\{f(x,t)\, |\, t \mapsto f(x, t) \in \mathcal{P}_{N+1}(\omega), f(x, 0) = 0\}$ if $u\in\mathcal{P}_{N}(\omega)$, which means that for function $v$, in general, there exist Taylor series expansion terms at least up to the $N+1$ order in $\Dtn{n}$. Substituting the estimates (\ref{eq:errs_ests_temp}) and using the additional function $v$ (\ref{eq:add_function_v}) in the second term of the expression (\ref{eq:ader_dg_corr}) leads to the following expression:
\begin{equation}
\begin{split}
&\sum\limits_{s = 0}^{N} \frac{(-1)^{s}}{(s+1)!}\left(\frac{a\Dtn{n}}{\Dx}\right)^{s+1}
\frac{\uphi_{k}}{w_{k}} \left(\frac{d^{s}u^{-,n}}{d\xi^{s}}(1) - \frac{d^{s}u^{n}}{d\xi^{s}}(0)\right)\\ & =
\frac{\uphi_{k}}{w_{k}}\sum\limits_{s = 0}^{N} \frac{1}{(s+1)!}\left(\frac{a\Dtn{n}}{\Dx}\right)^{s+1}\left(\frac{\Dx}{a}\right)^{s}
\left[\pdern{u}{t}{s}(x_{i}^{-}, t^{n}) - \pdern{u}{t}{s}(x_{i}^{+}, t^{n})\right] + O\left(\Dx^{N+1}\right)\\ & =
\frac{\uphi_{k}}{w_{k}}\sum\limits_{s = 0}^{N} \frac{1}{(s+1)!}\left(\frac{a\Dtn{n}}{\Dx}\right)^{s+1}\left(\frac{\Dx}{a}\right)^{s}
\left[\pdern{v}{t}{s+1}(x_{i}^{-}, t^{n}) - \pdern{v}{t}{s+1}(x_{i}^{+}, t^{n})\right] + O\left(\Dx^{N+1}\right)\\ & =
-\frac{1}{\Dx}\frac{\uphi_{k}}{w_{k}}\sum\limits_{s = 0}^{N+1}
\frac{\left(\Dtn{n}\right)^{s}}{s!}\pdernoper{t}{s}[\![a v]\!](x_{i}, t^{n}) + O\left(\Dx^{N+1}\right) =
-\frac{1}{\Dx}\frac{\uphi_{k}}{w_{k}}[\![a v]\!](x_{i}, t^{n+1})\\ & + \frac{1}{\Dx} O\left((\Dtn{n})^{N+2}\right)+\left(\Dx^{N+1}\right) =
-\frac{1}{\Dx}\frac{\uphi_{k}}{w_{k}}\int\limits_{t_{n}}^{t_{n+1}}dt\,[\![a u]\!](x_{i}, t) + O\left(\Dx^{N+1}\right),
\end{split}
\end{equation}
where $[\![f]\!](x_{i}, t) = f(x_{i}^{+}, t) - f(x_{i}^{-}, t)$ is the operator determining the magnitude of the discontinuity of the function $f = f(x, t)$ at point $x_{i}$, and constraint $v(x, t^{n}) = 0$ is assumed in the case $s = 0$ to complete the sum over $s$ to the first $N+2$ terms of the Taylor series of $[\![a u]\!](x, t^{n+1})$, and a well-known estimate of the remainder term of the expansion according to the Taylor formula
\begin{equation}
-\frac{1}{\Dx}\frac{\uphi_{k}}{w_{k}}[\![a v]\!](x_{i}, t^{n+1}) =
-\frac{1}{\Dx}\frac{\uphi_{k}}{w_{k}}\sum\limits_{s = 0}^{N+1}
\frac{\left(\Dtn{n}\right)^{s}}{s!}\pdernoper{t}{s}[\![a v]\!](x_{i}, t^{n}) + \frac{1}{\Dx} O\left((\Dtn{n})^{N+2}\right).
\end{equation}
A similar substitution of the estimates (\ref{eq:errs_ests_temp}) and using the additional function $v$ (\ref{eq:add_function_v}) in the third term of the expression (\ref{eq:ader_dg_corr}) leads to the following expression:
\begin{equation}
\begin{split}
&\sum\limits_{s = 0}^{N} \frac{(-1)^{s}}{(s+1)!}\left(\frac{a\Dtn{n}}{\Dx}\right)^{s+1}
\frac{\ophi_{k}}{w_{k}} \left(\frac{d^{s}u^{n}}{d\xi^{s}}(1) - \frac{d^{s}u^{+,n}}{d\xi^{s}}(0)\right)\\ & =
\frac{\ophi_{k}}{w_{k}}\sum\limits_{s = 0}^{N} \frac{1}{(s+1)!}\left(\frac{a\Dtn{n}}{\Dx}\right)^{s+1}\left(\frac{\Dx}{a}\right)^{s}
\left[\pdern{u}{t}{s}(x_{i+1}^{-}, t^{n}) - \pdern{u}{t}{s}(x_{i+1}^{+}, t^{n})\right] + O\left(\Dx^{N+1}\right)\\ & =
\frac{\ophi_{k}}{w_{k}}\sum\limits_{s = 0}^{N} \frac{1}{(s+1)!}\left(\frac{a\Dtn{n}}{\Dx}\right)^{s+1}\left(\frac{\Dx}{a}\right)^{s}
\left[\pdern{v}{t}{s+1}(x_{i+1}^{-}, t^{n}) - \pdern{v}{t}{s+1}(x_{i+1}^{+}, t^{n})\right] + O\left(\Dx^{N+1}\right)\\ & =
-\frac{1}{\Dx}\frac{\ophi_{k}}{w_{k}}\sum\limits_{s = 1}^{N+1}
\frac{\left(\Dtn{n}\right)^{s}}{s!}\pdernoper{t}{s}[\![a v]\!](x_{i+1}, t^{n}) + O\left(\Dx^{N+1}\right) =
-\frac{1}{\Dx}\frac{\ophi_{k}}{w_{k}}[\![a v]\!](x_{i+1}, t^{n+1})\\ & + \frac{1}{\Dx}O\left((\Dtn{n})^{N+2}\right)+\left(\Dx^{N+1}\right) =
-\frac{1}{\Dx}\frac{\ophi_{k}}{w_{k}}\int\limits_{t_{n}}^{t_{n+1}}dt\,[\![a u]\!](x_{i+1}, t) + O\left(\Dx^{N+1}\right),
\end{split}
\end{equation}
where the constraint $\Dtn{n} \propto \Dx$ is used. The final form of the expression (\ref{eq:ader_dg_corr}) for the numerical solution (\ref{eq:u_n_def}) is as follows:
\begin{equation}\label{eq:errs_ests_final}
\begin{split}
&u^{n+1}(\xi) - u^{n}(\xi) = \sum\limits_{k = 0}^{N} \left(\hat{u}^{n+1}_{k} - \hat{u}^{n}_{k}\right)\varphi_{k}(\xi) =
  \sum\limits_{s = 1}^{N} \frac{(-1)^{s}}{s!}\left(\frac{a\Dtn{n}}{\Dx}\right)^{s} \left[\sum\limits_{k = 0}^{N}\hat{u}^{s, n}_{k}\varphi_{k}(\xi)\right]\\ & +
  \theta( a) \sum\limits_{s = 0}^{N} \frac{(-1)^{s}}{(s+1)!}\left(\frac{a\Dtn{n}}{\Dx}\right)^{s+1}
  \left(\frac{d^{s}u^{-,n}}{d\xi^{s}}(1) - \frac{d^{s}u^{n}}{d\xi^{s}}(0)\right)\left[\sum\limits_{k = 0}^{N} \frac{\uphi_{k}}{w_{k}}\varphi_{k}(\xi)\right]\\ & +
  \theta(-a) \sum\limits_{s = 0}^{N} \frac{(-1)^{s}}{(s+1)!}\left(\frac{a\Dtn{n}}{\Dx}\right)^{s+1}
  \left(\frac{d^{s}u^{n}}{d\xi^{s}}(1) - \frac{d^{s}u^{+,n}}{d\xi^{s}}(0)\right)\left[\sum\limits_{k = 0}^{N} \frac{\ophi_{k}}{w_{k}}\varphi_{k}(\xi)\right]\\ & =
  u^{n}(\xi(x - a\Dtn{n}), t^{n}) - u^{n}(\xi(x)) - \Bigg[
  \theta( a) \frac{\delta_{\mathcal{P}_{N}}(\xi)}{\Dx}\int\limits_{t_{n}}^{t_{n+1}}dt\,[\![a u]\!](x_{i}, t)\\ & -
  \theta(-a) \frac{\delta_{\mathcal{P}_{N}}(\xi-1)}{\Dx}\int\limits_{t_{n}}^{t_{n+1}}dt\,[\![a u]\!](x_{i+1}, t) + O\left(\Dx^{N+1}\right)\Bigg] =
  u(x(\xi), t^{n+1}) - u(x(\xi), t^{n}) + O\left(\Dx^{N+1}\right),
\end{split}
\end{equation}
where the expressions for $\delta_{\mathcal{P}_{N}}$ (\ref{eq:delta_func_varphi_coeffs}) and (\ref{eq:delta_func_varphi_action}) are used. The expression (\ref{eq:errs_ests_final}) shows that the approximation order of the ADER-DG numerical method with LST-DG predictor is $p = N+1$.

\section{Stability analysis}
\label{sec:stab_anal}

This Section is devoted to the stability analysis of the ADER-DG numerical method with the LST-DG predictor. Subsection~\ref{sec:stab_anal:qual_cons} ``Qualitative considerations'' presents a qualitative analysis of the ADER-DG numerical method's stability and obtains an asymptotic estimate of the boundary Courant number $\mathrm{CFL}_{\rm max}(N)$ for high degree $N \gg 1$ of basis polynomials. Subsection~\ref{sec:stab_anal:statment_quant_cons} ``Statement of the problem of quantitative considerations'' presents a detailed derivation of the relations necessary for a rigorous stability analysis of the ADER-DG numerical method with the LST-DG predictor within the framework of the Von Neumann criterion (spectral stability criterion). Subsection~\ref{sec:stab_anal:empir_cons} ``Empirical considerations'' presents a comprehensive set of empirical results from analyzing the spectrum of the evolution operator matrix for a single time step. Based on this analysis, several assumptions are made and substantiated regarding the localization of the eigenvalue boundary upon reaching which the stability of the ADER-DG numerical method with the LST-DG predictor is lost. Subsection~\ref{sec:stab_anal:quant_cons} ``Rigorous quantitative considerations'', based on the obtained empirical results, significantly simplifies the general stability problem of the ADER-DG numerical method with the LST-DG predictor. This results in a set of polynomials whose roots determine the boundary Courant number $\mathrm{CFL}_{\rm max}(N)$.

\subsection{Qualitative considerations}
\label{sec:stab_anal:qual_cons}

After presenting the approximation analysis of the ADER-DG numerical method with LST-DG predictor and calculating the order of approximation, a qualitative (but not strictly quantitative) analysis of the stability of the ADER-DG numerical method with LST-DG predictor is presented. A rigorous quantitative analysis of the stability of the ADER-DG numerical method is quite extensive and complex, and will be presented after the qualitative stability analysis. The obtained expression (\ref{eq:ader_dg_corr}) allows us to carry out a non-rigorous, but qualitatively correct analysis of the stability of the numerical method ADER-DG with the LST-DG predictor, which will allow us to identify the leading asymptotics of the maximum Courant number $\mathrm{CFL}_{\rm max}$ in the case of sufficiently large polynomial degrees $N \gg 1$.

The obtained expression (\ref{eq:ader_dg_corr}) qualitatively demonstrates that the ``characteristic advection velocity'' for each individual harmonic $u_{k}$ of the numerical solution (\ref{eq:u_n_def}) of the ADER-DG numerical method is transported not with the advection velocity $a$, but with a different effective velocities $a^{\rm eff, +}_{k}$ and $a^{\rm eff, -}_{k}$:
\begin{equation}\label{eq:a_eff_pm}
a^{\rm eff, +}_{k} = \frac{a \ophi_{k}}{w_{k}},\quad
a^{\rm eff, -}_{k} = \frac{a \uphi_{k}}{w_{k}},
\end{equation}
which, of course, is not quantitatively rigorous. Therefore, the Courant number $\mathrm{CFL}_{\rm max} \propto 1/a^{\rm eff}$ must be quantitatively estimated from the value of the maximum effective velocity $a^{\rm eff}$ inherent in the advective transport of individual harmonics $\hat{u}_{k}\varphi_{k}(\xi)$ of the solution (\ref{eq:u_n_def}):
\begin{equation}
\begin{split}\label{eq:a_eff}
a^{\rm eff} = \max\limits_{k}\left[\left|a^{\rm eff, +}_{k}\right|\right]
            = \max\limits_{k}\left[\left|a^{\rm eff, -}_{k}\right|\right]
            = a \cdot \max\limits_{k}\left[\left|\frac{\ophi_{k}}{w_{k}}\right|\right]
            = a \cdot \max\limits_{k}\left[\left|\frac{\uphi_{k}}{w_{k}}\right|\right],
\end{split}
\end{equation}
where the symmetry properties $\varphi_{k}(\xi) = \varphi_{N-k}(1 - \xi)$ of the basis polynomials $\{\varphi_{k}\}$ (\ref{eq:phi_def}) are taken into account when calculating the maximum values, leading to the relationships $\ophi_{k} = \uphi_{N - k}$, $w_{k} = w_{N-k}$. It should be noted that these properties apply only to the selected nodal basis $\{\varphi_{k}\}$ (\ref{eq:phi_def}) of Lagrange interpolation polynomials with nodal points at the roots of shifted Legendre polynomials $\tilde{L}_{N+1}$, and in the case of a different basis, a similar analysis must be performed taking into account the properties of the other basis. However, taking into account the properties of quadrature rules of the highest algebraic order~\cite{Szego_1075, Abramowitz_Stegun}, to which the GL quadrature rule belongs, no significant differences in terms of the results of stability analysis in the case of other nodal bases, with nodes at the roots of polynomials, are expected. The fact that the signs of the effective harmonic advection velocities (\ref{eq:a_eff}) are not taken into account in this case also speaks in favor of the ``purely qualitative'' nature of this analysis.

To obtain estimated values $a^{\rm eff}$, it is convenient to consider the left boundary $\xi = 0$ of the reference domain $\omega$ and the values $\uphi_{k}$. The expression for the basis functions $\{\varphi_{p}\}$ is conveniently chosen in the form through the original shifted Legendre polynomial $\tilde{L}_{N+1}(\xi)$~\cite{Szego_1075}:
\begin{equation}\label{eq:stab_anal_estim_temp_1}
\varphi_{p}(\xi) = \frac{\tilde{L}_{N+1}(\xi)}{(\xi - \xi_{p})\tilde{L}_{N+1}'(\xi_{p})},\quad\Rightarrow\quad
\uphi_{p} = \frac{(-1)^{N}}{\xi_{p}\tilde{L}_{N+1}'(\xi_{p})},
\end{equation}
where the defining property $\tilde{L}_{n}(0) = (-1)^{n}$ is used. The expression for the weights $w_{p}$ of the GL quadrature rule can be conveniently written in the form~\cite{Abramowitz_Stegun}, also rewritten in terms of the shifted Legendre polynomial $\tilde{L}_{N+1}(\xi)$:
\begin{equation}\label{eq:stab_anal_estim_temp_2}
w_{p} = \frac{1}{\xi_{p}(1-\xi_{p})[\tilde{L}_{N+1}'(\xi_{p})]^{2}},\quad\Rightarrow\quad
\left|\frac{\uphi_{p}}{w_{p}}\right| = (1-\xi_{p})|\tilde{L}_{N+1}'(\xi_{p})|.
\end{equation}
For a quantitative estimate of the nodal point $\{\xi_{p}\}$ asymptotics $N\rightarrow\infty$, it is convenient to use the well-known Mehler-Heine formula:
\begin{equation}
\lim\limits_{n\rightarrow\infty} L_{n}\left(\cos\left(\frac{\theta}{n}\right)\right) = J_{0}(\theta),\quad
\cos\left(\frac{\theta}{n}\right) \approx 1 - \frac{\theta^{2}}{2n^{2}},\quad\Rightarrow\quad
\theta_{p}^{2} \approx j_{0, p+1}^{2},
\end{equation}
where $J_{0}$ is the Bessel function of the first kind and order $0$, $j_{0, p+1}$ is the $p+1$ root of the Bessel function $J_{0}$. Taking into account the definition of shifted Legendre polynomials $\tilde{L}_{n}(\xi) = L_{n}(2\xi-1) = (-1)^{n}L_{n}(1-2\xi)$, the following expression for the nodal point asymptotics $\{\xi_{p}\}$ is obtained for higher degrees $N \gg 1$:
\begin{equation}\label{eq:stab_anal_estim_temp_3}
1 - 2\xi_{p} = 1 - \frac{\theta_{p}^{2}}{2(N+1)^{2}},\quad\Rightarrow\quad
\xi_{p} \approx \frac{j_{0,p+1}^{2}}{4(N+1)^{2}}.
\end{equation}
To quantitatively estimate the asymptotic behavior of the derivatives $|\tilde{L}_{N+1}'(\xi_{p})|$ of the shifted Legendre polynomials at the nodes $\{\xi_{p}\}$ that appear in the expression (\ref{eq:stab_anal_estim_temp_2}), a corollary from the Mehler-Heine formula is used:
\begin{equation}\label{eq:stab_anal_estim_temp_4}
\lim\limits_{n\rightarrow\infty} \left[\frac{1}{n^{2}} L_{n}'\left(\cos\left(\frac{\theta}{n}\right)\right)\right] = \frac{J_{1}(\theta)}{\theta},\quad\Rightarrow\quad
|\tilde{L}_{N+1}'(\xi_{p})| \approx 4(N+1)^{2}\frac{|J_{1}(j_{0,p+1})|}{j_{0,p+1}},
\end{equation}
where the definition of shifted Legendre polynomials is also used, and $J_{1}$ is the Bessel function of the first kind and order $1$. The resulting expression (\ref{eq:stab_anal_estim_temp_2}), after substitution (\ref{eq:stab_anal_estim_temp_3}) and (\ref{eq:stab_anal_estim_temp_4}), took the following form:
\begin{equation}\label{eq:stab_anal_estim_temp_5}
\begin{split}
\left|\frac{\uphi_{p}}{w_{p}}\right| \approx 4(N+1)^{2}\frac{|J_{1}(j_{0,p+1})|}{j_{0,p+1}},\ \Rightarrow\ &
\max\limits_{p}\left[\left|\frac{\uphi_{p}}{w_{p}}\right|\right] =
  4(N+1)^{2}\max\limits_{p}\left[\frac{|J_{1}(j_{0, p+1})|}{j_{0,p+1}^{2}}\right] =
  4\frac{|J_{1}(j_{0, 1})|}{j_{0,1}^{2}}(N+1)^{2} \approx 0.8635 (N+1)^{2} \propto (N+1)^{2},
\end{split}
\end{equation}
from which follows an asymptotic estimate of the effective advection velocity $a^{\rm eff}$ (\ref{eq:a_eff}) and the maximum Courant number $\mathrm{CFL}_{\rm max}$:
\begin{equation}\label{eq:cfls_asymp_est}
\begin{split}
a^{\rm eff} \propto (N+1)^{2},\ \Rightarrow\ \mathrm{CFL}_{\rm max} \propto \frac{1}{(N+1)^{2}}.
\end{split}
\end{equation}
It should be noted immediately that this estimate $a^{\rm eff}$ overestimates (by $21$--$47$\%) the original exact expression (\ref{eq:a_eff}), which is, in general, only an estimate and should not be expected to provide even satisfactory quantitative correspondence to the exact values $\mathrm{CFL}_{\rm max}$. However, the resulting expression (\ref{eq:cfls_asymp_est}) correctly expresses the asymptotic behavior for large polynomial degrees $N \gg 1$, which may be important for preliminary estimates of computational complexity and computational costs of the ADER-DG numerical method with LST-DG predictor.

It is important to note that the obtained qualitative estimates (\ref{eq:cfls_asymp_est}) differ significantly from those widely used estimates $\mathrm{CFL}_{\rm max} \propto 1/(2N+1)$ in existing studies~\cite{ader_dg_ideal_flows, ader_dg_ale, ader_dg_grmhd, ader_dg_gr_prd, ader_dg_simple_mod_2016, fron_phys, ader_dg_axioms, exahype, ader_stiff_3, ader_stiff_4, ader_eff_blas} and are relatively consistent with more accurate and in-depth estimates~\cite{PNPM_DG_2008, ader_dg_PNPM, ader_dg_stab}, which may indicate the importance of the obtained results. It is clear that the expressions for limiting the CFL criterion for a time step $\Dtn{n}$, typically formulated in existing studies~\cite{ader_dg_ideal_flows, exahype, ader_stiff_3, ader_stiff_4, ader_eff_blas}, are expressed in a form
\begin{equation}\label{eq:old_eff_cfl}
\begin{split}
\mathrm{CFL}_{\rm max} \propto \frac{\mathrm{C}}{2N+1},
\end{split}
\end{equation}
that always allows one to select an ``effective Courant number'' $\mathrm{C} \leq 1$ such that the calculations are stable.

\subsection{Statement of the problem of quantitative considerations}
\label{sec:stab_anal:statment_quant_cons}

A rigorous quantitative analysis of stability of the ADER-DG numerical method with LST-DG predictor in this work is carried out on the basis of the use of the Von Neumann criterion  of stability (spectral criterion), which has proven itself well for studying the stability of linear numerical methods for linear equations, and is based on the use of the previously obtained expression (\ref{eq:ader_dg_corr}), which is rewritten in the following form:
\begin{equation}\label{eq:evol_oper_prop}
u(x, t^{n}) = \exp(ikx),\qquad u(x, t^{n+1}) = R\exp(ikx) = R \cdot u(x, t^{n}),
\end{equation}
where the following notations are introduced:
\begin{equation}\label{eq:evol_oper_def}
R = R(\mathrm{CFL}, \theta),\quad \mathrm{CFL} = \frac{a\Dtn{n}}{\Dx},\quad \theta = k\Dx,
\end{equation}
where $R(\mathrm{CFL}, \theta)$ is the evolution operator (or the transition operator) for one time step $\Dtn{n}$, and the dependence on $a\Dtn{n}/\Dx$ and $k\Dx$ obviously follows from considerations of similarity and the dimensionality of the problem: $\mathrm{CFL}$ and $\theta$ are conveniently chosen two unique dimensionless parameters of the problem. The use of the Von Neumann criterion  of stability (spectral criterion) presupposes consideration of the evolution over one time step $\Dtn{n}$ of one characteristic harmonic of the numerical solution, which is chosen in the form of a plane wave $\exp(ikx)$. Representing the solution (\ref{eq:u_n_def}) as an expansion in basis polynomials $\{\varphi_{k}\}_{k}$ is a linear operation, therefore, the transition to the solution $\hat{u}^{\pm}(\xi)$ in neighboring cells $\Omega_{i \pm 1}\equiv\Omega_{\pm}$ in the form of multiplication by the ``shift exponential factor'' $\exp(\pm ik\Dx)$ has a similar form for individual expansion coefficients $\hat{u}_{k}$ of the numerical solution (\ref{eq:u_n_def}). In this case, the solution in neighboring coordinate cells $\hat{u}^{\pm}(\xi)$ is represented in the following form:
\begin{equation}\label{eq:r_matrix_def}
\hat{u}^{\pm, n}_{k} = \exp(\pm ik\Dx)\cdot\hat{u}^{n}_{k},\quad
\hat{u}^{n+1}_{k} = \sum\limits_{l = 0}^{N} R_{kl} \hat{u}^{n}_{l},\quad
\mathrm{R}(\mathrm{CFL}, \theta) = \{R_{kl}(\mathrm{CFL}, \theta)\}_{k,l},
\end{equation}
and the evolution operator $R(\mathrm{CFL}, \theta)$ is conveniently represented in the form of a matrix $\{R_{kl}(\mathrm{CFL}, \theta)\}_{k,l}$ defining a linear operator (the ADER-DG numerical method ((\ref{eq:ader_dg_corr_temp_1}) and (\ref{eq:errs_ests_final})) and the original problem (\ref{eq:adv_eq_src}) are linear) in the finite-dimensional representation space of the numerical solution (\ref{eq:u_n_def}). The local discrete space-time solution $q(\tau, \xi)$ (\ref{eq:local_sol_def}) obtained as a result of solving the system of algebraic equations of the LST-DG predictor (\ref{eq:lstdg_pred_eq_weak}) (see the expression (\ref{eq:lstdg_pred_eq_sol_temp_1_ind_form})) also depends linearly on the numerical solution $\hat{u}(\xi)$ (\ref{eq:u_n_def}), therefore it also has the following expression for the expansion coefficients $\{\hat{q}_{p_{0}p_{1}}\}_{p_{0}p_{1}}$ of the solution in neighboring coordinate cells $\Omega_{\pm}$:
\begin{equation}\label{eq:q_exp_shift}
\hat{q}^{\pm}_{p_{0}p_{1}} = \exp(\pm ik\Dx)\hat{q}_{p_{0}p_{1}}.
\end{equation}
Using the expression (\ref{eq:ader_dg_corr_temp_1}) for the ADER-DG numerical method and the exact Riemann solver (\ref{eq:f_rp}) for the problem (\ref{eq:adv_eq_src}) leads to the following expression for the coefficients of the numerical solution at the new time step:
\begin{equation}
\begin{split}
\hat{u}^{n+1}_{k} & =
  \hat{u}^{n}_{k} + \frac{\mathrm{CFL}}{w_{k}}\Bigg\{\intrefdom{\tau}\intrefdom{\xi}\,\varphi_{k}'(\xi) q(\tau, \xi) -
  \intrefdom{\tau}\,\Bigg[\ophi_{k} \Big(\theta(a)q(\tau, 1) + \theta(-a)q^{+}(\tau, 0)\Big)\\ & -
  \uphi_{k} \Big(\theta(a)q^{-}(\tau, 1) + \theta(-a)q(\tau, 0)\Big)\Bigg]\Bigg\} =
  \hat{u}^{n}_{k} + \frac{\mathrm{CFL}}{w_{k}}\Bigg\{\intrefdom{\tau}\intrefdom{\xi}\,\varphi_{k}'(\xi) q(\tau, \xi)\\ & -
  \intrefdom{\tau}\,\Bigg[\theta(a) \Big(\ophi_{k} - \uphi_{k}\exp(-i\theta)\Big)q(\tau, 1) +
  \theta(-a)\Big(\ophi_{k}\exp(i\theta) - \uphi_{k}\Big)q(\tau, 0)\Bigg]\Bigg\},
\end{split}
\end{equation}
where the properties (\ref{eq:phi_def}) and (\ref{eq:gl_rule}) of the chosen functional basis $\{\varphi_{k}\}_{k}$ and the expression (\ref{eq:q_exp_shift}) are used. Substituting the representation of the local discrete space-time solution $q(\tau, \xi)$ in the form (\ref{eq:local_sol_def}) of an expansion in basis functions $\{\varphi_{p_{0}}\varphi_{p_{1}}\}_{p_{0}p_{1}}$ leads to the following expression:
\begin{equation}\label{eq:r_matrix_deriv_temp_1}
\begin{split}
&\hat{u}^{n+1}_{k} =
  \hat{u}^{n}_{k} + \frac{\mathrm{CFL}}{w_{k}}\Bigg\{
      \sum\limits_{p_{0} = 0}^{N}\sum\limits_{p_{1} = 0}^{N} \intrefdom{\tau}\,\varphi_{p_{0}}(\tau)
      \intrefdom{\xi}\,\varphi_{k}'(\xi)\varphi_{p_{1}}(\xi)\,\hat{q}_{p_{0}p_{1}}\\ & - \sum\limits_{p_{0} = 0}^{N}\sum\limits_{p_{1} = 0}^{N}
      \intrefdom{\tau}\,\varphi_{p_{0}}(\tau)\Bigg[
      \theta(a) \ophi_{p_{1}}\Big(\ophi_{k} - \uphi_{k}\exp(-i\theta)\Big)\,\hat{q}_{p_{0}p_{1}} + \theta(-a)\uphi_{p_{1}}\Big(\ophi_{k}\exp(i\theta) - \uphi_{k}\Big)\,\hat{q}_{p_{0}p_{1}} 
      \Bigg]\Bigg\}\\ & =
  \hat{u}^{n}_{k} + \frac{\mathrm{CFL}}{w_{k}} \sum\limits_{p_{0} = 0}^{N}\sum\limits_{p_{1} = 0}^{N} w_{p_{0}} \Bigg\{
      \varrho_{p_{1}k} - \Bigg[\theta(a) \ophi_{p_{1}}\Big(\ophi_{k} - \uphi_{k}\exp(-i\theta)\Big)
  + \theta(-a)\uphi_{p_{1}}\Big(\ophi_{k}\exp(i\theta) - \uphi_{k}\Big)\Bigg]
      \Bigg\}\hat{q}_{p_{0}p_{1}},
\end{split}
\end{equation}
where the definition (\ref{eq:matrices_def}) of matrix $\varrho$ is used. Substituting expression (\ref{eq:lstdg_pred_eq_sol_temp_1_ind_form}) into (\ref{eq:r_matrix_deriv_temp_1}) yields the following expression:
\begin{equation}\label{eq:r_matrix_deriv_temp_2}
\begin{split}
\hat{u}^{n+1}_{k} & =
  \sum\limits_{l = 0}^{N} \left[\vphantom{\intrefdom{\tau}}\right. \delta_{kl} +
  \frac{\mathrm{CFL}}{w_{k}} \sum\limits_{p_{0} = 0}^{N}\sum\limits_{p_{1} = 0}^{N} w_{p_{0}} \Bigg\{
      \varrho_{p_{1}k} - \Bigg[\theta(a) \ophi_{p_{1}}\Big(\ophi_{k} - \uphi_{k}\exp(-i\theta)\Big)\\
  & + \theta(-a)\uphi_{p_{1}}\Big(\ophi_{k}\exp(i\theta) - \uphi_{k}\Big)\Bigg]
      \Bigg\}\sigma_{p_{0}p_{1}, k}\left.\vphantom{\intrefdom{\tau}}\right] \hat{u}^{n}_{l}\\ & =
  \sum\limits_{l = 0}^{N} \left[\vphantom{\intrefdom{\tau}}\right. \delta_{kl} +
    \frac{\mathrm{CFL}}{w_{k}} \sum\limits_{p_{0} = 0}^{N}\sum\limits_{p_{1} = 0}^{N} w_{p_{0}} \Bigg\{
        \varrho_{p_{1}k} - \Bigg[\theta(a) \ophi_{p_{1}}\Big(\ophi_{k} - \uphi_{k}\exp(-i\theta)\Big)\\
    & + \theta(-a)\uphi_{p_{1}}\Big(\ophi_{k}\exp(i\theta) - \uphi_{k}\Big)\Bigg]
        \Bigg\} \Bigg\{\sum\limits_{s = 0}^{N} (-1)^{s} \mathrm{CFL}^{s} \frac{\tau_{p_{0}}^{s}}{s!}
        \left[\mathrm{D}^{s}\right]_{p_{1}k}\Bigg\} \left.\vphantom{\intrefdom{\tau}}\right] \hat{u}^{n}_{l}\\ & =
  \sum\limits_{l = 0}^{N} \left[\vphantom{\intrefdom{\tau}}\right. \delta_{kl} +
    \sum\limits_{p_{1} = 0}^{N} \frac{1}{w_{k}}\Bigg\{
        w_{p_{1}}D_{p_{1}k} - \Bigg[\theta(a) \ophi_{p_{1}}\Big(\ophi_{k} - \uphi_{k}\exp(-i\theta)\Big)\\
    & + \theta(-a)\uphi_{p_{1}}\Big(\ophi_{k}\exp(i\theta) - \uphi_{k}\Big)\Bigg]
        \Bigg\} \Bigg\{\sum\limits_{s = 0}^{N} \frac{(-1)^{s}\mathrm{CFL}^{s+1}}{(s+1)!} \left[\mathrm{D}^{s}\right]_{p_{1}k}\Bigg\}
        \left.\vphantom{\intrefdom{\tau}}\right] \hat{u}^{n}_{l}.
\end{split}
\end{equation}
The last expression (\ref{eq:r_matrix_deriv_temp_2}) allowed us to explicitly obtain an expression for the elements of matrix $\mathrm{R}$ (\ref{eq:r_matrix_def}) of the evolution operator $R(\mathrm{CFL}, \theta)$:
\begin{equation}\label{eq:r_matrix_elems_expr}
\begin{split}
R_{kl}(\mathrm{CFL}, \theta) = \delta_{kl} +
    \sum\limits_{p_{1} = 0}^{N} \frac{1}{w_{k}}\Bigg[&
        w_{p_{1}}D_{p_{1}k} - \theta(a) \ophi_{p_{1}}\Big(\ophi_{k} - \uphi_{k}\exp(-i\theta)\Big)
      + \theta(-a)\uphi_{p_{1}}\Big(\ophi_{k}\exp(i\theta) - \uphi_{k}\Big)
        \Bigg] \Bigg[\sum\limits_{s = 0}^{N} \frac{(-1)^{s}\mathrm{CFL}^{s+1}}{(s+1)!} \left[\mathrm{D}^{s}\right]_{p_{1}k}\Bigg],
\end{split}
\end{equation}
which is conveniently rewritten in pure matrix form:
\begin{equation}\label{eq:r_matrix_in_matrix_form}
\mathrm{R}(\mathrm{CFL}, \theta) = I + \mathrm{A}(\theta)\mathrm{B}(\mathrm{CFL}),\quad
\mathrm{A}(\theta) = \{A_{kl}(\theta)\}_{k,l},\quad
\mathrm{B}(\mathrm{CFL}) = \{B_{kl}(\mathrm{CFL})\}_{k,l},
\end{equation}
where $I = \{\delta_{kl}\}_{kl}$ is the identity matrix, and the matrices $\mathrm{A}(\theta)$ and $\mathrm{B}(\mathrm{CFL})$ are introduced with following elements:
\begin{equation}\label{eq:a_and_b_matrices_def}
\begin{split}
A_{kl}(\theta) = \frac{1}{w_{k}}\Bigg[w_{l}D_{lk} - \Big[
\theta(a) \ophi_{l}\Big(\ophi_{k} - \uphi_{k}\exp(-i\theta)\Big) +
\theta(-a)\uphi_{l}\Big(\ophi_{k}\exp(i\theta) - \uphi_{k}\Big)\Big]\Bigg],\quad
B_{kl}(\mathrm{CFL}) = \sum\limits_{s = 0}^{N} \frac{(-1)^{s}\mathrm{CFL}^{s+1}}{(s+1)!} \left[\mathrm{D}^{s}\right]_{kl}.
\end{split}
\end{equation}

The expressions (\ref{eq:r_matrix_elems_expr}), (\ref{eq:r_matrix_in_matrix_form}) and (\ref{eq:a_and_b_matrices_def}) obtained above for the matrix elements $R_{kl}$ the matrix $\mathrm{R}$ (\ref{eq:r_matrix_def}) of the evolution operator $R(\mathrm{CFL}, \theta)$ can be obtained by using pure matrix operations, which is of interest from the perspective of developing a high-performance software implementation of the ADER-DG numerical method with the LST-DG predictor using the BLAS interface~\cite{ader_eff_blas}. Below is presented the derivation of the matrix $\mathrm{R}$ of the evolution operator $R(\mathrm{CFL}, \theta)$, for which the expression (\ref{eq:lstdg_pred_eq_sol_temp_1}) for the vector of expansion coefficients of the local discrete space-time solution is used. The following vector and matrix notations are introduced:
\begin{equation}
\boldsymbol{\Theta}(\tau, \xi) = \{\Theta_{\mathrm{p}}(\tau, \xi)\}_{\mathrm{p}},\quad
\boldsymbol{\varphi}(\xi) = \{\varphi_{k}(\xi)\}_{k},\quad
\boldsymbol{\Theta}(\tau, \xi) = \boldsymbol{\varphi}(\tau)\otimes\boldsymbol{\varphi}(\xi),
\end{equation}
using which the local discrete space-time solution $q(\tau, \xi)$ (\ref{eq:local_sol_def}) and solutions $u^{n}(\xi)$ and $u^{n+1}(\xi)$ (\ref{eq:u_n_def}) are expressed in convolution forms:
\begin{equation}
q(\tau, \xi) = \boldsymbol{\Theta}(\tau, \xi)^{T}\hat{\mathbf{q}} = \hat{\mathbf{q}}^{T}\boldsymbol{\Theta}(\tau, \xi),\quad
u^{n}(\xi) = \boldsymbol{\varphi}(\xi)^{T}\hat{\mathbf{u}}^{n} = [\hat{\mathbf{u}}^{n}]^{T}\boldsymbol{\varphi}(\xi),\quad
u^{n+1}(\xi) = \boldsymbol{\varphi}(\xi)^{T}\hat{\mathbf{u}}^{n+1} = [\hat{\mathbf{u}}^{n+1}]^{T}\boldsymbol{\varphi}(\xi).
\end{equation}
Substituting the introduced matrix form of the solutions $q(\tau, \xi)$ (\ref{eq:local_sol_def}), $u^{n}(\xi)$ and $u^{n+1}(\xi)$ (\ref{eq:u_n_def}) into the expression for the ADER-DG numerical method (\ref{eq:ader_dg_corr_temp_1}) leads to the following expression
\begin{equation}
\begin{split}
\hat{\mathbf{u}}^{n+1} & =
  \hat{\mathbf{u}}^{n} + \mathrm{CFL} \cdot m^{-1}\Bigg\{\intrefdom{\tau}\intrefdom{\xi}\,\boldsymbol{\varphi}'(\xi) q(\tau, \xi)
  - \intrefdom{\tau}\,\Bigg[\theta(a) \Big(\ovecphi - \uvecphi\exp(-i\theta)\Big)q(\tau, 1)\\
  & + \theta(-a)\Big(\ovecphi\exp(i\theta) - \uvecphi\Big)q(\tau, 0)\Bigg]\Bigg\} =
  \hat{\mathbf{u}}^{n} + \mathrm{CFL} \cdot m^{-1}\Bigg\{\intrefdom{\tau}\intrefdom{\xi}\,\boldsymbol{\varphi}'(\xi) [\boldsymbol{\varphi}(\tau)\otimes\boldsymbol{\varphi}(\xi)]^{T}\\ 
  & - \intrefdom{\tau}\,\Bigg[\theta(a) \Big(\ovecphi - \uvecphi\exp(-i\theta)\Big)[\boldsymbol{\varphi}(\tau)\otimes\ovecphi]^{T}
  + \theta(-a)\Big(\ovecphi\exp(i\theta) - \uvecphi\Big)[\boldsymbol{\varphi}(\tau)\otimes\uvecphi]^{T}\Bigg]\Bigg\}\\
  & \hphantom{-\intrefdom{\tau}\,\Bigg[}\cdot \Big(I \otimes I + \mathrm{CFL} \cdot \chi \otimes \mathrm{D}\Big)^{-1}\Big(\mathbf{1} \otimes \hat{\mathbf{u}}^{n}\Big),
\end{split}
\end{equation}
where expressions (\ref{eq:matrices_K_def}) are used to represent the transformation matrices in the form of Kronecker products of the main matrices of the functional basis $\{\varphi_{k}\}_{k}$. Explicit transformations were implemented in the resulting expression, leading to the following expression:
\begin{equation}
\begin{split}
\hat{\mathbf{u}}^{n+1} =
  \Bigg[I + \mathrm{CFL} \cdot m^{-1}\Bigg\{&\intrefdom{\tau}\intrefdom{\xi}\,\boldsymbol{\varphi}'(\xi) [\boldsymbol{\varphi}(\tau)\otimes\boldsymbol{\varphi}(\xi)]^{T}
  - \intrefdom{\tau}\,\Bigg[\theta(a) \Big(\ovecphi - \uvecphi\exp(-i\theta)\Big)[\boldsymbol{\varphi}(\tau)\otimes\ovecphi]^{T}\\
  & + \theta(-a)\Big(\ovecphi\exp(i\theta) - \uvecphi\Big)[\boldsymbol{\varphi}(\tau)\otimes\uvecphi]^{T}\Bigg]\Bigg\}
    \cdot \left\{\sum\limits_{s = 0}^{N} \frac{(-1)^{s} \mathrm{CFL}^{s}}{s!}\, \boldsymbol{\tau}_{s} \otimes \mathrm{D}^{s}\right\}\Bigg]\hat{\mathbf{u}}^{n},
\end{split}
\end{equation}
where is used the following property of the Kronecker product
\begin{equation}
(\mathrm{G}\otimes\mathrm{H})(\mathbf{a}\otimes\mathbf{b}) = (\mathrm{G}\mathbf{a})\otimes(\mathrm{H}\mathbf{b}) =
(\mathrm{G}\mathbf{a})\otimes(\mathrm{H}I\mathbf{b}) = ((\mathrm{G}\mathbf{a})\otimes(\mathrm{H}I))(1\otimes\mathbf{b}) =
((\mathrm{G}\mathbf{a})\otimes\mathrm{H})\mathbf{b},
\end{equation}
which holds for any matrices $\mathrm{G}$, $\mathrm{H} \in \mathcal{R}^{n}\times\mathcal{R}^{n}$ and vectors $\mathbf{a}$, $\mathbf{b} \in \mathcal{R}^{n}$, the following notation for property (\ref{eq:chi_int_prop}) is introduced:
\begin{equation}
\chi\boldsymbol{\tau}_{s} = \frac{\boldsymbol{\tau}_{s+1}}{s+1},\quad
\chi\mathbf{1} = \boldsymbol{\tau},\quad \boldsymbol{\tau}_{s} = \{\tau_{p}^{s}\}_{p},
\end{equation}
as well as the nilpotency $\mathrm{D}^{N+1} = 0$ of the differentiation matrix $\mathrm{D}$ is used (see also (\ref{eq:local_sol_inv_matrix})). From the last expression obtained and the definition (\ref{eq:r_matrix_def}), the expression for the matrix $\mathrm{R}(\mathrm{CFL}, \theta)$ follows:
\begin{equation}\label{eq:r_matrix_deriv_temp_3}
\begin{split}
\mathrm{R}(\mathrm{CFL}, \theta) = I + \mathrm{CFL} \cdot m^{-1}&\Bigg\{\intrefdom{\tau}\intrefdom{\xi}\,\boldsymbol{\varphi}'(\xi)
    [\boldsymbol{\varphi}(\tau)\otimes\boldsymbol{\varphi}(\xi)]^{T}
    - \intrefdom{\tau}\,\Bigg[\theta(a) \Big(\ovecphi - \uvecphi\exp(-i\theta)\Big)[\boldsymbol{\varphi}(\tau)\otimes\ovecphi]^{T}\\
  & + \theta(-a)\Big(\ovecphi\exp(i\theta) - \uvecphi\Big)[\boldsymbol{\varphi}(\tau)\otimes\uvecphi]^{T}\Bigg]\Bigg\}
    \cdot \left\{\sum\limits_{s = 0}^{N} \frac{(-1)^{s} \mathrm{CFL}^{s}}{s!}\, \boldsymbol{\tau}_{s} \otimes \mathrm{D}^{s}\right\}.
\end{split}
\end{equation}
The first term in the first brackets of the last expression (\ref{eq:r_matrix_deriv_temp_3}) is expressed in the following form
\begin{equation}\label{eq:r_matrix_deriv_temp_4}
\begin{split}
\intrefdom{\tau}\intrefdom{\xi}&\,\boldsymbol{\varphi}'(\xi) [\boldsymbol{\varphi}(\tau)\otimes\boldsymbol{\varphi}(\xi)]^{T} =
\intrefdom{\xi}\intrefdom{\xi}\,\boldsymbol{\varphi}'(\xi)\otimes\boldsymbol{\varphi}^{T}(\tau)\otimes\boldsymbol{\varphi}^{T}(\xi)\\ & =
\intrefdom{\xi}\,\boldsymbol{\varphi}'(\xi)\otimes\left[\intrefdom{\tau}\,\boldsymbol{\varphi}^{T}(\tau)\right]\otimes\boldsymbol{\varphi}^{T}(\xi) =
\intrefdom{\xi}\,\boldsymbol{\varphi}'(\xi)\otimes\mathbf{w}^{T}\otimes\boldsymbol{\varphi}^{T}(\xi) \equiv
\mathbf{w}^{T} \otimes \intrefdom{\xi}\,(\boldsymbol{\varphi}'(\xi) \otimes \boldsymbol{\varphi}^{T}(\xi)) = \mathbf{w}^{T} \otimes \varrho^{T},
\end{split}
\end{equation}
where the properties $\mathbf{a}^{T}\otimes\mathbf{b} = \mathbf{b}\otimes\mathbf{a}^{T}$, $\mathbf{a}(\mathbf{b}\otimes\mathbf{c})^{T} = \mathbf{a}\otimes\mathbf{b}^{T}\otimes\mathbf{c}^{T}$ of the Kronecker product are used. The second terms in the first brackets of the expression (\ref{eq:r_matrix_deriv_temp_3}) are transformed similarly (\ref{eq:r_matrix_deriv_temp_4}):
\begin{equation}
\begin{split}
&\intrefdom{\tau}\,\ovecphi\,[\boldsymbol{\varphi}(\tau)\otimes\ovecphi]^{T} \equiv
\left[\intrefdom{\tau}\,\boldsymbol{\varphi}(\tau)^{T}\right] \otimes \ovecphi \otimes \ovecphi^{T} = \mathbf{w}^{T} \otimes \ovecphi \otimes \ovecphi^{T},\\
&\intrefdom{\tau}\,\uvecphi\,[\boldsymbol{\varphi}(\tau)\otimes\ovecphi]^{T} = \mathbf{w}^{T} \otimes \uvecphi \otimes \ovecphi^{T},\quad
\intrefdom{\tau}\,\ovecphi\,[\boldsymbol{\varphi}(\tau)\otimes\uvecphi]^{T} = \mathbf{w}^{T} \otimes \ovecphi \otimes \uvecphi^{T},\\
&\intrefdom{\tau}\,\uvecphi\,[\boldsymbol{\varphi}(\tau)\otimes\uvecphi]^{T} = \mathbf{w}^{T} \otimes \uvecphi \otimes \uvecphi^{T}.
\end{split}
\end{equation}
Full expansion of the brackets in the expression (\ref{eq:r_matrix_deriv_temp_3}) required the following transformation:
\begin{equation}\label{eq:r_matrix_deriv_temp_5_1}
\begin{split}
&(\mathbf{w}^{T}\otimes\varrho^{T})(\boldsymbol{\tau}_{s}\otimes\mathrm{D}) = (\mathbf{w}^{T}\boldsymbol{\tau}_{s})\otimes(\varrho^{T}\,\mathrm{D}) \equiv
(\mathbf{w}^{T}\boldsymbol{\tau}_{s})(\varrho^{T}\,\mathrm{D}) = \cfrac{\varrho^{T}\,\mathrm{D}}{s+1},
\end{split}
\end{equation}
where the property (\ref{eq:b_prop_ode}) is used, and four more similar transformations:
\begin{equation}\label{eq:r_matrix_deriv_temp_5_2}
\begin{split}
&(\mathbf{w}^{T}\otimes\ovecphi\otimes\ovecphi^{T})(\boldsymbol{\tau}_{s}\otimes\mathrm{D}) = \cfrac{(\ovecphi\otimes\ovecphi^{T})\,\mathrm{D}}{s+1},\qquad
 (\mathbf{w}^{T}\otimes\uvecphi\otimes\ovecphi^{T})(\boldsymbol{\tau}_{s}\otimes\mathrm{D}) = \cfrac{(\uvecphi\otimes\ovecphi^{T})\,\mathrm{D}}{s+1},\\
&(\mathbf{w}^{T}\otimes\ovecphi\otimes\uvecphi^{T})(\boldsymbol{\tau}_{s}\otimes\mathrm{D}) = \cfrac{(\ovecphi\otimes\uvecphi^{T})\,\mathrm{D}}{s+1},\qquad
 (\mathbf{w}^{T}\otimes\uvecphi\otimes\uvecphi^{T})(\boldsymbol{\tau}_{s}\otimes\mathrm{D}) = \cfrac{(\uvecphi\otimes\uvecphi^{T})\,\mathrm{D}}{s+1}.
\end{split}
\end{equation}
Substituting the resulting expressions (\ref{eq:r_matrix_deriv_temp_5_1}) and (\ref{eq:r_matrix_deriv_temp_5_2}) into the expression (\ref{eq:r_matrix_deriv_temp_3}) yielded the following expression:
\begin{equation}\label{eq:r_matrix_elems_expr_dup}
\begin{split}
\mathrm{R}(\mathrm{CFL}, \theta) & = I + \mathrm{CFL} \cdot m^{-1}\Bigg[\varrho^{T} - \Big[\theta(a) \Big(\ovecphi - \uvecphi\exp(-i\theta)\Big)\otimes\ovecphi^{T}\\
  & + \theta(-a)\Big(\ovecphi\exp(i\theta) - \uvecphi\Big)\otimes\uvecphi^{T}\Big]\Bigg]
  \Bigg[\sum\limits_{s = 0}^{N} \frac{(-1)^{s} \mathrm{CFL}^{s}}{s!}\, \frac{\mathrm{D}^{s}}{s+1}\Bigg]\\
  & = I + m^{-1}\Bigg[\mathrm{D}^{T} m - \Big[\theta(a) \Big(\ovecphi - \uvecphi\exp(-i\theta)\Big)\otimes\ovecphi^{T}\\
  & + \theta(-a)\Big(\ovecphi\exp(i\theta) - \uvecphi\Big)\otimes\uvecphi^{T}\Big]\Bigg]
  \Bigg[\sum\limits_{s = 0}^{N} \frac{(-1)^{s} \mathrm{CFL}^{s+1}}{(s+1)!}\, \mathrm{D}^{s}\Bigg],
\end{split}
\end{equation}
which, taking into account the definition of matrix $\mathrm{R}(\mathrm{CFL}, \theta) = I + \mathrm{A}(\theta)\mathrm{B}(\mathrm{CFL})$ (\ref{eq:r_matrix_in_matrix_form}) in its representation form, allowed us to determine expressions for matrices $\mathrm{A}(\theta)$ and $\mathrm{B}(\mathrm{CFL})$:
\begin{equation}\label{eq:a_and_b_matrices_def_dup}
\begin{split}
\mathrm{A}(\theta) = m^{-1}\Bigg[\mathrm{D}^{T} m - 
\Big[\theta(a) \Big(\ovecphi - \uvecphi\exp(-i\theta)\Big)\otimes\ovecphi^{T} + \theta(-a)\Big(\ovecphi\exp(i\theta) - \uvecphi\Big)\otimes\uvecphi^{T}\Big]\Bigg],\quad
\mathrm{B}(\mathrm{CFL}) = \sum\limits_{s = 0}^{N} \frac{(-1)^{s} \mathrm{CFL}^{s+1}}{(s+1)!}\, \mathrm{D}^{s},
\end{split}
\end{equation}
which is of course consistent with expression (\ref{eq:a_and_b_matrices_def}).

The work~\cite{ader_dg_stab} notes the property of the matrix $\mathrm{R}(\mathrm{CFL}, \theta)$ (\ref{eq:r_matrix_in_matrix_form}) of the evolution operator $R$, which consists in the correct evolution of a completely constant solution $u(\xi) \equiv \mathrm{const}$ (for a linear uniform problem (\ref{eq:adv_eq_src}), it is clear that this is reduced to the correct reproduction of a ``unit solution'' $u(\xi) \equiv 1$, sometimes called ``constant-preserving property'') independent of the Courant number $\mathrm{CFL}$. In this work, the constant-preserving property is rigorously proven from the obtained expression for the matrix $\mathrm{R}(\mathrm{CFL}, \theta)$ of the evolution operator $R$ for one time step $\Dtn{n}$. For a constant unit solution $u^{n}(\xi) \equiv 1$, the wave vector $k = 0$ is chosen, as a result of which the phase $\theta = 0$ is chosen. The expansion coefficients $\mathbf{u}^{n}$ of the numerical solution at the previous time step are calculated trivially:
\begin{equation}
u^{n}(\xi) \equiv 1,\ \Rightarrow\ 
\hat{u}_{n}^{k} = \frac{1}{w_{k}}\intrefdom{\xi}\, u^{n}(\xi)\varphi_{k}(\xi) \equiv u^{n}(\xi_{k}) = 1,\ \Rightarrow\ 
\mathbf{u}^{n} = \mathbf{1}.
\end{equation}
The expansion coefficients of the numerical solution $\mathbf{u}^{n+1}$ at the new time step are determined by the following expression:
\begin{equation}
\mathbf{u}^{n+1} = \mathrm{R}(\mathrm{CFL}, 0)\mathbf{u}^{n} = \mathrm{R}(\mathrm{CFL}, 0)\mathbf{1} = I\mathbf{1} + \mathrm{A}(0)(\mathrm{B}(\mathrm{CFL})\mathbf{1}).
\end{equation}
Then, the constant-preserving property reduces to requirement $\mathrm{A}(0)(\mathrm{B}(\mathrm{CFL})\mathbf{1}) = \mathbf{0}$. For matrix $\mathrm{B}(\mathrm{CFL})$, an obvious relation holds
\begin{equation}
\mathrm{B}(\mathrm{CFL})\mathbf{1} = \sum\limits_{s = 0}^{N} \frac{(-1)^{s} \mathrm{CFL}^{s+1}}{(s+1)!}\, (\mathrm{D}^{s}\mathbf{1}) = \mathbf{1},
\end{equation}
since the derivative of the constant is zero: $\mathrm{D}^{s} \mathbf{1} = 0$ if $s > 0$. For the action of matrix $\mathrm{A}(0)$ on the vector of ones $\mathbf{1}$, the following chain of relations is obtained:
\begin{equation}
\begin{split}
\mathrm{A}(0)\mathbf{1} = m^{-1}\Big[\mathrm{D}^{T}m\mathbf{1} - 
\Big[\ovecphi - \uvecphi\Big]\otimes\Big[\theta(a) \ovecphi^{T} + \theta(-a)\otimes\uvecphi^{T}\Big]\Big]\mathbf{1} & =
m^{-1}\Big[\mathrm{D}^{T}m\mathbf{1} - \Big[\ovecphi - \uvecphi\Big]\otimes\Big[\theta(a) (\ovecphi^{T}\mathbf{1}) + \theta(-a)(\uvecphi^{T}\mathbf{1})\Big]\Big]\\ & =
m^{-1}\Big[\mathrm{D}^{T}m\mathbf{1} - \Big[\ovecphi - \uvecphi\Big]\Big[\theta(a) + \theta(-a)\Big]\Big] =
m^{-1}\Big[\mathrm{D}^{T}m\mathbf{1} - \Big[\ovecphi - \uvecphi\Big]\Big].
\end{split}
\end{equation}
Direct calculation of the first term in the brackets of the last expression obtained is accomplished by the following chain of transformations the $p$-th component of the resulting vector:
\begin{equation}
\begin{split}
\big[\mathrm{D}^{T} m \mathbf{1}\big]_{p} & = \sum\limits_{q = 0}^{N} w_{q} D_{qp} = \sum\limits_{q = 0}^{N} w_{q} \frac{{\varrho}_{qp}}{w_{q}} =
\sum\limits_{q = 0}^{N} {\varrho}_{qp} = \sum\limits_{q = 0}^{N} \intrefdom{\xi}\, \varphi_{q}(\xi) \varphi_{p}'(\xi) =
\intrefdom{\xi}\, \left[\sum\limits_{q = 0}^{N} \varphi_{q}(\xi)\right] \varphi_{p}'(\xi) =
\intrefdom{\xi}\, \varphi_{p}'(\xi) = \ophi_{p} - \uphi_{p},
\end{split}
\end{equation}
which led to the following expression and the chain of implications
\begin{equation}\label{eq:cont_presiv_prop}
\mathrm{D}^{T} m \mathbf{1} = \ovecphi - \uvecphi,\quad\Rightarrow\quad
\mathrm{A}(0)\mathbf{1} = \mathbf{0},\quad\Rightarrow\quad
\mathrm{R}(\mathrm{CFL}, 0)\mathbf{1} = \mathbf{1},
\end{equation}
which rigorously proves the constant-preserving property of the ADER-DG method with LST-DG predictor.

It is clear that the Von Neumann stability criterion (spectral criterion) consists of the requirement on the Courant number $\mathrm{CFL}$ that the spectral radius $\rho[\mathrm{R}(\mathrm{CFL}, \theta)]$ of the matrix $\mathrm{R}(\mathrm{CFL}, \theta)$ (\ref{eq:r_matrix_in_matrix_form}) of the evolution operator $R$ for one time step $\Dtn{n}$ be no greater than one for any value of phase $\theta$:
\begin{equation}\label{eq:cfl_max_base_def}
\mathrm{CFL}_{\rm max} = \sup\Big\{\mathrm{CFL}\,\Big|\, \forall \theta \in [0, 2\pi] \subset \mathcal{R},\, \rho[\mathrm{R}(\mathrm{CFL}, \theta)] \leqslant 1\Big\},
\end{equation}
where the supremum is chosen from considerations of the presence of an upper limit of the time step $\Dtn{n}$ for a fixed coordinate step $\Dx$ ($\mathrm{CFL}_{\rm max} \in [0, 1]$ for explicit methods, which include the ADER-DG numerical method with the LST-DG predictor) and the continuity of the set of values of the time step $\Dtn{n}$.

\begin{figure}[h!]
\centering
\includegraphics[width=0.028125\textwidth]{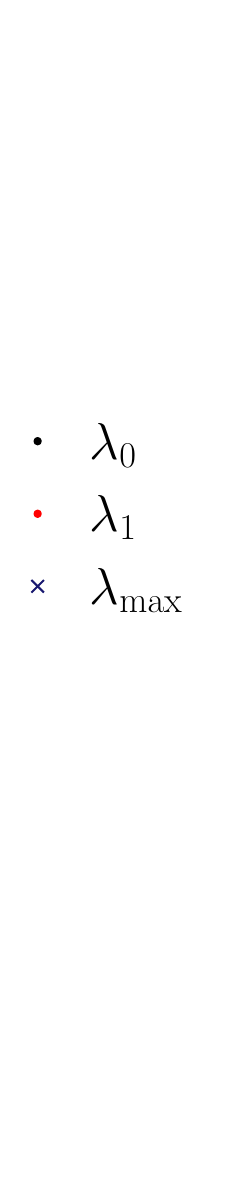}
\includegraphics[width=0.15\textwidth]{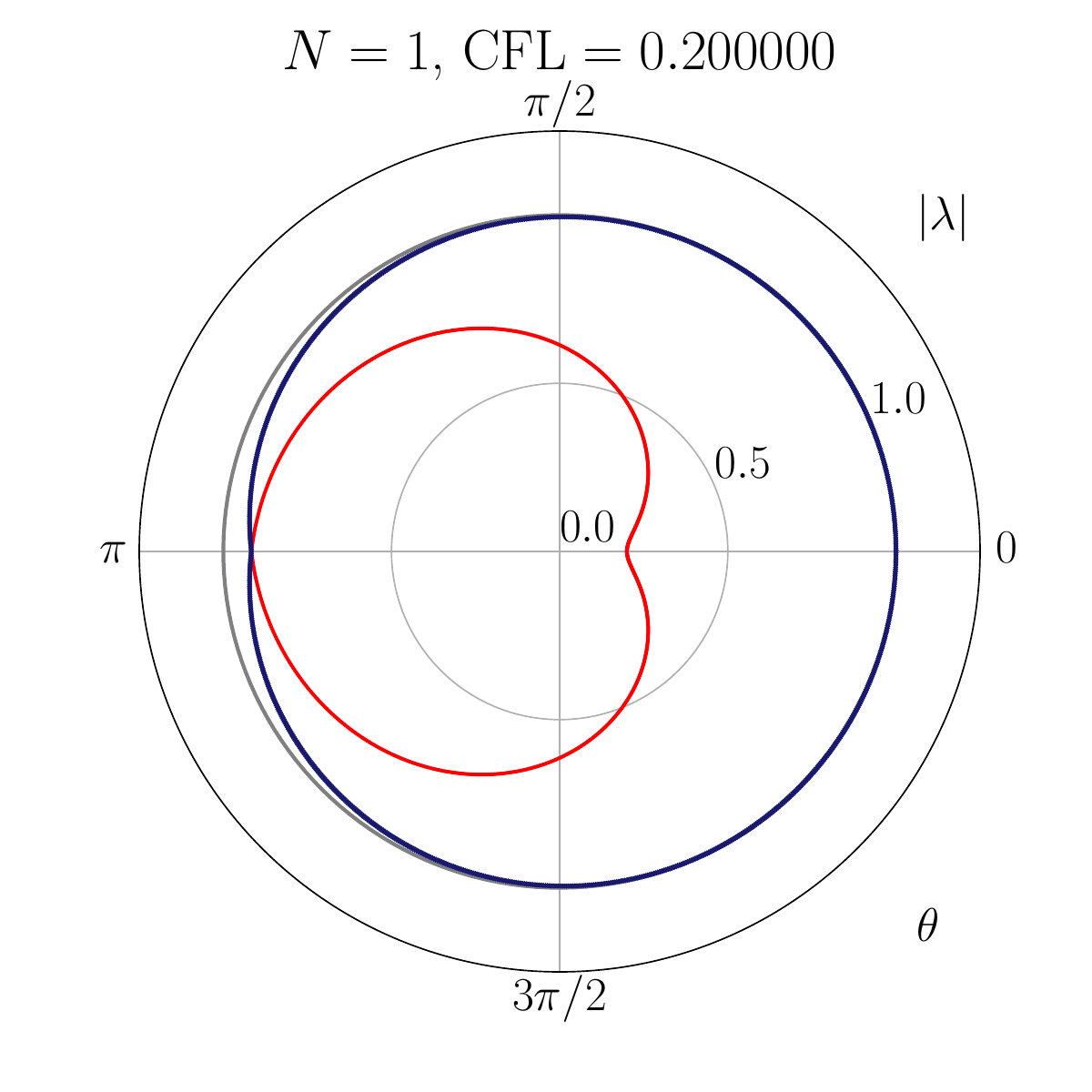}
\includegraphics[width=0.15\textwidth]{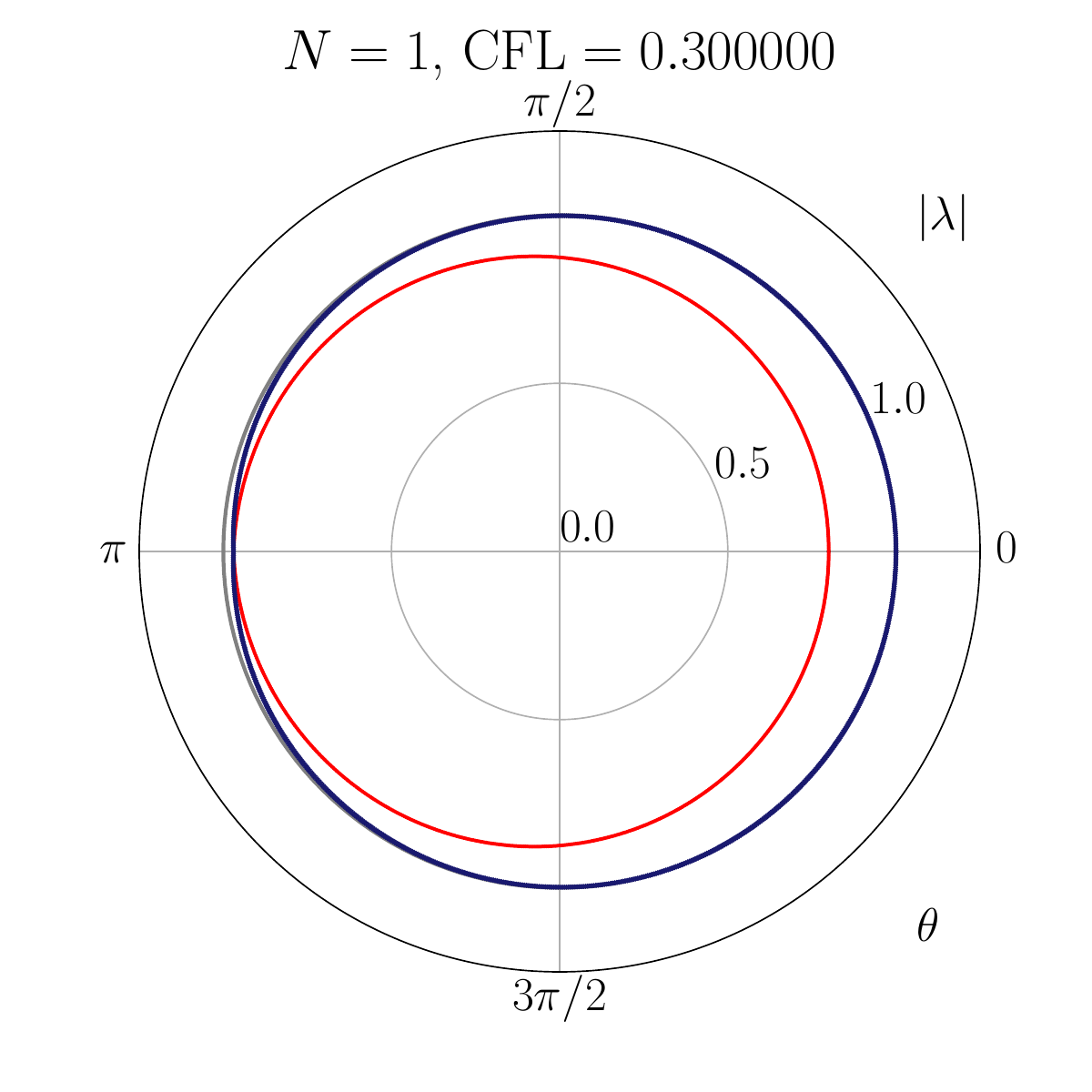}
\includegraphics[width=0.15\textwidth]{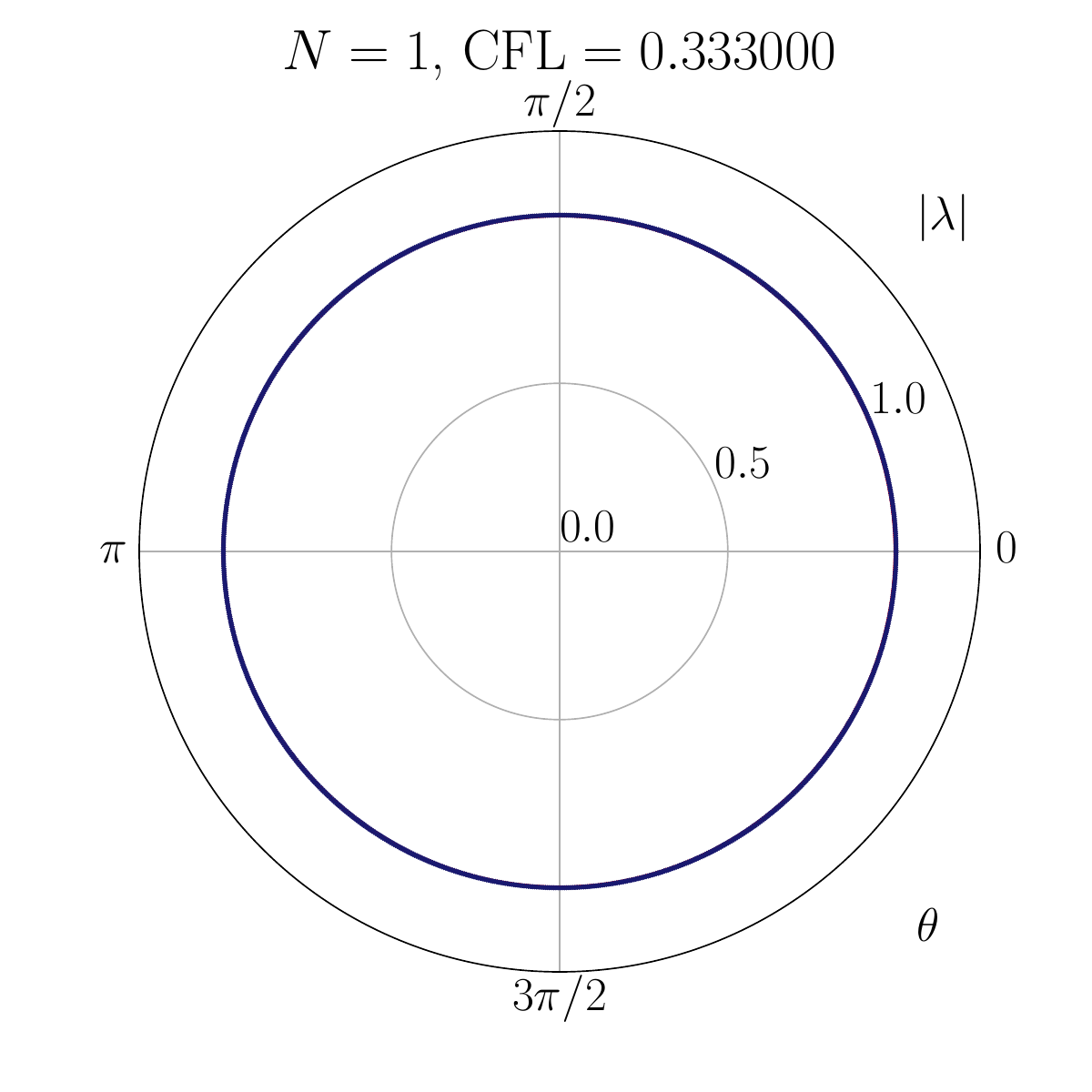}
\includegraphics[width=0.15\textwidth]{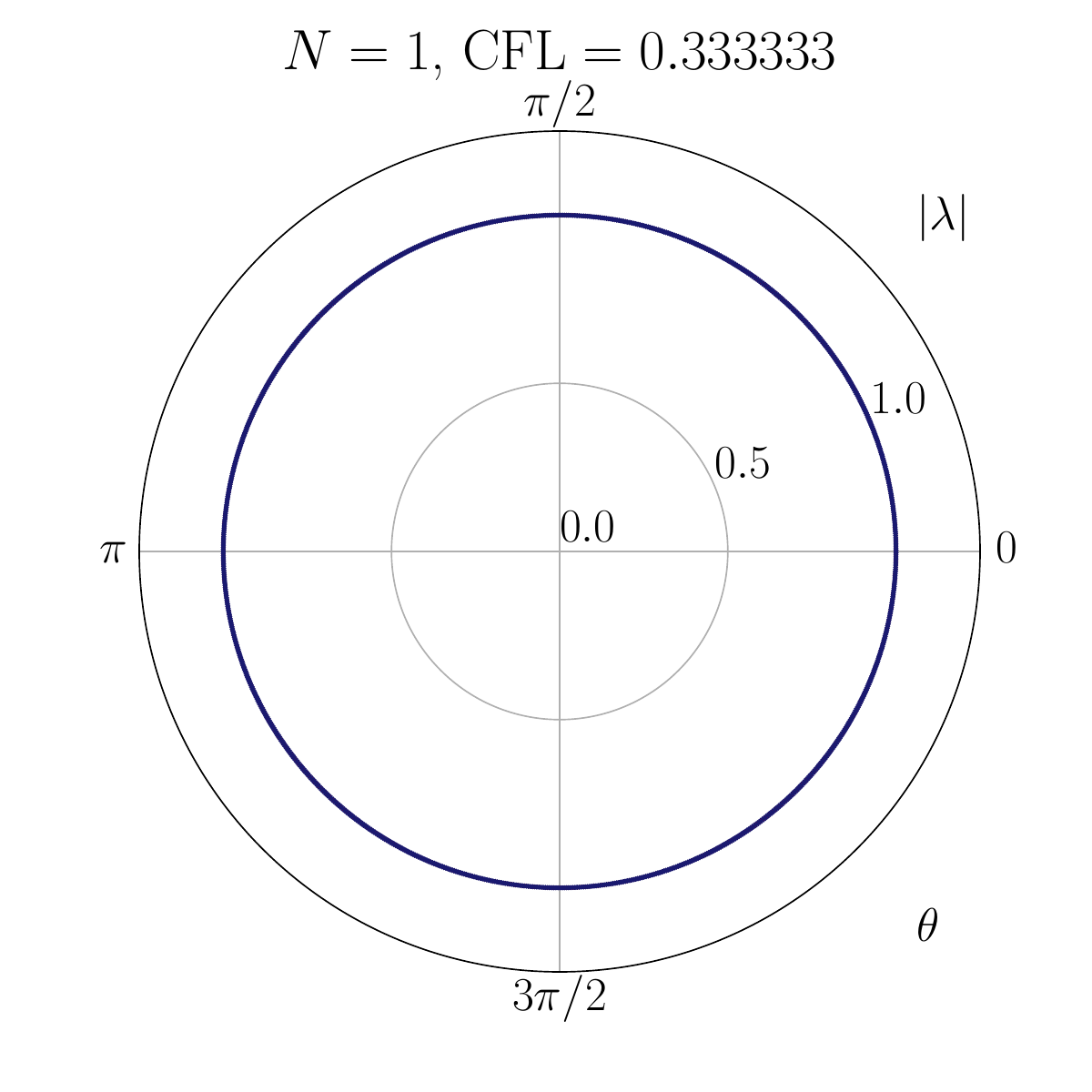}
\includegraphics[width=0.15\textwidth]{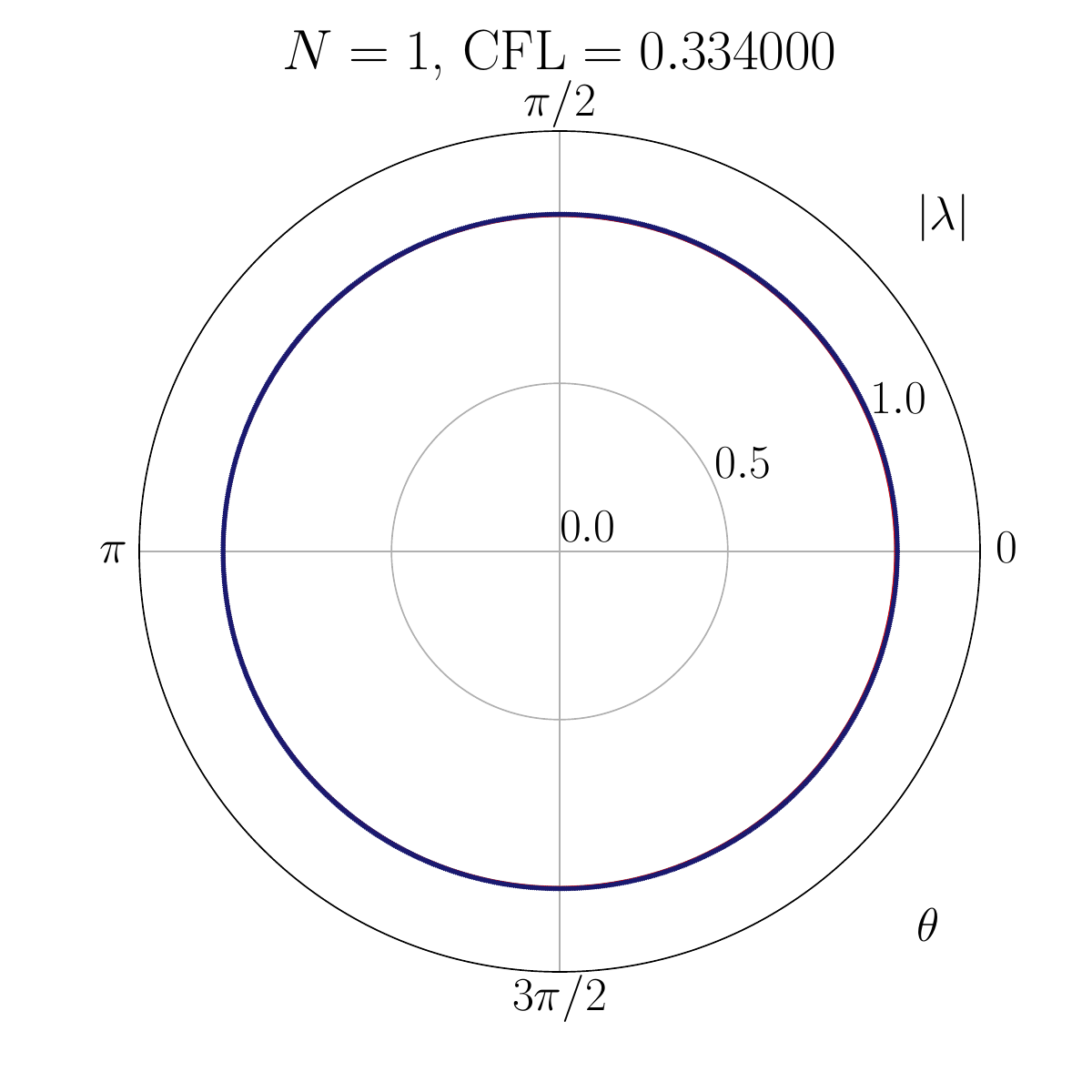}
\includegraphics[width=0.15\textwidth]{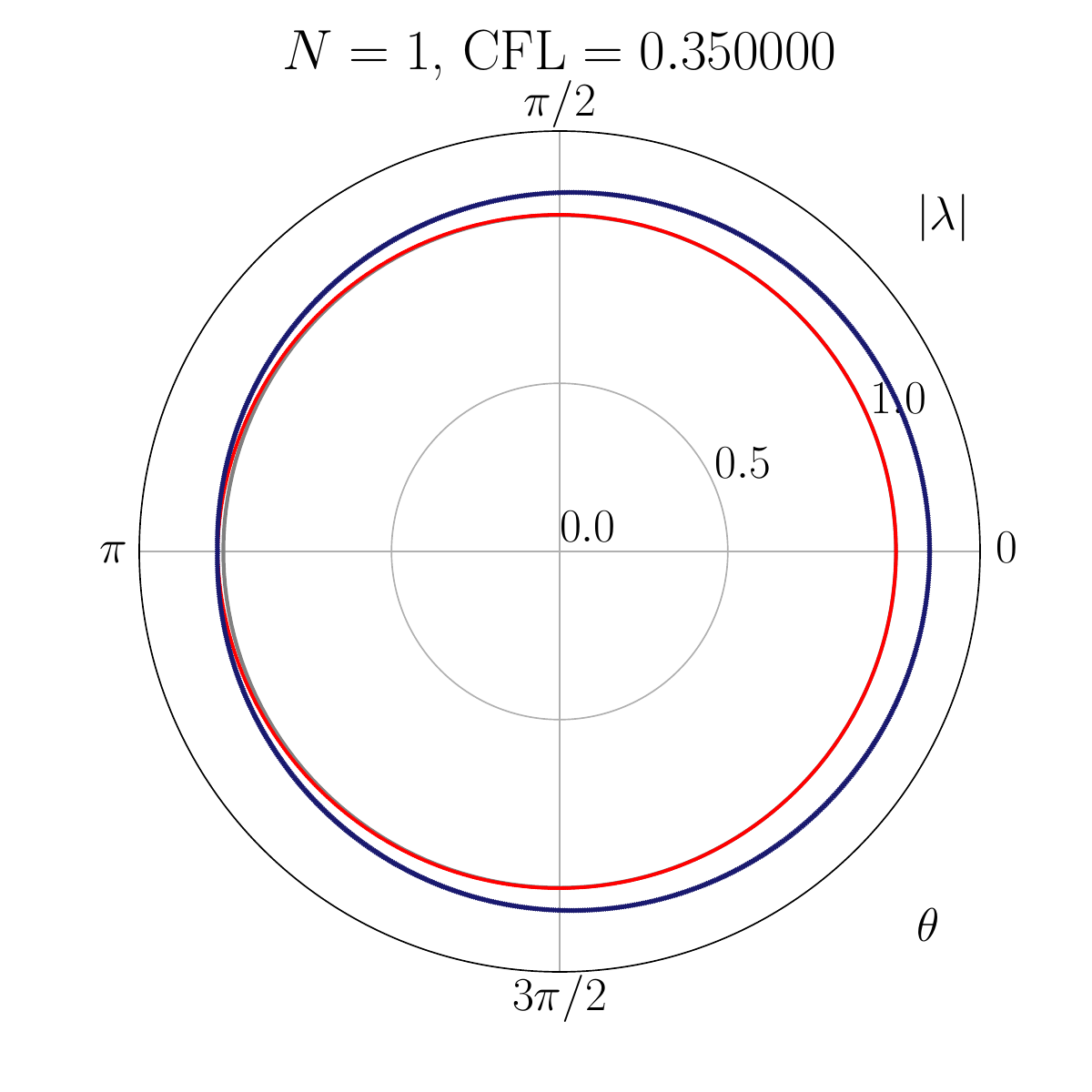}\\
\includegraphics[width=0.028125\textwidth]{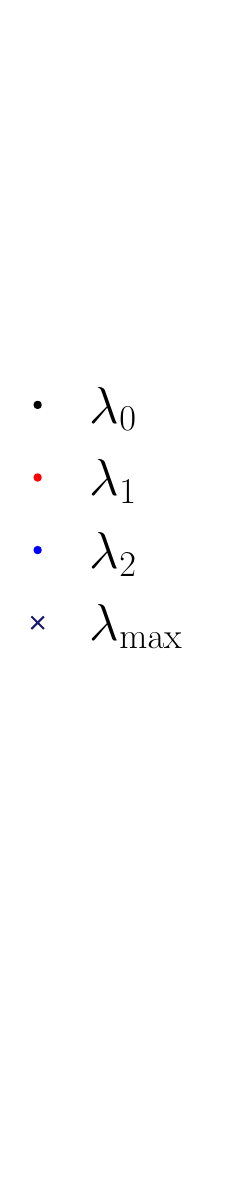}
\includegraphics[width=0.15\textwidth]{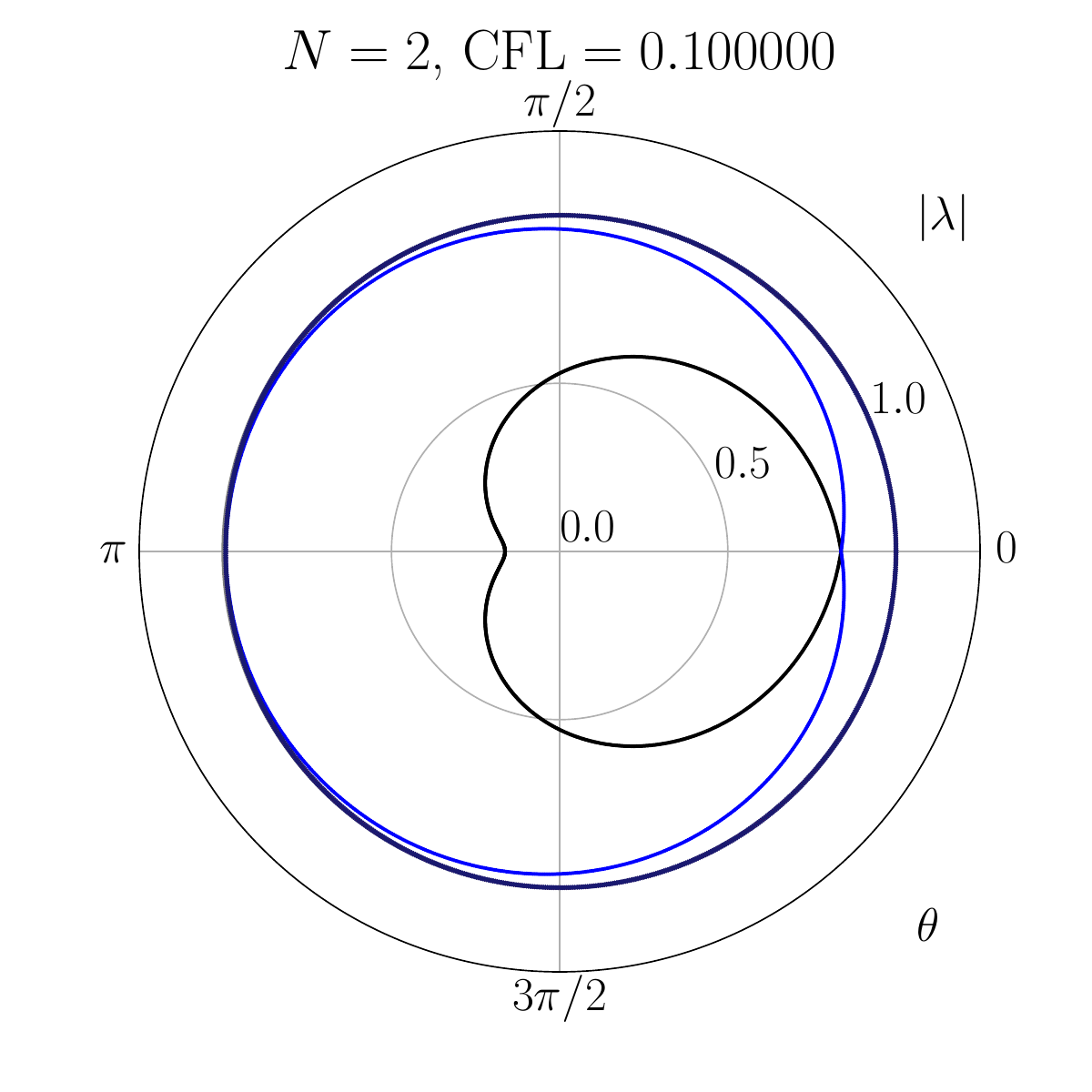}
\includegraphics[width=0.15\textwidth]{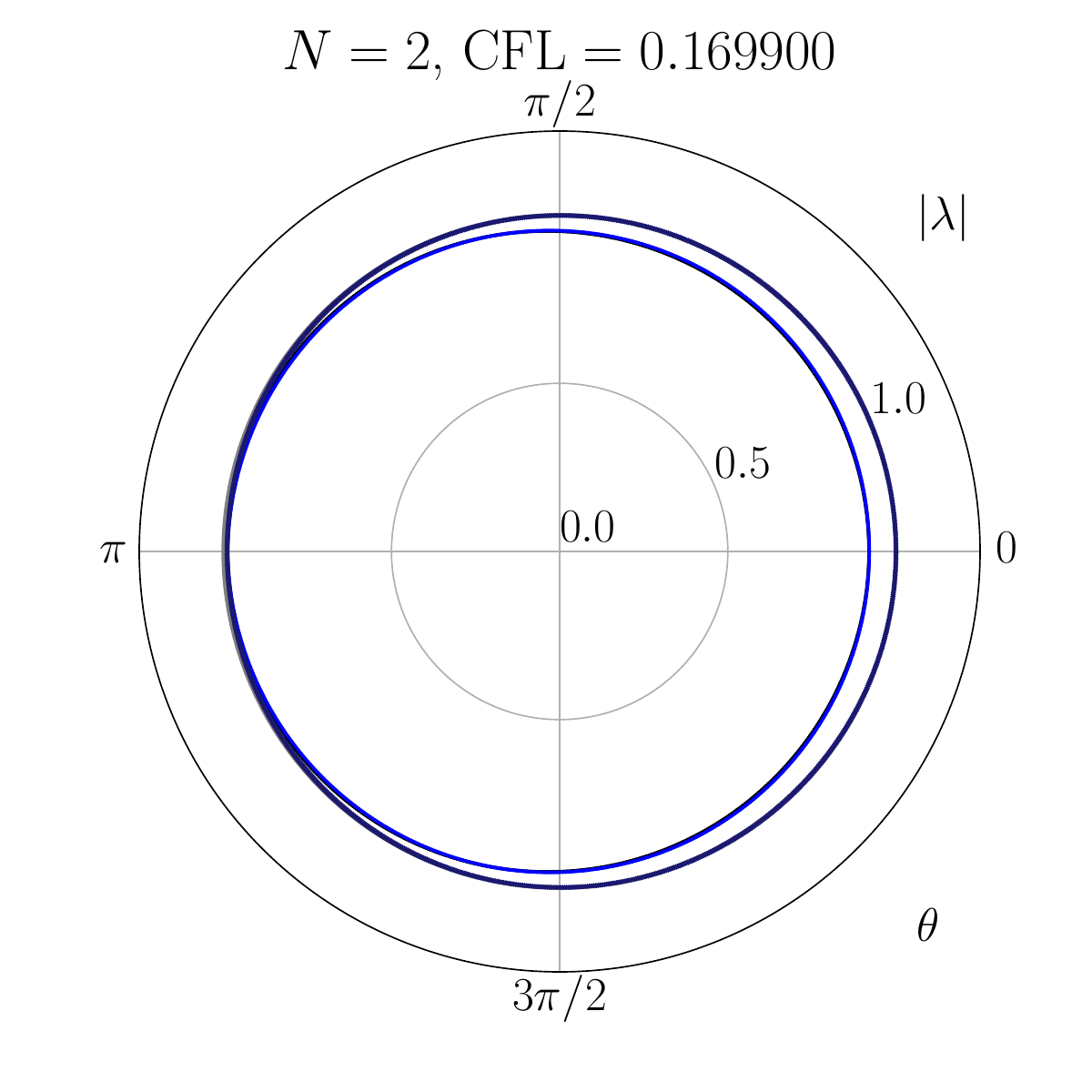}
\includegraphics[width=0.15\textwidth]{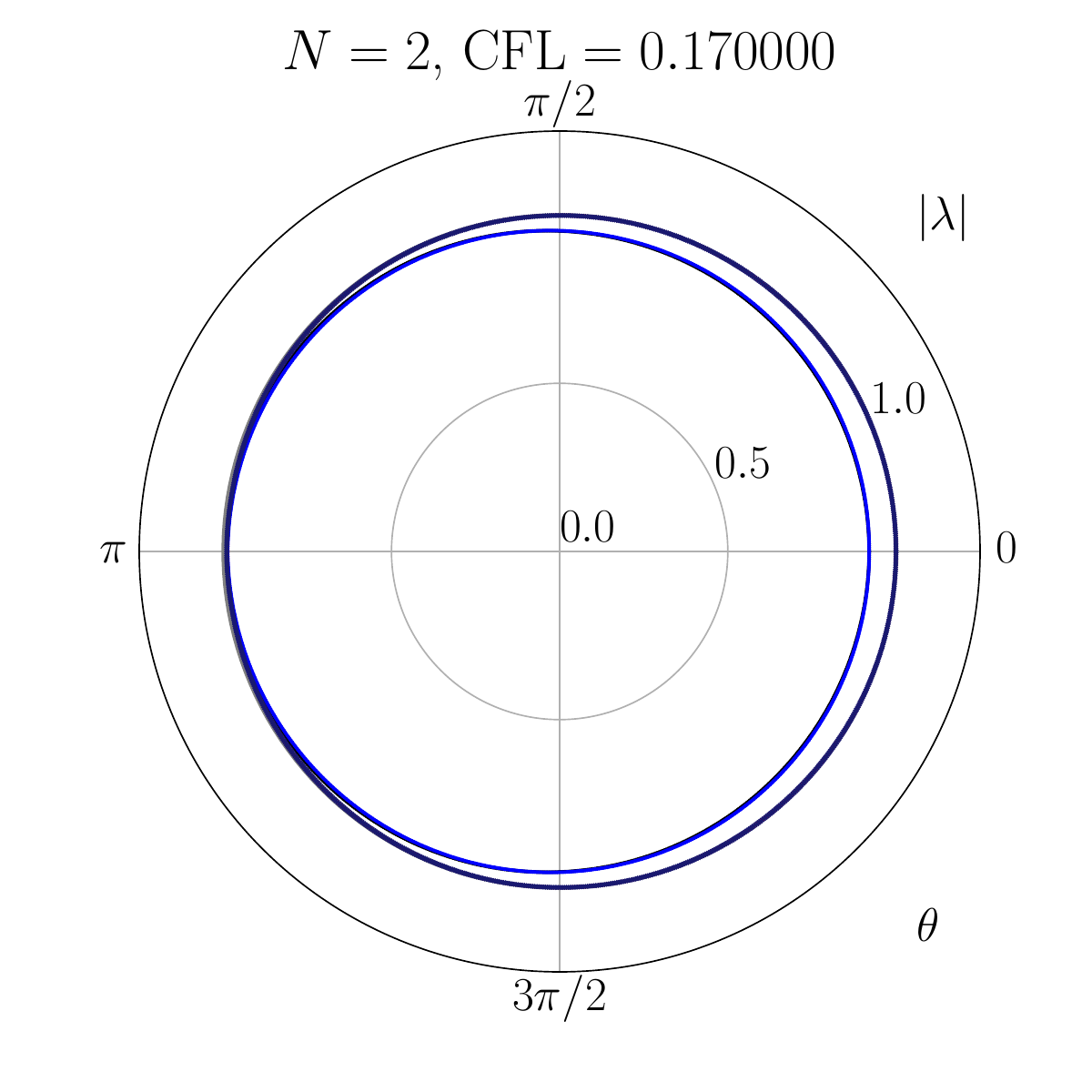}
\includegraphics[width=0.15\textwidth]{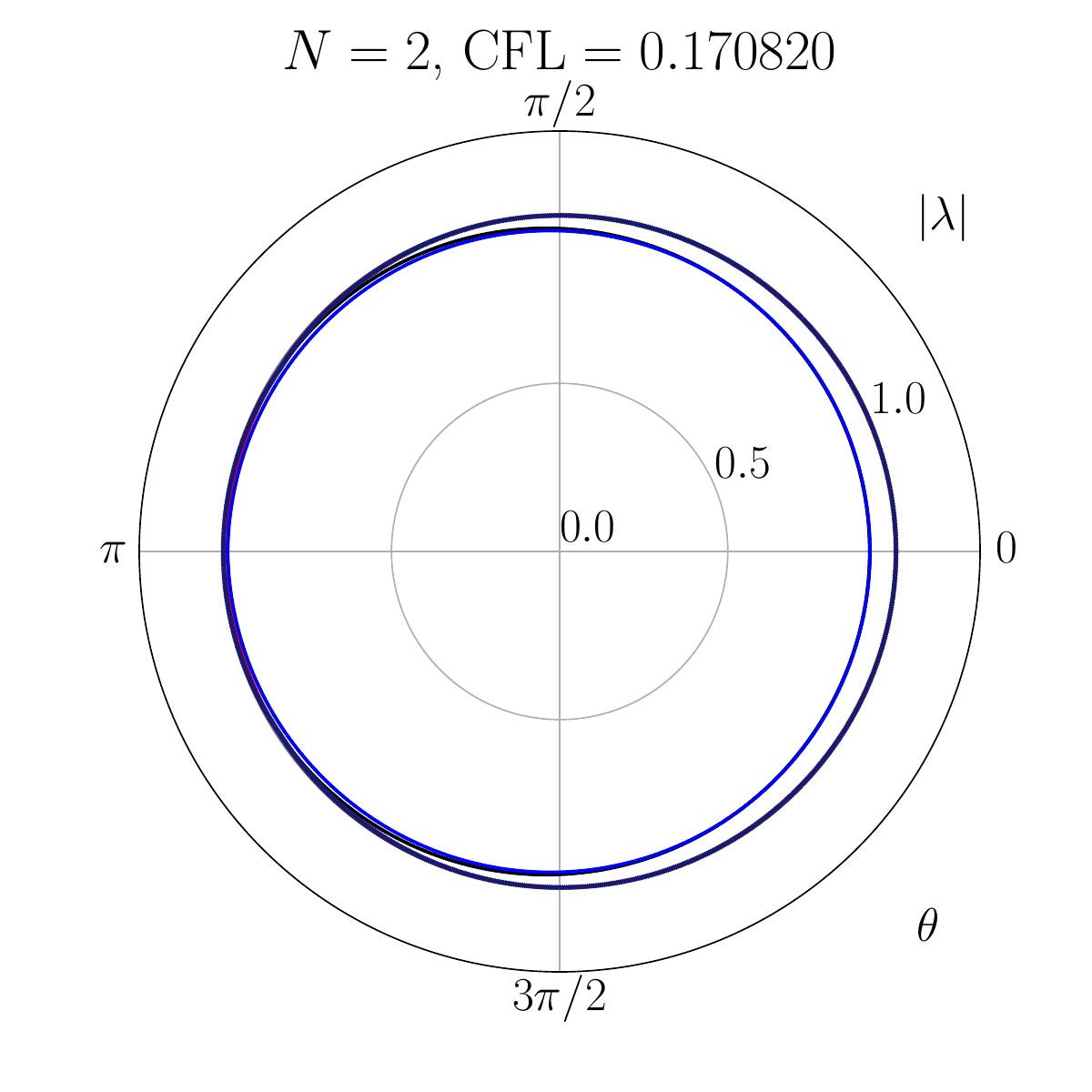}
\includegraphics[width=0.15\textwidth]{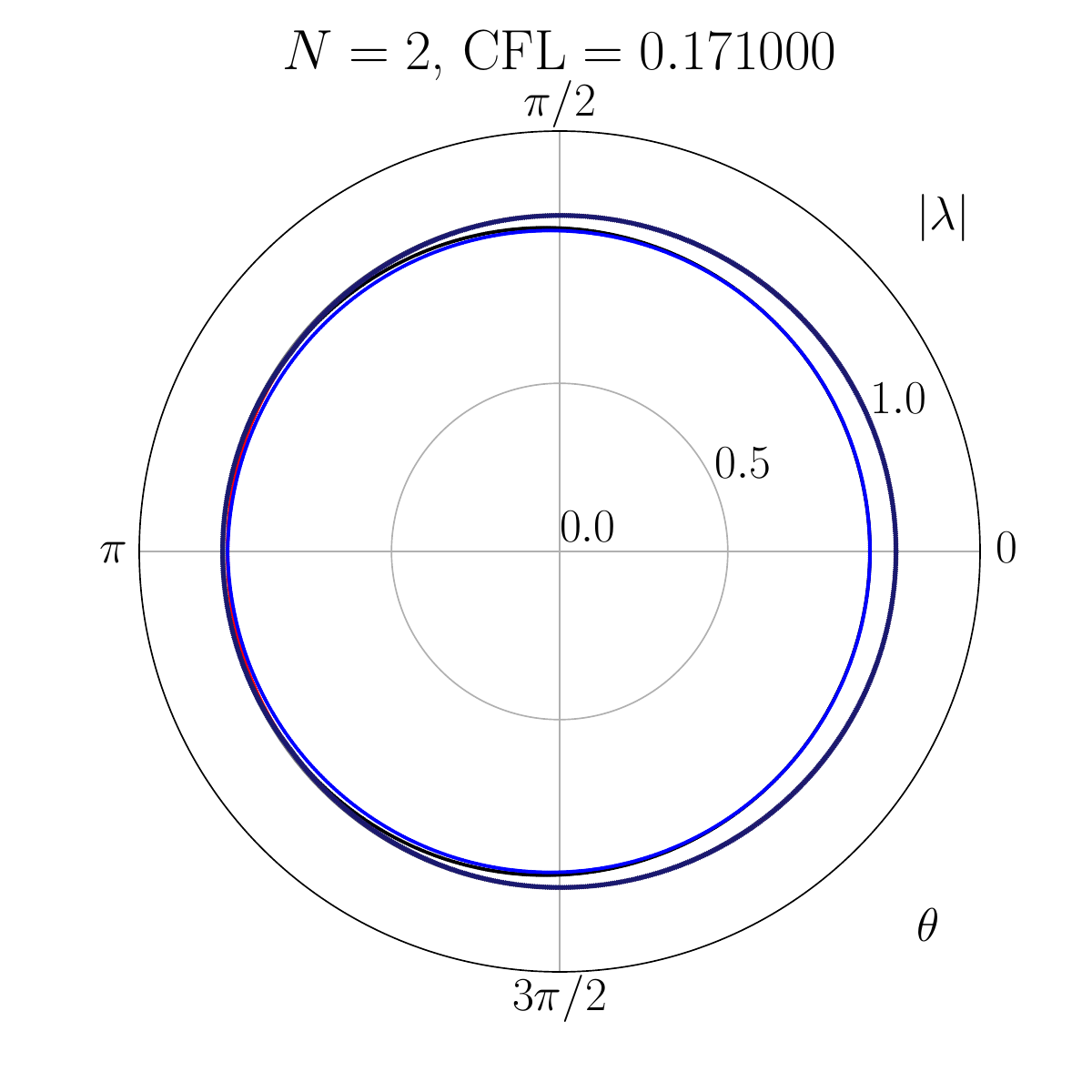}
\includegraphics[width=0.15\textwidth]{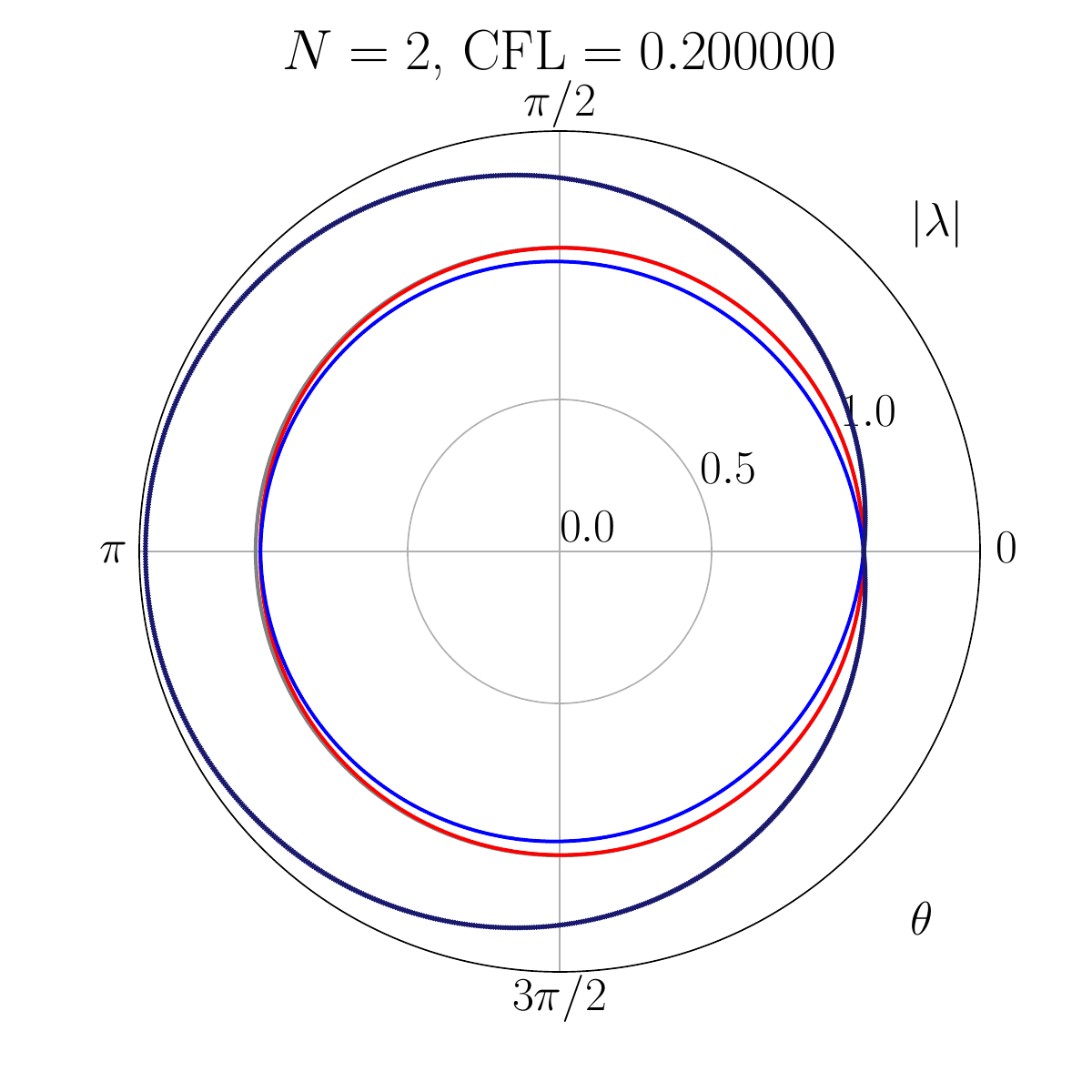}\\
\includegraphics[width=0.028125\textwidth]{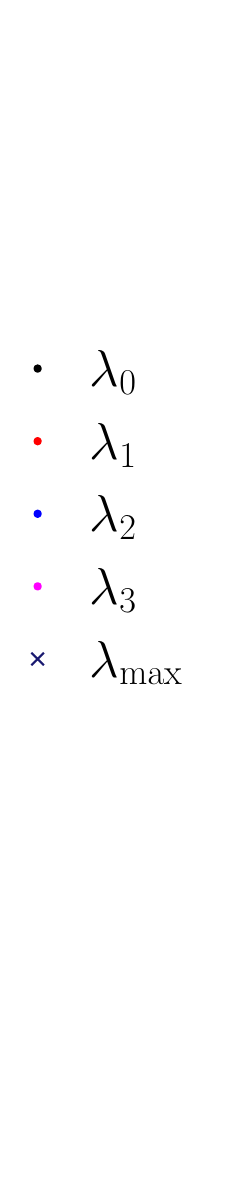}
\includegraphics[width=0.15\textwidth]{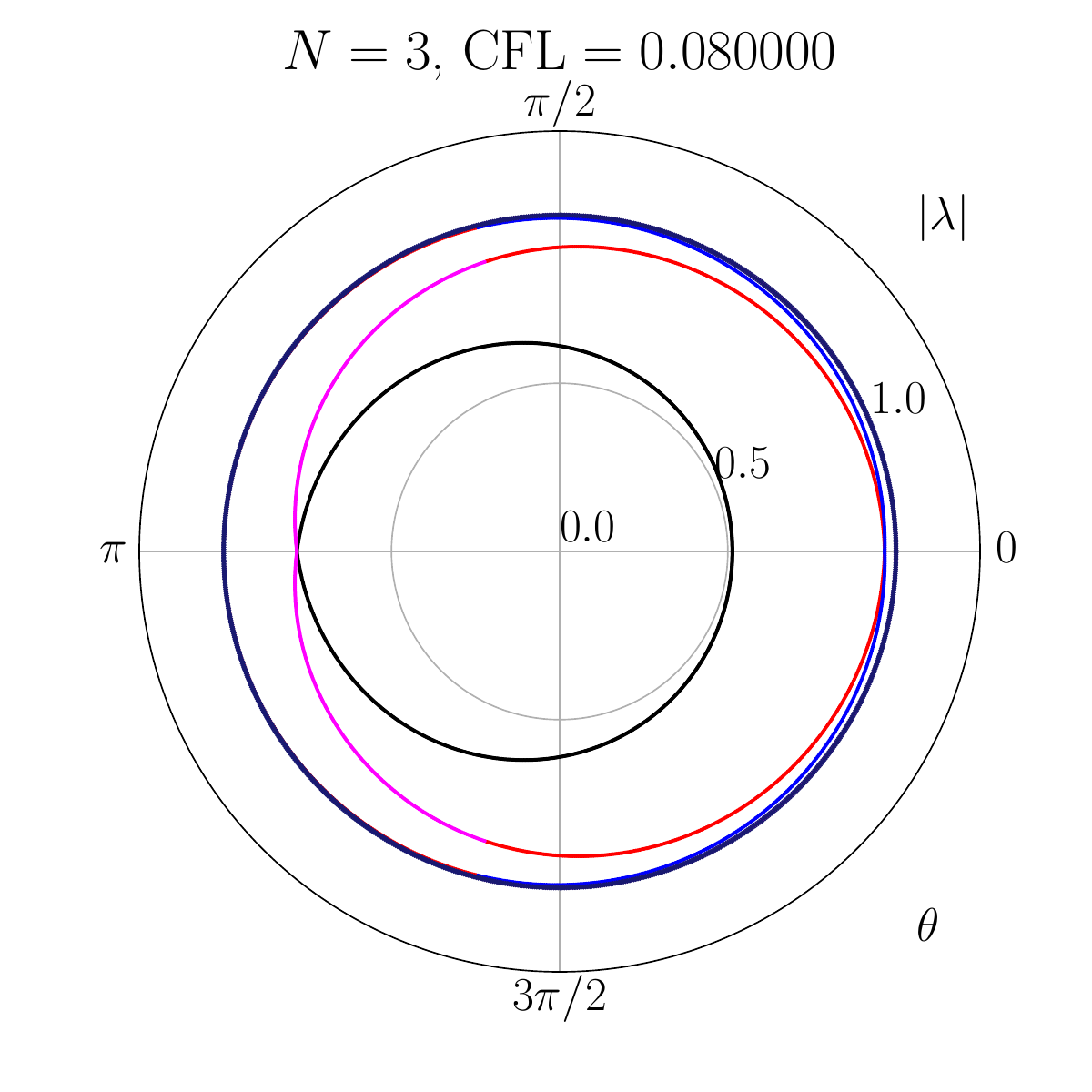}
\includegraphics[width=0.15\textwidth]{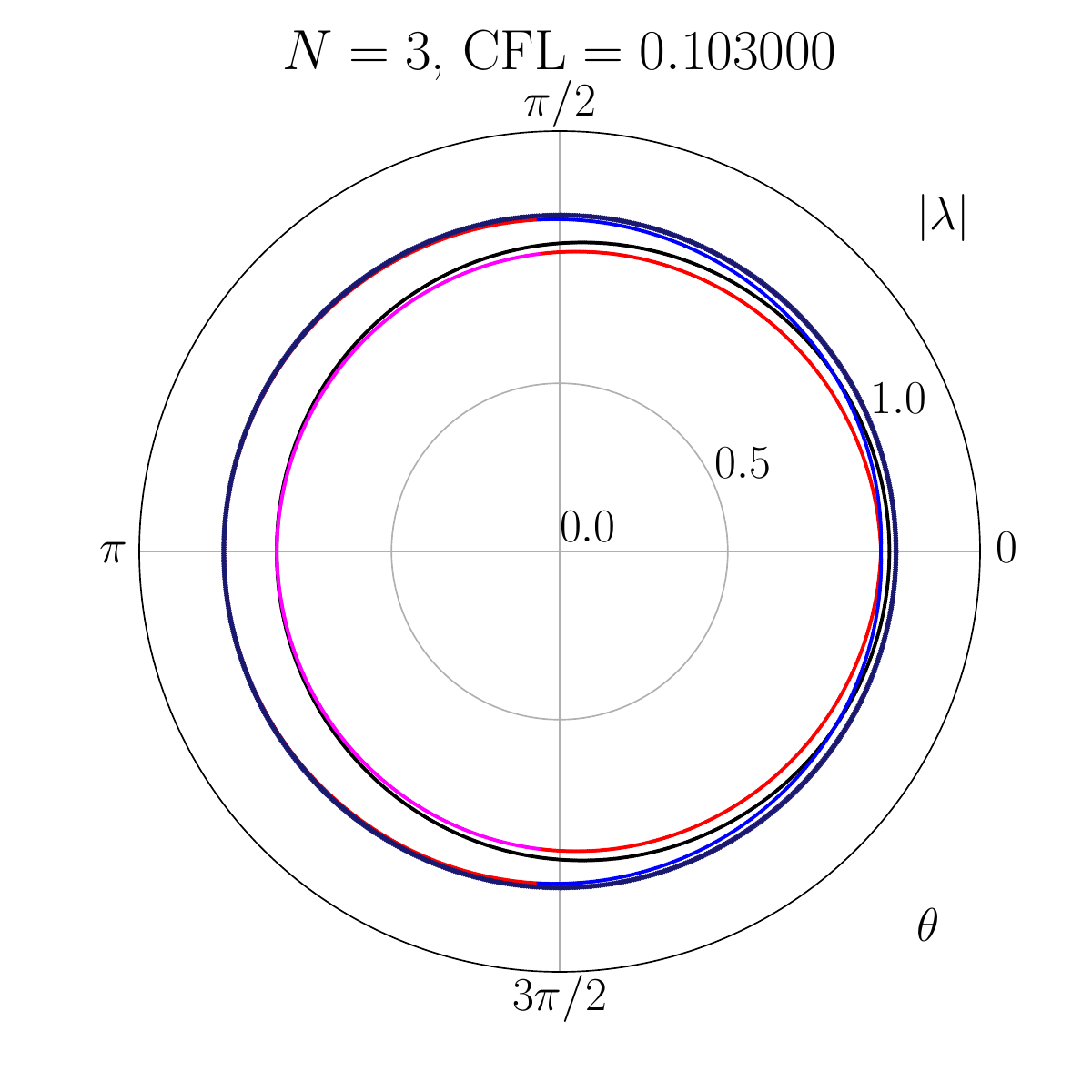}
\includegraphics[width=0.15\textwidth]{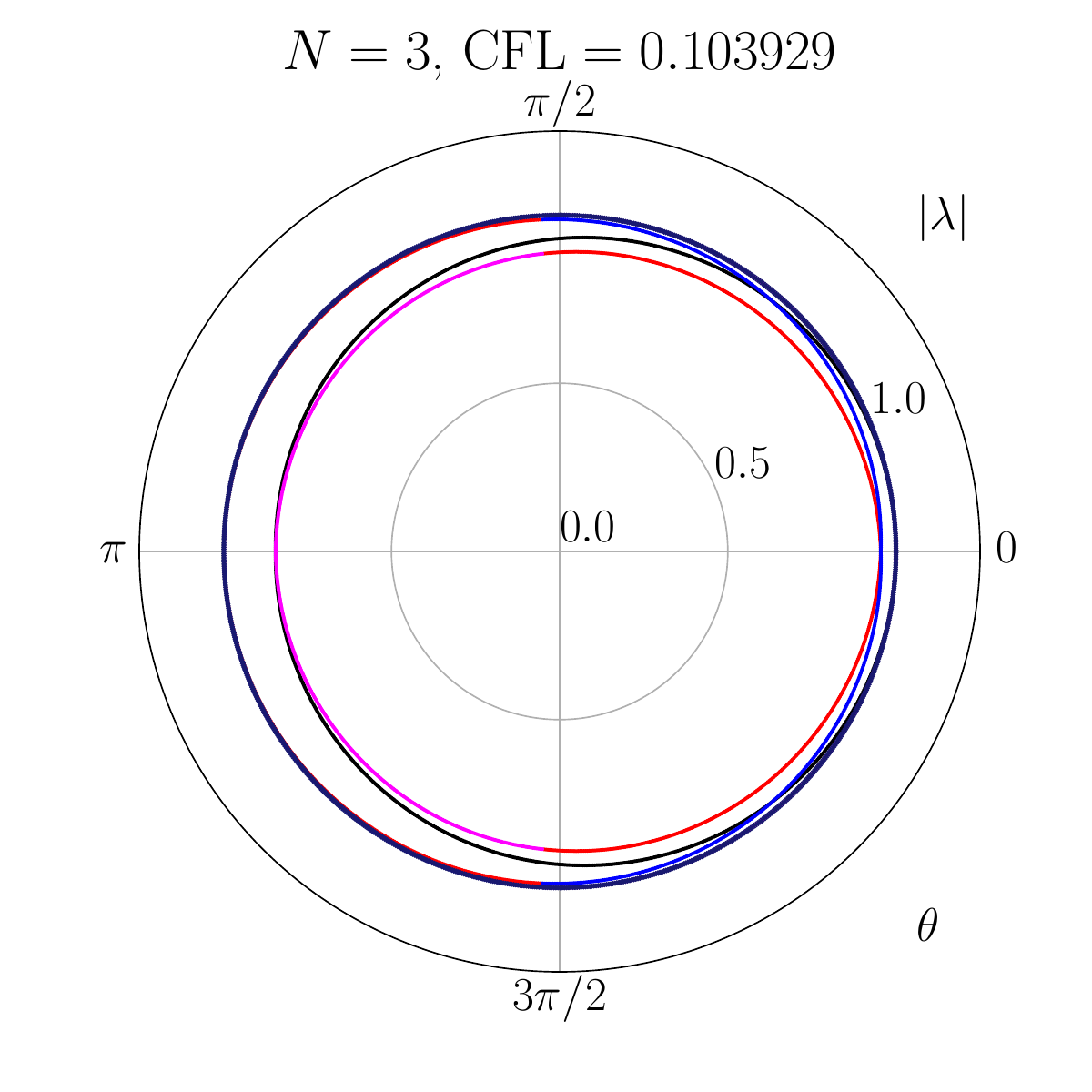}
\includegraphics[width=0.15\textwidth]{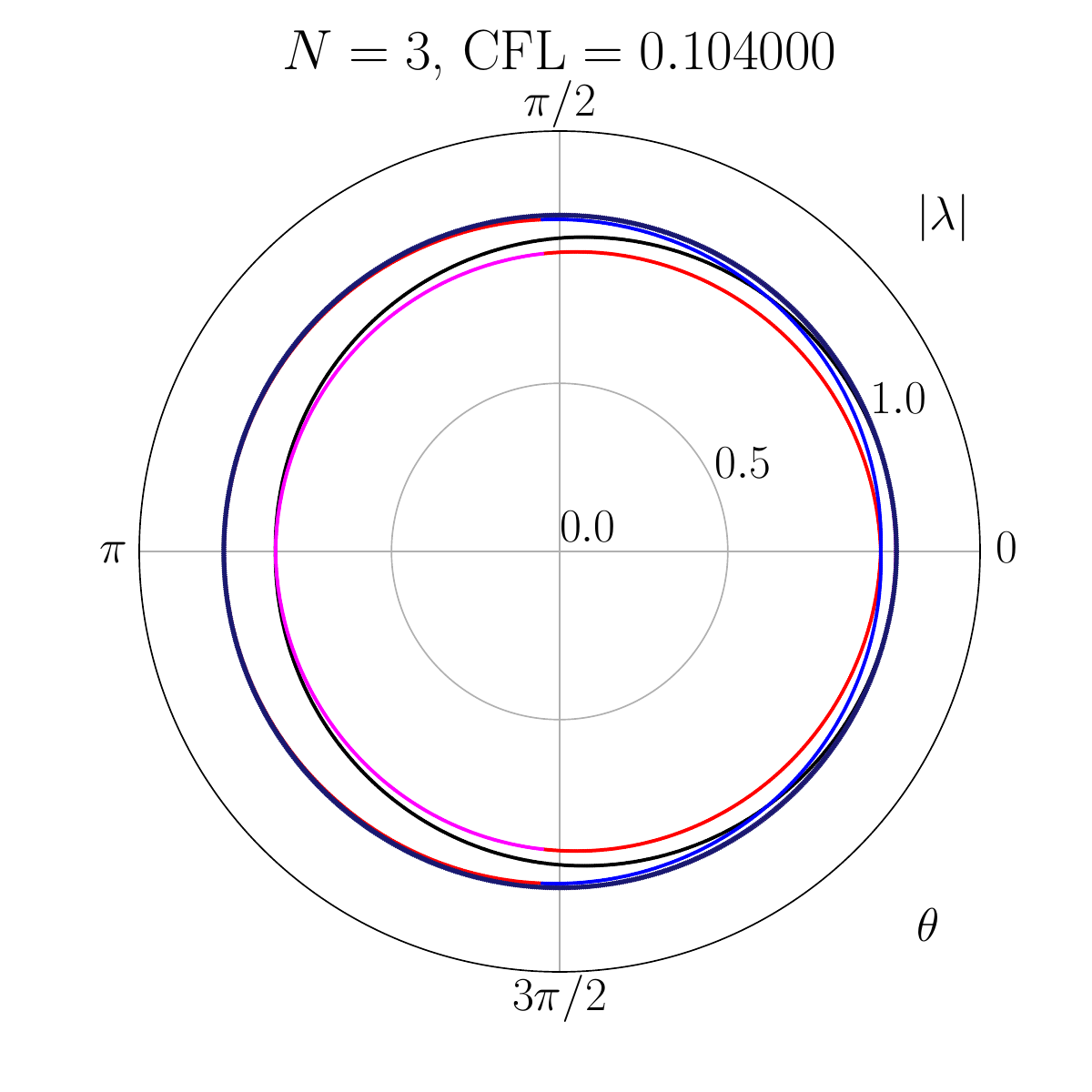}
\includegraphics[width=0.15\textwidth]{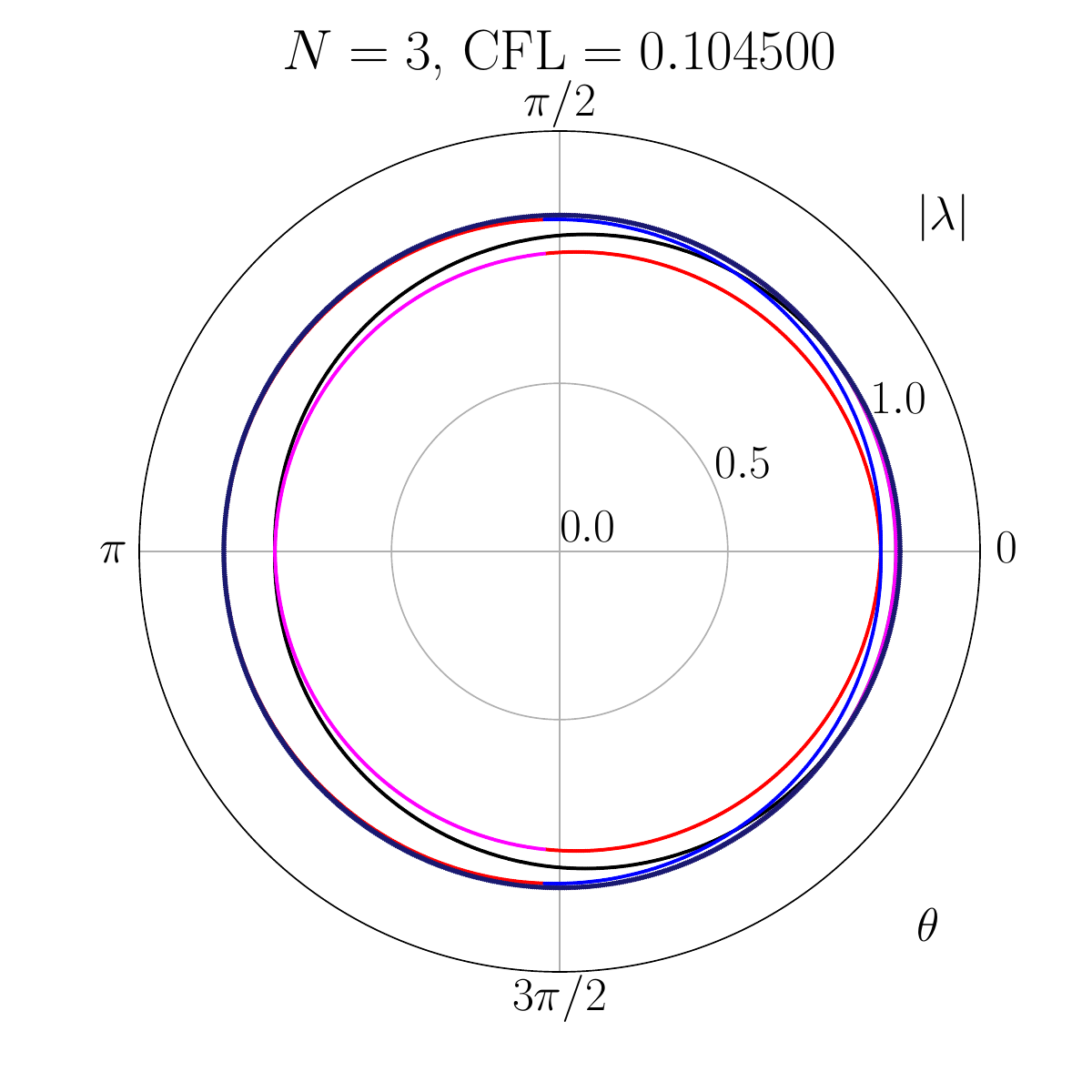}
\includegraphics[width=0.15\textwidth]{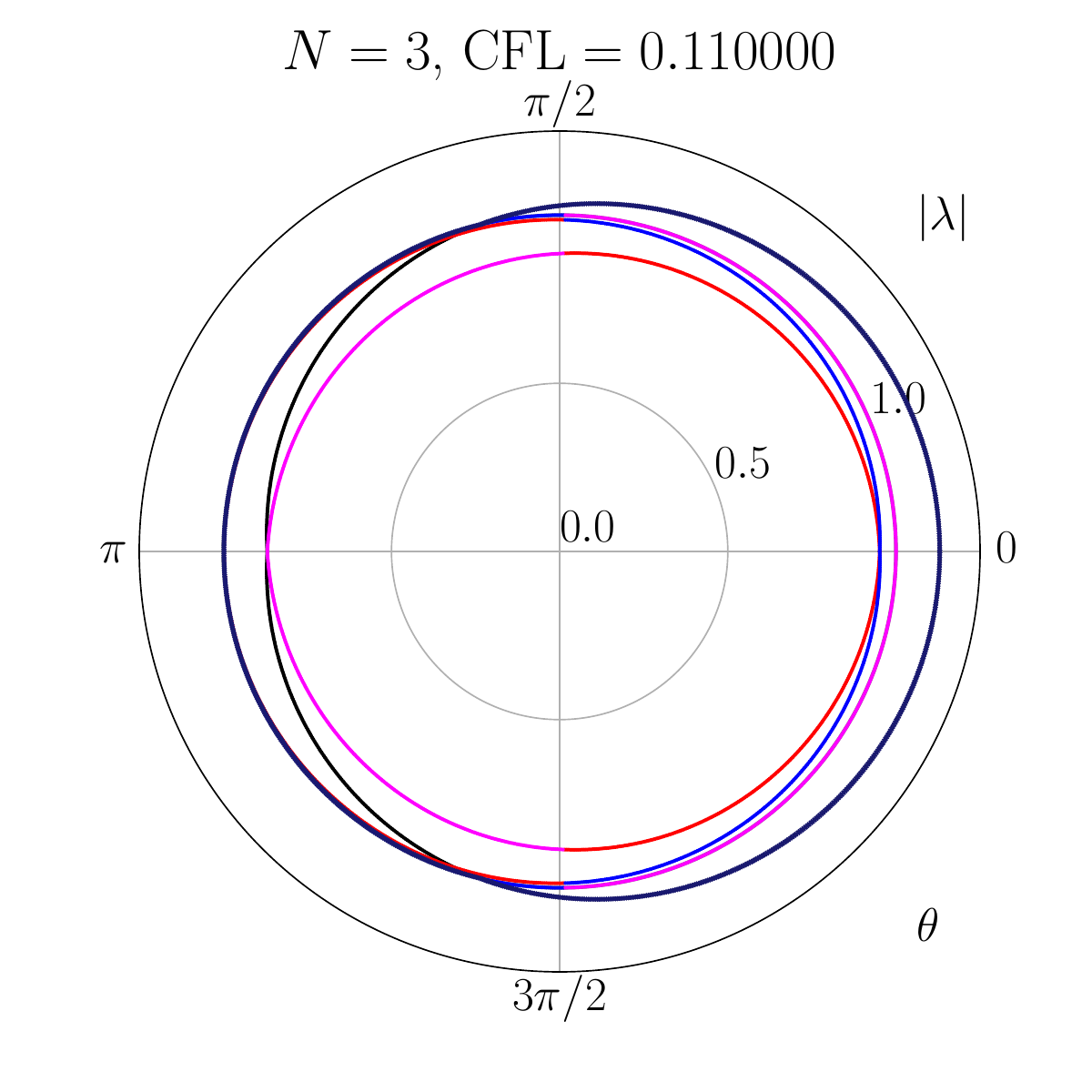}\\
\includegraphics[width=0.028125\textwidth]{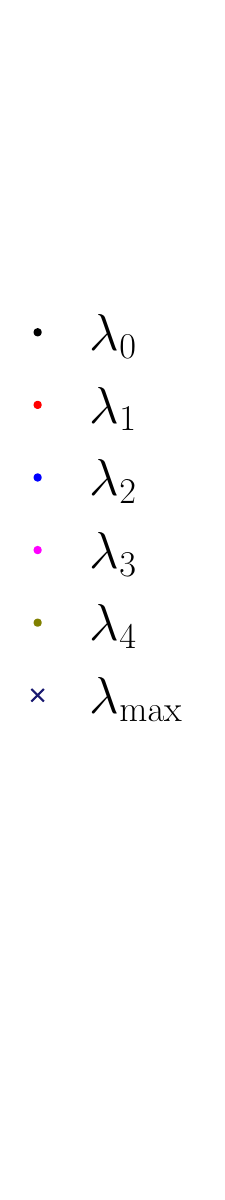}
\includegraphics[width=0.15\textwidth]{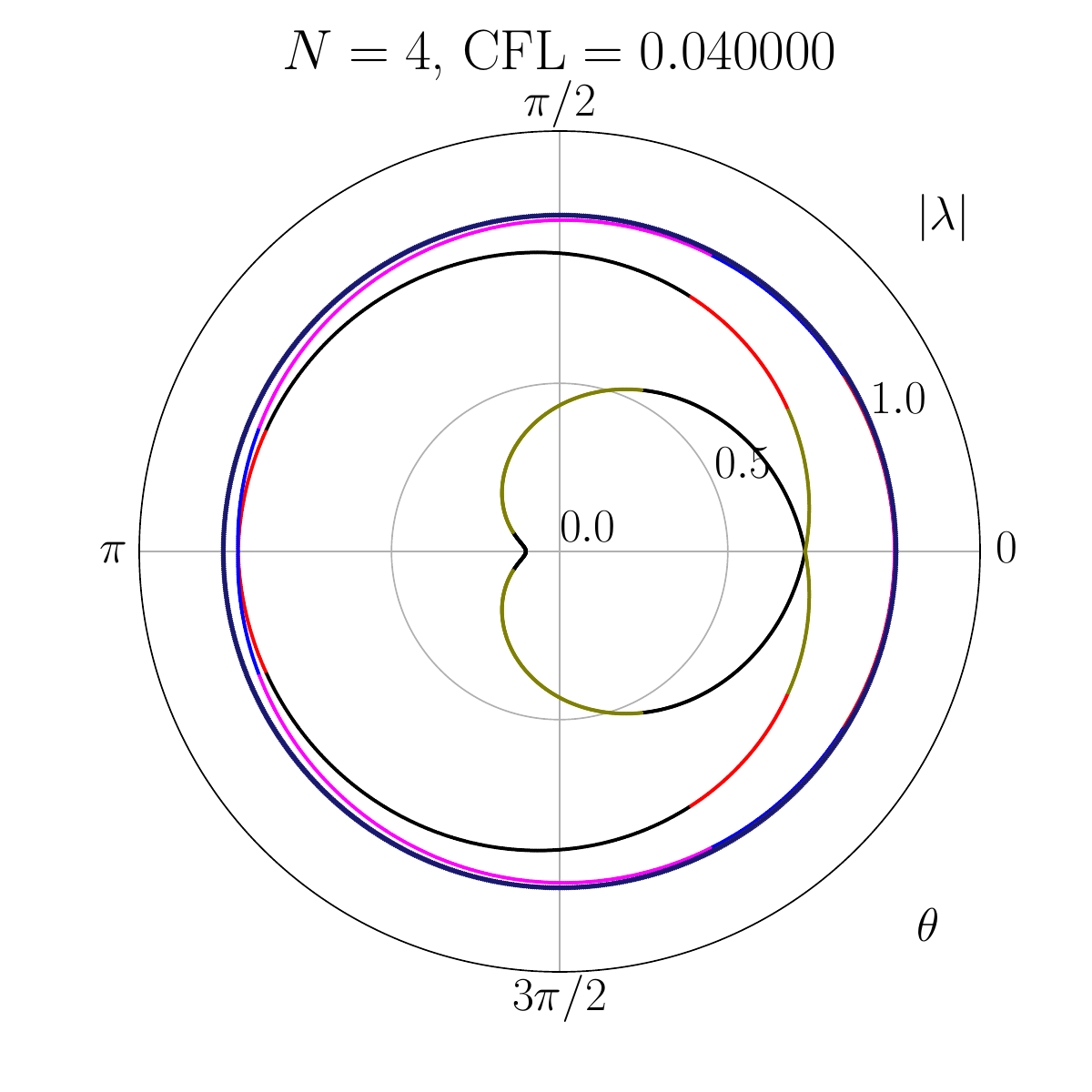}
\includegraphics[width=0.15\textwidth]{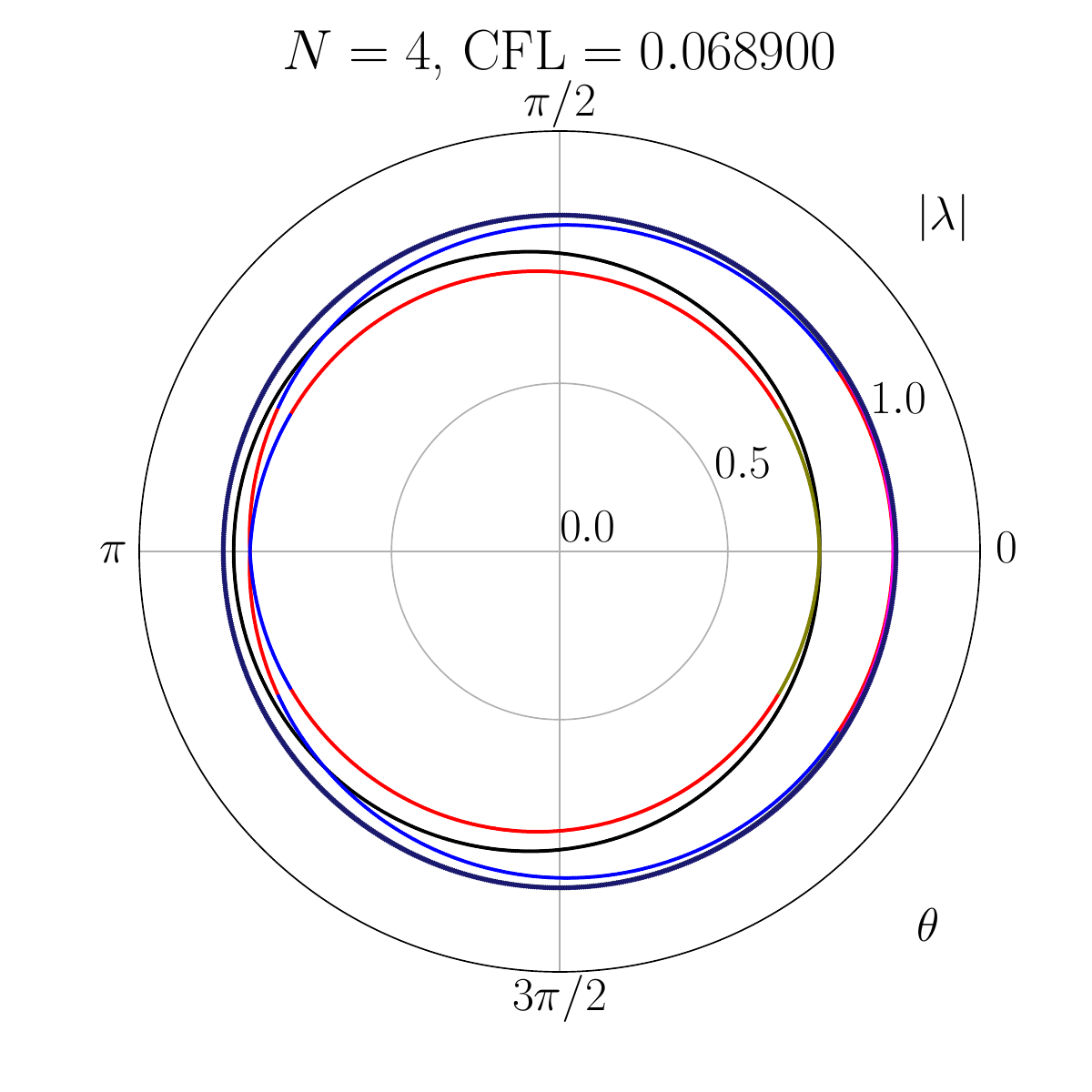}
\includegraphics[width=0.15\textwidth]{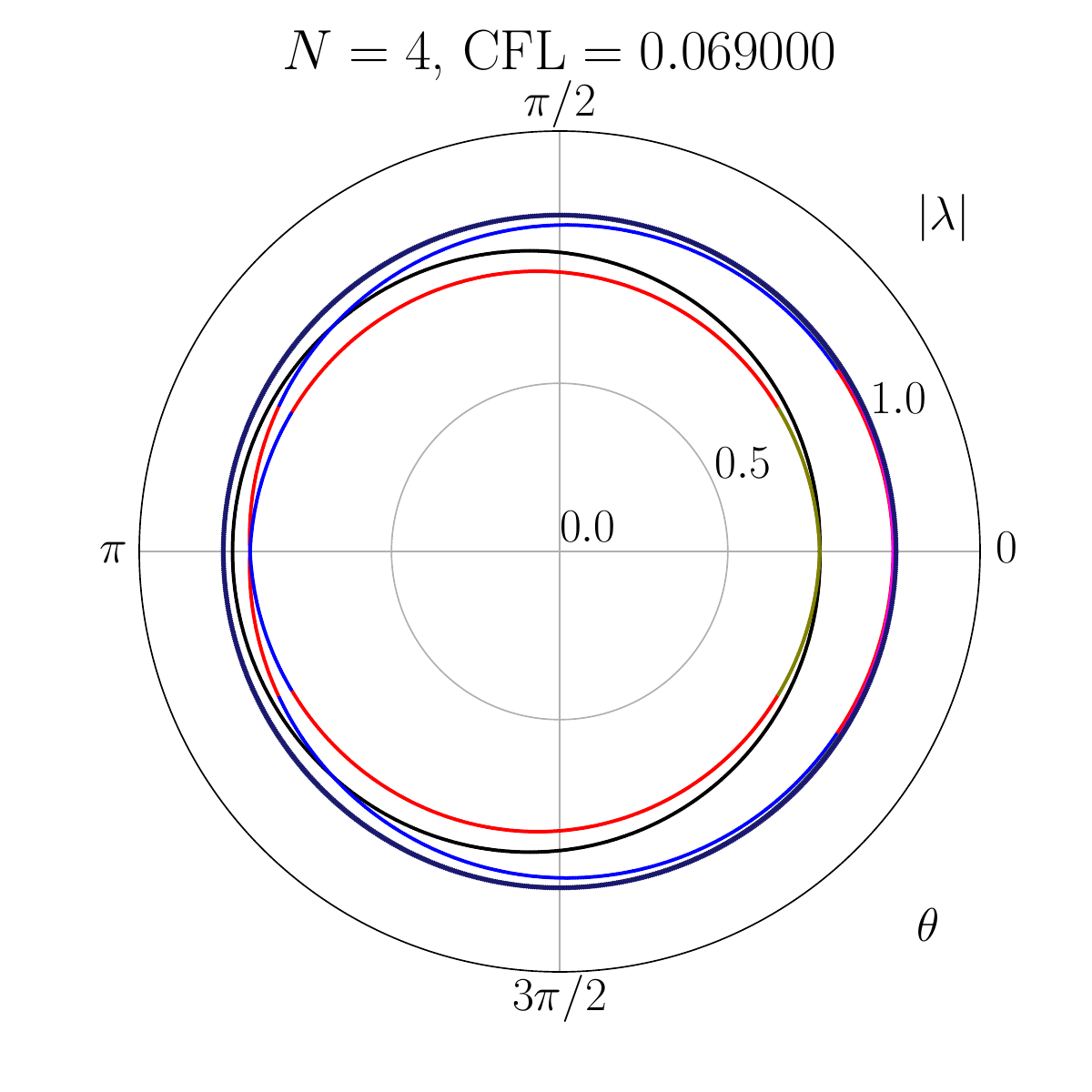}
\includegraphics[width=0.15\textwidth]{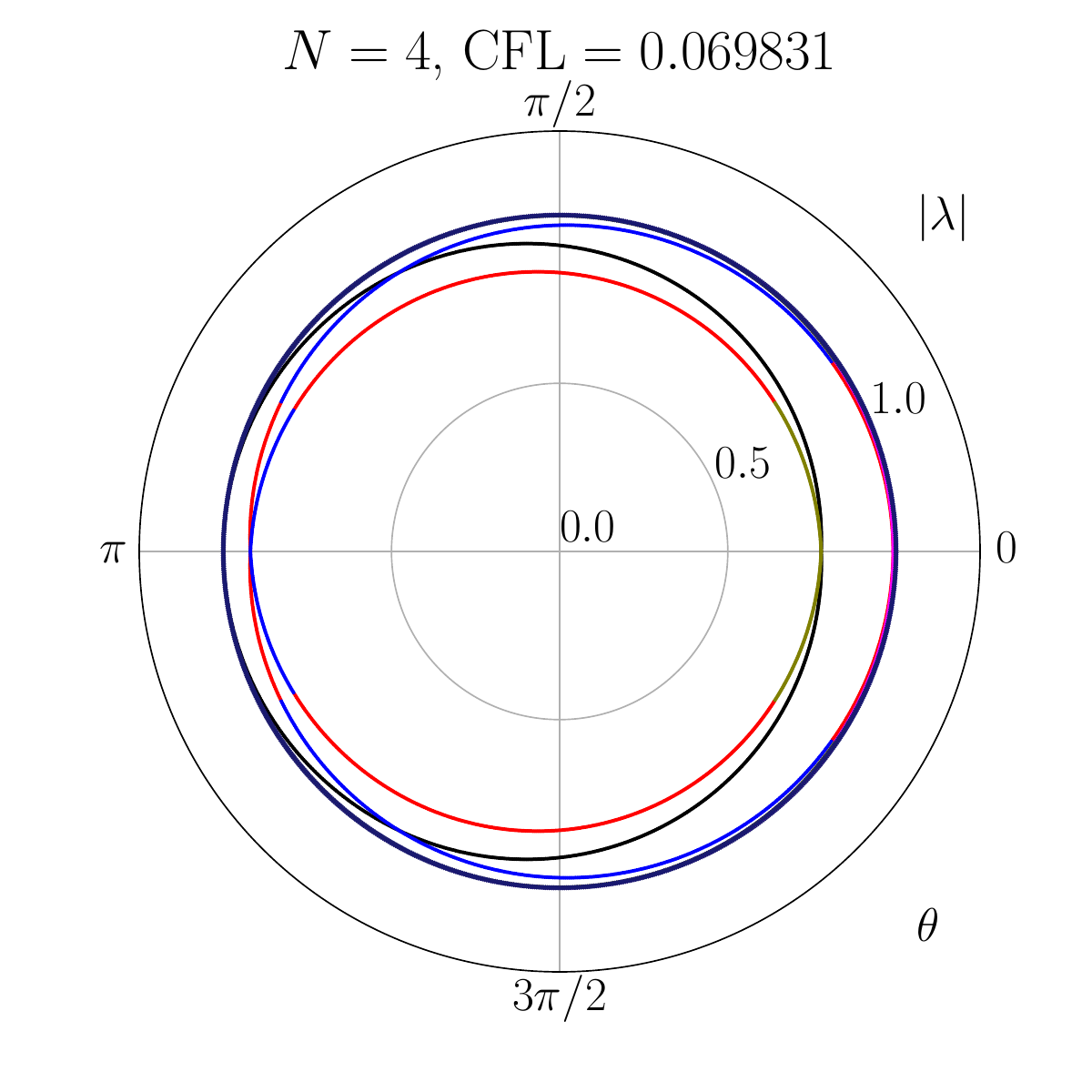}
\includegraphics[width=0.15\textwidth]{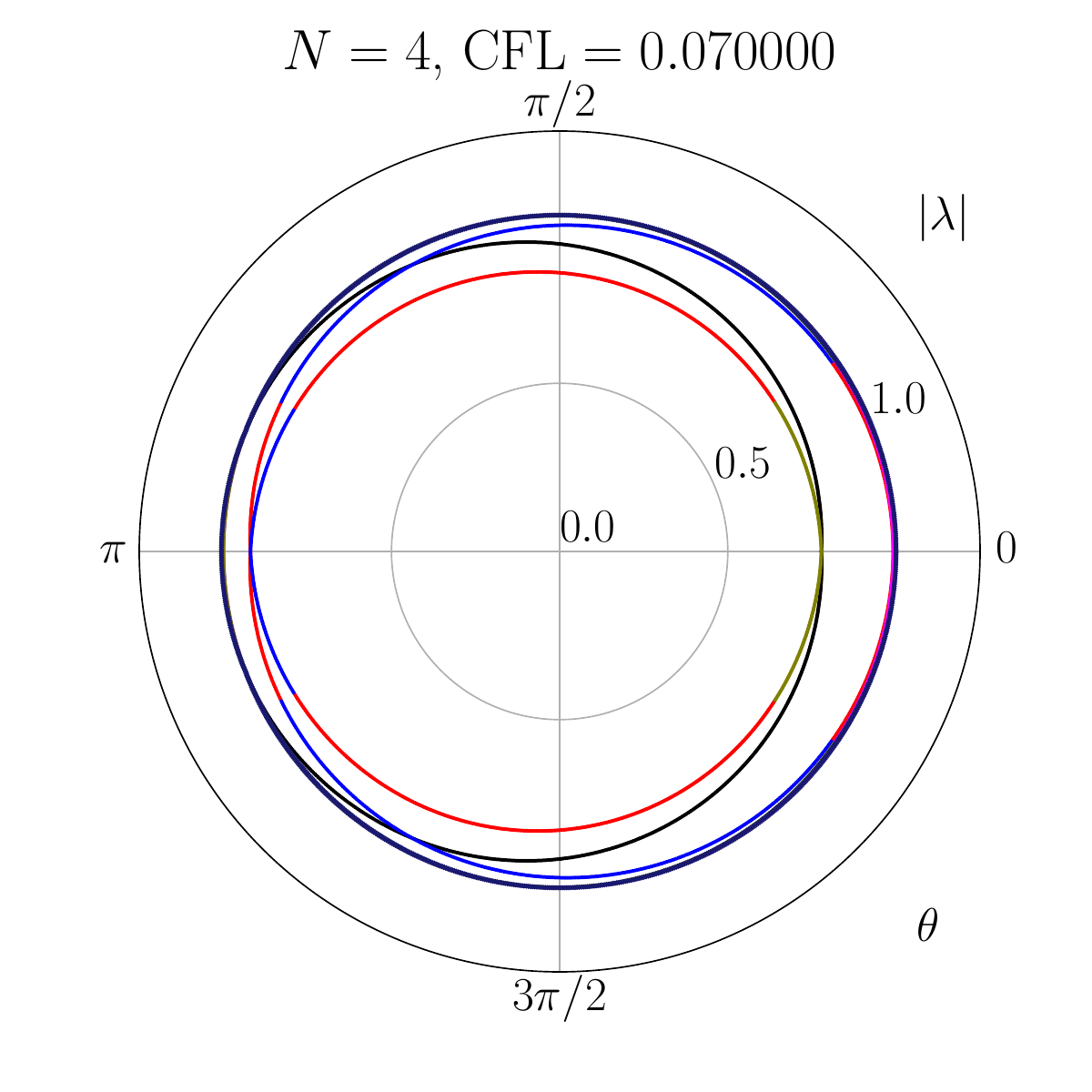}
\includegraphics[width=0.15\textwidth]{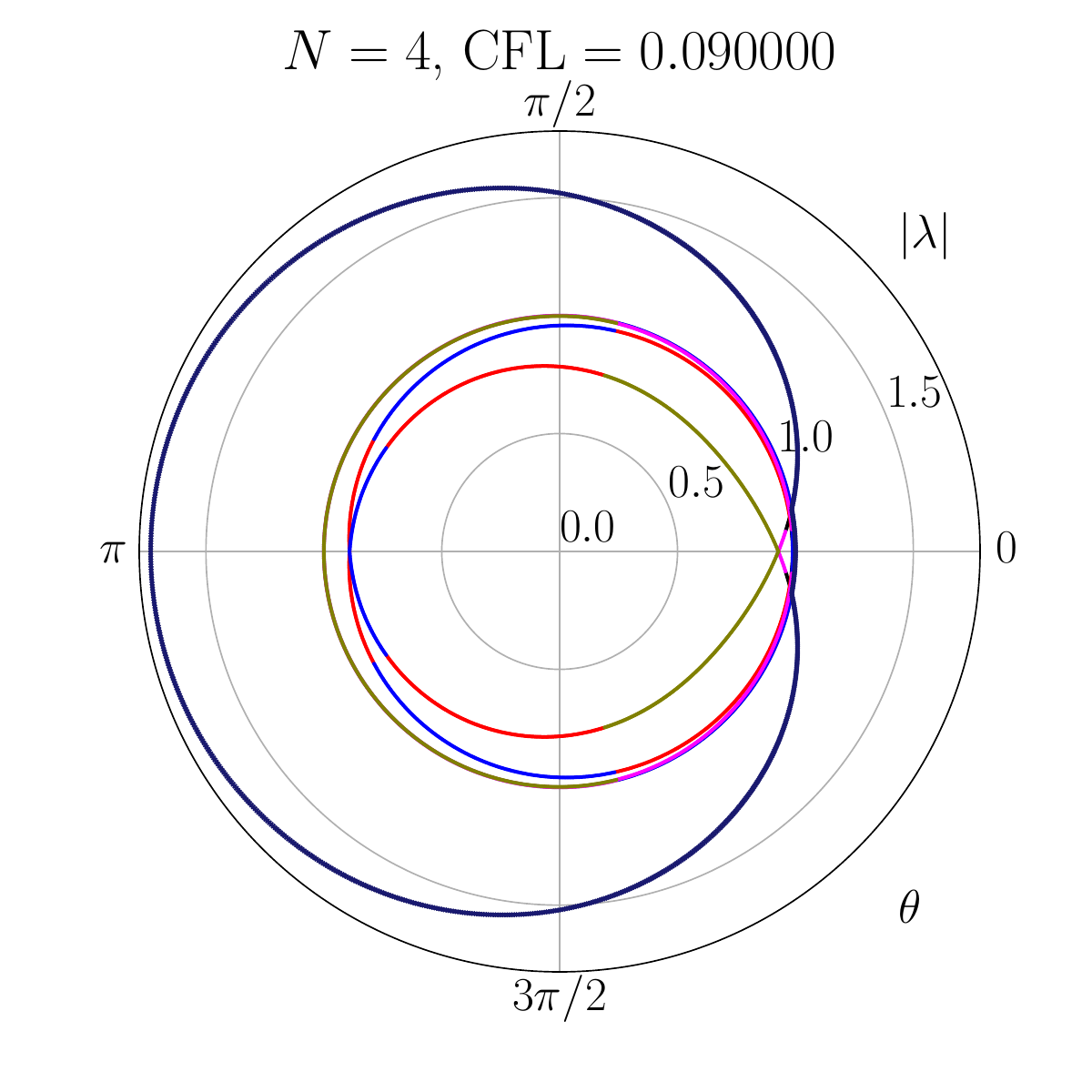}\\
\includegraphics[width=0.028125\textwidth]{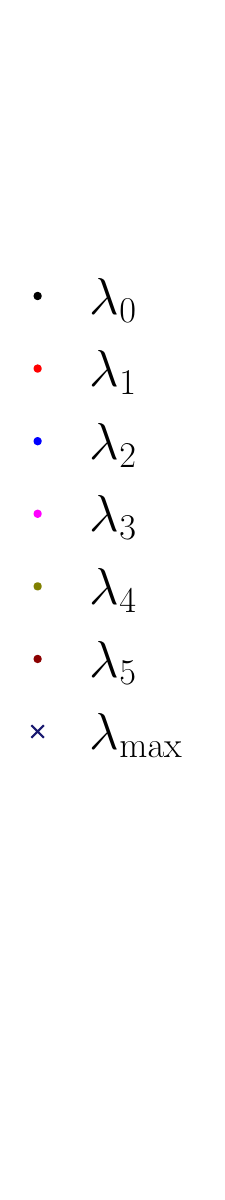}
\includegraphics[width=0.15\textwidth]{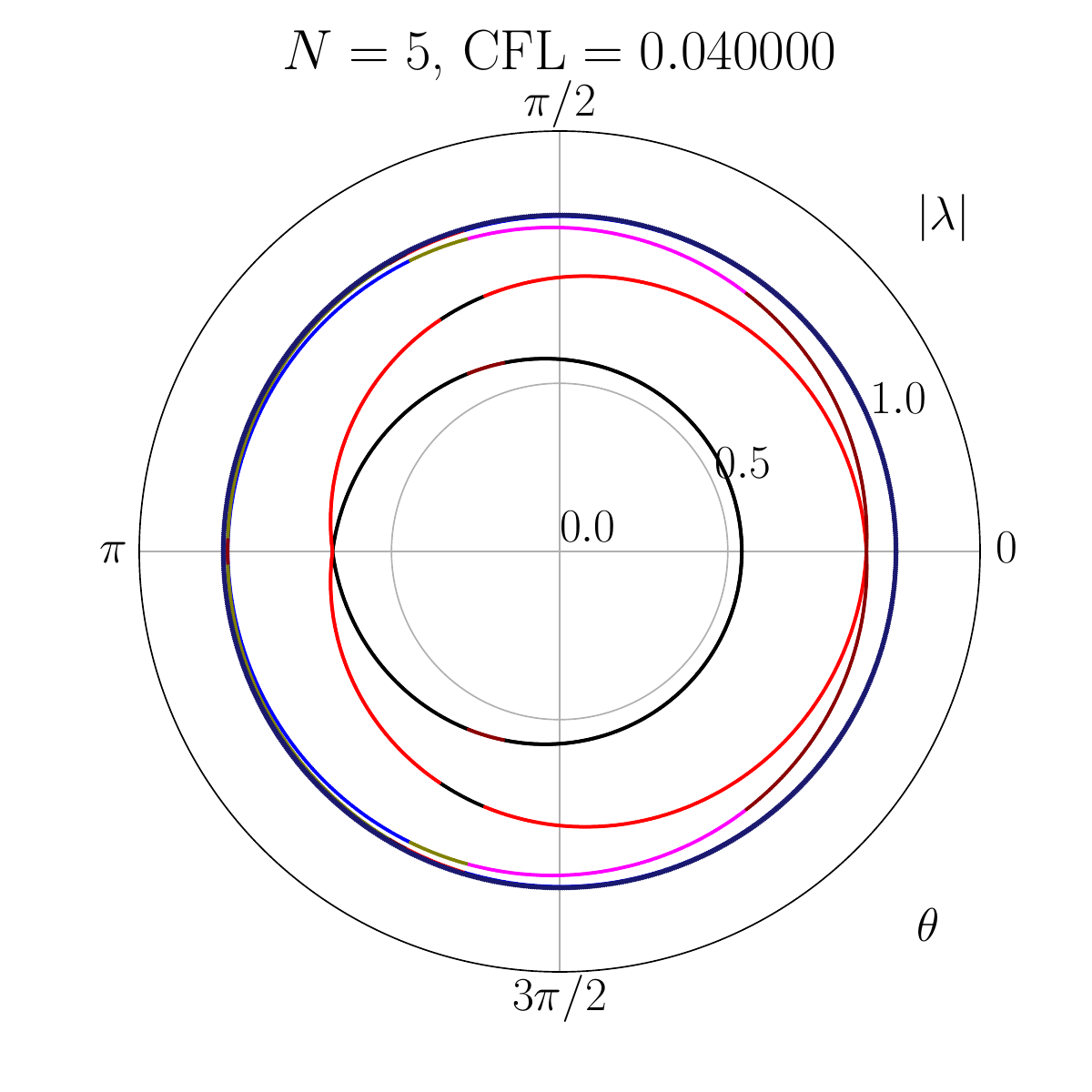}
\includegraphics[width=0.15\textwidth]{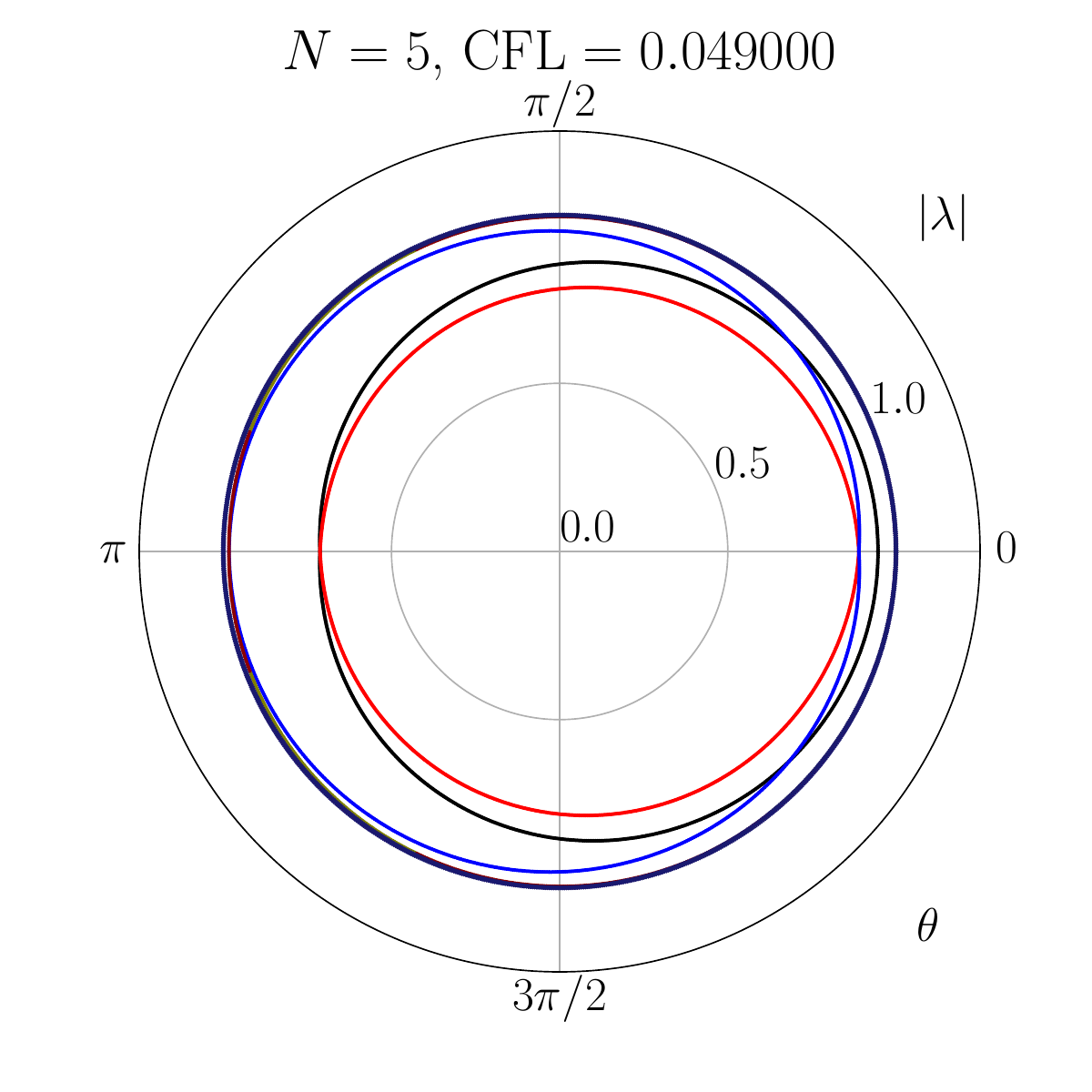}
\includegraphics[width=0.15\textwidth]{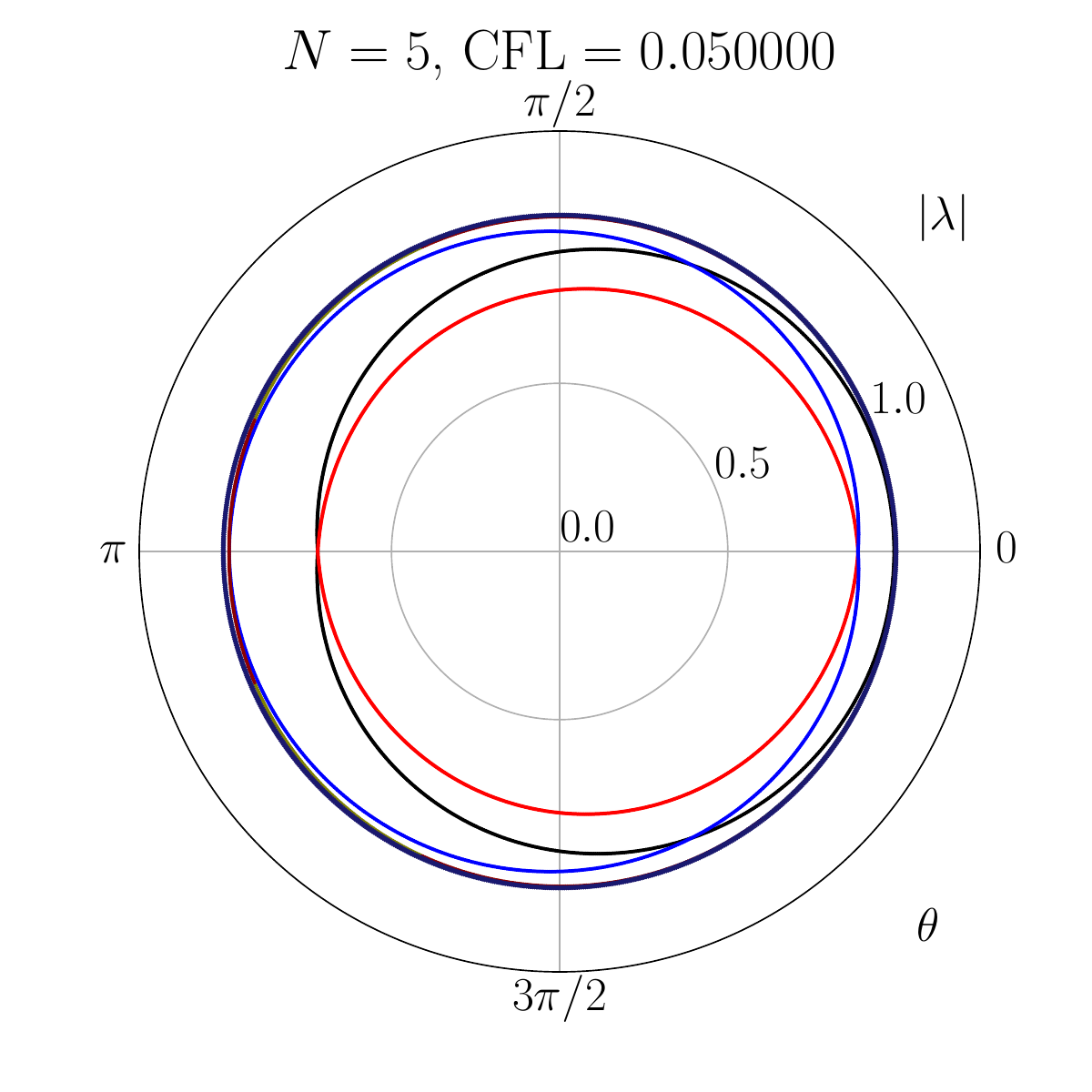}
\includegraphics[width=0.15\textwidth]{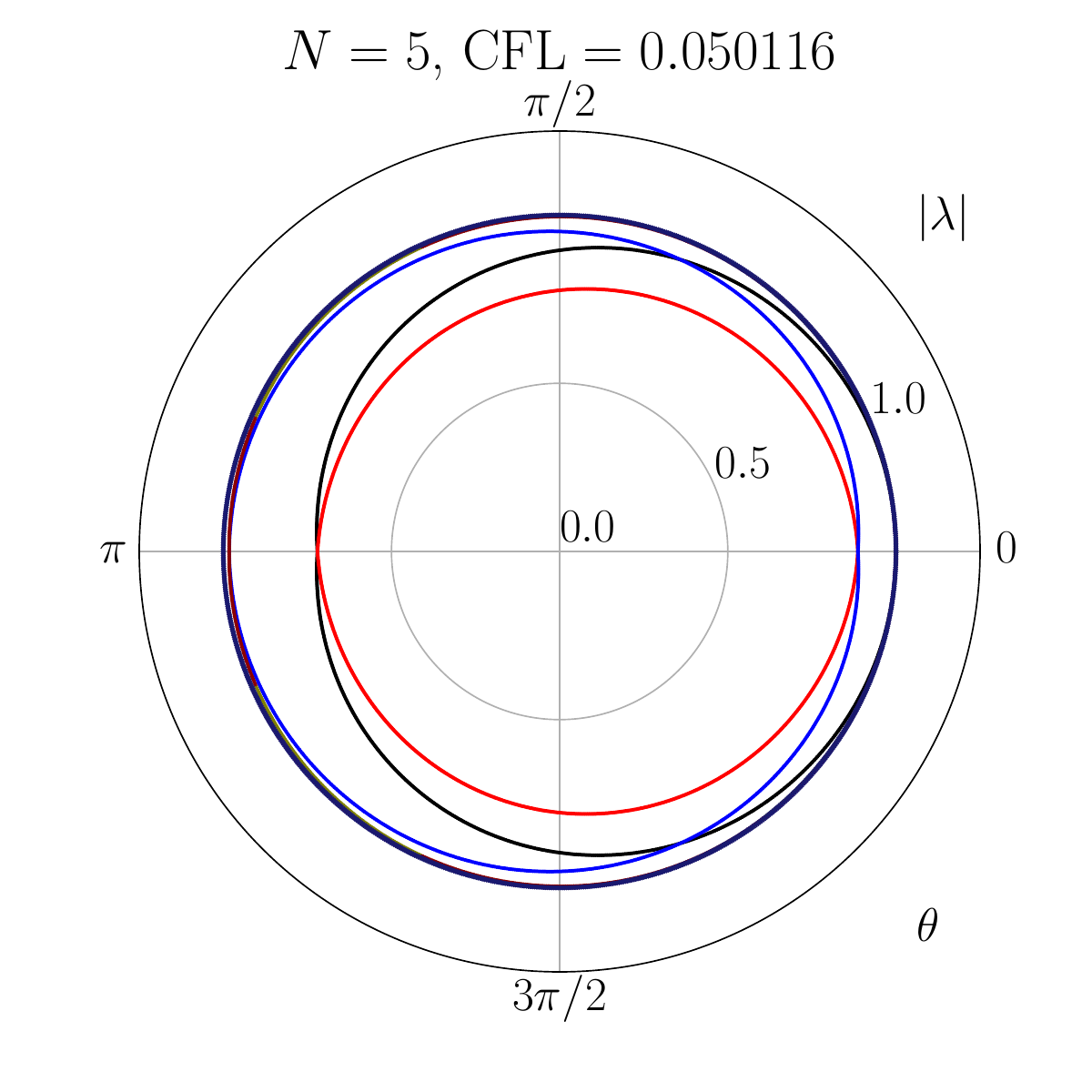}
\includegraphics[width=0.15\textwidth]{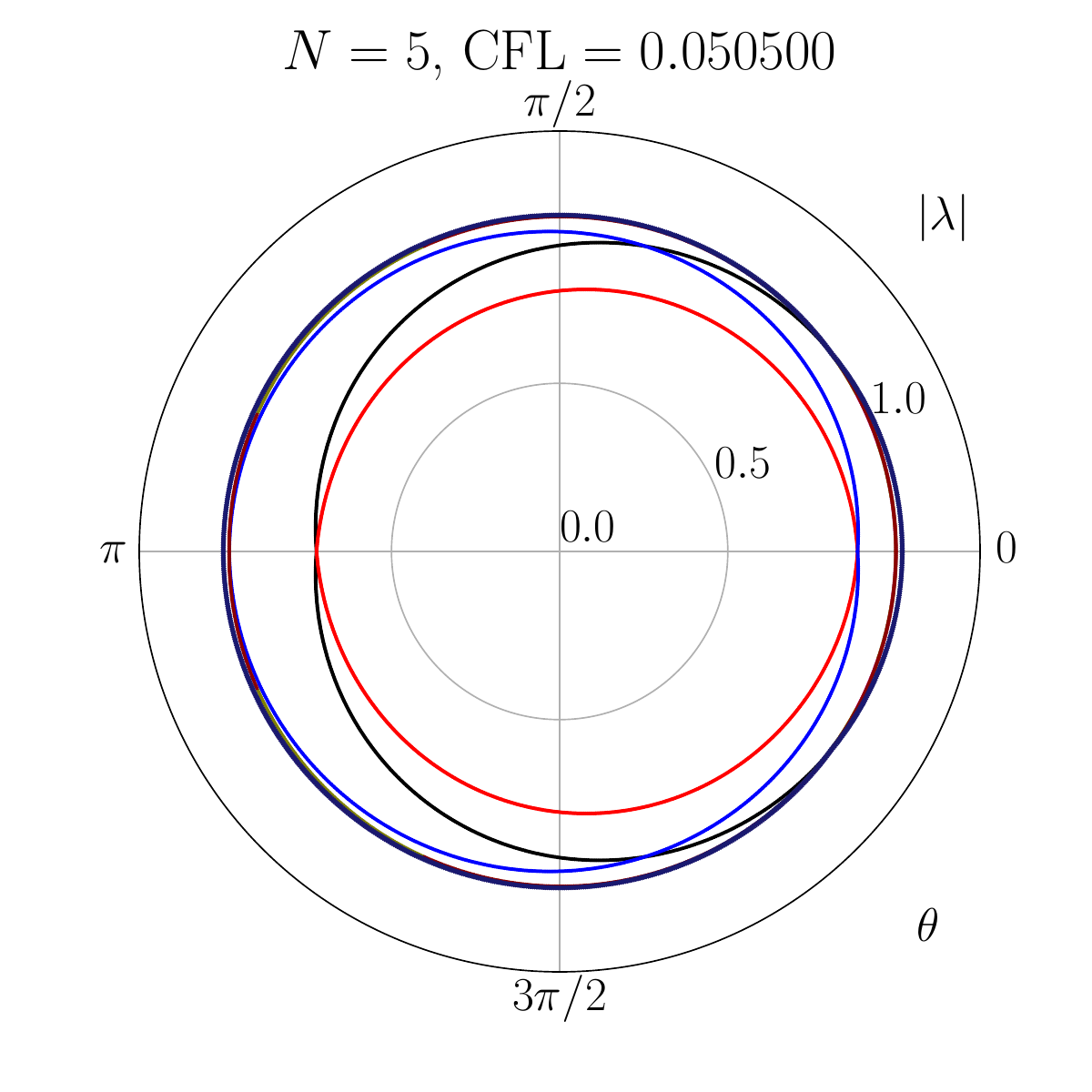}
\includegraphics[width=0.15\textwidth]{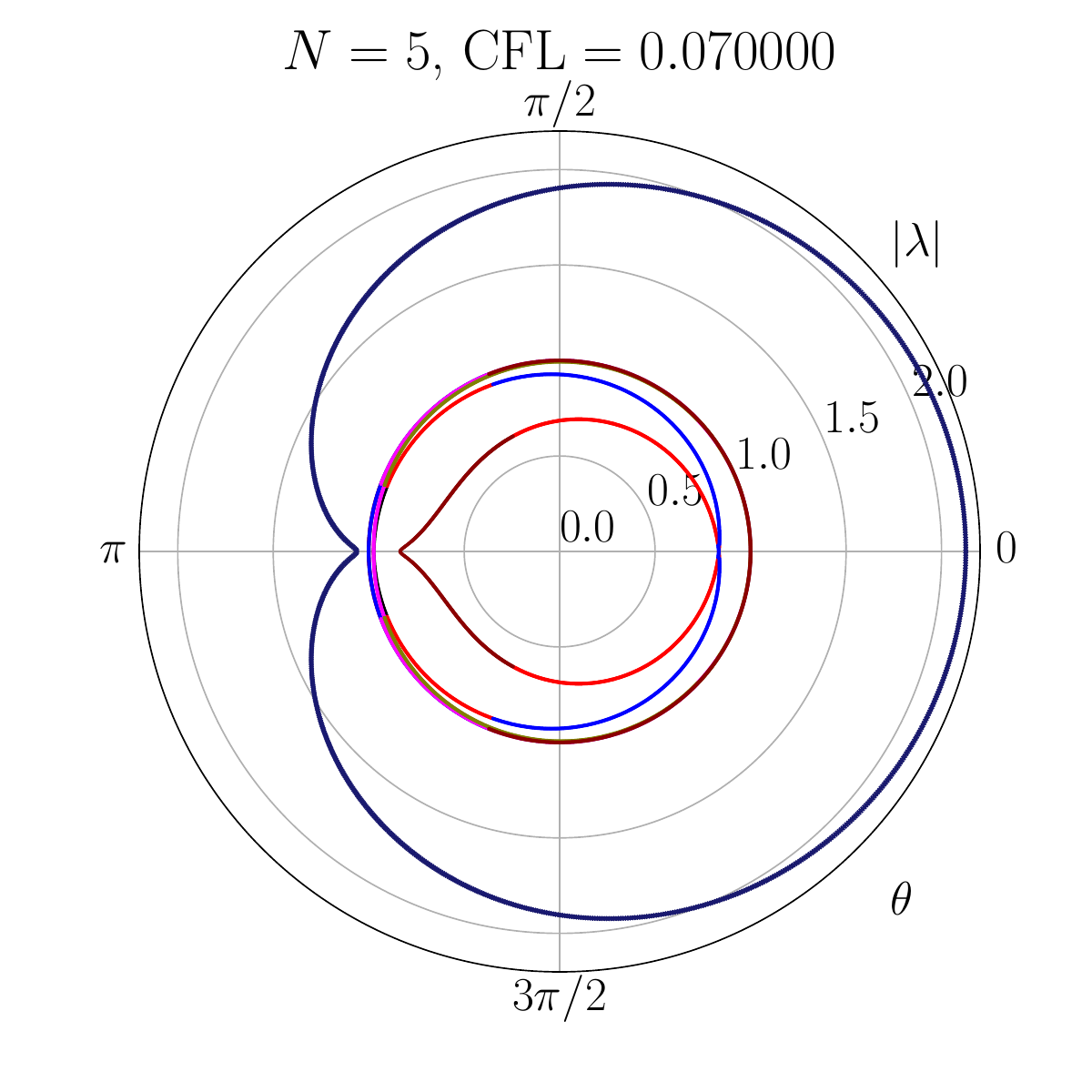}\\
\includegraphics[width=0.028125\textwidth]{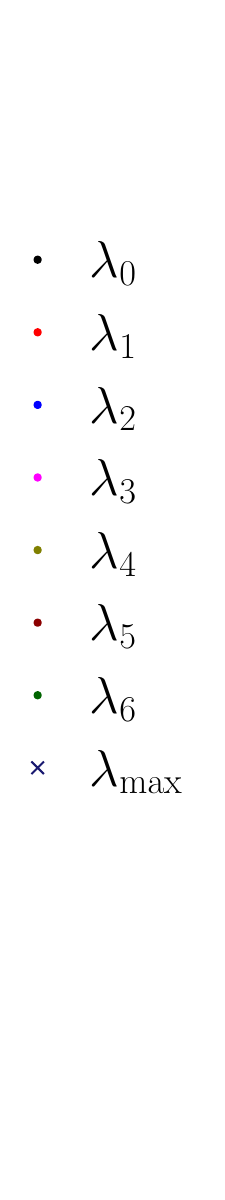}
\includegraphics[width=0.15\textwidth]{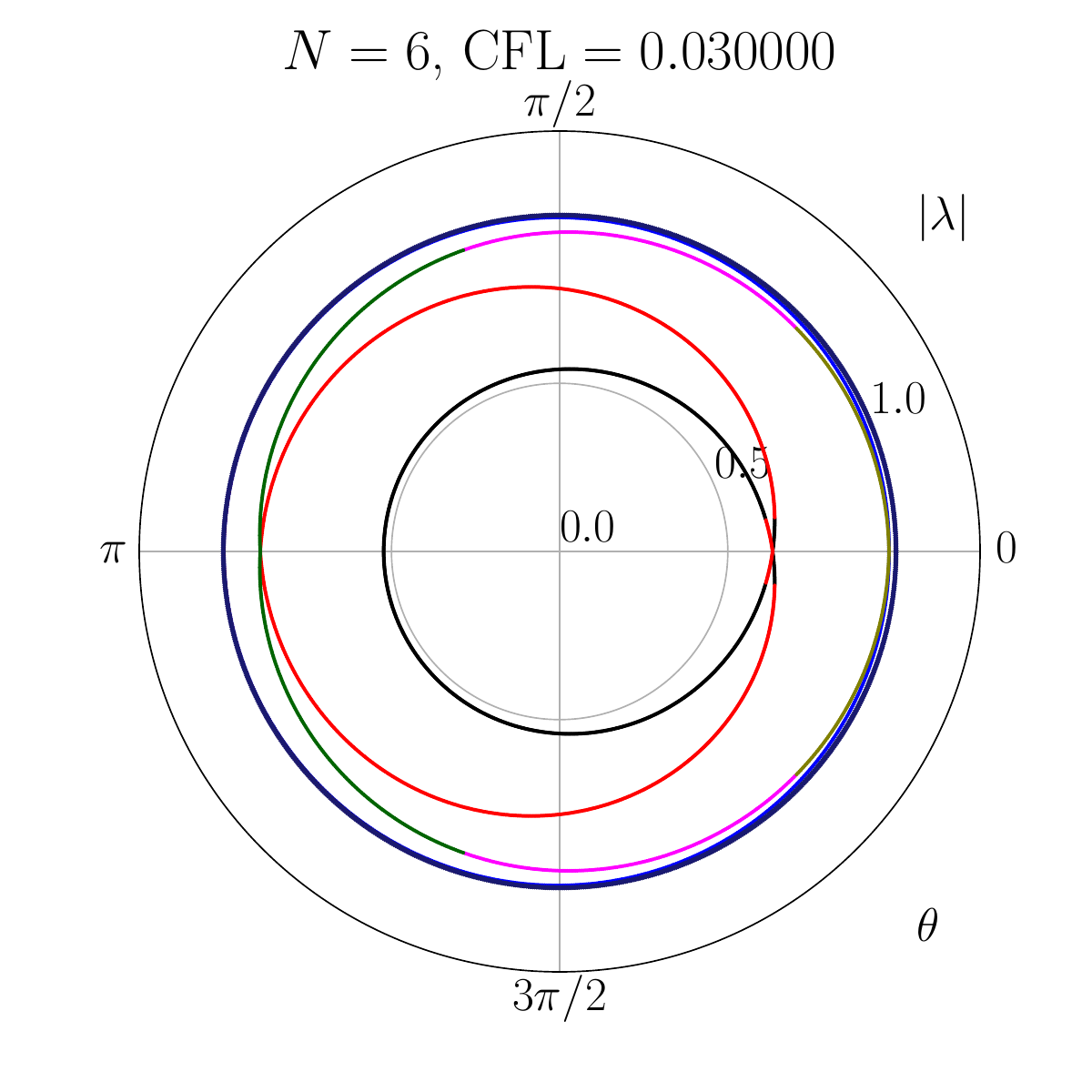}
\includegraphics[width=0.15\textwidth]{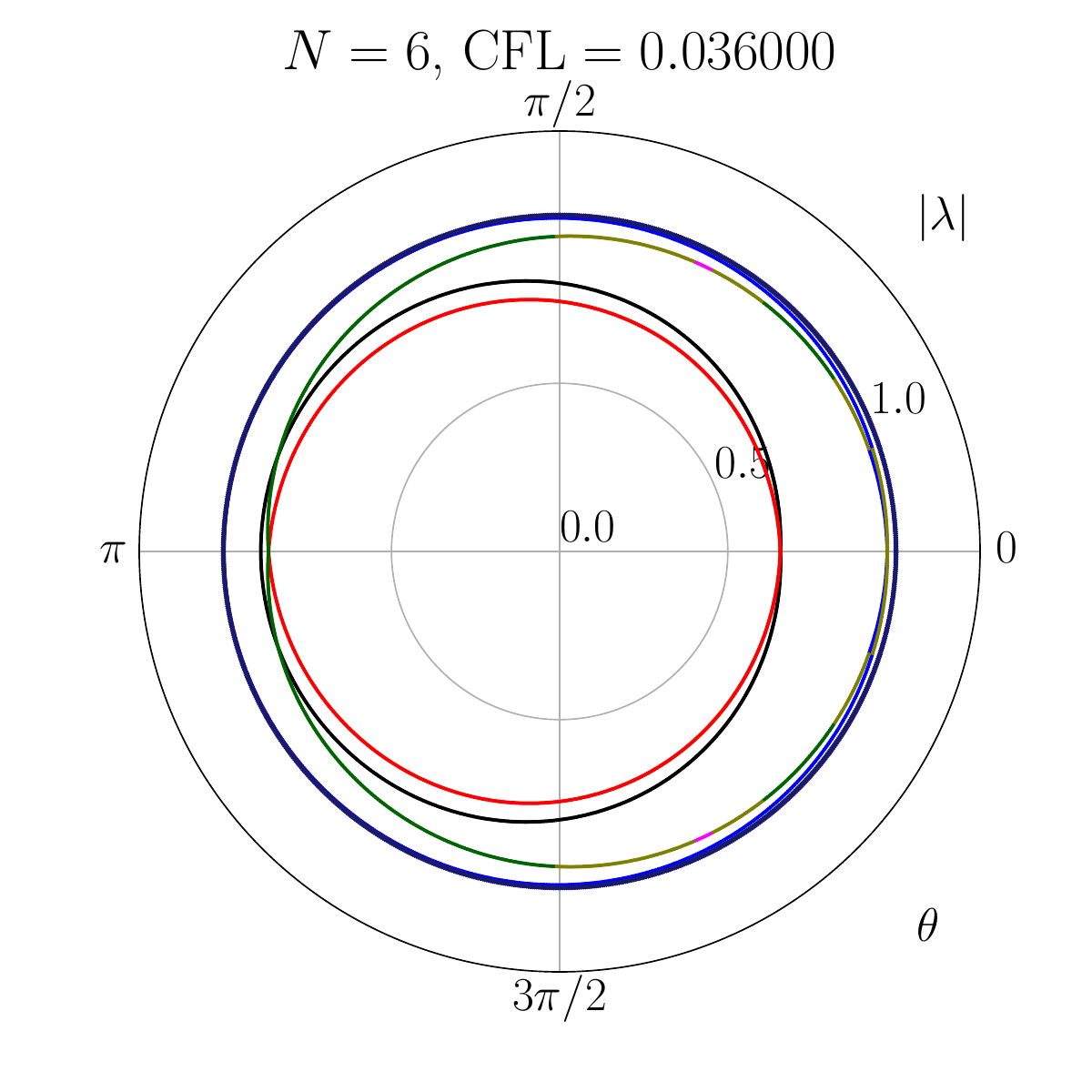}
\includegraphics[width=0.15\textwidth]{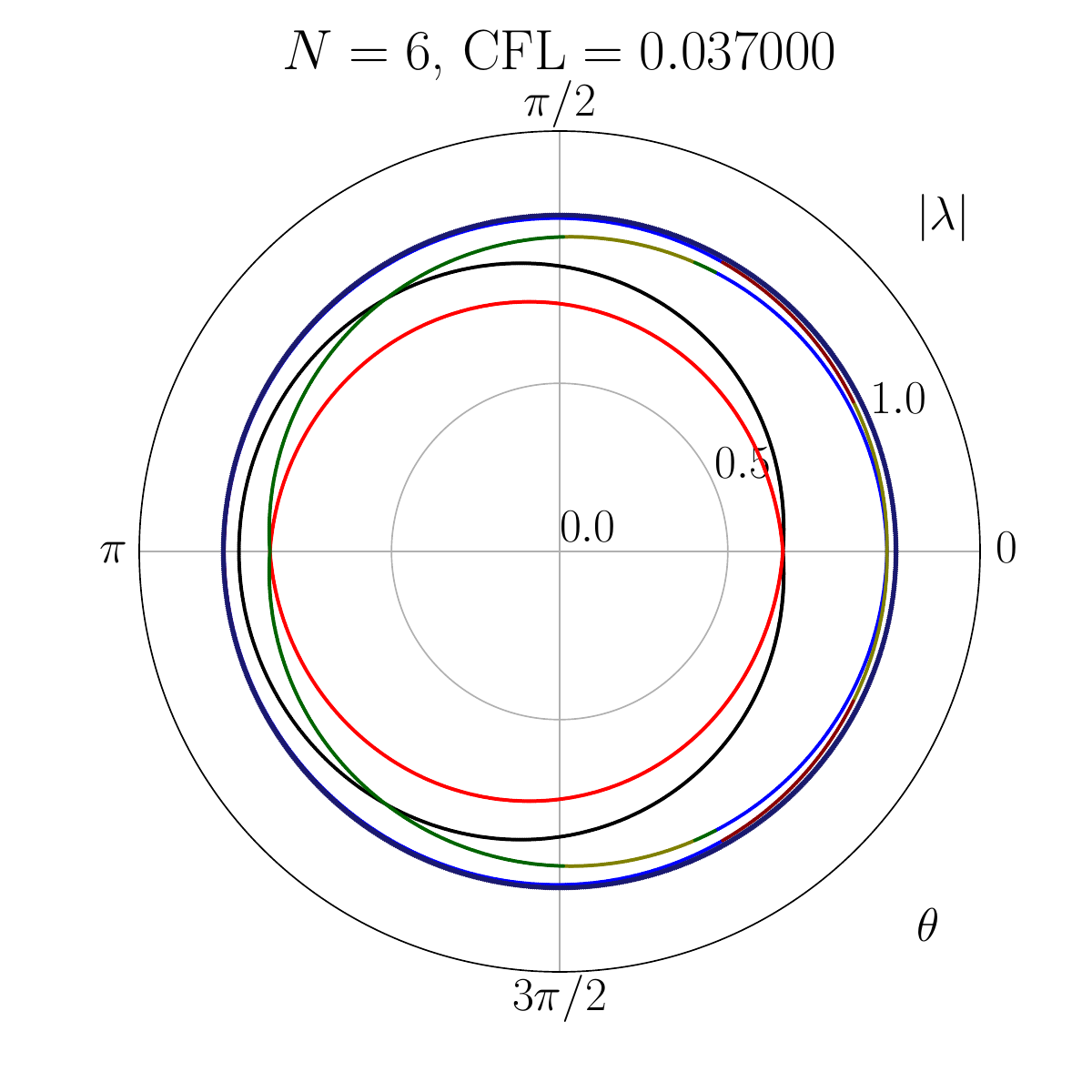}
\includegraphics[width=0.15\textwidth]{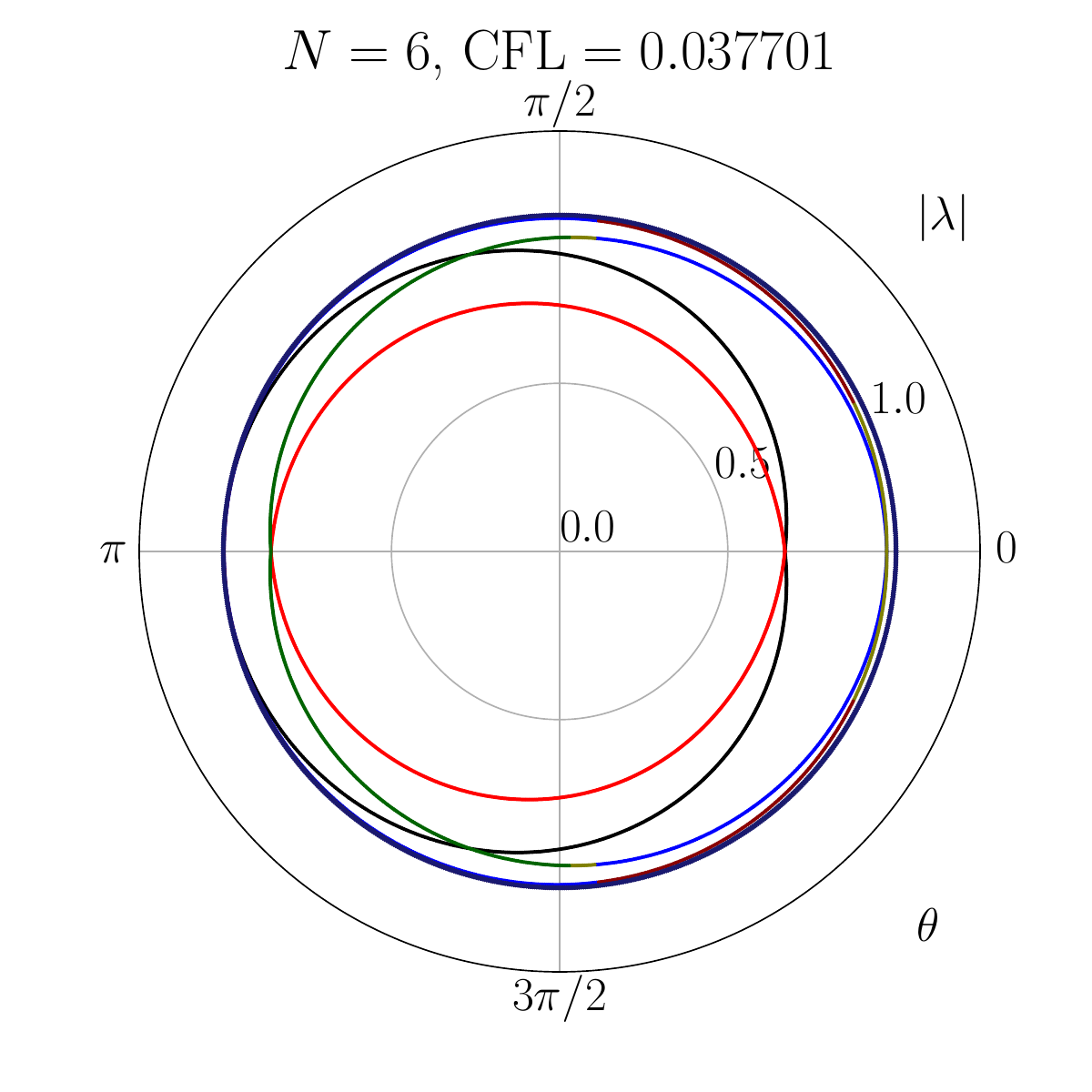}
\includegraphics[width=0.15\textwidth]{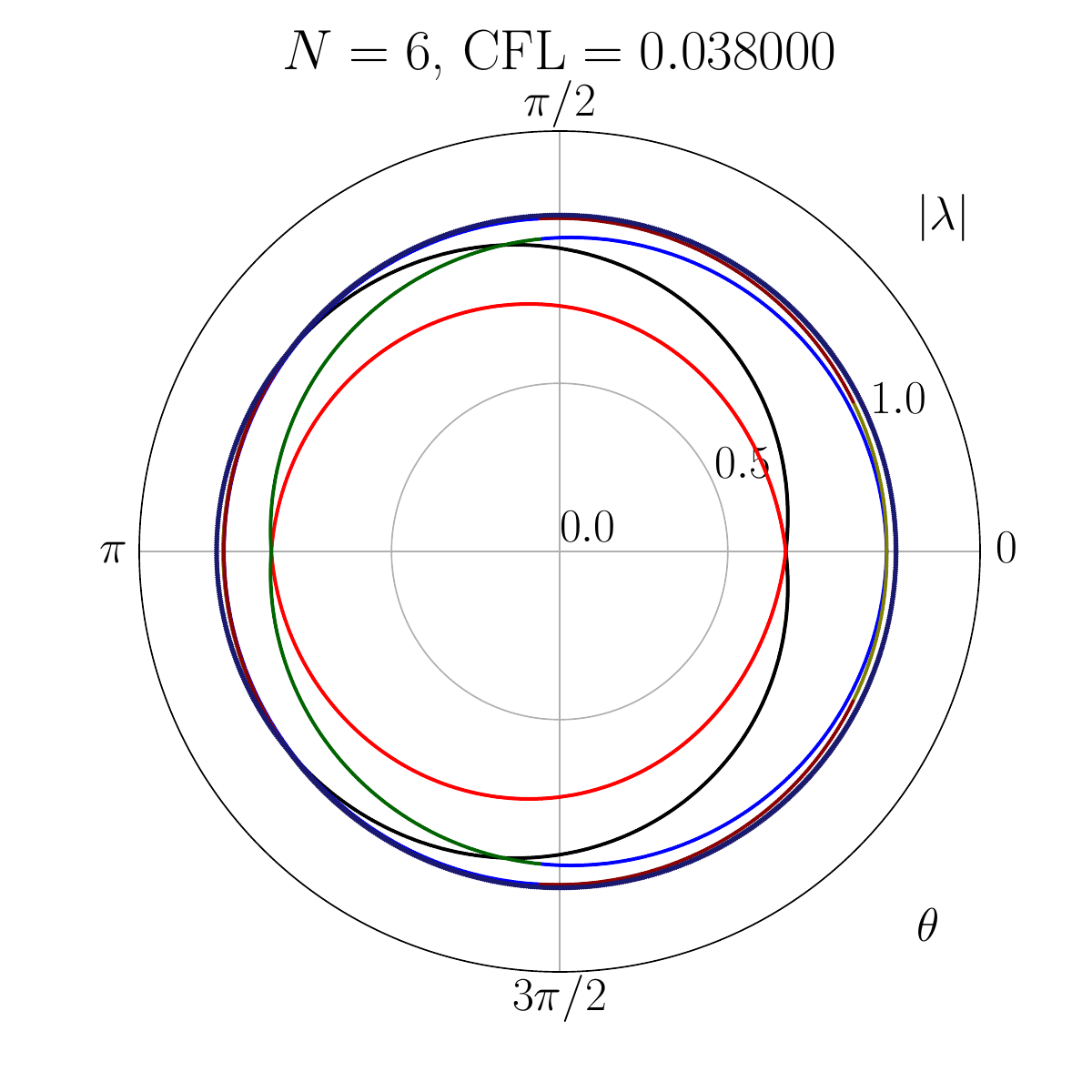}
\includegraphics[width=0.15\textwidth]{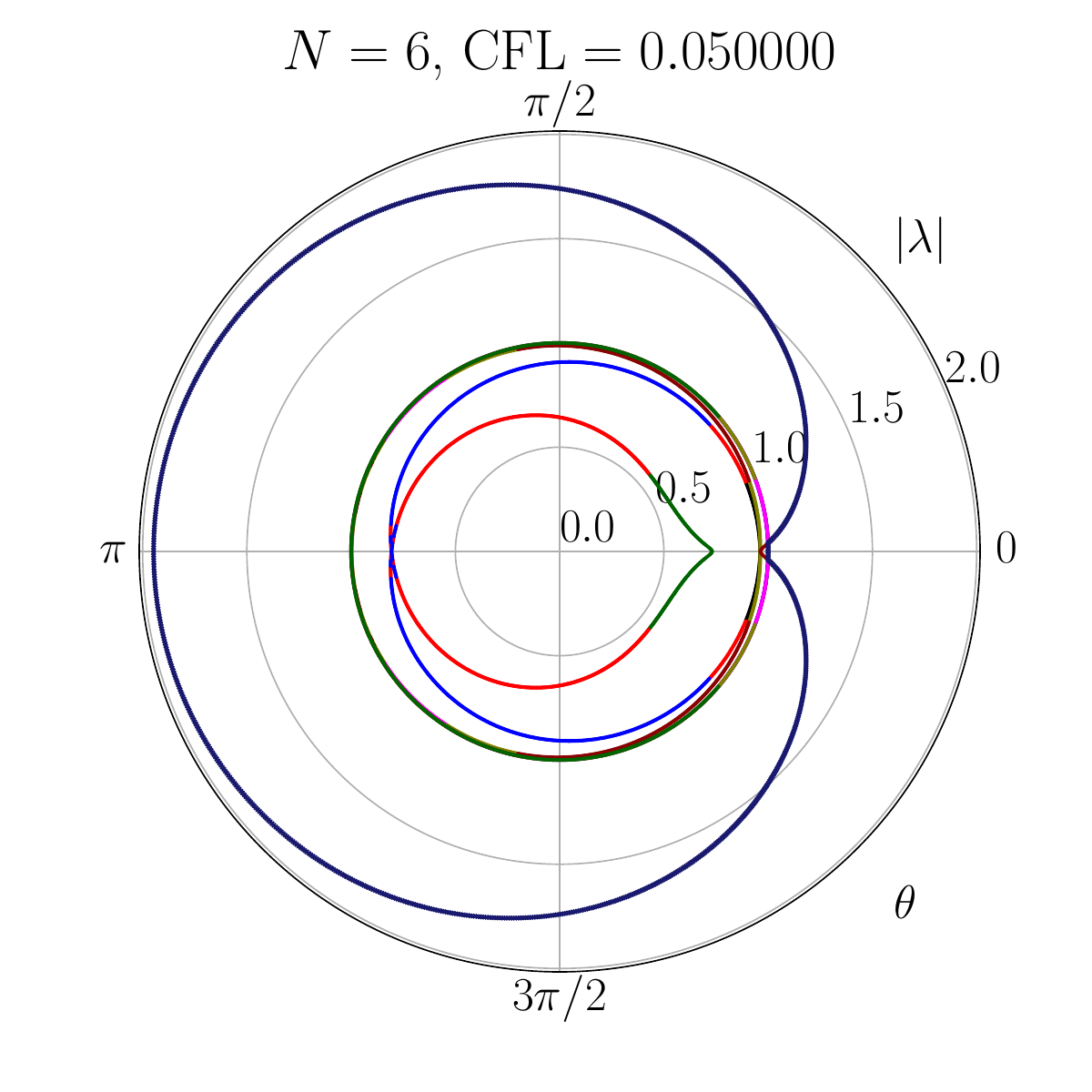}\\
\caption{%
The dependence of the absolute values $|\lambda_{k}|$ of the spectrum of the matrix $\mathrm{R}(\mathrm{CFL}, \theta)$ (\ref{eq:r_matrix_elems_expr_dup}) (eigenvalues $\lambda_{k} = \lambda_{k}(\mathrm{CFL}, \theta)$, $k = 0, \ldots, N$) of the evolution operator $R$ (\ref{eq:evol_oper_prop}) for a single time step $\Dtn{n}$ on phase $\theta = k\Dx$ for several selected values of the Courant number $\mathrm{CFL}$ for polynomial degrees $N = 1, \ldots, 6$ --- the polar plot with phase $\theta$ as angle and absolute value $|\lambda|$ as radius. The range of phase $\theta\in[0, 2\pi)$ is sampled on a uniform grid of $1000$ nodes. Legends for each row of the graphs are located on the left. The gray circle of unit radius defines the stability boundary. The Courant numbers $\mathrm{CFL}$ are taken deep inside the stability region $\mathrm{CFL} \in [0, \mathrm{CFL}_{\rm max}]$ (left column), in the region of guaranteed instability $\mathrm{CFL} > \mathrm{CFL}_{\rm max}$ (right column) and near the boundary of the stability region $\mathrm{CFL}_{\rm max}$ (the values $\mathrm{CFL}_{\rm max}$ are selected from work~\cite{ader_dg_stab}, as well as the values $\mathrm{CFL}_{\rm max}$ calculated further in this work in Table~\ref{tab:cfls_max_data} and in Figure~\ref{fig:cfls_max_data}). \textit{Note}: the phase $\theta$ is not the phase $\arg \lambda_{k}$ of eigenvalue $\lambda_{k}$; the dependence of eigenvalues $\lambda_{k}$ of the matrix $\mathrm{R}(\mathrm{CFL}, \theta)$ (\ref{eq:r_matrix_elems_expr_dup}) for these polynomial degrees $N$ is presented in Figure~\ref{fig:spectrum_exact_eigvals_degrees_1_6}.
}
\label{fig:rhos_on_theta_degrees_1_6}
\end{figure}

\subsection{Empirical considerations}
\label{sec:stab_anal:empir_cons}

\begin{figure}[h!]
\centering
\includegraphics[width=0.028125\textwidth]{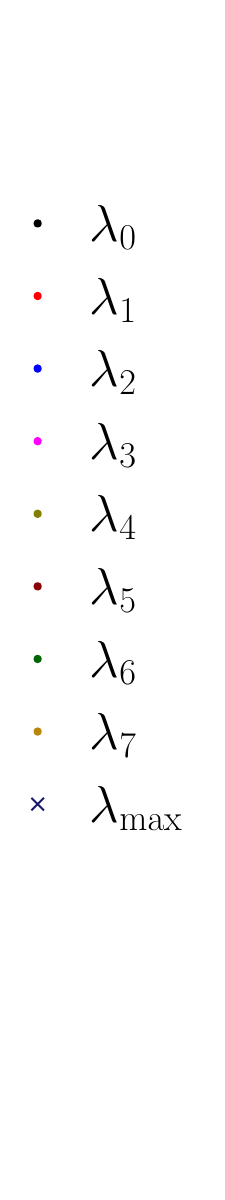}
\includegraphics[width=0.15\textwidth]{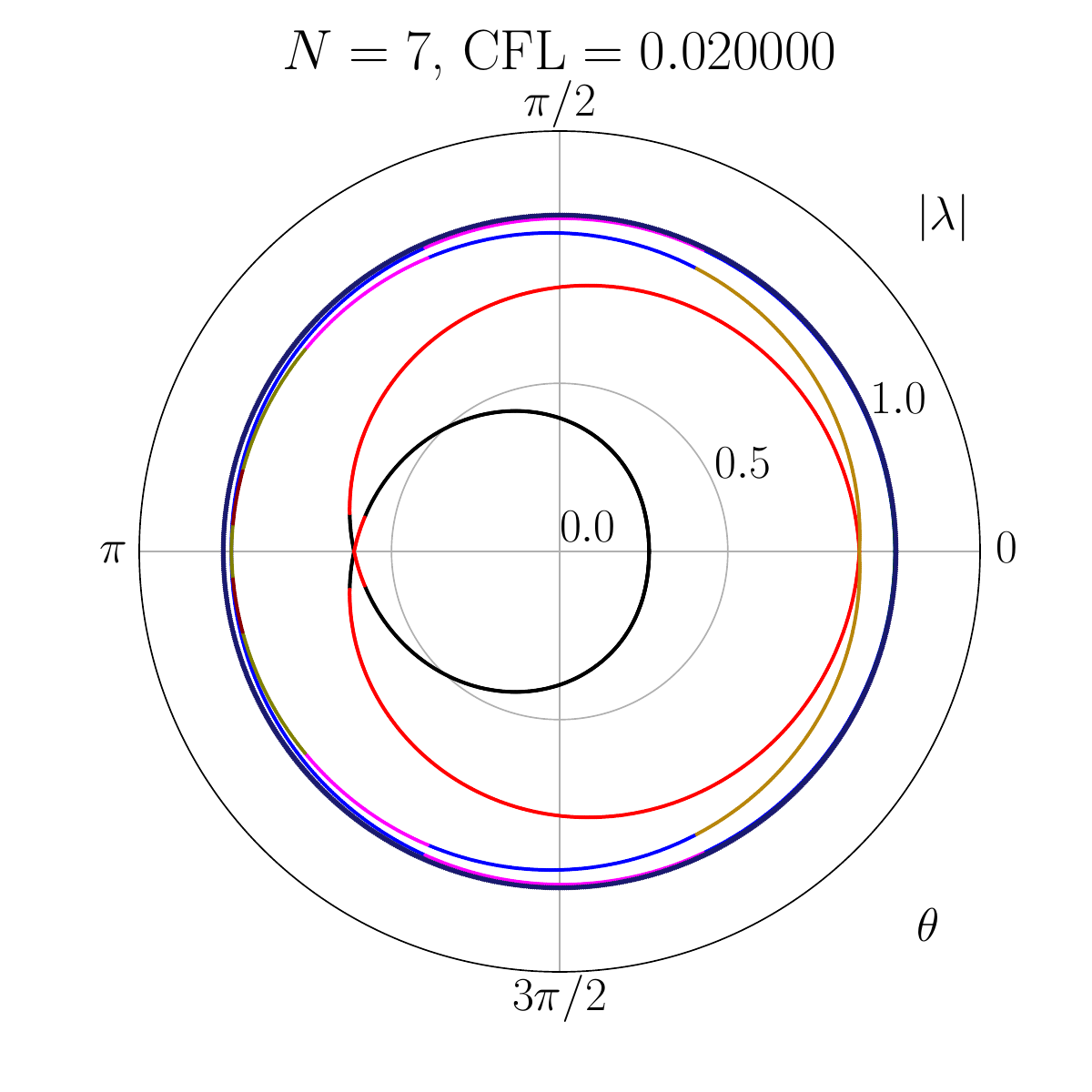}
\includegraphics[width=0.15\textwidth]{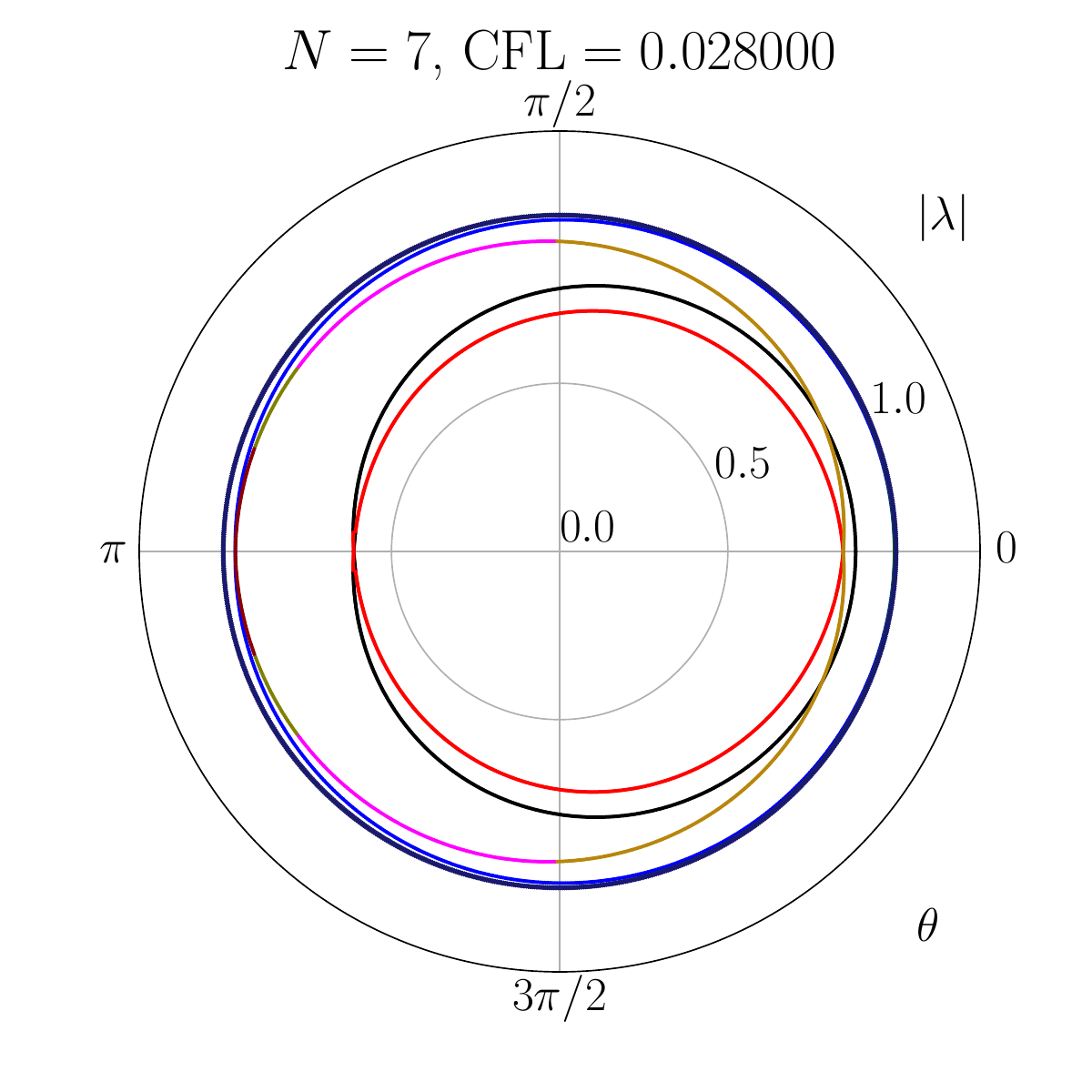}
\includegraphics[width=0.15\textwidth]{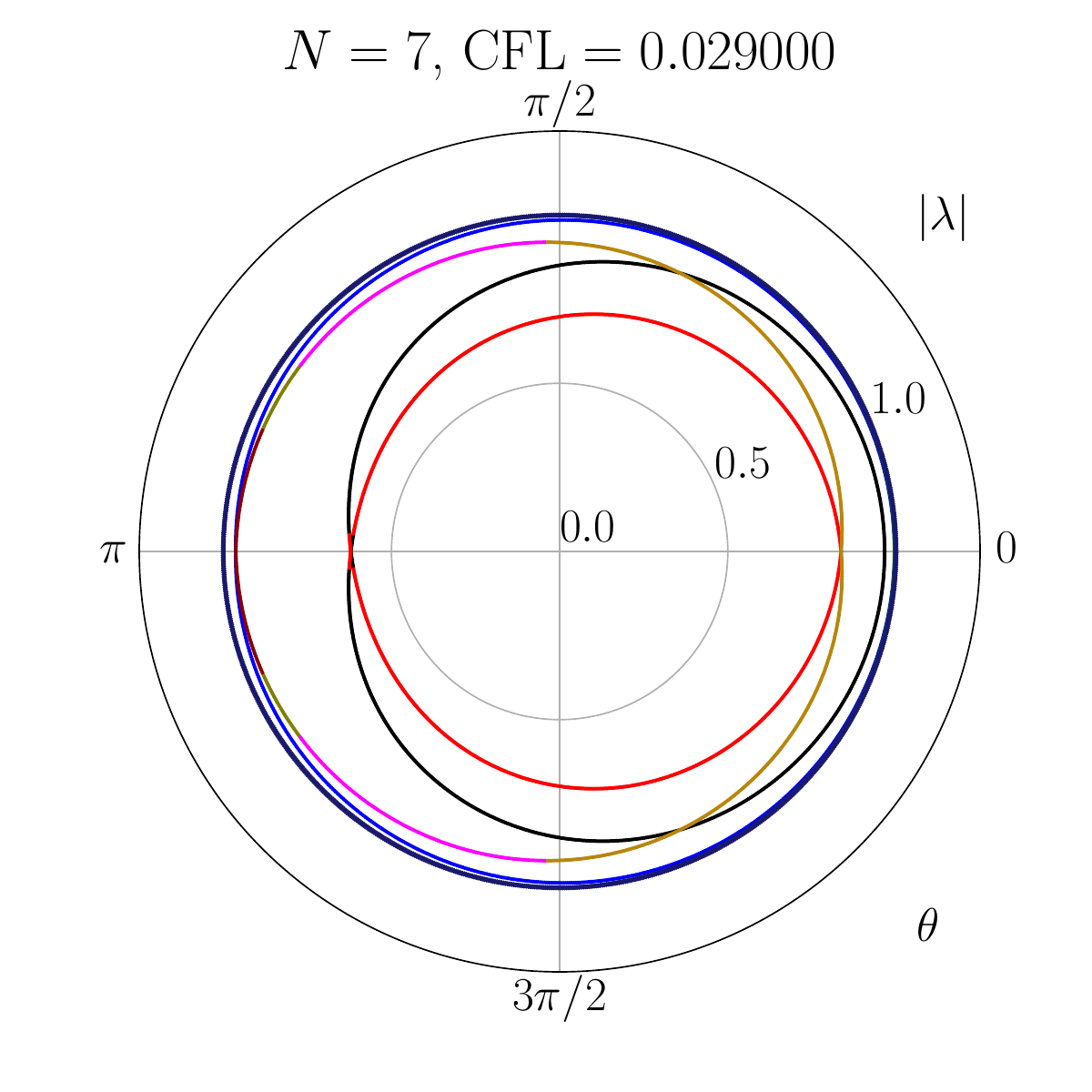}
\includegraphics[width=0.15\textwidth]{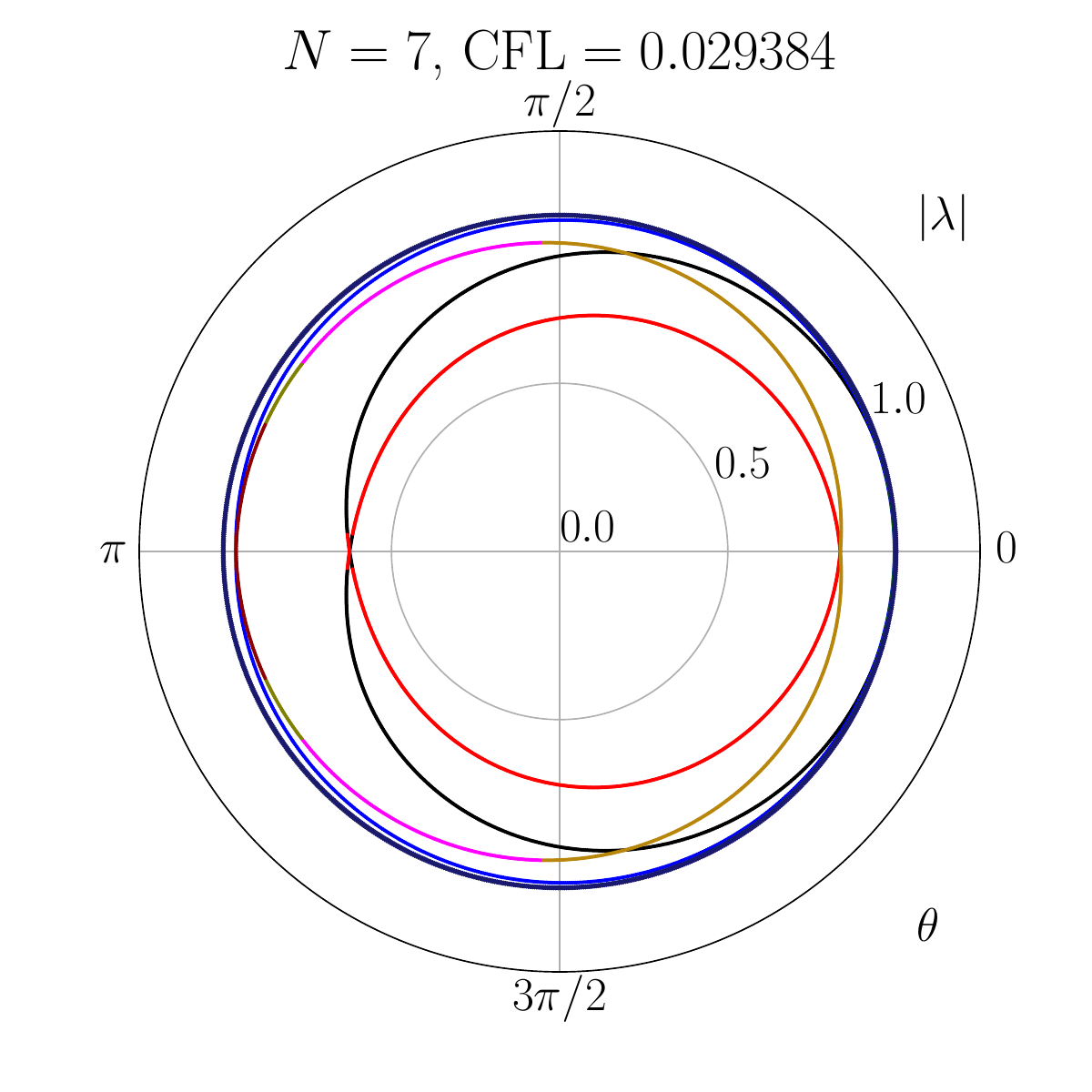}
\includegraphics[width=0.15\textwidth]{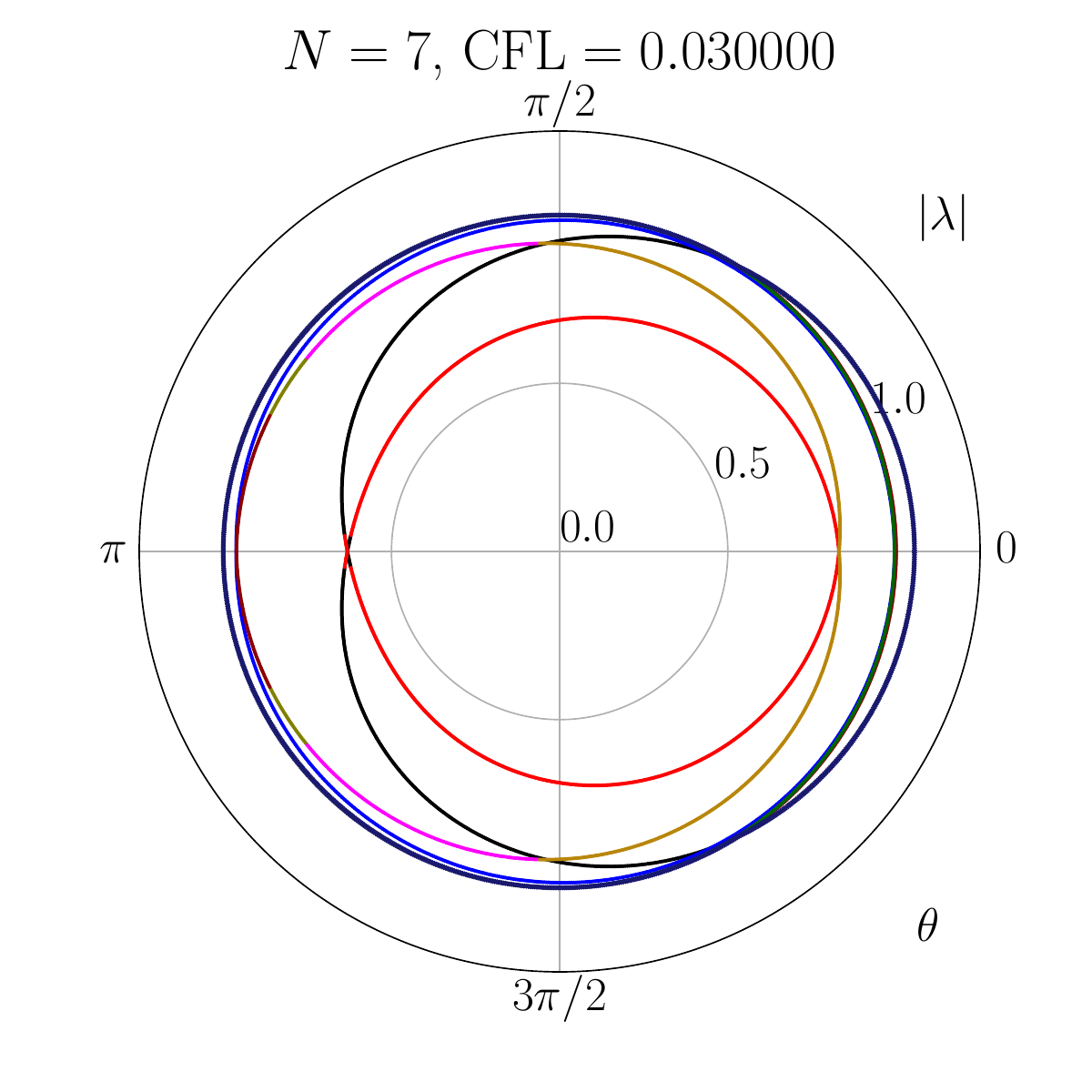}
\includegraphics[width=0.15\textwidth]{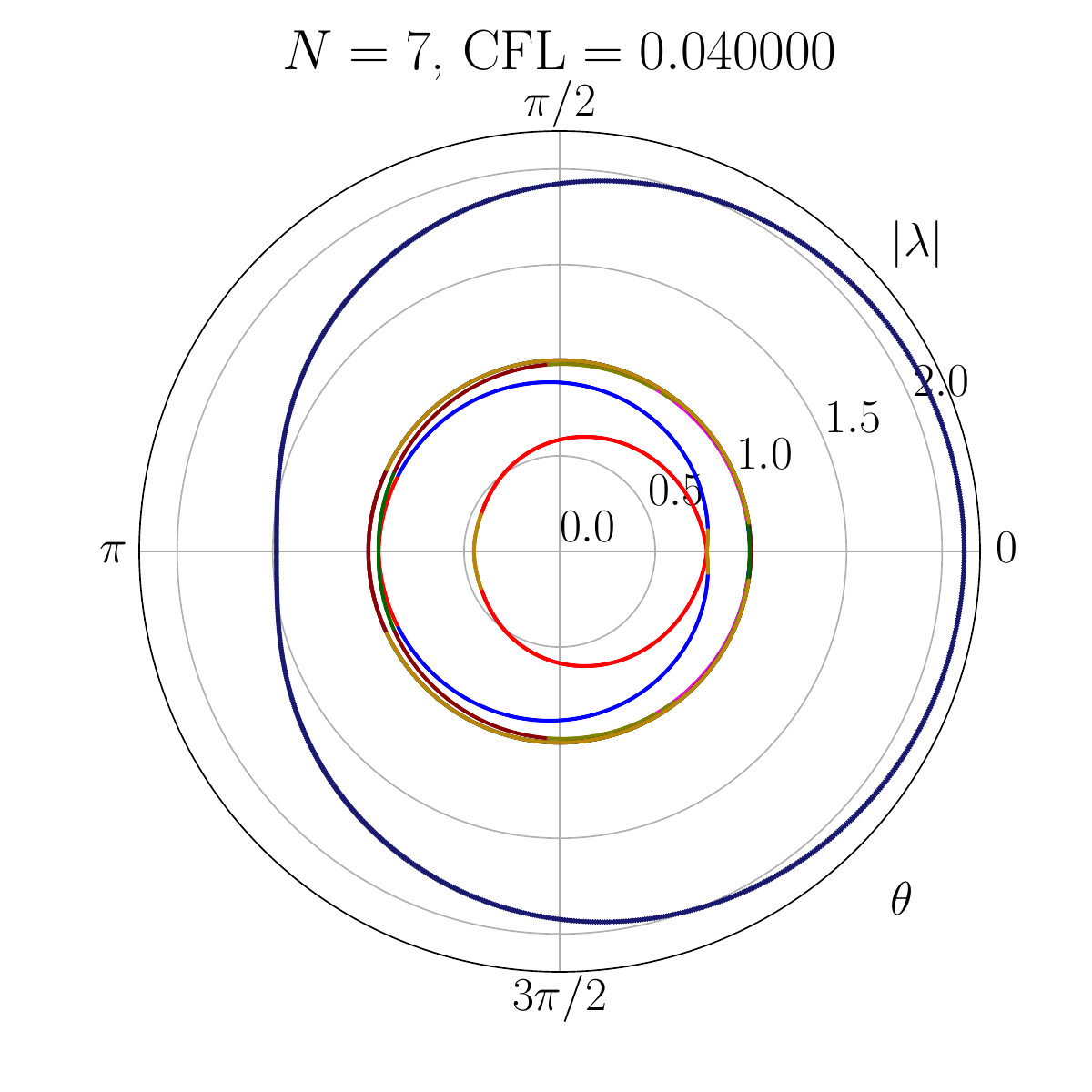}\\
\includegraphics[width=0.028125\textwidth]{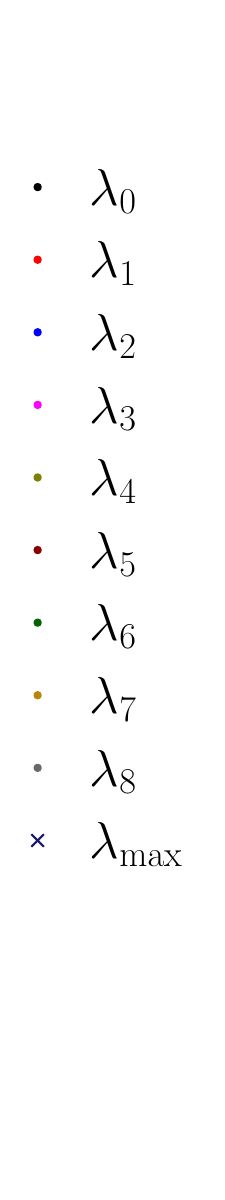}
\includegraphics[width=0.15\textwidth]{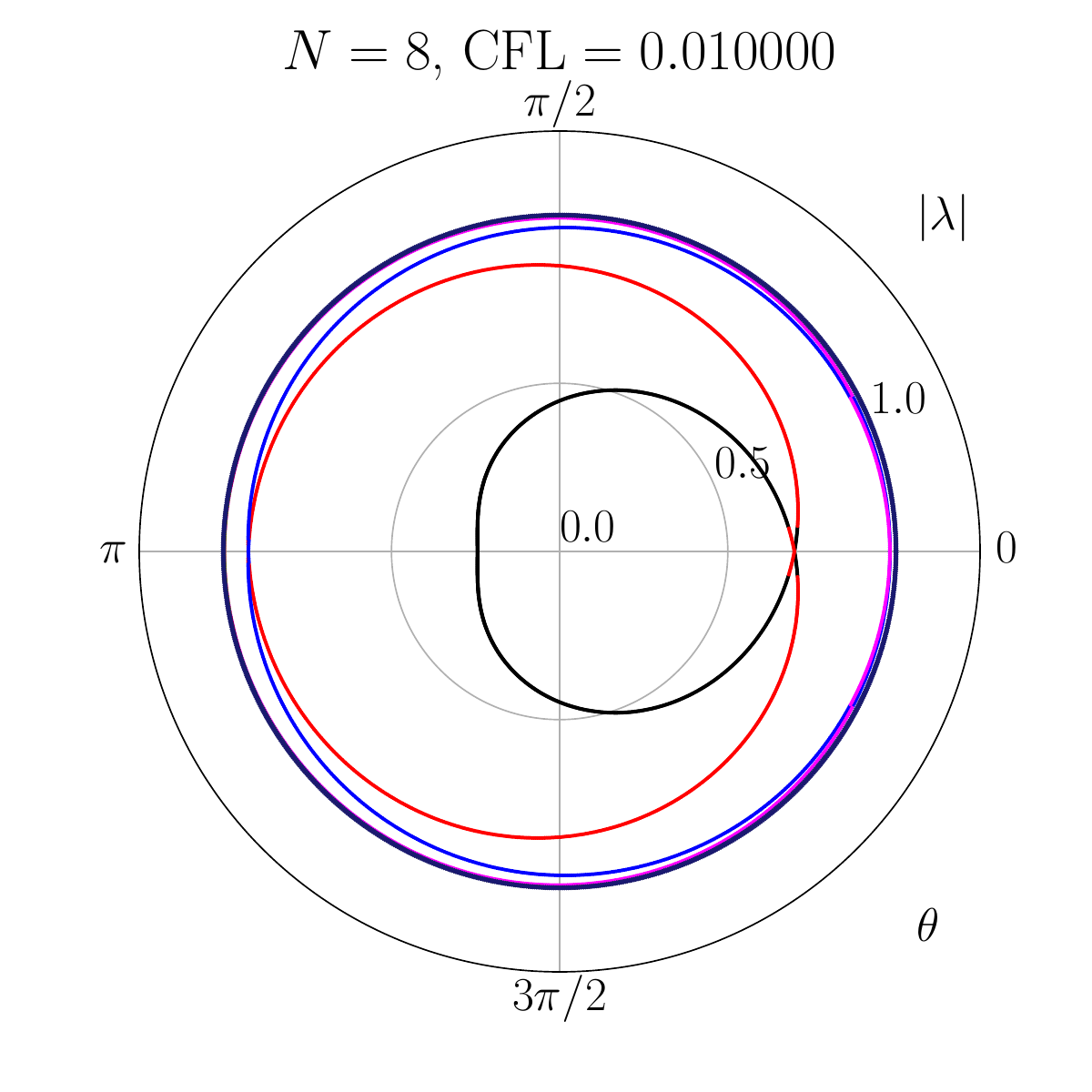}
\includegraphics[width=0.15\textwidth]{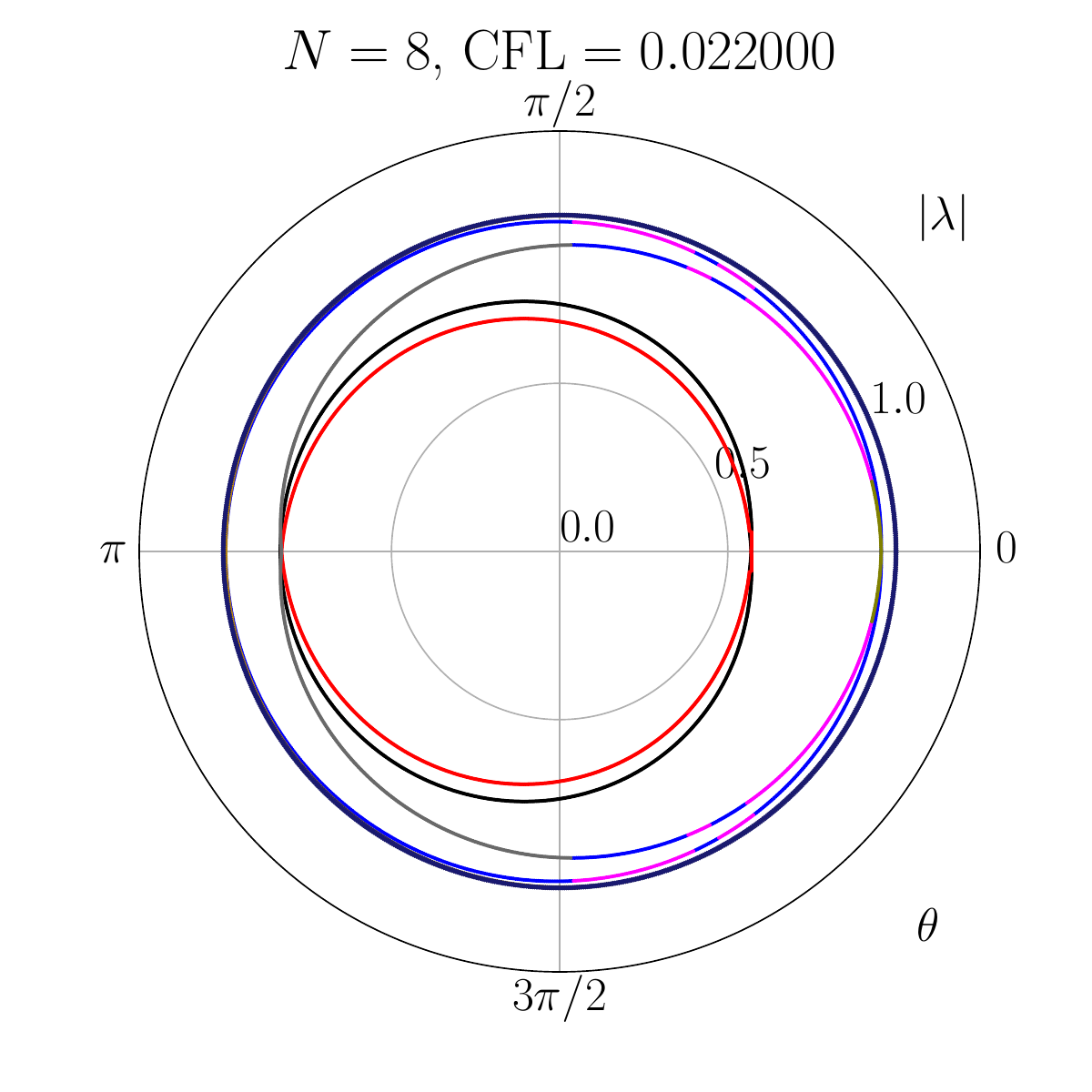}
\includegraphics[width=0.15\textwidth]{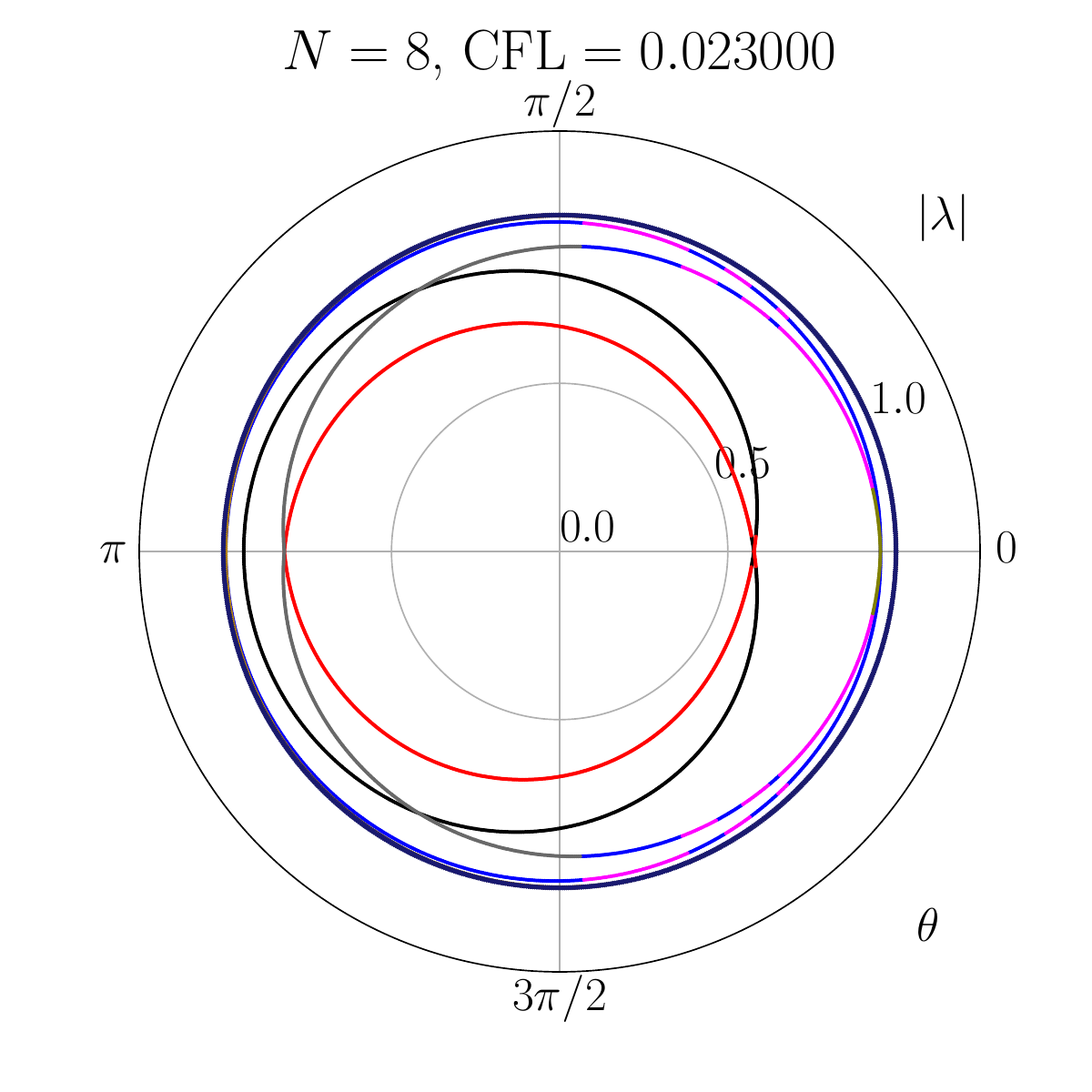}
\includegraphics[width=0.15\textwidth]{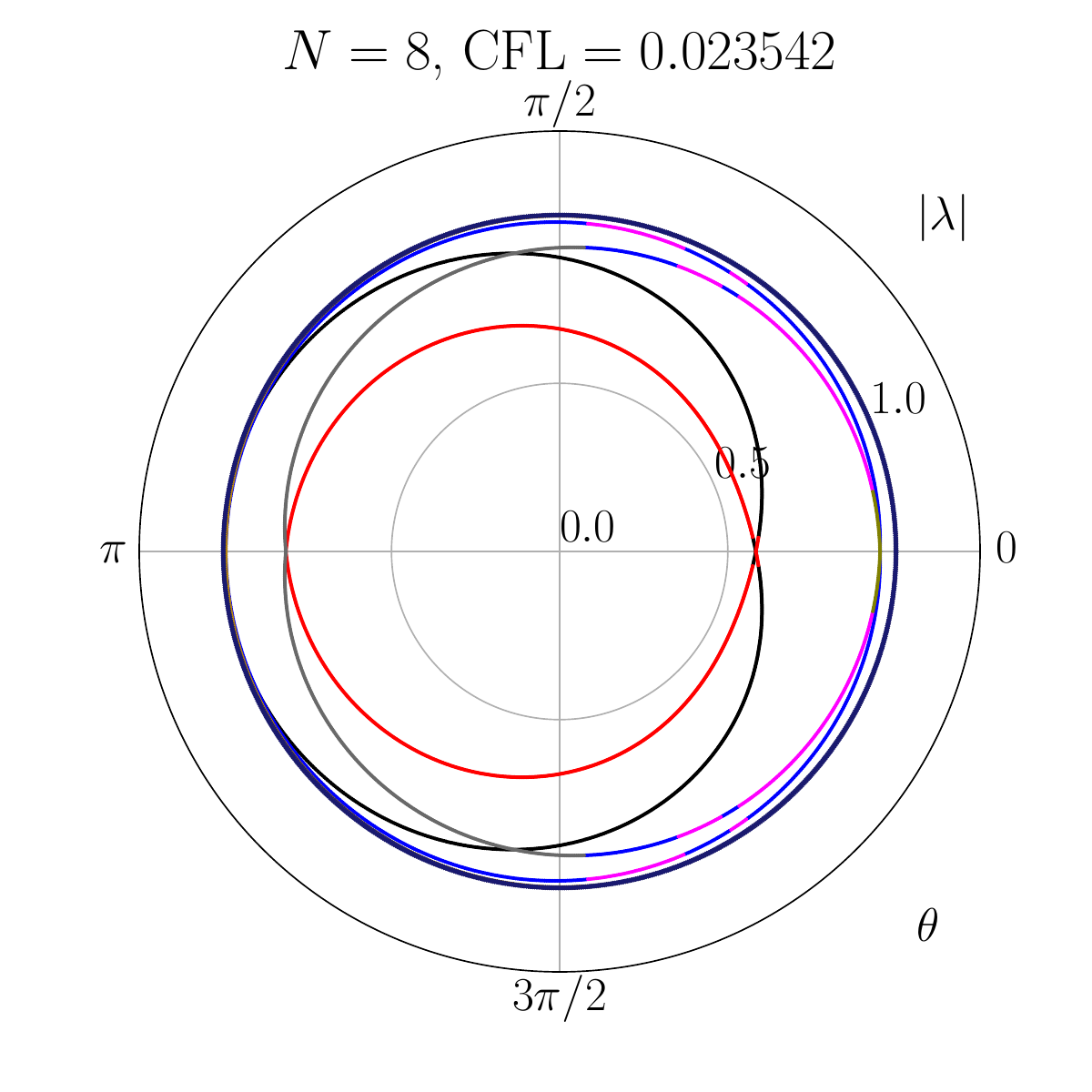}
\includegraphics[width=0.15\textwidth]{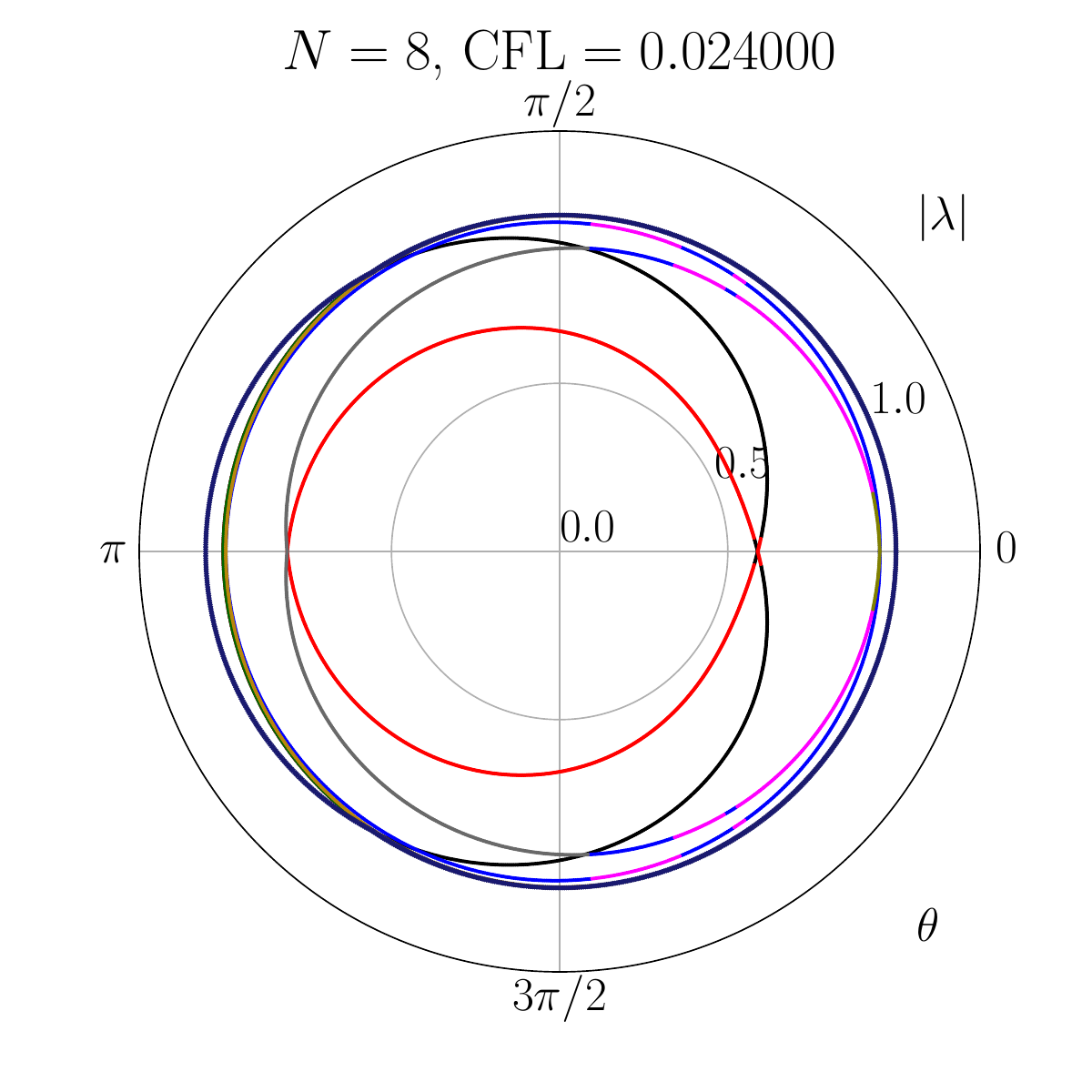}
\includegraphics[width=0.15\textwidth]{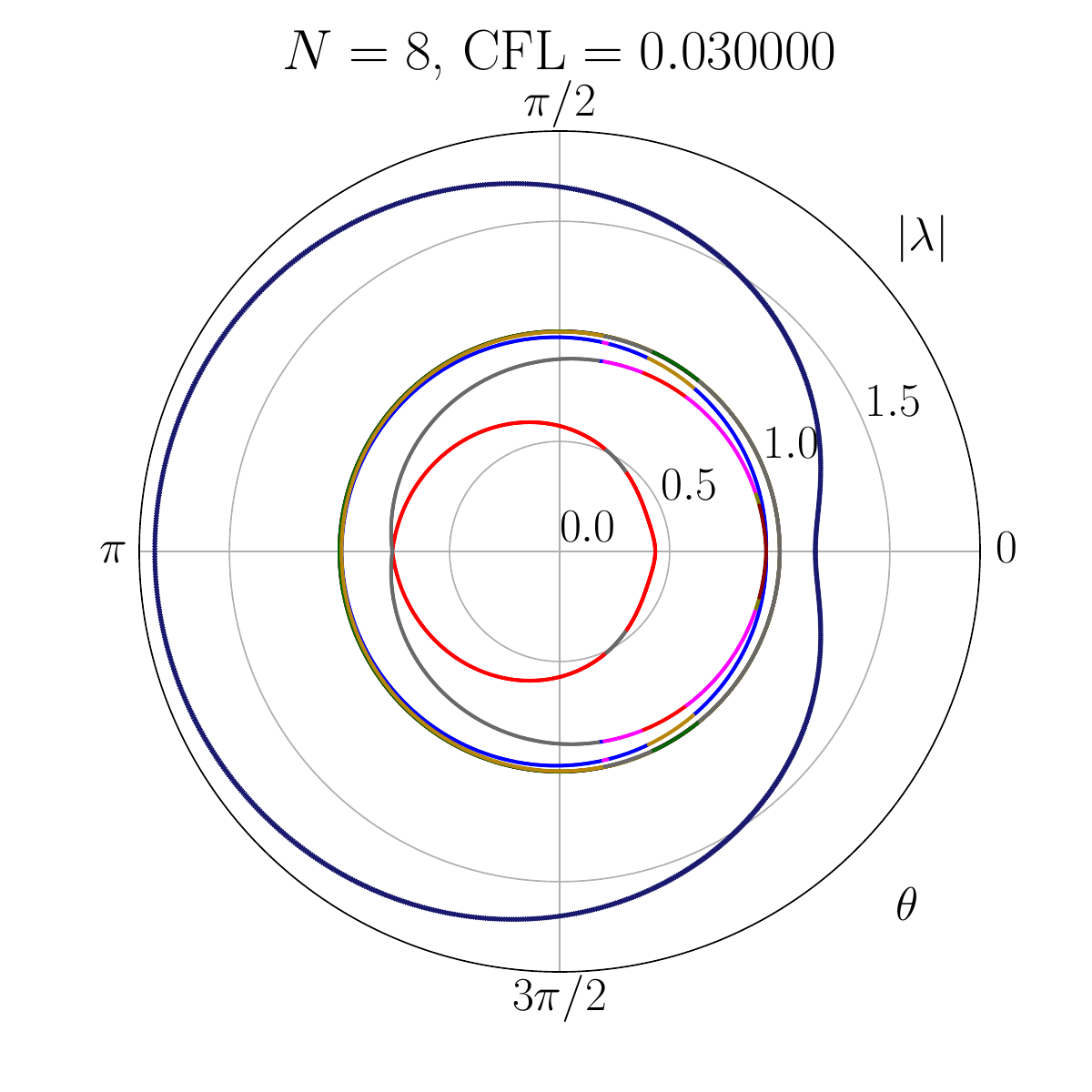}\\
\includegraphics[width=0.028125\textwidth]{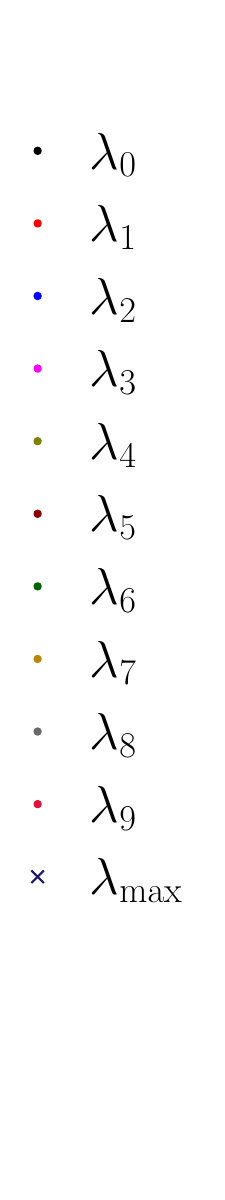}
\includegraphics[width=0.15\textwidth]{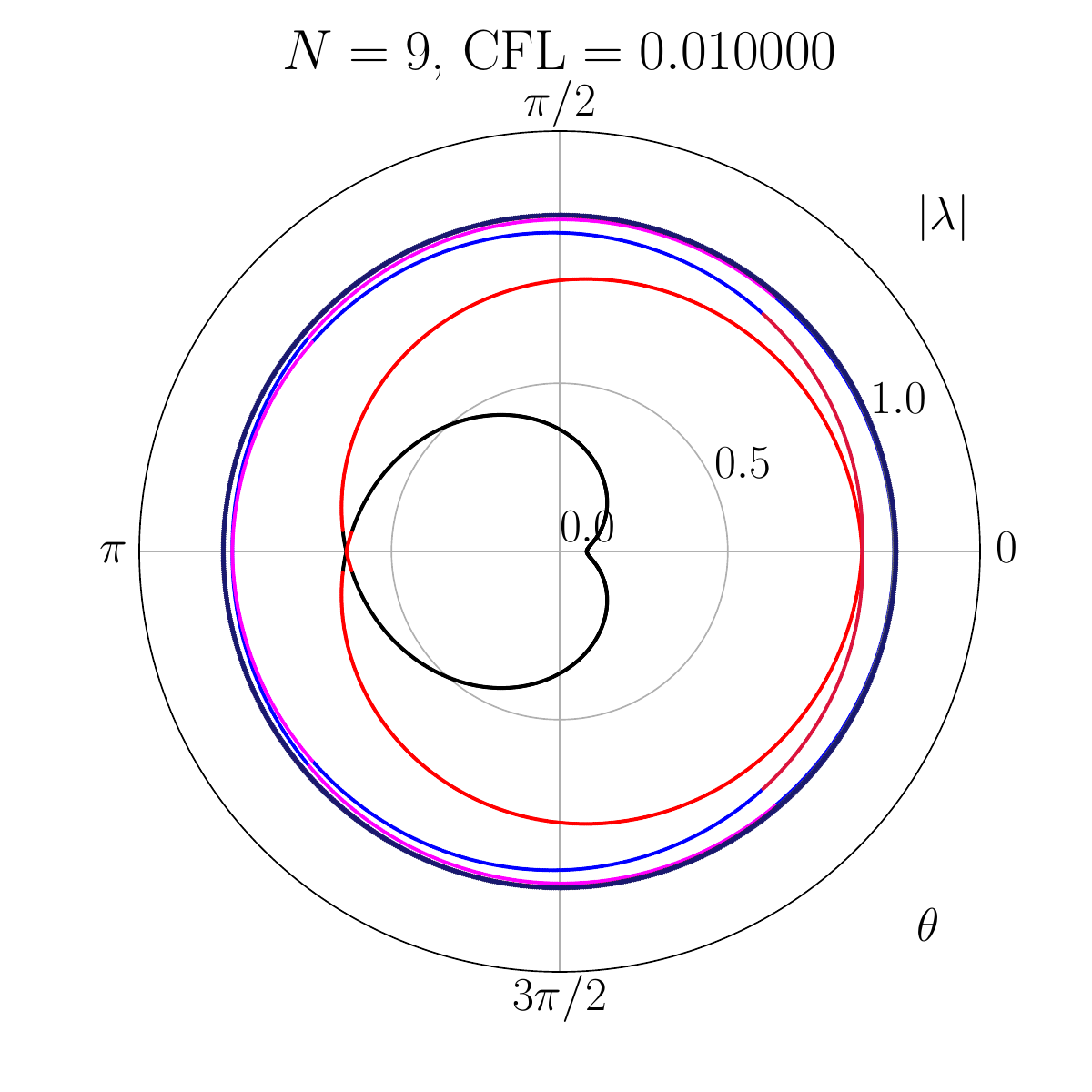}
\includegraphics[width=0.15\textwidth]{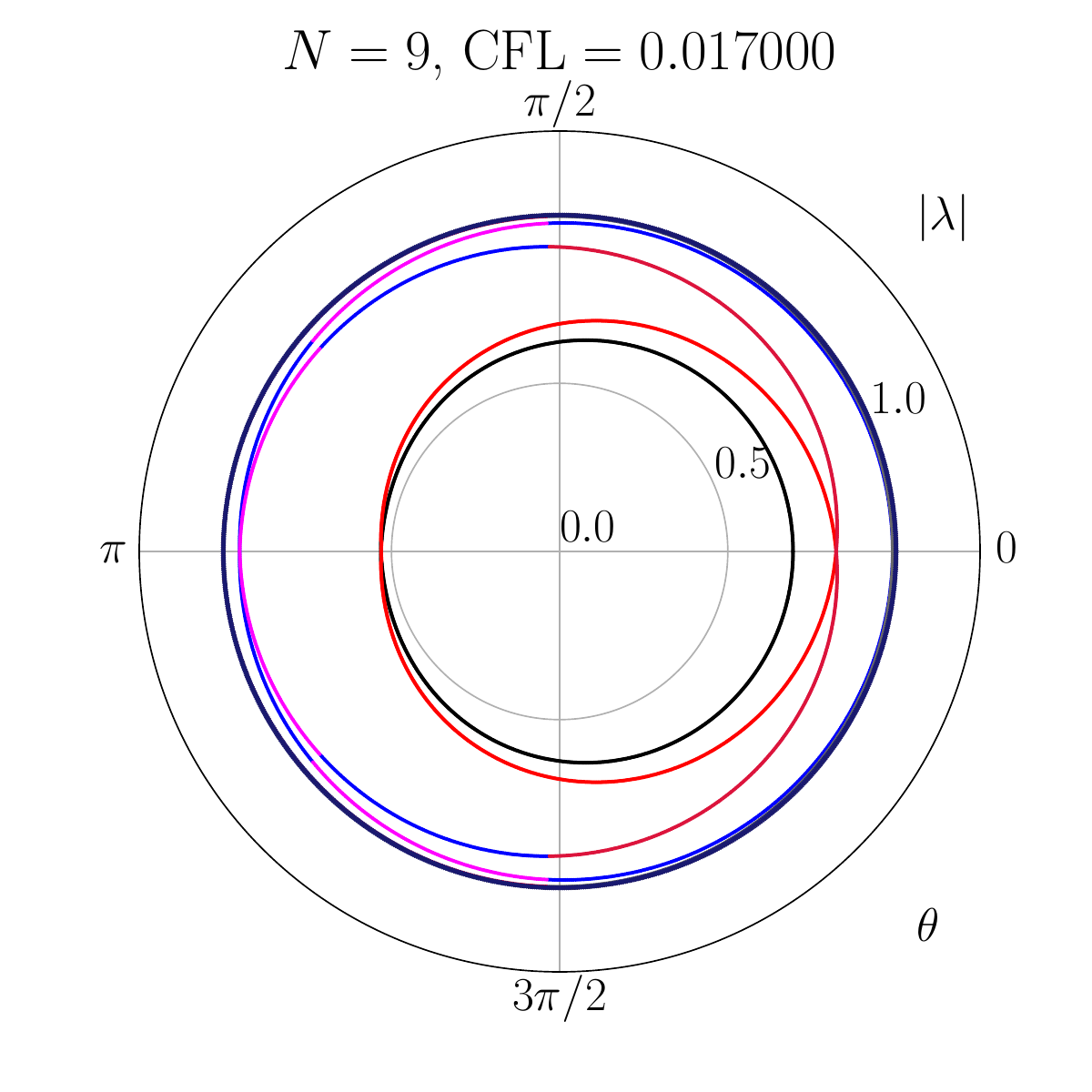}
\includegraphics[width=0.15\textwidth]{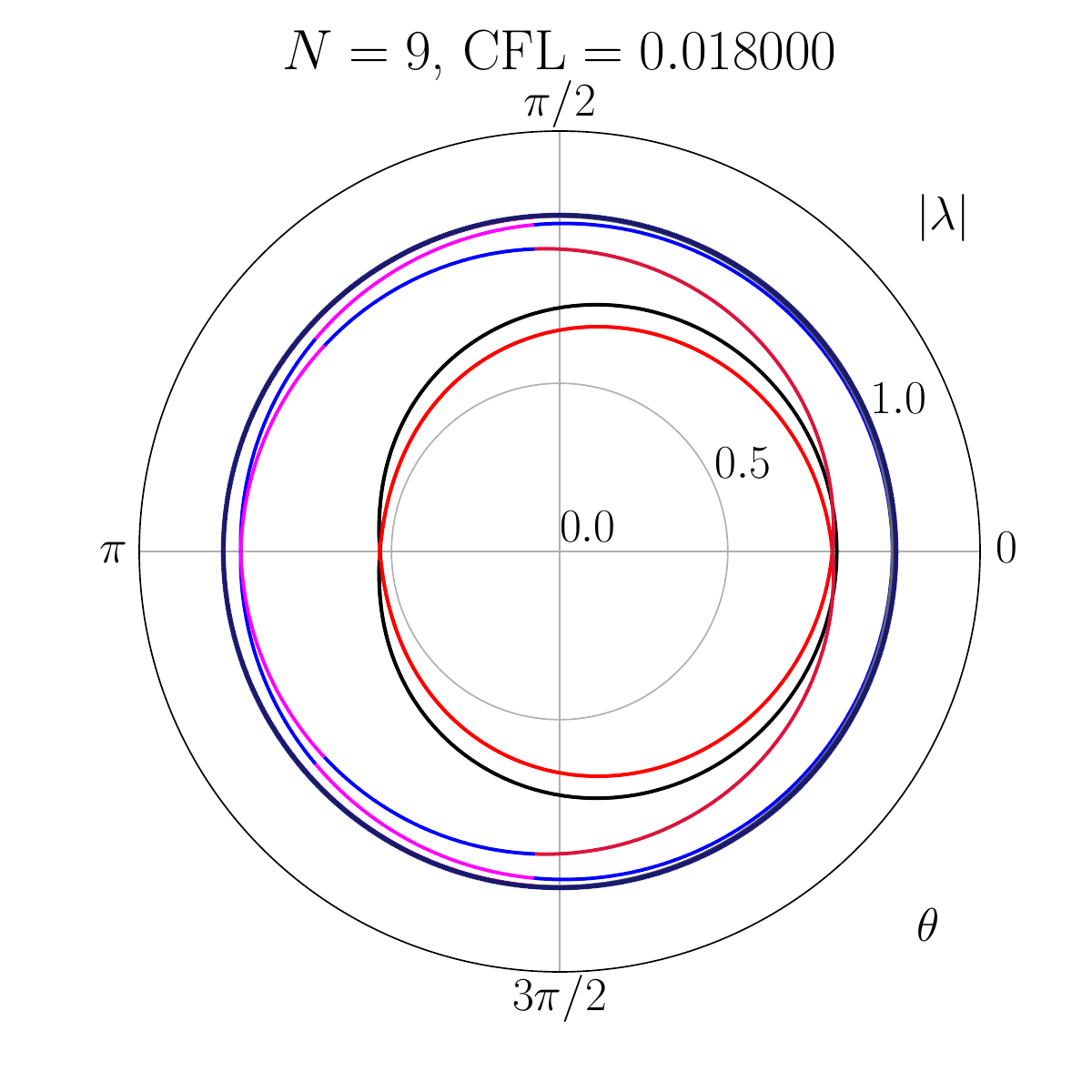}
\includegraphics[width=0.15\textwidth]{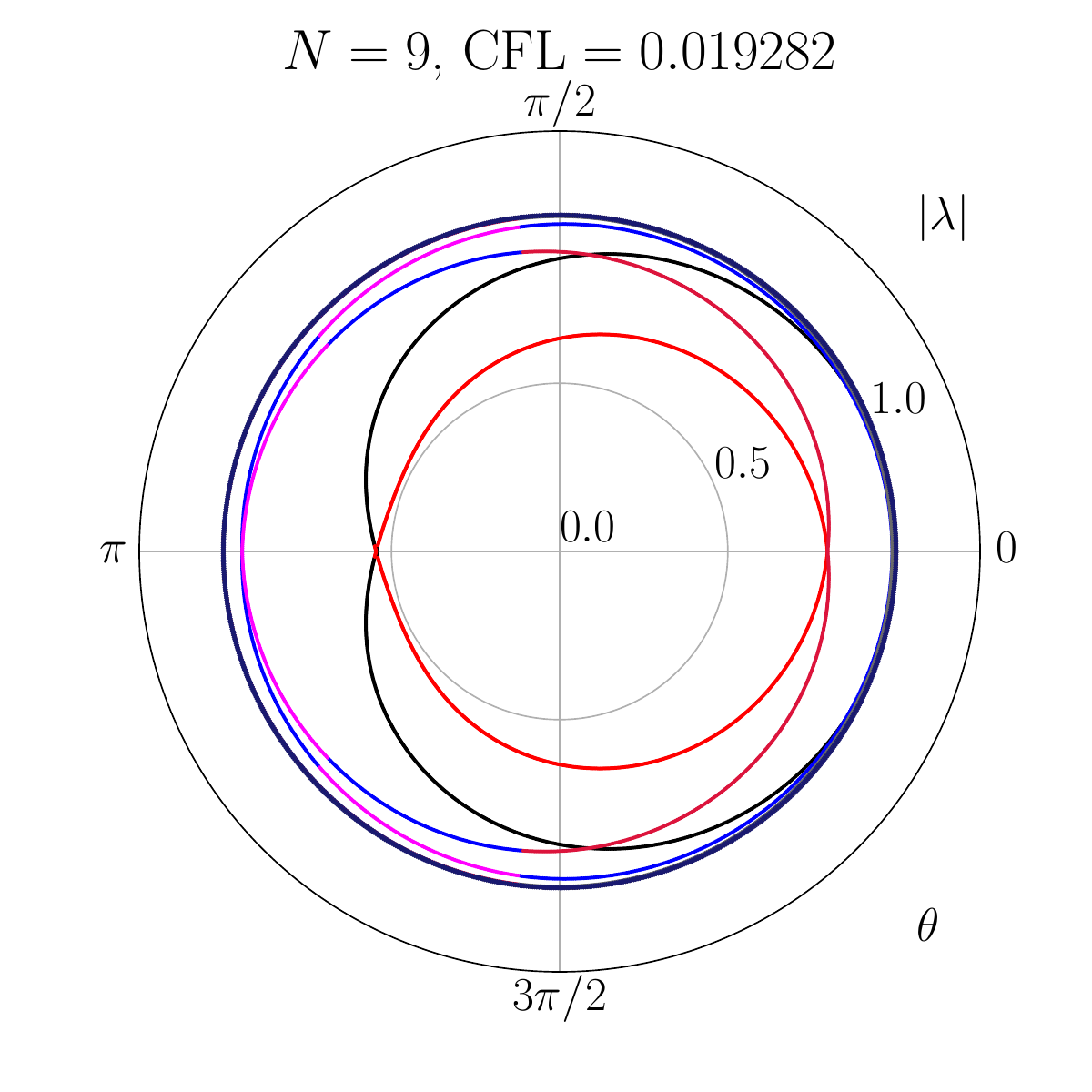}
\includegraphics[width=0.15\textwidth]{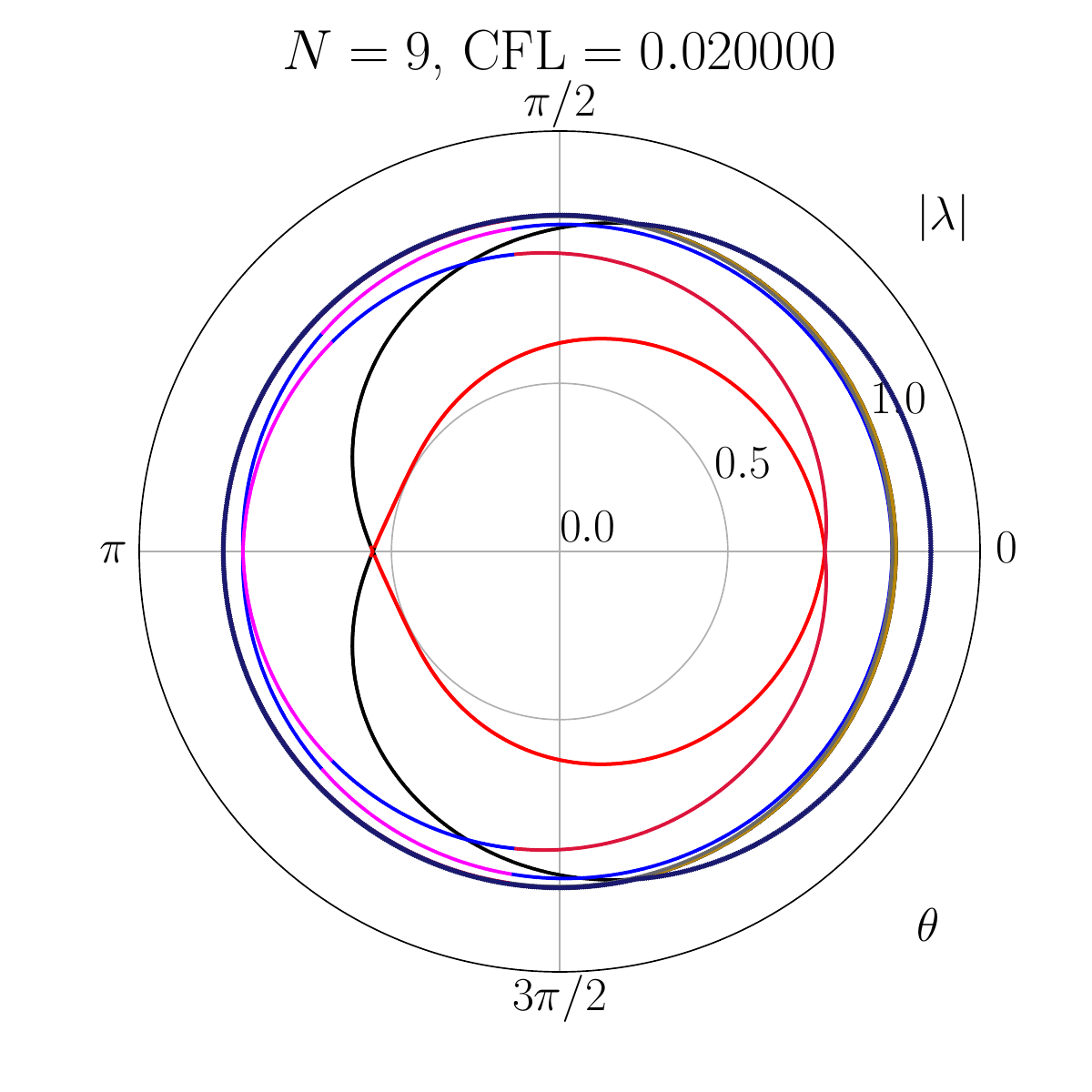}
\includegraphics[width=0.15\textwidth]{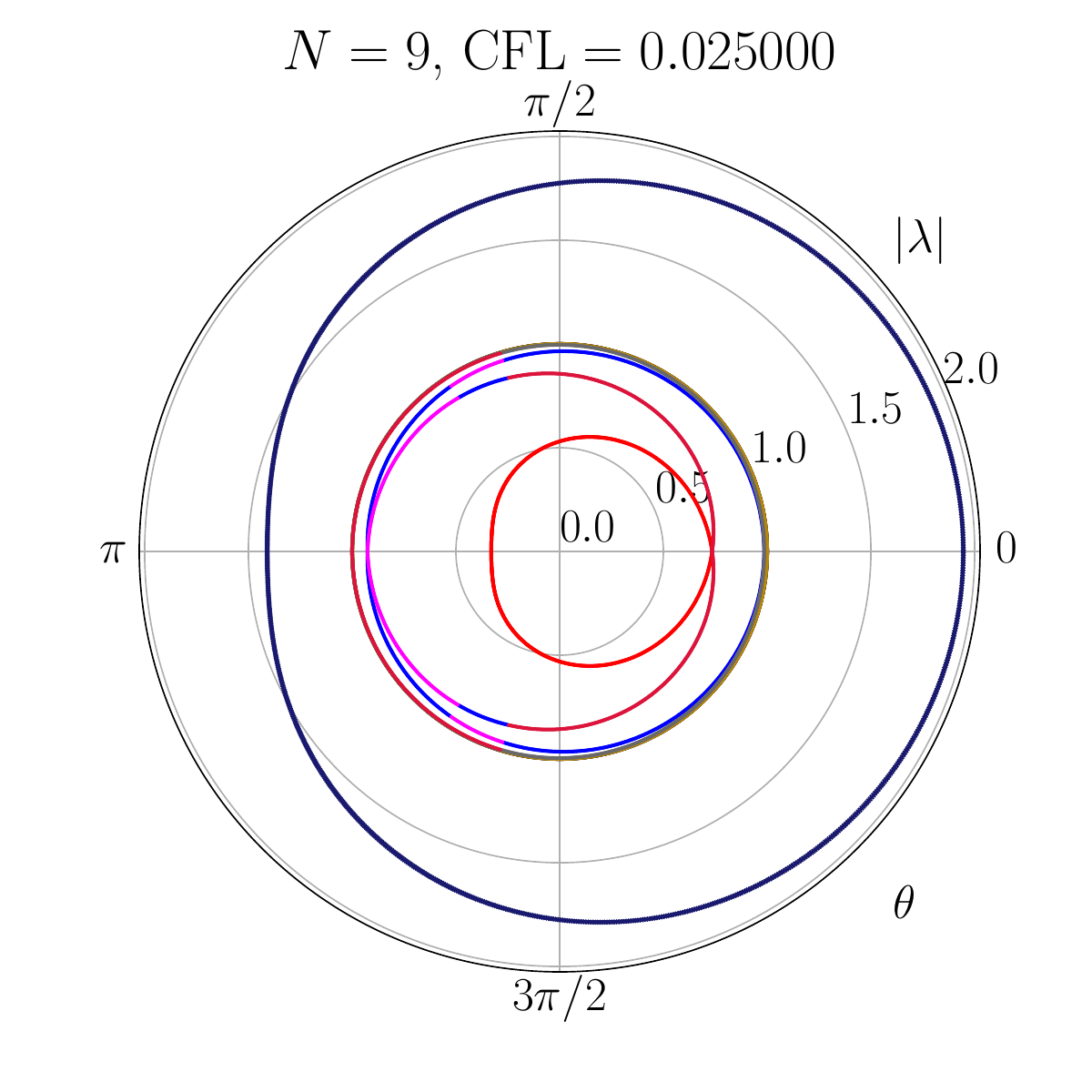}\\
\includegraphics[width=0.028125\textwidth]{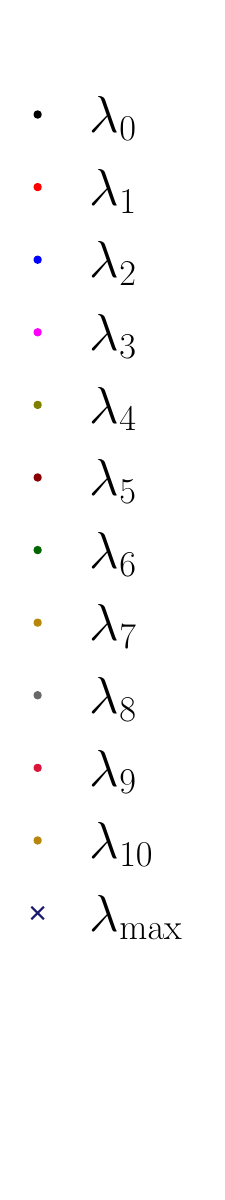}
\includegraphics[width=0.15\textwidth]{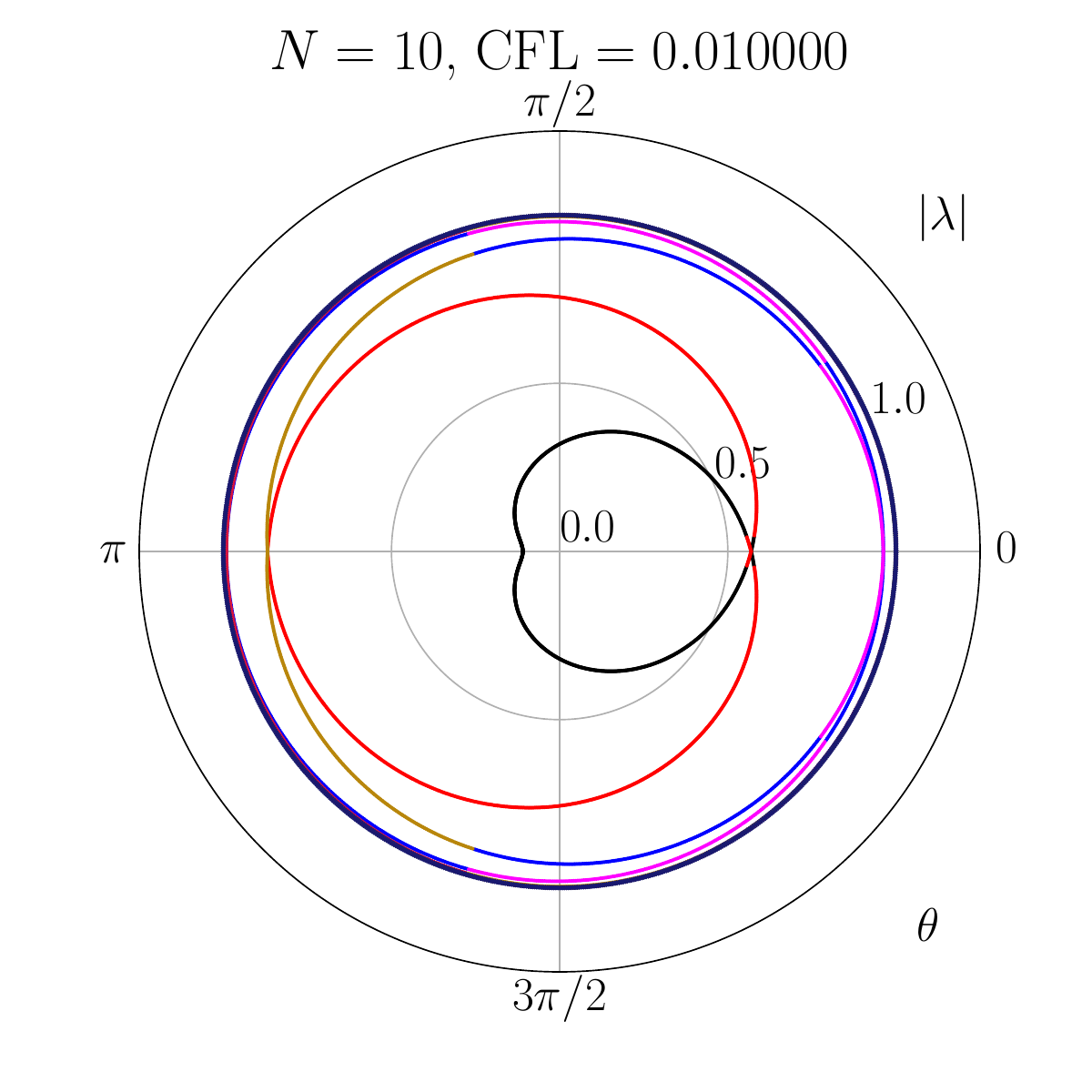}
\includegraphics[width=0.15\textwidth]{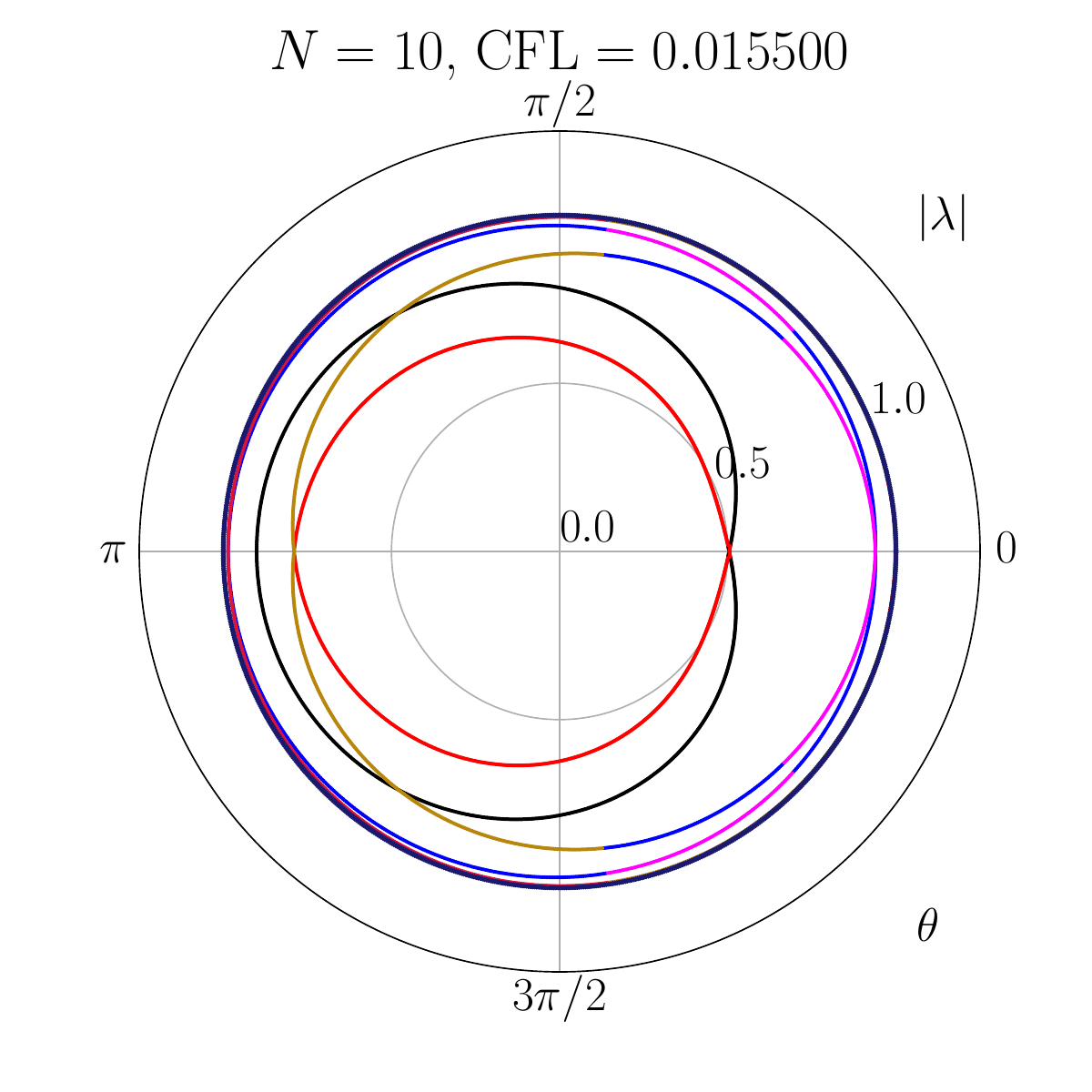}
\includegraphics[width=0.15\textwidth]{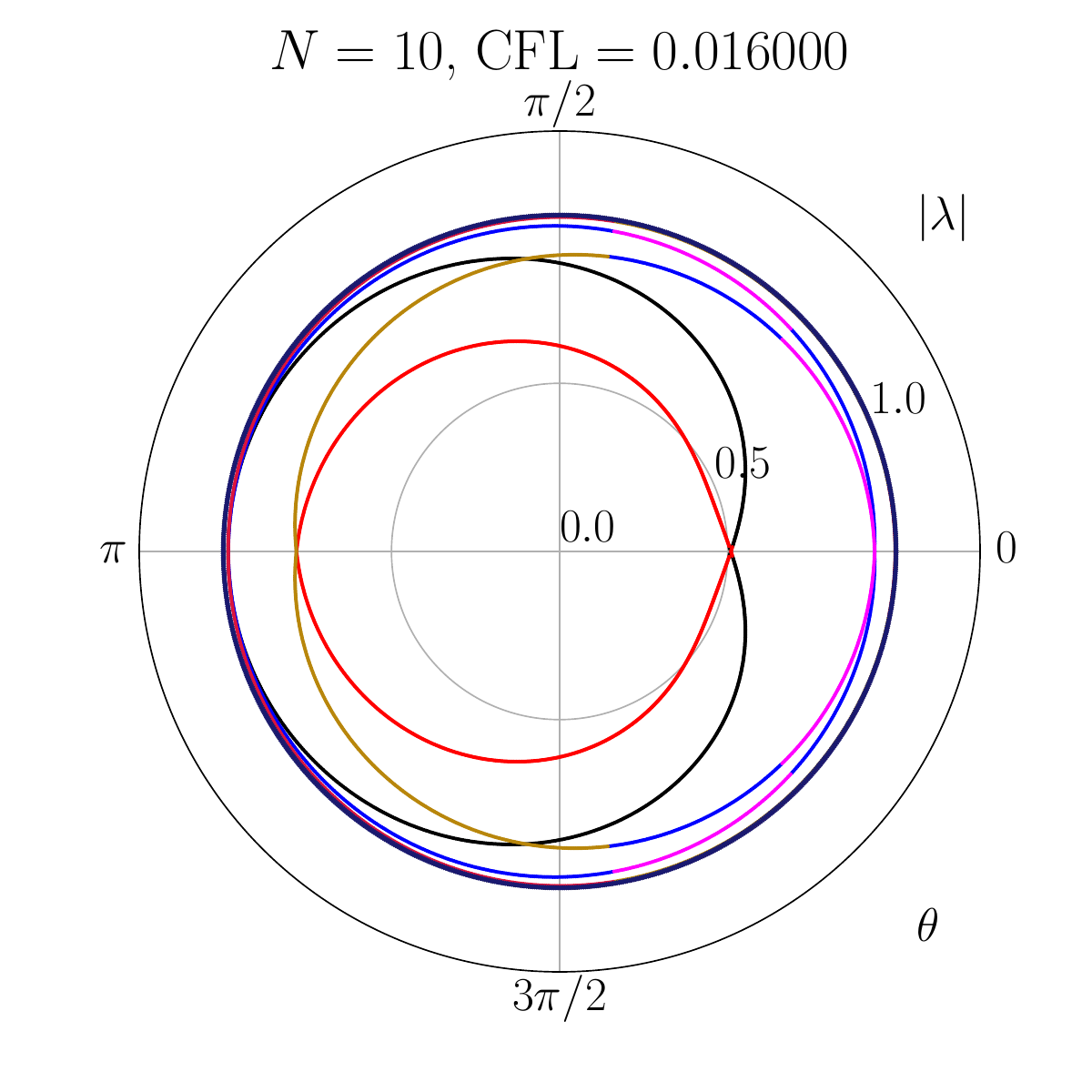}
\includegraphics[width=0.15\textwidth]{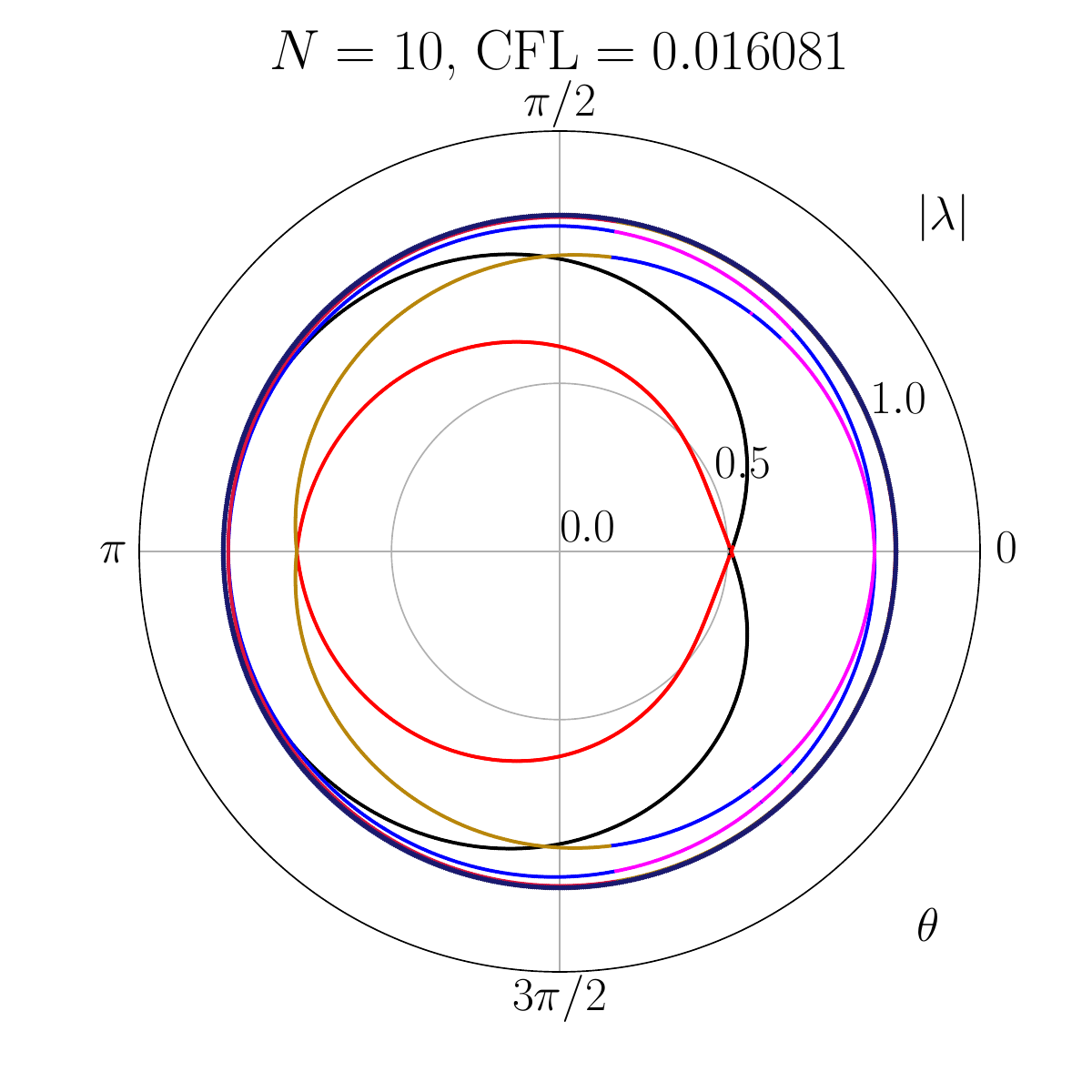}
\includegraphics[width=0.15\textwidth]{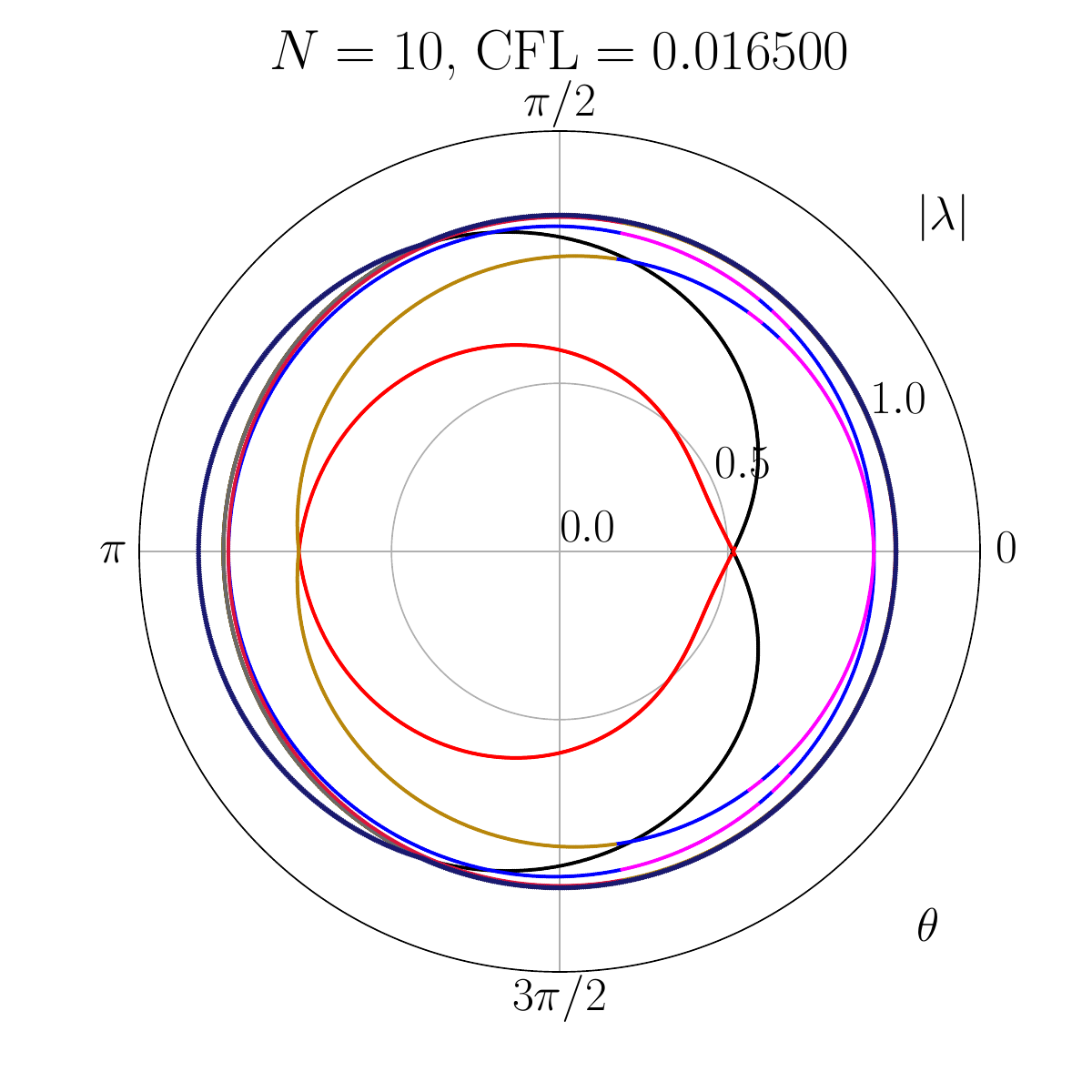}
\includegraphics[width=0.15\textwidth]{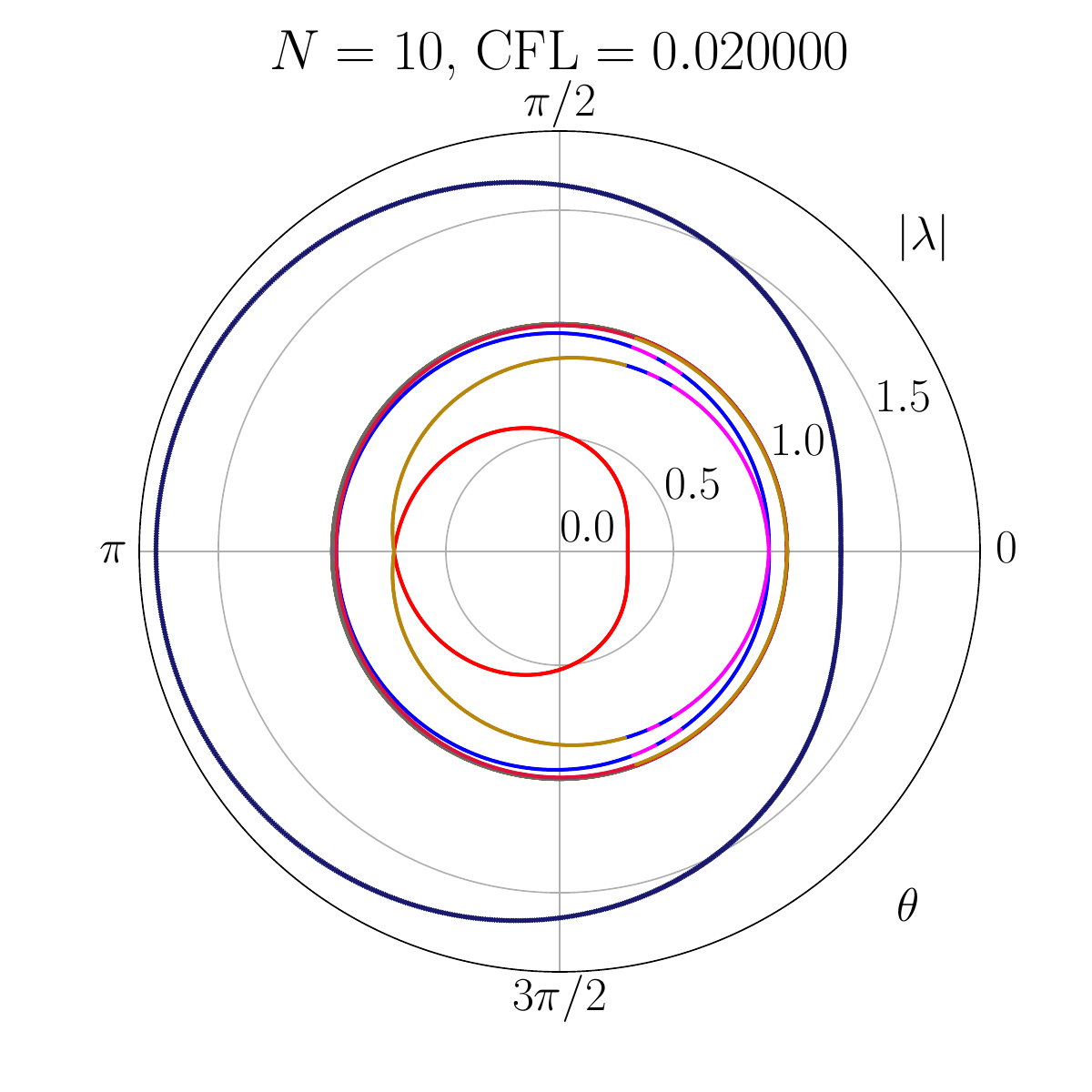}\\
\includegraphics[width=0.028125\textwidth]{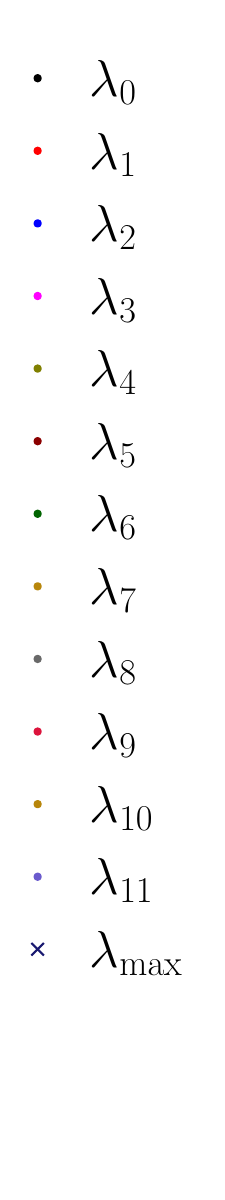}
\includegraphics[width=0.15\textwidth]{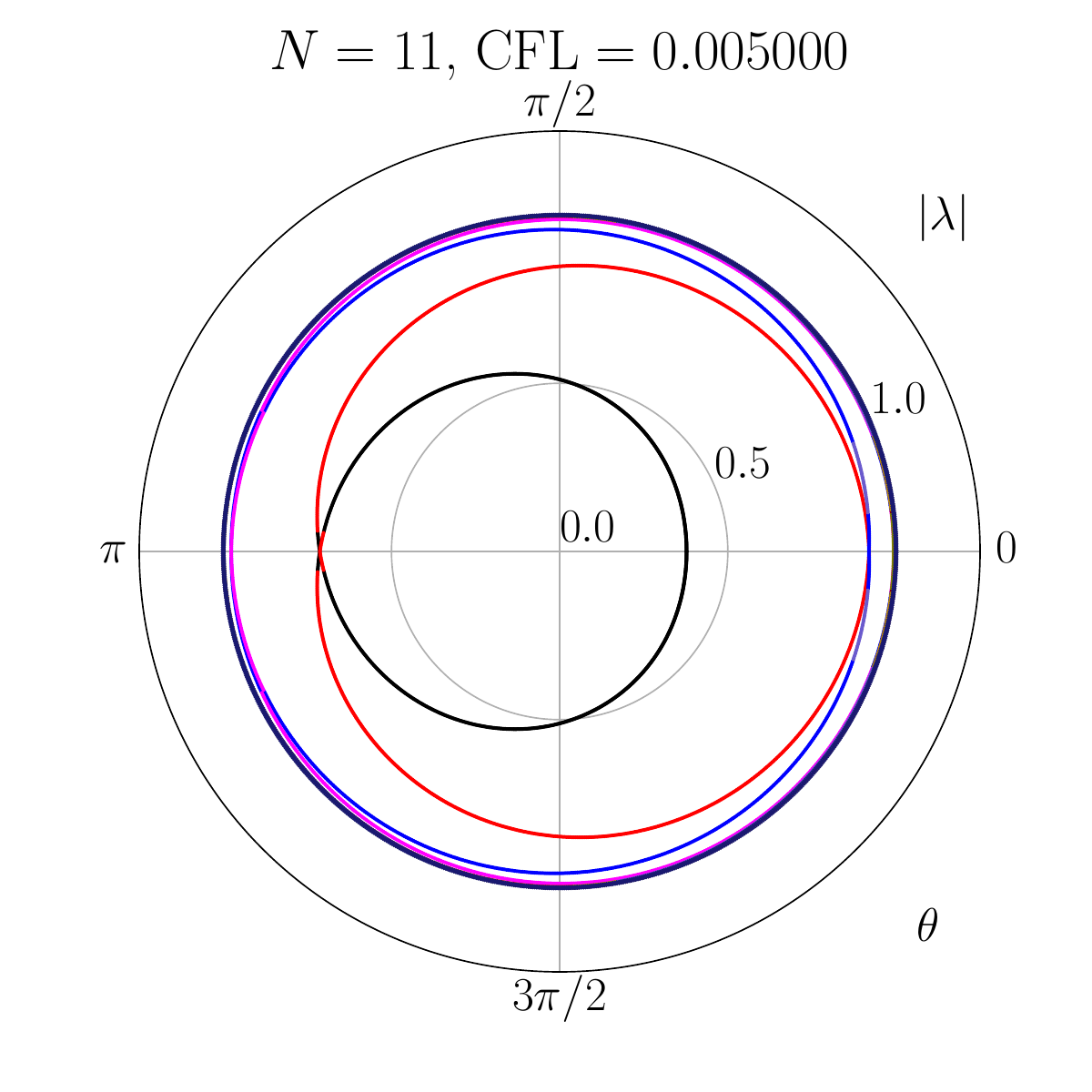}
\includegraphics[width=0.15\textwidth]{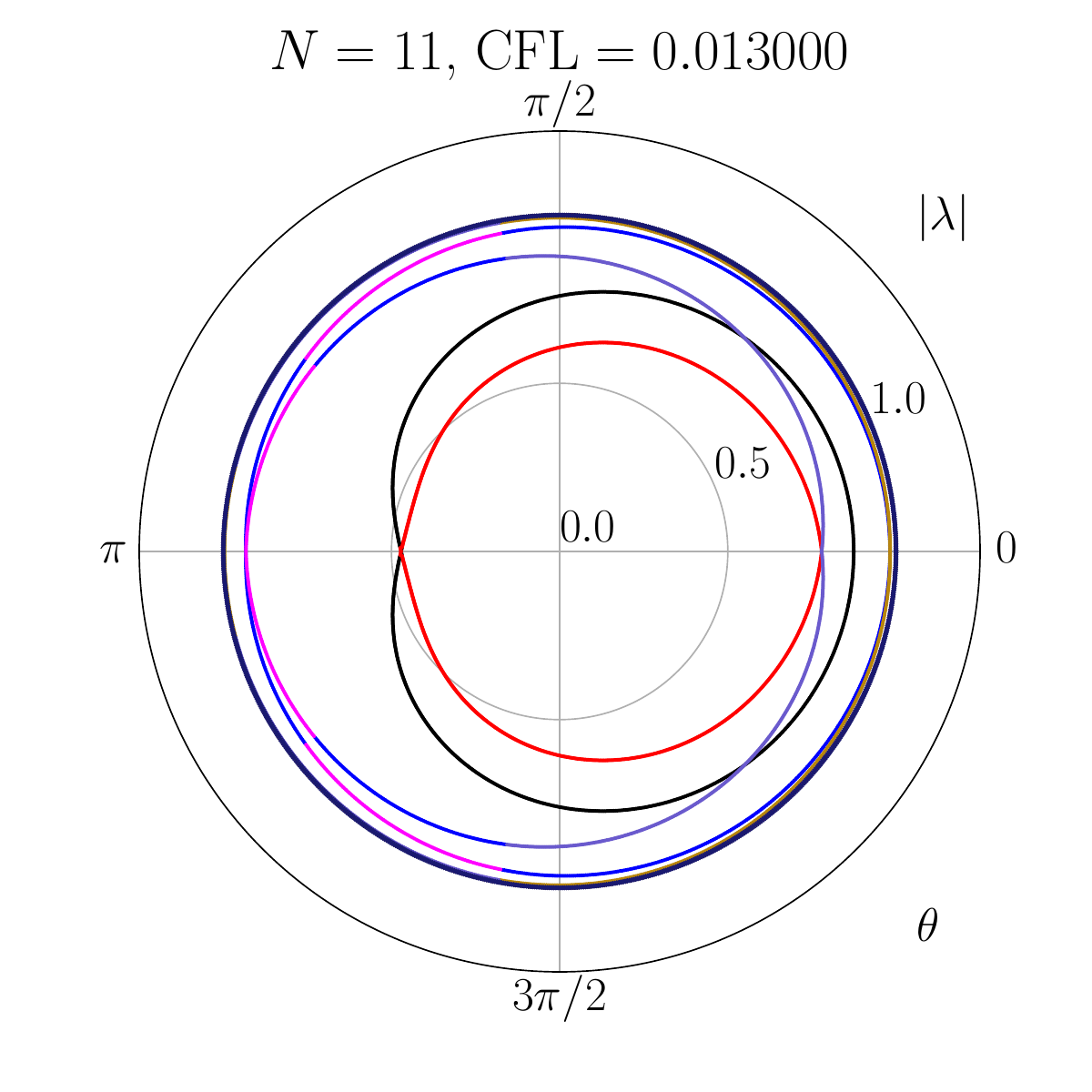}
\includegraphics[width=0.15\textwidth]{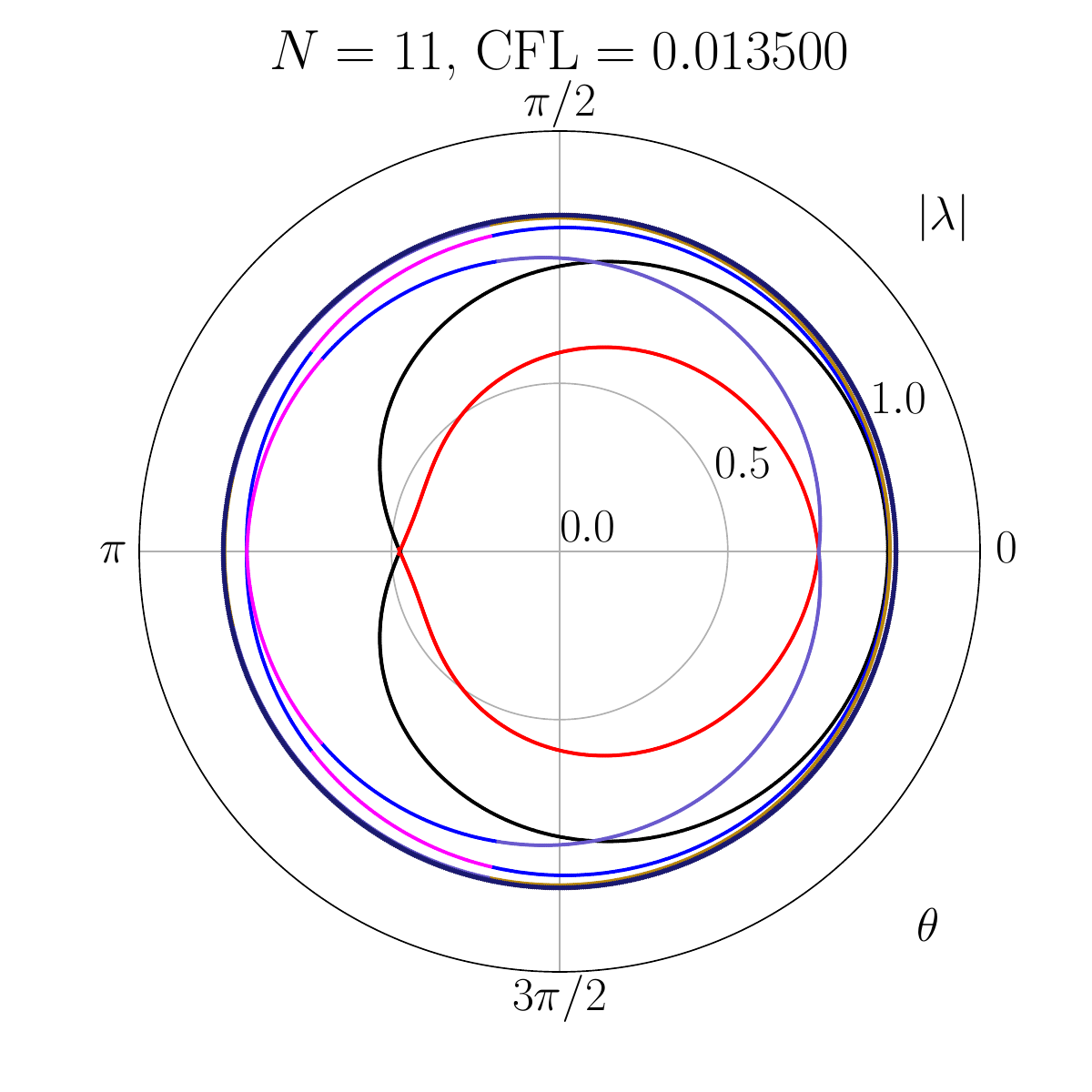}
\includegraphics[width=0.15\textwidth]{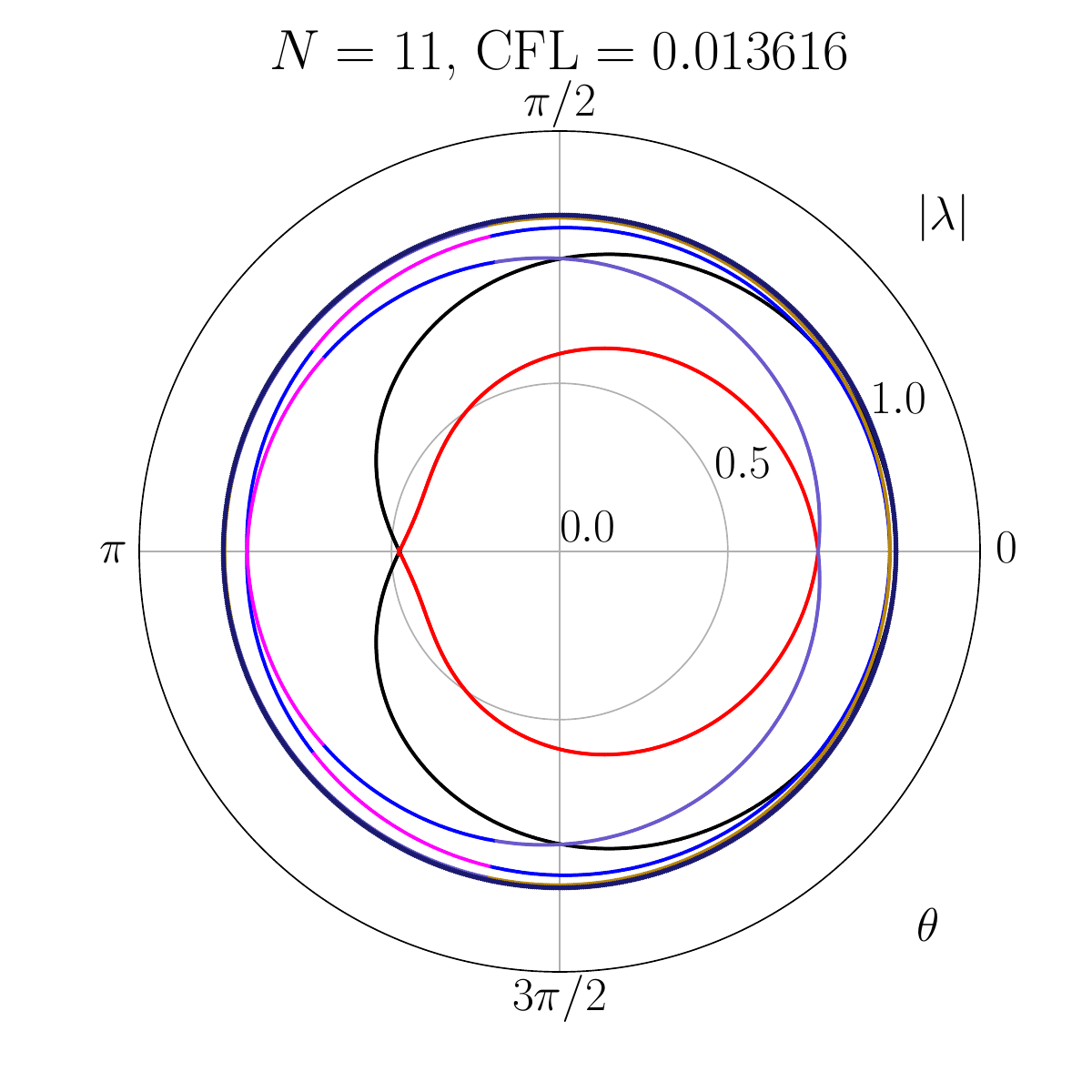}
\includegraphics[width=0.15\textwidth]{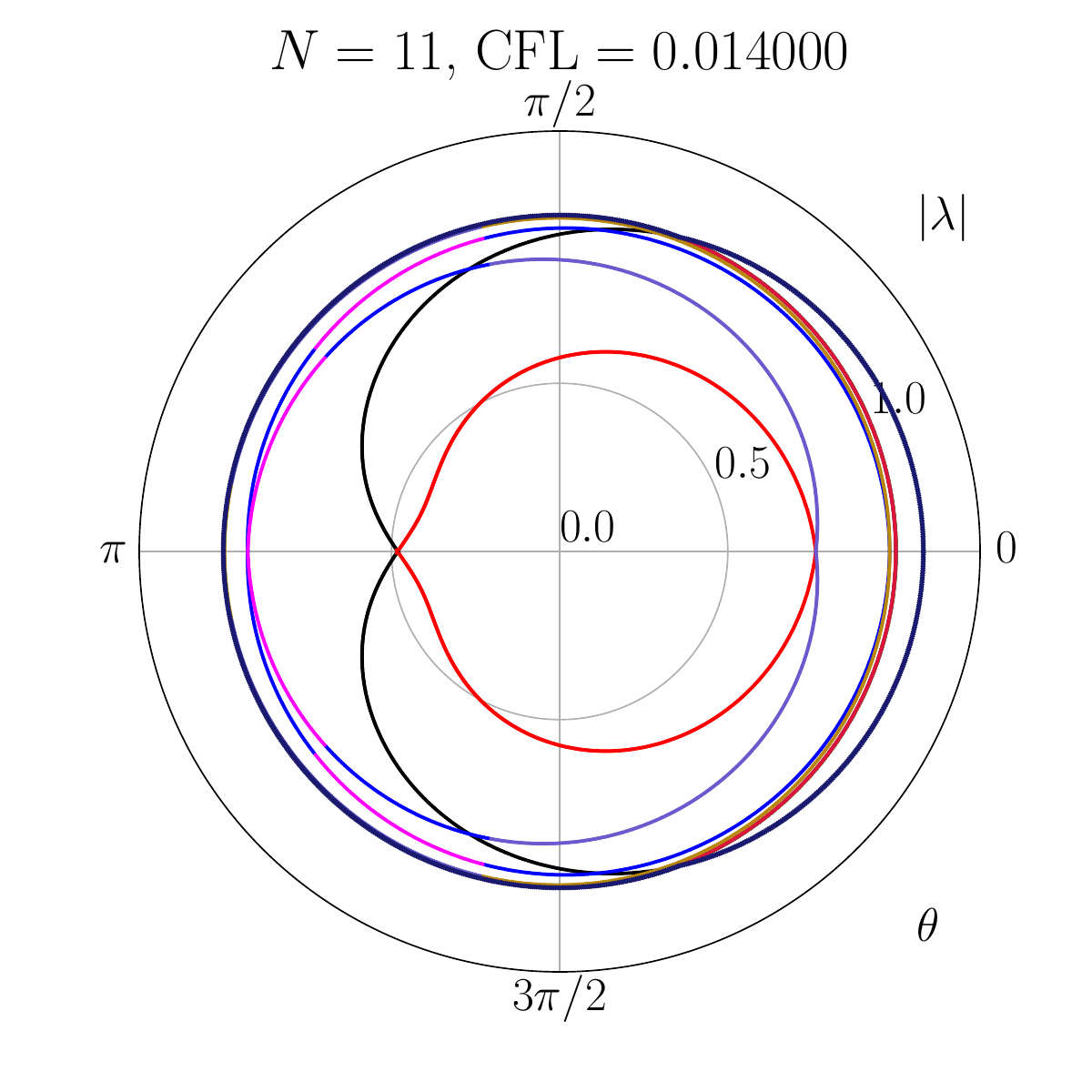}
\includegraphics[width=0.15\textwidth]{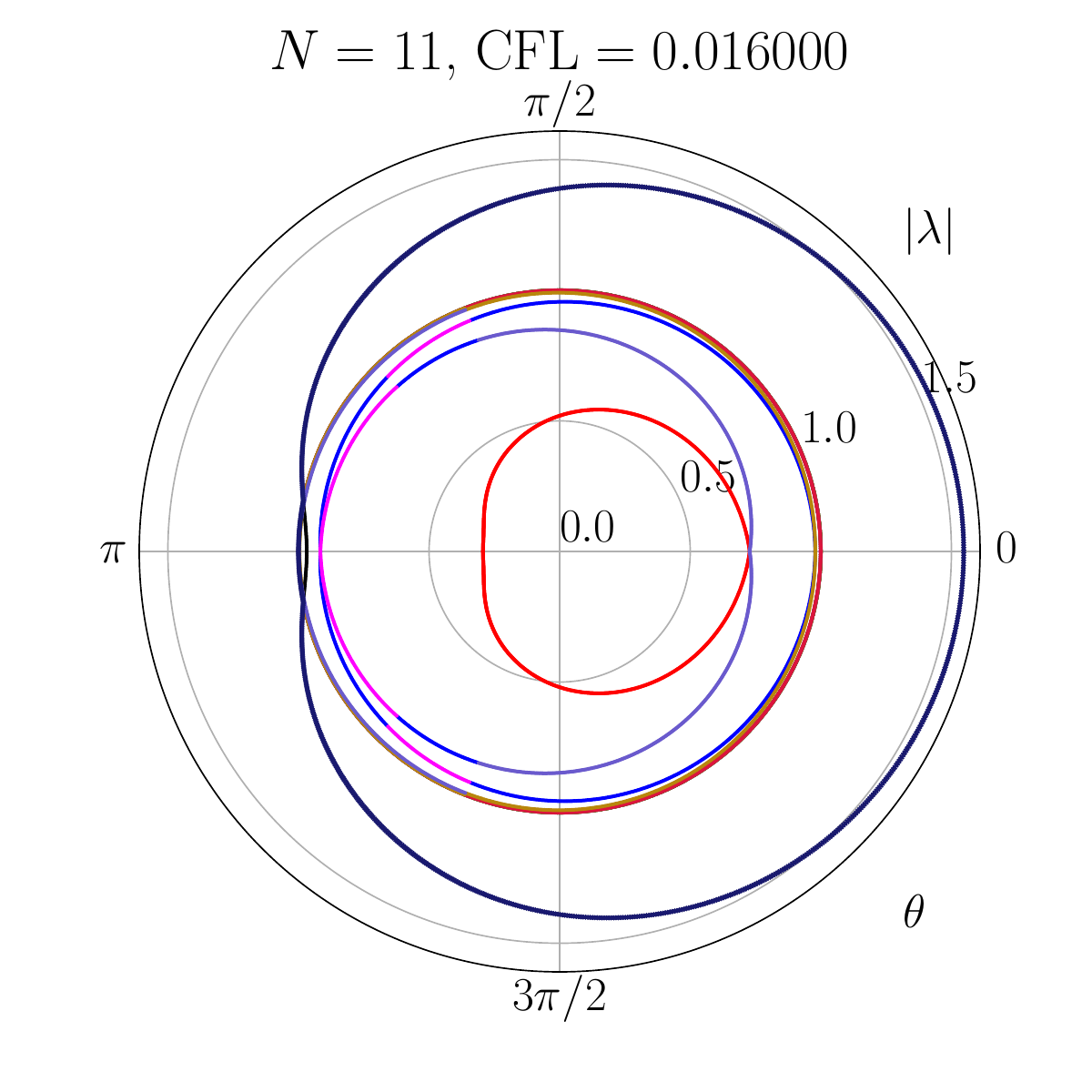}\\
\includegraphics[width=0.028125\textwidth]{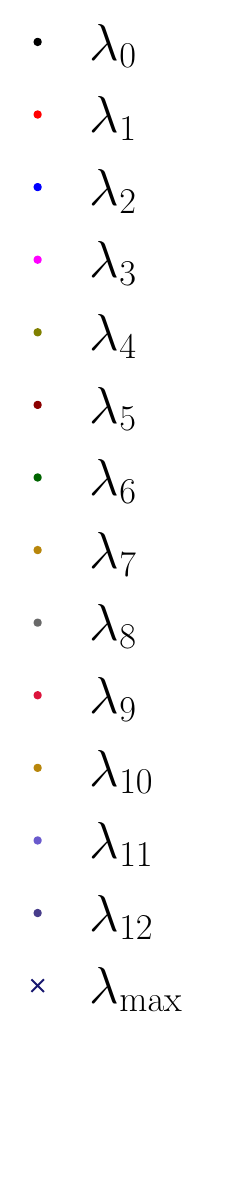}
\includegraphics[width=0.15\textwidth]{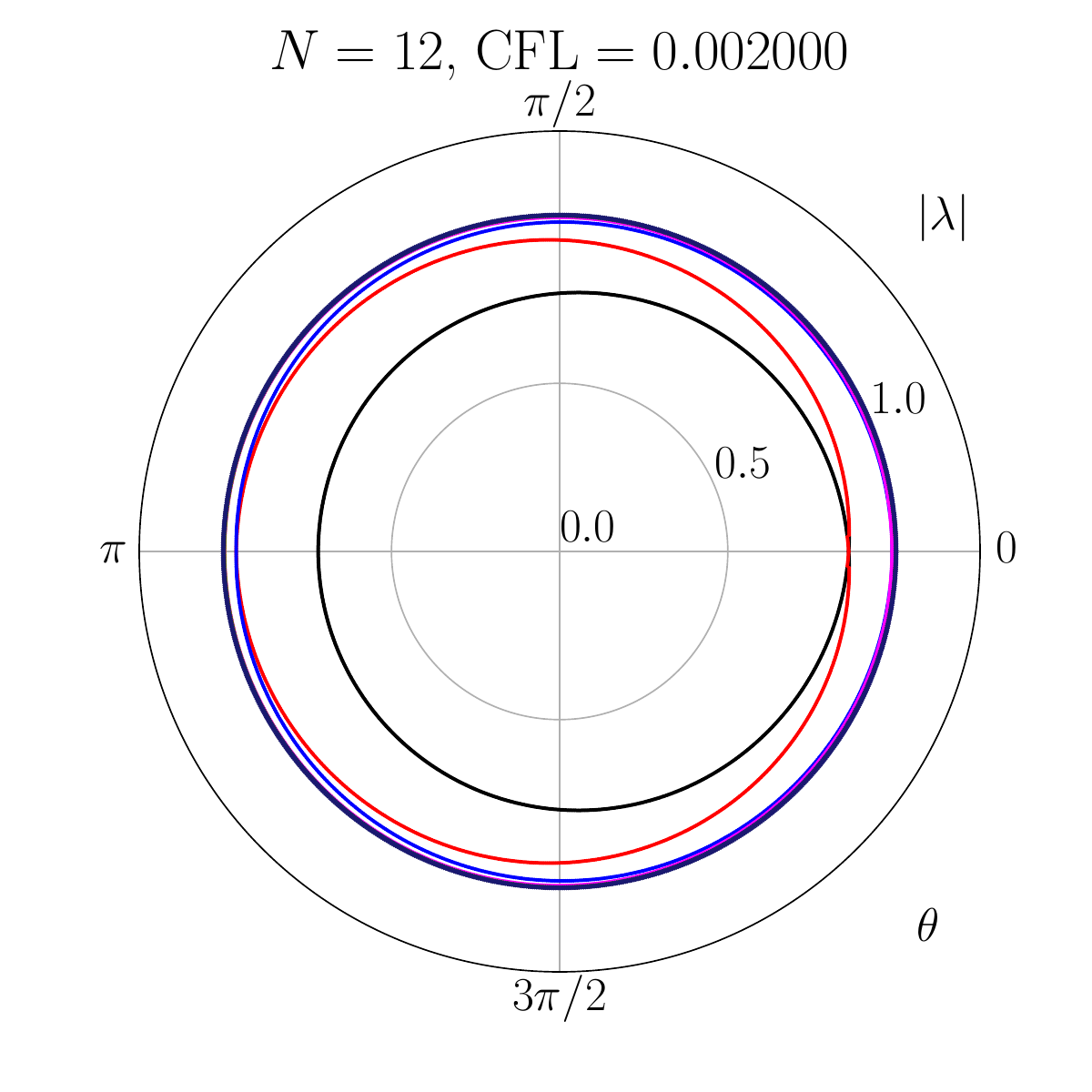}
\includegraphics[width=0.15\textwidth]{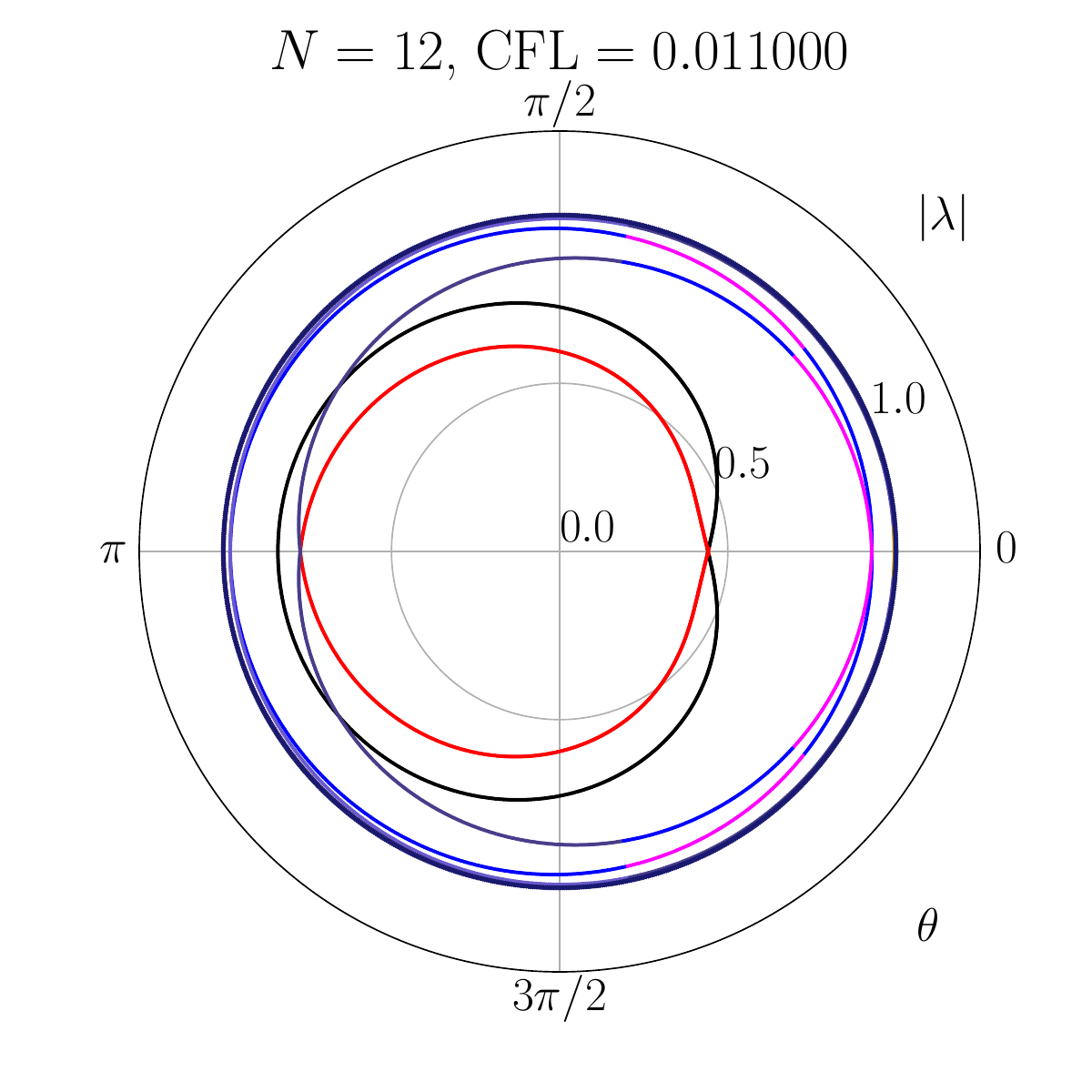}
\includegraphics[width=0.15\textwidth]{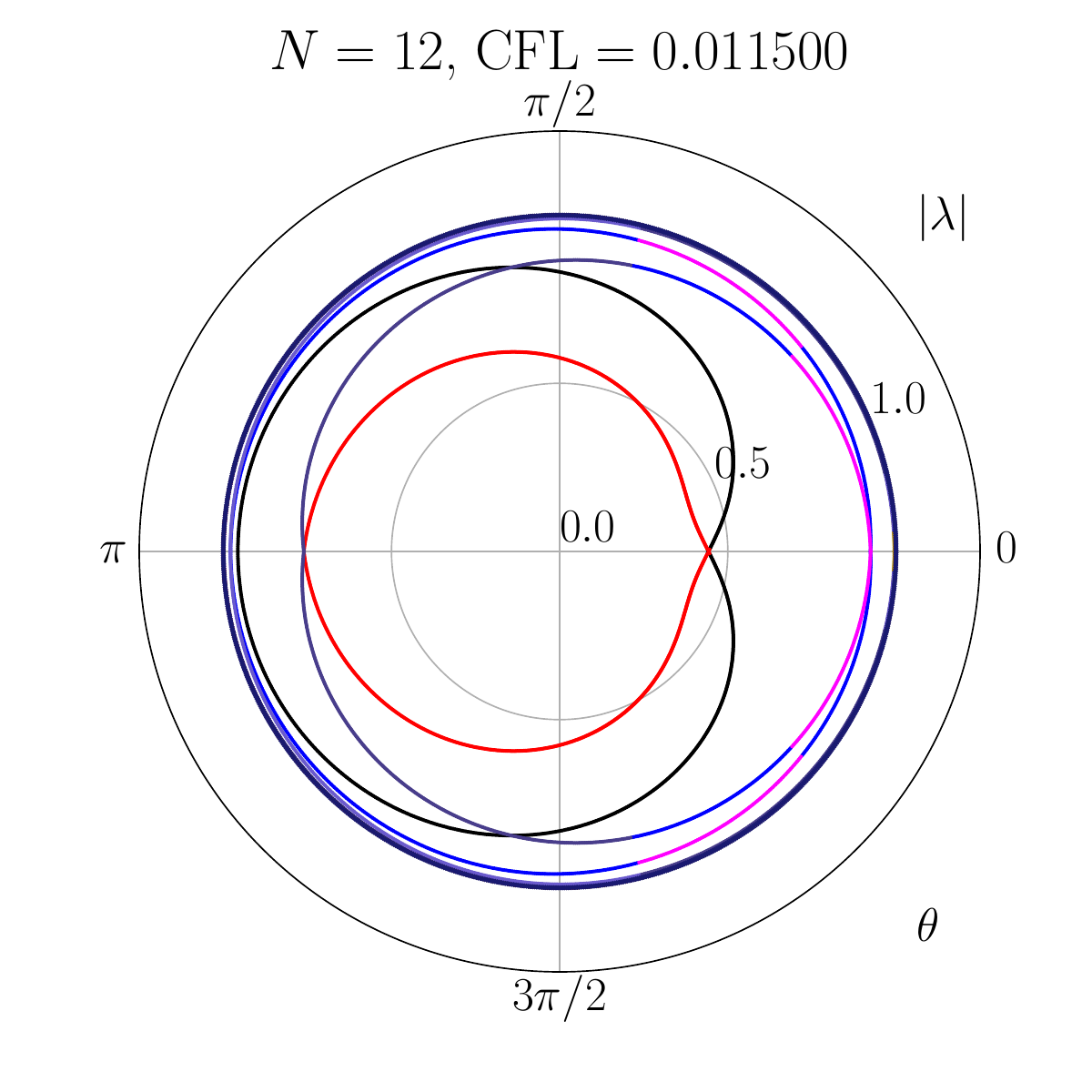}
\includegraphics[width=0.15\textwidth]{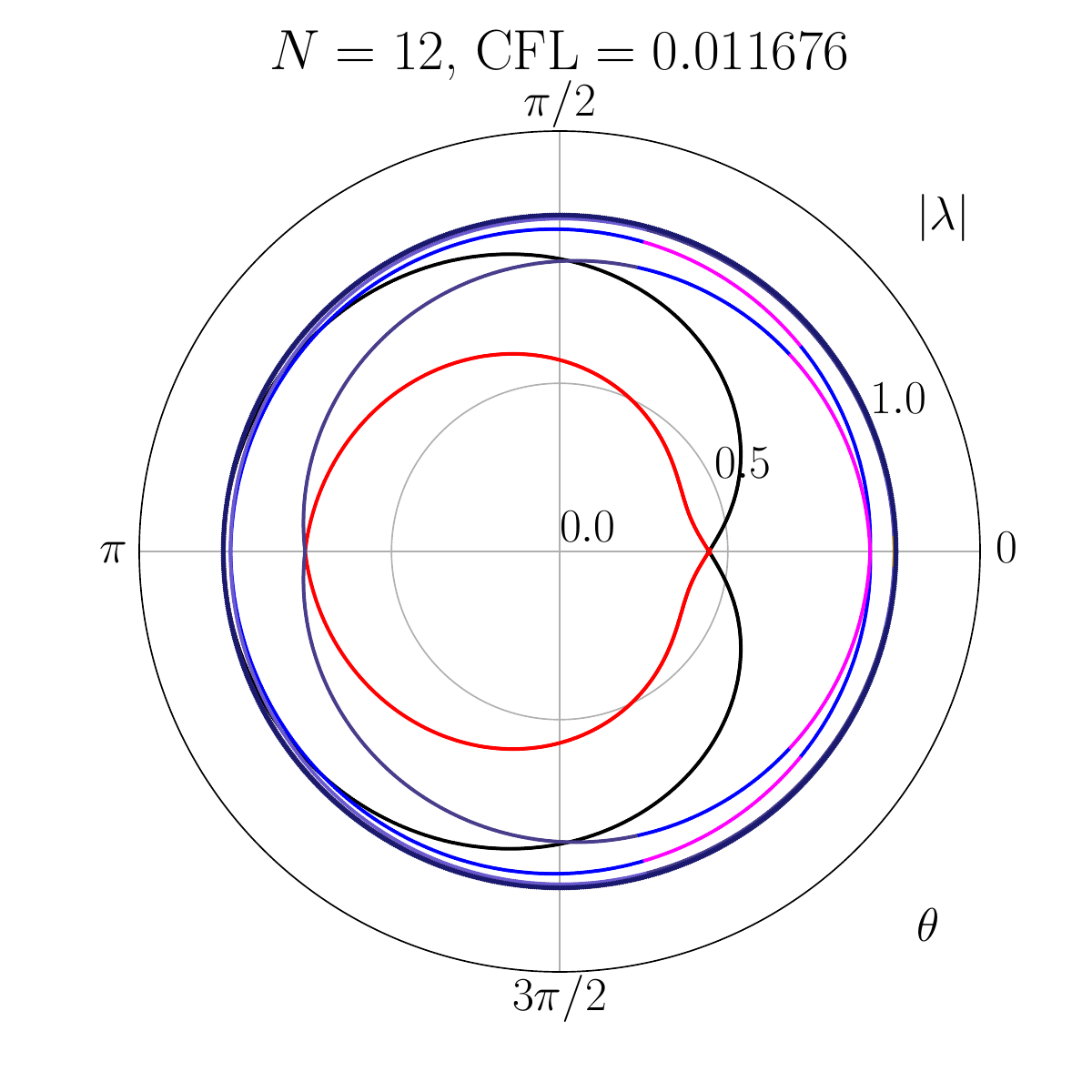}
\includegraphics[width=0.15\textwidth]{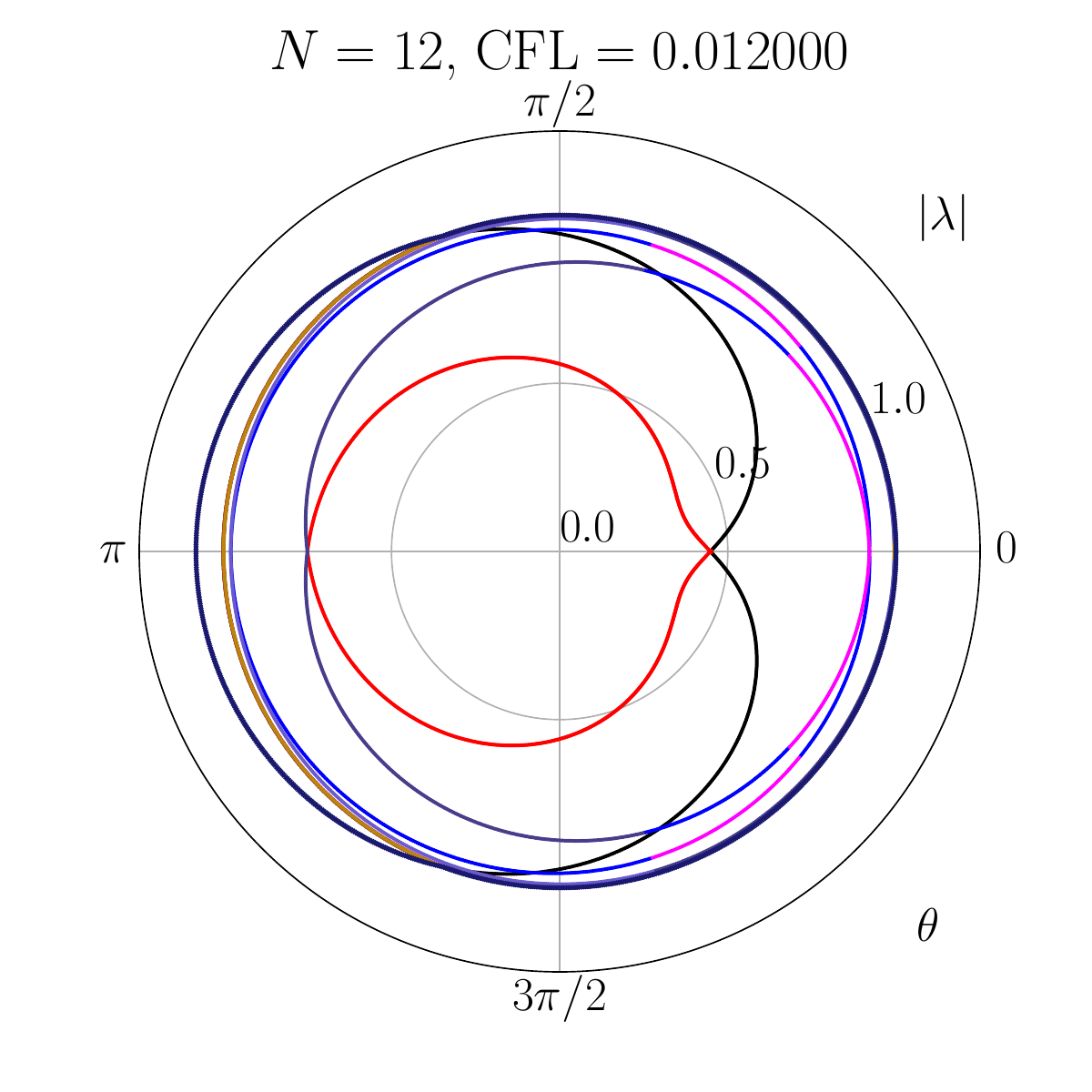}
\includegraphics[width=0.15\textwidth]{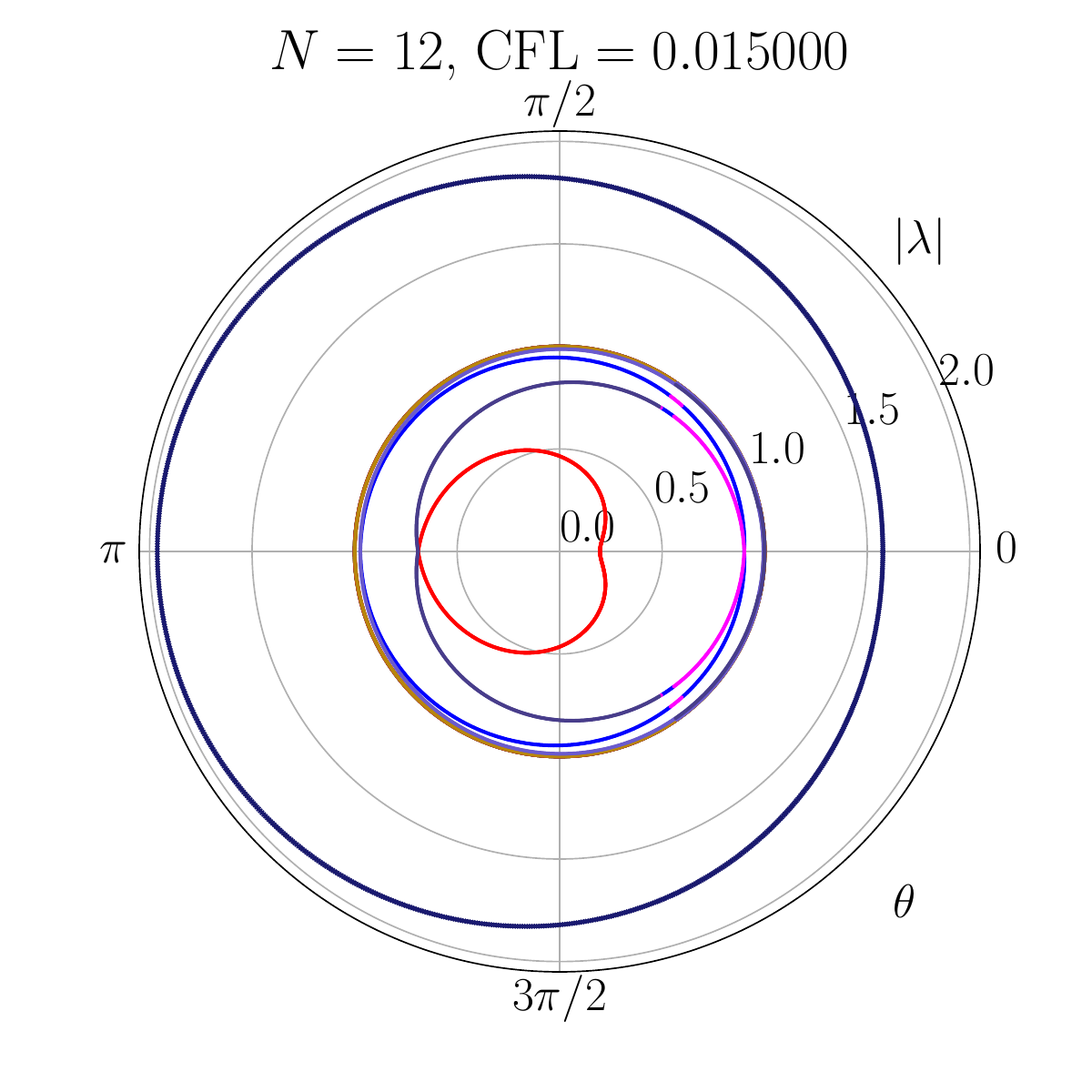}\\
\caption{%
The dependence of the absolute values $|\lambda_{k}|$ of the spectrum of the matrix $\mathrm{R}(\mathrm{CFL}, \theta)$ (\ref{eq:r_matrix_elems_expr_dup}) (eigenvalues $\lambda_{k} = \lambda_{k}(\mathrm{CFL}, \theta)$, $k = 0, \ldots, N$) of the evolution operator $R$ (\ref{eq:evol_oper_prop}) for a single time step $\Dtn{n}$ on phase $\theta = k\Dx$ for several selected values of the Courant number $\mathrm{CFL}$ for polynomial degrees $N = 1, \ldots, 6$ --- the polar plot with phase $\theta$ as angle and absolute value $|\lambda|$ as radius. The range of phase $\theta\in[0, 2\pi)$ is sampled on a uniform grid of $1000$ nodes. Legends for each row of the graphs are located on the left. The gray circle of unit radius defines the stability boundary. The Courant numbers $\mathrm{CFL}$ are taken deep inside the stability region $\mathrm{CFL} \in [0, \mathrm{CFL}_{\rm max}]$ (left column), in the region of guaranteed instability $\mathrm{CFL} > \mathrm{CFL}_{\rm max}$ (right column) and near the boundary of the stability region $\mathrm{CFL}_{\rm max}$ (the values $\mathrm{CFL}_{\rm max}$ are selected from work~\cite{ader_dg_stab} for $N = 7$, $8$, $9$, as well as the values $\mathrm{CFL}_{\rm max}$ calculated further in this work in Table~\ref{tab:cfls_max_data} and in Figure~\ref{fig:cfls_max_data}). \textit{Note}: the phase $\theta$ is not the phase $\arg \lambda_{k}$ of eigenvalue $\lambda_{k}$; the dependence of eigenvalues $\lambda_{k}$ of the matrix $\mathrm{R}(\mathrm{CFL}, \theta)$ (\ref{eq:r_matrix_elems_expr_dup}) for these polynomial degrees $N$ is presented in Figure~\ref{fig:spectrum_exact_eigvals_degrees_7_12}.
}
\label{fig:rhos_on_theta_degrees_7_12}
\end{figure}

\begin{figure}[h!]
\centering
\includegraphics[width=0.028125\textwidth]{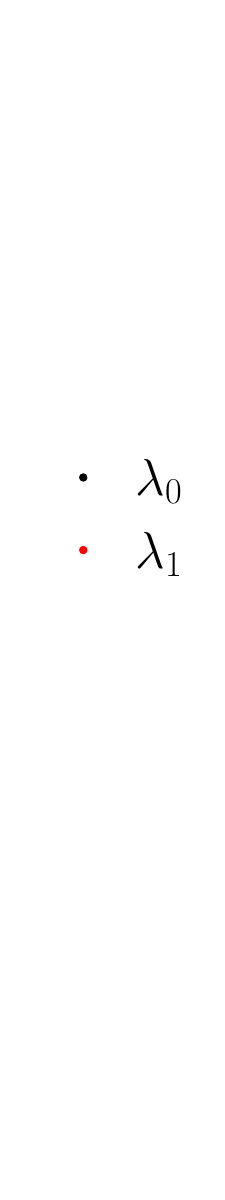}
\includegraphics[width=0.15\textwidth]{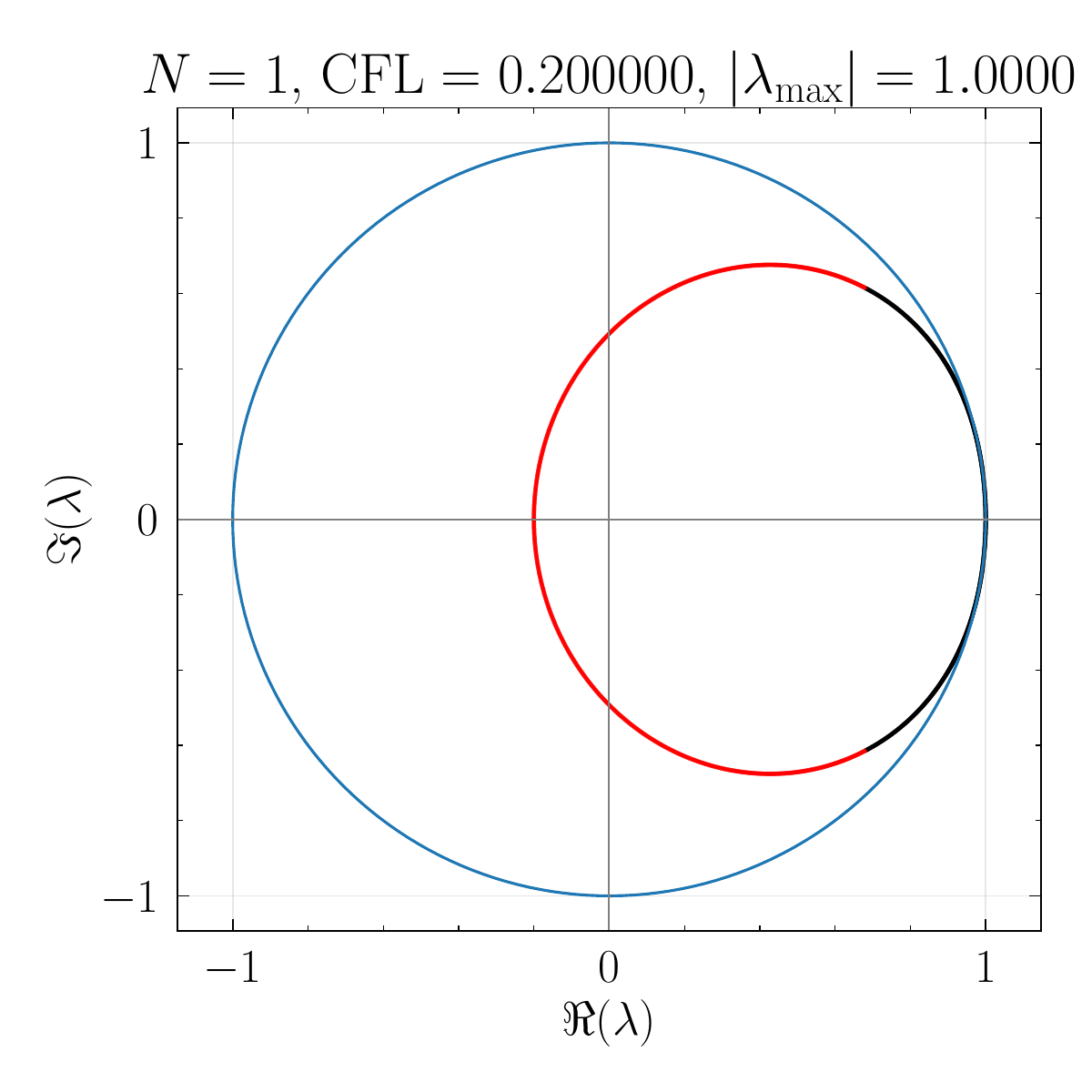}
\includegraphics[width=0.15\textwidth]{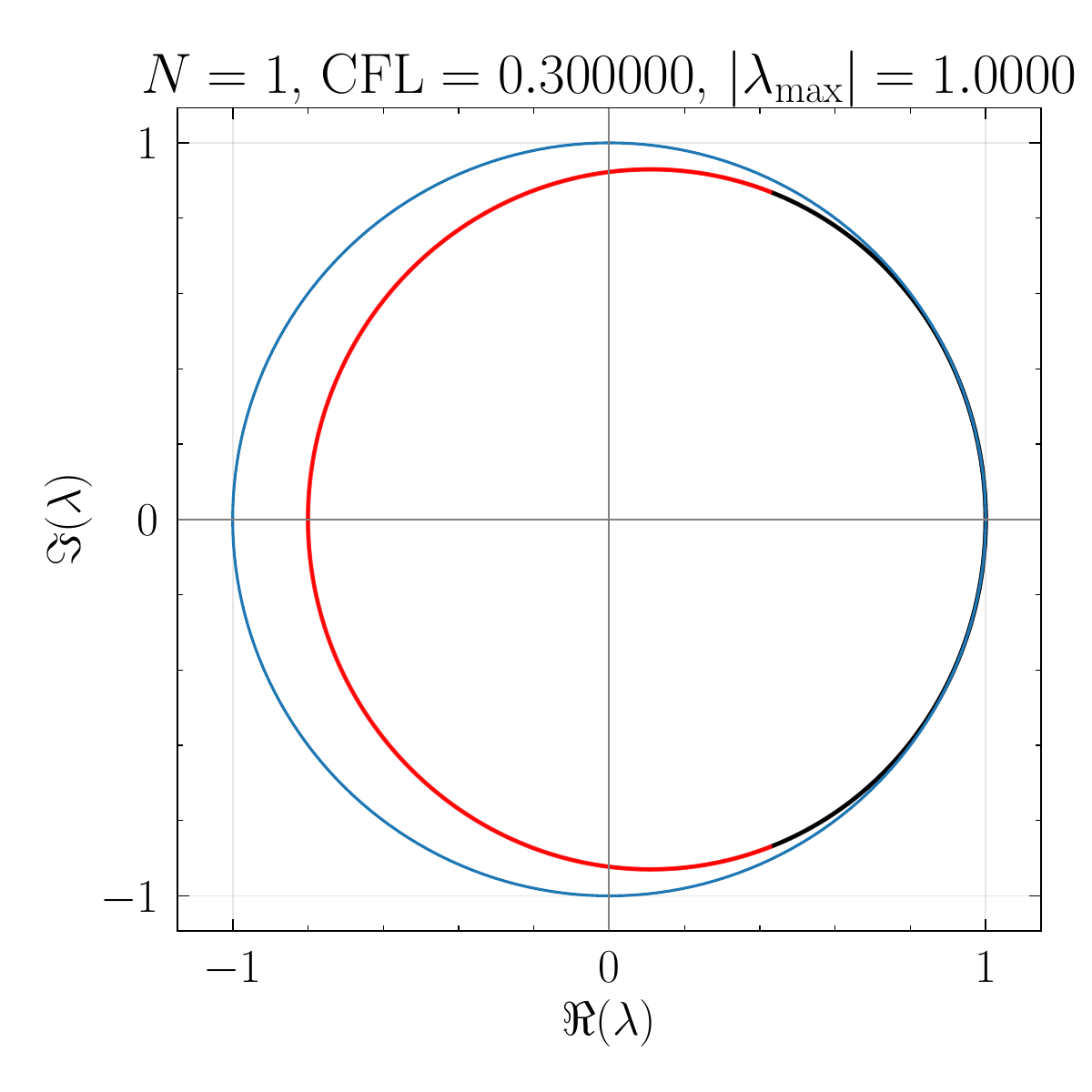}
\includegraphics[width=0.15\textwidth]{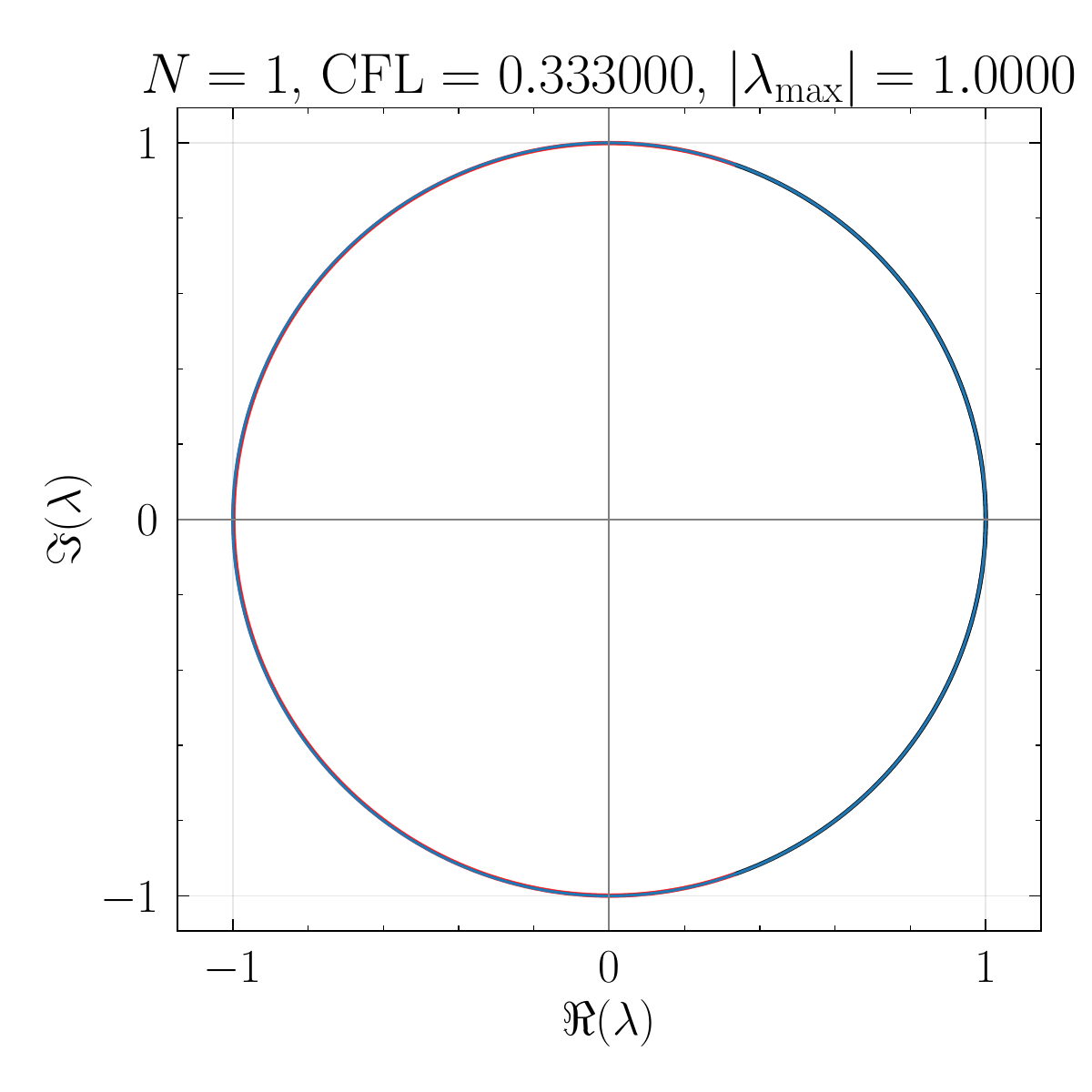}
\includegraphics[width=0.15\textwidth]{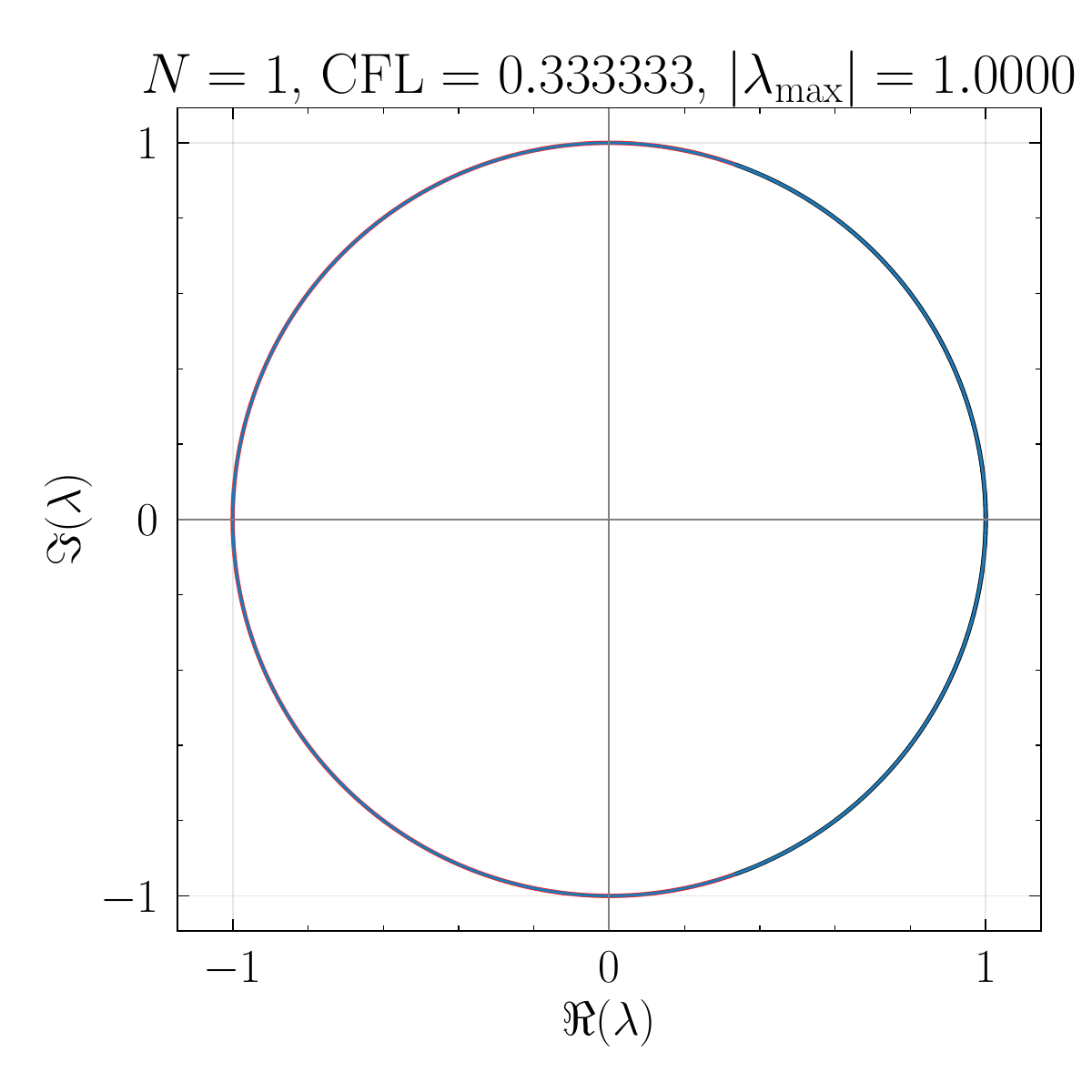}
\includegraphics[width=0.15\textwidth]{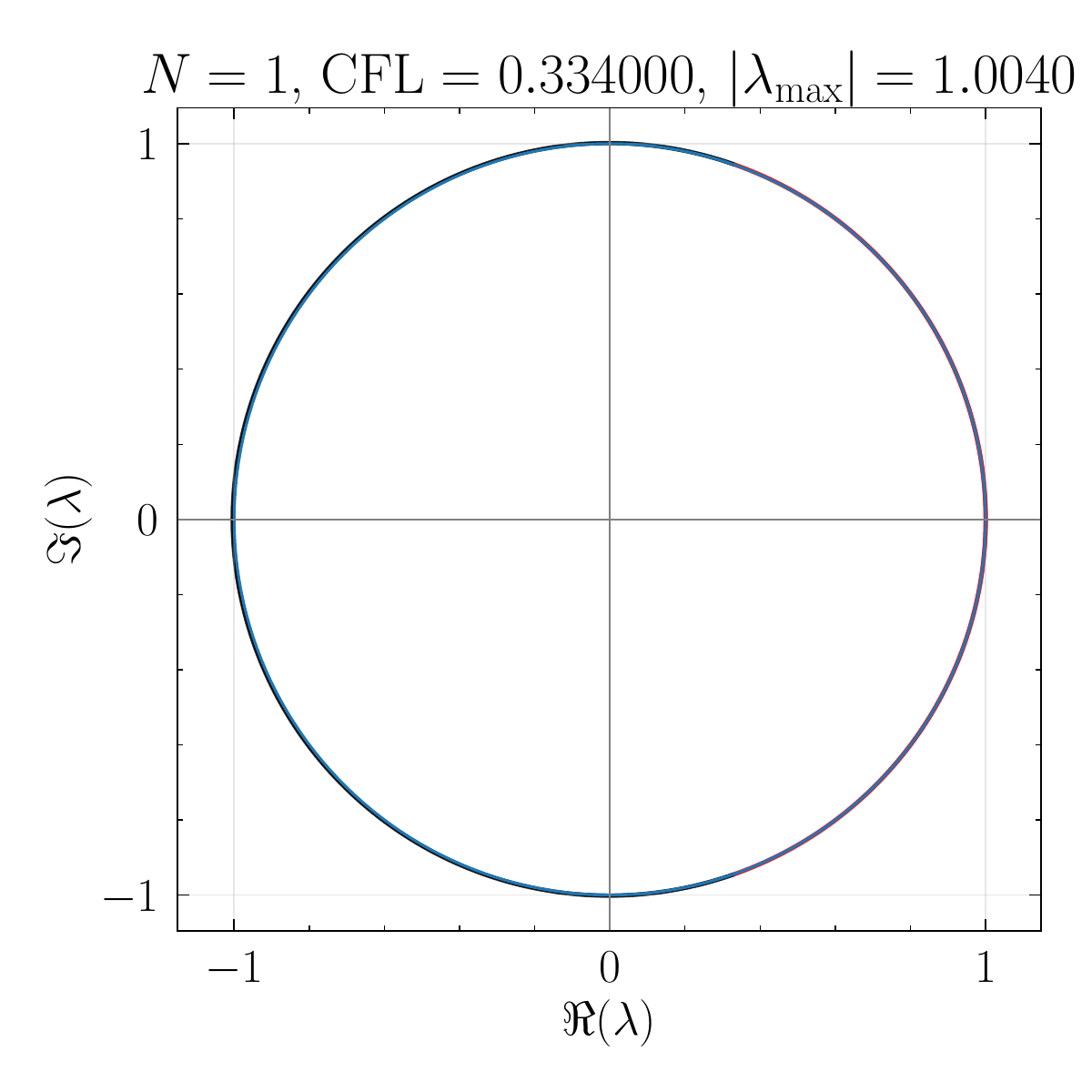}
\includegraphics[width=0.15\textwidth]{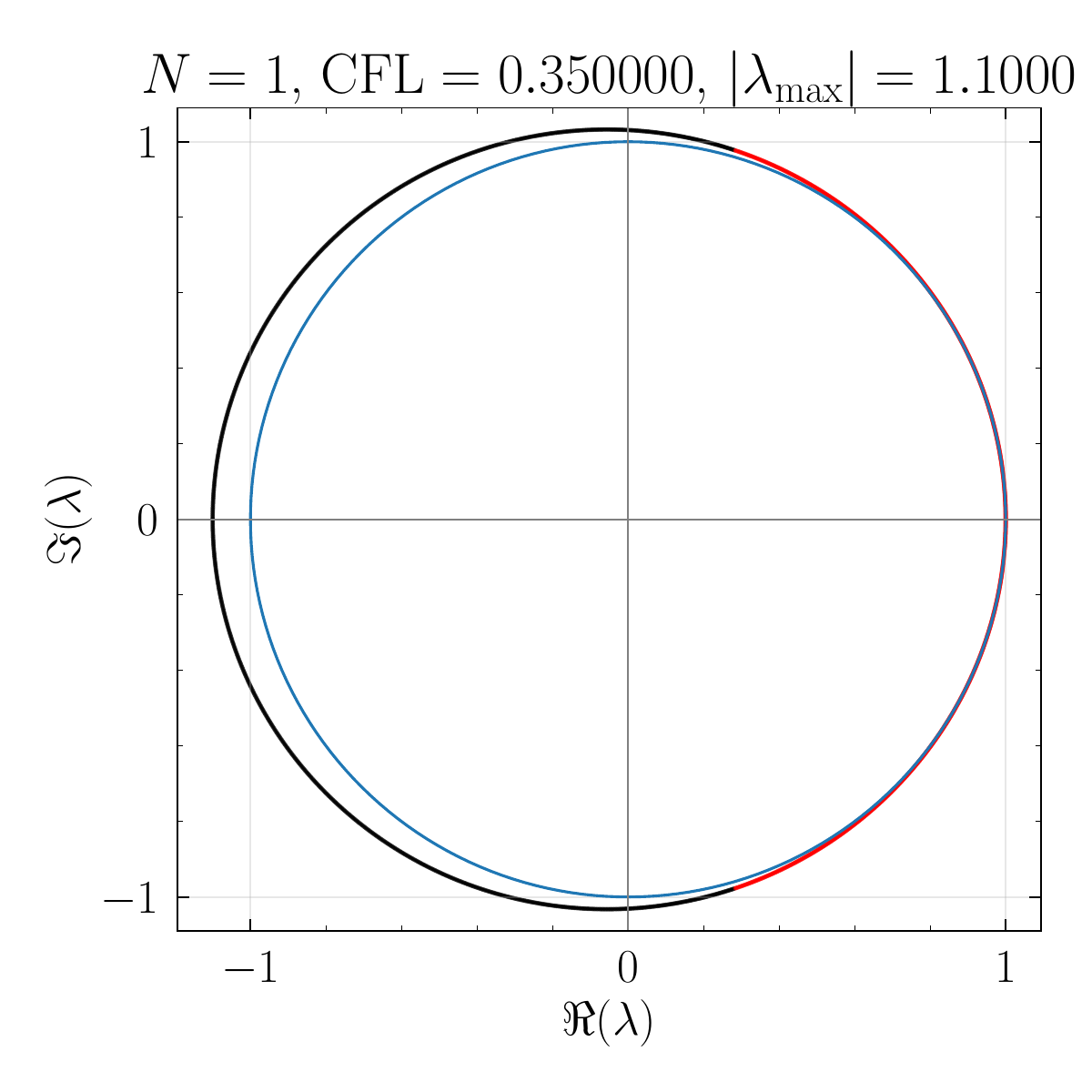}\\
\includegraphics[width=0.028125\textwidth]{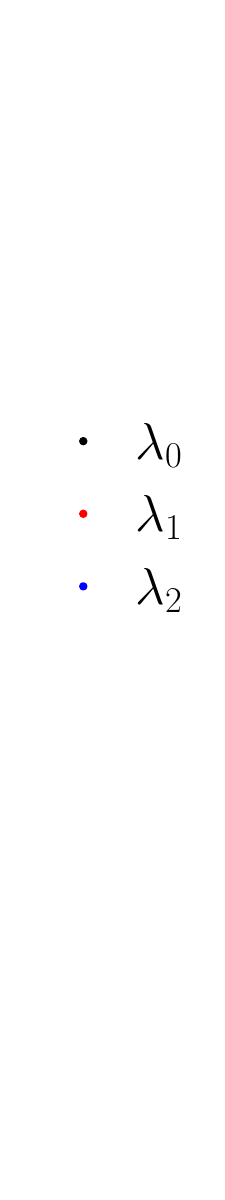}
\includegraphics[width=0.15\textwidth]{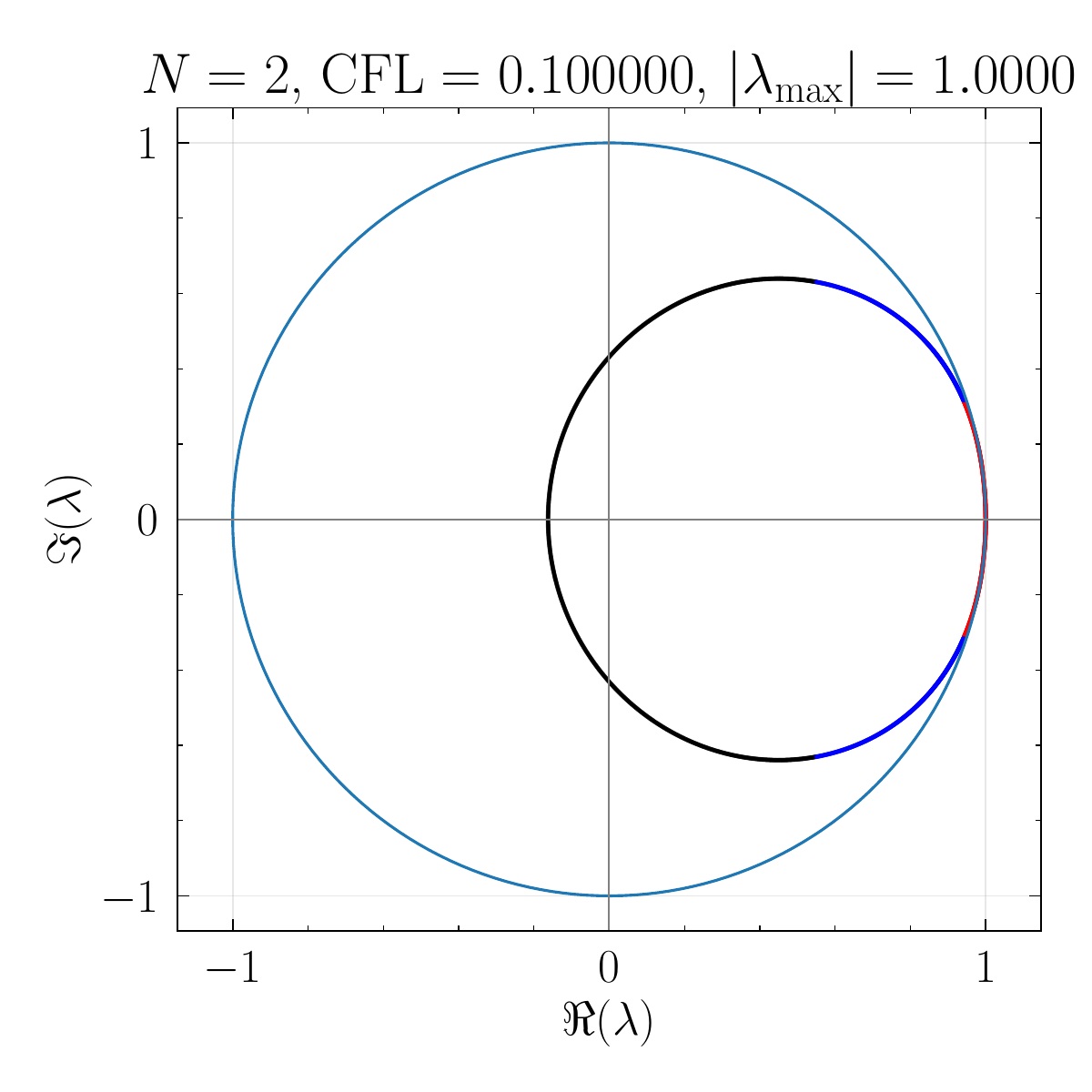}
\includegraphics[width=0.15\textwidth]{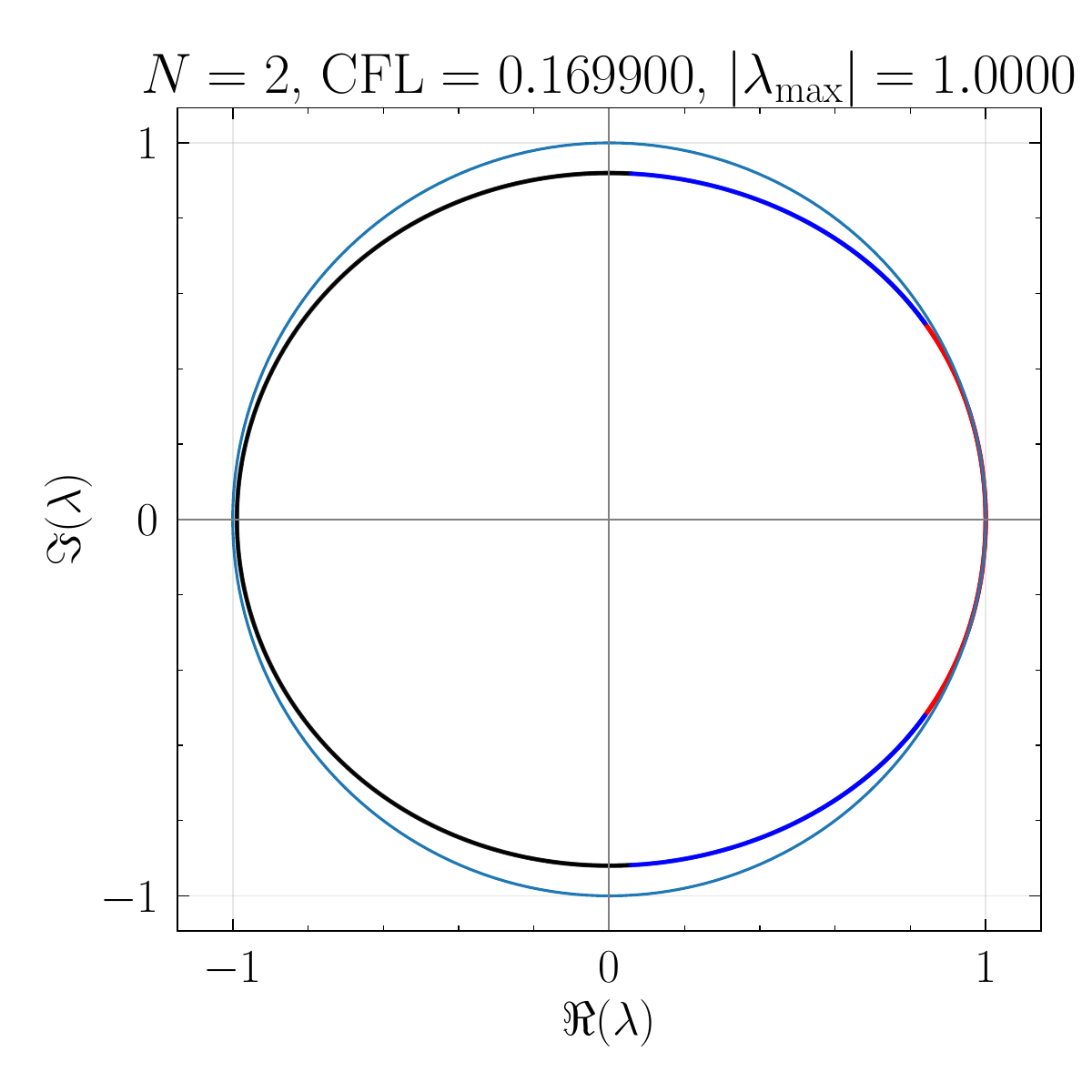}
\includegraphics[width=0.15\textwidth]{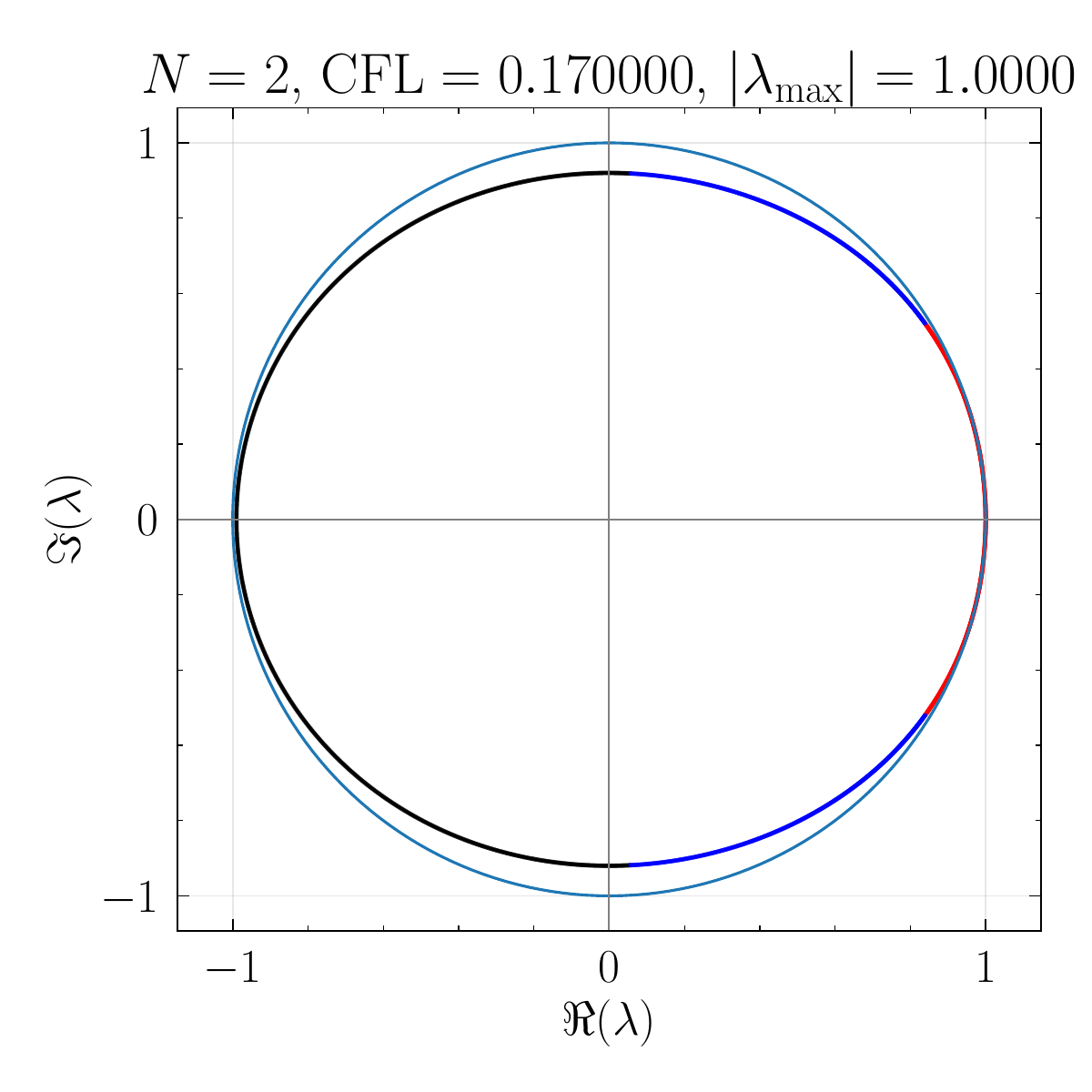}
\includegraphics[width=0.15\textwidth]{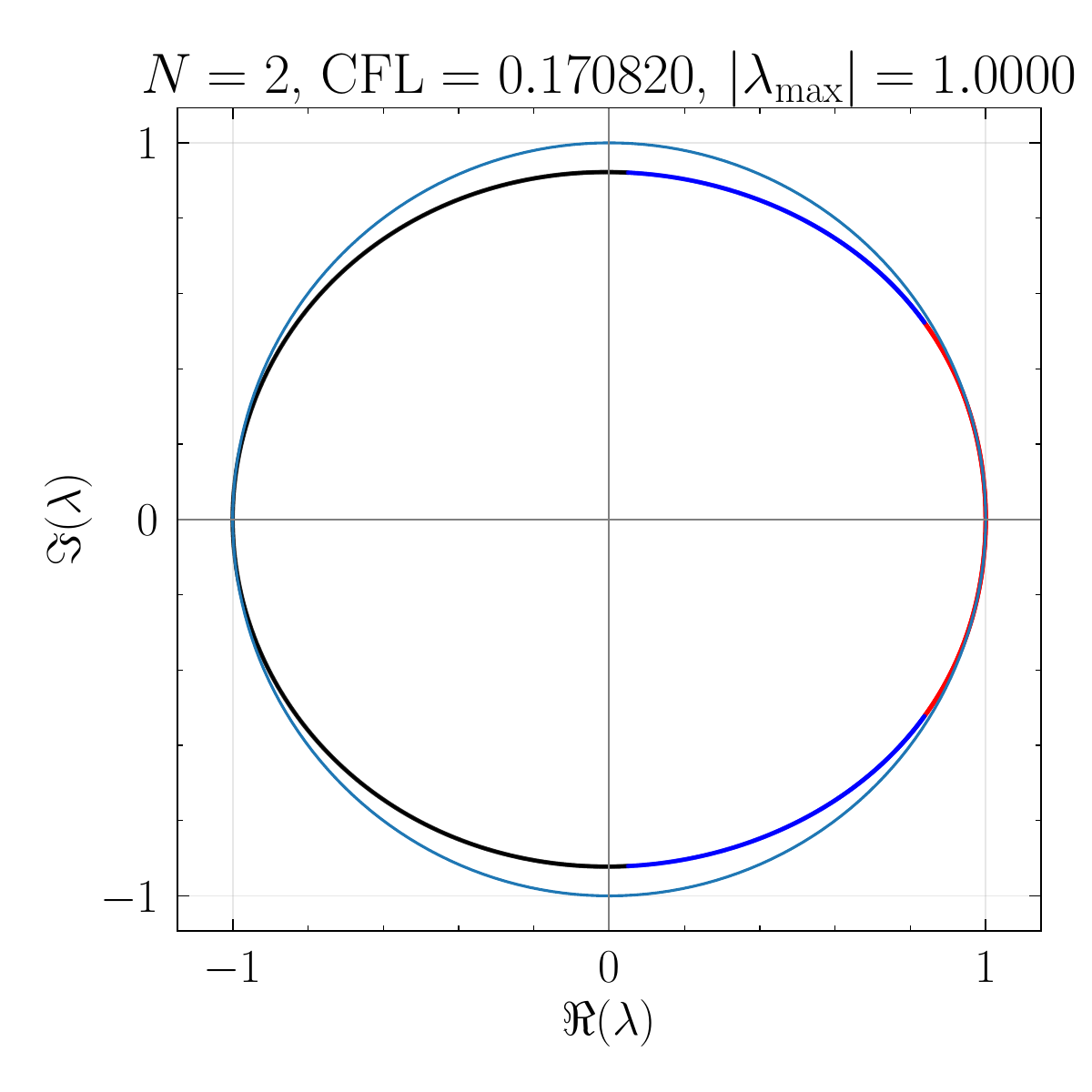}
\includegraphics[width=0.15\textwidth]{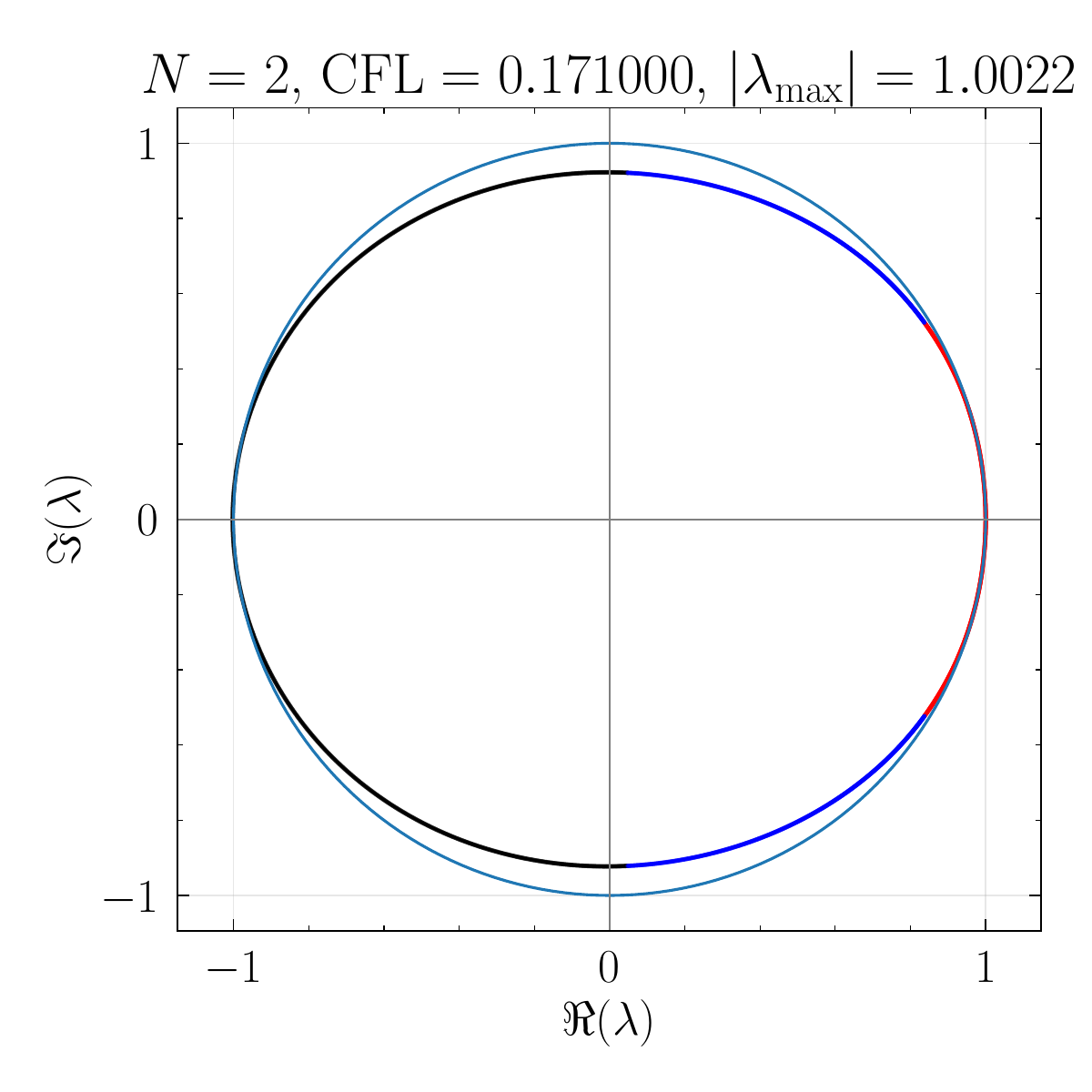}
\includegraphics[width=0.15\textwidth]{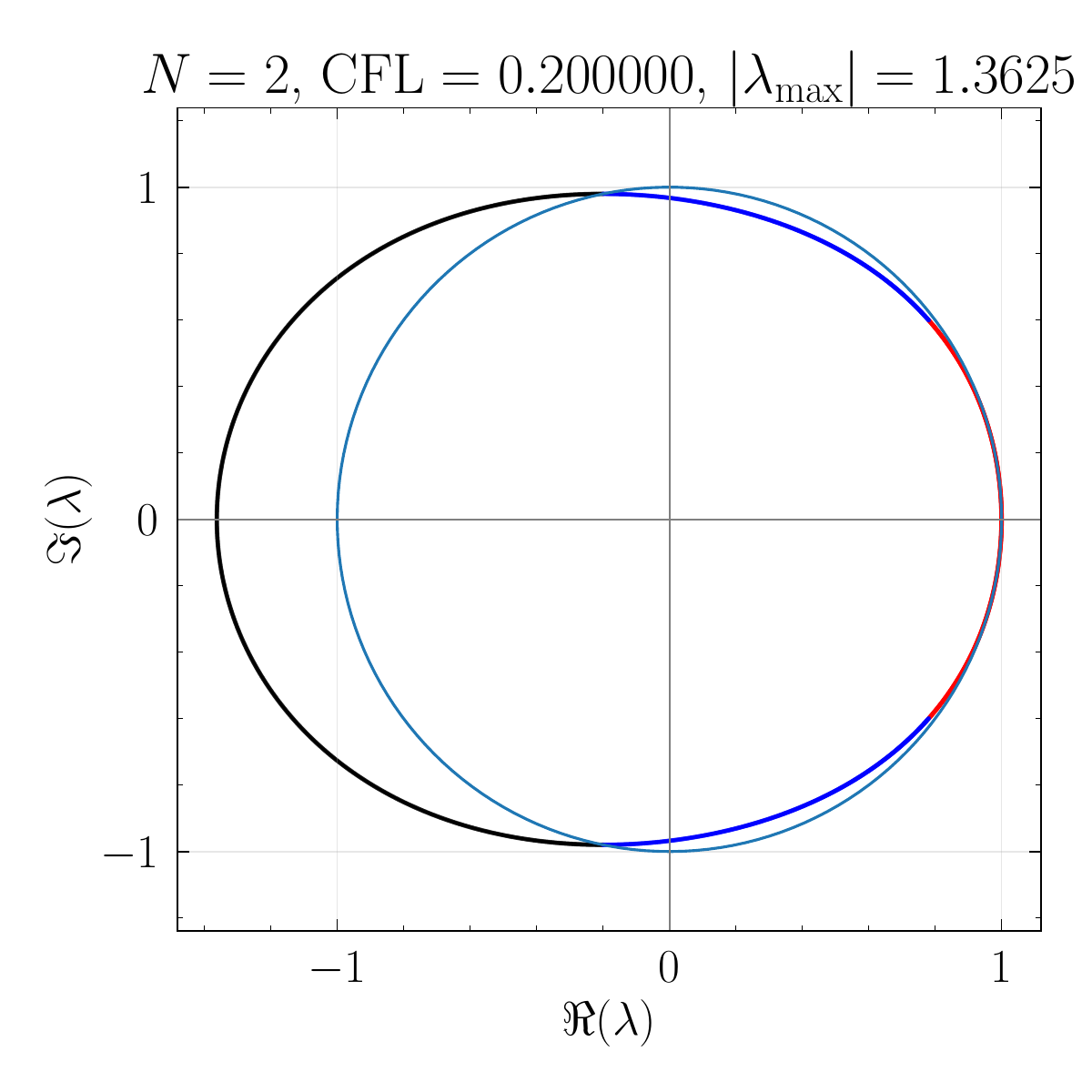}\\
\includegraphics[width=0.028125\textwidth]{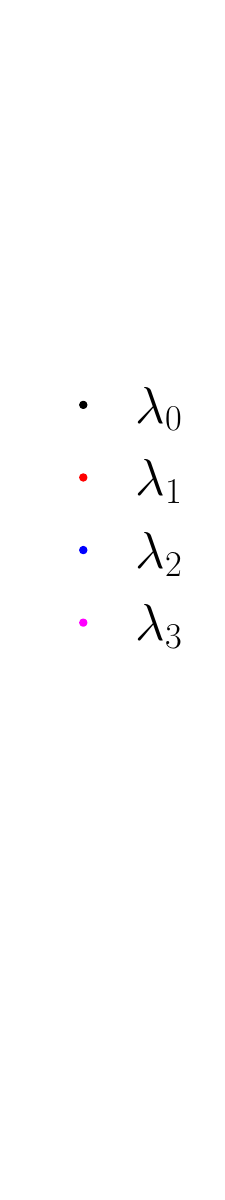}
\includegraphics[width=0.15\textwidth]{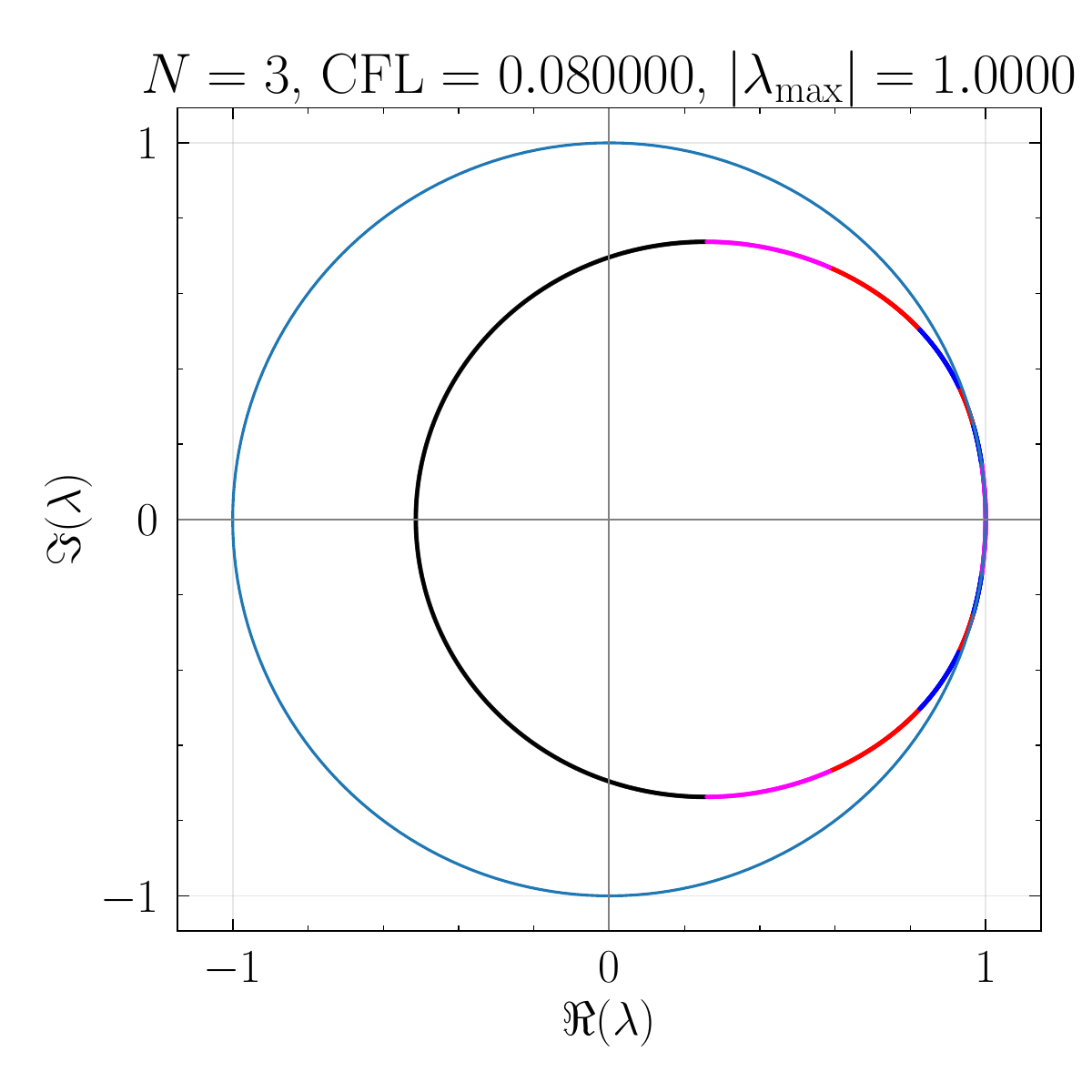}
\includegraphics[width=0.15\textwidth]{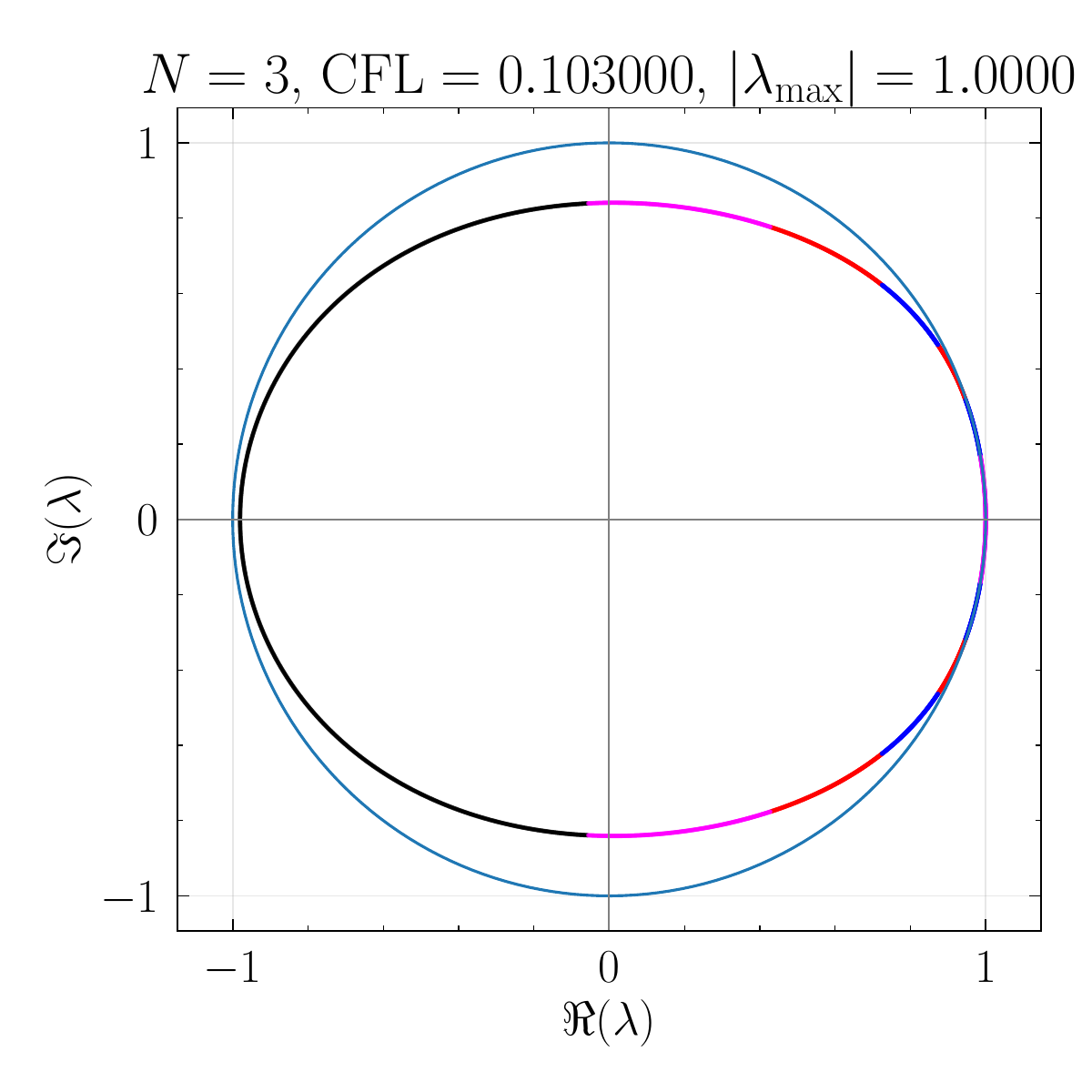}
\includegraphics[width=0.15\textwidth]{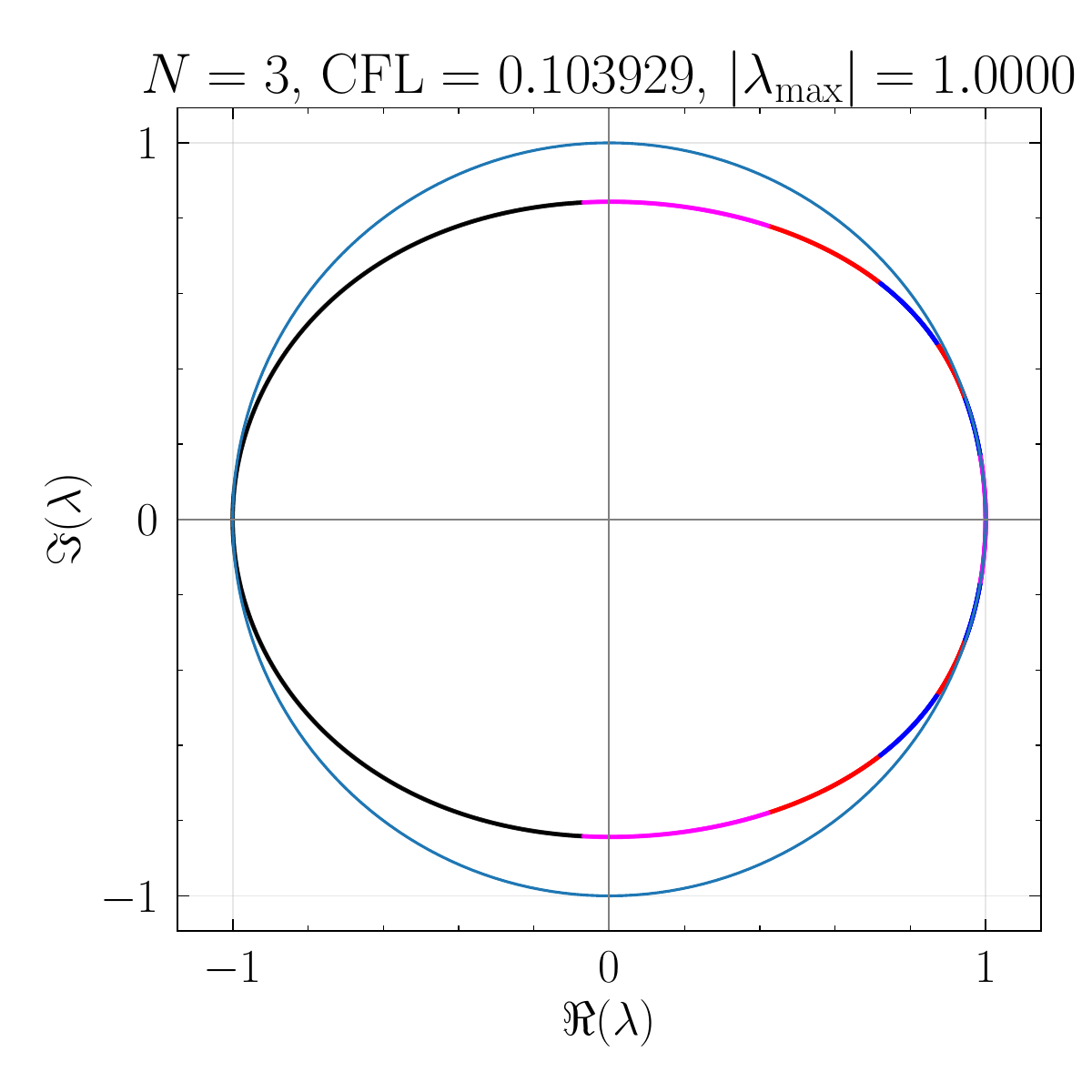}
\includegraphics[width=0.15\textwidth]{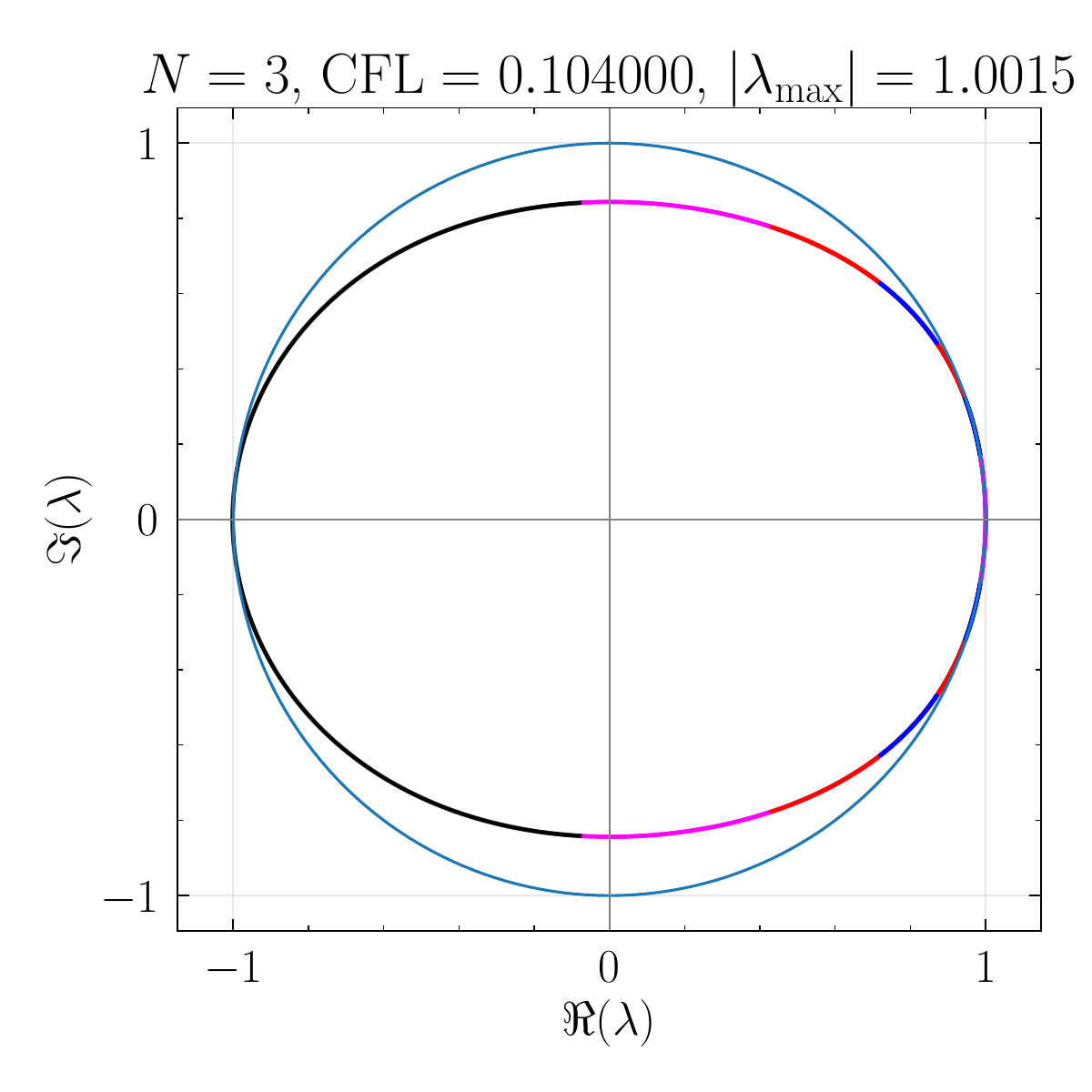}
\includegraphics[width=0.15\textwidth]{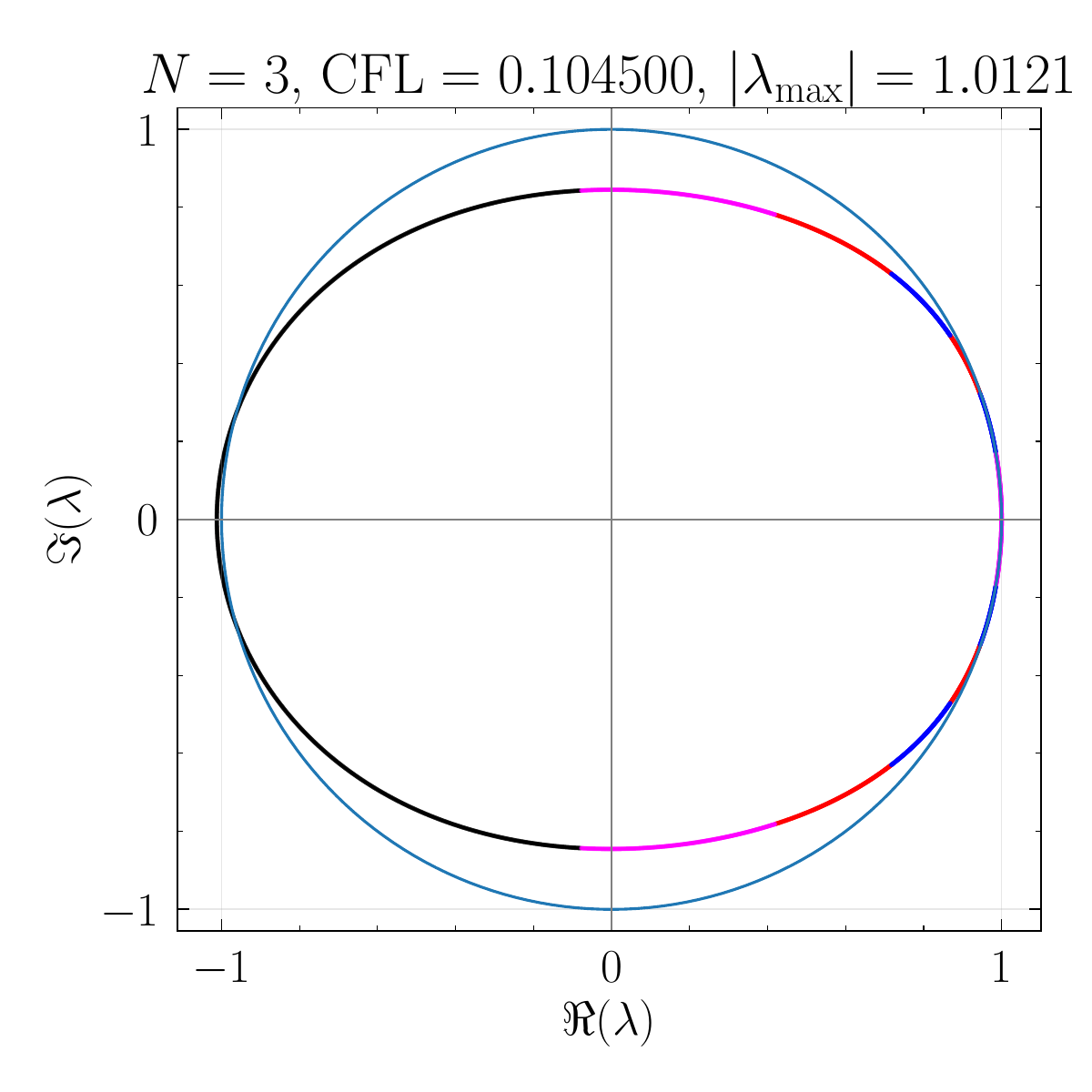}
\includegraphics[width=0.15\textwidth]{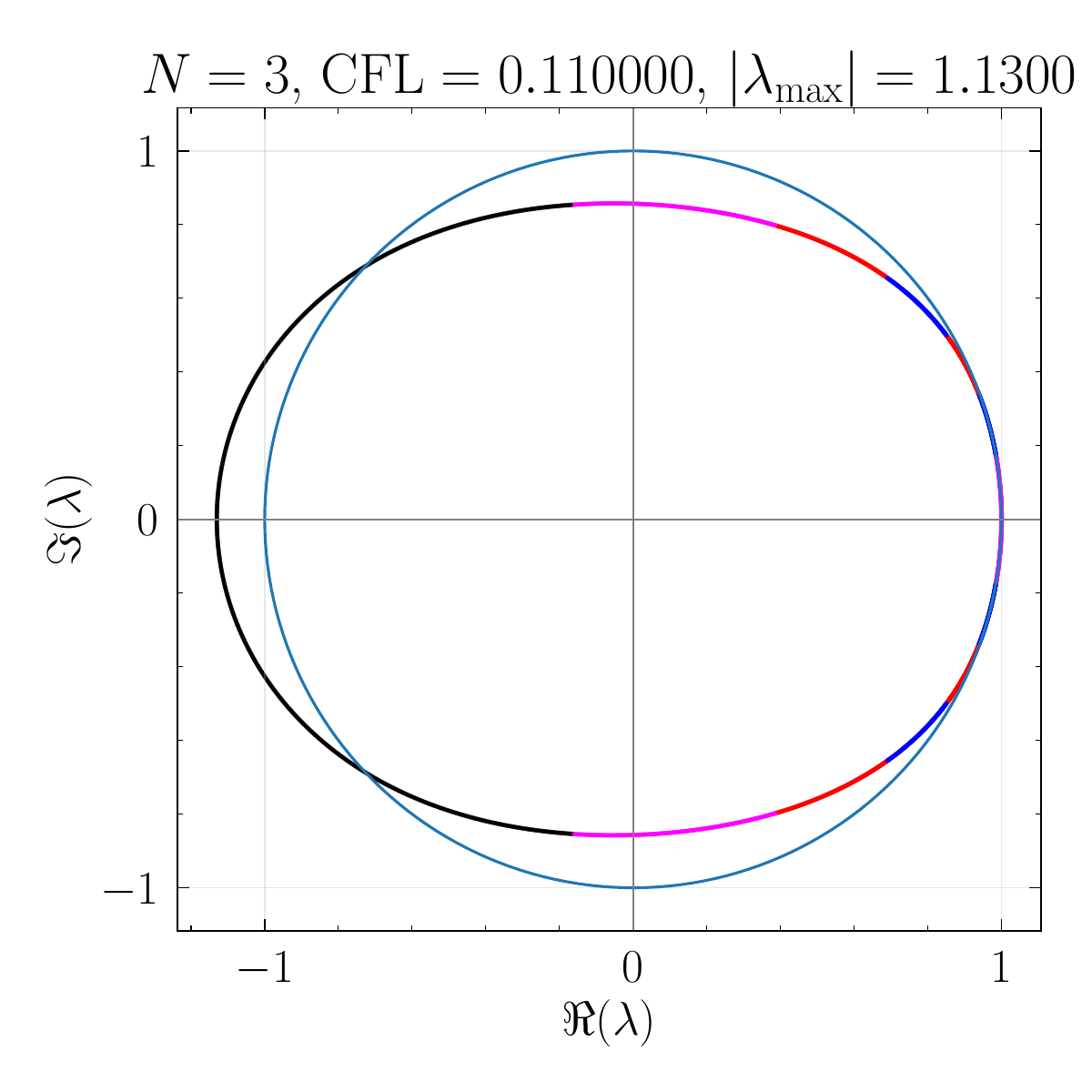}\\
\includegraphics[width=0.028125\textwidth]{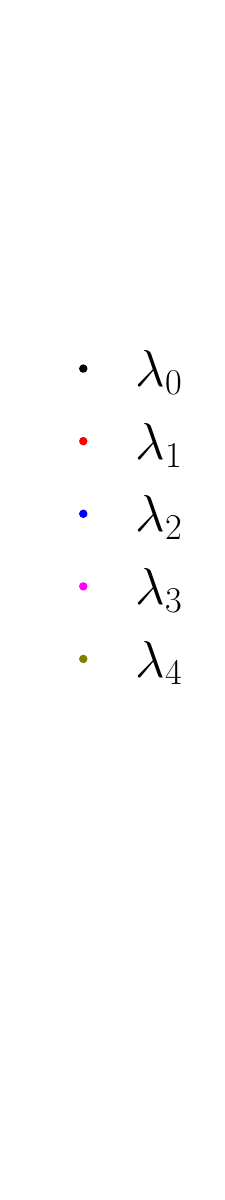}
\includegraphics[width=0.15\textwidth]{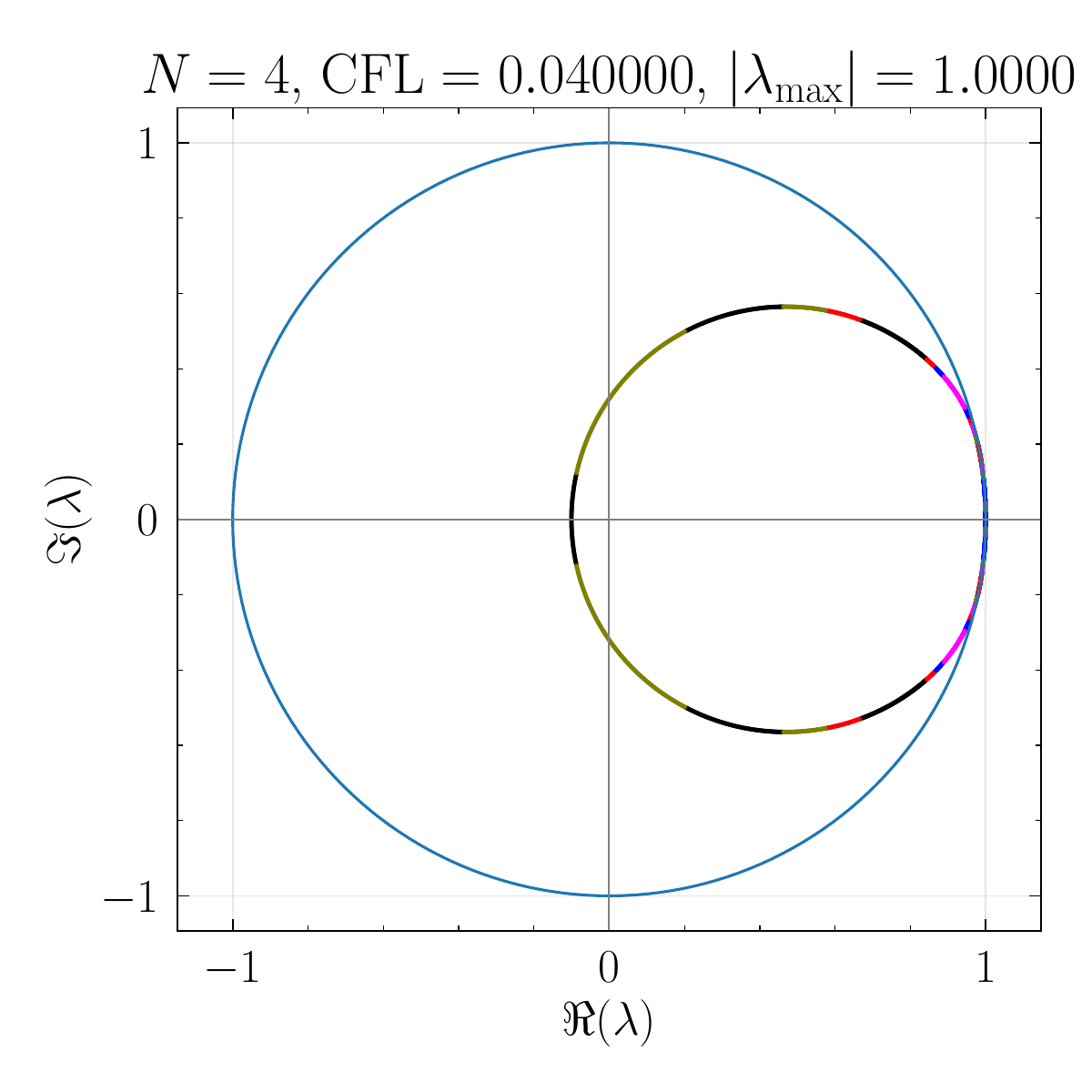}
\includegraphics[width=0.15\textwidth]{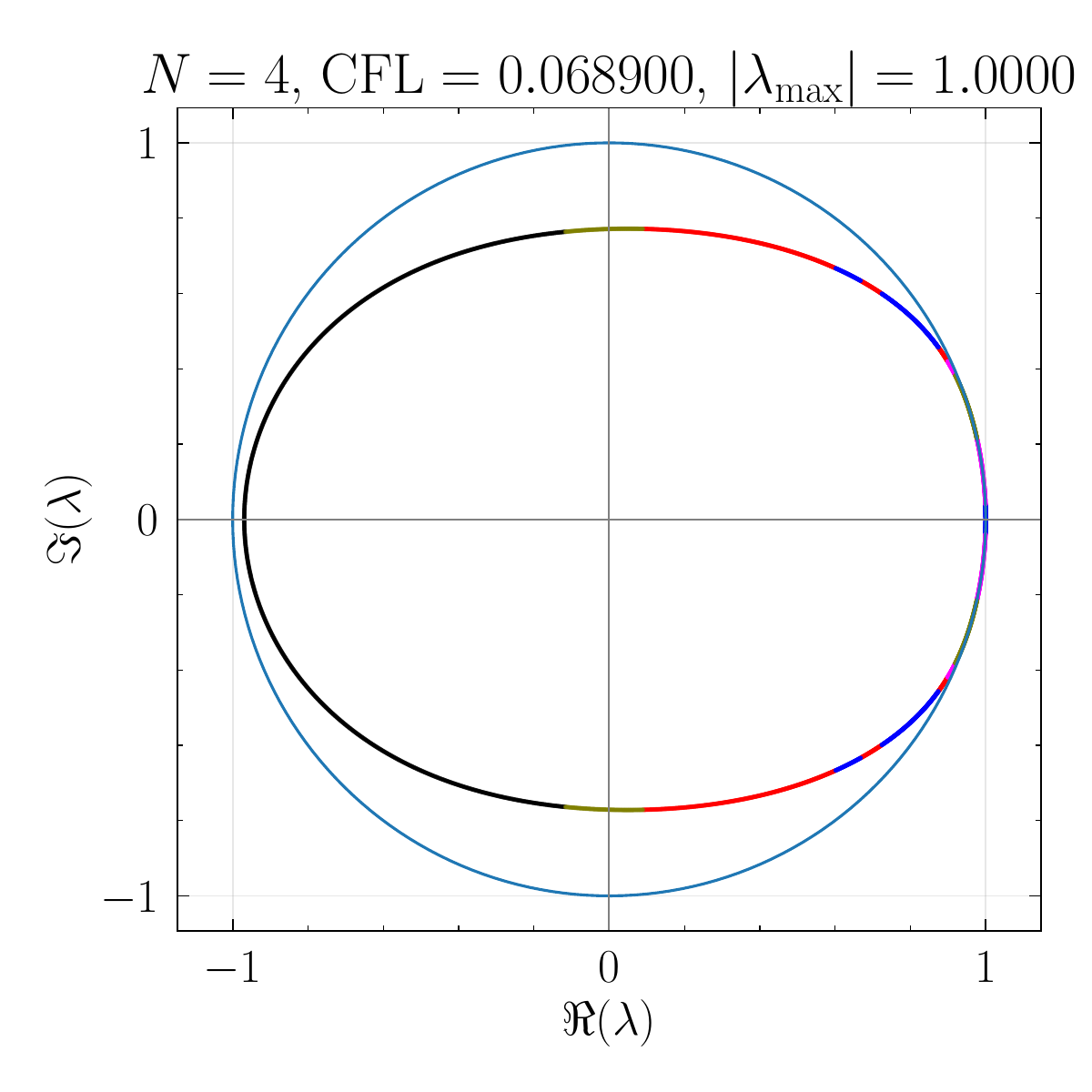}
\includegraphics[width=0.15\textwidth]{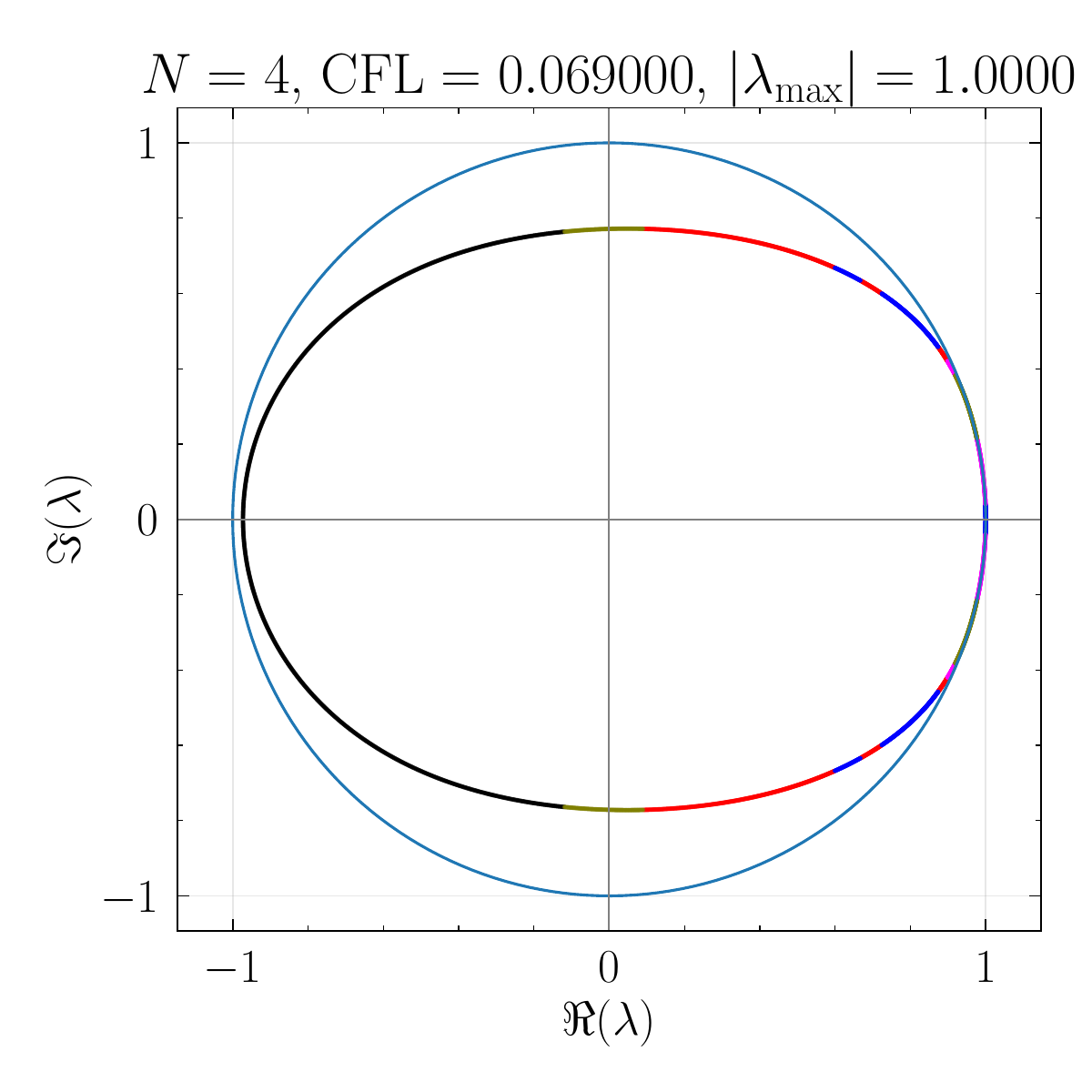}
\includegraphics[width=0.15\textwidth]{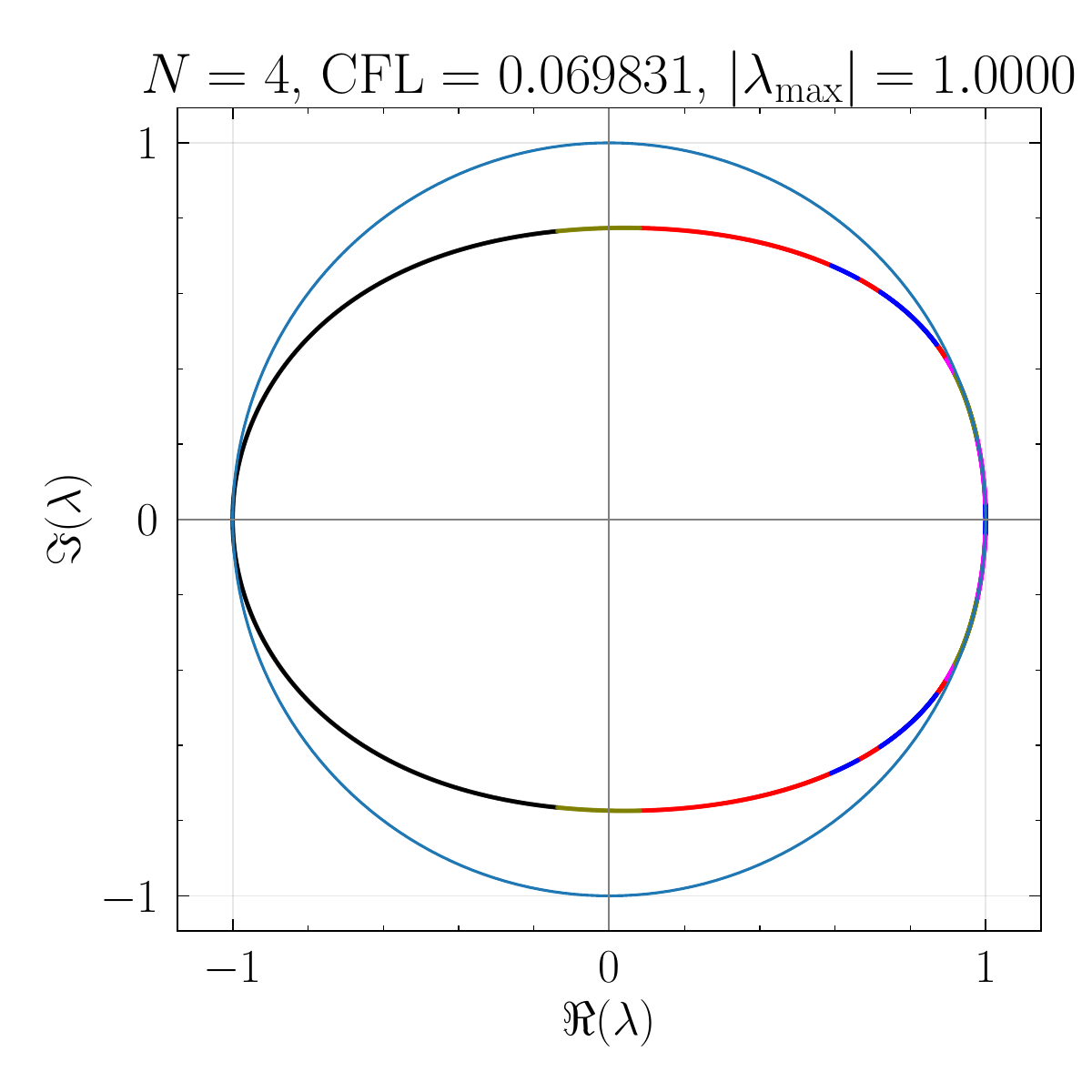}
\includegraphics[width=0.15\textwidth]{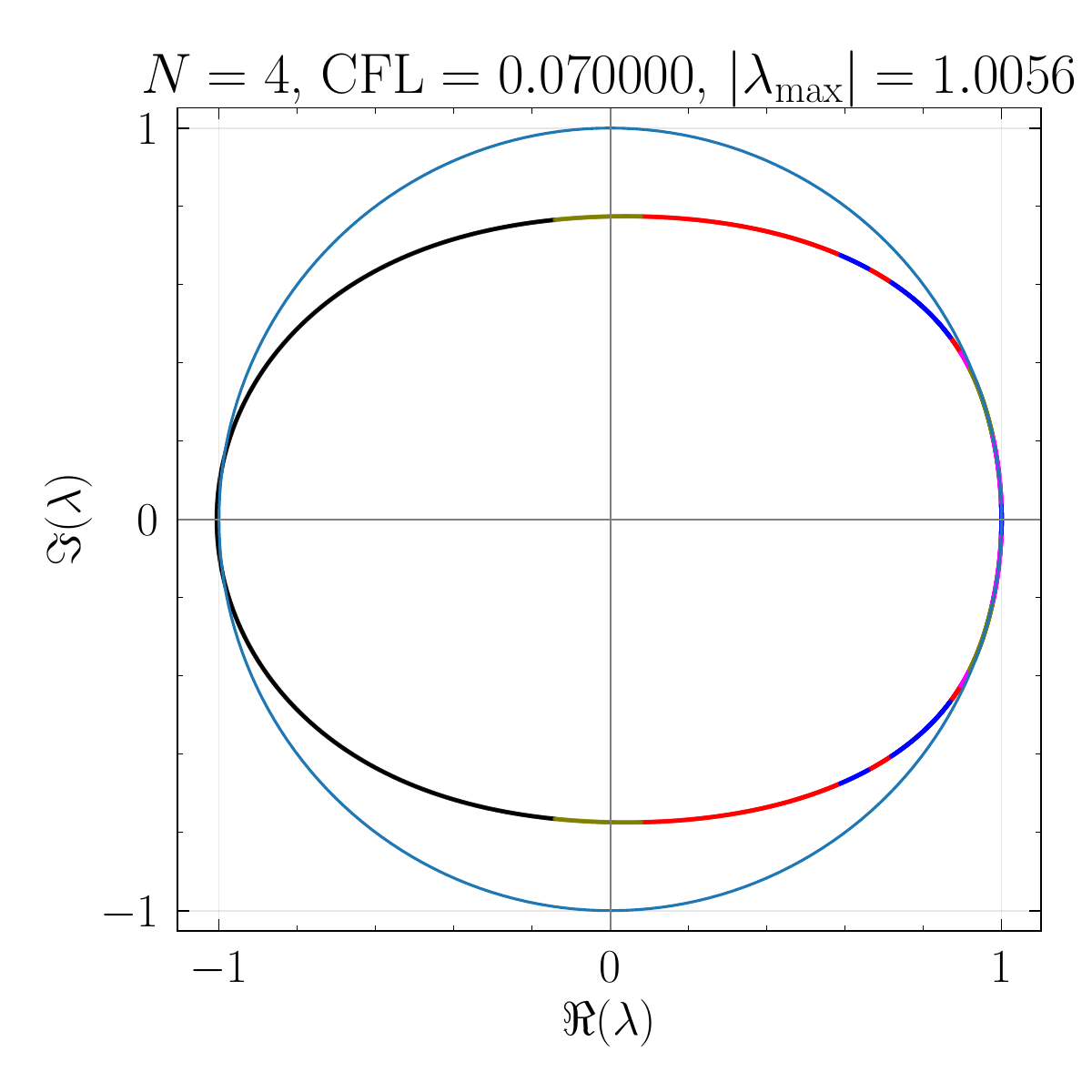}
\includegraphics[width=0.15\textwidth]{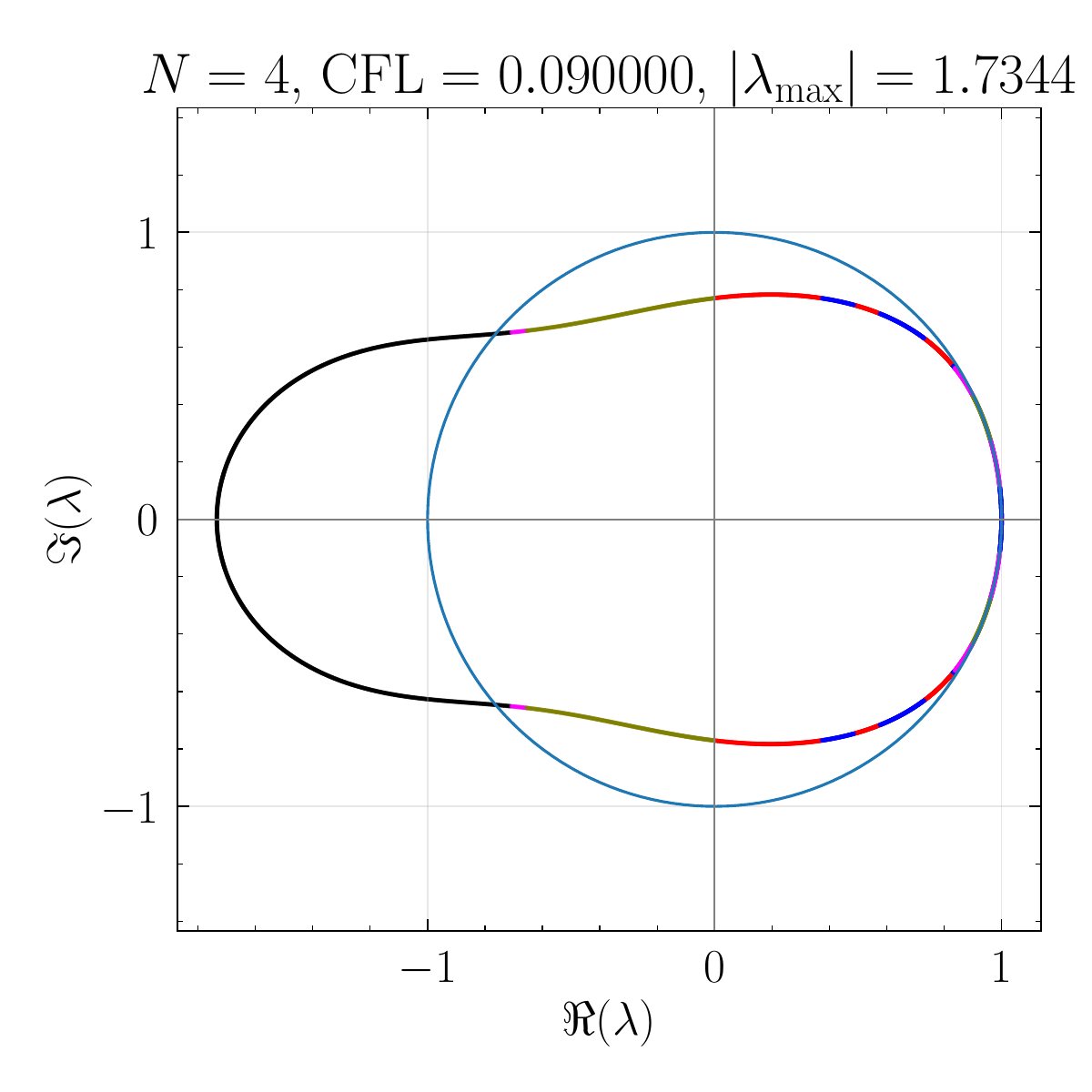}\\
\includegraphics[width=0.028125\textwidth]{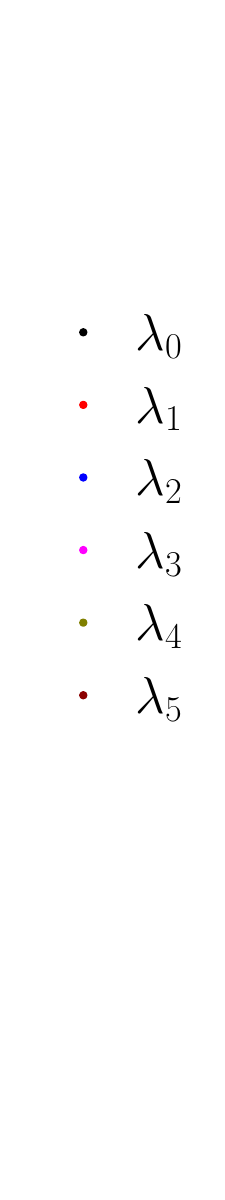}
\includegraphics[width=0.15\textwidth]{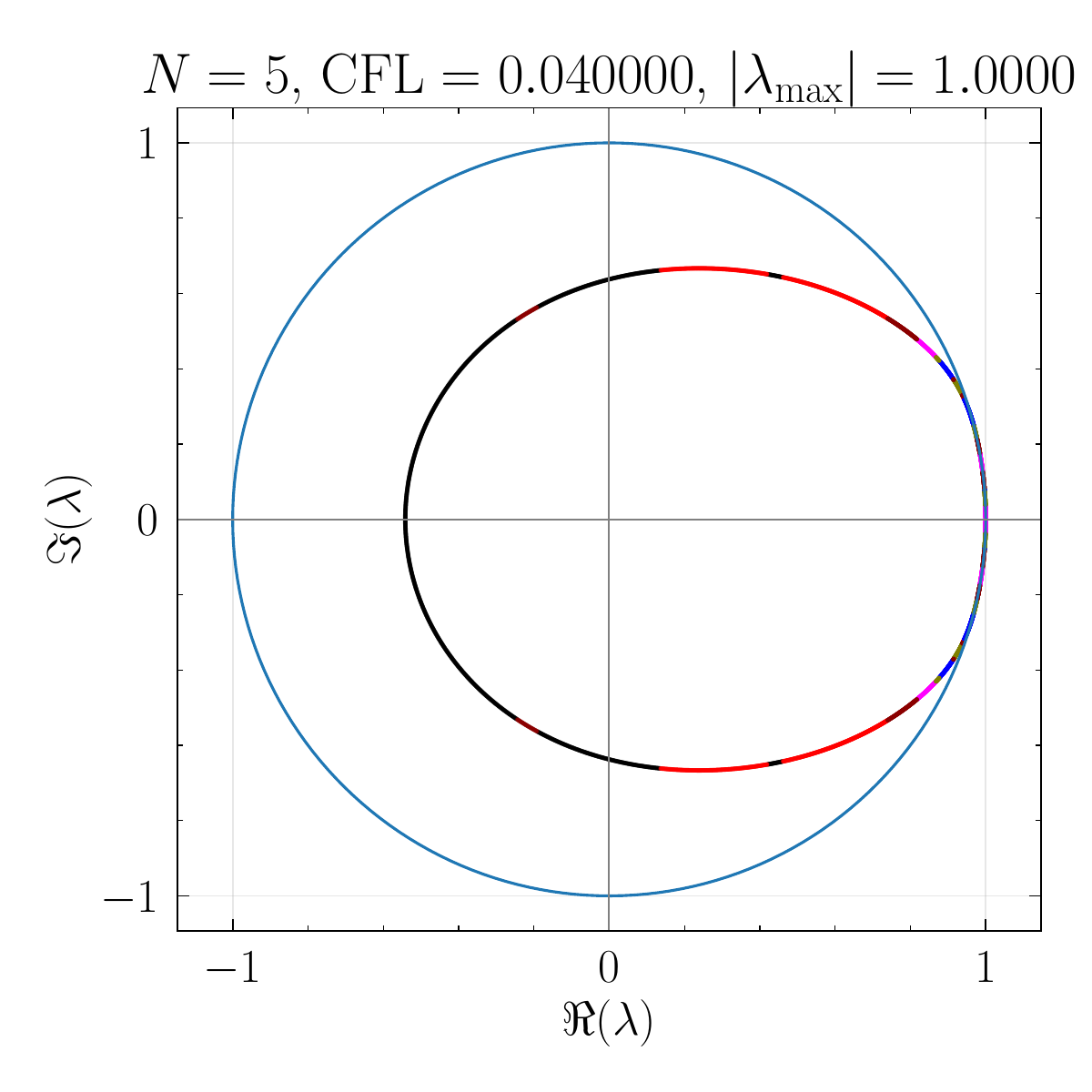}
\includegraphics[width=0.15\textwidth]{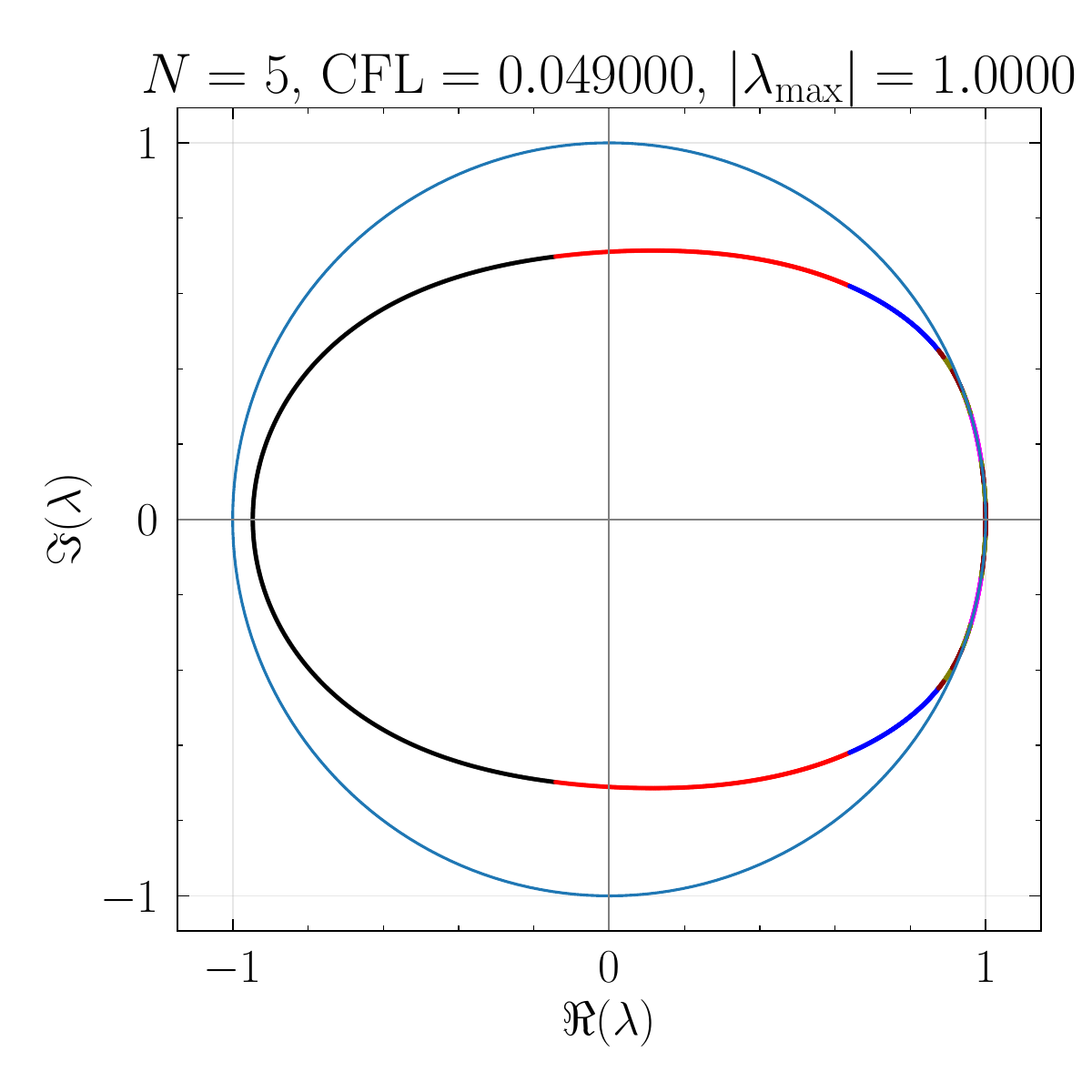}
\includegraphics[width=0.15\textwidth]{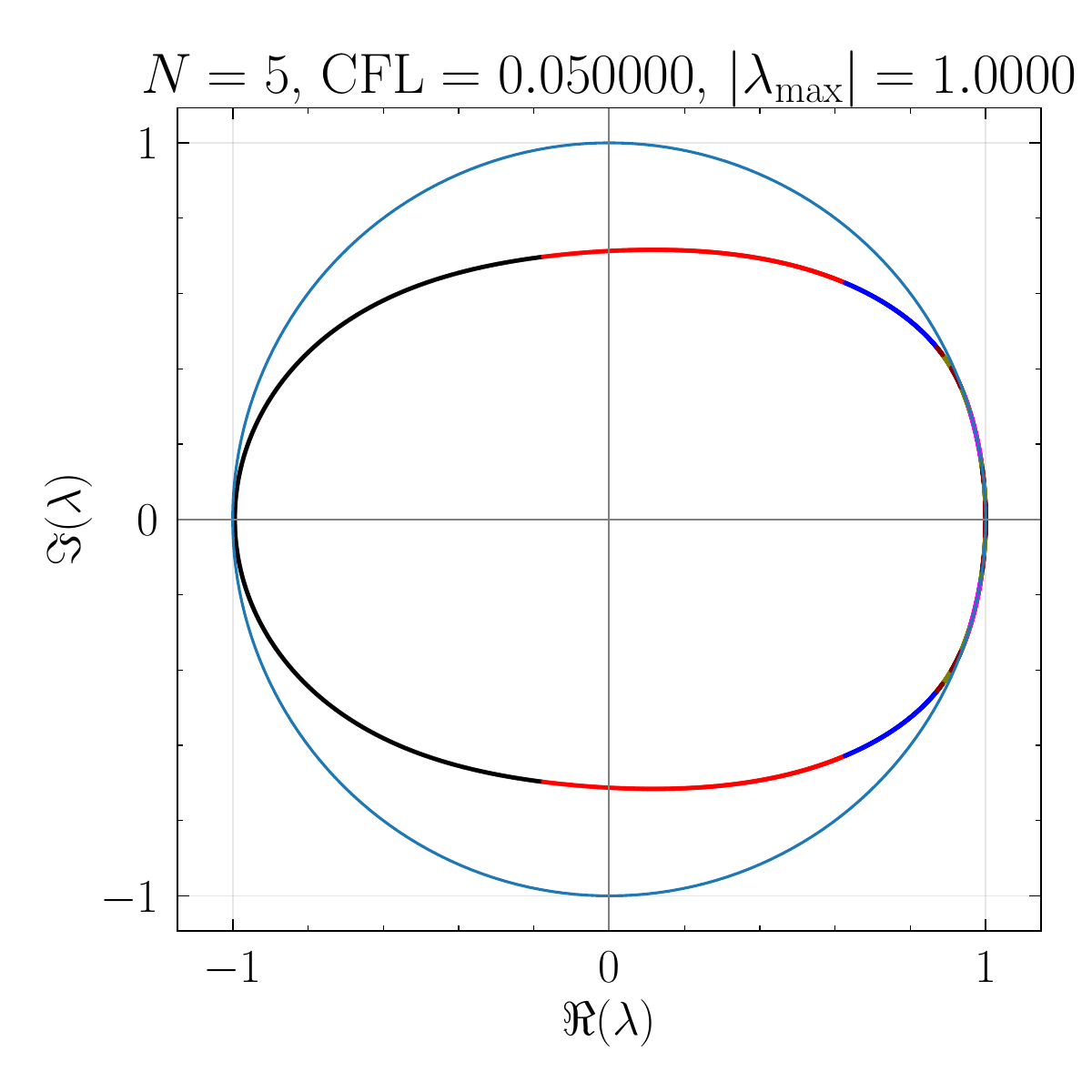}
\includegraphics[width=0.15\textwidth]{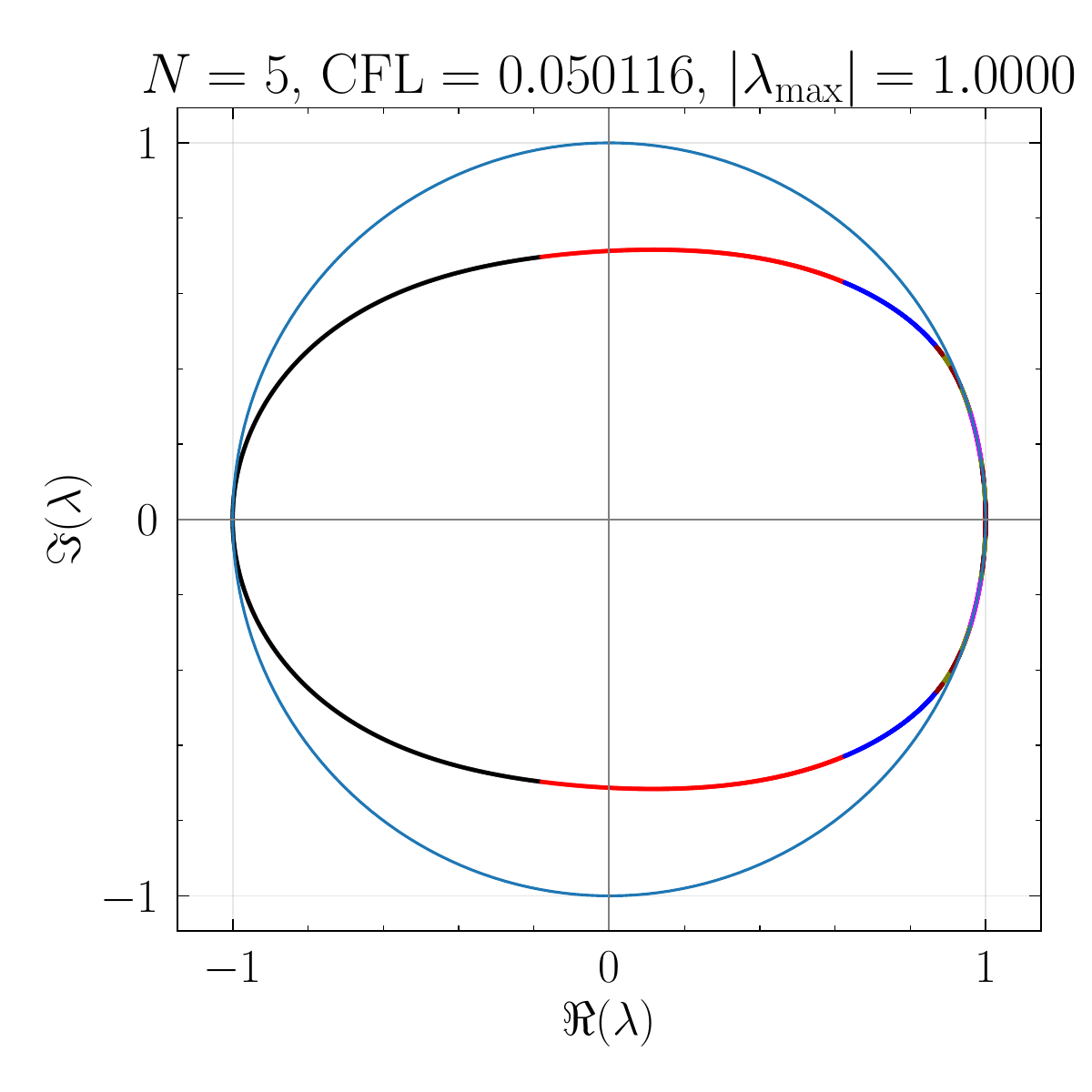}
\includegraphics[width=0.15\textwidth]{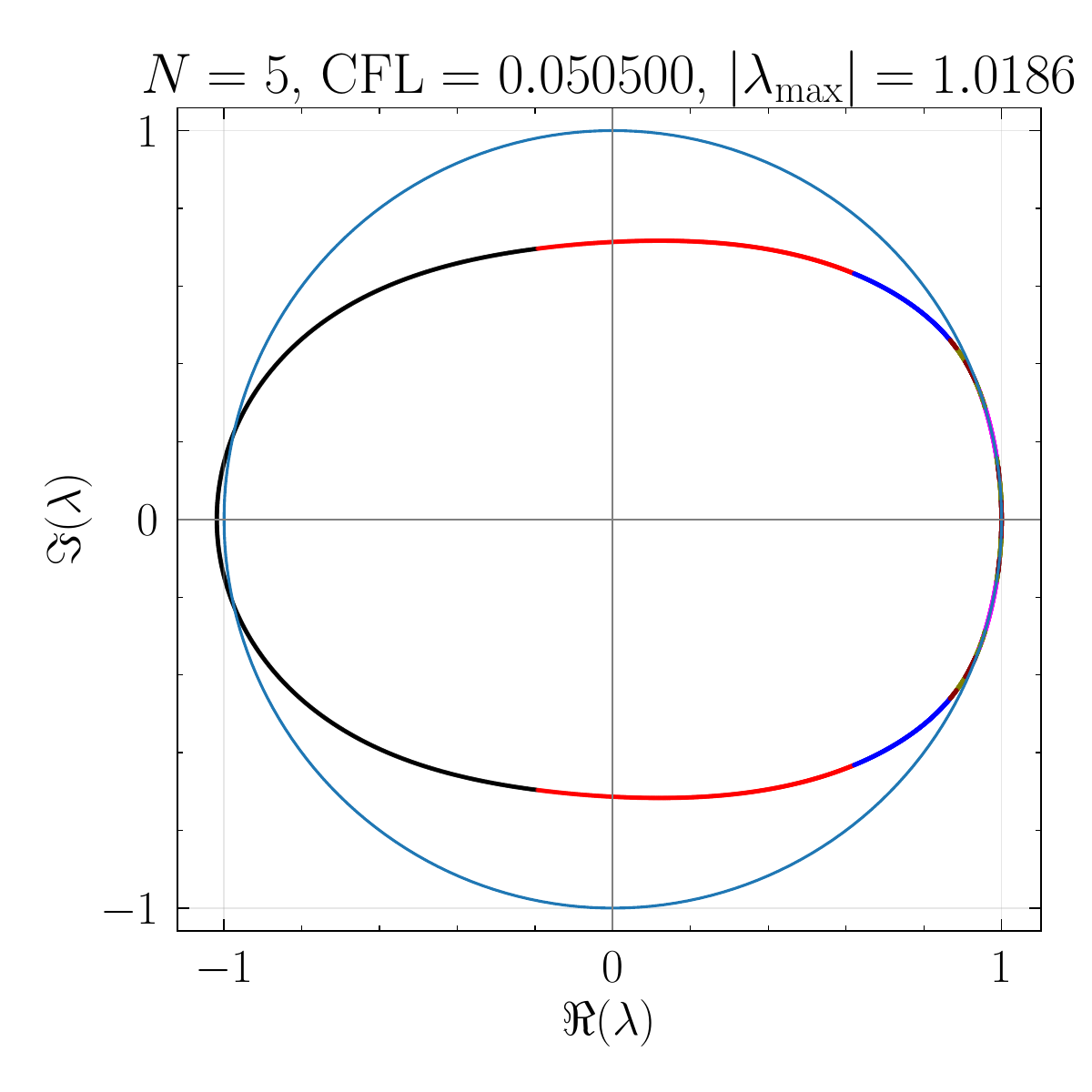}
\includegraphics[width=0.15\textwidth]{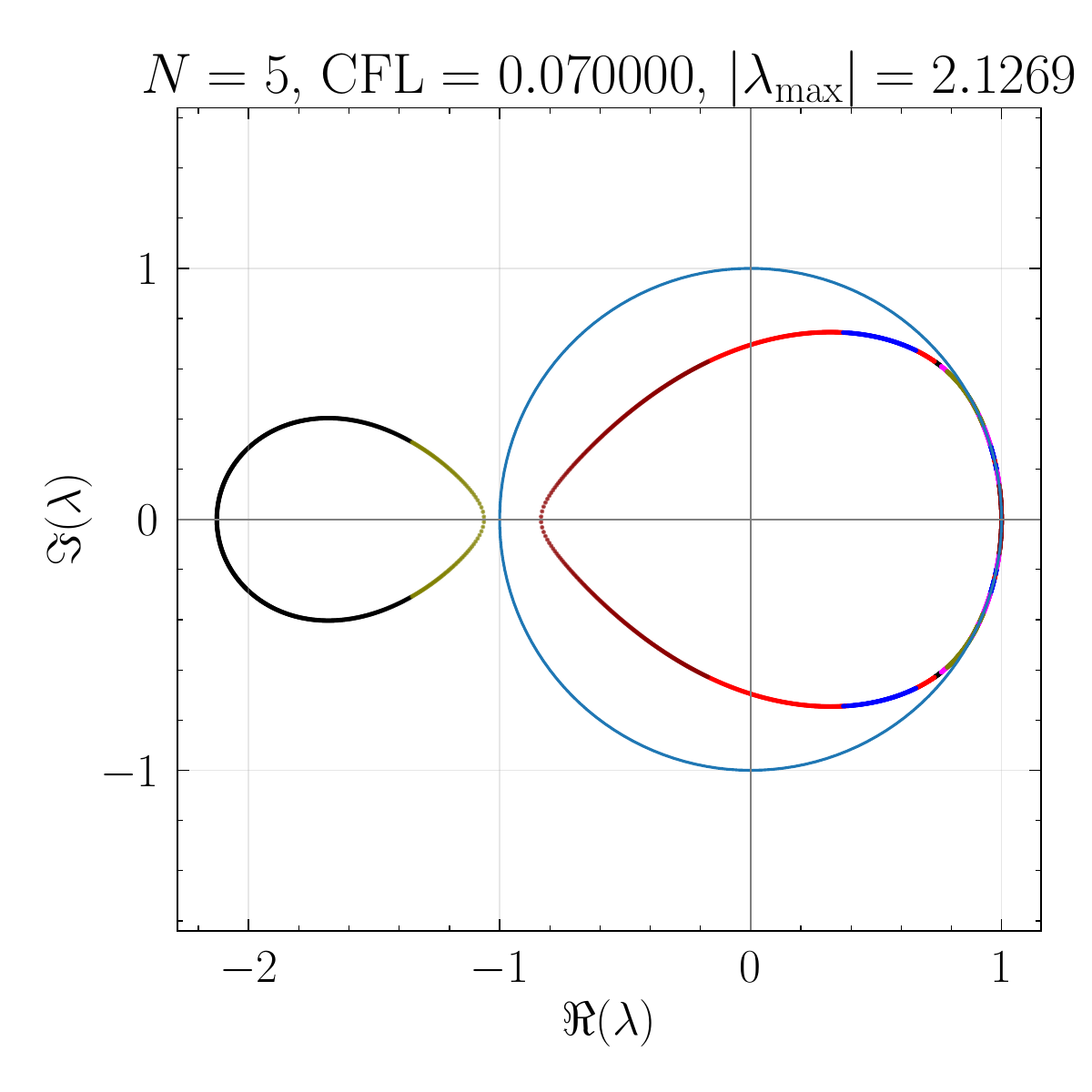}\\
\includegraphics[width=0.028125\textwidth]{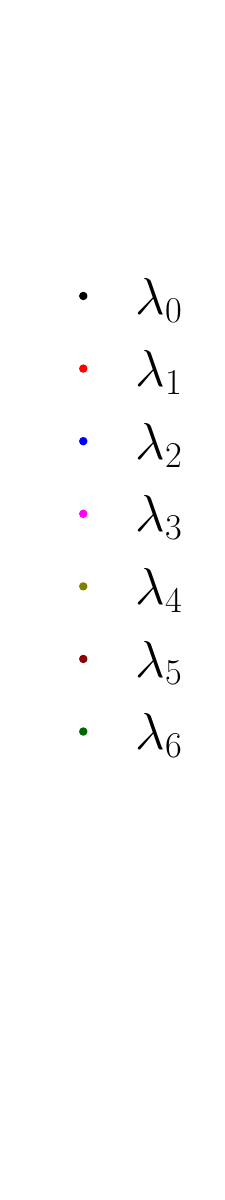}
\includegraphics[width=0.15\textwidth]{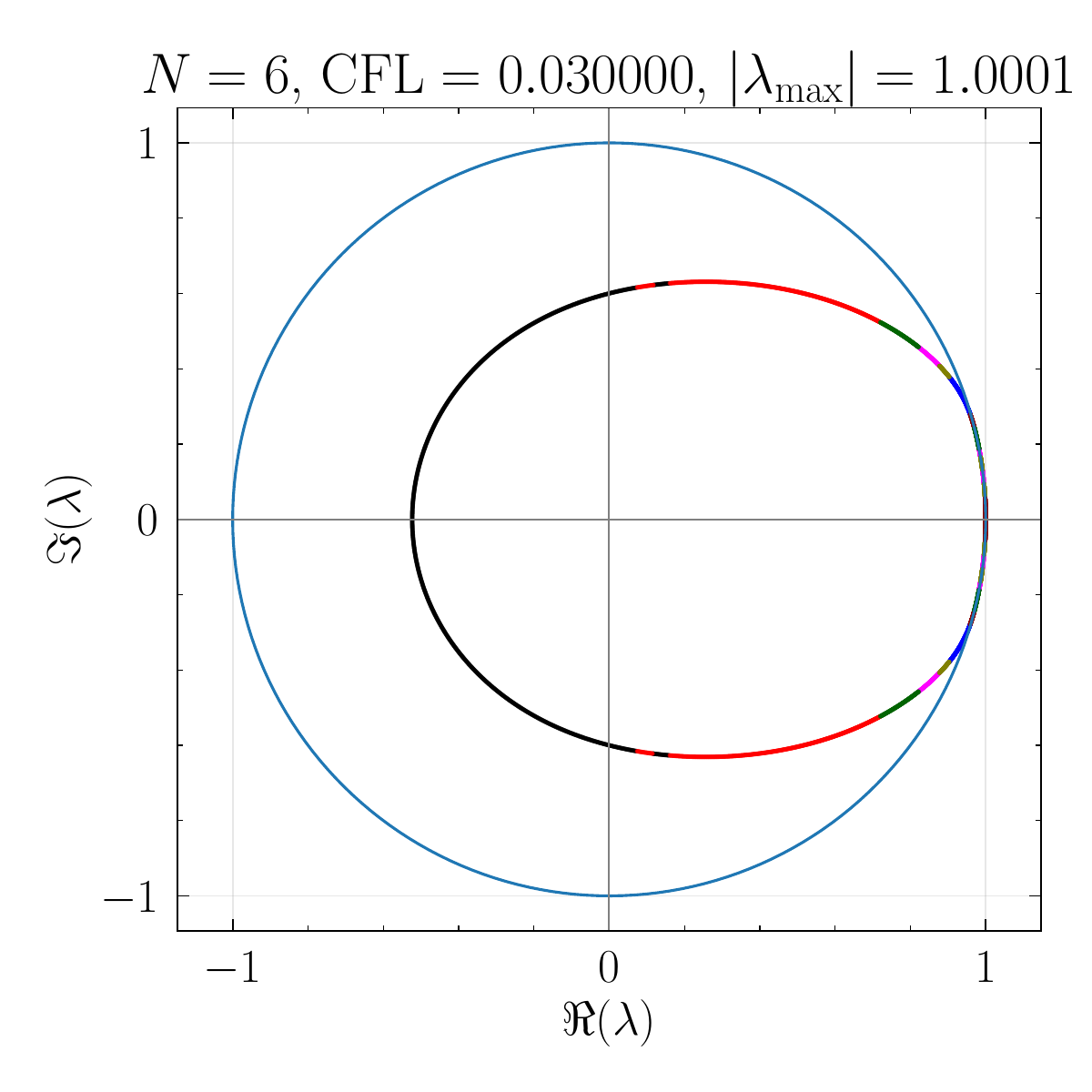}
\includegraphics[width=0.15\textwidth]{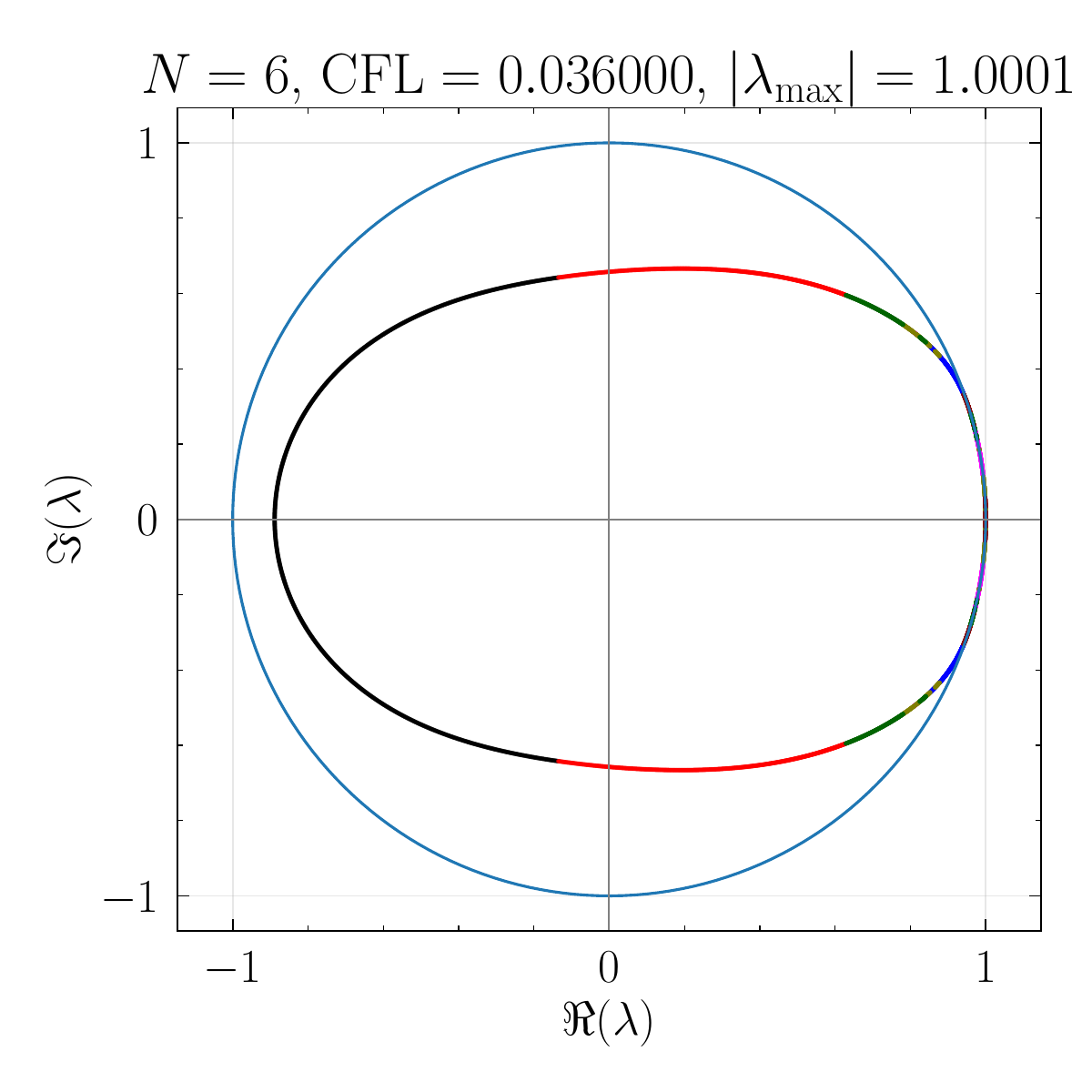}
\includegraphics[width=0.15\textwidth]{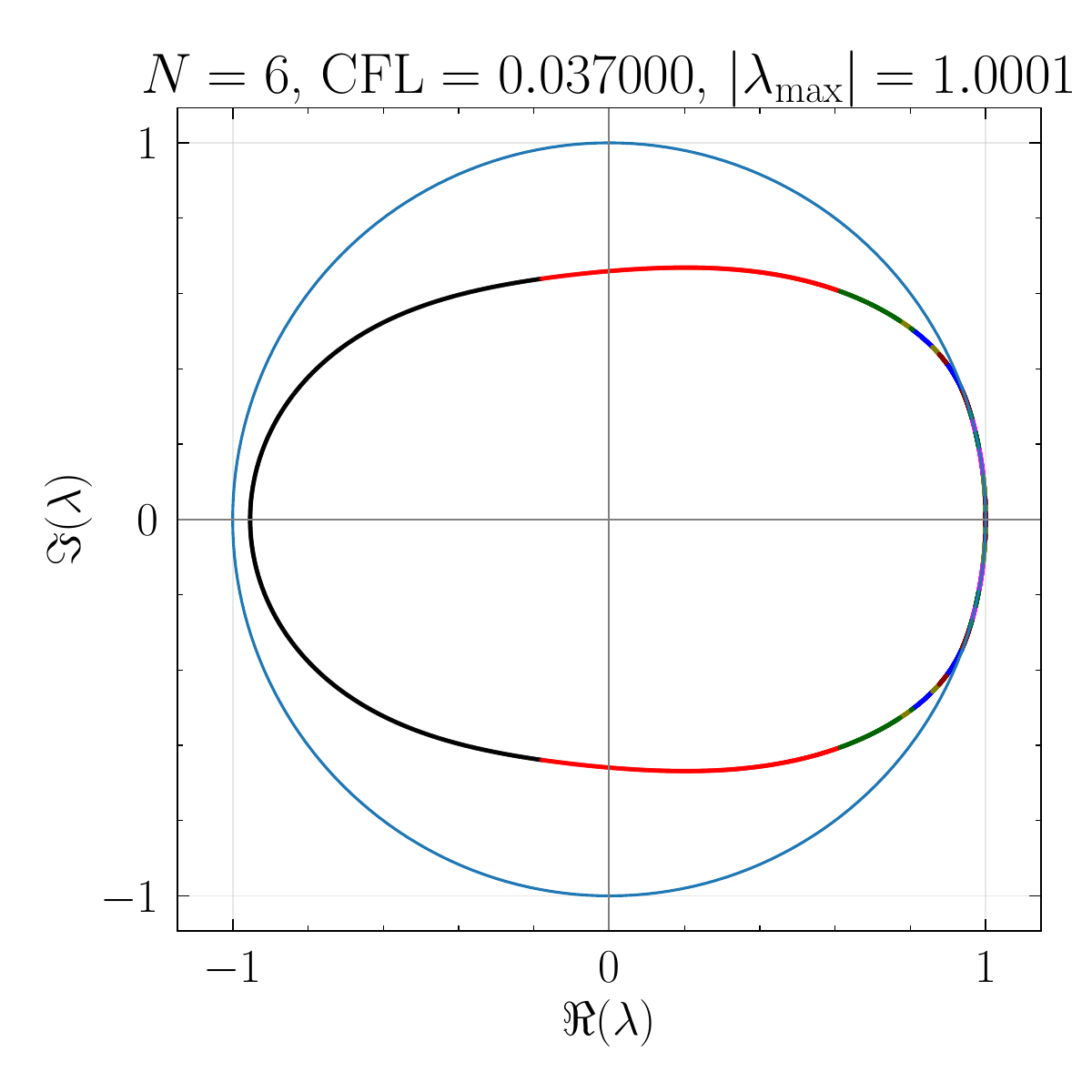}
\includegraphics[width=0.15\textwidth]{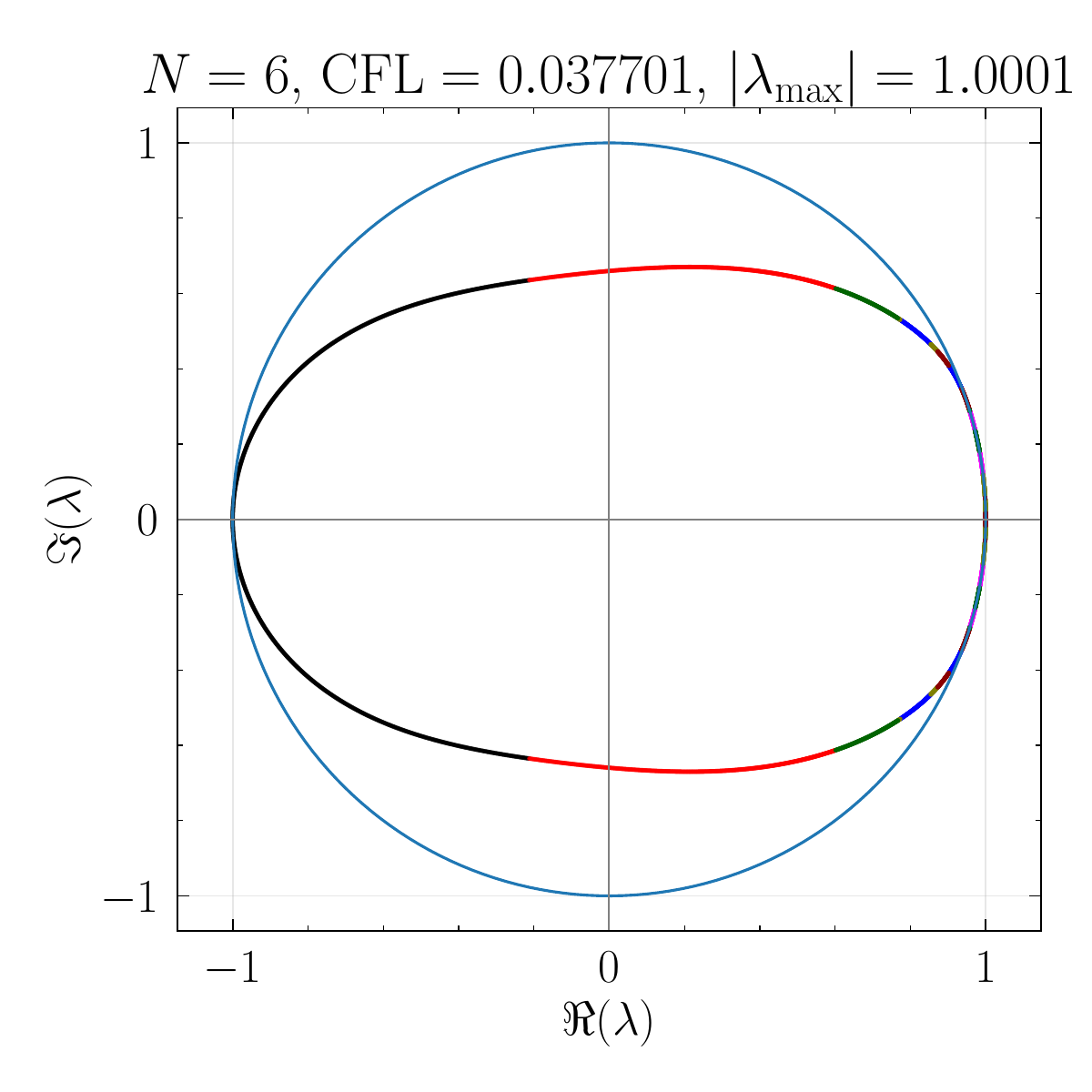}
\includegraphics[width=0.15\textwidth]{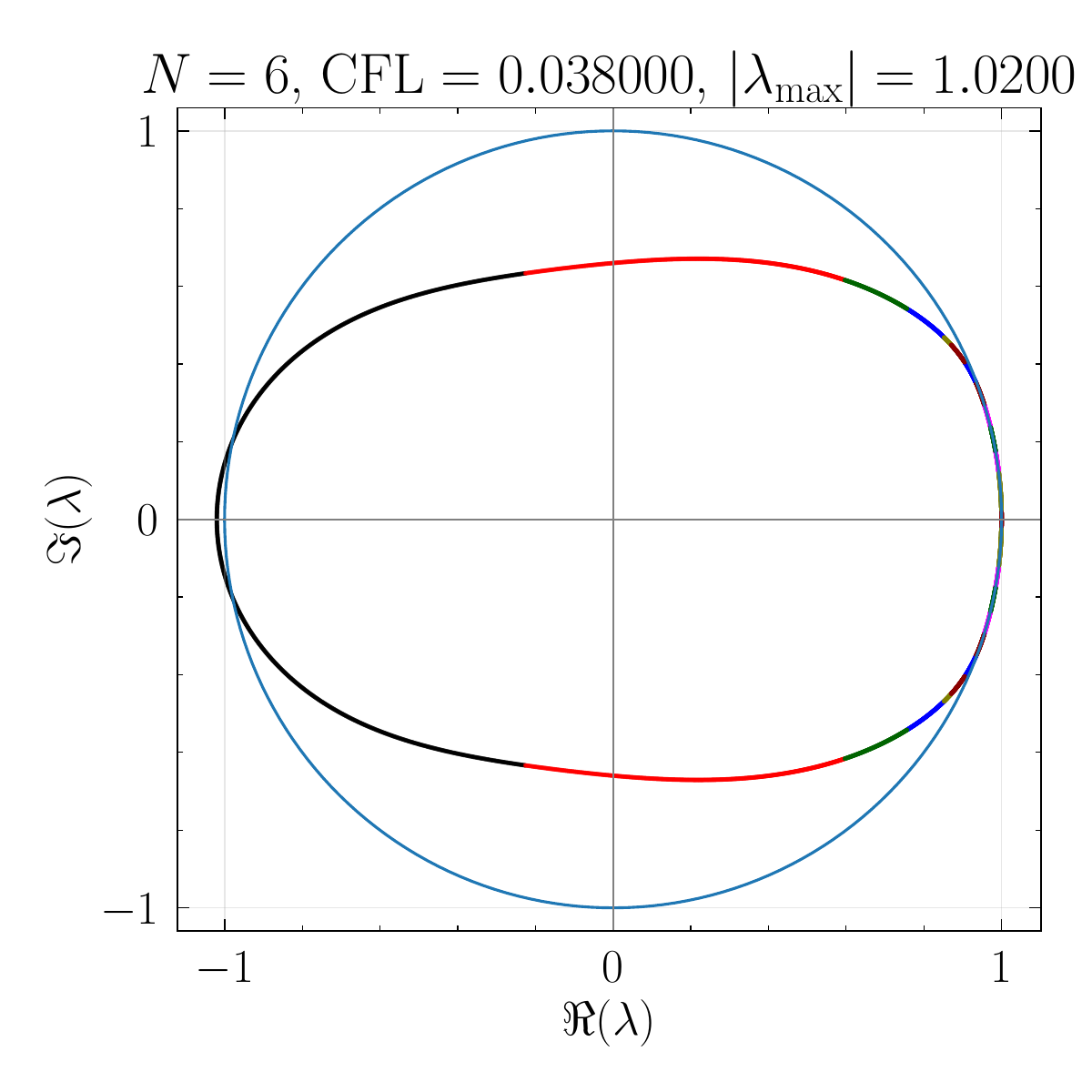}
\includegraphics[width=0.15\textwidth]{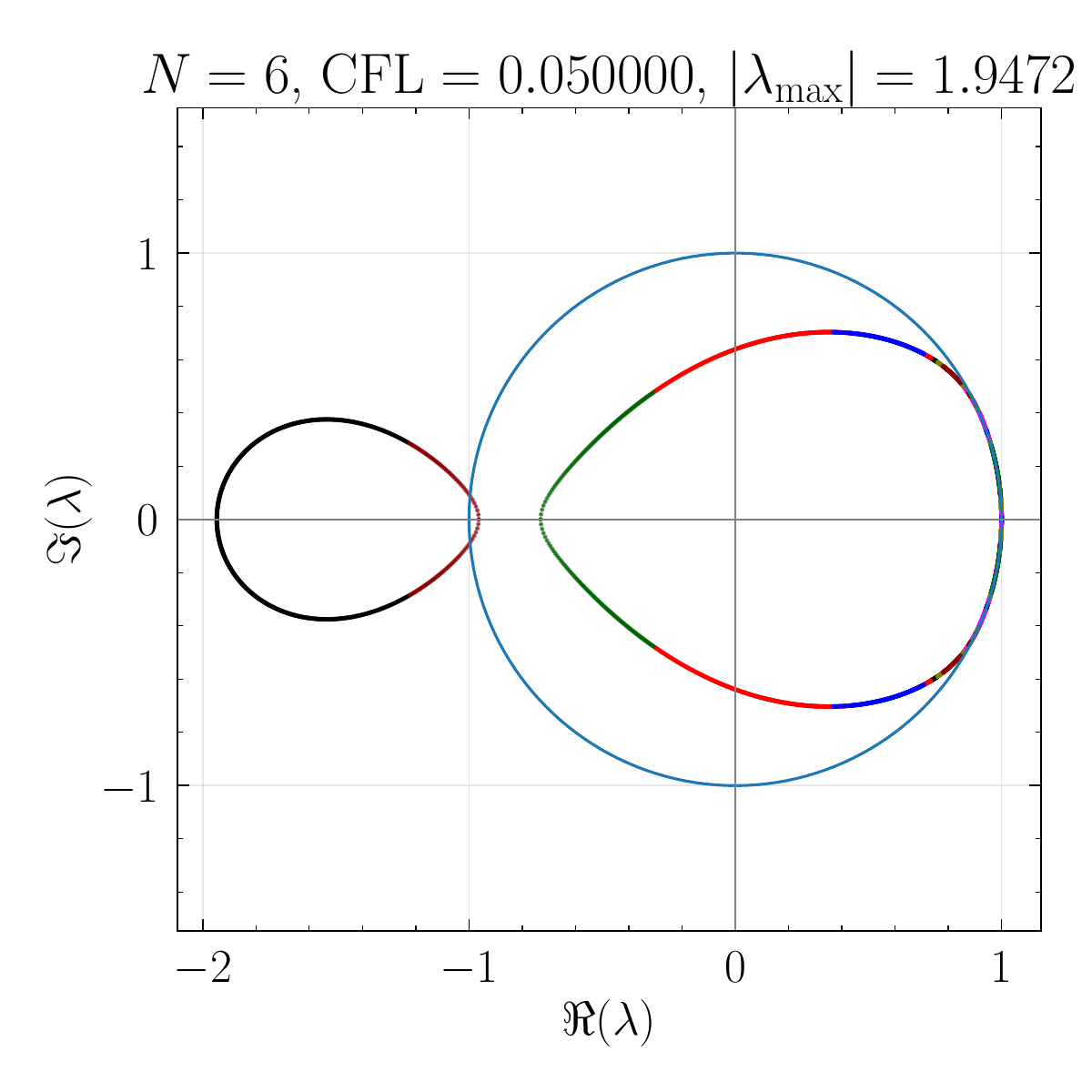}\\
\caption{%
Spectrum of the matrix $\mathrm{R}(\mathrm{CFL}, \theta)$ (\ref{eq:r_matrix_elems_expr_dup}) (eigenvalues $\lambda_{k} = \lambda_{k}(\mathrm{CFL}, \theta)$, $k = 0, \ldots, N$) of the evolution operator $R$ (\ref{eq:evol_oper_prop}) for a single time step $\Dtn{n}$ for several selected values of the Courant number $\mathrm{CFL}$ for polynomial degrees $N = 1, \ldots, 6$. The range of phase $\theta\in[0, 2\pi)$ is sampled on a uniform grid of $1000$ nodes. Legends for each row of the graphs are located on the left. $|\lambda_{\rm max}|$ is the absolute value of the largest eigenvalue $\lambda_{k}(\mathrm{CFL}, \theta)$ in a graph. The gray circle of unit radius defines the stability boundary. The Courant numbers $\mathrm{CFL}$ are taken deep inside the stability region $\mathrm{CFL} \in [0, \mathrm{CFL}_{\rm max}]$ (left column), in the region of guaranteed instability $\mathrm{CFL} > \mathrm{CFL}_{\rm max}$ (right column) and near the boundary of the stability region $\mathrm{CFL}_{\rm max}$ (the values $\mathrm{CFL}_{\rm max}$ are selected from work~\cite{ader_dg_stab}, as well as the values $\mathrm{CFL}_{\rm max}$ calculated further in this work in Table~\ref{tab:cfls_max_data} and in Figure~\ref{fig:cfls_max_data}). \textit{Note}: the phase $\arg \lambda_{k}$ of eigenvalue $\lambda_{k}$ is not the phase $\theta$; the dependence of the absolute value $|\lambda_{k}|$ on phase $\theta$ for these polynomial degrees $N$ is presented in Figure~\ref{fig:rhos_on_theta_degrees_1_6}.
}
\label{fig:spectrum_exact_eigvals_degrees_1_6}
\end{figure}

\begin{figure}[h!]
\centering
\includegraphics[width=0.028125\textwidth]{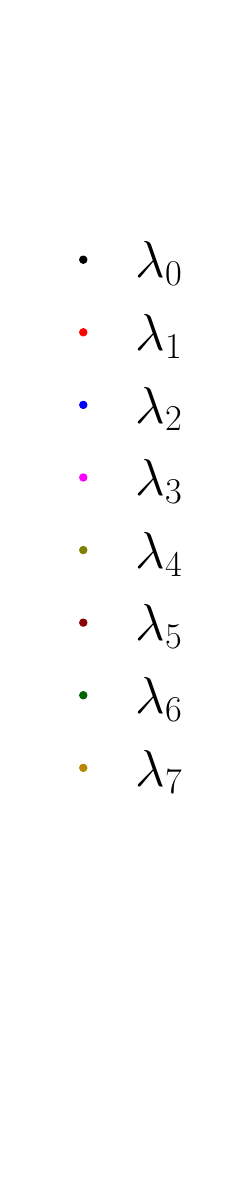}
\includegraphics[width=0.15\textwidth]{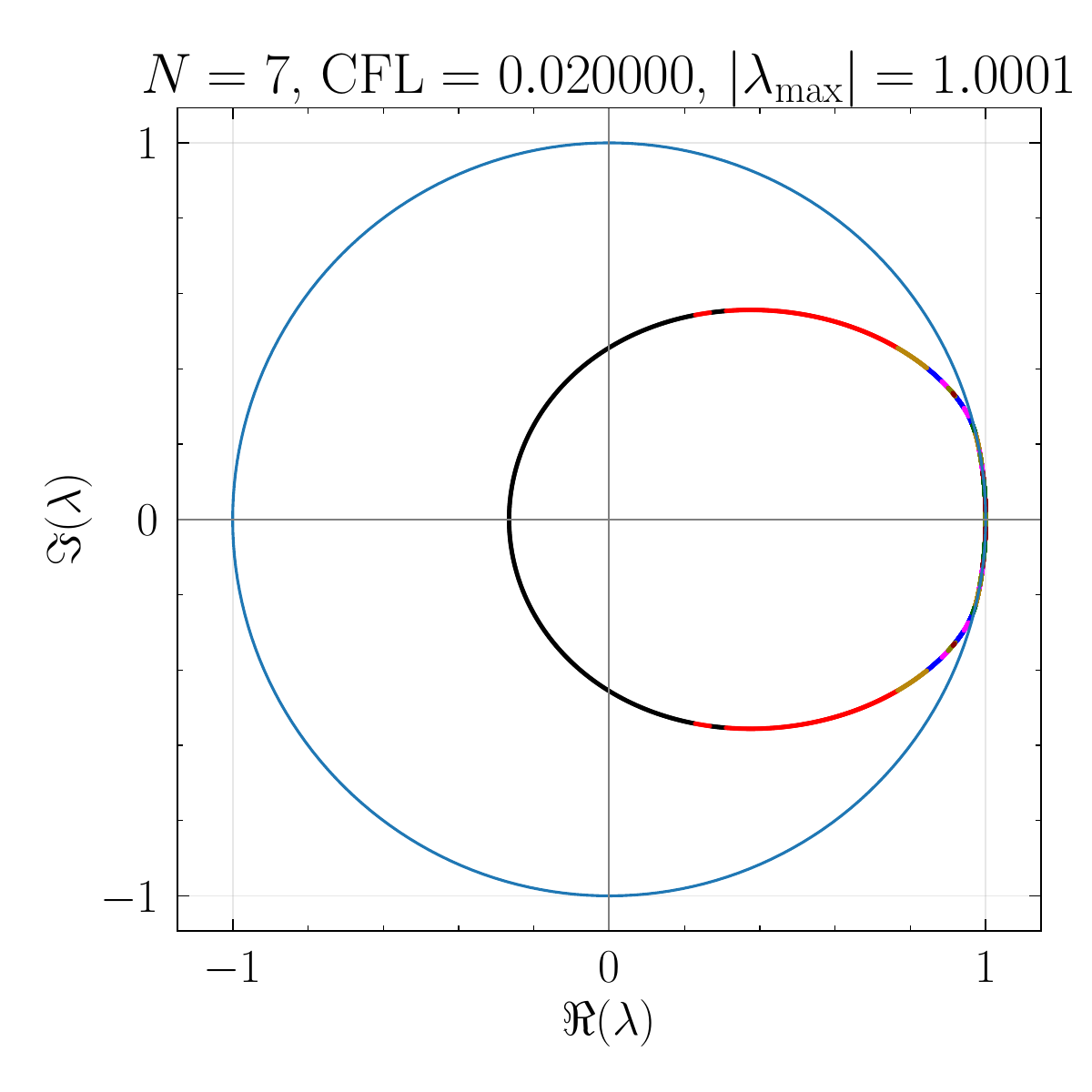}
\includegraphics[width=0.15\textwidth]{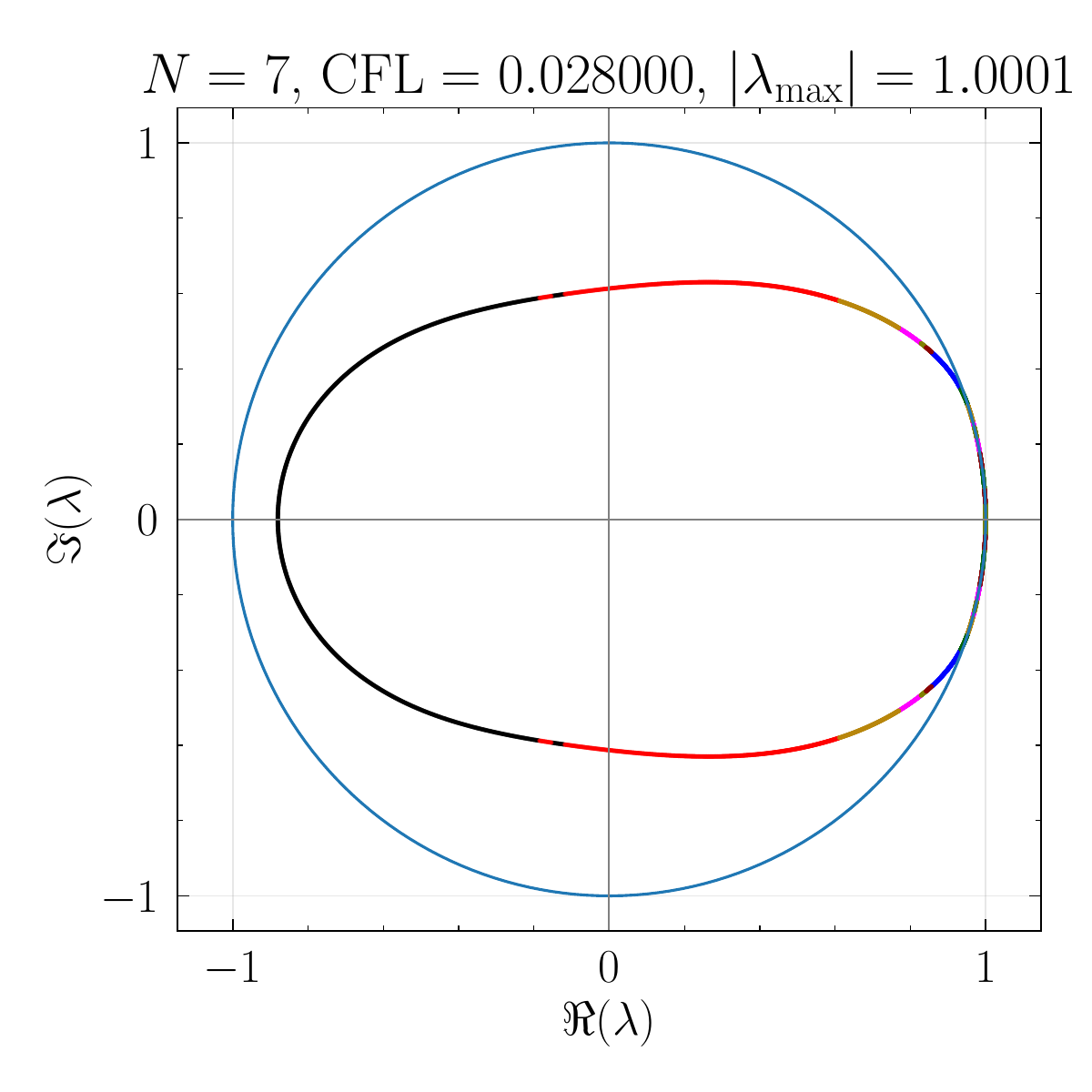}
\includegraphics[width=0.15\textwidth]{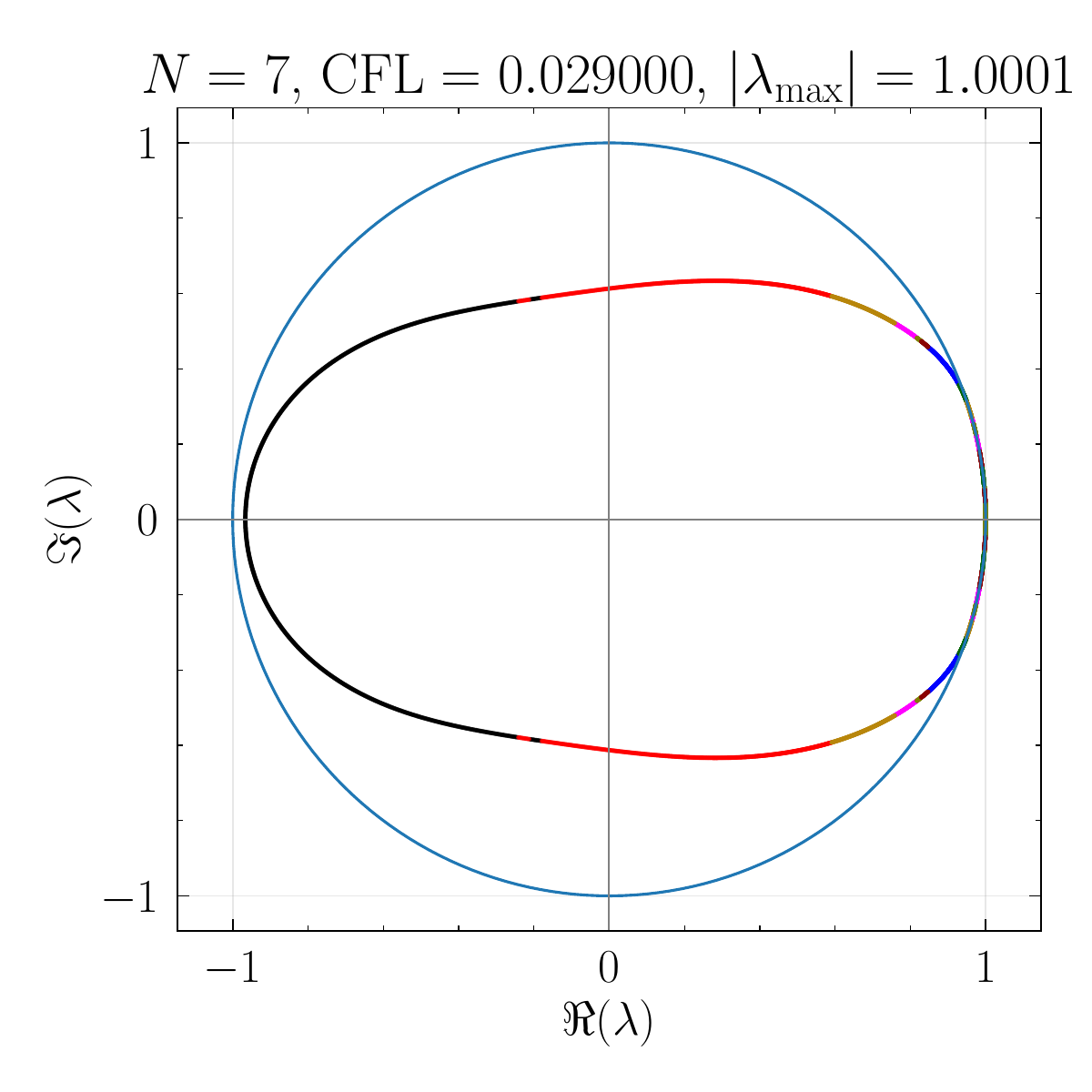}
\includegraphics[width=0.15\textwidth]{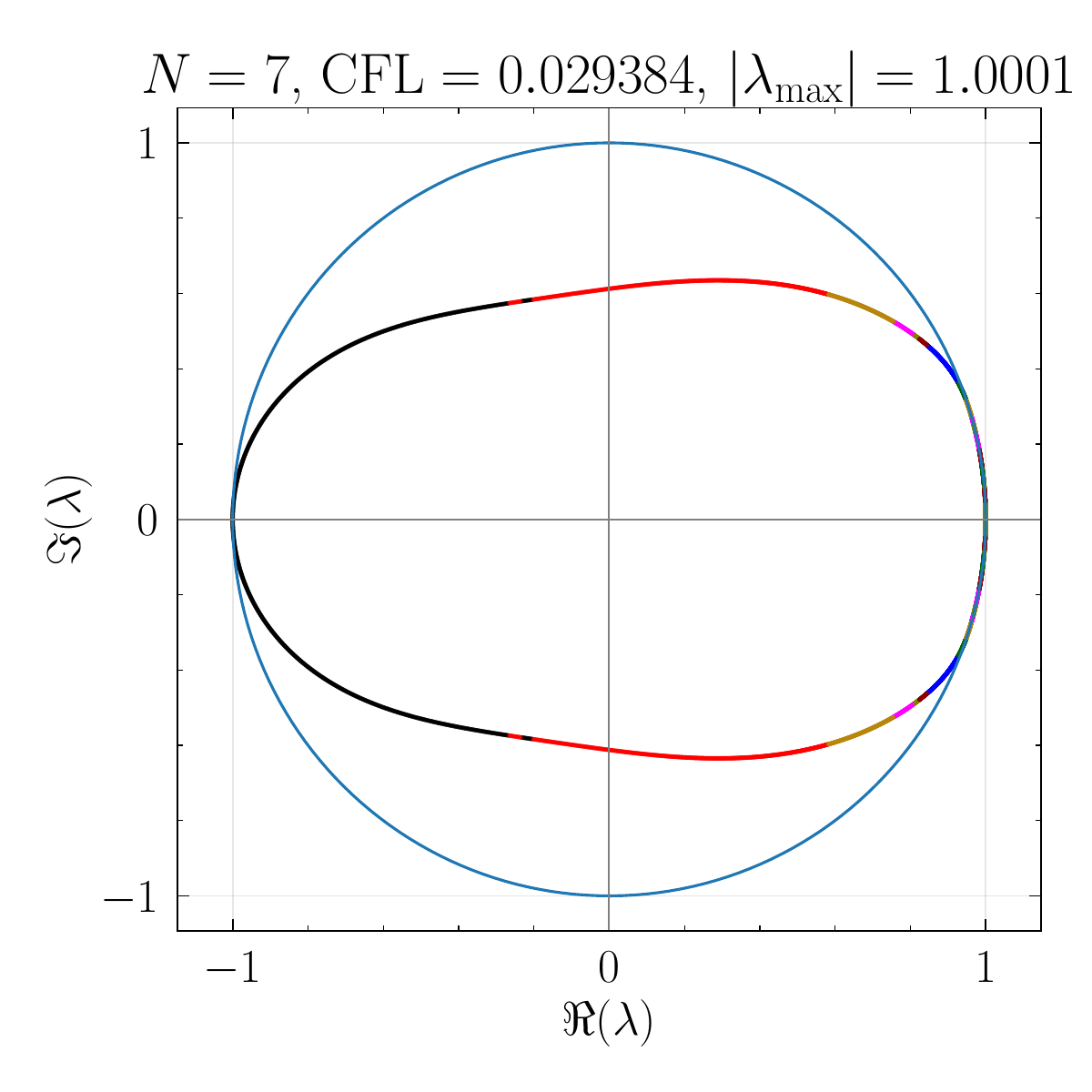}
\includegraphics[width=0.15\textwidth]{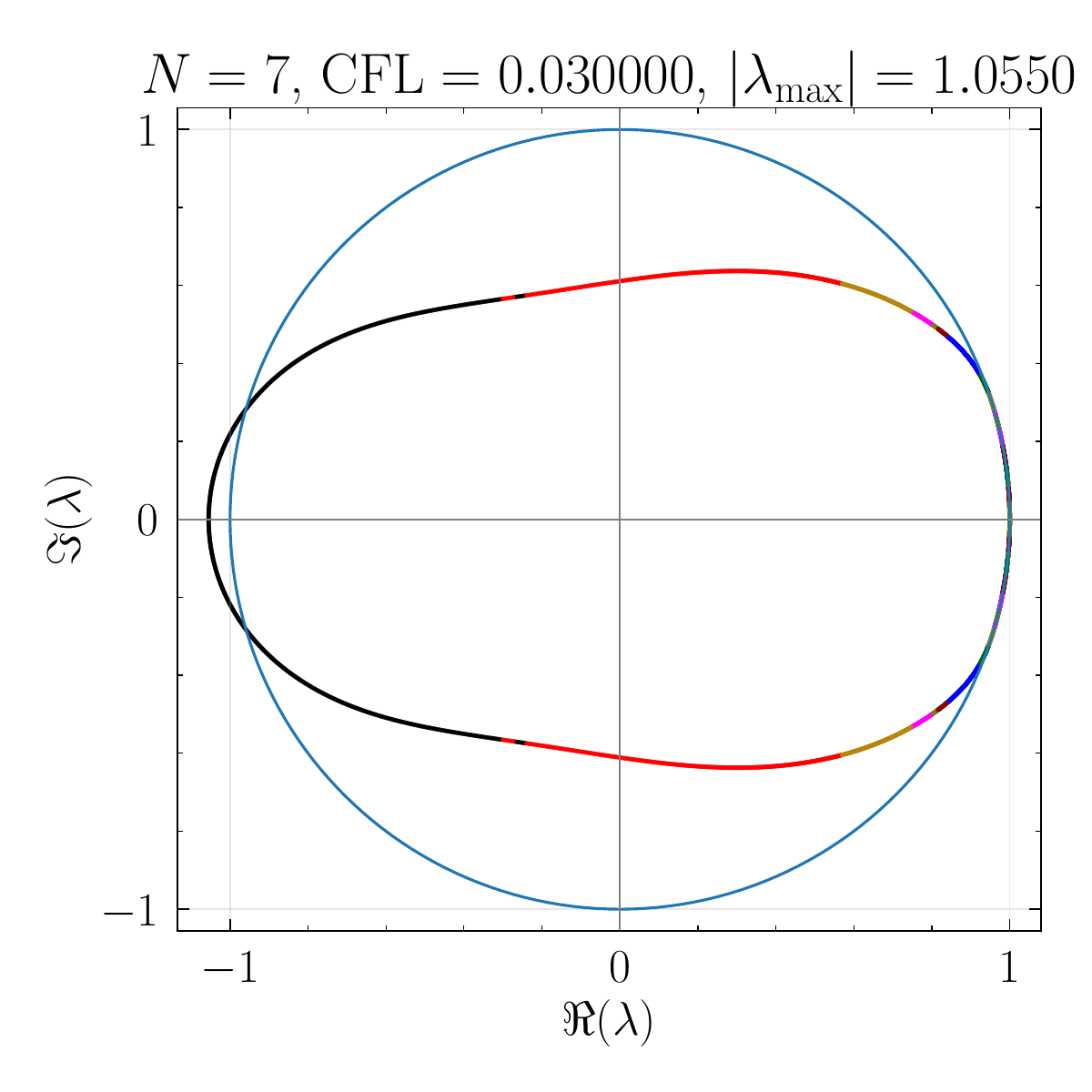}
\includegraphics[width=0.15\textwidth]{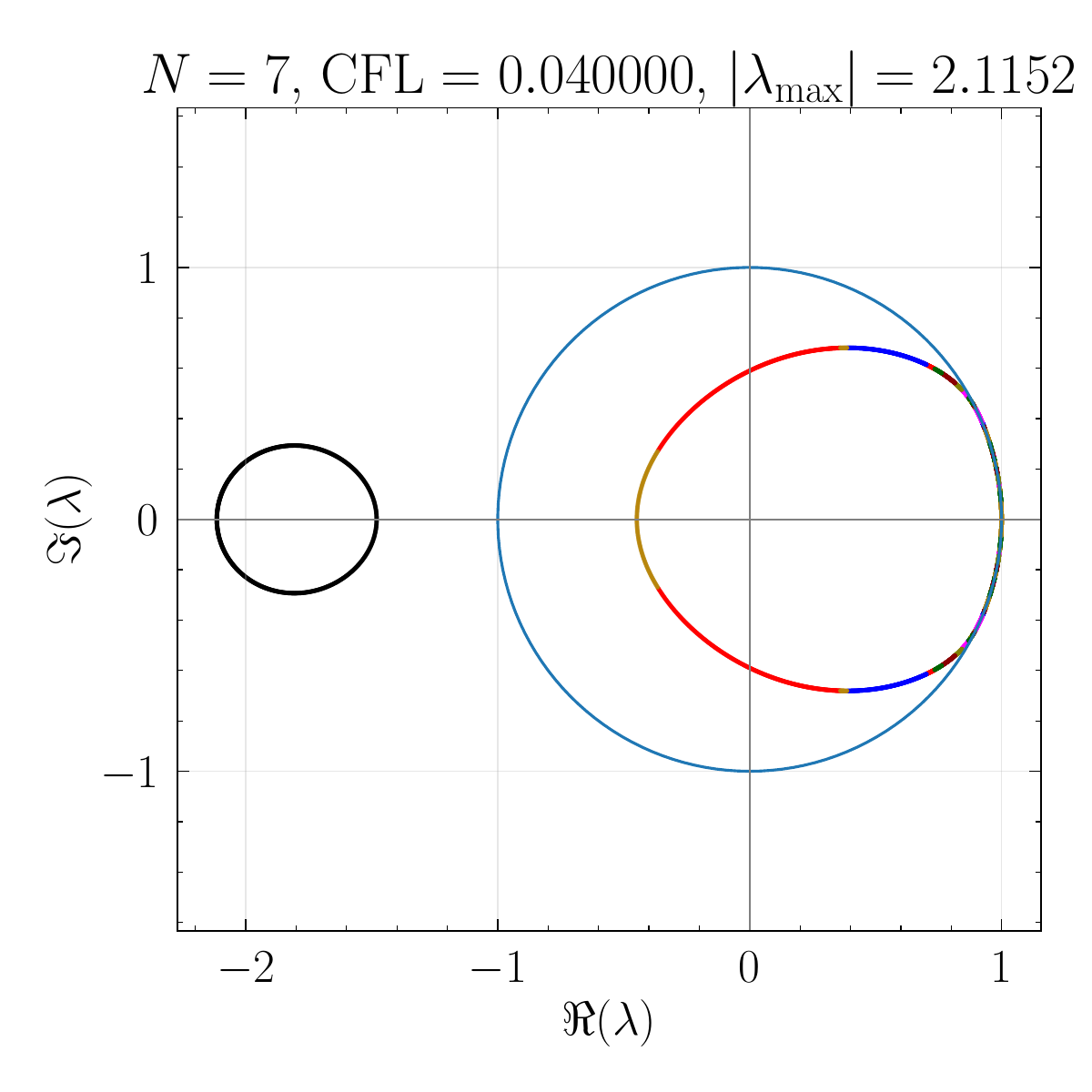}\\
\includegraphics[width=0.028125\textwidth]{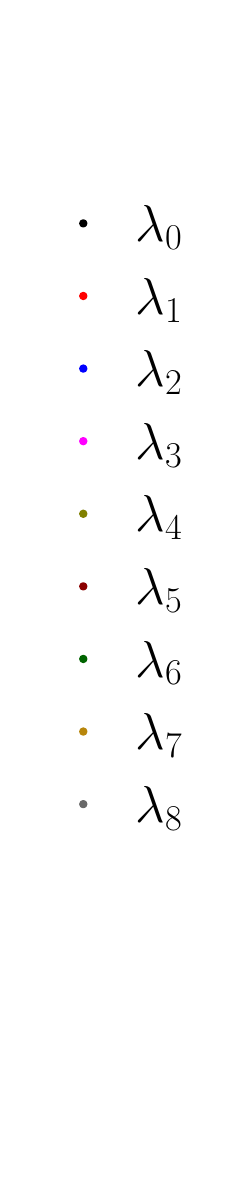}
\includegraphics[width=0.15\textwidth]{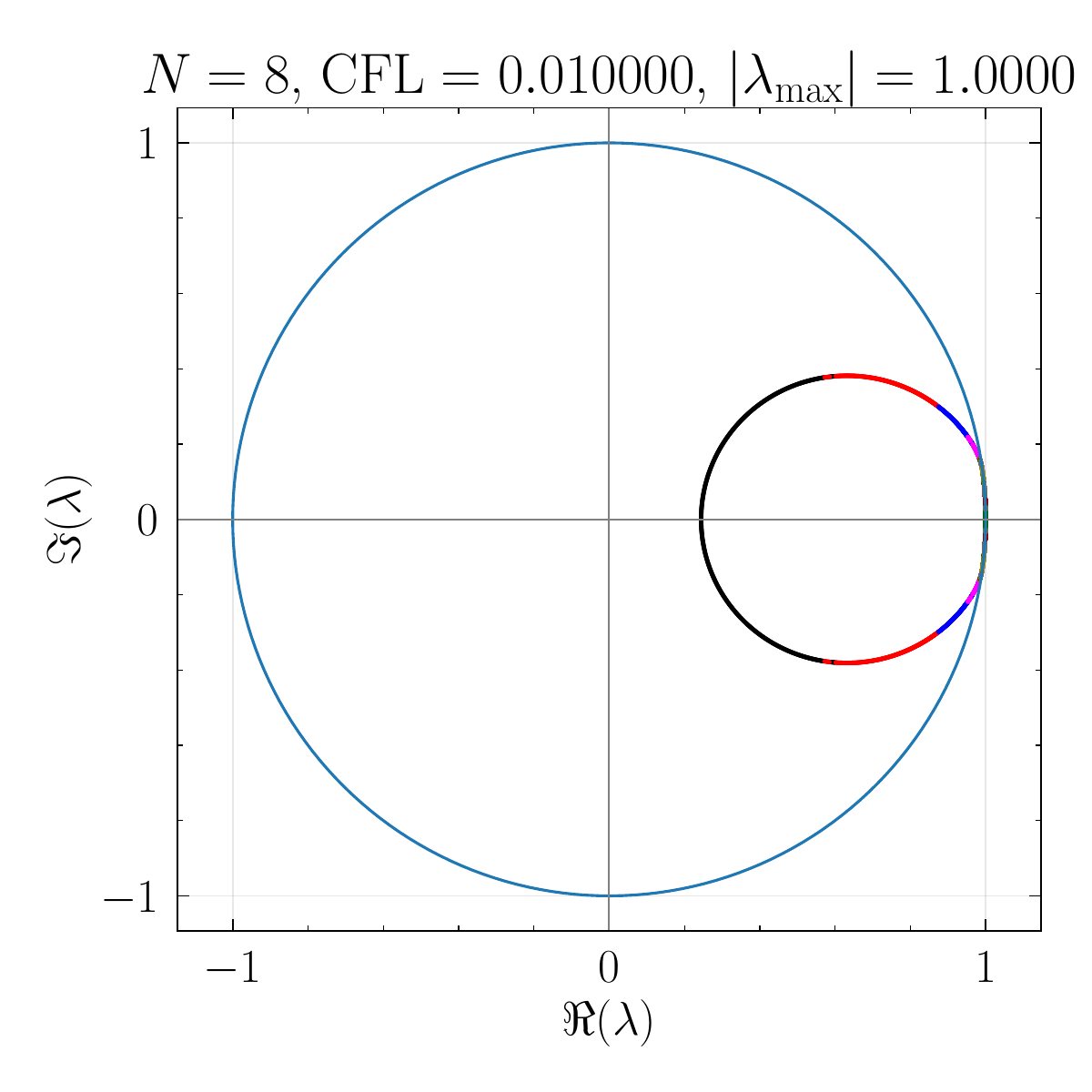}
\includegraphics[width=0.15\textwidth]{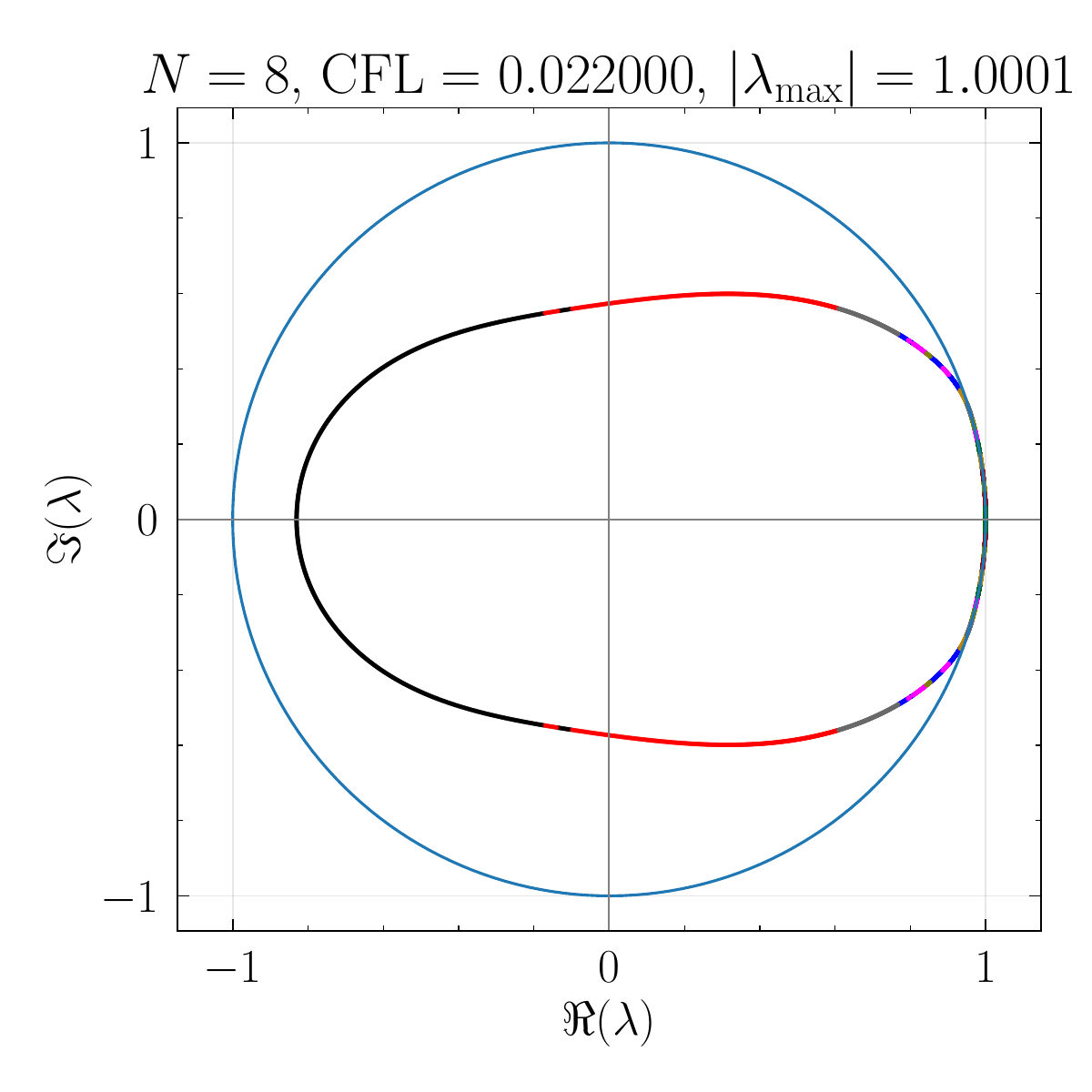}
\includegraphics[width=0.15\textwidth]{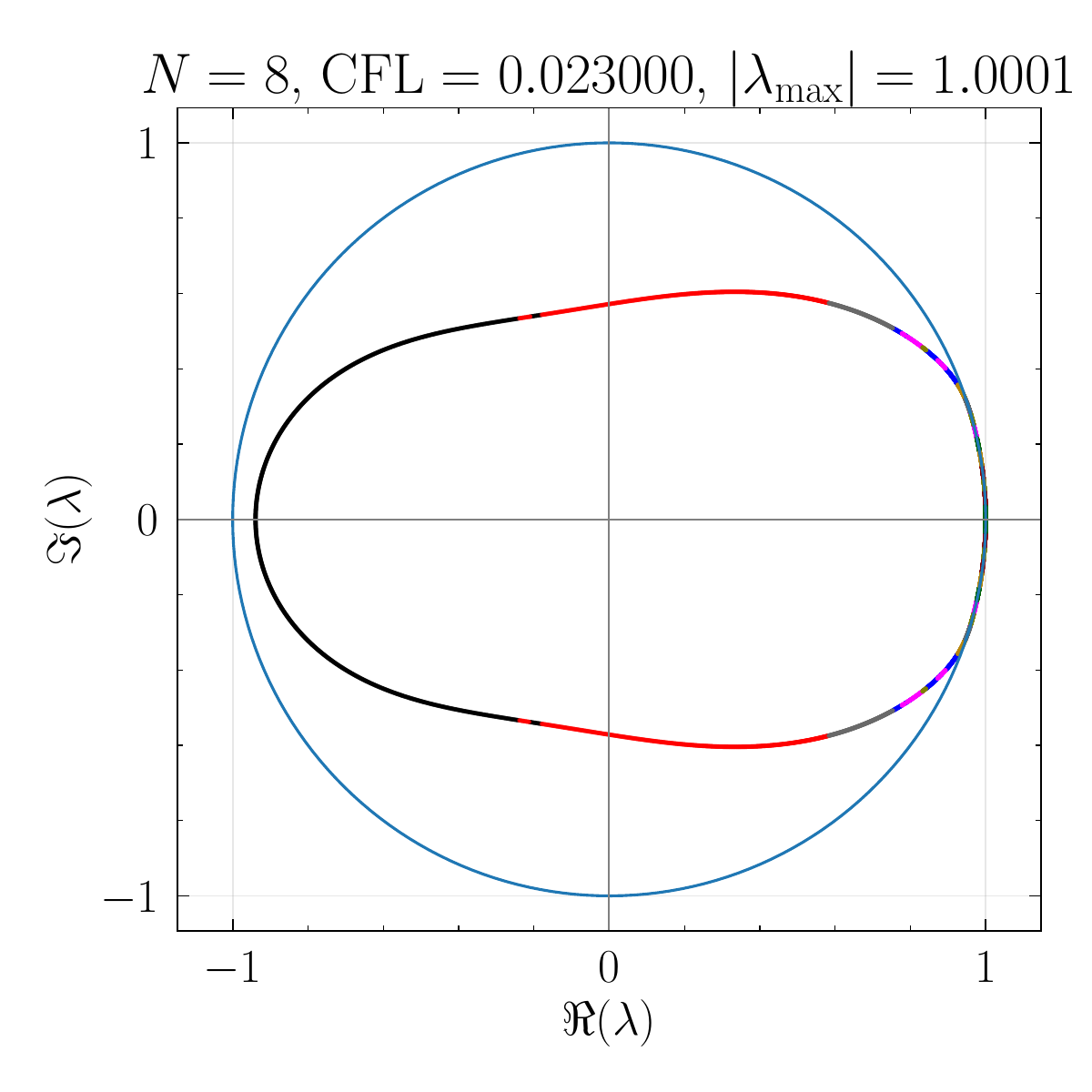}
\includegraphics[width=0.15\textwidth]{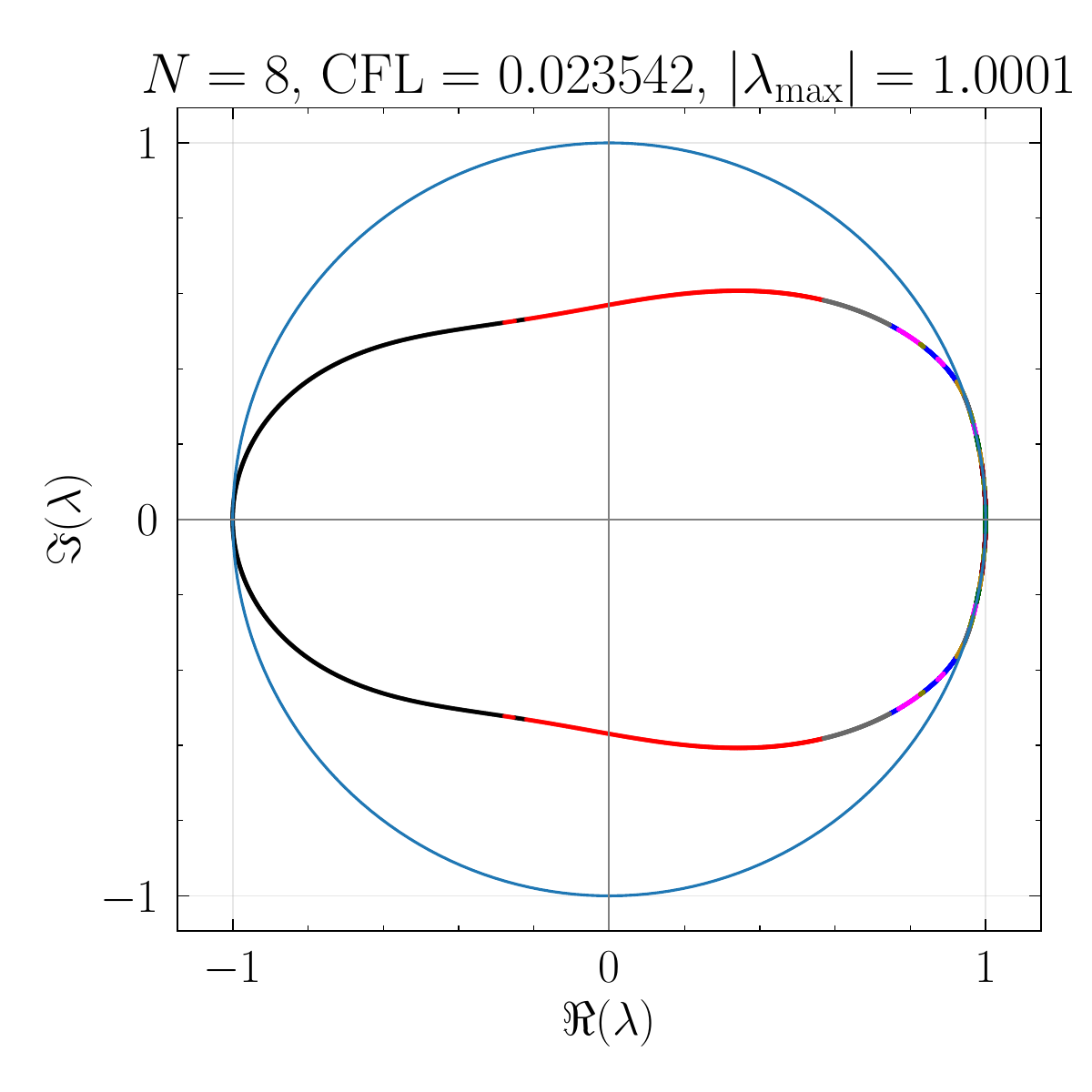}
\includegraphics[width=0.15\textwidth]{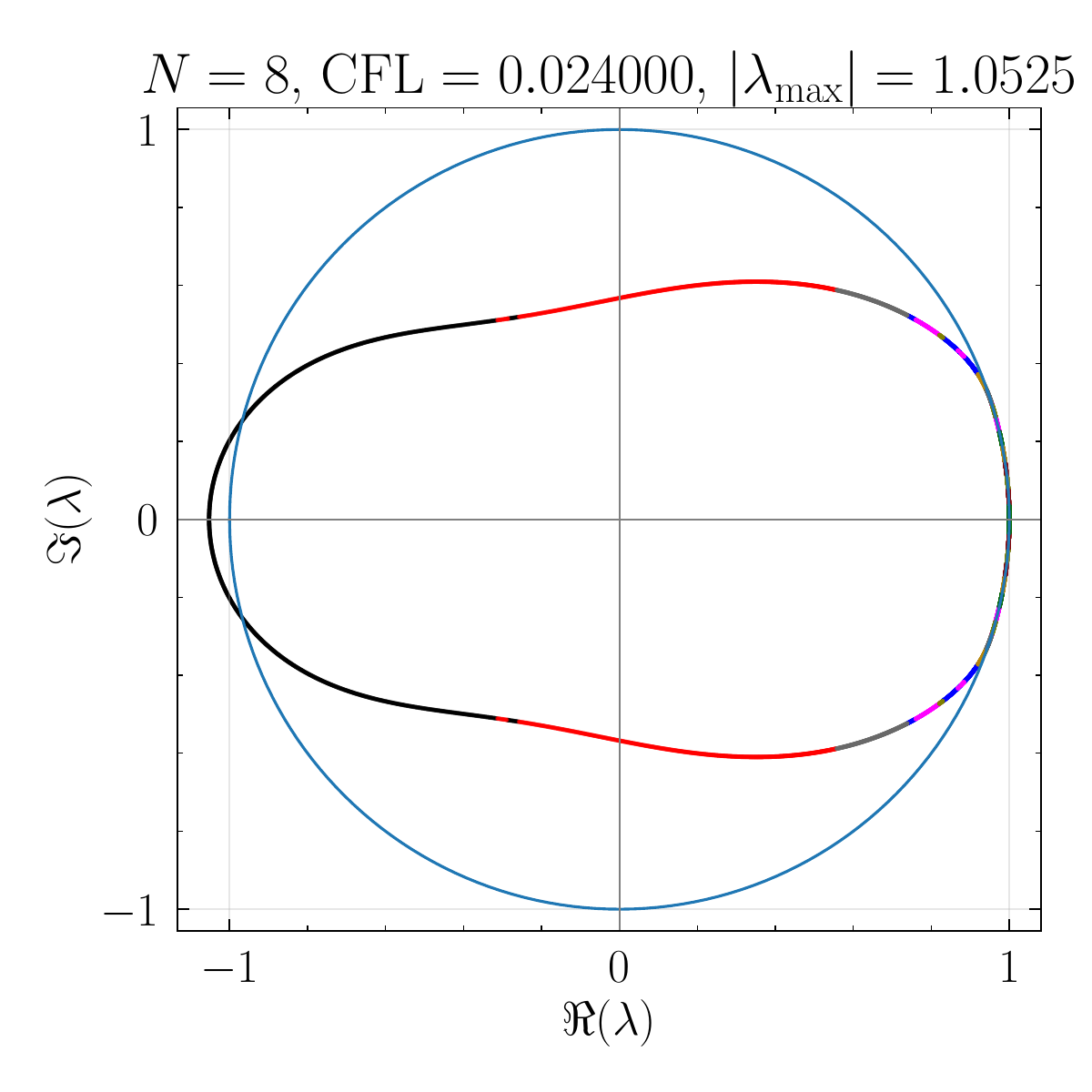}
\includegraphics[width=0.15\textwidth]{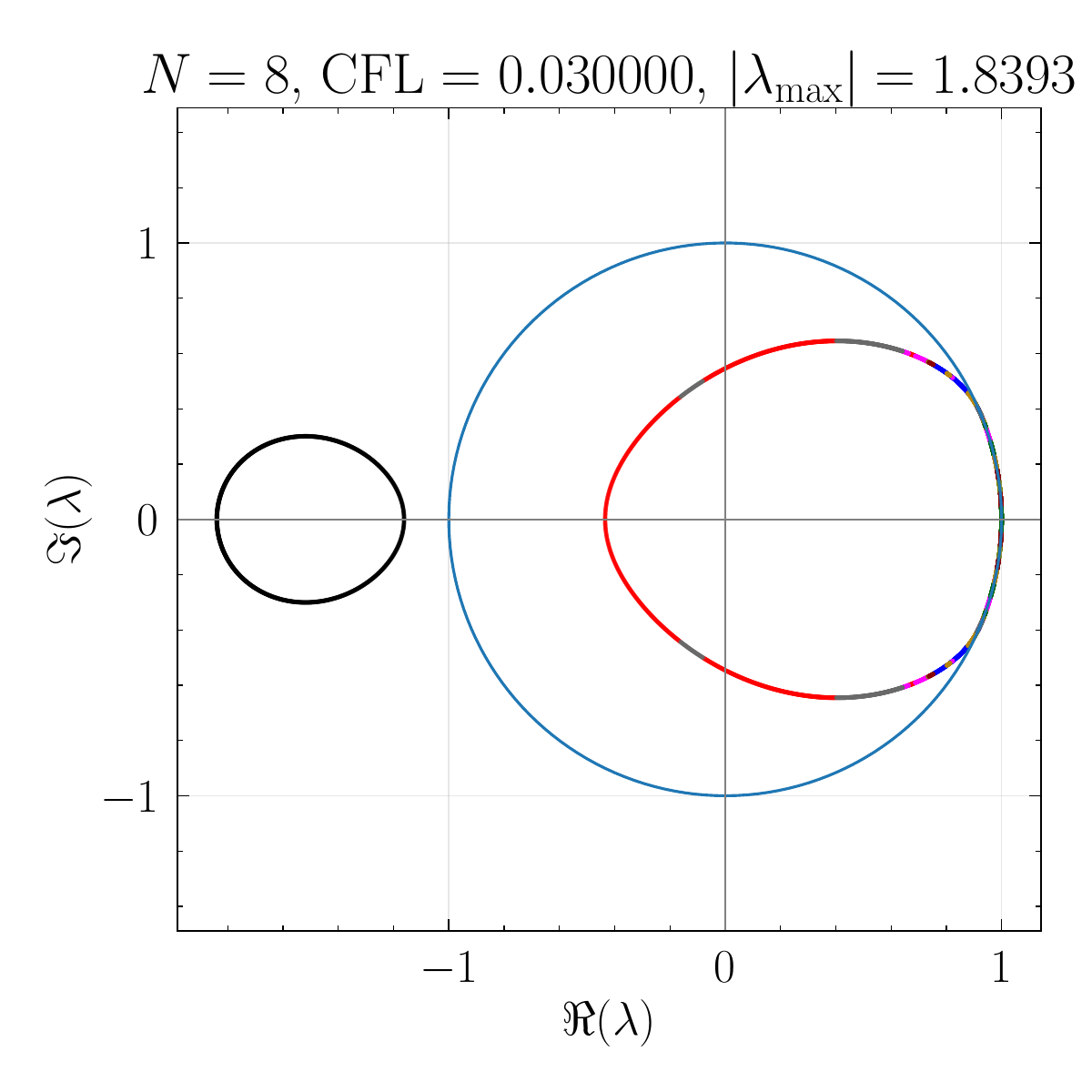}\\
\includegraphics[width=0.028125\textwidth]{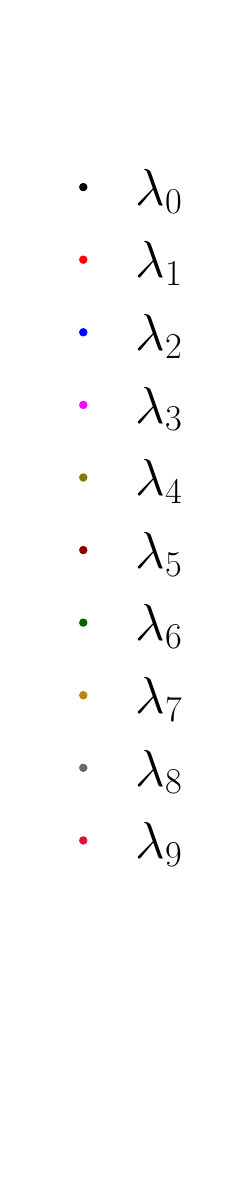}
\includegraphics[width=0.15\textwidth]{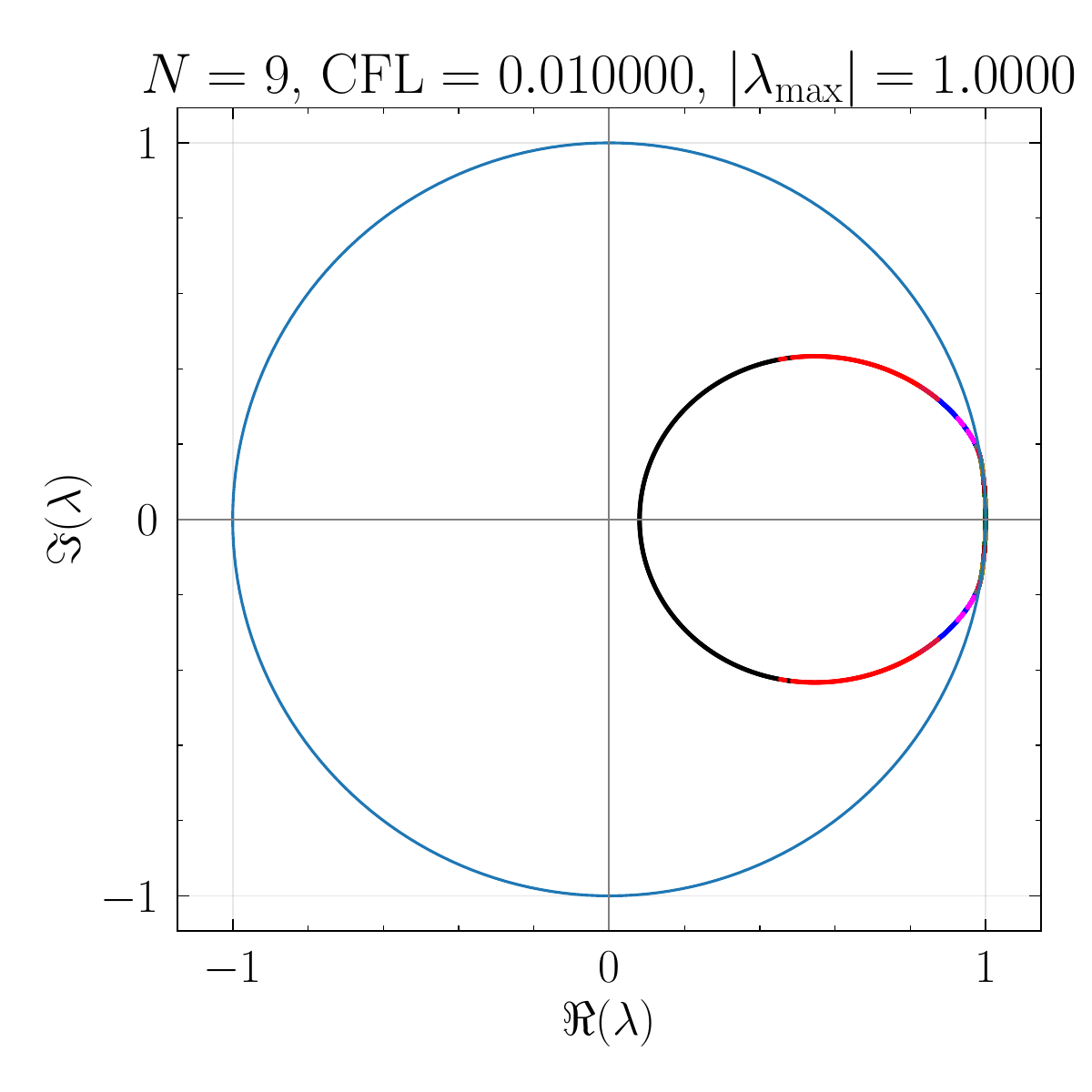}
\includegraphics[width=0.15\textwidth]{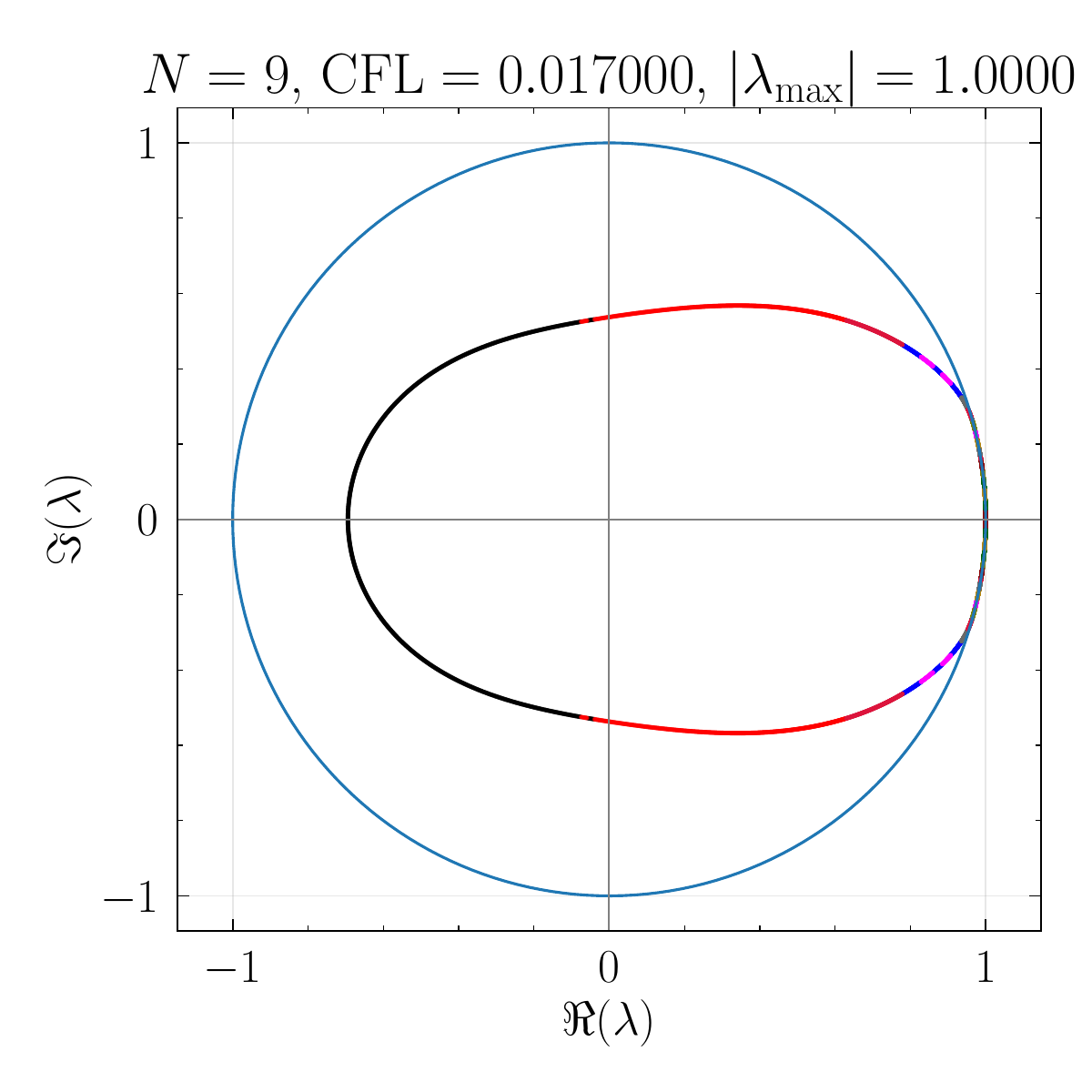}
\includegraphics[width=0.15\textwidth]{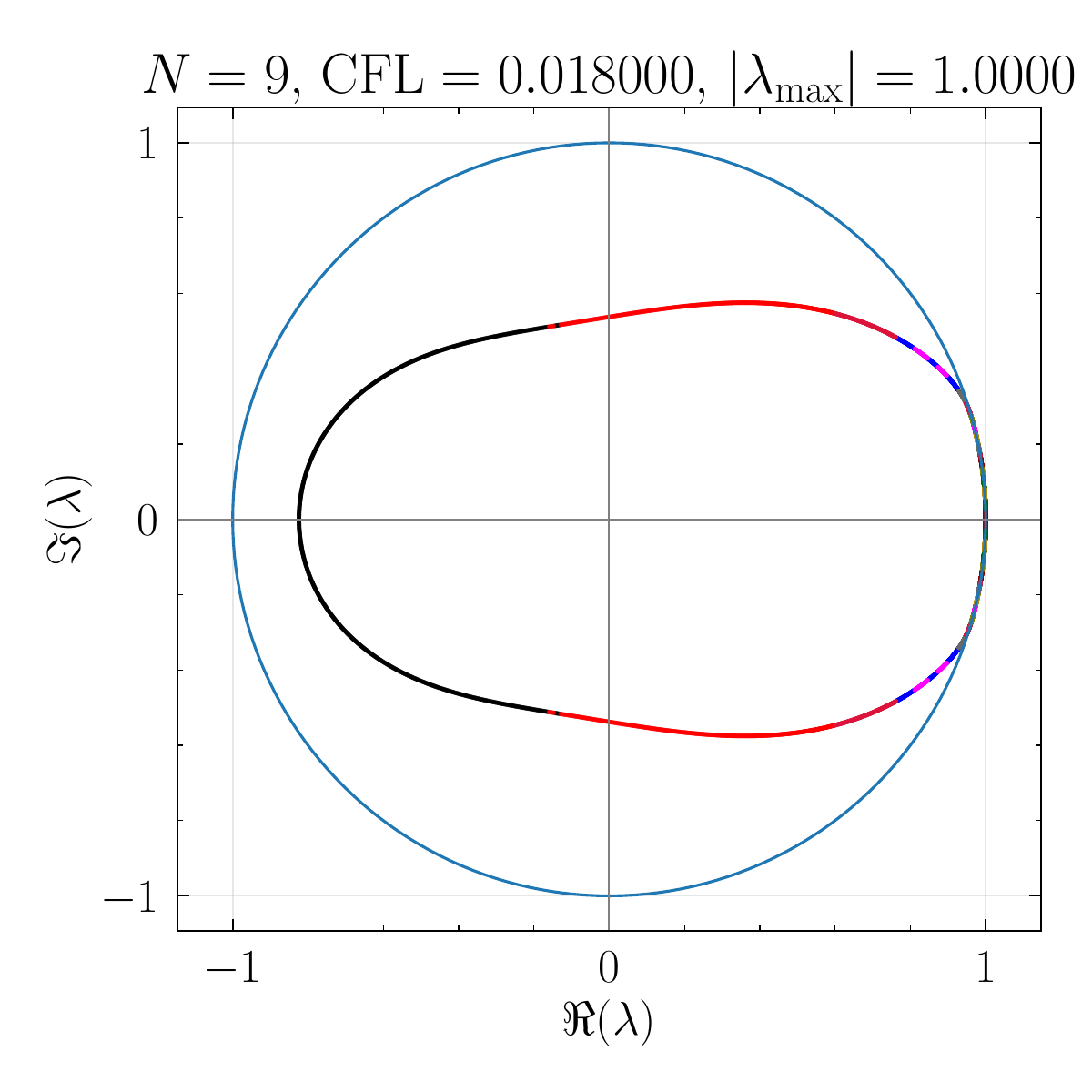}
\includegraphics[width=0.15\textwidth]{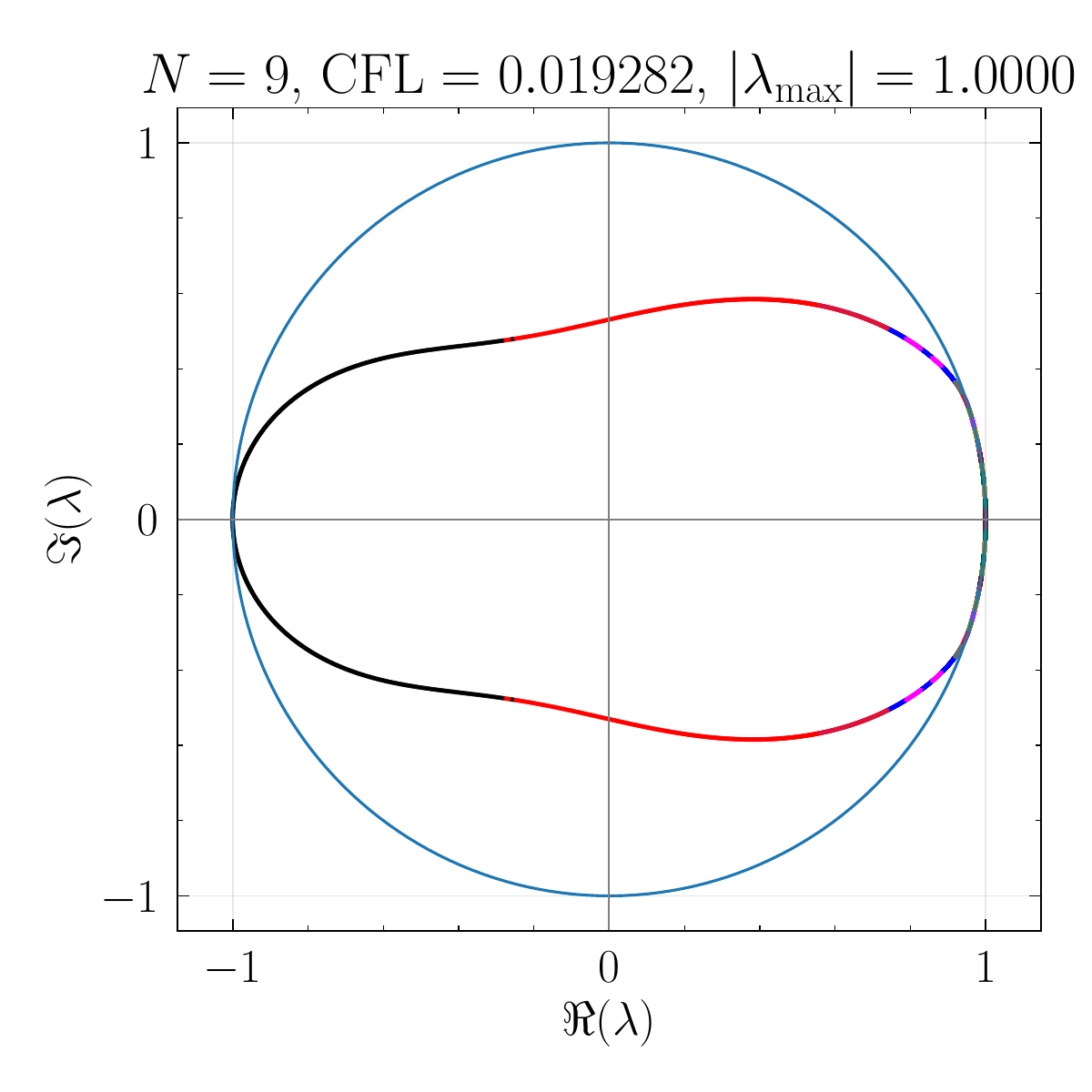}
\includegraphics[width=0.15\textwidth]{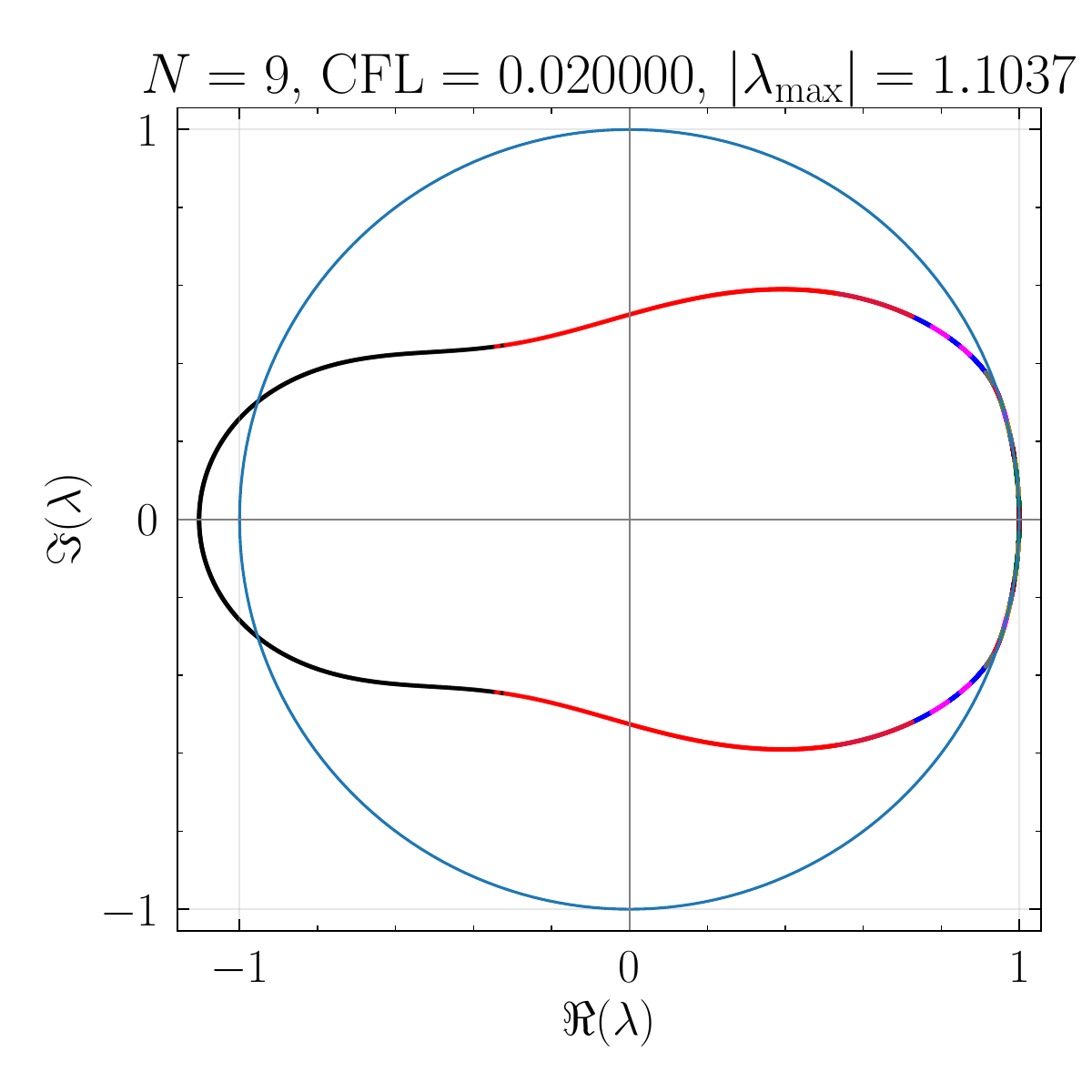}
\includegraphics[width=0.15\textwidth]{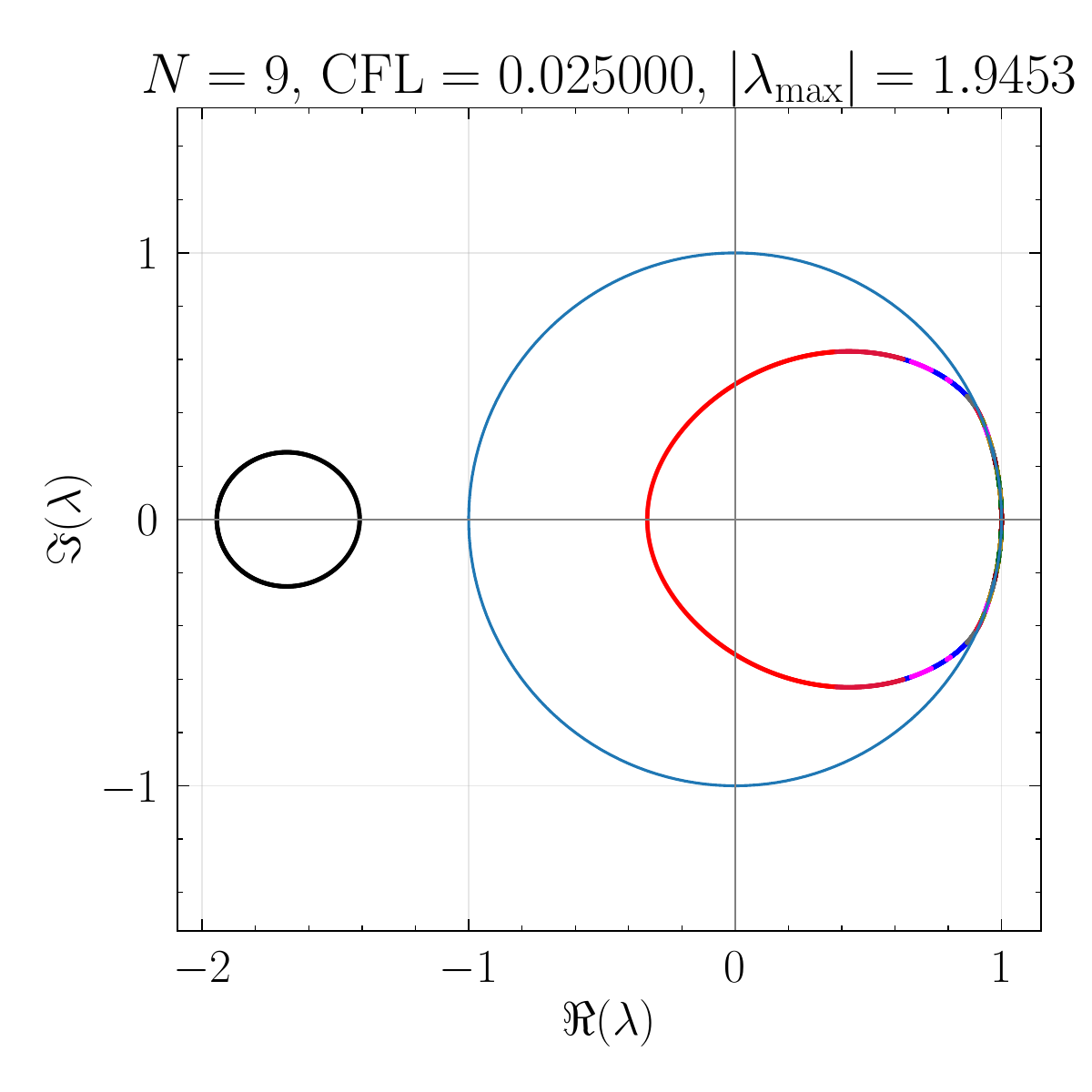}\\
\includegraphics[width=0.028125\textwidth]{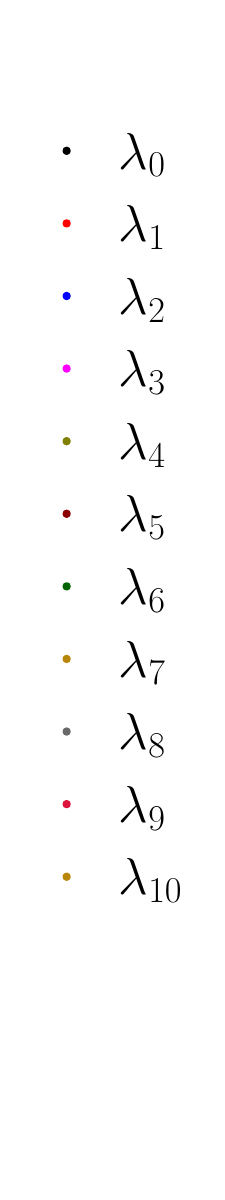}
\includegraphics[width=0.15\textwidth]{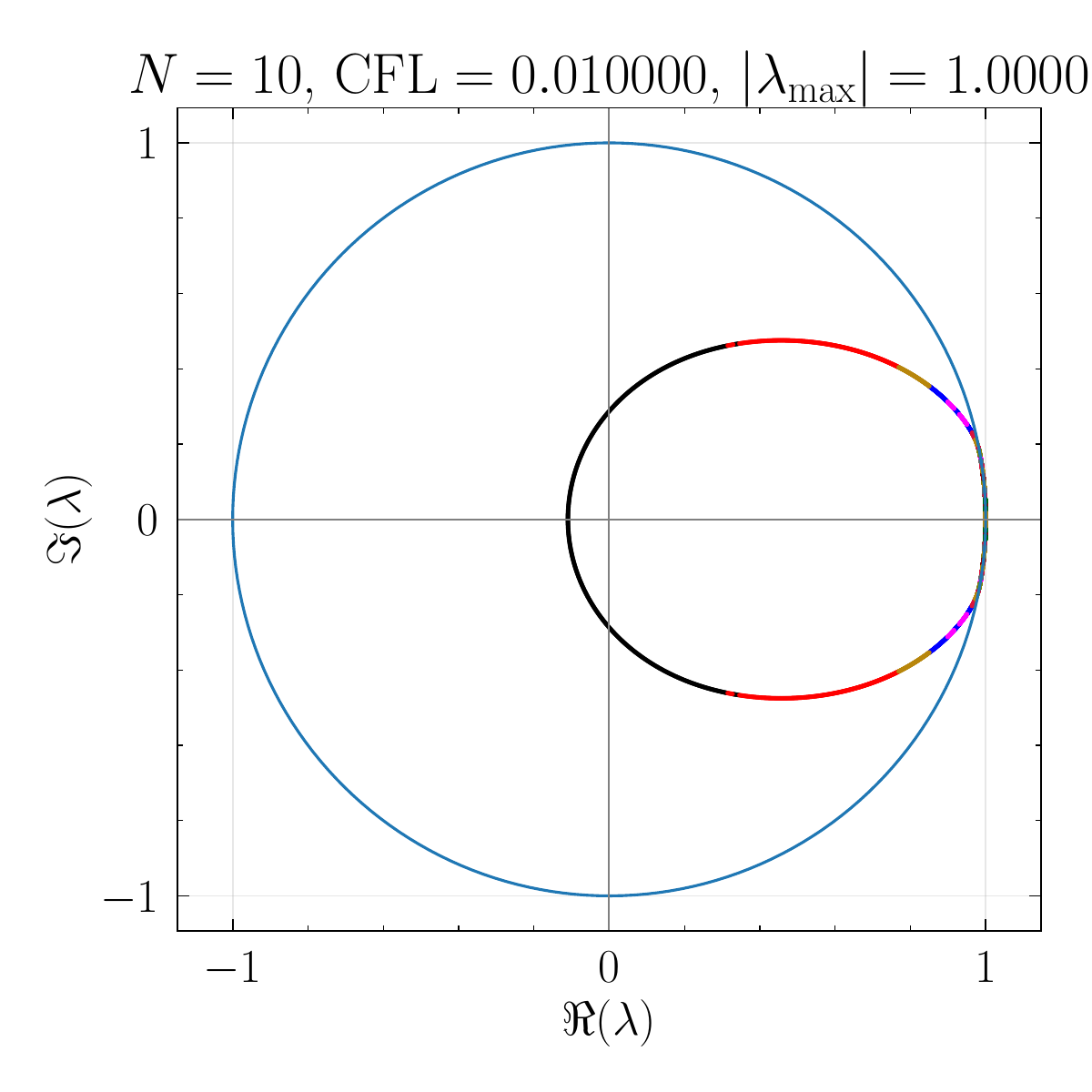}
\includegraphics[width=0.15\textwidth]{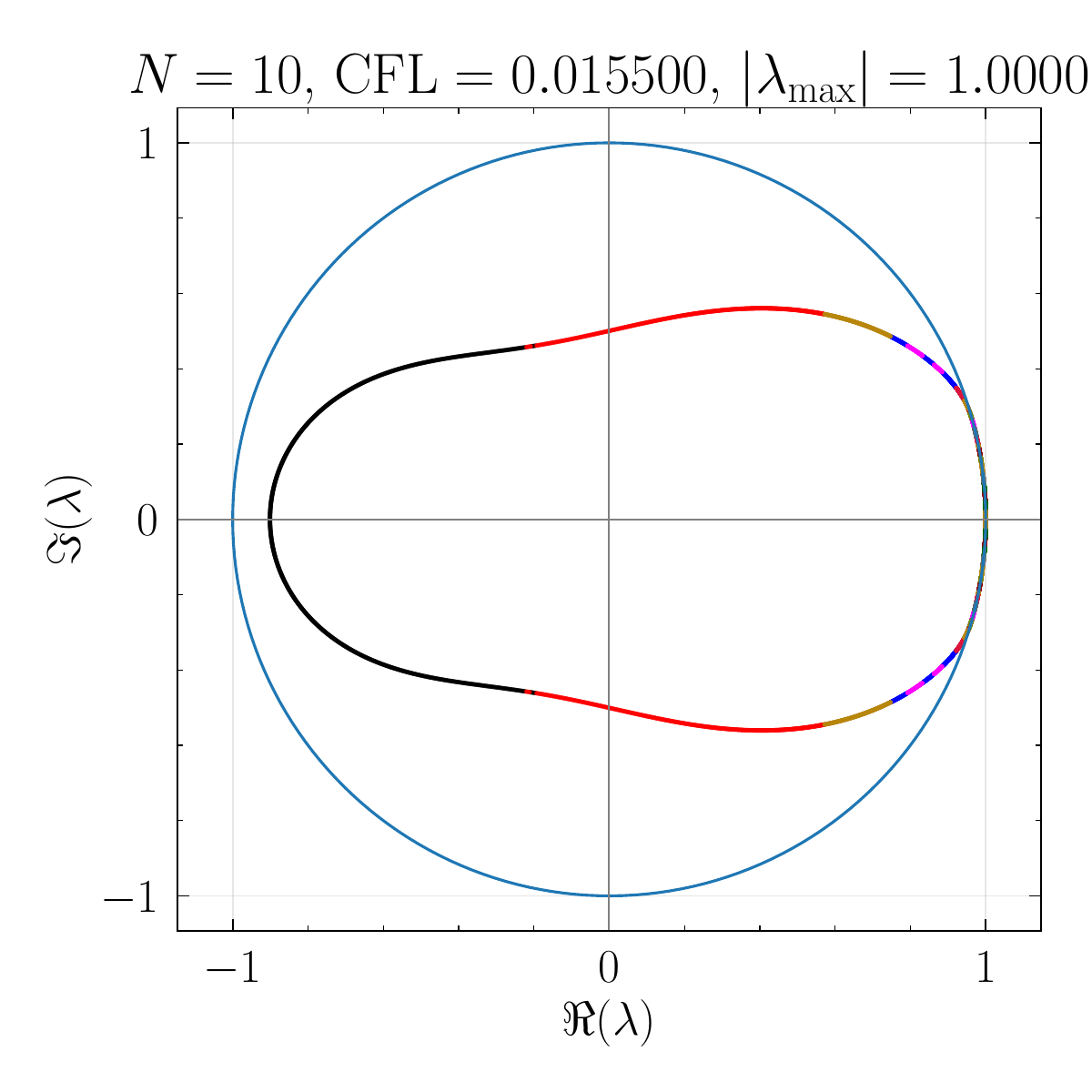}
\includegraphics[width=0.15\textwidth]{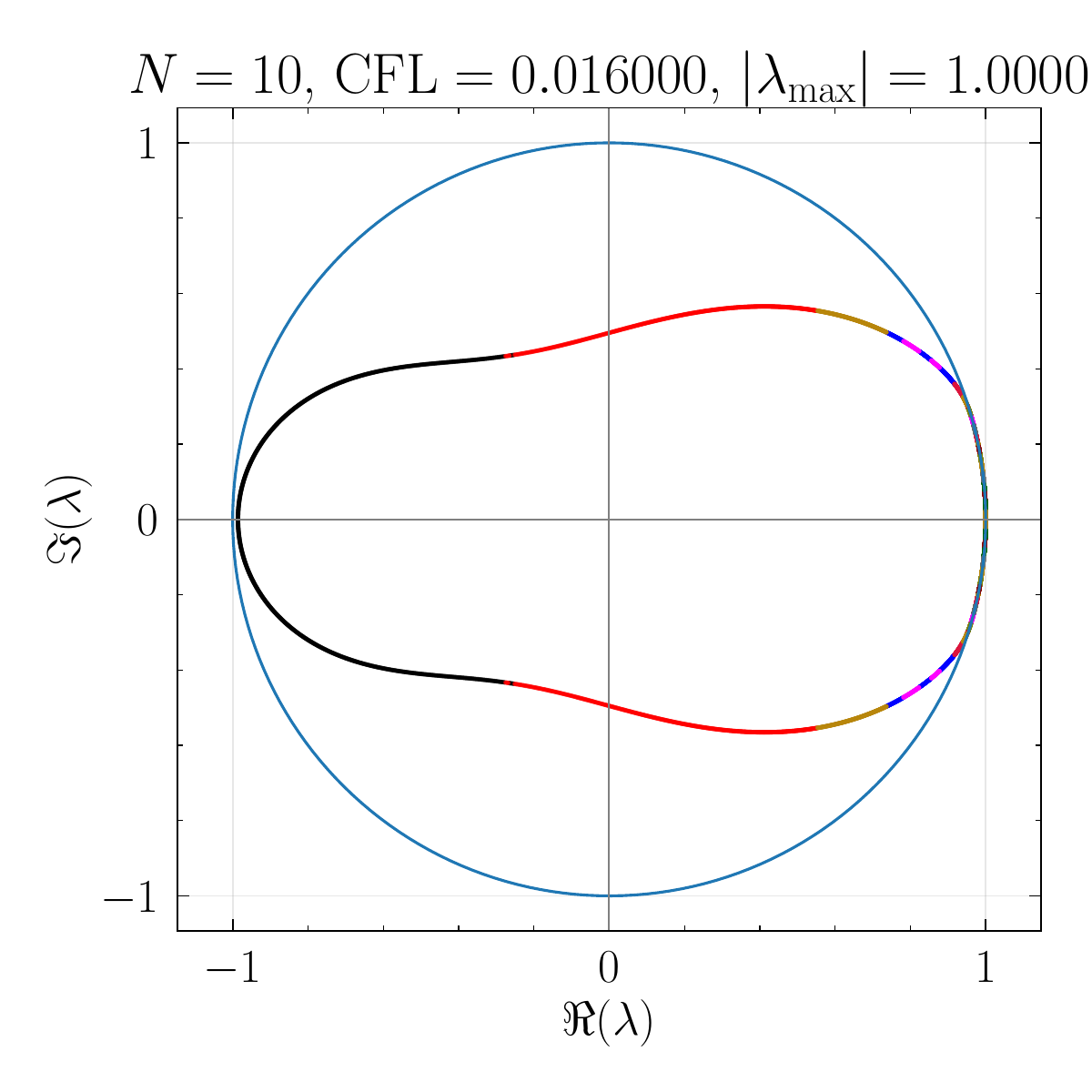}
\includegraphics[width=0.15\textwidth]{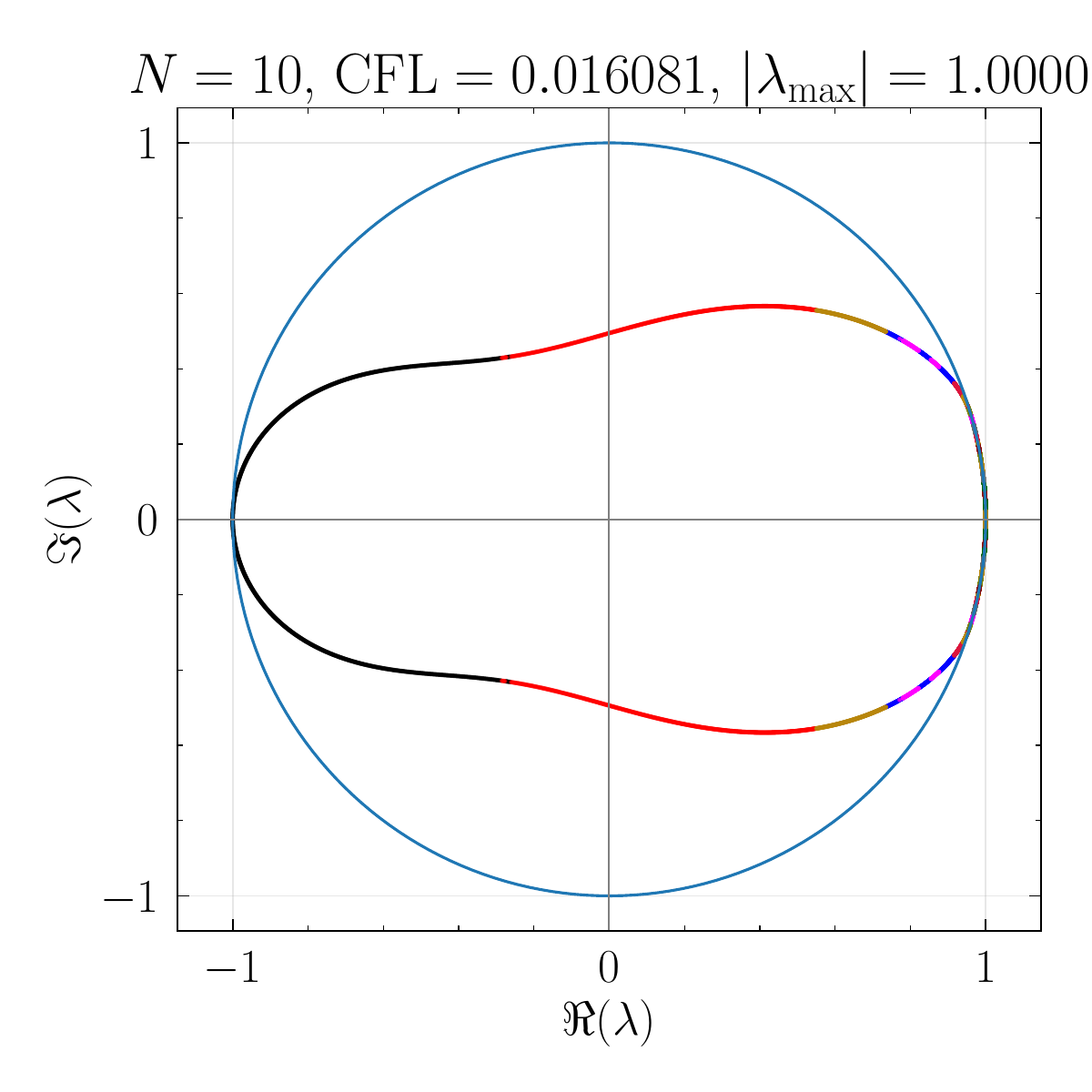}
\includegraphics[width=0.15\textwidth]{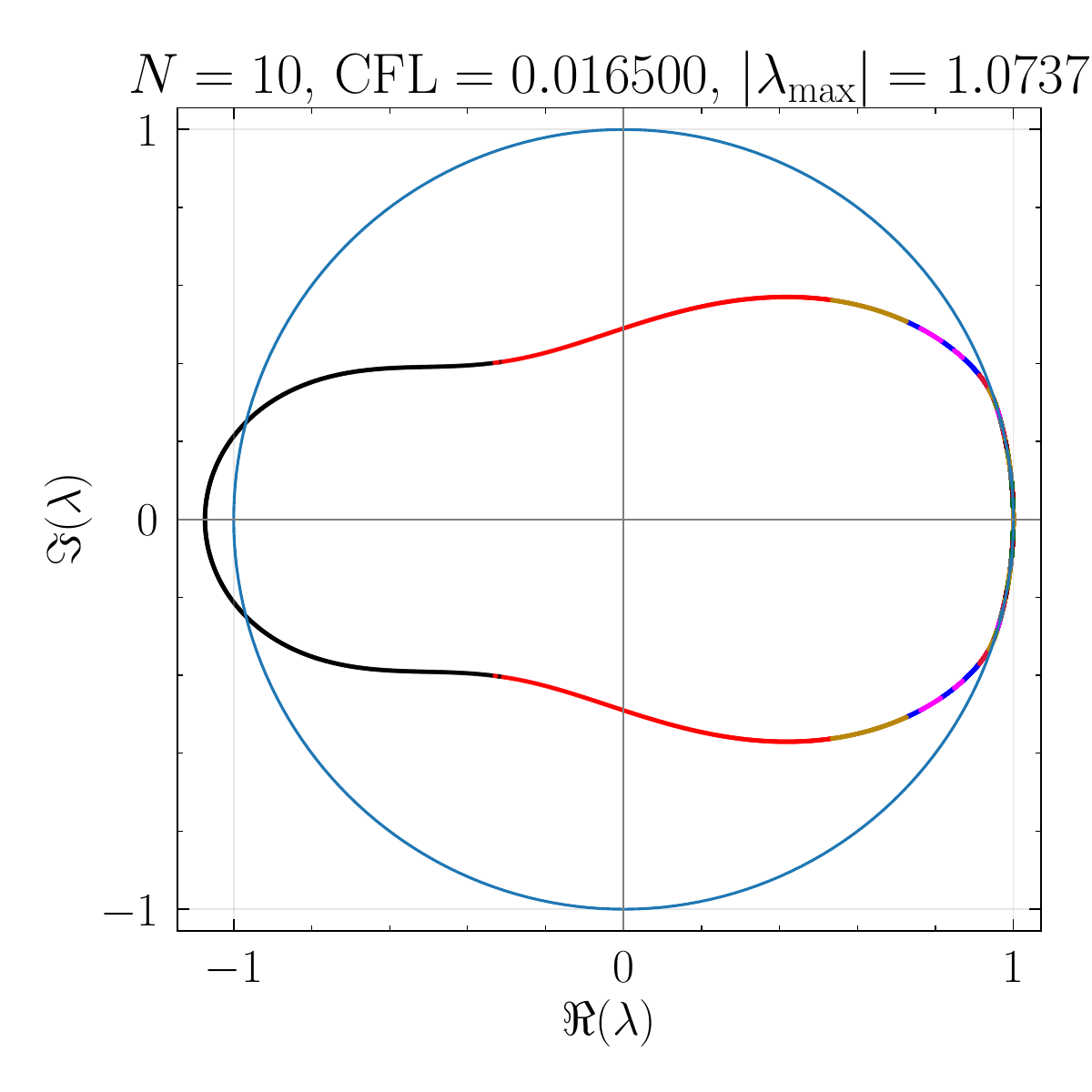}
\includegraphics[width=0.15\textwidth]{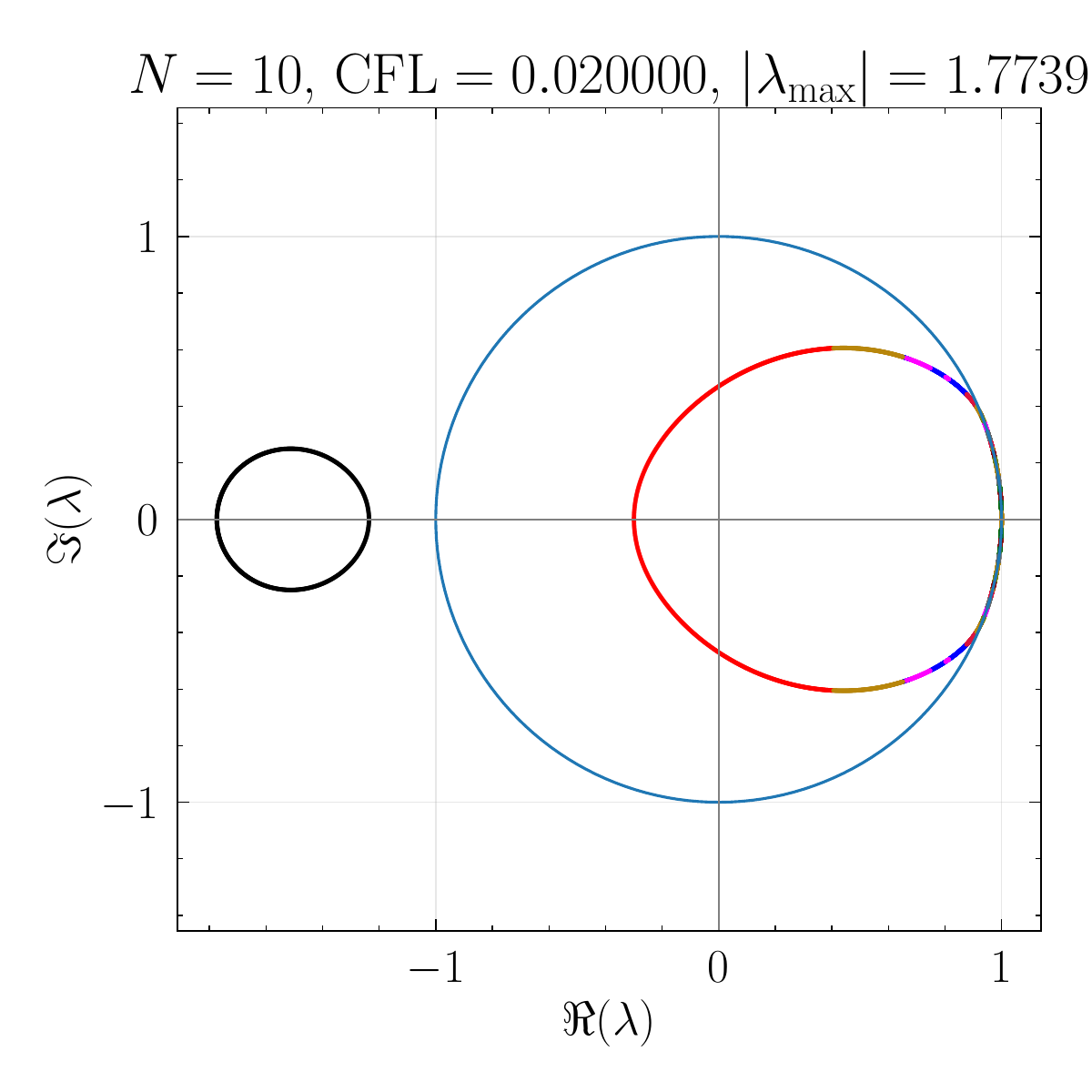}\\
\includegraphics[width=0.028125\textwidth]{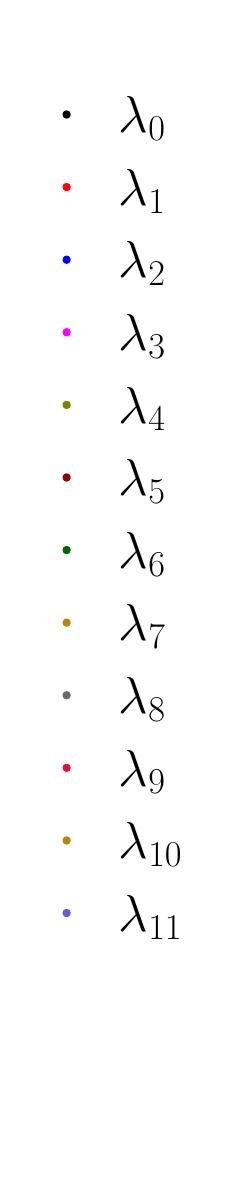}
\includegraphics[width=0.15\textwidth]{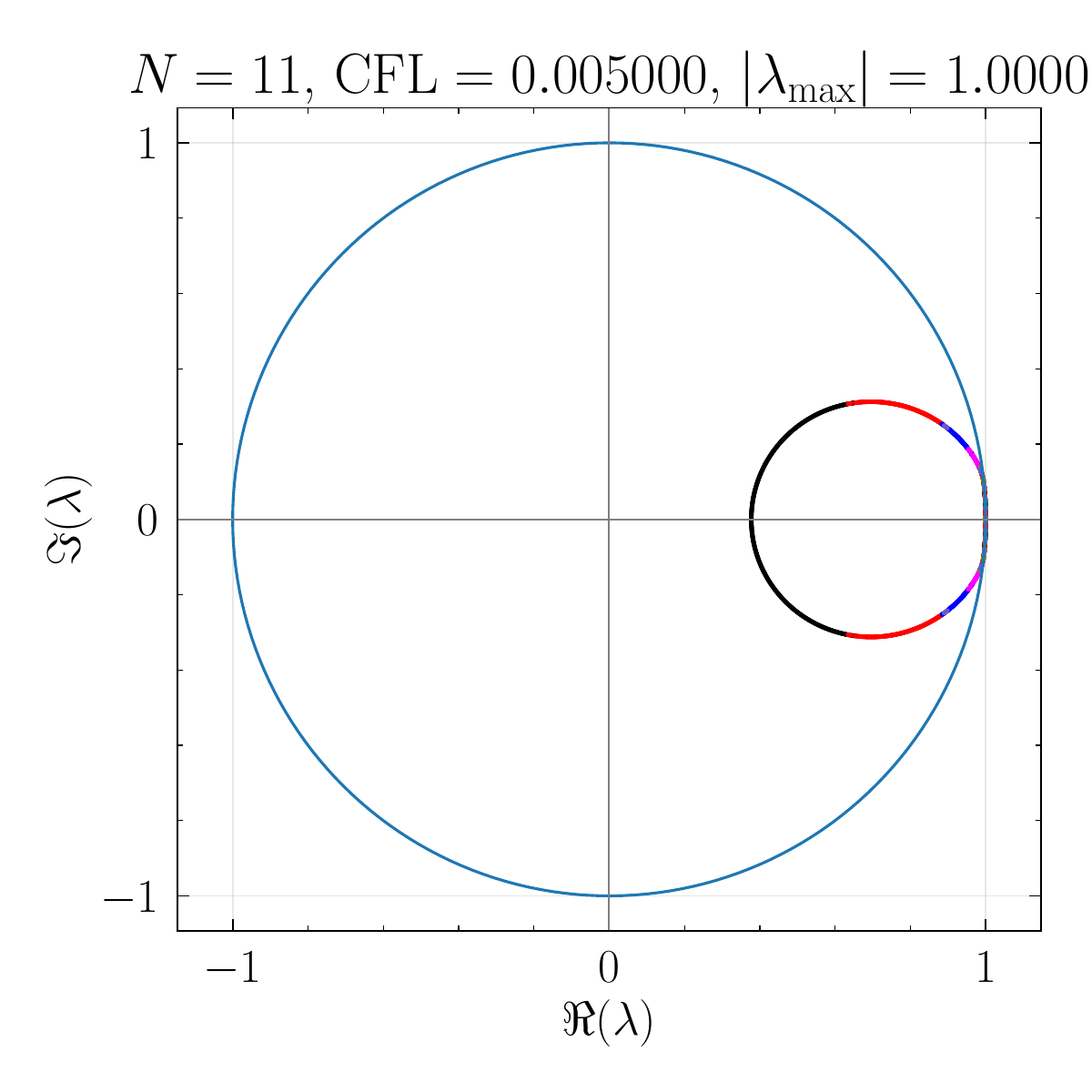}
\includegraphics[width=0.15\textwidth]{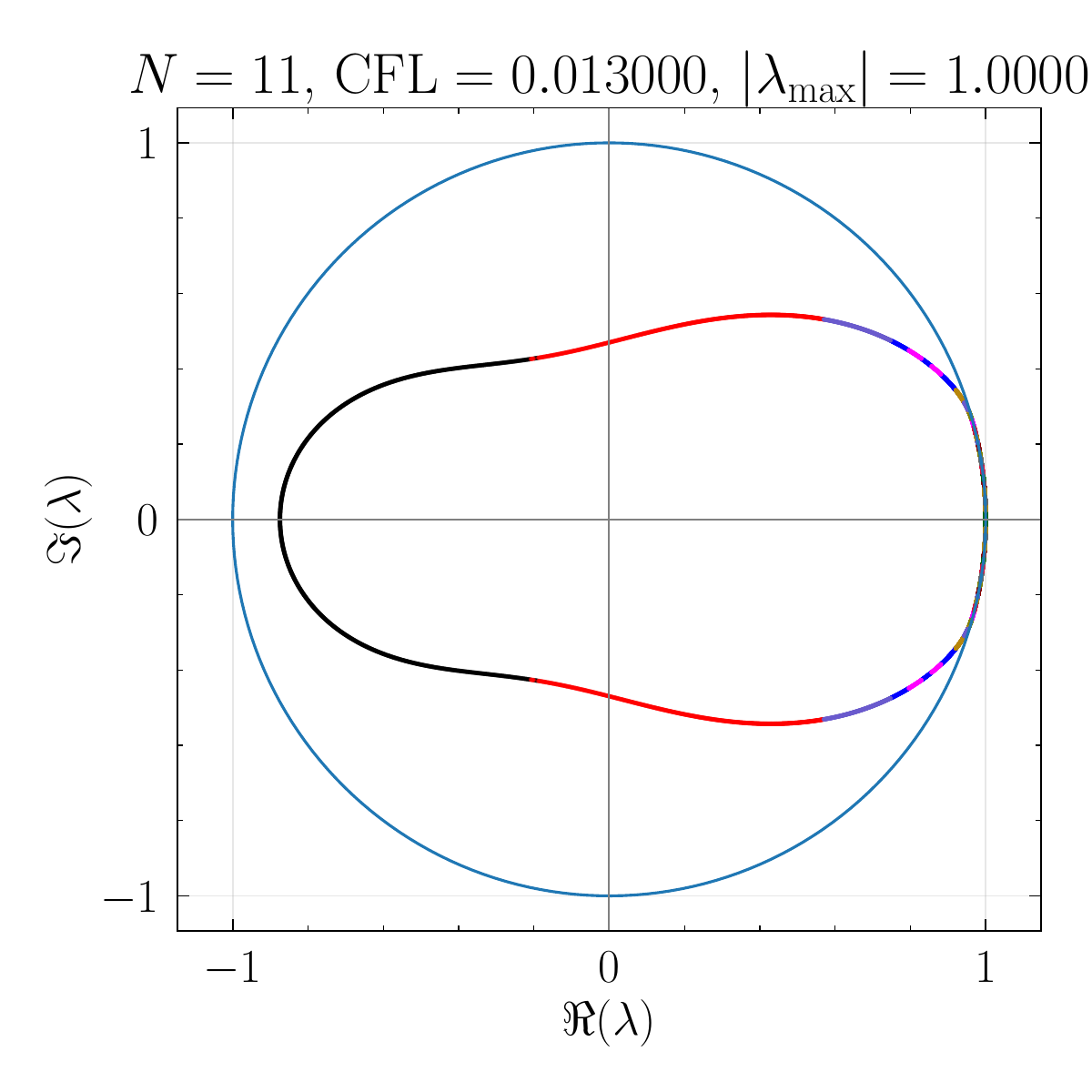}
\includegraphics[width=0.15\textwidth]{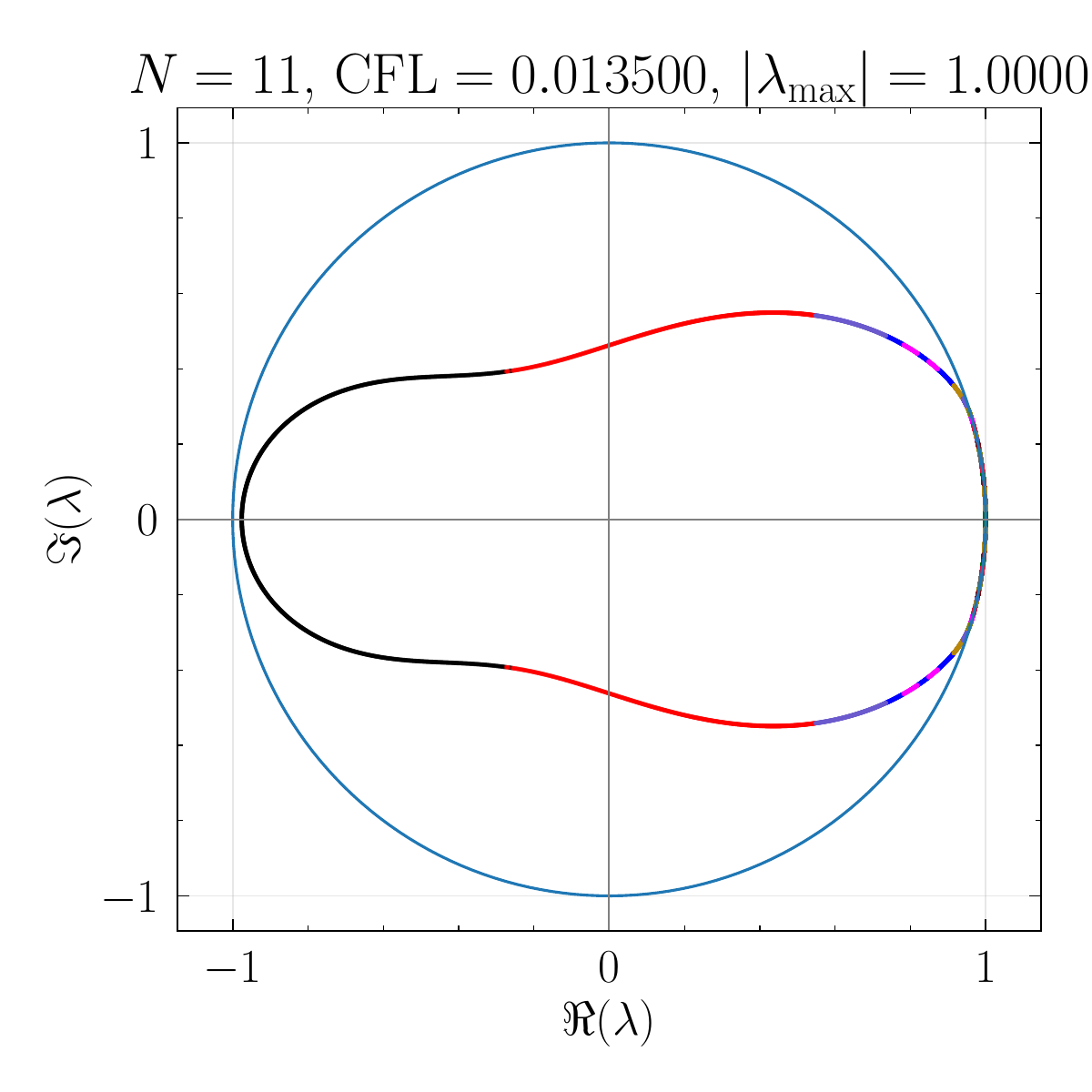}
\includegraphics[width=0.15\textwidth]{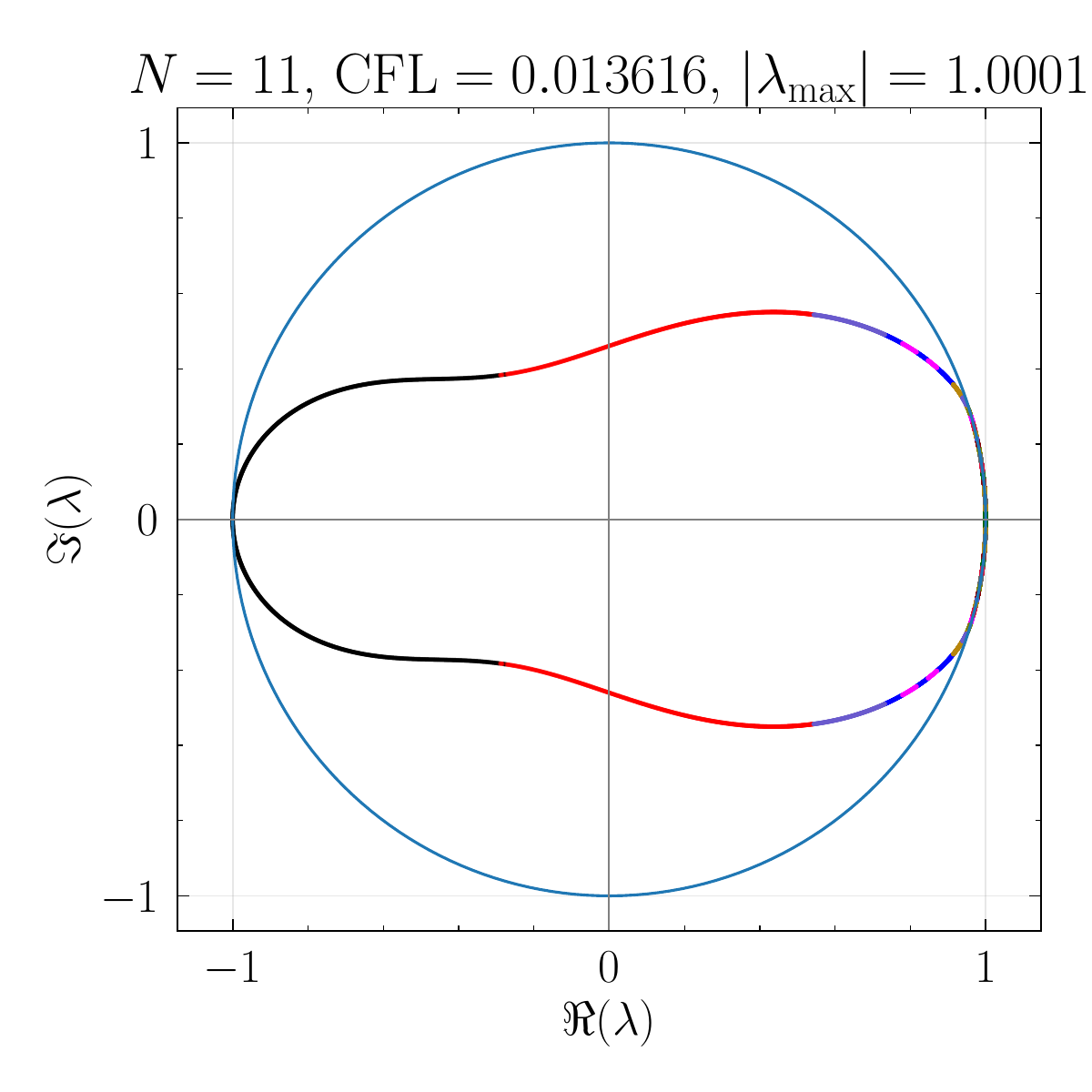}
\includegraphics[width=0.15\textwidth]{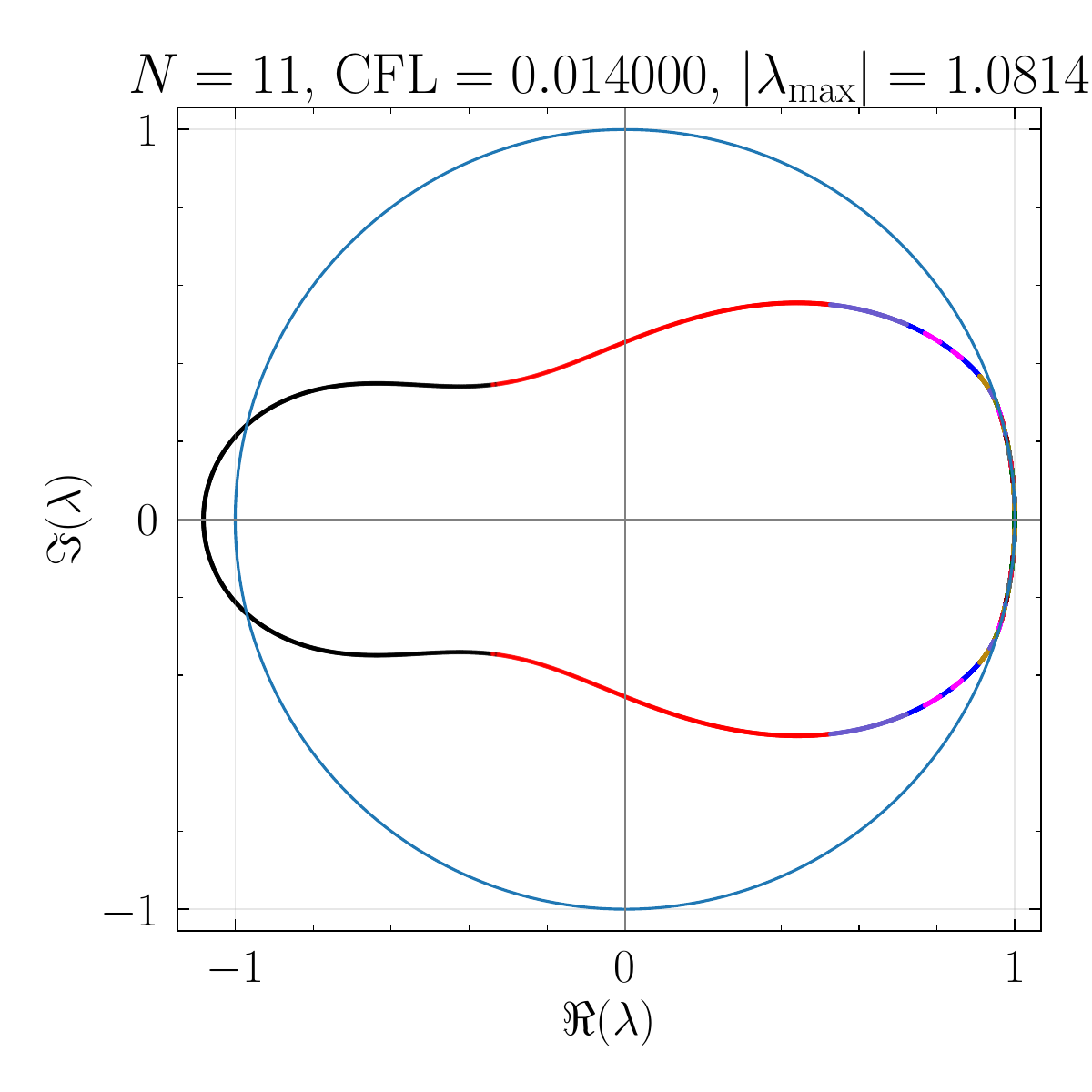}
\includegraphics[width=0.15\textwidth]{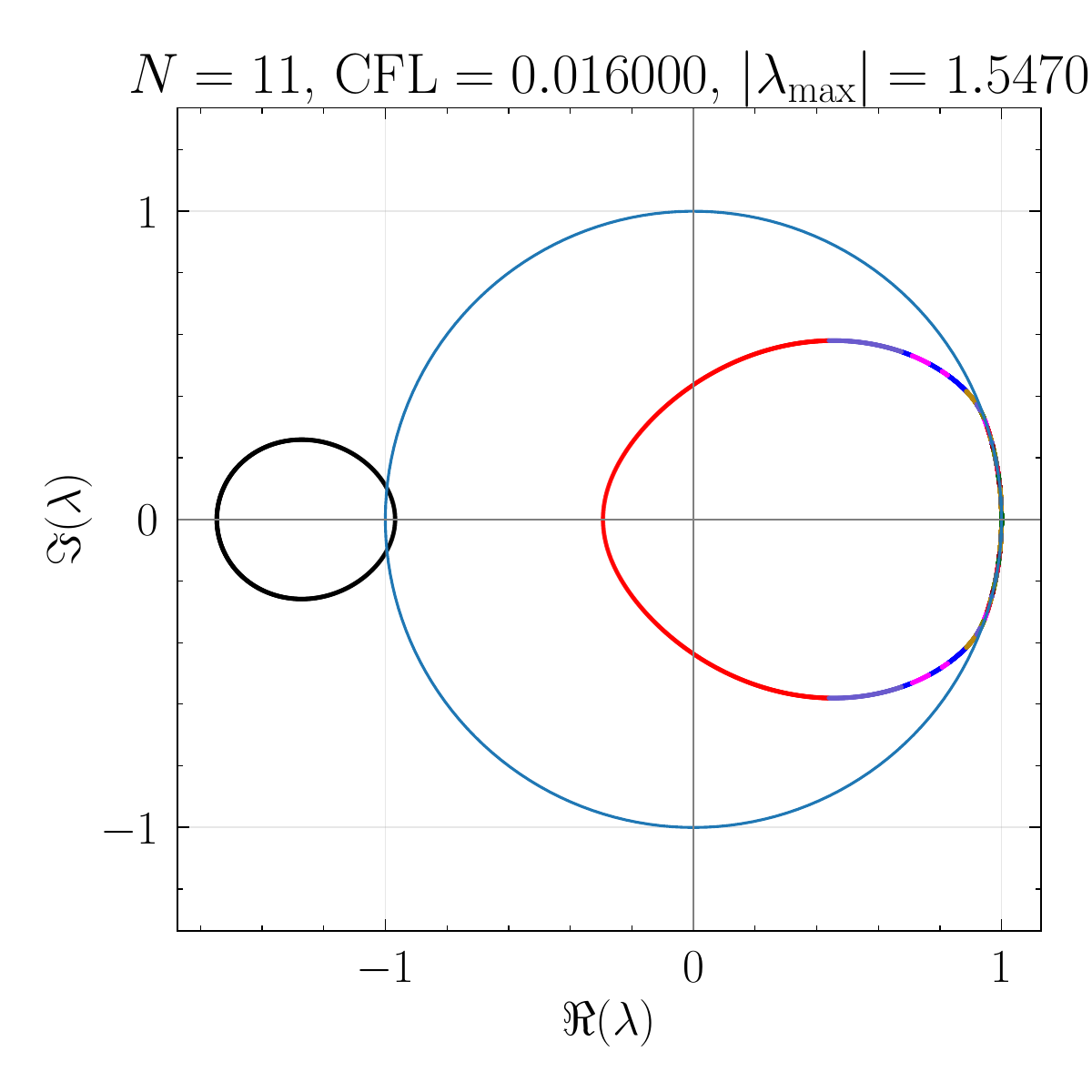}\\
\includegraphics[width=0.028125\textwidth]{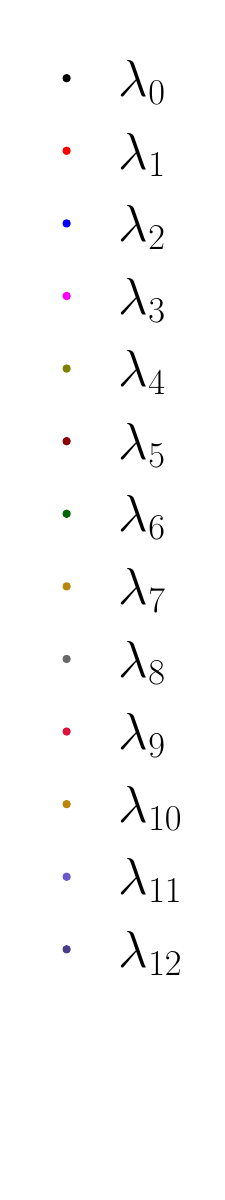}
\includegraphics[width=0.15\textwidth]{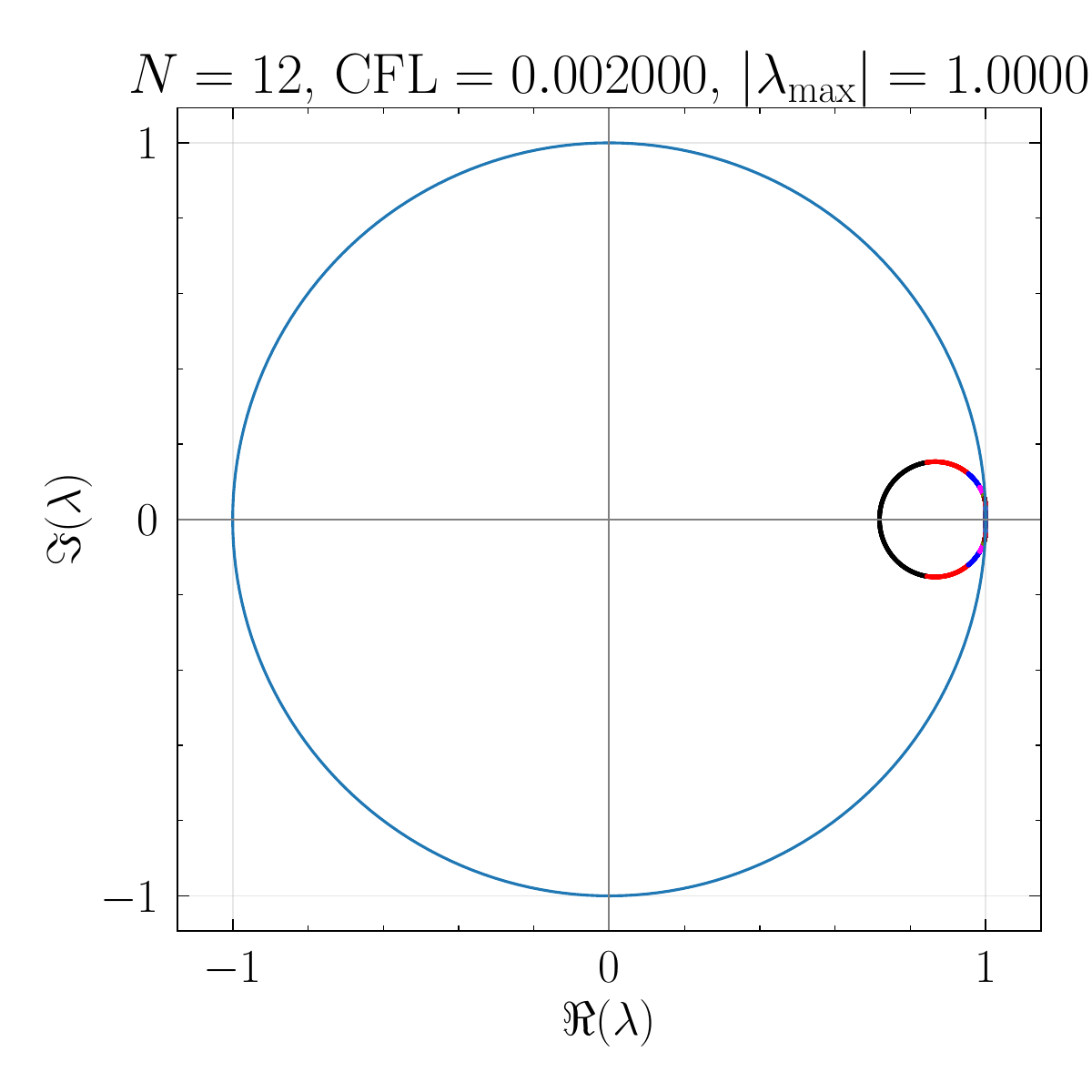}
\includegraphics[width=0.15\textwidth]{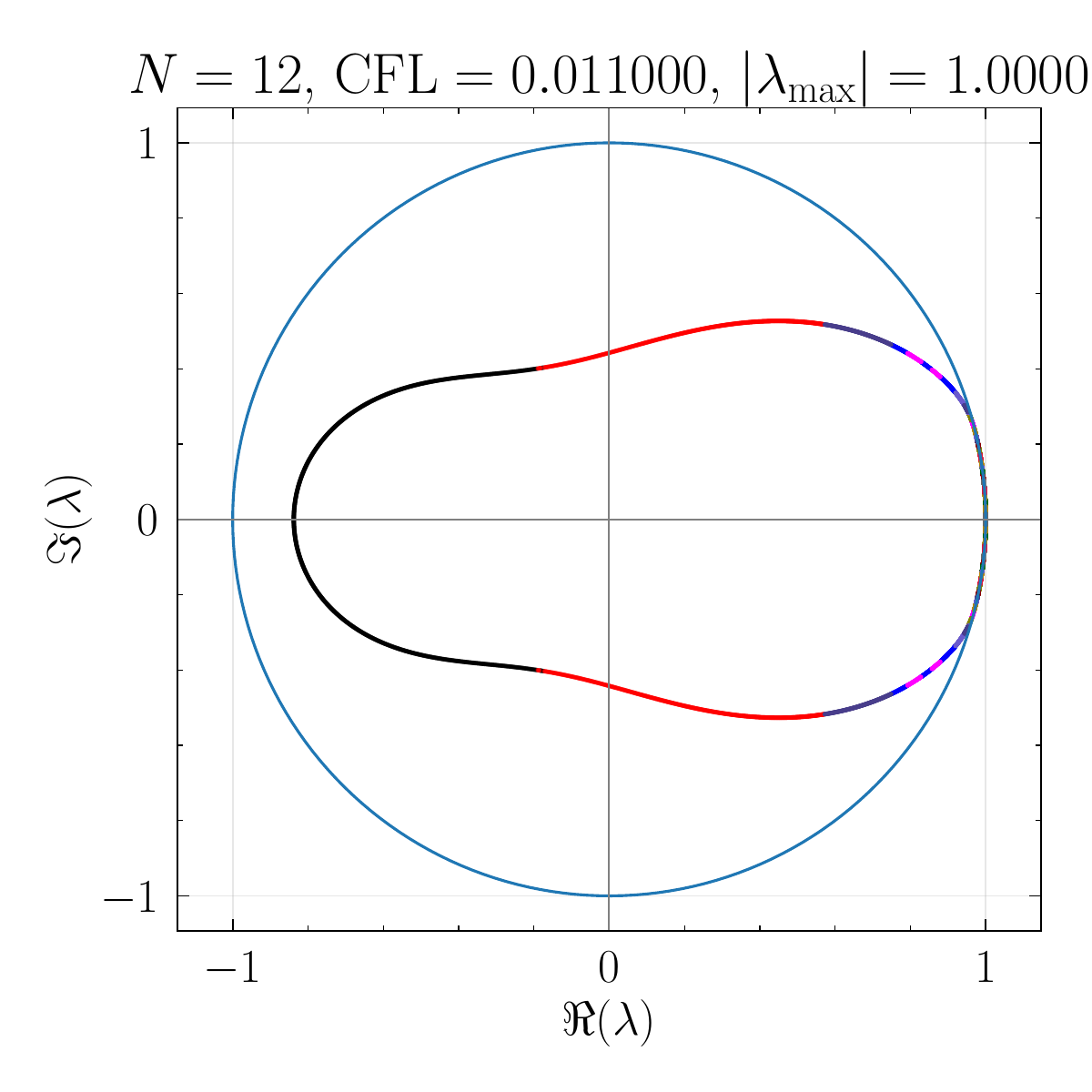}
\includegraphics[width=0.15\textwidth]{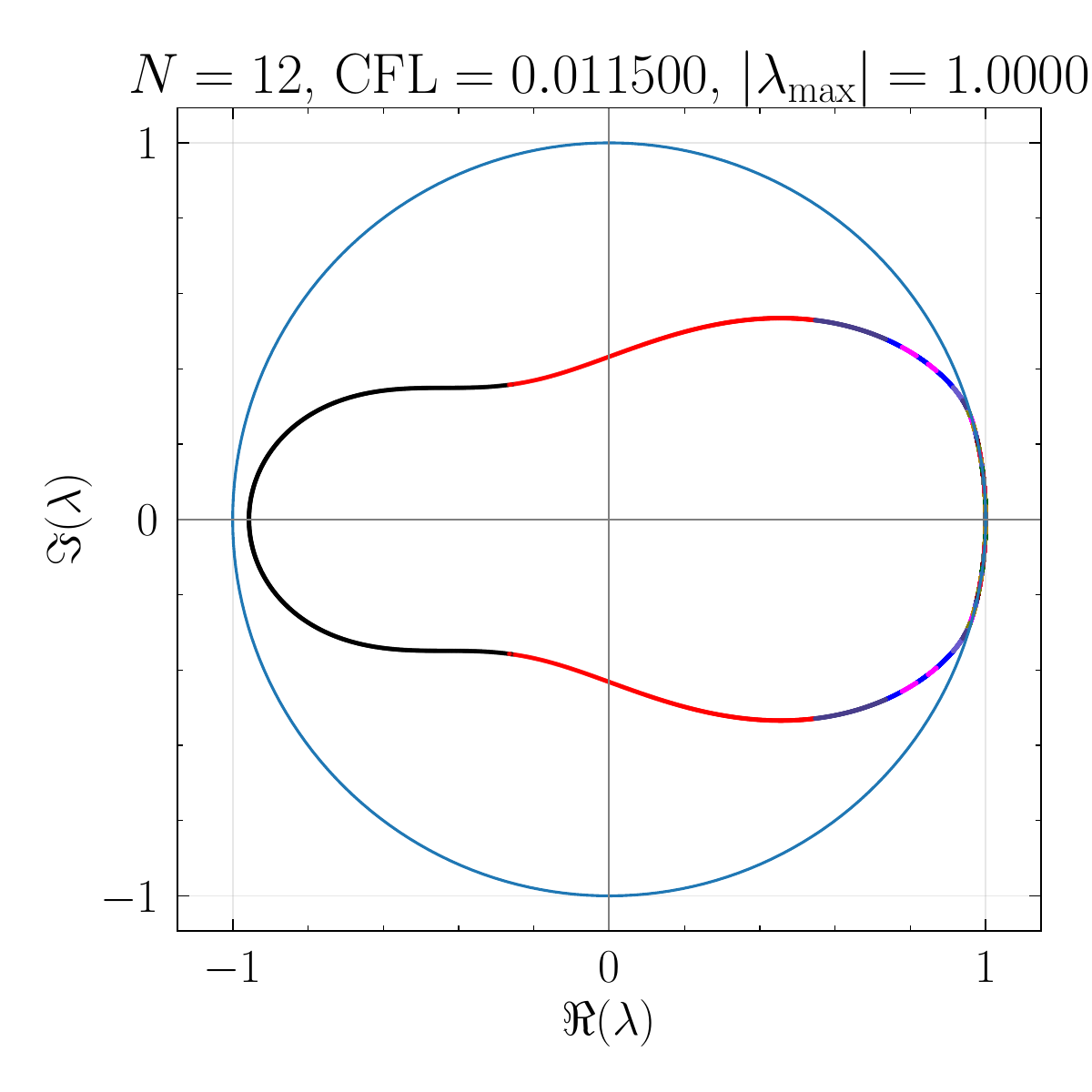}
\includegraphics[width=0.15\textwidth]{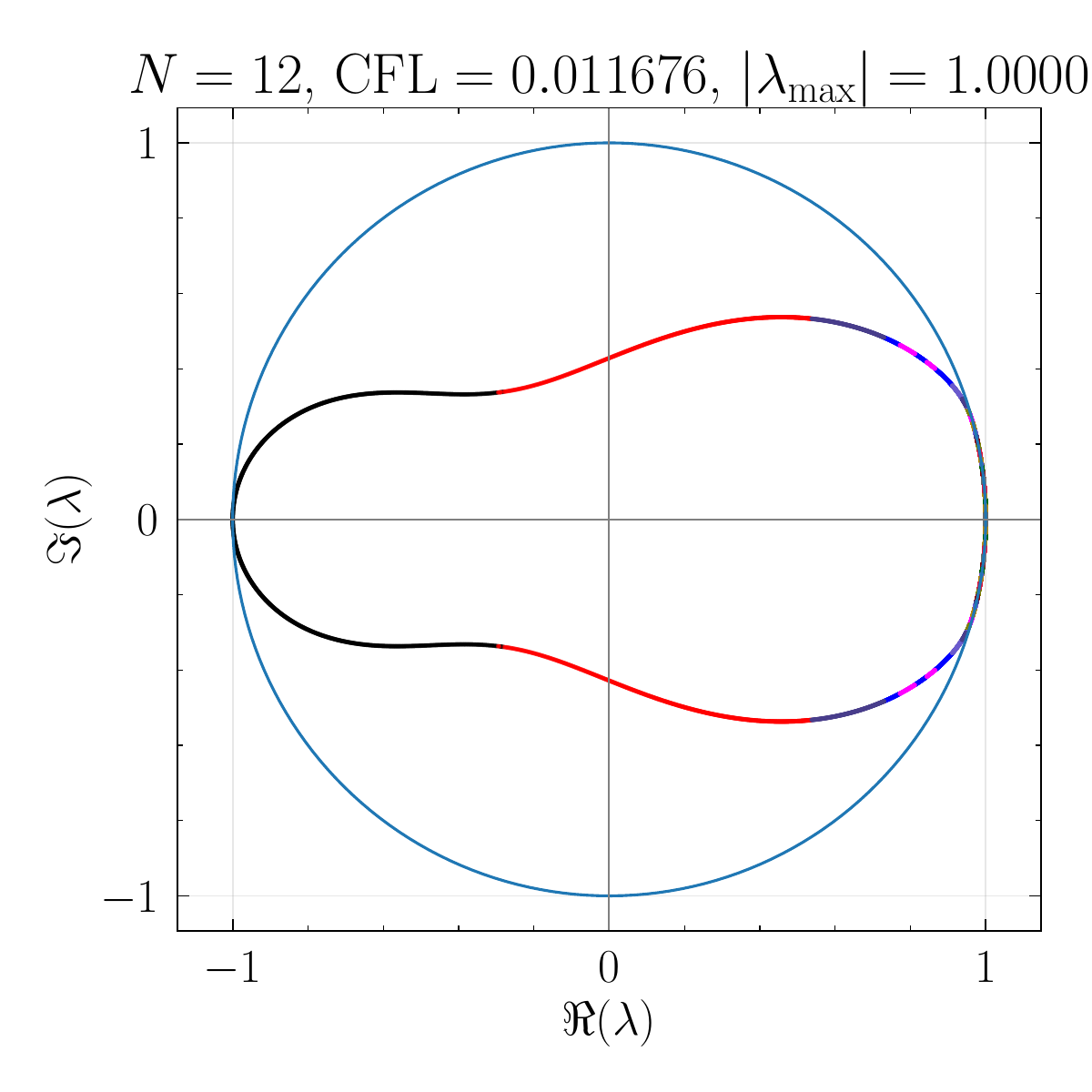}
\includegraphics[width=0.15\textwidth]{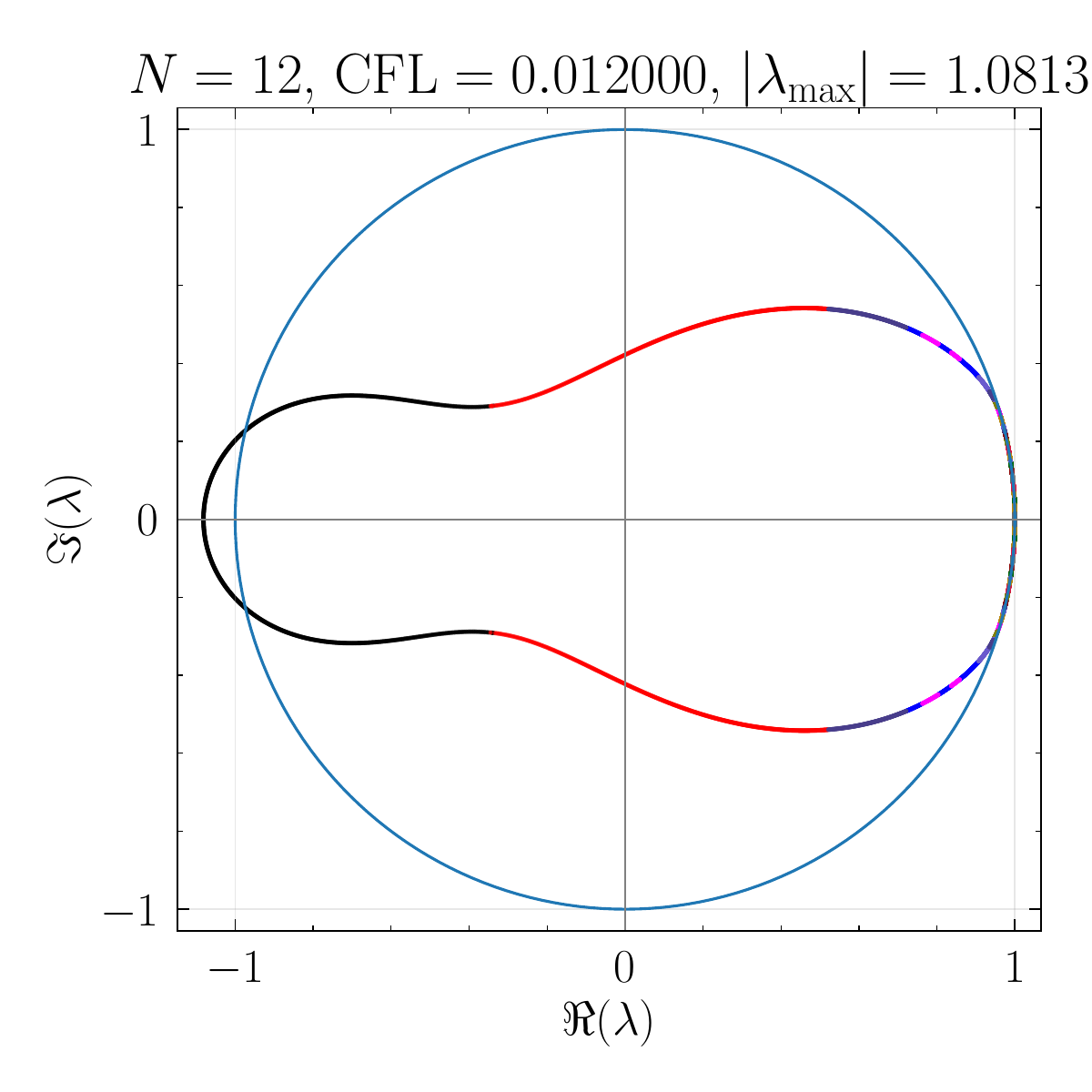}
\includegraphics[width=0.15\textwidth]{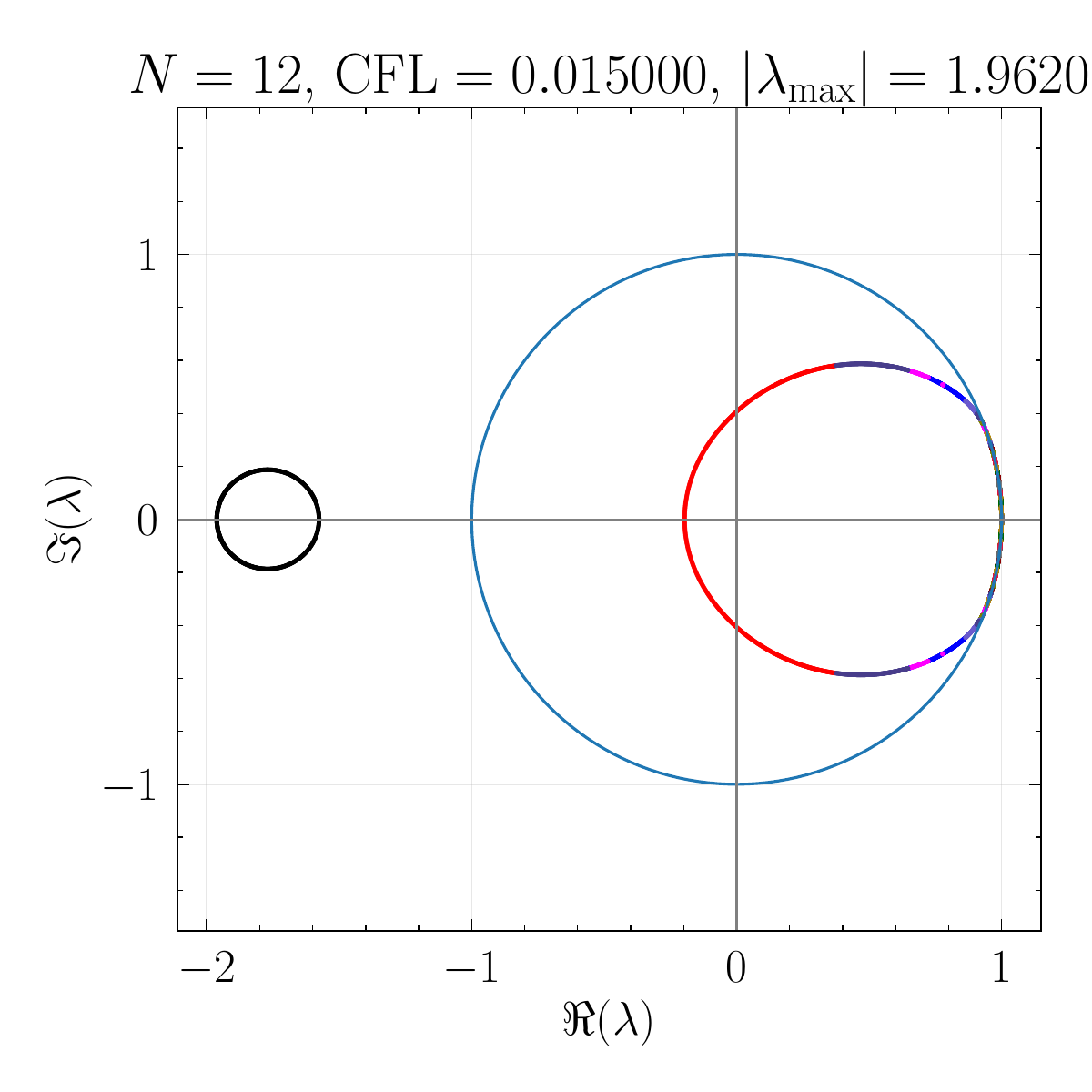}\\
\caption{%
Spectrum of the matrix $\mathrm{R}(\mathrm{CFL}, \theta)$ (\ref{eq:r_matrix_elems_expr_dup}) (eigenvalues $\lambda_{k} = \lambda_{k}(\mathrm{CFL}, \theta)$, $k = 0, \ldots, N$) of the evolution operator $R$ (\ref{eq:evol_oper_prop}) for a single time step $\Dtn{n}$ for several selected values of the Courant number $\mathrm{CFL}$ for polynomial degrees $N = 7, \ldots, 12$. The range of phase $\theta\in[0, 2\pi)$ is sampled on a uniform grid of $1000$ nodes. Legends for each row of the graphs are located on the left. $|\lambda_{\rm max}|$ is the absolute value of the largest eigenvalue $\lambda_{k}(\mathrm{CFL}, \theta)$ in a graph. The gray circle of unit radius defines the stability boundary. The Courant numbers $\mathrm{CFL}$ are taken deep inside the stability region $\mathrm{CFL} \in [0, \mathrm{CFL}_{\rm max}]$ (left column), in the region of guaranteed instability $\mathrm{CFL} > \mathrm{CFL}_{\rm max}$ (right column) and near the boundary of the stability region $\mathrm{CFL}_{\rm max}$ (the values $\mathrm{CFL}_{\rm max}$ are selected from work~\cite{ader_dg_stab} for $N = 7$, $8$, $9$, as well as the values $\mathrm{CFL}_{\rm max}$ calculated further in this work in Table~\ref{tab:cfls_max_data} and in Figure~\ref{fig:cfls_max_data}). \textit{Note}: the phase $\arg \lambda_{k}$ of eigenvalue $\lambda_{k}$ is not the phase $\theta$; the dependence of the absolute value $|\lambda_{k}|$ on phase $\theta$ for these polynomial degrees $N$ is presented in Figure~\ref{fig:rhos_on_theta_degrees_7_12}.
}
\label{fig:spectrum_exact_eigvals_degrees_7_12}
\end{figure}

In general, it is impossible to analytically rigorously analyze the matrix spectrum $\rho[\mathrm{R}(\mathrm{CFL}, \theta)]$ to determine the stability region $[0, \mathrm{CFL}_{\rm max}]$ for arbitrary degree $N$ of polynomials $\{\varphi_{p}\}$. In this work, a rigorous study combines both the rigorous derivation of algebraic equations (or, more precisely, polynomials) for the boundary value of the Courant number $\mathrm{CFL}_{\rm max}$ (\ref{eq:cfl_max_base_def}) and the derivation and substantiation of some purely empirical considerations necessary for obtaining these algebraic equations. The equation for the boundary Courant number $\mathrm{CFL}_{\rm max}$ of the stability region is quite complex for at least two reasons: it is an inequality $\rho[\mathrm{R}(\mathrm{CFL}, \theta)] \leqslant 1$ rather than an equation in the strict sense, since the problem involves an additional parameter --- the phase $\theta$, and the solution requires that the inequality be satisfied for any phase $\theta \in [0, 2\pi]$; and the eigenvalues $\lambda_{k}$ themselves are complex numbers --- $\lambda_{k} \in \mathcal{C}$. Therefore, before solving the problem of determining the stability region, graphical representations of the eigenvalue dependencies were constructed, which are presented in Figures~\ref{fig:rhos_on_theta_degrees_1_6},~\ref{fig:rhos_on_theta_degrees_7_12},~\ref{fig:spectrum_exact_eigvals_degrees_1_6} and~\ref{fig:spectrum_exact_eigvals_degrees_7_12}. All calculations to obtain these results are performed using a developed program in the \texttt{python} programming language. Floating-point calculations are performed using the \texttt{mpmath} module (with \texttt{gmp2} module), which enables calculations with arbitrarily high precision representation of real numbers as floating-point numbers. The presented results are calculated using \texttt{mpmath.mp.dps = 1000} (for comparison, for double-precision floating-point numbers \texttt{double}, \texttt{mpmath.mp.dps = 15}). The phase $\theta \in [0, 2\pi)$ is discretized using a uniform grid with $1000$ nodes. The calculation of the eigenvalues is additionally controlled by verification the control error $\|(\mathrm{R} - \lambda_{k}I)\mathbf{r}_{k}\|_{\infty}$, where $\mathbf{r}_{k}$ is the right eigenvector, and the control error did not exceed $\sim 10^{-996}$--$10^{-1000}$ for degrees $N \leqslant 4$ and up to $\sim 10^{-994}$--$10^{-997}$ for degree $N = 12$. It is important to note that the results presented in Figures~\ref{fig:rhos_on_theta_degrees_1_6},~\ref{fig:rhos_on_theta_degrees_7_12},~\ref{fig:spectrum_exact_eigvals_degrees_1_6} and~\ref{fig:spectrum_exact_eigvals_degrees_7_12} include the Courant numbers $\mathrm{CFL}_{\rm max}$ (\ref{eq:cfl_max_base_def}) obtained in the work~\cite{ader_dg_stab}, as well as those obtained in this work, which will be presented later in the text in Table~\ref{tab:cfls_max_data} and in Figure~\ref{fig:cfls_max_data}.

Figures~\ref{fig:rhos_on_theta_degrees_1_6} and~\ref{fig:rhos_on_theta_degrees_7_12} demonstrate the dependence of the absolute value $|\lambda_{k}|$ of each eigenvalue $\lambda_{k}$ of the matrix $\mathrm{R}(\mathrm{CFL}, \theta)$ (\ref{eq:r_matrix_in_matrix_form}) of the evolution operator $R$ on the phase $\theta$ for several selected values of the Courant number $\mathrm{CFL}$ for polynomial degrees $N = 1, \ldots, 12$: for polynomial degrees $N = 1, \ldots, 6$ (Figure~\ref{fig:rhos_on_theta_degrees_1_6}) and polynomial degrees $N = 7, \ldots, 12$ (Figure~\ref{fig:rhos_on_theta_degrees_7_12}). The Courant numbers $\mathrm{CFL}$ are taken deep inside the stability region $\mathrm{CFL} \in [0, \mathrm{CFL}_{\rm max}]$, in the region of guaranteed instability $\mathrm{CFL} > \mathrm{CFL}_{\rm max}$ and near the boundary of the stability region $\mathrm{CFL}_{\rm max}$ (\ref{eq:cfl_max_base_def}) --- the values $\mathrm{CFL}_{\rm max}$ are selected from work~\cite{ader_dg_stab}, as well as the values $\mathrm{CFL}_{\rm max}$ calculated further in this work. The presented dependencies $|\lambda_{k}| = |\lambda_{k}|(\theta)$ clearly demonstrate the complex structure of the phase $\theta$ dependence, especially for high-degree polynomials $N \geqslant 4$ (for low-degree polynomials, the boundary values of the Courant $\mathrm{CFL}_{\rm max}$ (\ref{eq:cfl_max_base_def}) number are simply calculated precisely analytically), which requires a very high degree of precision in resolving the intersection points of the curves of the maximum eigenvalue versus phase with the unit circle. Therefore, these results ``seem to directly suggest'' that it is desirable to eliminate the phase $\theta$ dependence in the form of a complex exponent $\exp(\pm i\theta)$ within the framework of a rigorous analysis, for example, by isolating the complex exponent $\exp(\pm i\theta)$ as a common factor on one side of the equation and then taking the absolute value of both sides of the equation (and using the property $|\exp(\pm i\theta)| = 1$ for $\theta\in[0, 2\pi]\subset\mathcal{R}$).

Figures~\ref{fig:spectrum_exact_eigvals_degrees_1_6} and~\ref{fig:spectrum_exact_eigvals_degrees_7_12} demonstrate the set of eigenvalues $\lambda_{k}$ of the matrix $\mathrm{R}(\mathrm{CFL}, \theta)$ (\ref{eq:r_matrix_in_matrix_form}) of the evolution operator $R$ for all phase values $\theta\in[0, 2\pi]$ in the complex plane for polynomial degrees $N = 1, \ldots, 12$: for polynomial degrees $N = 1, \ldots, 6$ (Figure~\ref{fig:spectrum_exact_eigvals_degrees_1_6}) and polynomial degrees $N = 7, \ldots, 12$ (Figure~\ref{fig:spectrum_exact_eigvals_degrees_7_12}). The presented dependencies $\lambda_{k} = \lambda_{k}(\mathrm{CFL}, \theta)$ also clearly demonstrate the complex structure of the dependency, especially for high polynomial degrees $N$. However, in this case, the phase $\theta$ dependence is not evident and does not ``obscure the overall picture'' of the studied dependency. The most important conclusion that clearly follows from the presented dependencies is that the intersection of the unit circle by the dependency $|\lambda_{k}| = |\lambda_{k}|(\mathrm{CFL}, \theta)$ with increasing Courant number occurs in the vicinity of the point $\lambda = -1$, therefore, it is expected that it is precisely upon reaching this eigenvalue $\lambda_{k} = -1$ with increasing Courant number $\mathrm{CFL}$ that the stability of the numerical method is lost. In general, this result could be considered expected, since significant loss of stability by high-order numerical methods usually occurs when oscillations of the initially sufficiently smooth solution occur already at the first step of time integration, and for such oscillations to occur, it is necessary for the eigenvalue $\lambda$ to have a negative real part (which leads to a characteristic factor $(-1)^{n}$ for one or more harmonics $u^{n+1}_{k}$ of the numerical solution $u^{n+1}(\xi)$); and the location on the real axis may have symmetry reasons. It is important to note the following: the phase of an eigenvalue $\arg \lambda_{k}$ and the phase $\theta$ are different quantities --- the $\arg \lambda_{k}$ is the result of a calculation, and the phase $\theta$ is an independent argument in the problem of calculating the eigenvalues of the matrix $\mathrm{R}(\mathrm{CFL}, \theta)$ (\ref{eq:r_matrix_in_matrix_form}) of the evolution operator $R$. Therefore, the loss of stability in the vicinity of an eigenvalue $\lambda = -1$ does not clearly demonstrate anything about the phase $\theta$ for which this loss of stability occurs (of course, the phase $\theta$ can be calculated separately). The obtained results showed that it is better to analyze the stability boundary $\mathrm{CFL}_{\rm max}$ (\ref{eq:cfl_max_base_def}) as follows: first, transform the original problem $\rho[\mathrm{R}(\mathrm{CFL}, \theta)] \leqslant 1$ in order to exclude the phase $\theta$ included in the analyzed expression in the form $\exp(\pm i\theta)$, and then try to simplify this expression as much as possible under the assumption of searching for the boundary value of the Courant number $\mathrm{CFL}_{\rm max}$ for which one of the eigenvalues $\lambda_{k}$ of the matrix (for any phase $\theta$) reaches the value $\lambda = -1$.

One interesting detail revealed by analyzing the dependencies presented in Figures~\ref{fig:spectrum_exact_eigvals_degrees_1_6} and~\ref{fig:spectrum_exact_eigvals_degrees_7_12} should be noted. Although the stability $\rho[\mathrm{R}(\mathrm{CFL}, \theta)] \leqslant 1$ of the numerical method ADER-DG with LST-DG predictor is lost when one of the eigenvalues reaches $\lambda = -1$, for sufficiently large polynomial degrees $N$, even at lower Courant numbers $\mathrm{CFL} \ll \mathrm{CFL}_{\rm max}$ (\ref{eq:cfl_max_base_def}), some eigenvalues $\lambda_{k}$ of the matrix $\mathrm{R}(\mathrm{CFL}, \theta)$ (\ref{eq:r_matrix_in_matrix_form}) of the evolution operator $R$ with absolute values $|\lambda_{k}|$ greater than one remain in the vicinity of point $\lambda = +1$, approximately in the region of the eigenvalue phase $\arg\lambda_{k} \in [-\pi/4, +\pi/4]$. However, they exceed unity only by $\sim 10^{-5}$--$10^{-9}$, and the Courant numbers $\mathrm{CFL}$ for which the condition is achieved in this region are unusually small --- $1.5$--$4.0$ times smaller than the boundary value $\mathrm{CFL}_{\rm max}$. This phenomenon is not an error in the eigenvalue calculations (error $\sim 10^{-994}$--$10^{-997}$). The values $|\lambda_{\rm max}|$ presented in Figures~\ref{fig:spectrum_exact_eigvals_degrees_1_6} and~\ref{fig:spectrum_exact_eigvals_degrees_7_12} are derived with four decimal places, so this phenomenon is not observed in this representation. Many calculations performed numerically in all previous years (see~\cite{ader_dg_ideal_flows, ader_dg_ale, ader_dg_grmhd, ader_dg_gr_prd, ader_dg_simple_mod_2016, fron_phys, ader_dg_axioms, exahype, ader_stiff_3, ader_stiff_4, ader_eff_blas}) show that such a strong underestimation of the stability boundary $\mathrm{CFL}_{\rm max}$ is unnecessary. Otherwise, the studies would use ``effective Courant numbers'' (\ref{eq:old_eff_cfl}) $\mathrm{C} \sim 0.05$--$0.10$, which would be very significant but has never been observed in calculations. In ``real-world'' calculations, especially for quasilinear systems of equations, approximate Riemann solvers are used, characterized by a small but nonzero diffusivity that overcomes grows of the functions $\lambda_{k}^{n}$ with eigenvalues $\lambda_{k}$ greater than one by absolute value on $\sim 10^{-5}$--$10^{-9}$: $|\lambda_{k}^{n}| = |\exp(n\ln\lambda_{k})| \sim \exp(10^{-9}n)$--$\exp(10^{-5}n)$. This effect can be classified as practically unobservable which could only complicate the stability analysis of the numerical method ADER-DG with LST-DG predictor.

Mathematically, this phenomenon can be explained as follows. Despite the fact that the LST-DG predictor (\ref{eq:lstdg_pred_eq_src}) is implicit ((\ref{eq:lstdg_pred_eq_tmp_2}) and (\ref{eq:lstdg_pred_eq_tmp_3})), in the case of a linear advection equation (\ref{eq:adv_eq_src}), due to its linearity and the nilpotency $\mathrm{D}^{N+1} = 0$ of the differentiation matrix $\mathrm{D}$, the final solution $\hat{\mathbf{q}}$ of the system of algebraic equations of the LST-DG predictor represents a classical Taylor series expansion (\ref{eq:lstdg_pred_eq_sol_temp_2}) of the solution up to and including the $N$-degree term in $\tau$ (\ref{eq:mapping}). This is mathematically completely equivalent to the explicit method, whose ``stability function'' represents a classical Taylor series expansion of the evolution operator, and the approximation order (\ref{eq:local_sol_app_order}) is determined by the truncation error of the series (\ref{eq:local_sol_app_order}). In the case of the implicit method~\cite{ader_proofs_2025, ader_dg_ivp_ode_sinum}, this ``stability function'' would be a fractional polynomial function like the Pad\'{e} approximant. Explicit methods are characterized by precisely such peculiarities in the structure of the boundary of the stability region~\cite{Hairer_book_1, Hairer_book_2, Butcher_book_2016, Dekker_Verwer_1984}.

\subsection{Rigorous quantitative considerations}
\label{sec:stab_anal:quant_cons}

The eigenvalues of the matrix $\mathrm{R}(\mathrm{CFL}, \theta)$ (\ref{eq:r_matrix_in_matrix_form}) are determined by the roots of the characteristic equation, which, up to the sign on the left-hand side, has the form $\det[R(\mathrm{CFL}, \theta) - \lambda I] = 0$. This equation will then be subjected to a series of transformations that make the equation free of the phase factor $\exp(\pm i\theta)$, which contains the entire phase $\theta$ dependence of the characteristic equation. An equation will then be obtained for the boundary value of the Courant number $\mathrm{CFL}_{\rm max}$, which determines the stability boundary (\ref{eq:cfl_max_base_def}) of the ADER-DG numerical method with the LST-DG predictor. For the last transformation, it is also shown that the boundary value of the Courant number $\mathrm{CFL}_{\rm max}$ indeed corresponds to one of the eigenvalues $\lambda_{k}$ reaching the point $\lambda = -1$. It is also clarified how the above-described phenomenon of ``false'' violation of the stability condition manifests itself in this case. A detailed derivation depends significantly on the sign of the advection velocity $a$, so the derivation is first made for case $a > 0$. In this case, the expression for matrices $\mathrm{A}(\theta)$ and $\mathrm{B}(\mathrm{CFL})$ (\ref{eq:a_and_b_matrices_def_dup}), which define the matrix $\mathrm{R}(\mathrm{CFL}, \theta) = I + \mathrm{A}(\theta)\mathrm{B}(\mathrm{CFL})$ (\ref{eq:r_matrix_in_matrix_form}) of the evolution operator $R$, takes the following form:
\begin{equation}\label{eq:f_derivd_temp_1_a_plus}
\mathrm{A}(\theta) = m^{-1}\Big[\mathrm{D}^{T} m - \Big(\ovecphi - \uvecphi\exp(-i\theta)\Big)\otimes\ovecphi^{T}\Big],\qquad
\mathrm{B}(\mathrm{CFL}) = \sum\limits_{s = 0}^{N} \frac{(-1)^{s} \mathrm{CFL}^{s+1}}{(s+1)!}\, \mathrm{D}^{s}.
\end{equation}
Then, the left side of the characteristic equation $\det[R(\mathrm{CFL}, \theta) - \lambda I] = 0$ immediately admits the following chain of transformations:
\begin{equation}\label{eq:f_derivd_temp_2_a_plus}
\begin{split}
\det\Big[R(\mathrm{CFL}, \theta) - \lambda I\Big] = \det\big[(1 - \lambda)I + \mathrm{A}(\theta)\mathrm{B}(\mathrm{CFL})\Big] & =
\det\Big[(1 - \lambda)I + m^{-1}\mathrm{D}^{T} m \mathrm{B}(\mathrm{CFL}) - m^{-1}\Big[(\ovecphi - \exp(-i\theta)\uvecphi)\otimes\ovecphi^{T}\Big]\mathrm{B}(\mathrm{CFL})\Big]\\ &=
\det\Big[m^{-1}\Big]\det\Big[(1 - \lambda)m + \varrho^{T} \mathrm{B}(\mathrm{CFL}) - \Big[(\ovecphi - \exp(-i\theta)\uvecphi)\otimes\ovecphi^{T}\Big]\mathrm{B}(\mathrm{CFL})\Big],
\end{split}
\end{equation}
as a result of which the characteristic equation acquires a form that is relatively simpler to analyze:
\begin{equation}\label{eq:f_derivd_temp_3_a_plus}
\det\Big[(1 - \lambda)m + \varrho^{T} \mathrm{B}(\mathrm{CFL}) - \Big[(\ovecphi - \exp(-i\theta)\uvecphi)\otimes\ovecphi^{T}\Big]\mathrm{B}(\mathrm{CFL})\Big] = 0,
\end{equation}
in which the second term $\varrho^{T} \mathrm{B}(\mathrm{CFL})$ is needed in some transformation to reveal its direct connection with the known properties of matrix $\varrho$ and the differentiation matrix $\mathrm{D}$ included in it (\ref{eq:matrices_def}). To implement this transformation, an integration-by-parts identity is introduced for the chosen polynomial $f\in\mathcal{P}_{N}$ representation in the chosen functional basis $\{\varphi_{p}\}_{p}$:
\begin{equation}\label{eq:ibp_prop}
m\mathrm{D} + \mathrm{D}^{T}m = \ovecphi\otimes\ovecphi^{T} - \uvecphi\otimes\uvecphi^{T},\quad
\varrho + \varrho^{T} = \ovecphi\otimes\ovecphi^{T} - \uvecphi\otimes\uvecphi^{T},
\end{equation}
which is proved by simple integration-by-parts of the element $\varrho_{pq}$ of matrix $\varrho$ (\ref{eq:matrices_def}):
\begin{equation}
\begin{split}
&\intrefdom{\xi}\,\varphi_{p}(\xi)\varphi_{q}'(\xi) = \varphi_{p}(1)\varphi_{q}(1) - \varphi_{p}(0)\varphi_{q}(0) - \intrefdom{\xi}\,\varphi_{p}'(\xi)\varphi_{q}(\xi),\\
&\intrefdom{\xi}\,\varphi_{p}(\xi)\varphi_{q}'(\xi) + \intrefdom{\xi}\,\varphi_{p}'(\xi)\varphi_{q}(\xi) = \varphi_{p}(1)\varphi_{q}(1) - \varphi_{p}(0)\varphi_{q}(0).
\end{split}
\end{equation}
The resulting integration-by-parts identity (\ref{eq:ibp_prop}) is used to express the transposed matrix $\varrho^{T} = \ovecphi\otimes\ovecphi^{T} - \uvecphi\otimes\uvecphi^{T} - \varrho$, leading to the following chain of transformations:
\begin{equation}\label{eq:f_derivd_temp_4_a_plus}
\begin{split}
&(1 - \lambda)m + \varrho^{T} \mathrm{B}(\mathrm{CFL}) - \Big[(\ovecphi - \exp(-i\theta)\uvecphi)\otimes\ovecphi^{T}\Big]\mathrm{B}(\mathrm{CFL})\\ & =
(1 - \lambda)m + \ovecphi\otimes\ovecphi^{T}\mathrm{B}(\mathrm{CFL}) - \uvecphi\otimes\uvecphi^{T}\mathrm{B}(\mathrm{CFL}) -
\varrho\mathrm{B}(\mathrm{CFL}) - \Big[(\ovecphi - \exp(-i\theta)\uvecphi)\otimes\ovecphi^{T}\Big]\mathrm{B}(\mathrm{CFL})\\ & =
(1 - \lambda)m - \varrho\mathrm{B}(\mathrm{CFL}) - (\uvecphi\otimes\uvecphi^{T})\mathrm{B}(\mathrm{CFL}) + \exp(-i\theta)(\uvecphi\otimes\ovecphi^{T})\mathrm{B}(\mathrm{CFL}),
\end{split}
\end{equation}
the final result of which is that the second term now has a transparent meaning, which can be expressed by the following expression:
\begin{equation}\label{eq:f_derivd_temp_5_a_plus}
\varrho\mathrm{B}(\mathrm{CFL}) = -m\sum\limits_{s = 0}^{N}(-1)^{s+1}\frac{\mathrm{CFL}^{s+1}}{(s+1)!}\, \mathrm{D}^{s+1} =
m\left[I - \sum\limits_{s = 0}^{N}\frac{(-\mathrm{CFL})^{s}}{s!}\, \mathrm{D}^{s}\right] =
m\big[I - \exp\left(-\mathrm{CFL}\cdot\mathrm{D}\right)\big],
\end{equation}
where the nilpotency property $\mathrm{D}^{N+1} = 0$ of the differentiation matrix $\mathrm{D}$ is used. The subsequent chain of transformations involves identifying a convenient common factor in the expression under the determinant:
\begin{equation}\label{eq:f_derivd_temp_6_a_plus}
\begin{split}
&\det\Big[(1 - \lambda)m + \varrho^{T} \mathrm{B}(\mathrm{CFL}) - (\uvecphi\otimes\uvecphi^{T})\mathrm{B}(\mathrm{CFL}) +
\exp(-i\theta)(\uvecphi\otimes\ovecphi^{T})\mathrm{B}(\mathrm{CFL})\Big]\\ & =
\det\Big[m(1 - \lambda) - m\big[I - \exp\left(-\mathrm{CFL}\cdot\mathrm{D}\right)\big] - 
(\uvecphi\otimes\uvecphi^{T})\mathrm{B}(\mathrm{CFL}) + \exp(-i\theta)(\uvecphi\otimes\ovecphi^{T})\mathrm{B}(\mathrm{CFL})\Big]\\ & =
\det\Big[m(\exp\left(-\mathrm{CFL}\cdot\mathrm{D}\right) - \lambda I) -
(\uvecphi\otimes\uvecphi^{T})\mathrm{B}(\mathrm{CFL}) + \exp(-i\theta)(\uvecphi\otimes\ovecphi^{T})\mathrm{B}(\mathrm{CFL})\Big].
\end{split}
\end{equation}
To further simplify the resulting expression, it is assumed that $\lambda$ is not an eigenvalue of matrix $\exp(-\mathrm{CFL}\cdot\mathrm{D})$: $\lambda \not\in \sigma[\exp(-\mathrm{CFL}\cdot\mathrm{D})]$, which allows us to apply the well-known formula $\det(\mathrm{K}-\mathbf{a}\otimes\mathbf{b}^{T}) = \det{K}\cdot(1 - \mathbf{b}^{T}\mathrm{K}^{-1}\mathbf{a})$, where matrix $\mathrm{K}$ is assumed to be non-singular. Then, the characteristic equation immediately admits the following chain of transformations:
\begin{equation}\label{eq:f_derivd_temp_7_a_plus}
\begin{split}
&\det\Big[m(\exp(-\mathrm{CFL}\cdot\mathrm{D}) - \lambda I) -
(\uvecphi\otimes\uvecphi^{T})\mathrm{B}(\mathrm{CFL}) - \exp(-i\theta)(\uvecphi\otimes\ovecphi^{T})\mathrm{B}(\mathrm{CFL})\Big]\\ & =
\det\Big[m(\exp\left(-\mathrm{CFL}\cdot\mathrm{D}\right) - \lambda I) -
\uvecphi\otimes\Big(\uvecphi^{T}\mathrm{B}(\mathrm{CFL}) - \exp(-i\theta)\ovecphi^{T}\mathrm{B}(\mathrm{CFL})\Big)\Big]\\ & =
\det\Big[m(\exp\left(-\mathrm{CFL}\cdot\mathrm{D}\right) - \lambda I)\Big]
\Big[1 - \Big(\uvecphi^{T}\mathrm{B}(\mathrm{CFL})\\ & - \exp(-i\theta)\ovecphi^{T}\mathrm{B}(\mathrm{CFL})\Big)
\Big[m(\lambda I - \exp\left(-\mathrm{CFL}\cdot\mathrm{D}\right))\Big]^{-1}\uvecphi\Big] = 0.
\end{split}
\end{equation}
It turns out that the original characteristic equation $\det[R(\mathrm{CFL}, \theta) - \lambda I] = 0$ is reduced to the following scalar equation:
\begin{equation}\label{eq:f_derivd_temp_8_a_plus}
1 - \Big(\uvecphi^{T}\mathrm{B}(\mathrm{CFL}) - \exp(-i\theta)\ovecphi^{T}\mathrm{B}(\mathrm{CFL})\Big)
(\lambda I - \exp\left(-\mathrm{CFL}\cdot\mathrm{D}\right))^{-1}m^{-1}\uvecphi = 0.
\end{equation}
After introducing the following notation:
\begin{equation}\label{eq:upsilons_def_plus}
\begin{split}
&\Upsilon^{+}_{\mathrm{A}}(\mathrm{CFL}, \lambda) = \uvecphi^{T}\mathrm{B}(\mathrm{CFL})(\lambda I - \exp\left(-\mathrm{CFL}\cdot\mathrm{D}\right))^{-1}m^{-1}\uvecphi,\\
&\Upsilon^{+}_{\mathrm{B}}(\mathrm{CFL}, \lambda) = \ovecphi^{T}\mathrm{B}(\mathrm{CFL})(\lambda I - \exp\left(-\mathrm{CFL}\cdot\mathrm{D}\right))^{-1}m^{-1}\uvecphi,
\end{split}
\end{equation}
the equation (\ref{eq:f_derivd_temp_8_a_plus}) is transformed to the following form:
\begin{equation}\label{eq:upsilon_equation_plus_def_theta}
\exp(-i\theta)\Upsilon^{+}_{\mathrm{B}}(\mathrm{CFL}, \lambda) = \Upsilon^{+}_{\mathrm{A}}(\mathrm{CFL}, \lambda) - 1,
\end{equation}
or can be rewritten in the following form:
\begin{equation}\label{eq:upsilon_equation_plus_def_z}
z\Upsilon^{+}_{\mathrm{B}}(\mathrm{CFL}, \lambda) = \Upsilon^{+}_{\mathrm{A}}(\mathrm{CFL}, \lambda) - 1,\quad z\in\mathcal{C},\quad |z| = 1.
\end{equation}
The functions $\Upsilon^{+}_{\mathrm{A}}(\mathrm{CFL}, \lambda)$ and $\Upsilon^{+}_{\mathrm{B}}(\mathrm{CFL}, \lambda)$ (\ref{eq:upsilons_def_plus}) are conveniently defined by introducing the following polynomial representation:
\begin{equation}
\Upsilon^{+}_{\mathrm{A}}(\mathrm{CFL}, \lambda) = \uvecphi^{T}\mathbf{h}^{+}(\mathrm{CFL}, \lambda),\quad
\Upsilon^{+}_{\mathrm{B}}(\mathrm{CFL}, \lambda) = \ovecphi^{T}\mathbf{h}^{+}(\mathrm{CFL}, \lambda),
\end{equation}
where the polynomial coefficients $\mathbf{h}^{+} = \{h^{+}_{p}\}_{p}$ are defined as follows:
\begin{equation}\label{eq:h_plus_coeffs_def}
h^{+}_{p}(\mathrm{CFL}, \lambda) = \Big[\mathrm{B}(\mathrm{CFL})(\lambda I - \exp\left(-\mathrm{CFL}\cdot\mathrm{D}\right))^{-1}m^{-1}\uvecphi\Big]_{p},
\end{equation}
and determine the following polynomial:
\begin{equation}\label{eq:h_plus_def}
H^{+}(\xi;\, \mathrm{CFL}, \lambda) = \sum\limits_{p = 0}^{N} h^{+}_{p}(\mathrm{CFL}, \lambda) \varphi_{p}(\xi).
\end{equation}
Using this polynomial $H^{+}(\xi;\, \mathrm{CFL}, \lambda)$, the functions $\Upsilon^{+}_{\mathrm{A}}(\mathrm{CFL}, \lambda)$ and $\Upsilon^{+}_{\mathrm{B}}(\mathrm{CFL}, \lambda)$ (\ref{eq:upsilons_def_plus}) are defined as follows:
\begin{equation}\label{eq:upsilons_def_by_h_plus}
\Upsilon^{+}_{\mathrm{A}}(\mathrm{CFL}, \lambda) = H^{+}(0;\, \mathrm{CFL}, \lambda),\quad
\Upsilon^{+}_{\mathrm{B}}(\mathrm{CFL}, \lambda) = H^{+}(1;\, \mathrm{CFL}, \lambda),
\end{equation}
and the characteristic equation (\ref{eq:upsilon_equation_plus_def_z}) takes the following form:
\begin{equation}\label{eq:h_equation_plus_def}
z H^{+}(1;\, \mathrm{CFL}, \lambda) = H^{+}(0;\, \mathrm{CFL}, \lambda) - 1,\quad z\in\mathcal{C},\quad |z| = 1.
\end{equation}
Now, to free the characteristic equation $\det[R(\mathrm{CFL}, \theta) - \lambda I] = 0$ from phase $\theta$ dependence, it is necessary to simply take the absolute value of the left-hand and right-hand sides of the equation (\ref{eq:h_equation_plus_def}), which leads to the following equation:
\begin{equation}\label{eq:h_abs_equation_plus_def}
|H^{+}(1;\, \mathrm{CFL}, \lambda)| = |H^{+}(0;\, \mathrm{CFL}, \lambda) - 1|,
\end{equation}
the solution of which is significantly simpler than the original characteristic equation $\det[R(\mathrm{CFL}, \theta) - \lambda I] = 0$. This equation (\ref{eq:h_abs_equation_plus_def}) is then studied in the neighborhood of a point and is reduced to a purely polynomial form of the equation for the Courant number $\mathrm{CFL}$.

The simplification of the characteristic equation $\det[R(\mathrm{CFL}, \theta) - \lambda I] = 0$ in the case $a < 0$ is carried out similarly to the case $a > 0$, with only technical differences. The expression for matrix $\mathrm{B}(\mathrm{CFL})$ remained the same (\ref{eq:f_derivd_temp_1_a_plus}), while the expression for matrix $\mathrm{A}(\theta)$ took the following form (\ref{eq:a_and_b_matrices_def_dup}):
\begin{equation}\label{eq:f_derivd_temp_1_a_minus}
\mathrm{A}(\theta) = m^{-1}\Big[\mathrm{D}^{T} m - \Big(\ovecphi\exp(i\theta) - \uvecphi\Big)\otimes\uvecphi^{T}\Big].
\end{equation}
The characteristic equation $\det[R(\mathrm{CFL}, \theta) - \lambda I] = 0$ is rewritten as follows:
\begin{equation}
\begin{split}
&\det\Big[R(\mathrm{CFL}, \theta) - \lambda I\Big] = \det\big[(1 - \lambda)I + \mathrm{A}(\theta)\mathrm{B}(\mathrm{CFL})\Big]\\ & =
\det\Big[(1 - \lambda)I + m^{-1}\mathrm{D}^{T} m \mathrm{B}(\mathrm{CFL}) - m^{-1}\Big[(\exp(-i\theta)\ovecphi - \uvecphi)\otimes\uvecphi^{T}\Big]\mathrm{B}(\mathrm{CFL})\Big]\\ &=
\det\Big[m^{-1}\Big]\det\Big[(1 - \lambda)m + \varrho^{T} \mathrm{B}(\mathrm{CFL}) - \Big[(\exp(-i\theta)\ovecphi - \uvecphi)\otimes\uvecphi^{T}\Big]\mathrm{B}(\mathrm{CFL})\Big],
\end{split}
\end{equation}
resulting in the following notation:
\begin{equation}
\det\Big[(1 - \lambda)m + \varrho^{T} \mathrm{B}(\mathrm{CFL}) - \Big[(\exp(-i\theta)\ovecphi - \uvecphi)\otimes\uvecphi^{T}\Big]\mathrm{B}(\mathrm{CFL})\Big] = 0.
\end{equation}
Using the integration-by-parts identity (\ref{eq:ibp_prop}) to express the transposed matrix $\varrho^{T} = \ovecphi\otimes\ovecphi^{T} - \uvecphi\otimes\uvecphi^{T} - \varrho$, the last expression could be rewritten as follows
\begin{equation}
\begin{split}
&\det\Big[(1 - \lambda)m + \varrho^{T} \mathrm{B}(\mathrm{CFL}) + (\ovecphi\otimes\ovecphi^{T})\mathrm{B}(\mathrm{CFL}) -
\exp(-i\theta)(\ovecphi\otimes\uvecphi^{T})\mathrm{B}(\mathrm{CFL})\Big]\\ & =
\det\Big[m(1 - \lambda) - m\big[I - \exp\left(-\mathrm{CFL}\cdot\mathrm{D}\right)\big] + 
(\ovecphi\otimes\ovecphi^{T})\mathrm{B}(\mathrm{CFL}) - \exp(-i\theta)(\ovecphi\otimes\uvecphi^{T})\mathrm{B}(\mathrm{CFL})\Big]\\ & =
\det\Big[m(\exp\left(-\mathrm{CFL}\otimes\mathrm{D}\right) - \lambda I) +
(\ovecphi\otimes\ovecphi^{T})\mathrm{B}(\mathrm{CFL}) - \exp(-i\theta)(\ovecphi\otimes\uvecphi^{T})\mathrm{B}(\mathrm{CFL})\Big],
\end{split}
\end{equation}
for which, under assumption $\lambda \not\in \sigma[\exp(-\mathrm{CFL}\cdot\mathrm{D})]$, using the known relationship $\det(\mathrm{K}+\mathbf{a}\otimes\mathbf{b}^{T}) = \det{K}\cdot(1 + \mathbf{b}^{T}\mathrm{K}^{-1}\mathbf{a})$ for the non-singular matrix $K$, the following equation is obtained:
\begin{equation}
\begin{split}
&\det\Big[m(\exp(-\mathrm{CFL}\cdot\mathrm{D}) - \lambda I) +
(\ovecphi\otimes\ovecphi^{T})\mathrm{B}(\mathrm{CFL}) - \exp(-i\theta)(\ovecphi\otimes\uvecphi^{T})\mathrm{B}(\mathrm{CFL})\Big]\\ & =
\det\Big[m(\exp\left(-\mathrm{CFL}\cdot\mathrm{D}\right) - \lambda I) +
\ovecphi\otimes\Big(\ovecphi^{T}\mathrm{B}(\mathrm{CFL}) - \exp(-i\theta)\uvecphi^{T}\mathrm{B}(\mathrm{CFL})\Big)\Big]\\ & =
\det\Big[m(\exp\left(-\mathrm{CFL}\cdot\mathrm{D}\right) - \lambda I)\Big]
\Big[1 + \Big(\ovecphi^{T}\mathrm{B}(\mathrm{CFL})\\ & - \exp(-i\theta)\uvecphi^{T}\mathrm{B}(\mathrm{CFL})\Big)
\Big[m(\lambda I - \exp\left(-\mathrm{CFL}\cdot\mathrm{D}\right))\Big]^{-1}\ovecphi\Big] = 0,
\end{split}
\end{equation}
which resulted in an equation similar to (\ref{eq:f_derivd_temp_8_a_plus}):
\begin{equation}\label{eq:f_derivd_temp_8_a_minus}
1 + \Big(\ovecphi^{T}\mathrm{B}(\mathrm{CFL}) - \exp(-i\theta)\uvecphi^{T}\mathrm{B}(\mathrm{CFL})\Big)
(\lambda I - \exp\left(-\mathrm{CFL}\cdot\mathrm{D}\right))^{-1}m^{-1}\ovecphi = 0.
\end{equation}
After introducing the following notation:
\begin{equation}\label{eq:upsilons_def_minus}
\begin{split}
&\Upsilon^{-}_{\mathrm{A}}(\mathrm{CFL}, \lambda) = \ovecphi^{T}\mathrm{B}(\mathrm{CFL})(\lambda I - \exp\left(-\mathrm{CFL}\cdot\mathrm{D}\right))^{-1}m^{-1}\ovecphi,\\
&\Upsilon^{-}_{\mathrm{B}}(\mathrm{CFL}, \lambda) = \uvecphi^{T}\mathrm{B}(\mathrm{CFL})(\lambda I - \exp\left(-\mathrm{CFL}\cdot\mathrm{D}\right))^{-1}m^{-1}\ovecphi,
\end{split}
\end{equation}
the equation (\ref{eq:f_derivd_temp_8_a_minus}) is transformed to the following form similar to (\ref{eq:upsilon_equation_plus_def_theta}):
\begin{equation}
\exp(-i\theta)\Upsilon^{-}_{\mathrm{B}}(\mathrm{CFL}, \lambda) = \Upsilon^{-}_{\mathrm{A}}(\mathrm{CFL}, \lambda) + 1.
\end{equation}
The functions $\Upsilon^{-}_{\mathrm{A}}(\mathrm{CFL}, \lambda)$ and $\Upsilon^{-}_{\mathrm{B}}(\mathrm{CFL}, \lambda)$ (\ref{eq:upsilons_def_minus}) are conveniently defined by introducing the following polynomial representation:
\begin{equation}\label{eq:upsilons_def_by_h_minus}
\Upsilon_{\mathrm{A}}(\mathrm{CFL}, \lambda) = H^{-}(1;\, \mathrm{CFL}, \lambda),\quad
\Upsilon_{\mathrm{B}}(\mathrm{CFL}, \lambda) = H^{-}(0;\, \mathrm{CFL}, \lambda),
\end{equation}
where polynomial $H^{-}(\xi;\, \mathrm{CFL}, \lambda)$ is defined as follows:
\begin{equation}\label{eq:h_minus_def}
H^{-}(\xi;\, \mathrm{CFL}, \lambda) = \sum\limits_{p = 0}^{N} h^{-}_{p}(\mathrm{CFL}, \lambda) \varphi_{p}(\xi),
\end{equation}
and the coefficients $\{h^{-}_{p}(\mathrm{CFL}, \lambda)\}_{p}$ are given by the following relation:
\begin{equation}\label{eq:h_minus_coeffs_def}
h^{-}_{p}(\mathrm{CFL}, \lambda) = \Big[\mathrm{B}(\mathrm{CFL})(\lambda I - \exp\left(-\mathrm{CFL}\cdot\mathrm{D}\right))^{-1}m^{-1}\ovecphi\Big]_{p}.
\end{equation}
As a result, the characteristic equation $\det[R(\mathrm{CFL}, \theta) - \lambda I] = 0$ in the form (\ref{eq:f_derivd_temp_8_a_minus}) is transformed to the following form:
\begin{equation}\label{eq:h_equation_minus_def}
z H^{-}(0;\, \mathrm{CFL}, \lambda) = H^{-}(1;\, \mathrm{CFL}, \lambda) + 1,\quad z\in\mathcal{C},\quad |z| = 1,
\end{equation}
which, after taking the absolute parts of the left and right sides, takes the following form:
\begin{equation}\label{eq:h_abs_equation_minus_def}
|H^{-}(0;\, \mathrm{CFL}, \lambda)| = |H^{-}(1;\, \mathrm{CFL}, \lambda) + 1|.
\end{equation}
The resulting relations (\ref{eq:upsilons_def_minus}), (\ref{eq:upsilons_def_by_h_minus}), (\ref{eq:h_minus_def}), (\ref{eq:h_minus_coeffs_def}), (\ref{eq:h_equation_minus_def}) and (\ref{eq:h_abs_equation_minus_def}) in the case $a < 0$ are, in general, completely similar to the corresponding relations (\ref{eq:upsilons_def_plus}), (\ref{eq:upsilons_def_by_h_plus}), (\ref{eq:h_plus_def}), (\ref{eq:h_plus_coeffs_def}), (\ref{eq:h_equation_plus_def}) and (\ref{eq:h_abs_equation_plus_def}) in the case $a > 0$.

As a result, the equation for the Courant number $\mathrm{CFL}_{\rm max}$, which determines the stability boundary of the numerical method of ADER-DG with the LST-DG predictor, can be written in the following form:
\begin{equation}\label{eq:cfl_by_h_src}
\mathrm{CFL}_{\rm max} = \inf\left\{\mathrm{CFL}\, \left|
\begin{split}
&\mathrm{CFL}\in[0, 1]\subset\mathcal{R},\, \exists\, \lambda \in \mathcal{C},\, |\lambda| \geqslant 1:\\
&a > 0\,\wedge\,|H^{+}(1;\, \mathrm{CFL}, \lambda)| = |H^{+}(0;\, \mathrm{CFL}, \lambda) - 1|\ \vee\\
&a < 0\,\wedge\,|H^{-}(0;\, \mathrm{CFL}, \lambda)| = |H^{-}(1;\, \mathrm{CFL}, \lambda) + 1|
\end{split}
\right. \right\},
\end{equation}
where it is assumed that equation (\ref{eq:h_abs_equation_plus_def}) in case $a > 0$ or equation (\ref{eq:h_abs_equation_minus_def}) in case $a < 0$ will have at least some solution $\lambda \in \mathcal{C}$ that goes beyond the stability boundary $|\lambda| \geqslant 1$.

\begin{figure}[h!]
\centering
\includegraphics[width=0.24\textwidth]{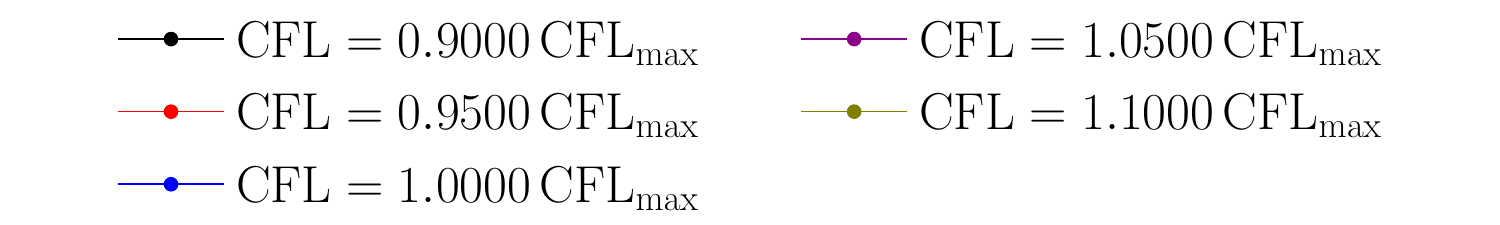}
\includegraphics[width=0.24\textwidth]{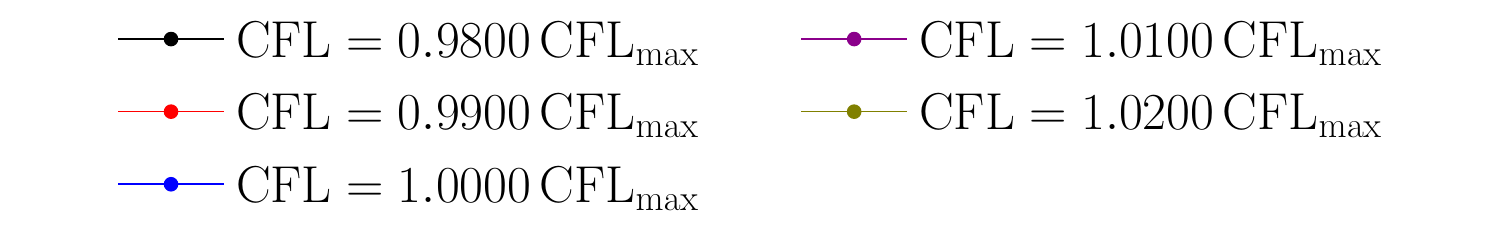}
\includegraphics[width=0.24\textwidth]{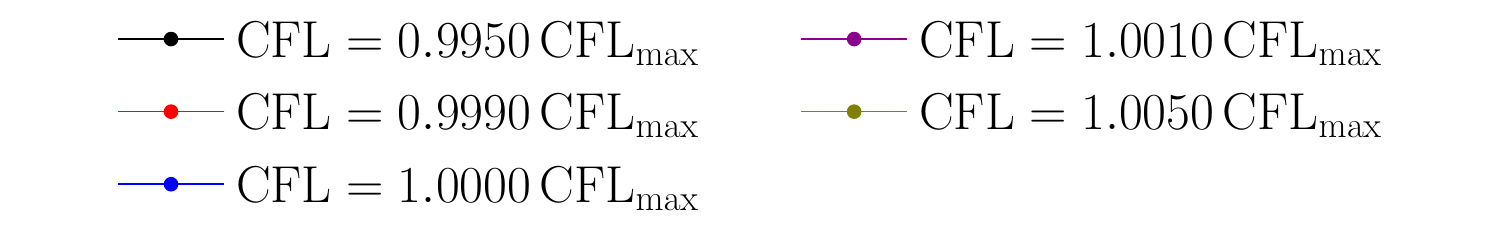}
\includegraphics[width=0.24\textwidth]{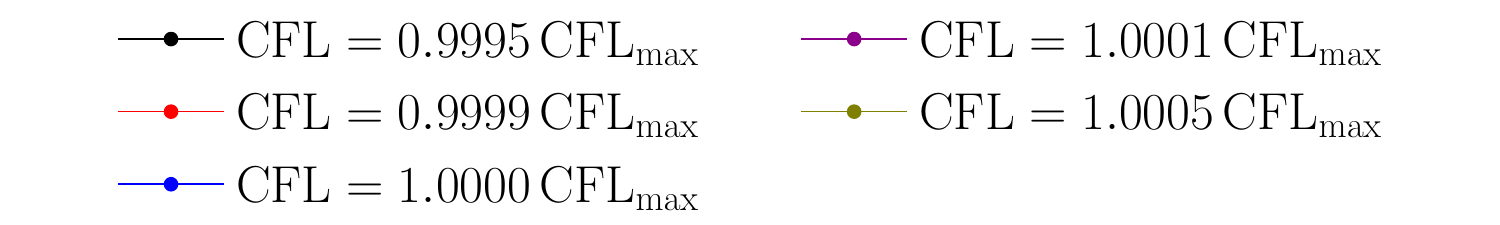}\\
\includegraphics[width=0.24\textwidth]{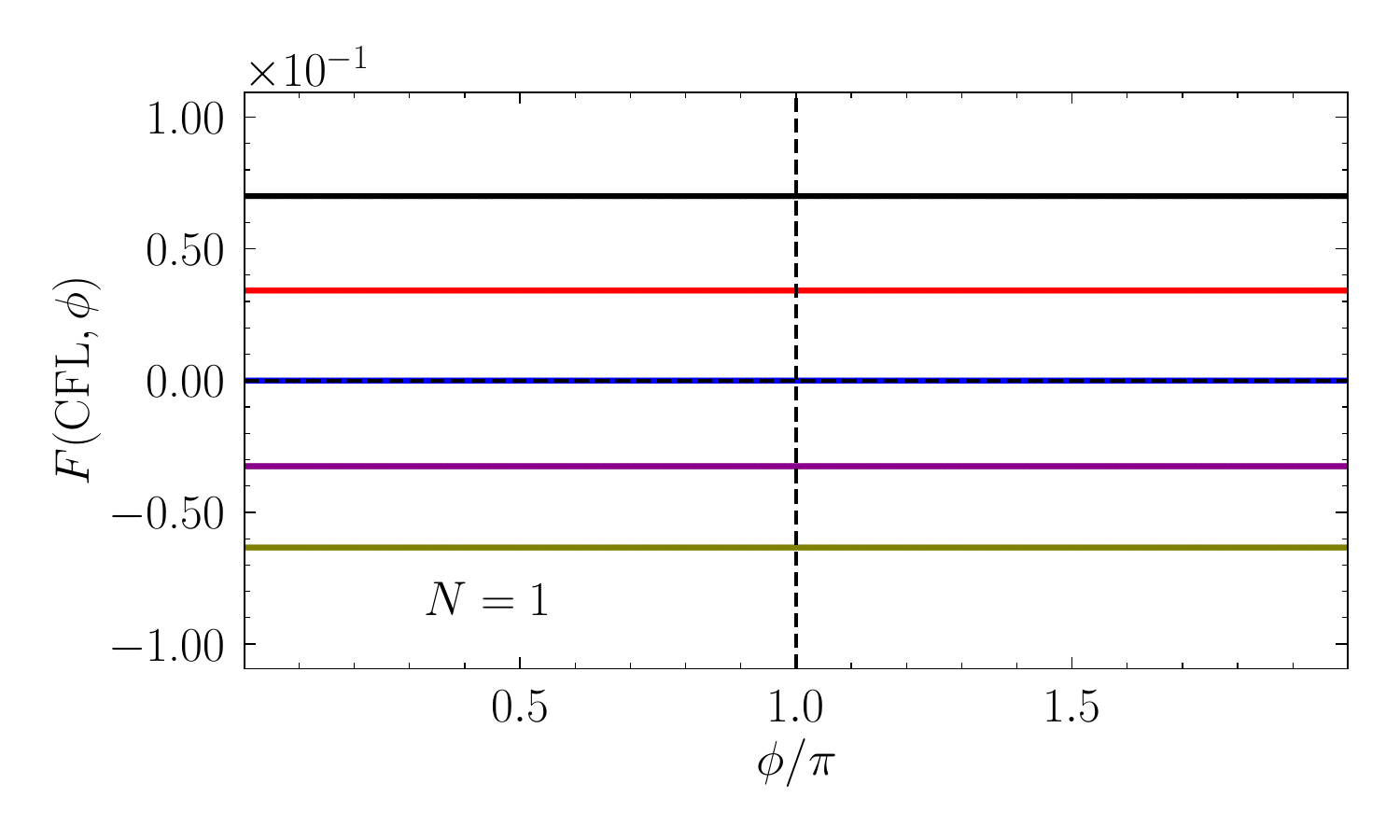}
\includegraphics[width=0.24\textwidth]{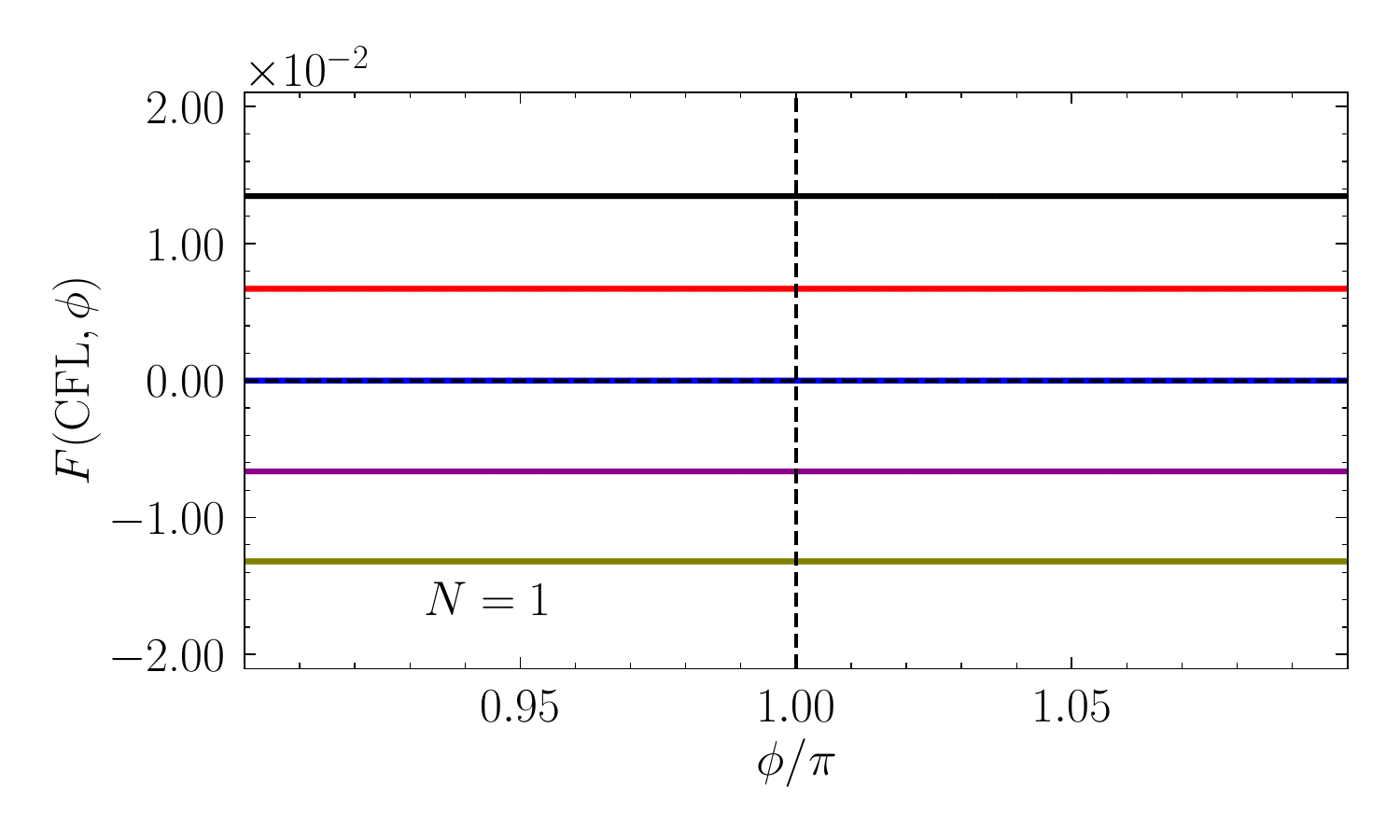}
\includegraphics[width=0.24\textwidth]{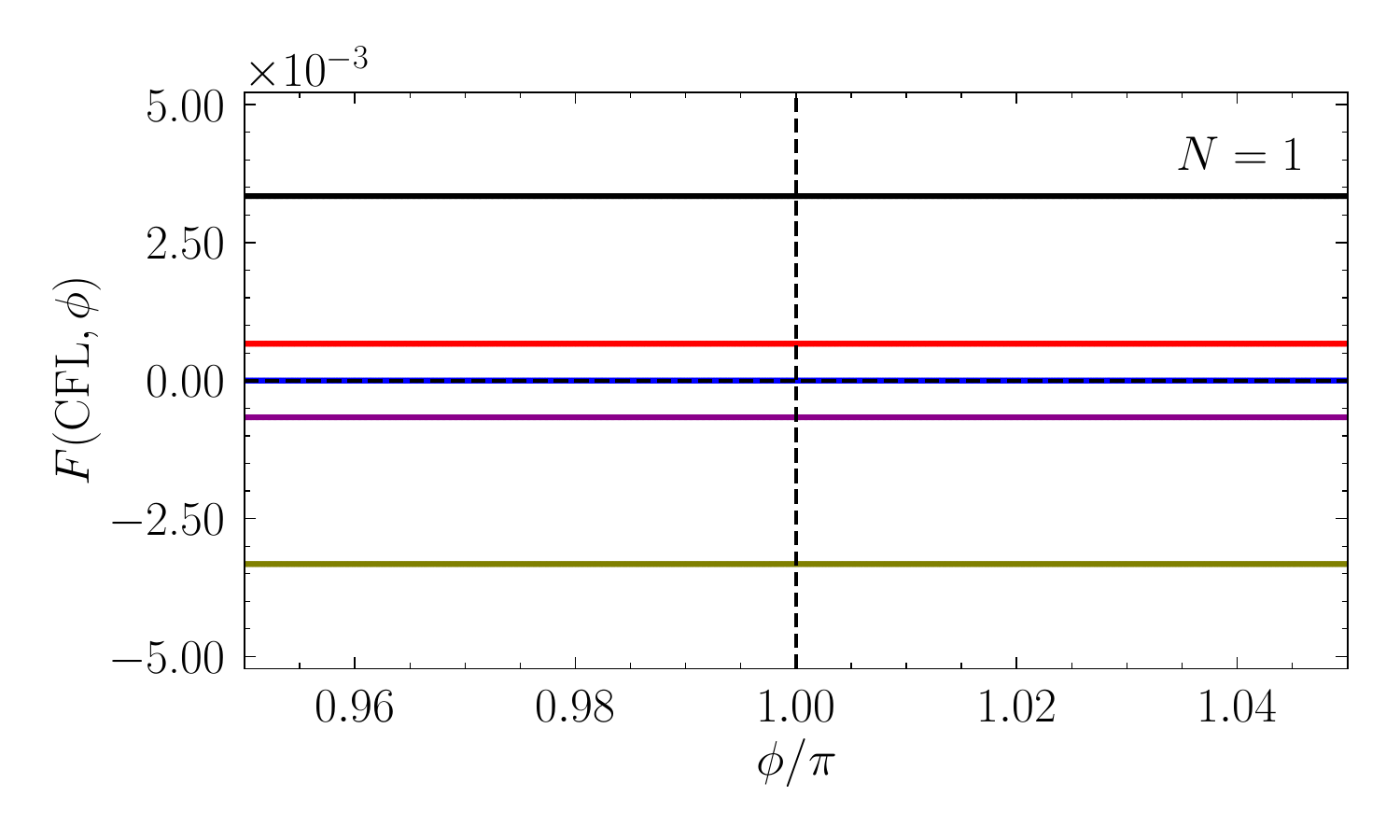}
\includegraphics[width=0.24\textwidth]{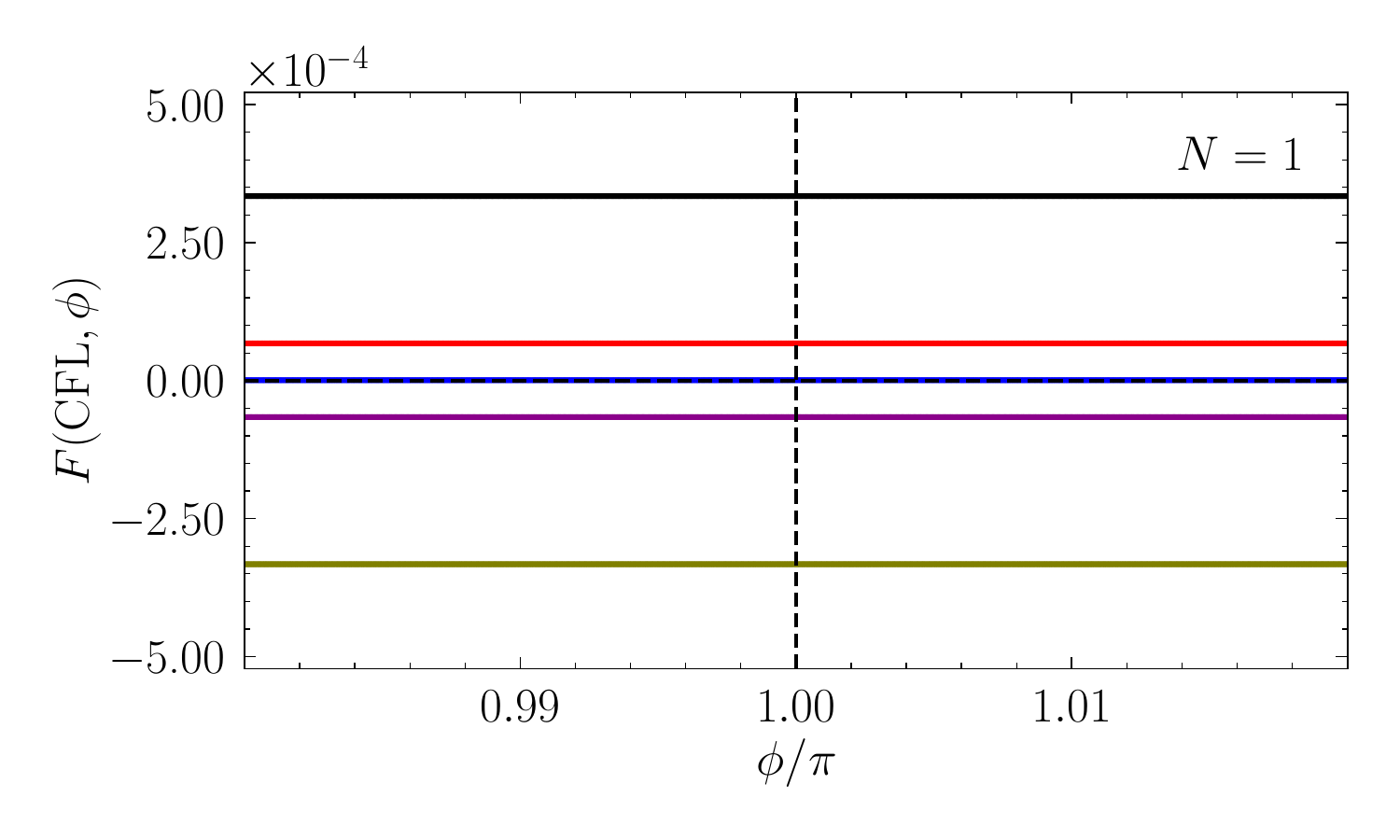}\\[-2mm]
\includegraphics[width=0.24\textwidth]{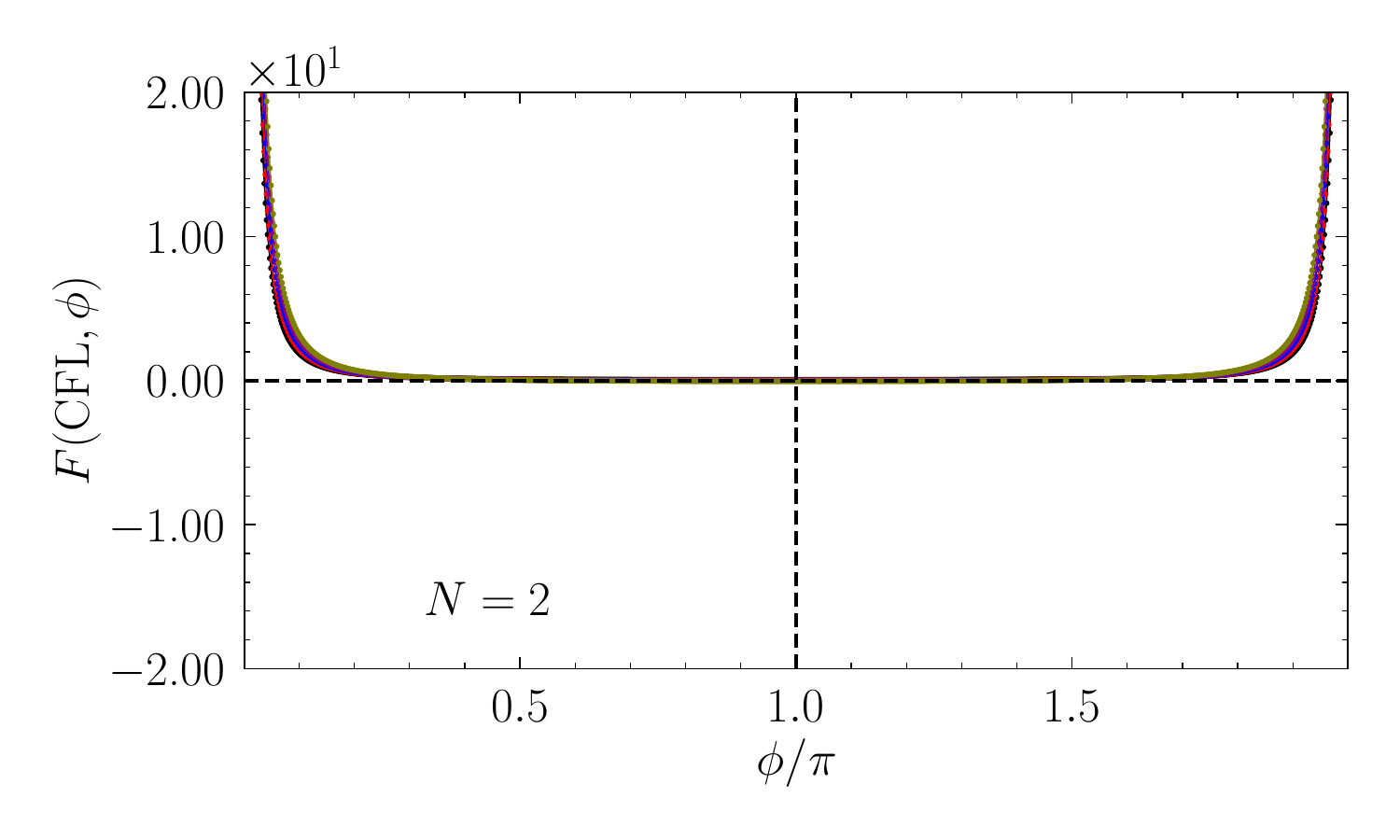}
\includegraphics[width=0.24\textwidth]{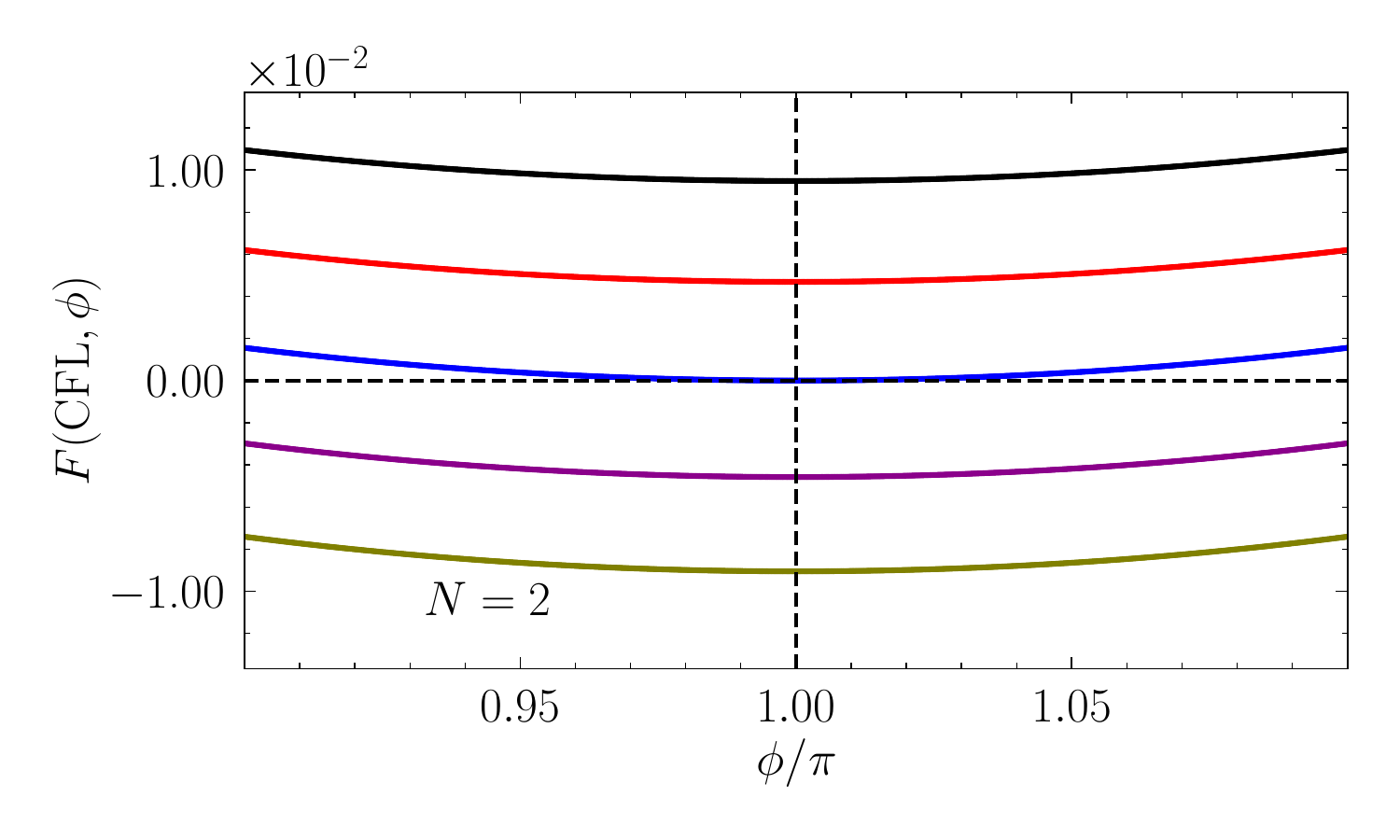}
\includegraphics[width=0.24\textwidth]{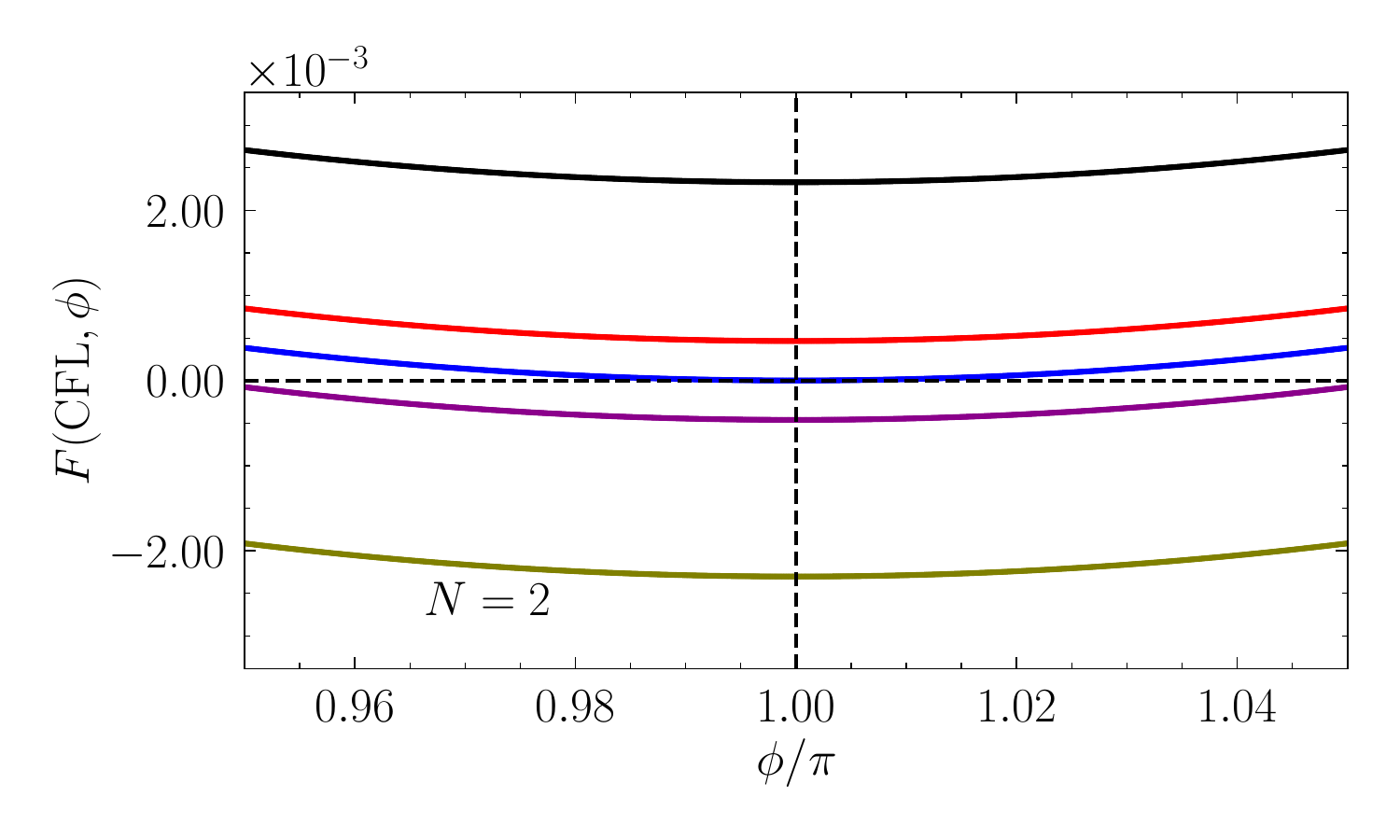}
\includegraphics[width=0.24\textwidth]{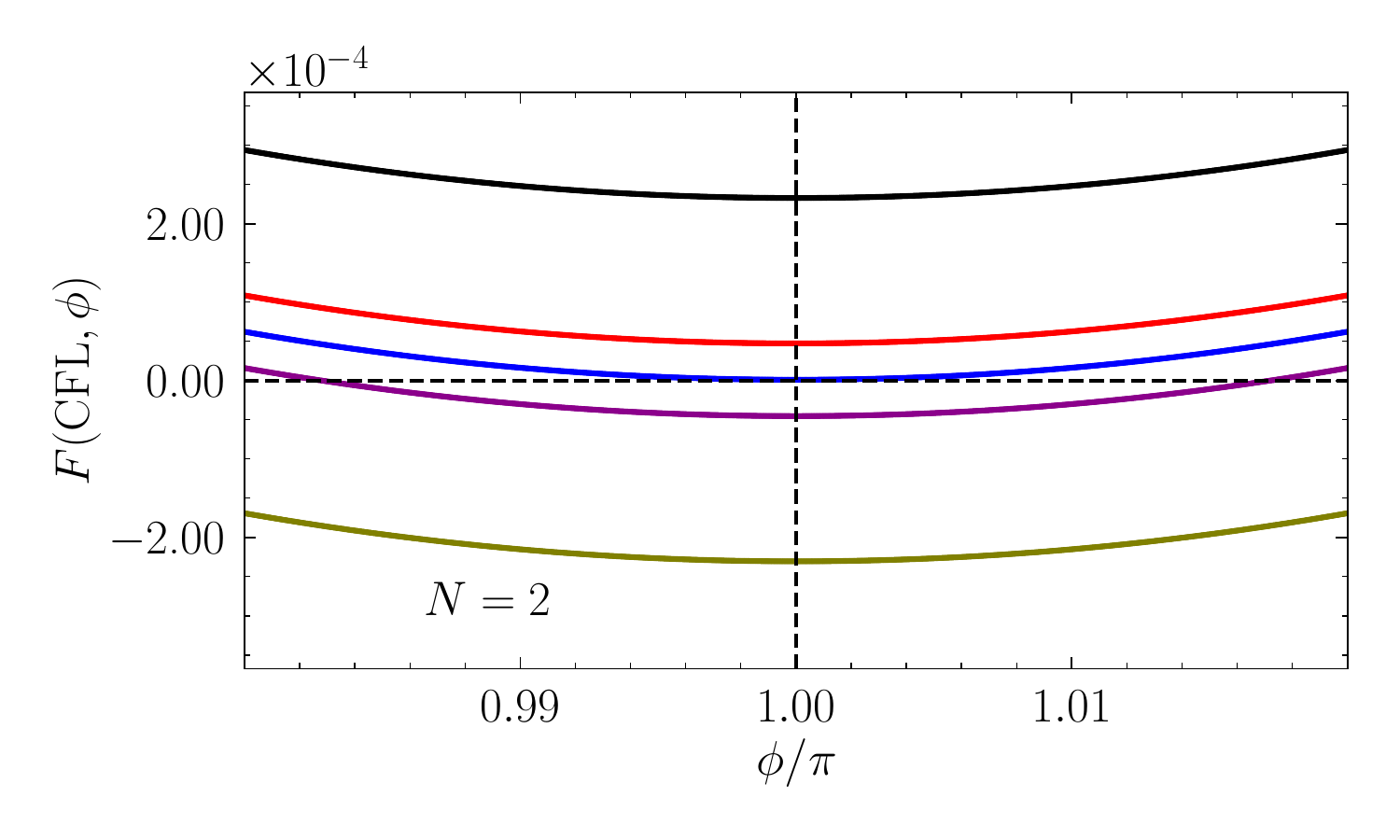}\\[-2mm]
\includegraphics[width=0.24\textwidth]{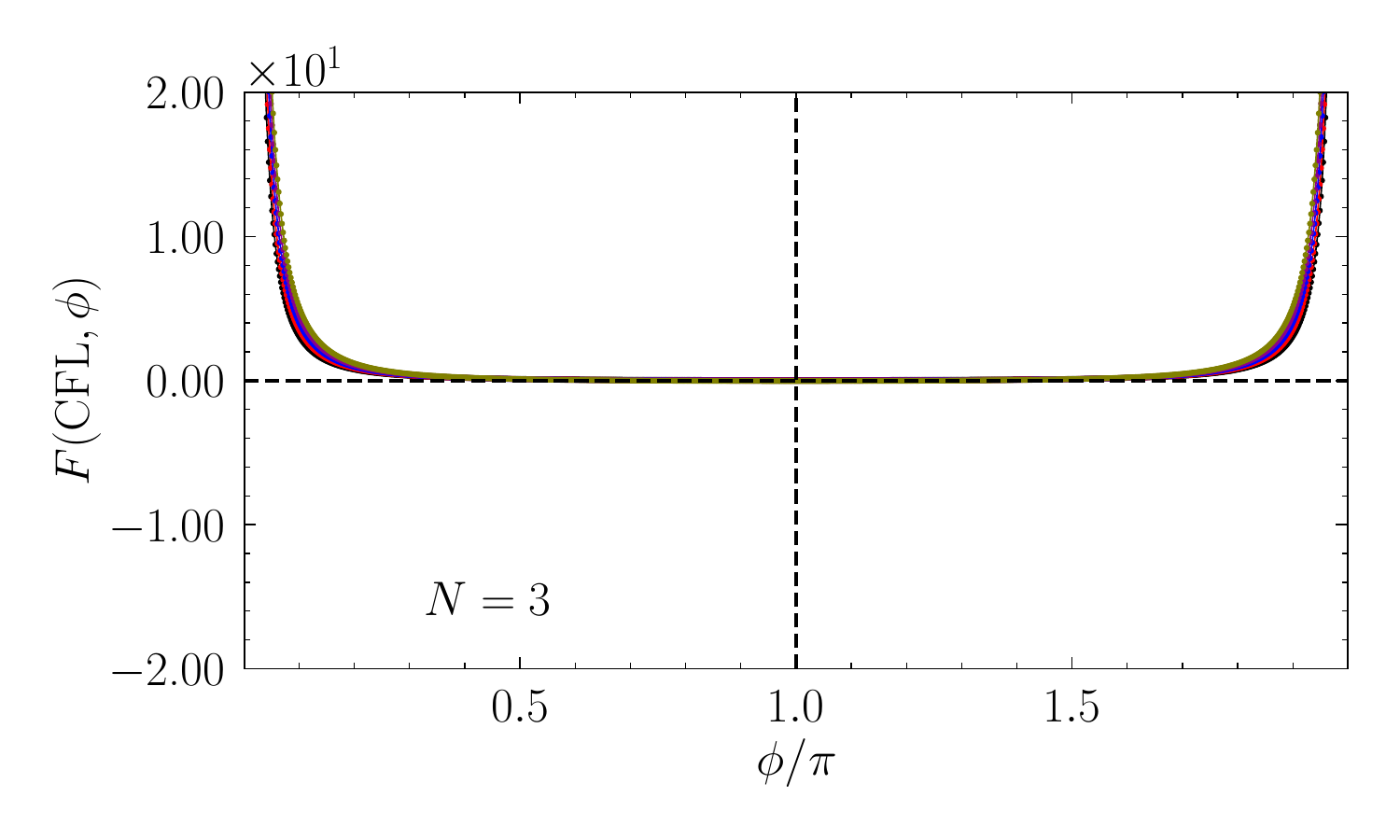}
\includegraphics[width=0.24\textwidth]{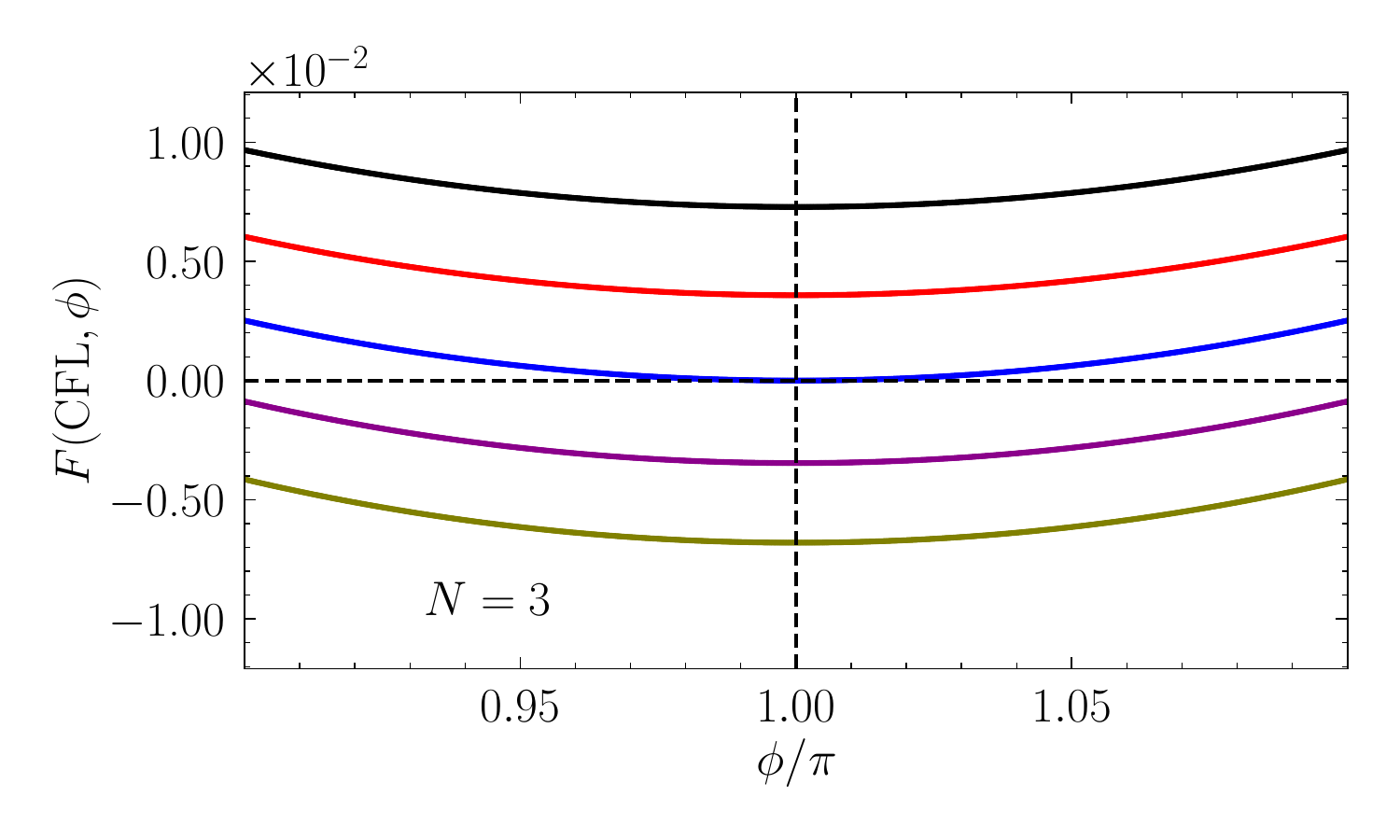}
\includegraphics[width=0.24\textwidth]{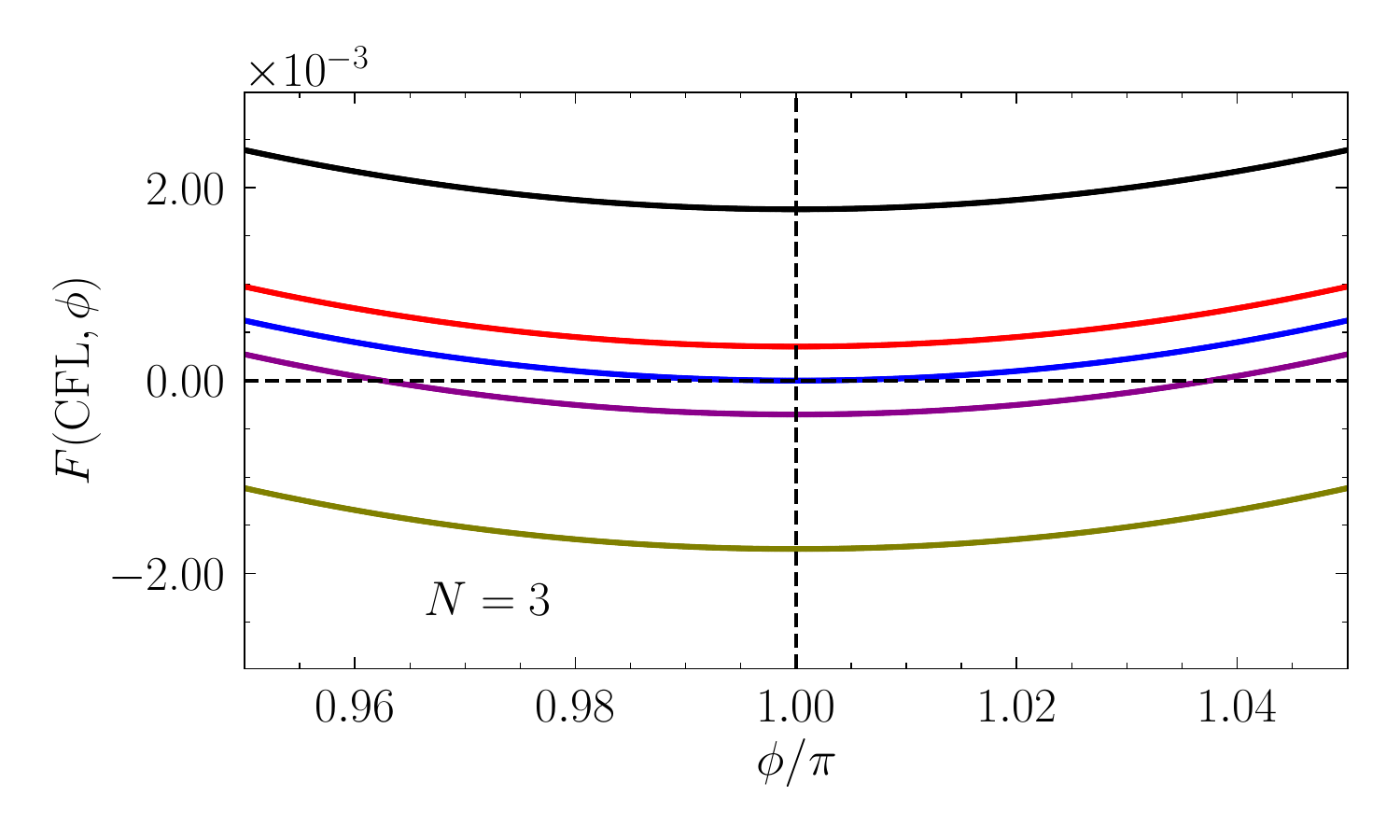}
\includegraphics[width=0.24\textwidth]{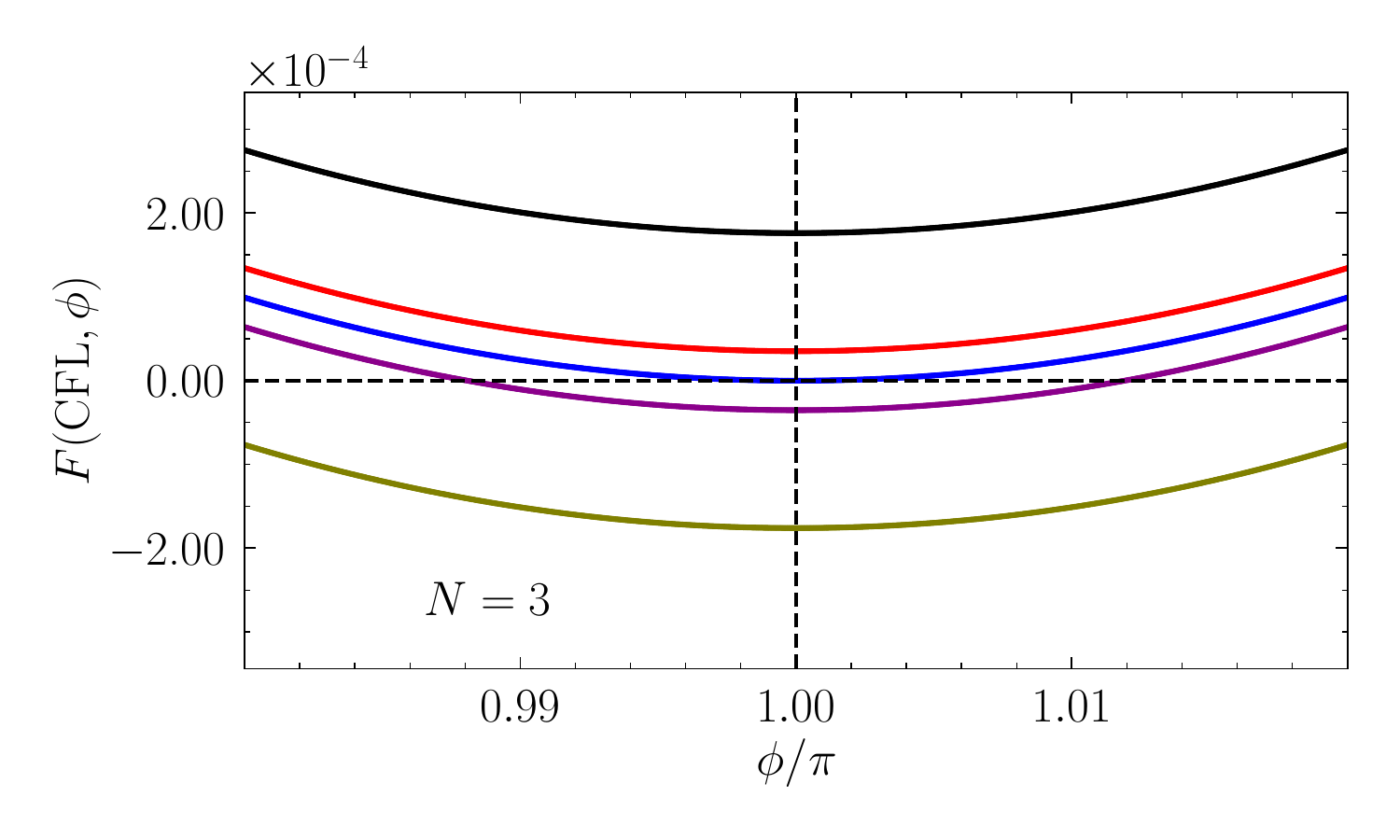}\\[-2mm]
\includegraphics[width=0.24\textwidth]{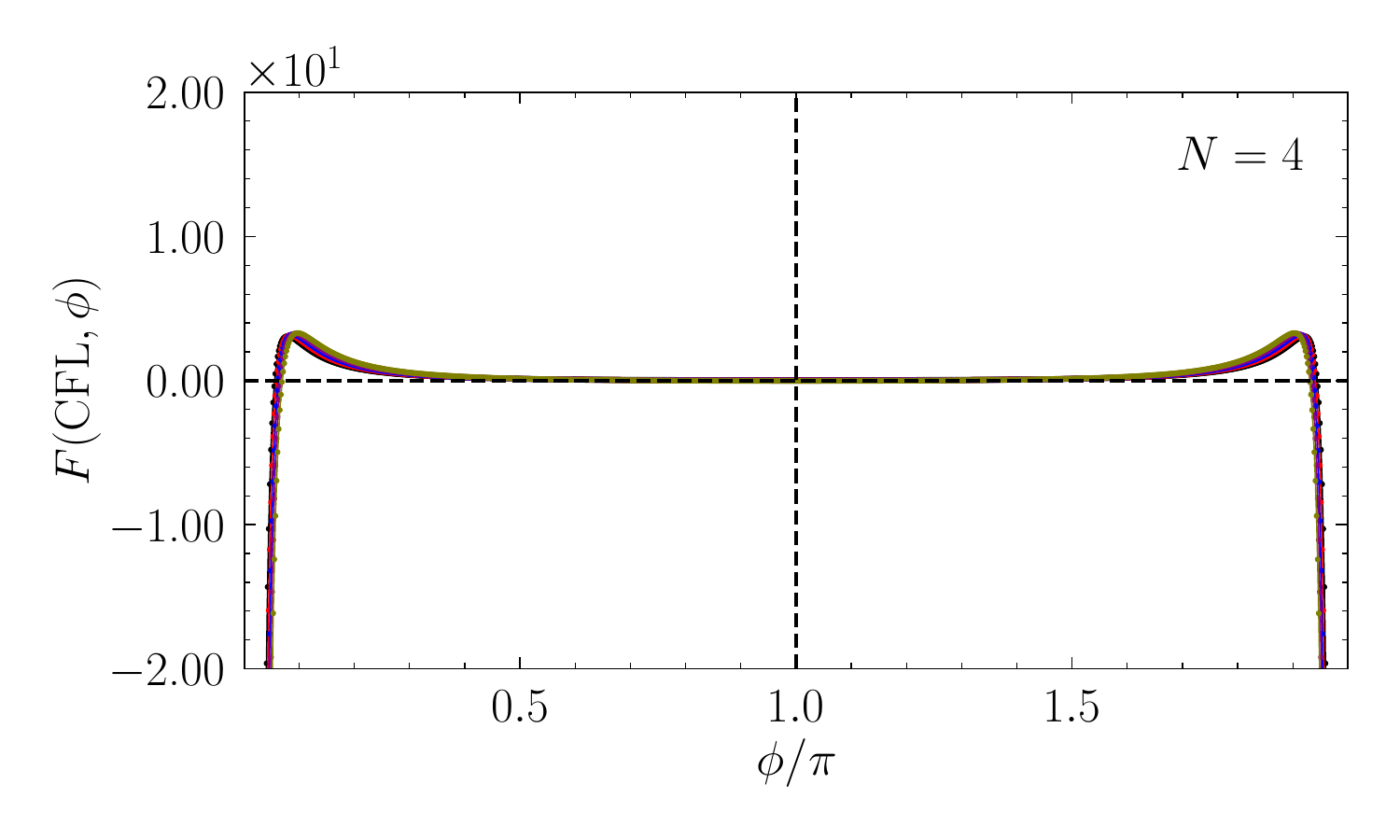}
\includegraphics[width=0.24\textwidth]{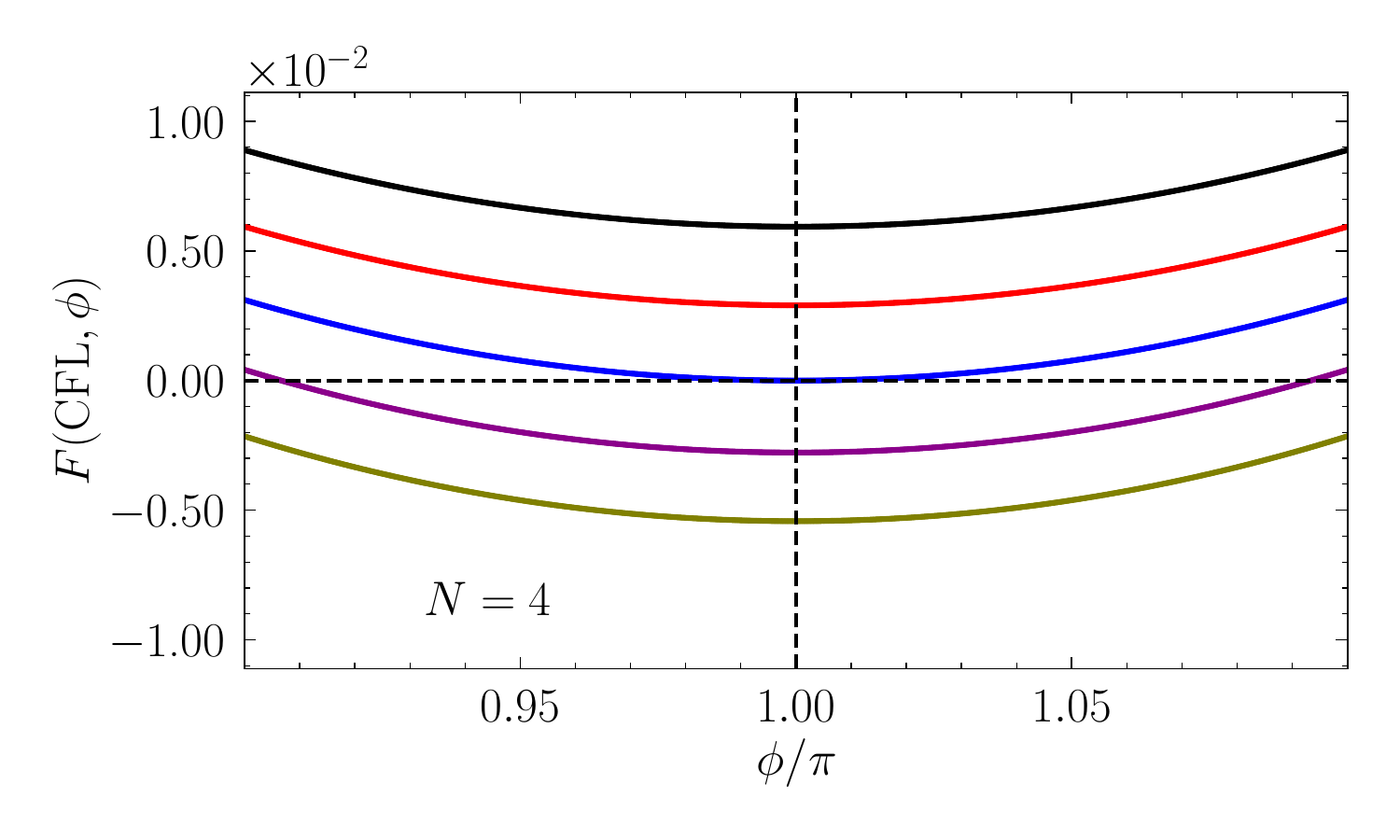}
\includegraphics[width=0.24\textwidth]{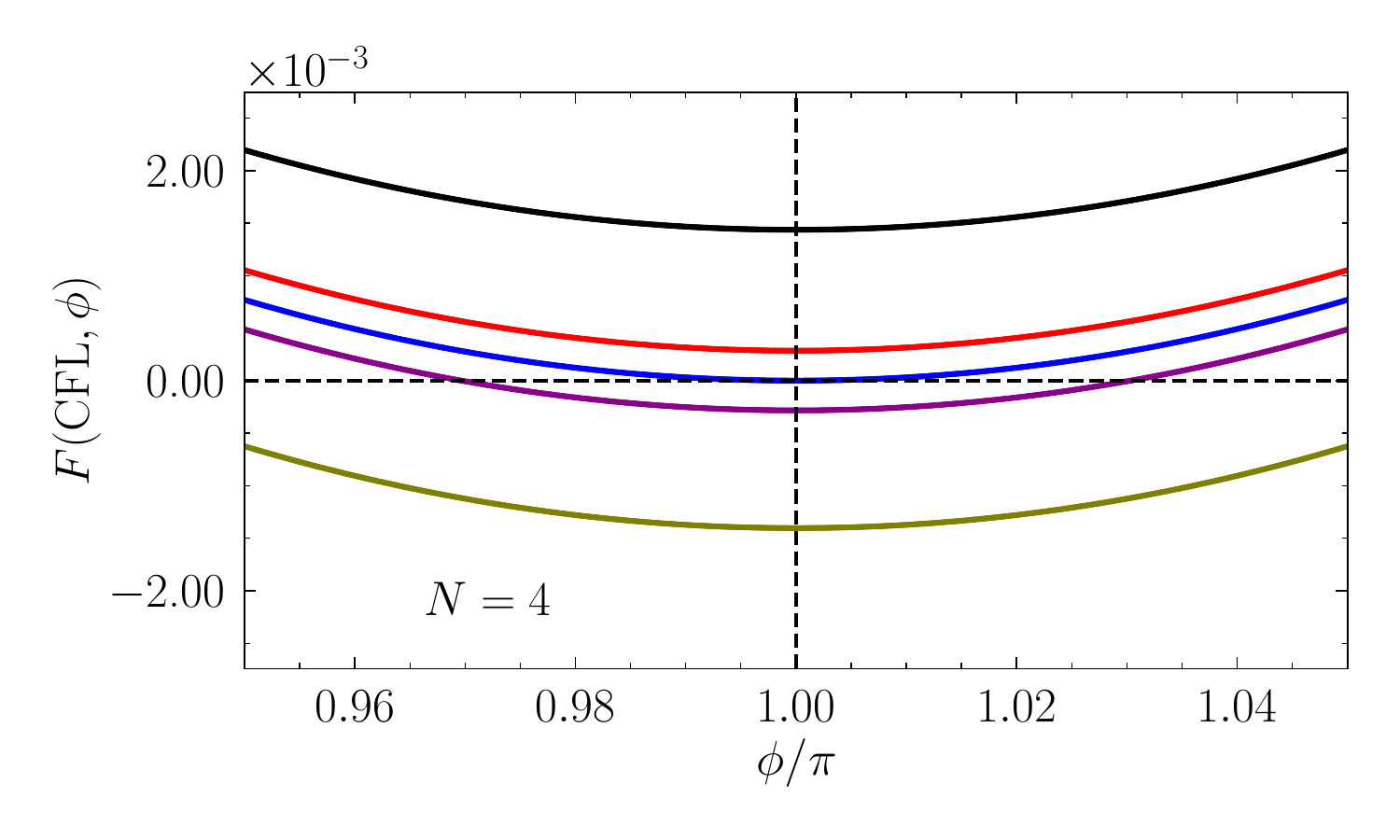}
\includegraphics[width=0.24\textwidth]{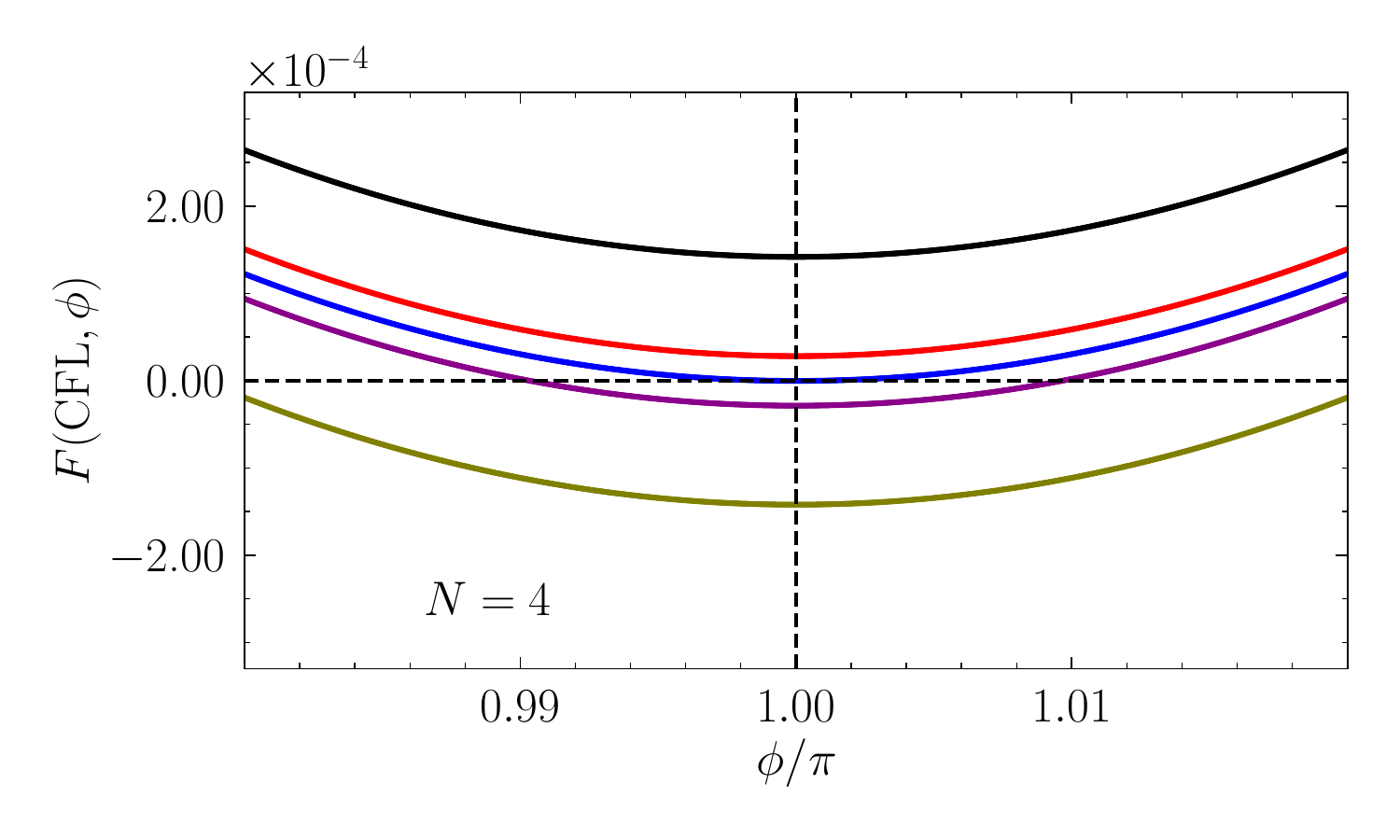}\\[-2mm]
\includegraphics[width=0.24\textwidth]{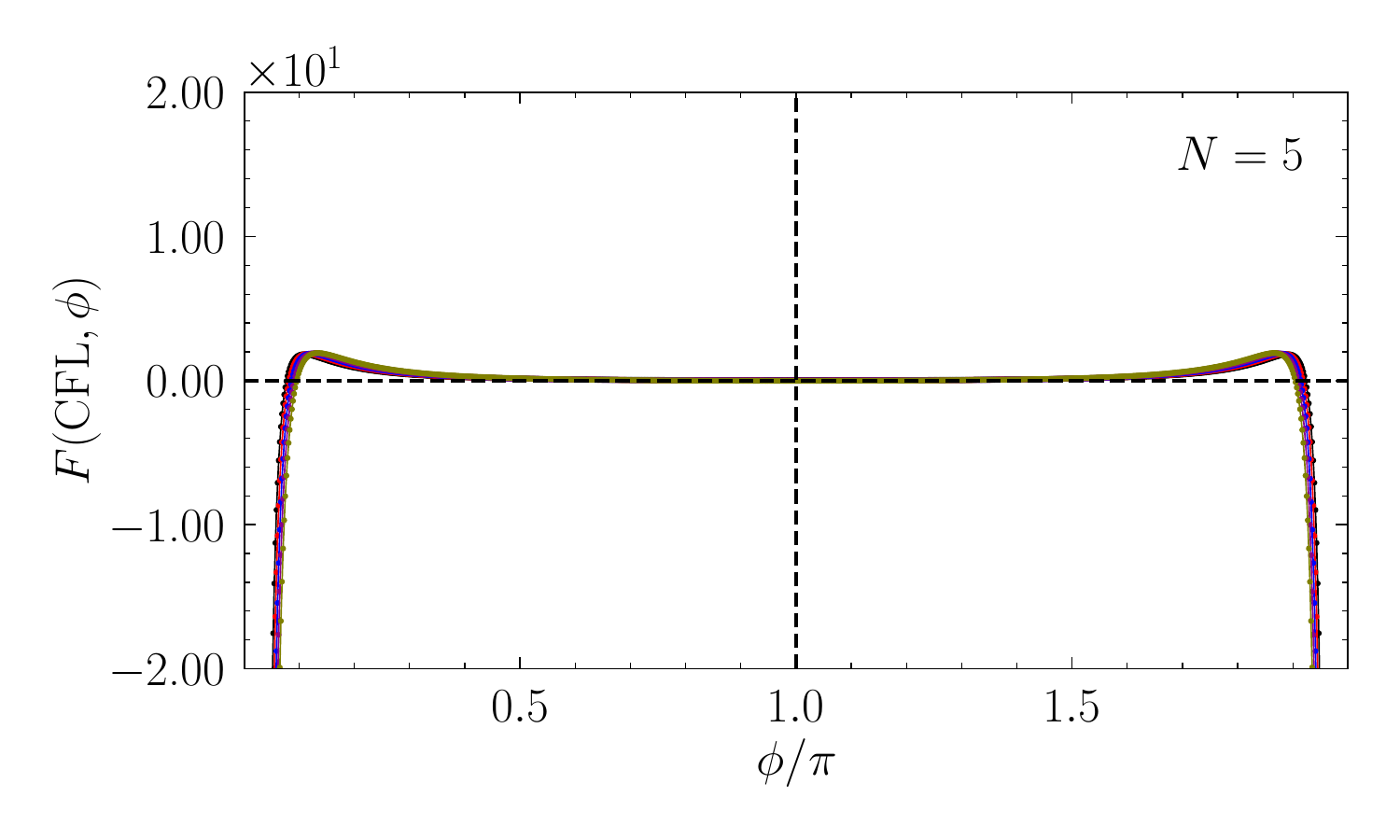}
\includegraphics[width=0.24\textwidth]{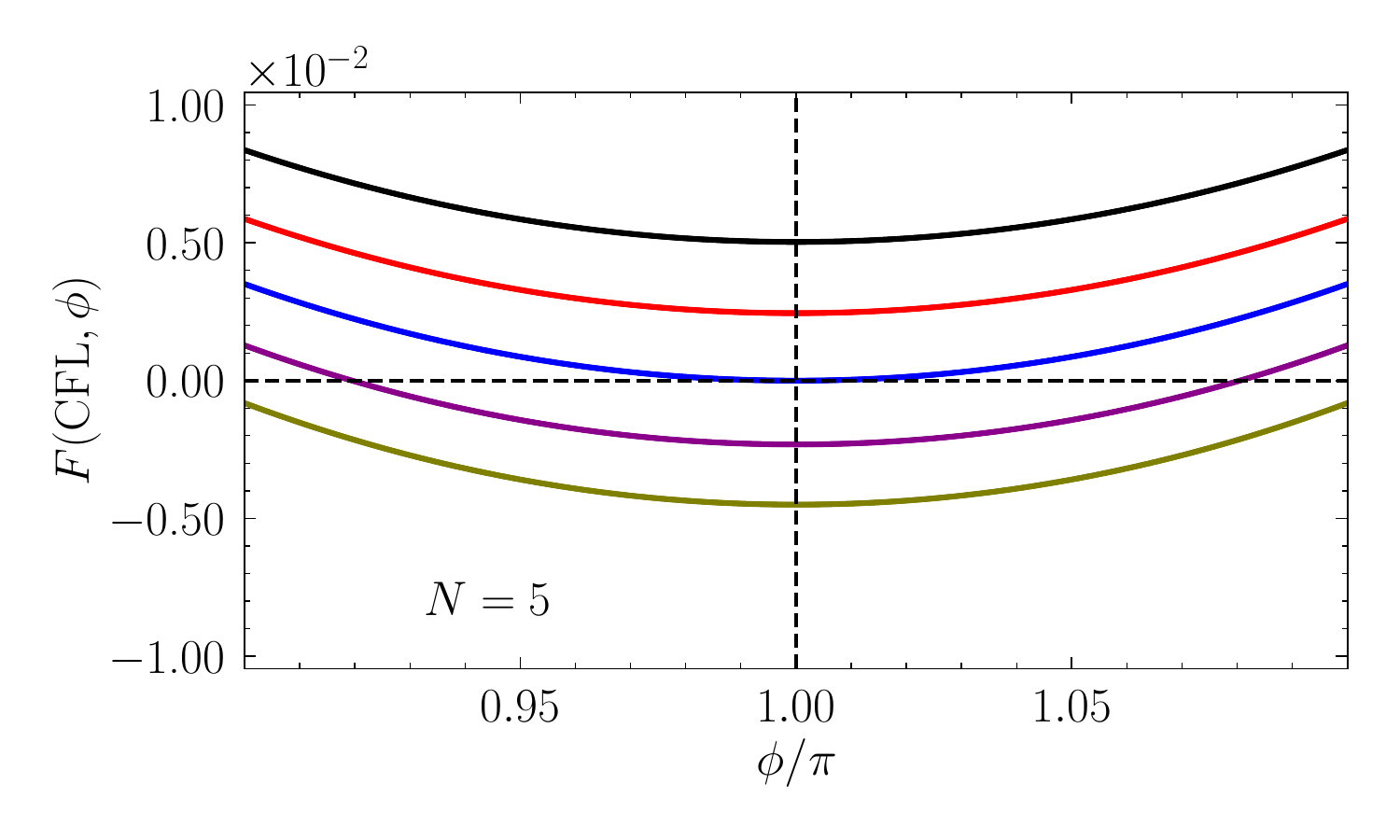}
\includegraphics[width=0.24\textwidth]{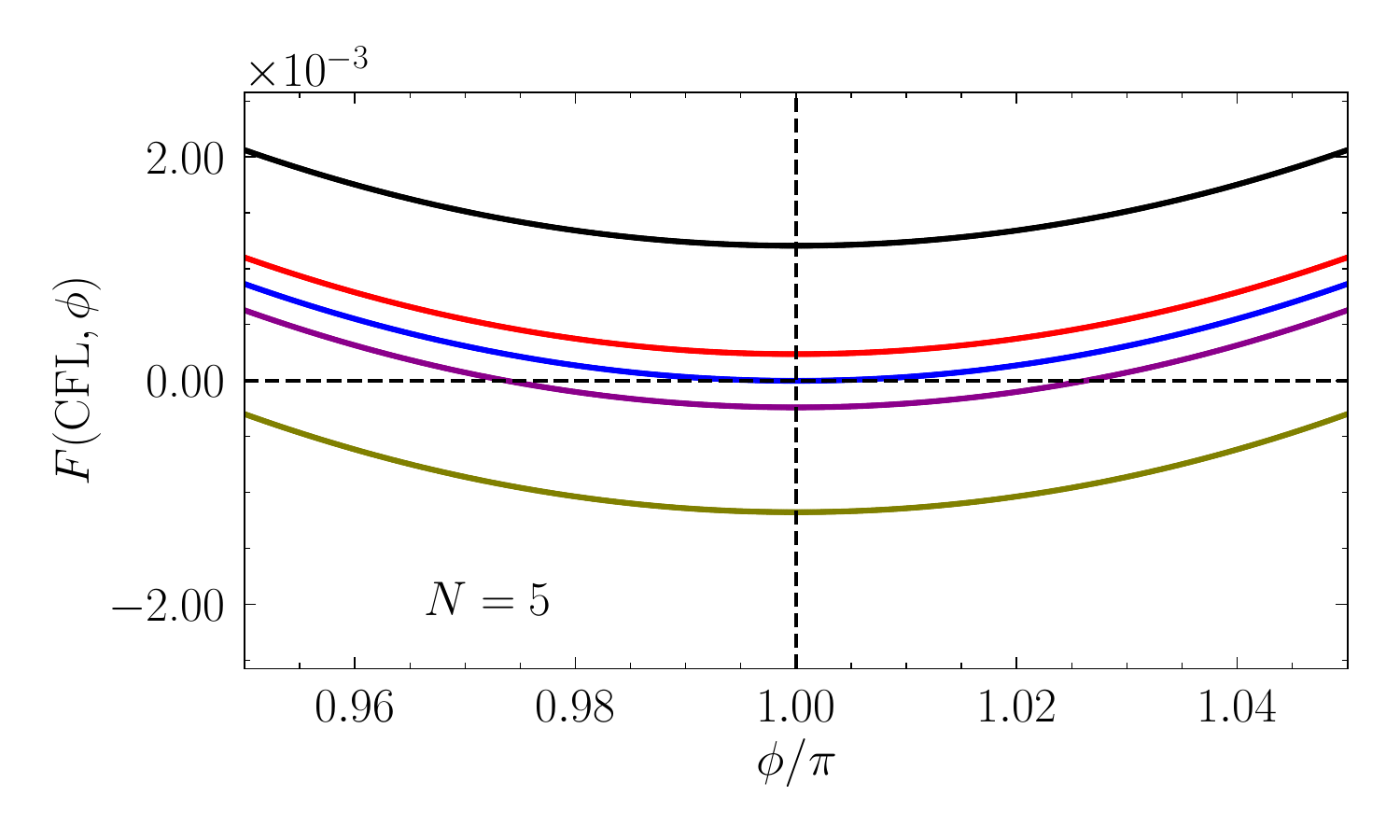}
\includegraphics[width=0.24\textwidth]{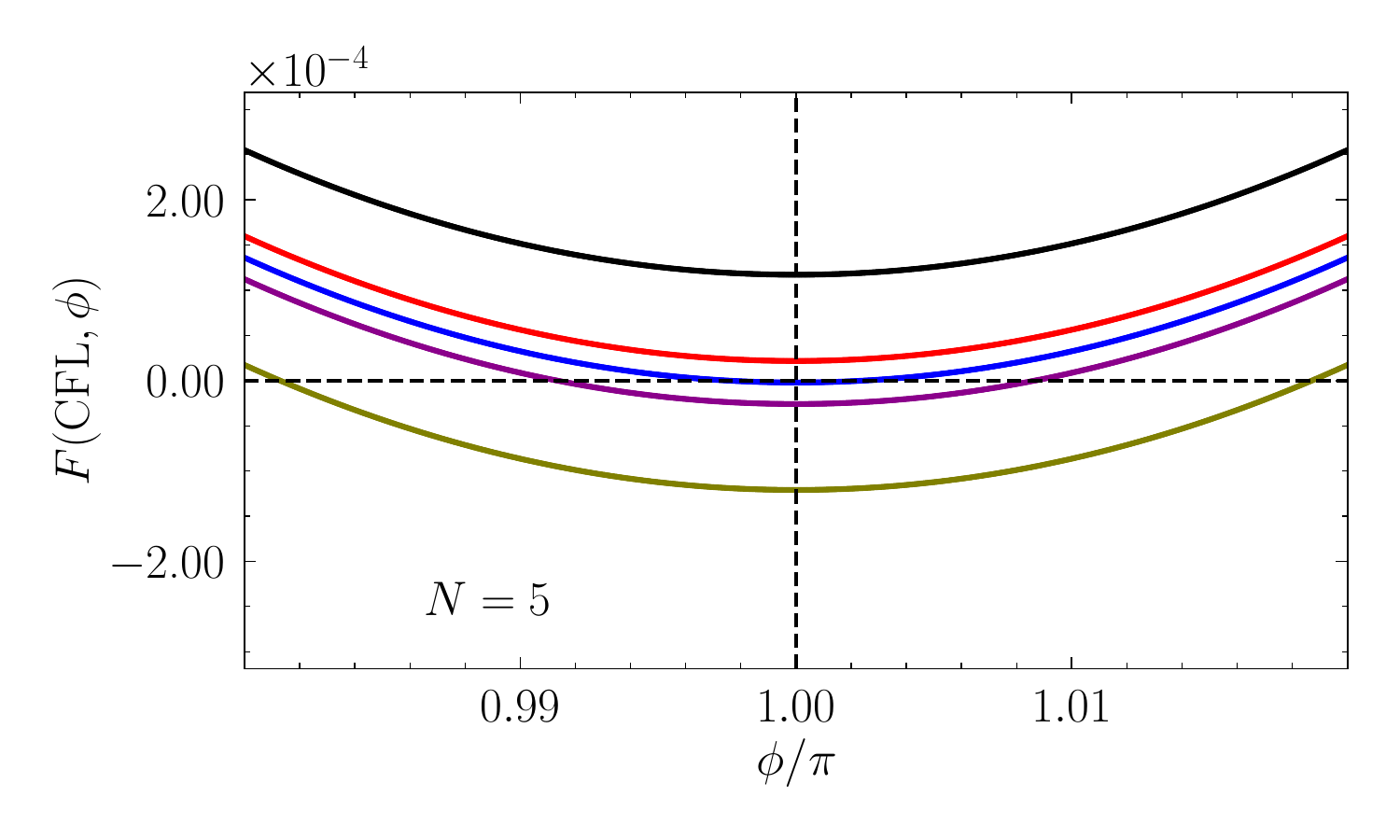}\\[-2mm]
\includegraphics[width=0.24\textwidth]{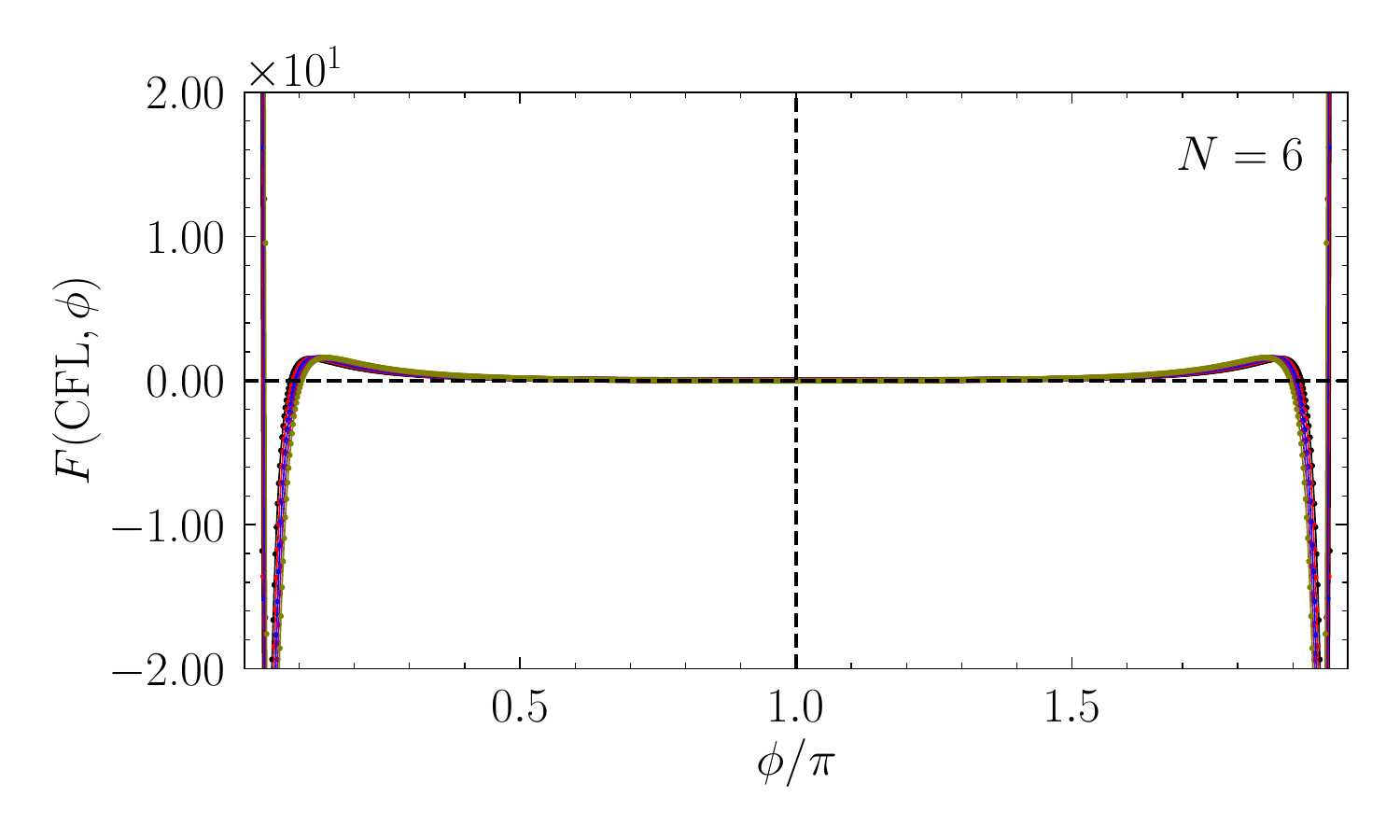}
\includegraphics[width=0.24\textwidth]{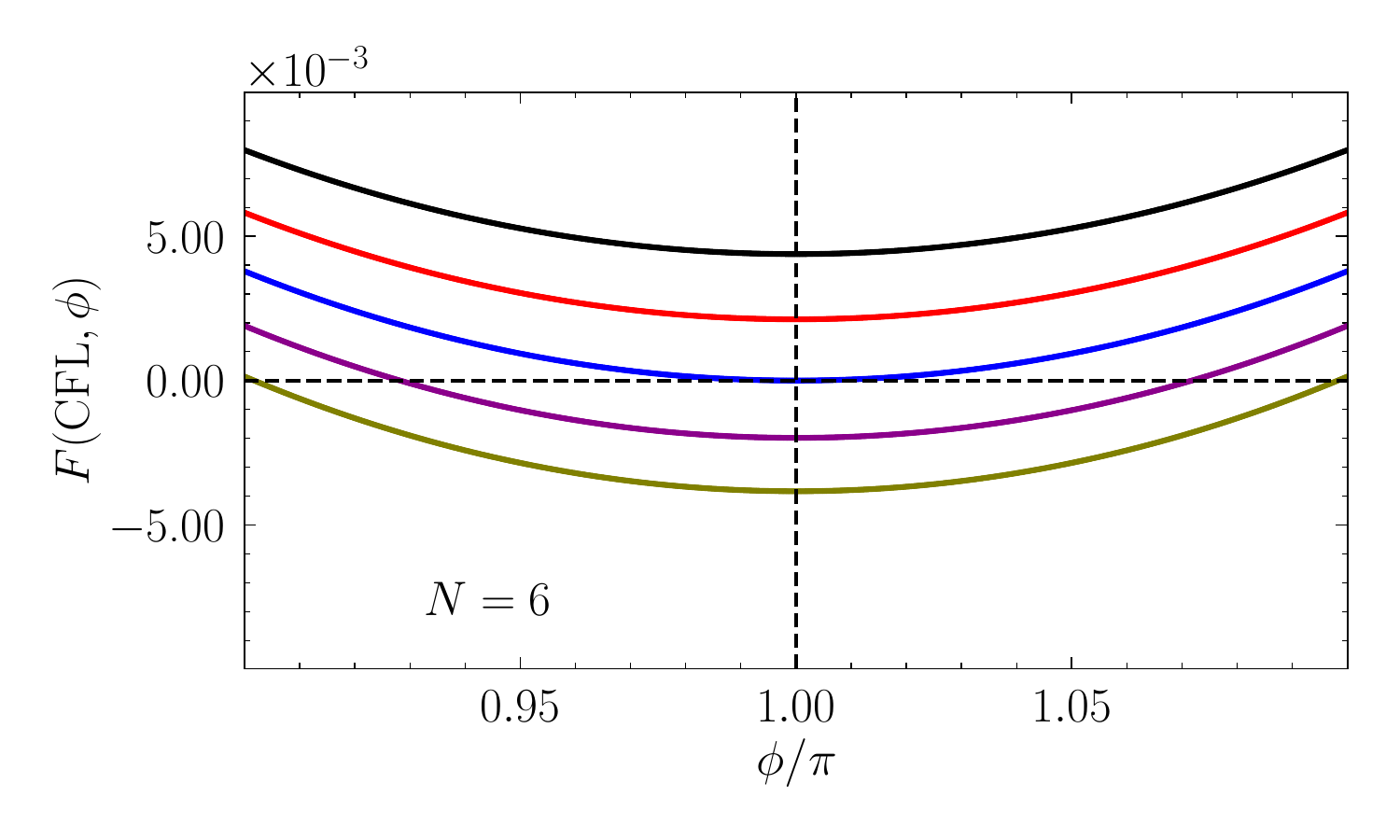}
\includegraphics[width=0.24\textwidth]{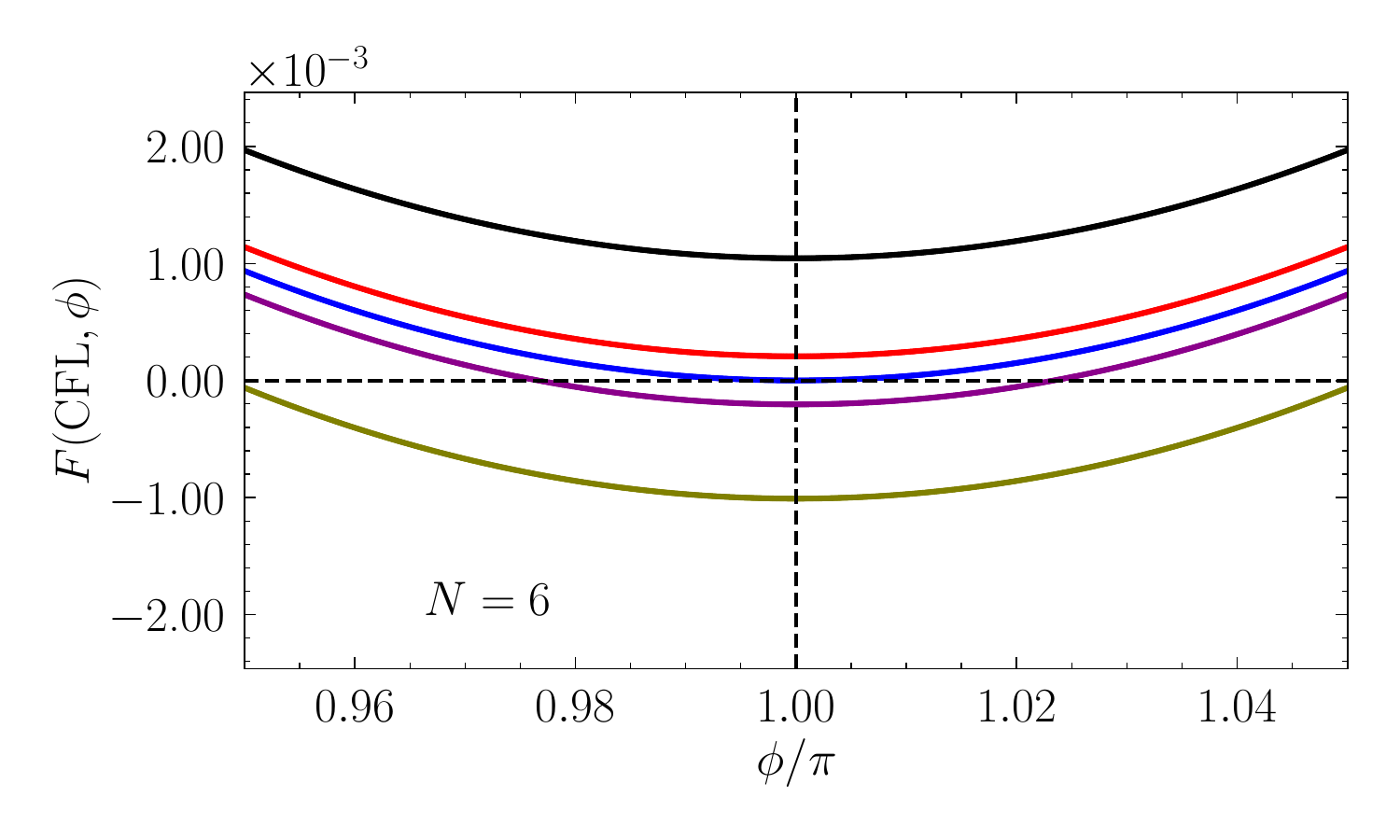}
\includegraphics[width=0.24\textwidth]{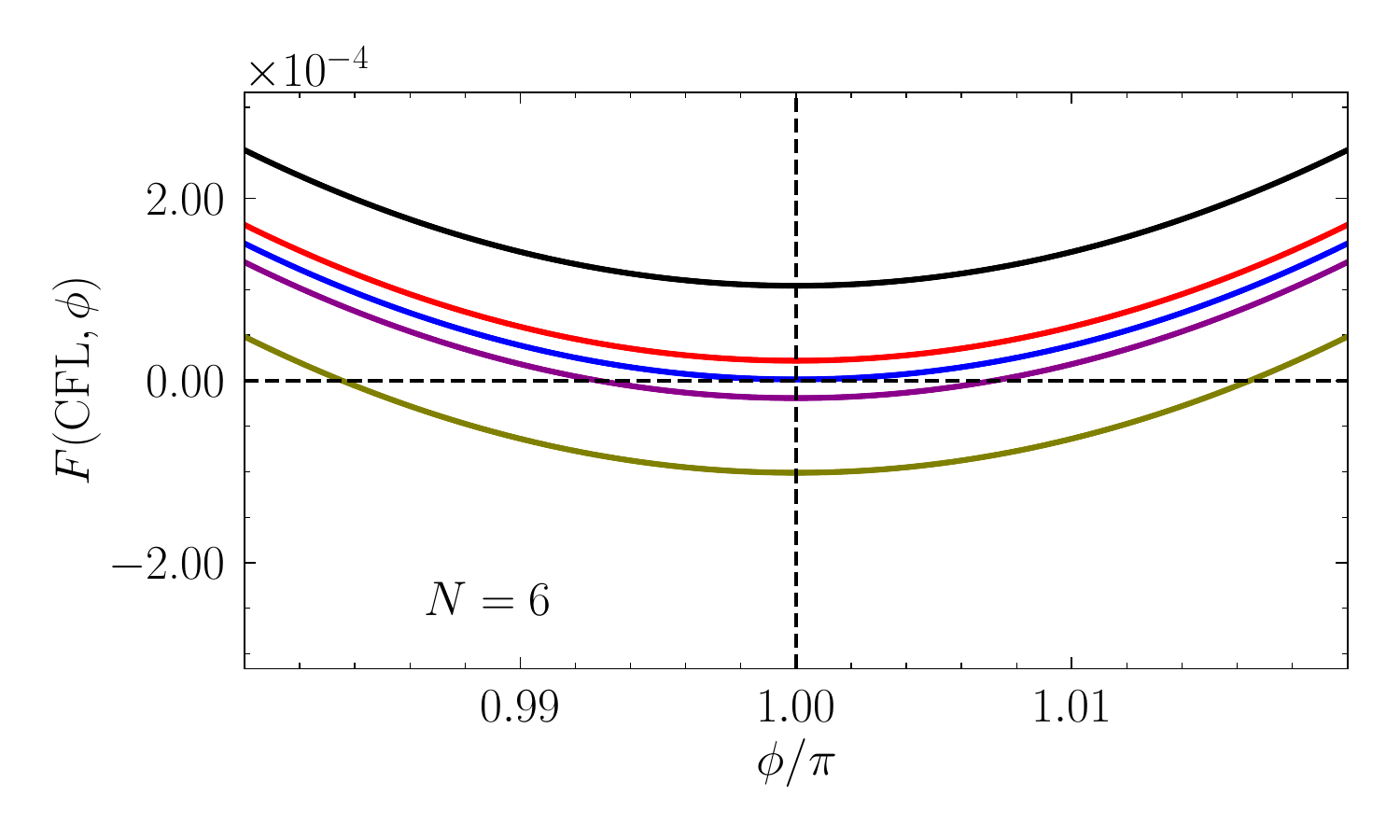}
\caption{%
The dependence of function $F(\mathrm{CFL}, \phi)$ (\ref{eq:f_def_plus}) on phase $\phi\in[0, 2\pi]$ for several values of the Courant number $\mathrm{CFL}$, selected in the vicinity of the stability boundary $\mathrm{CFL}_{\rm max}$, defined further in Table~\ref{tab:cfls_max_data} and in Figure~\ref{fig:cfls_max_data}, for polynomial degrees $N = 1, \ldots, 6$. The graph legends are distributed among columns and located at the top of the columns. From left to right, the graphs represent increasingly narrower neighborhoods of the stability boundary $\mathrm{CFL}_{\rm max}$, from $\mathrm{CFL} \in [0.90\, \mathrm{CFL}_{\rm max},\, 1.10\, \mathrm{CFL}_{\rm max}]$ in the left column to $\mathrm{CFL} \in [0.9995\, \mathrm{CFL}_{\rm max},\, 1.0005\, \mathrm{CFL}_{\rm max}]$ in the right column; the range of phase $\phi$ variation also narrows when moving from the left to the right columns of the graphs to ensure a correct and clear presentation of the root $\phi = \pi$ of the equation $F(\mathrm{CFL}, \pi) = 0$. For polynomial degrees $N \geqslant 4$, additional intersections of the dependencies of functions $F(\mathrm{CFL}, \phi)$ (\ref{eq:f_def_plus}) on phase $\phi$ are observed in the vicinity of points $0$ and $2\pi$, which corresponds to the ``false'' instability violations described in the main text. The horizontal dotted line indicates $F = 0$, the vertical dotted line indicates $\phi = \pi$ ($\lambda = \exp(\pm i\phi) = -1$). \textit{Note}: the phase $\phi$ of eigenvalue $\lambda_{k} = |\lambda_{k}|\exp(i\phi)$ is not the phase $\theta$.
}
\label{fig:dep_f_on_phi_diff_cfls_degrees_1_6}
\end{figure}

\begin{figure}[h!]
\centering
\includegraphics[width=0.24\textwidth]{dep_f_on_phi_diff_cfls_legend_Cidx_0.pdf}
\includegraphics[width=0.24\textwidth]{dep_f_on_phi_diff_cfls_legend_Cidx_1.pdf}
\includegraphics[width=0.24\textwidth]{dep_f_on_phi_diff_cfls_legend_Cidx_2.pdf}
\includegraphics[width=0.24\textwidth]{dep_f_on_phi_diff_cfls_legend_Cidx_3.pdf}\\
\includegraphics[width=0.24\textwidth]{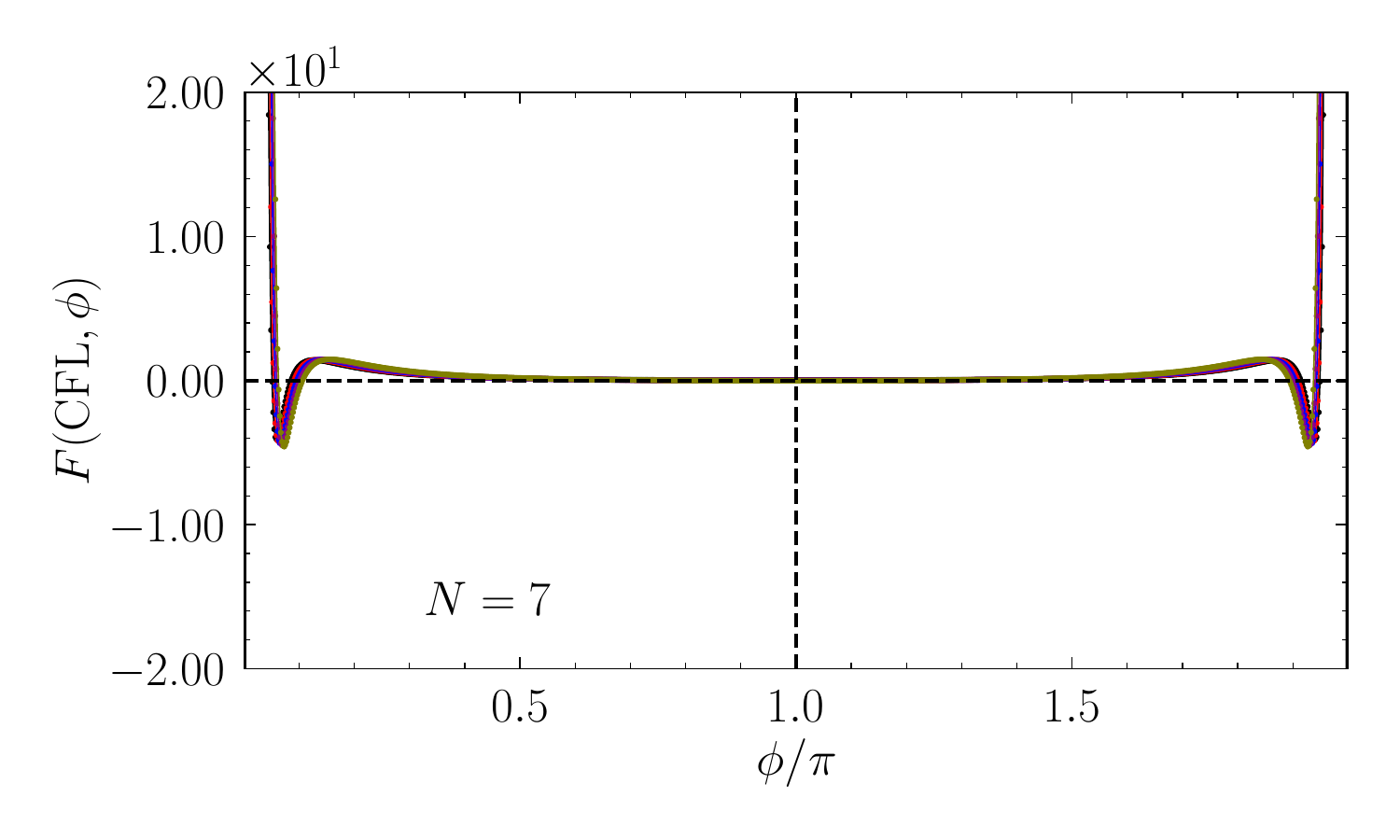}
\includegraphics[width=0.24\textwidth]{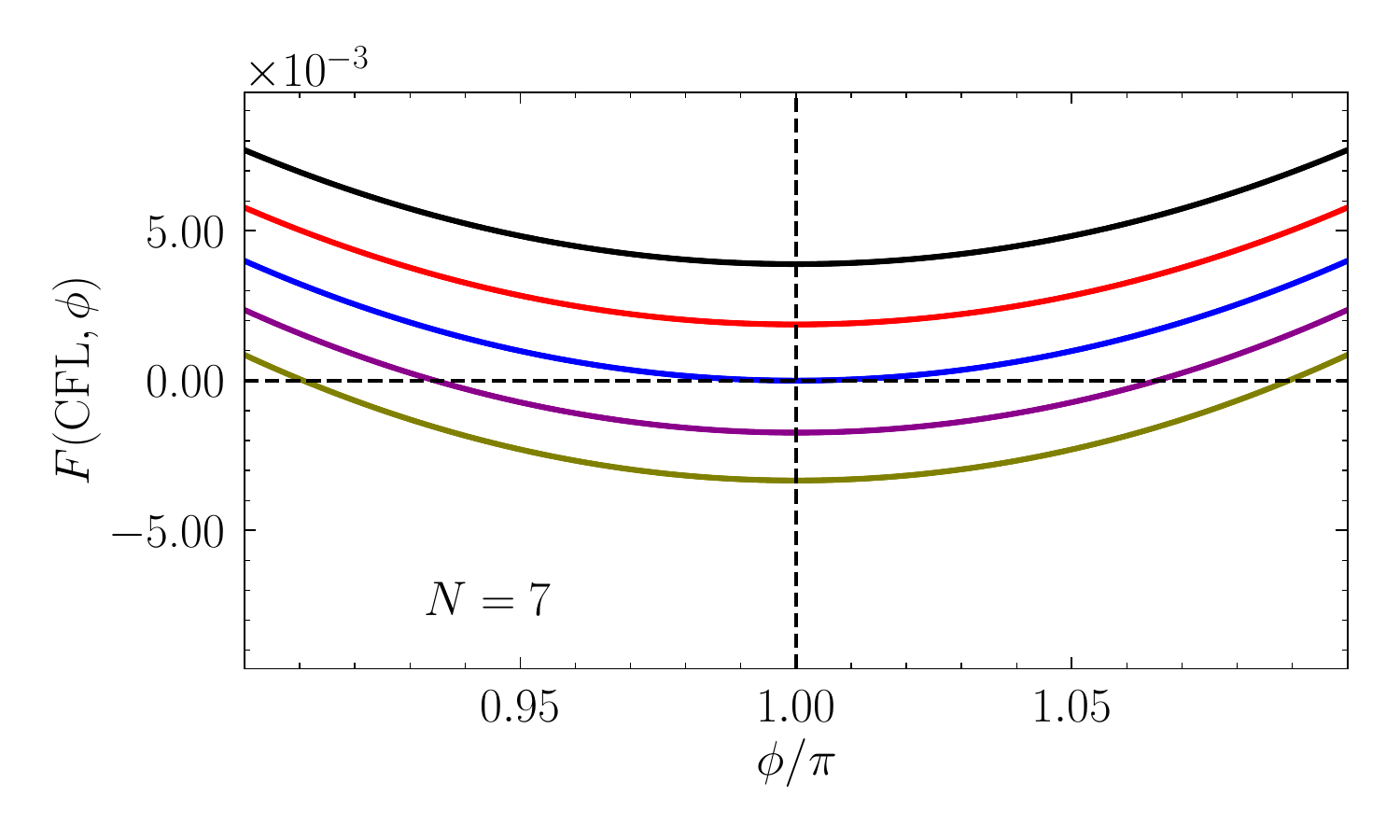}
\includegraphics[width=0.24\textwidth]{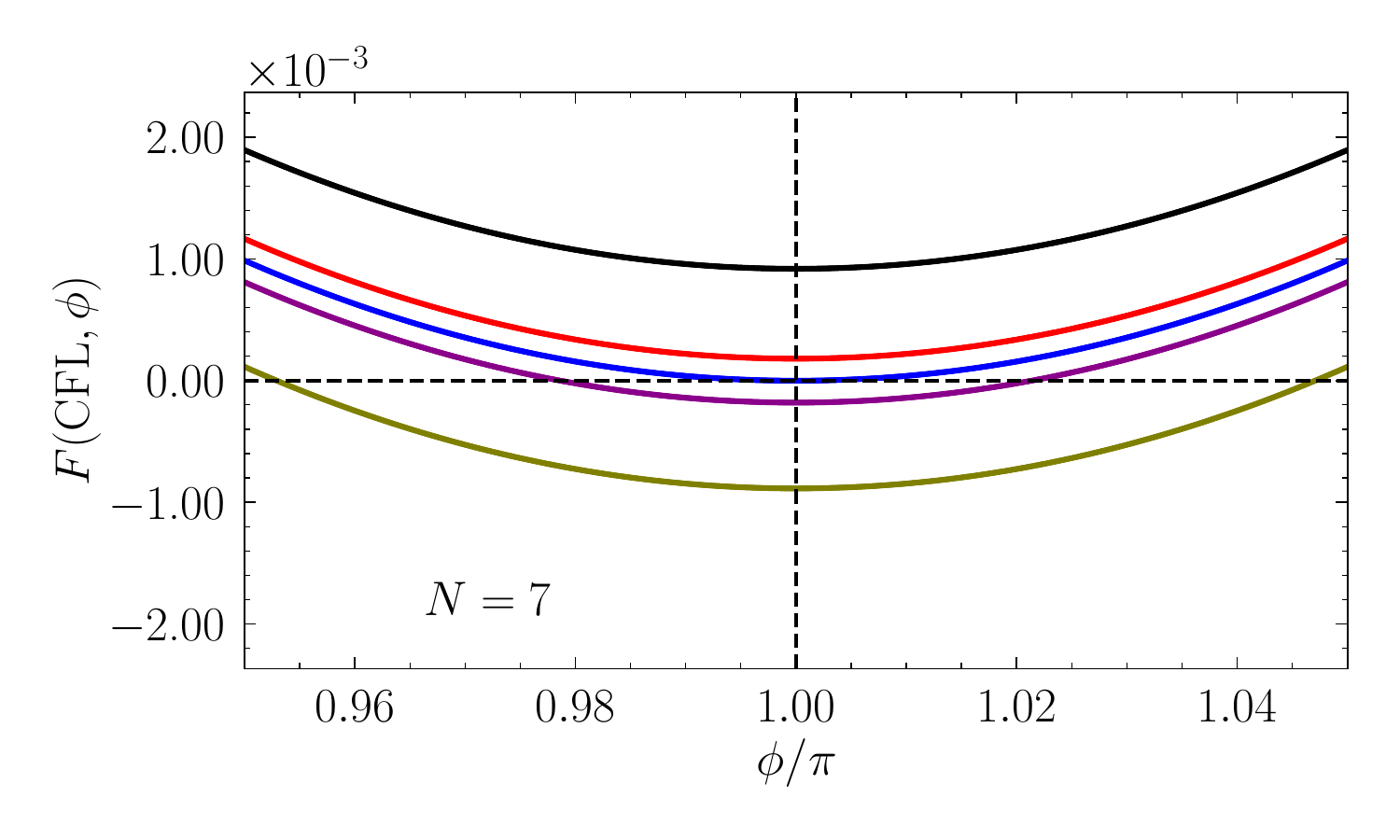}
\includegraphics[width=0.24\textwidth]{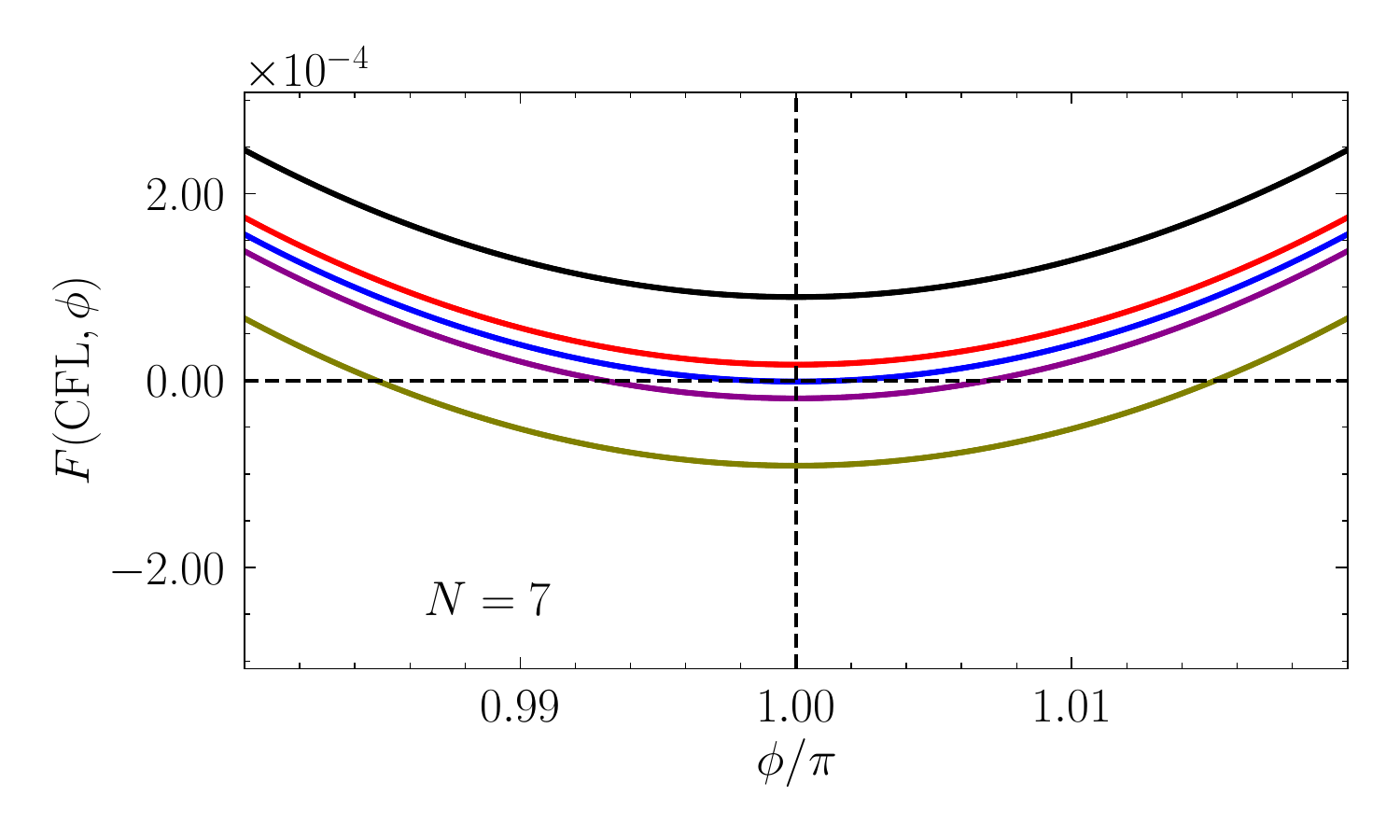}\\[-2mm]
\includegraphics[width=0.24\textwidth]{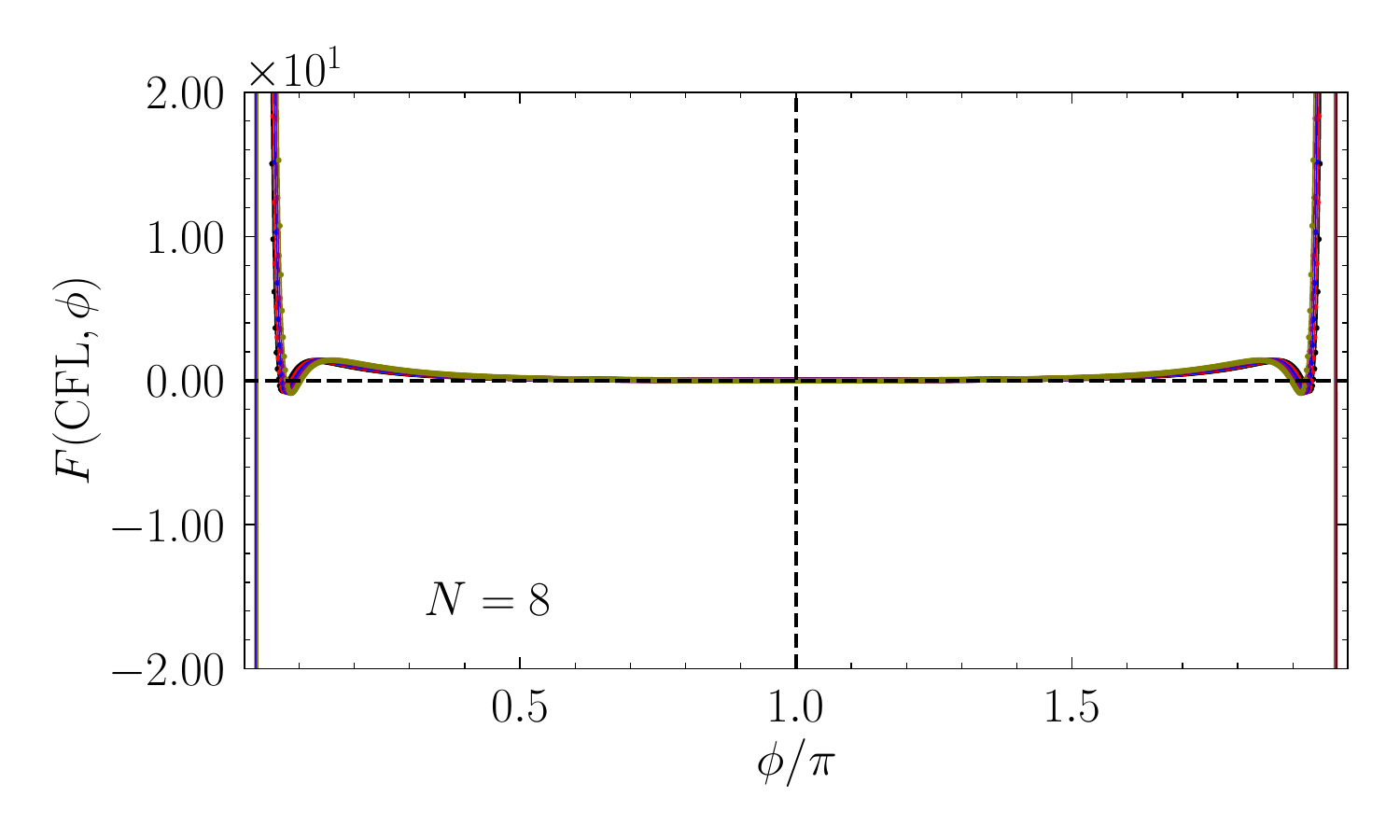}
\includegraphics[width=0.24\textwidth]{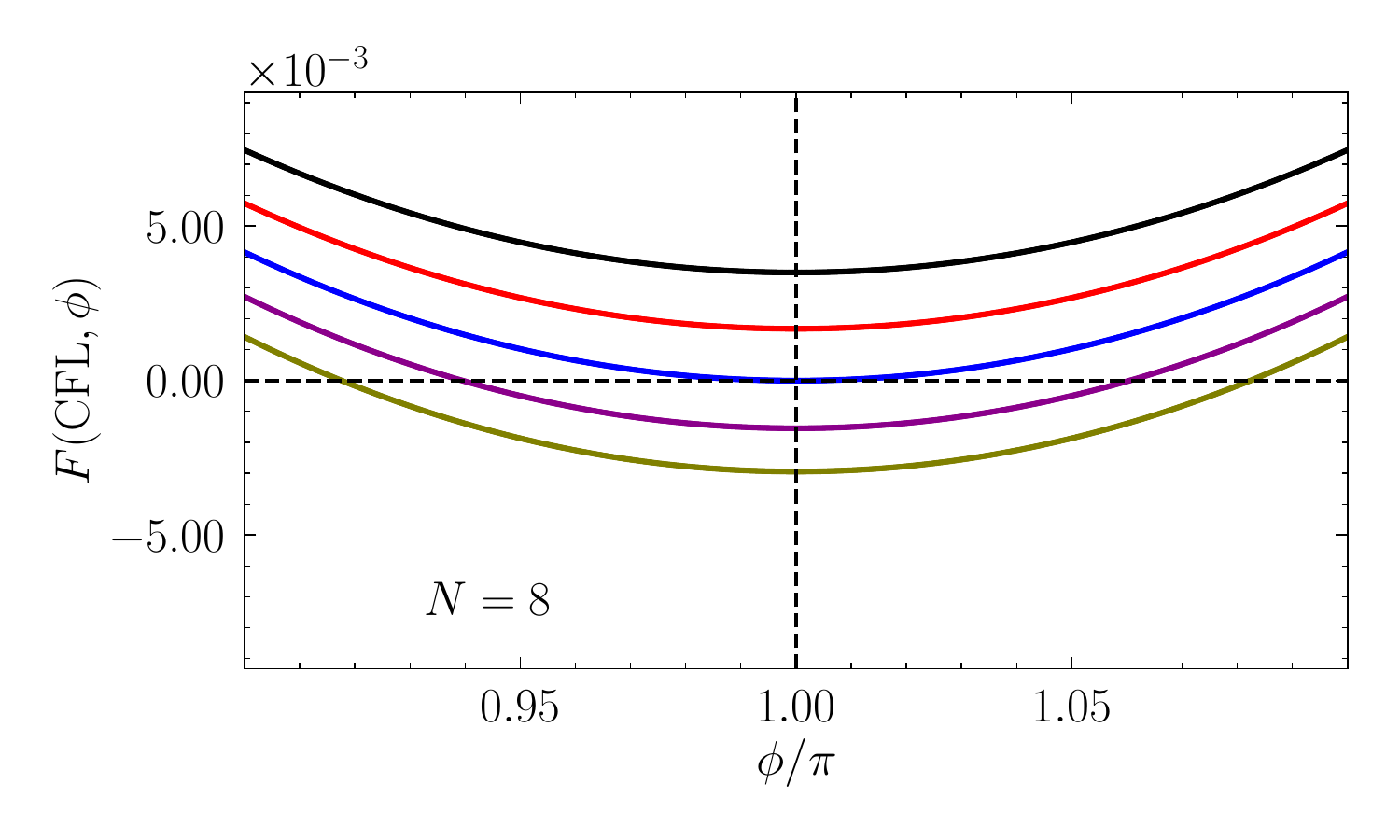}
\includegraphics[width=0.24\textwidth]{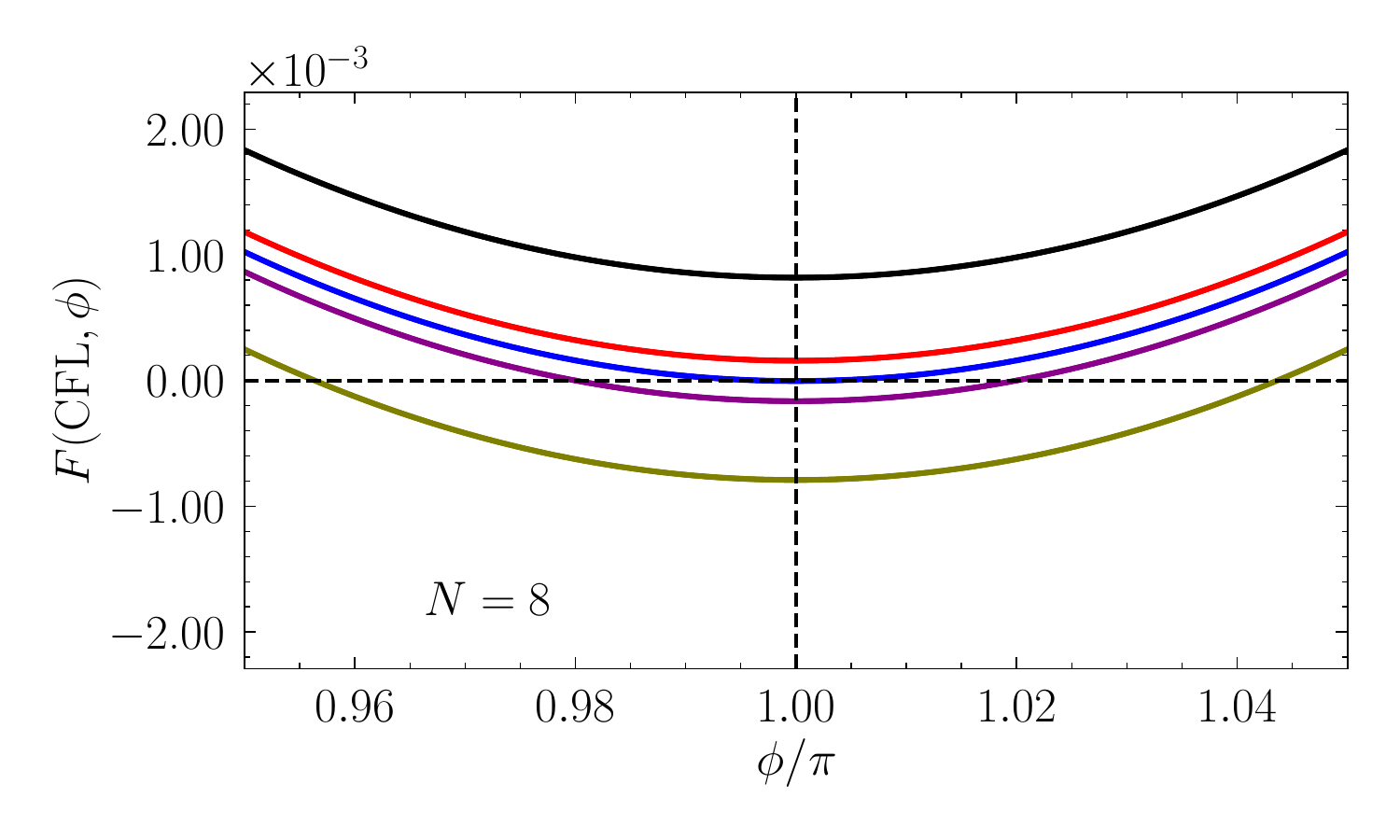}
\includegraphics[width=0.24\textwidth]{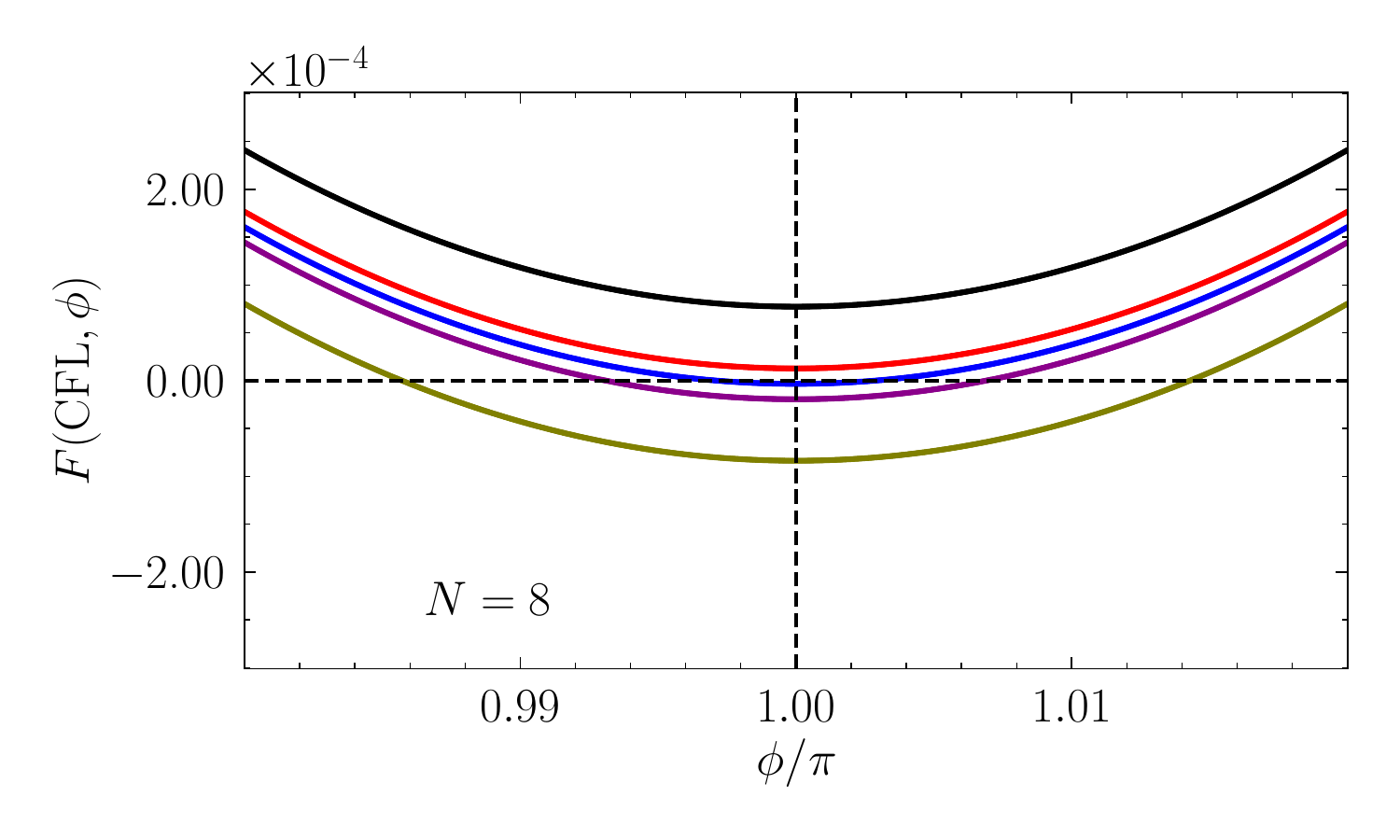}\\[-2mm]
\includegraphics[width=0.24\textwidth]{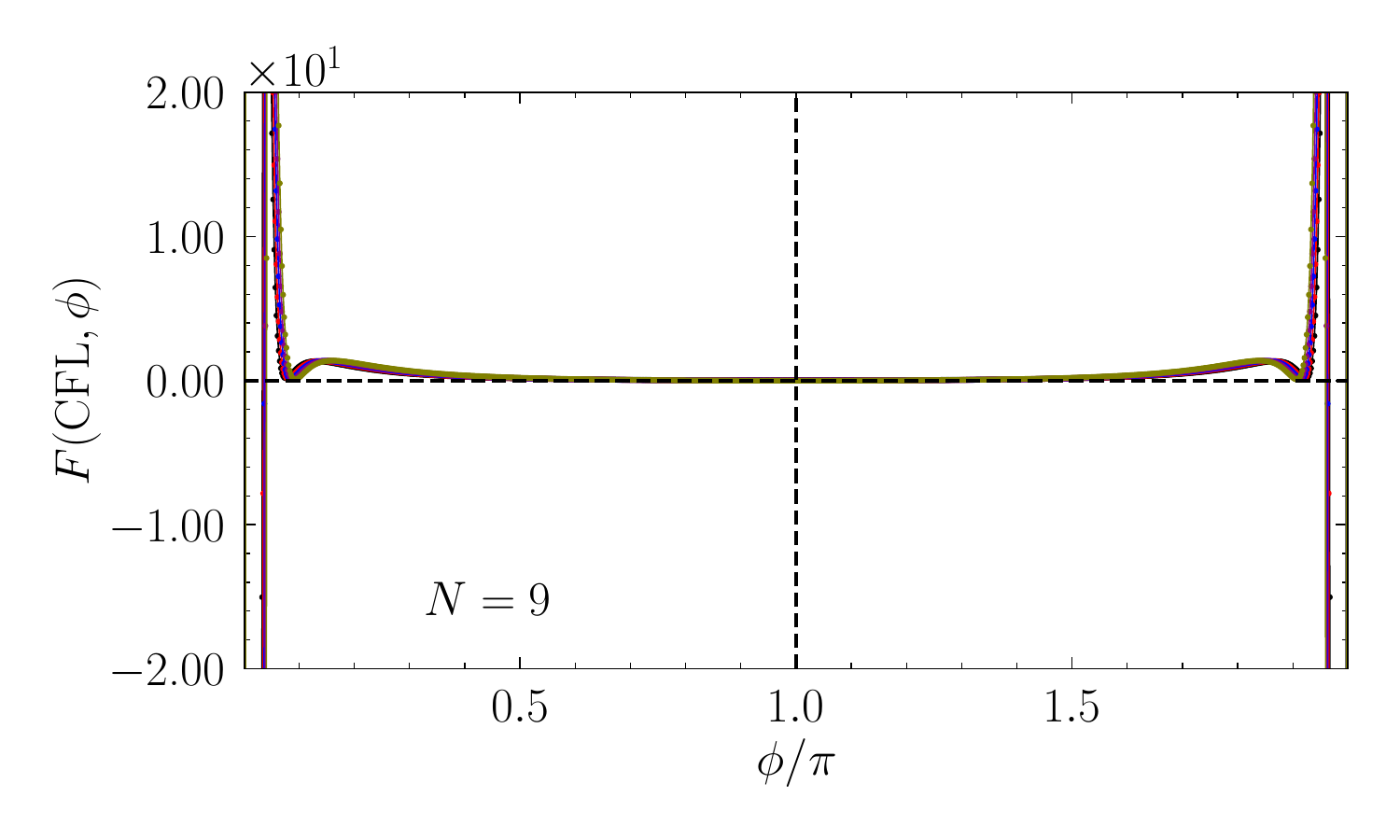}
\includegraphics[width=0.24\textwidth]{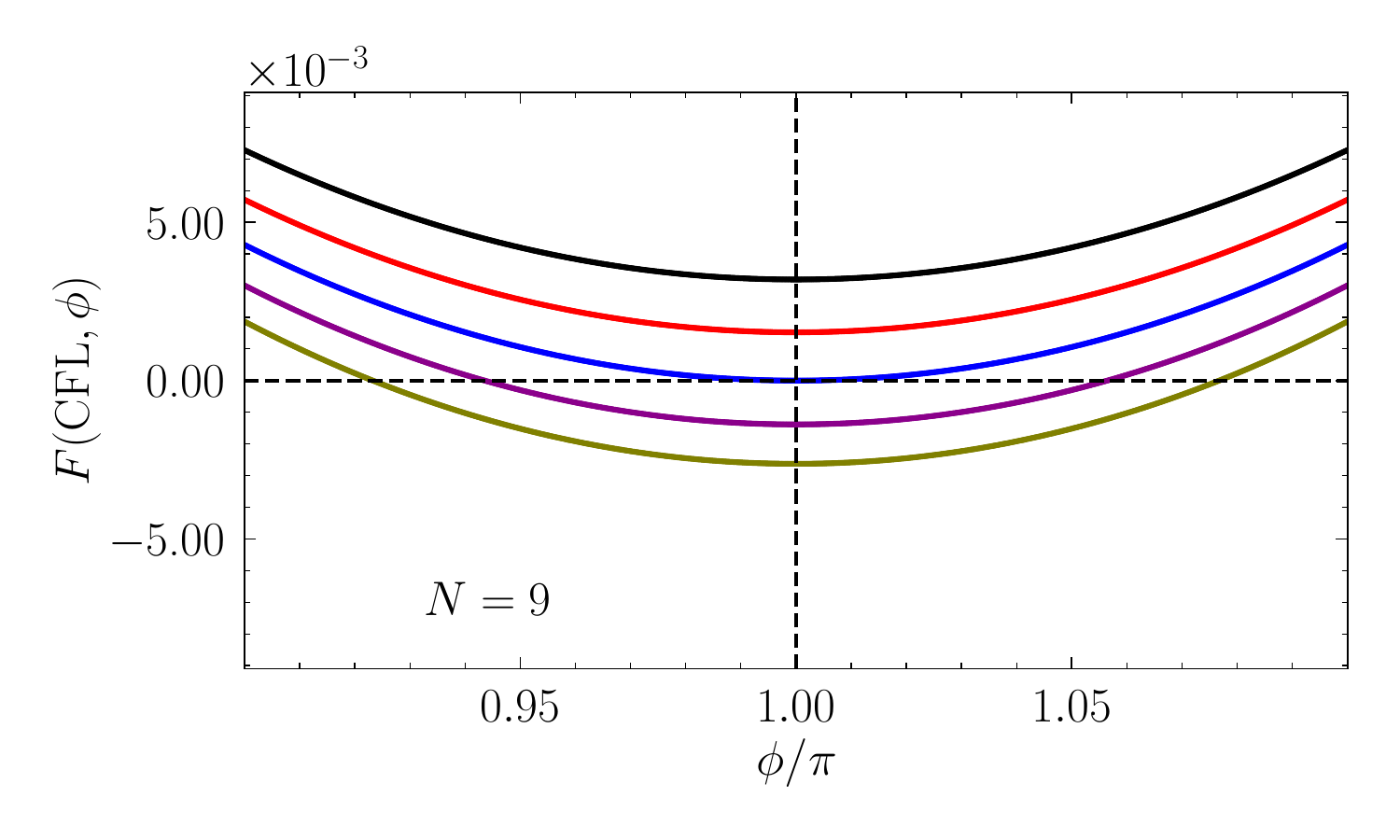}
\includegraphics[width=0.24\textwidth]{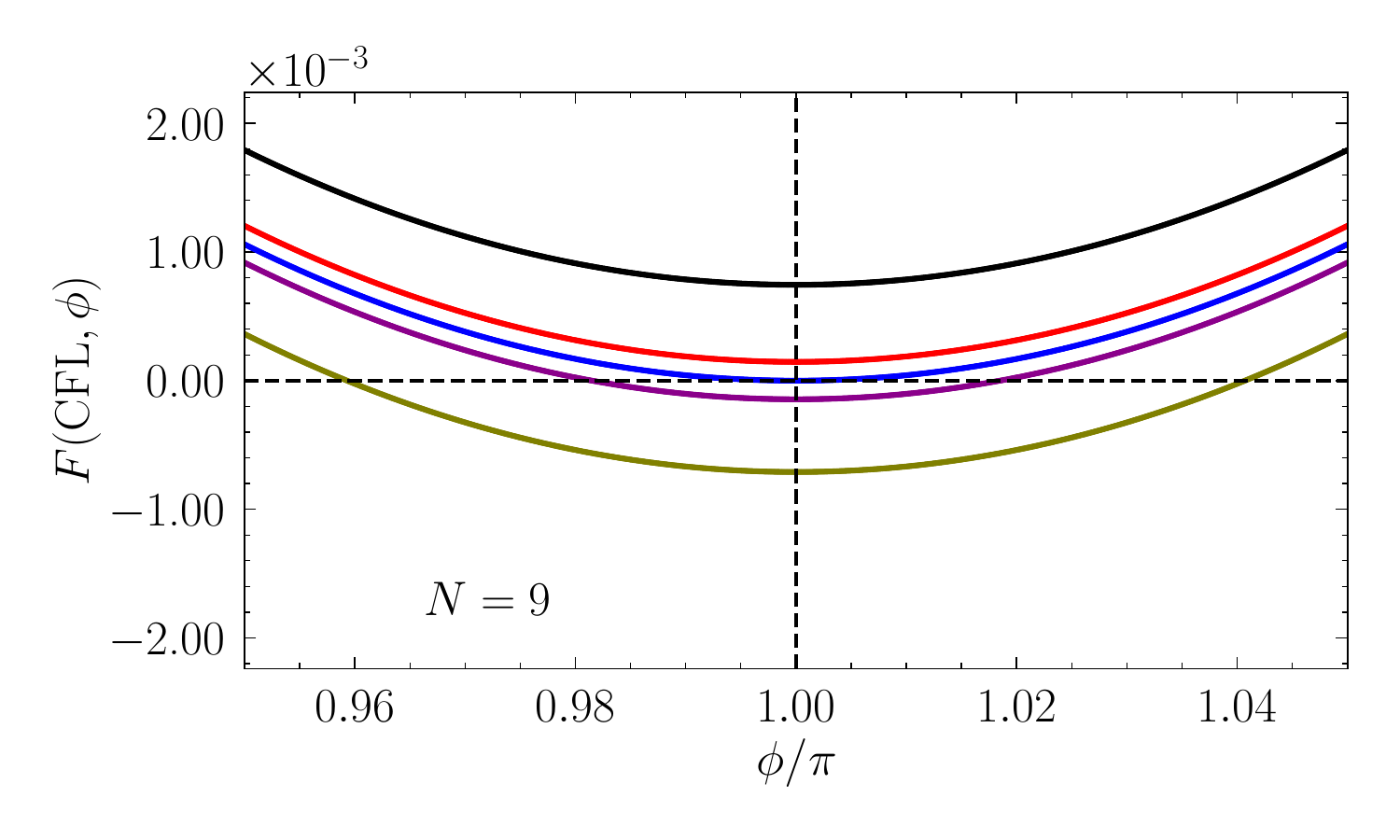}
\includegraphics[width=0.24\textwidth]{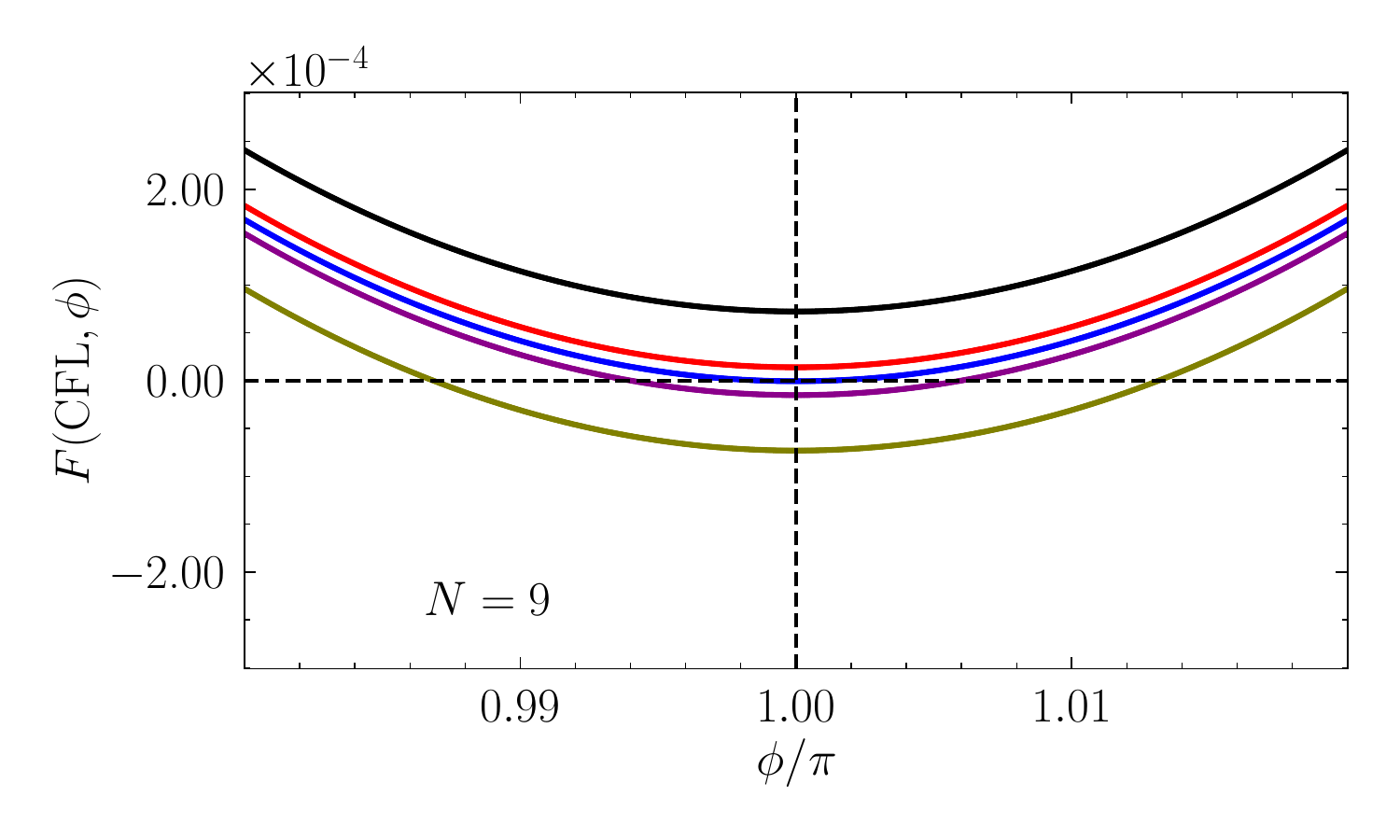}\\[-2mm]
\includegraphics[width=0.24\textwidth]{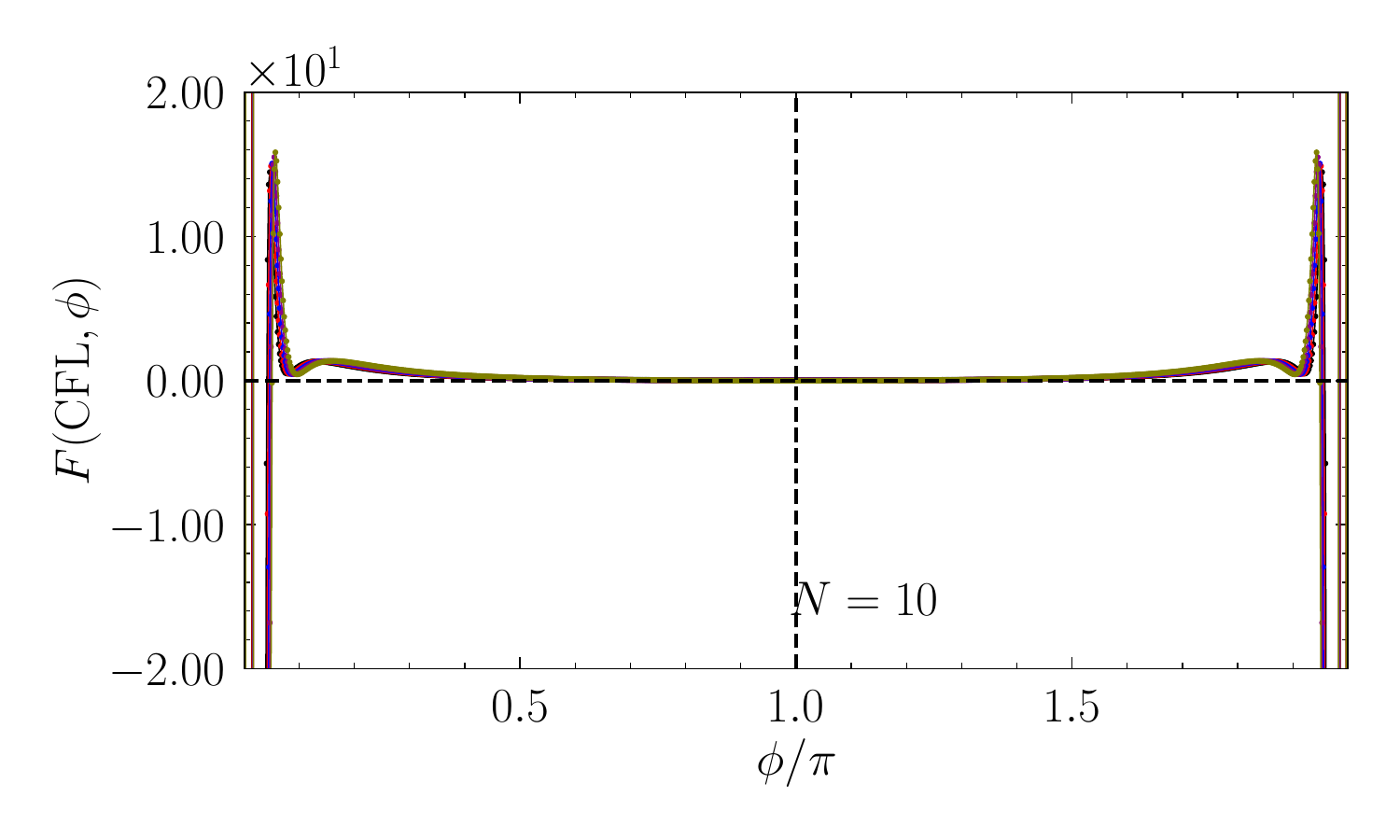}
\includegraphics[width=0.24\textwidth]{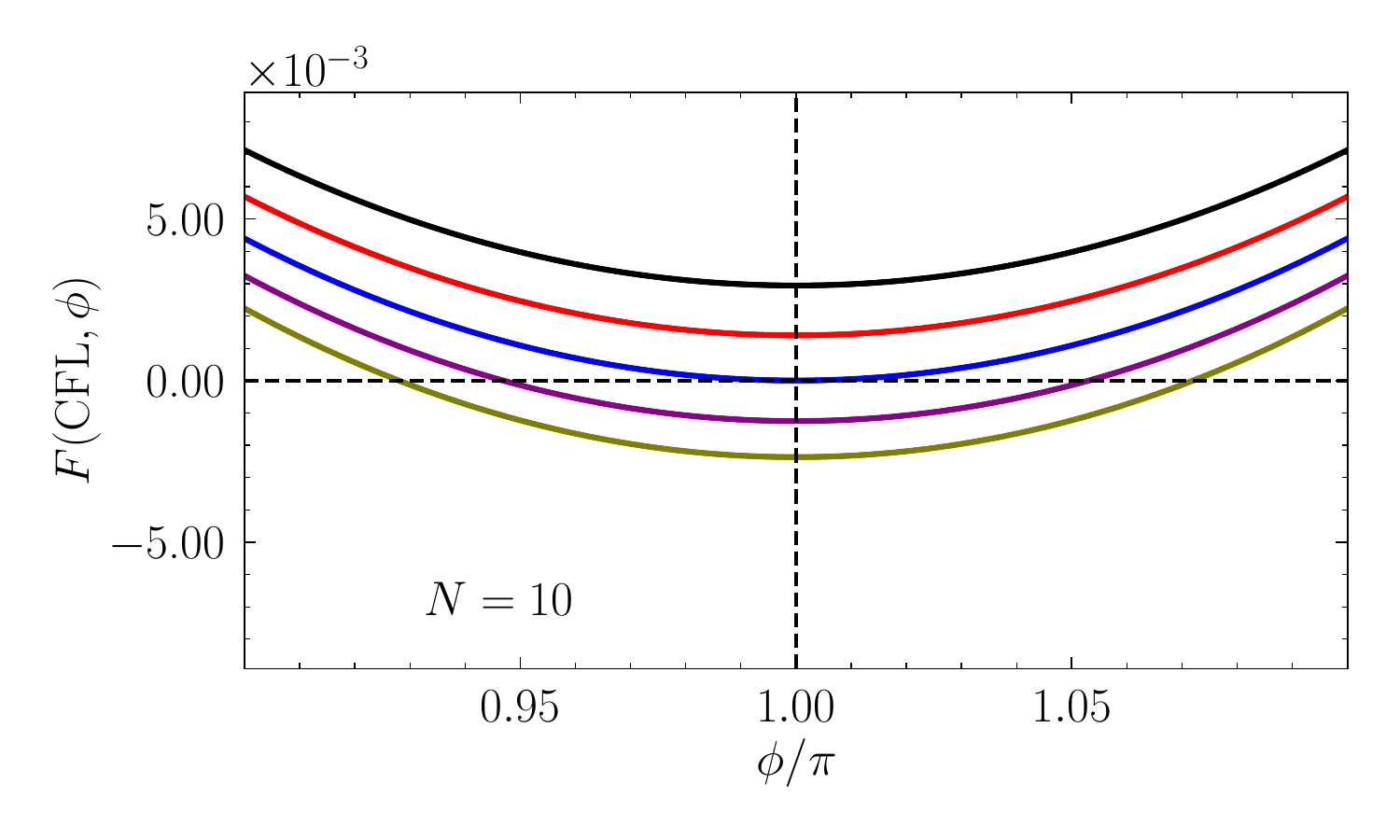}
\includegraphics[width=0.24\textwidth]{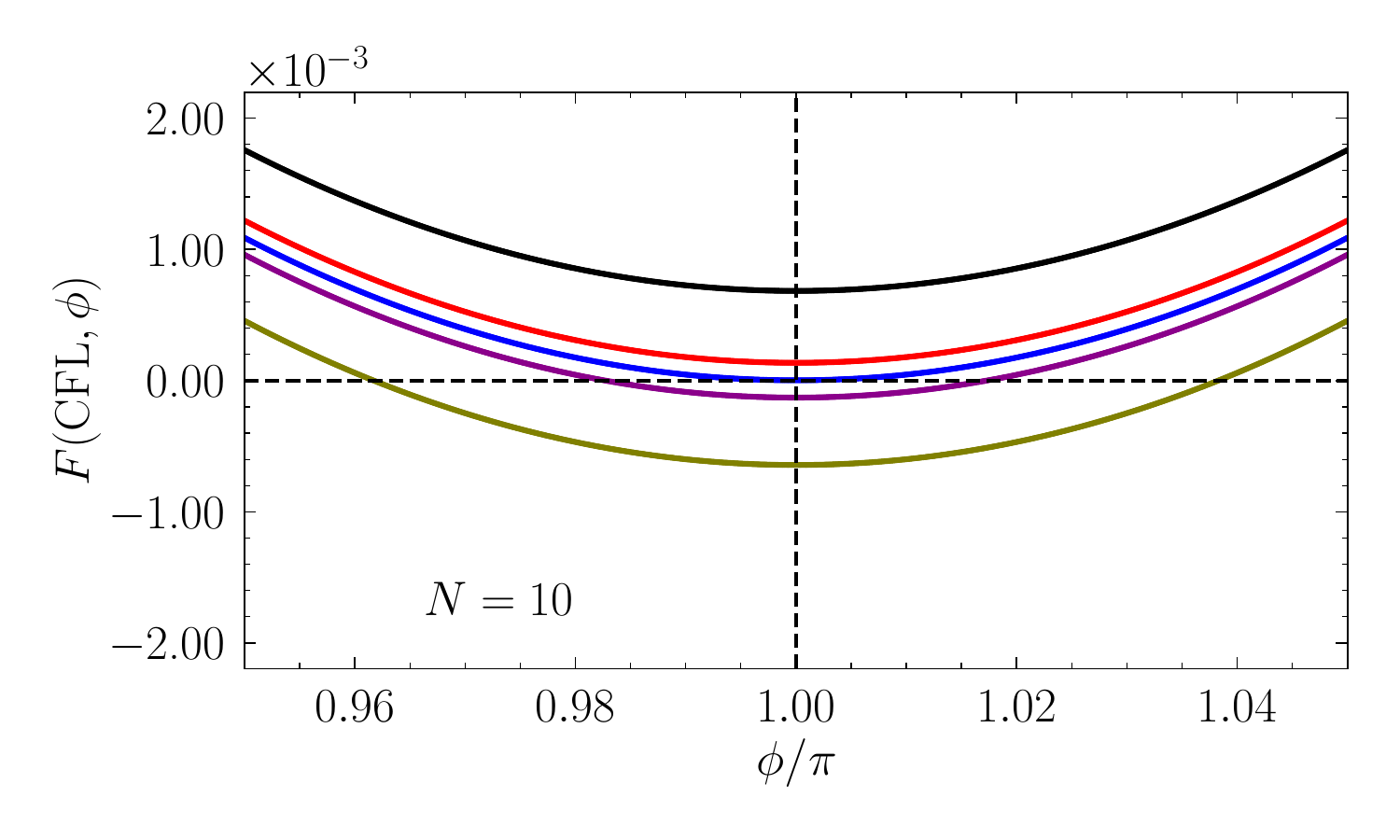}
\includegraphics[width=0.24\textwidth]{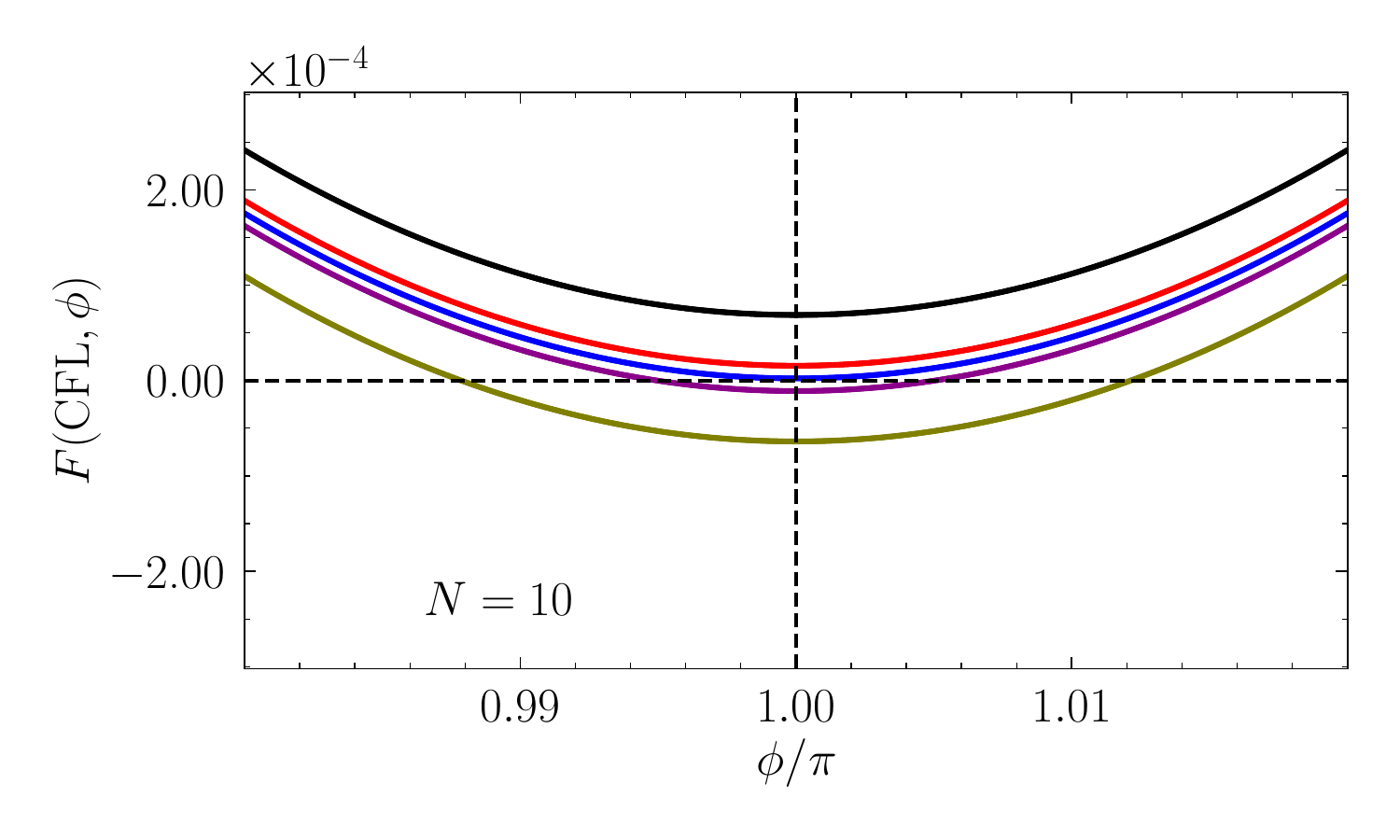}\\[-2mm]
\includegraphics[width=0.24\textwidth]{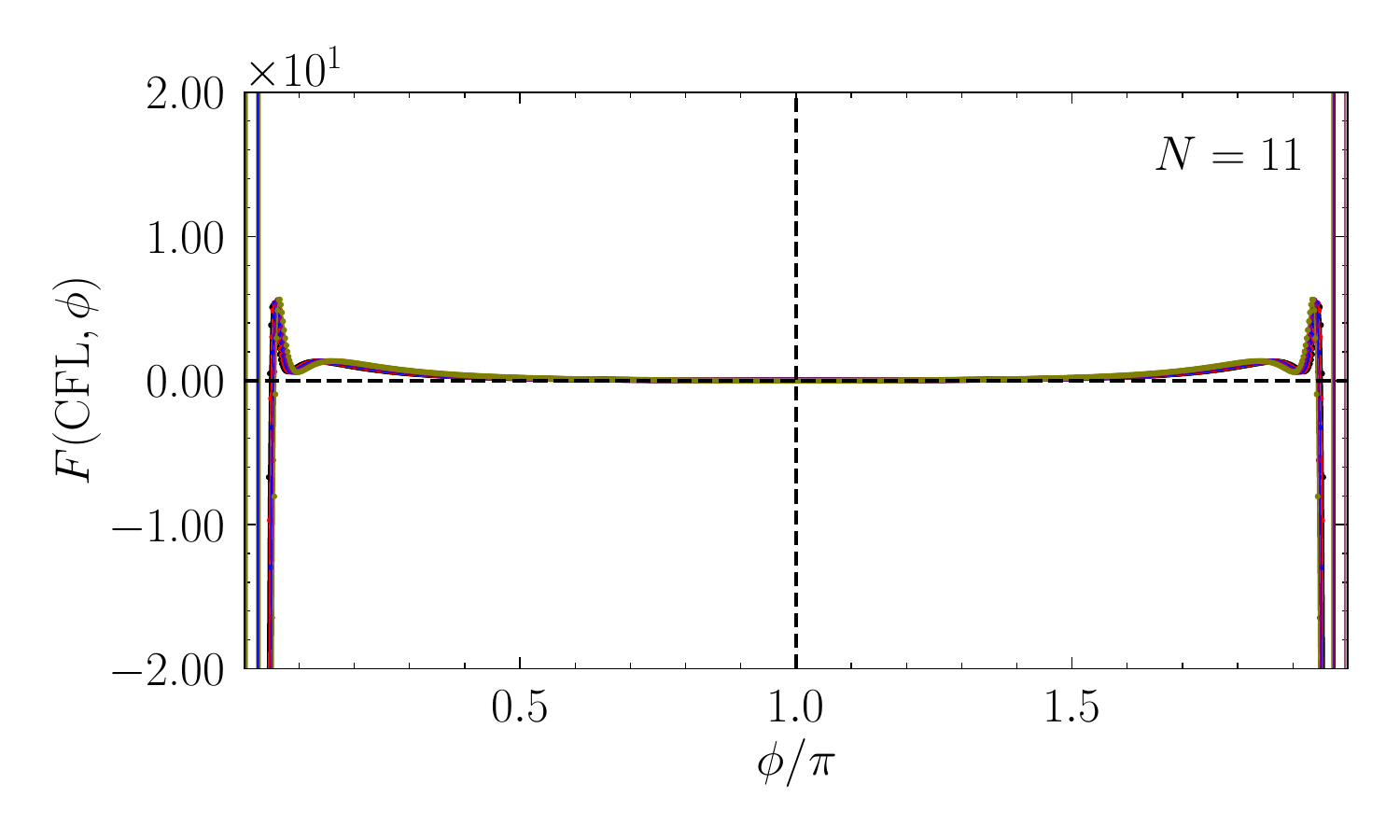}
\includegraphics[width=0.24\textwidth]{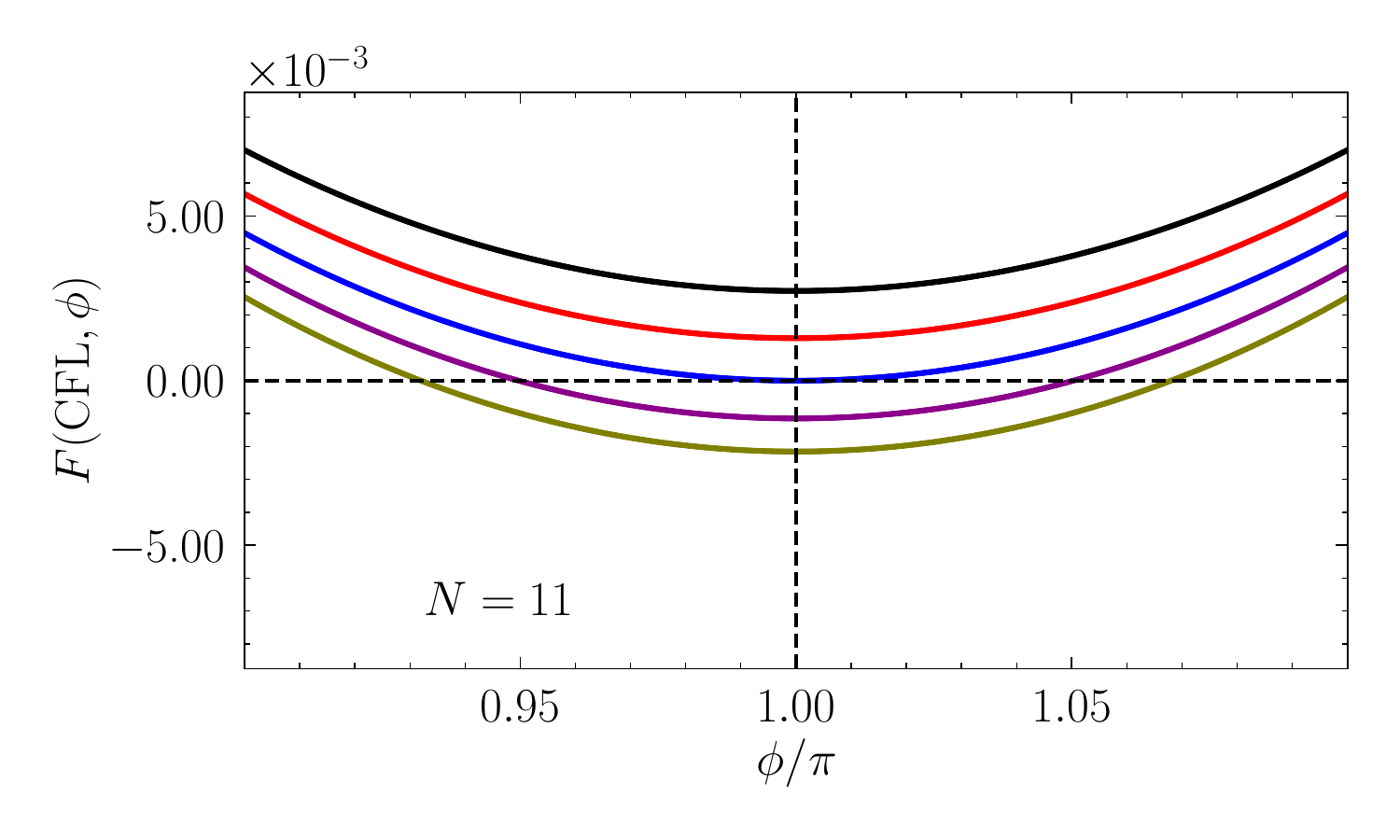}
\includegraphics[width=0.24\textwidth]{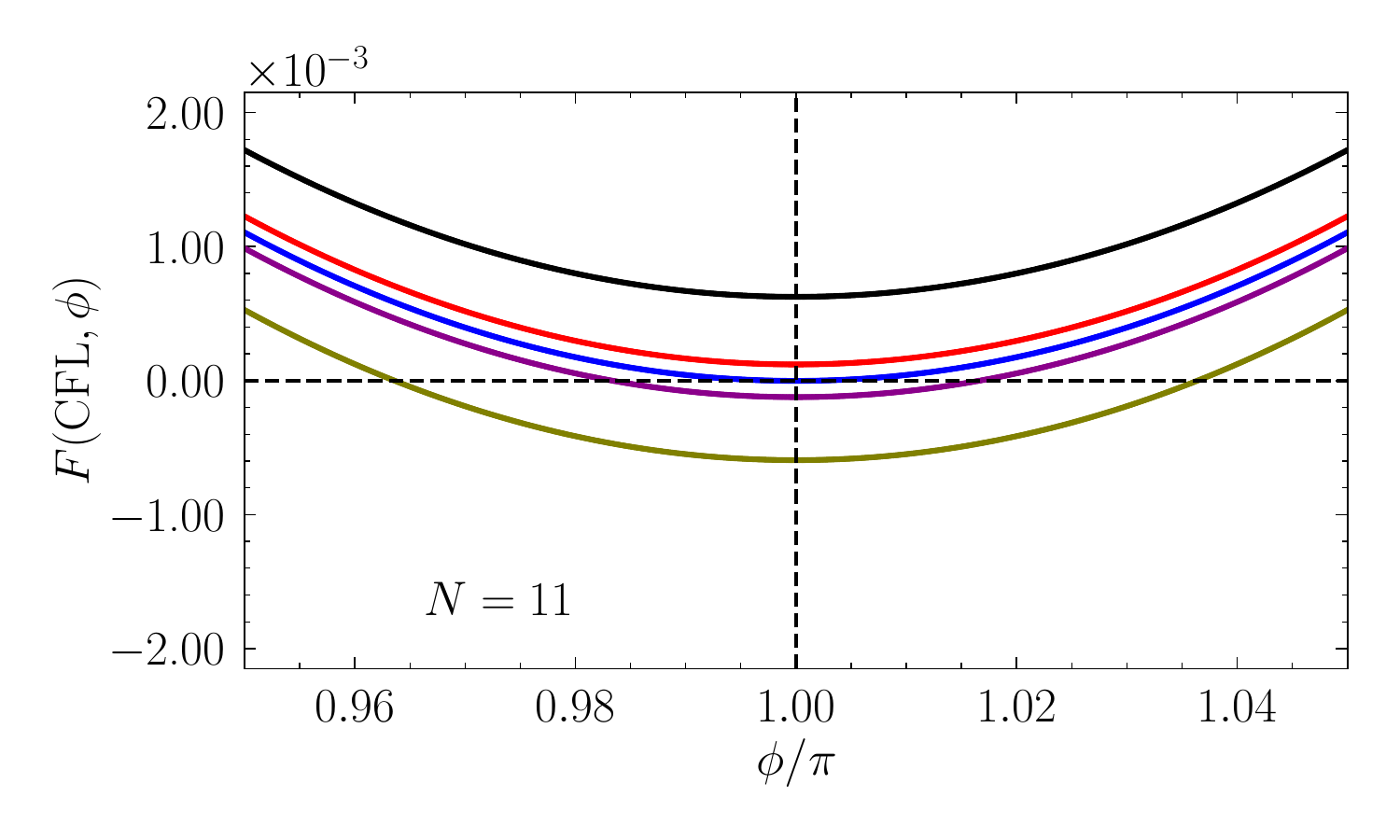}
\includegraphics[width=0.24\textwidth]{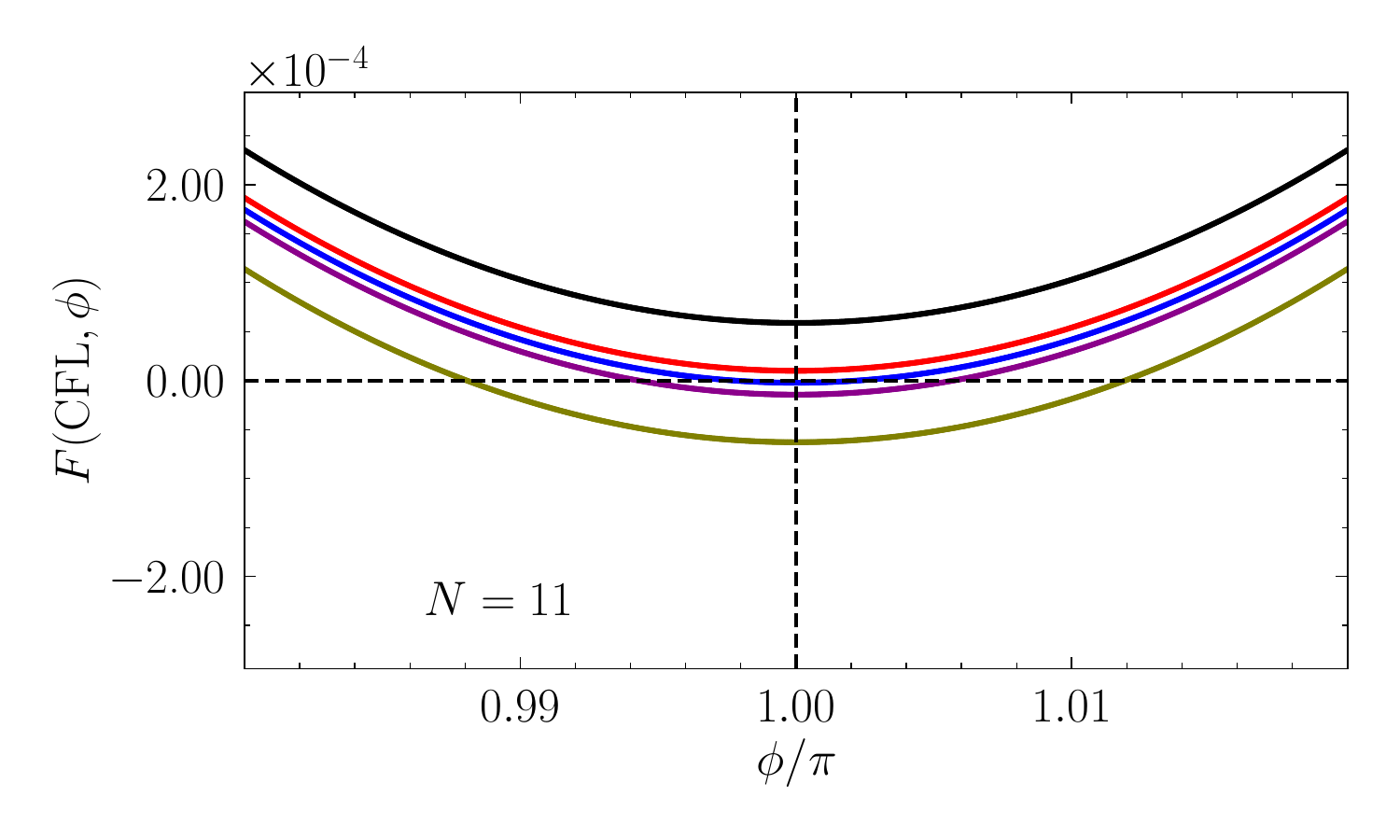}\\[-2mm]
\includegraphics[width=0.24\textwidth]{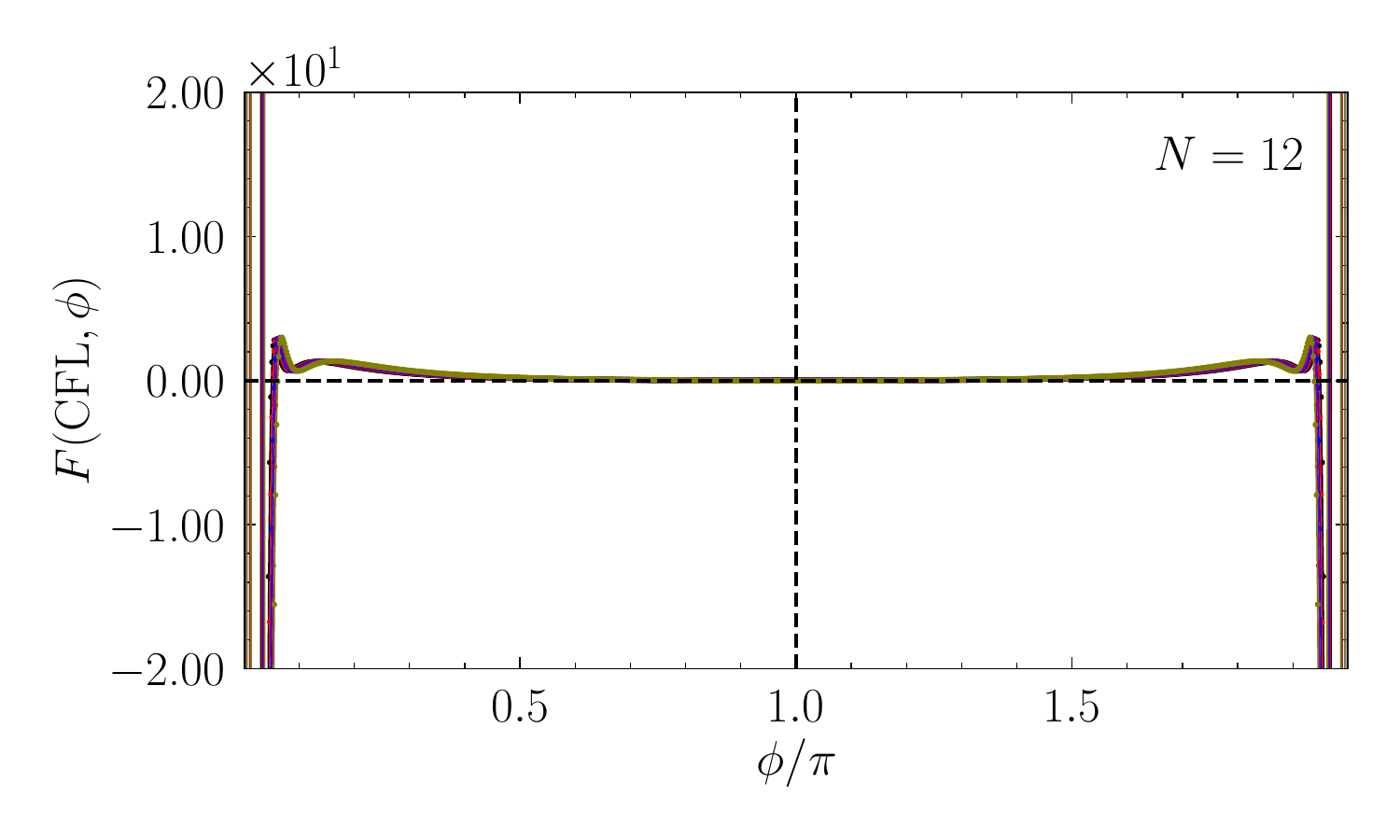}
\includegraphics[width=0.24\textwidth]{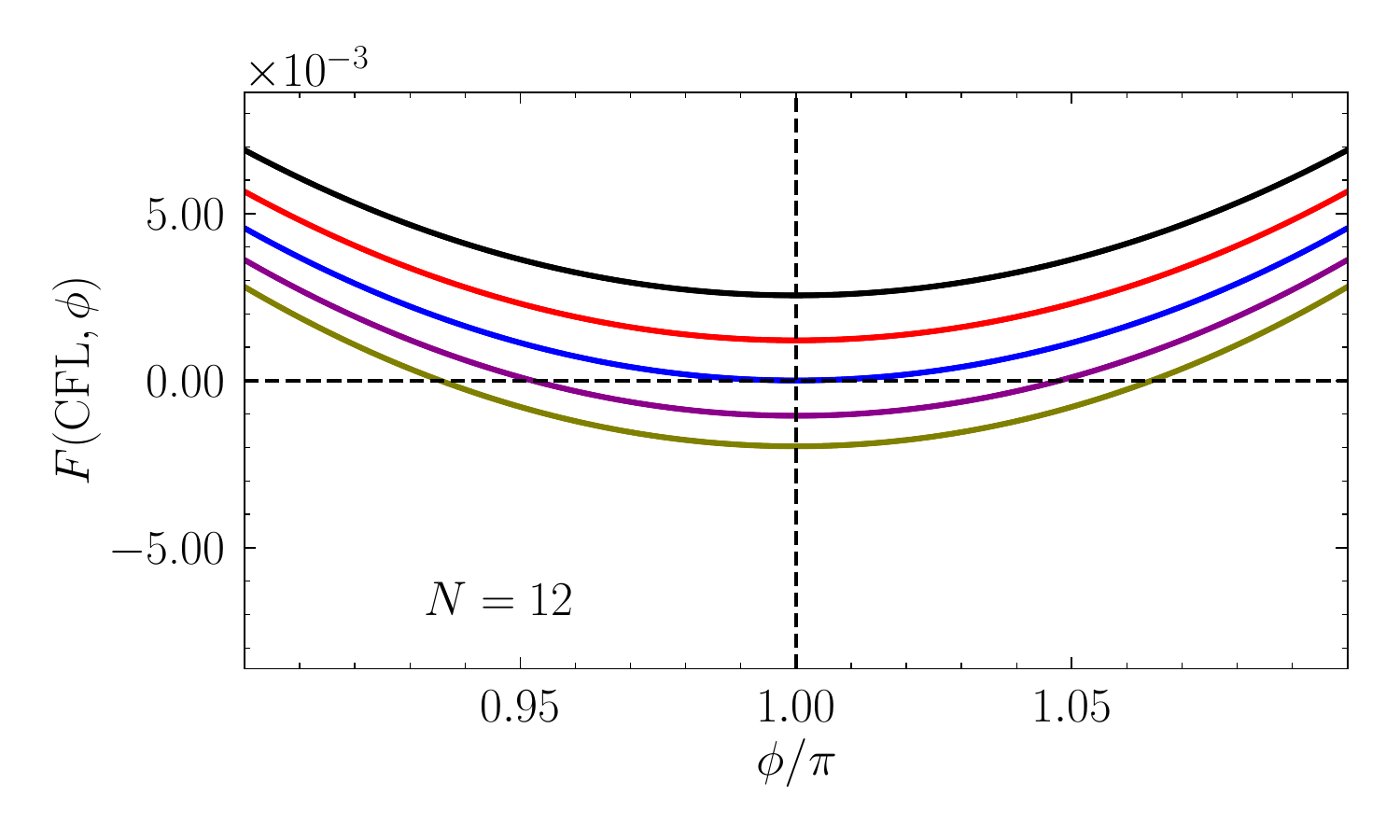}
\includegraphics[width=0.24\textwidth]{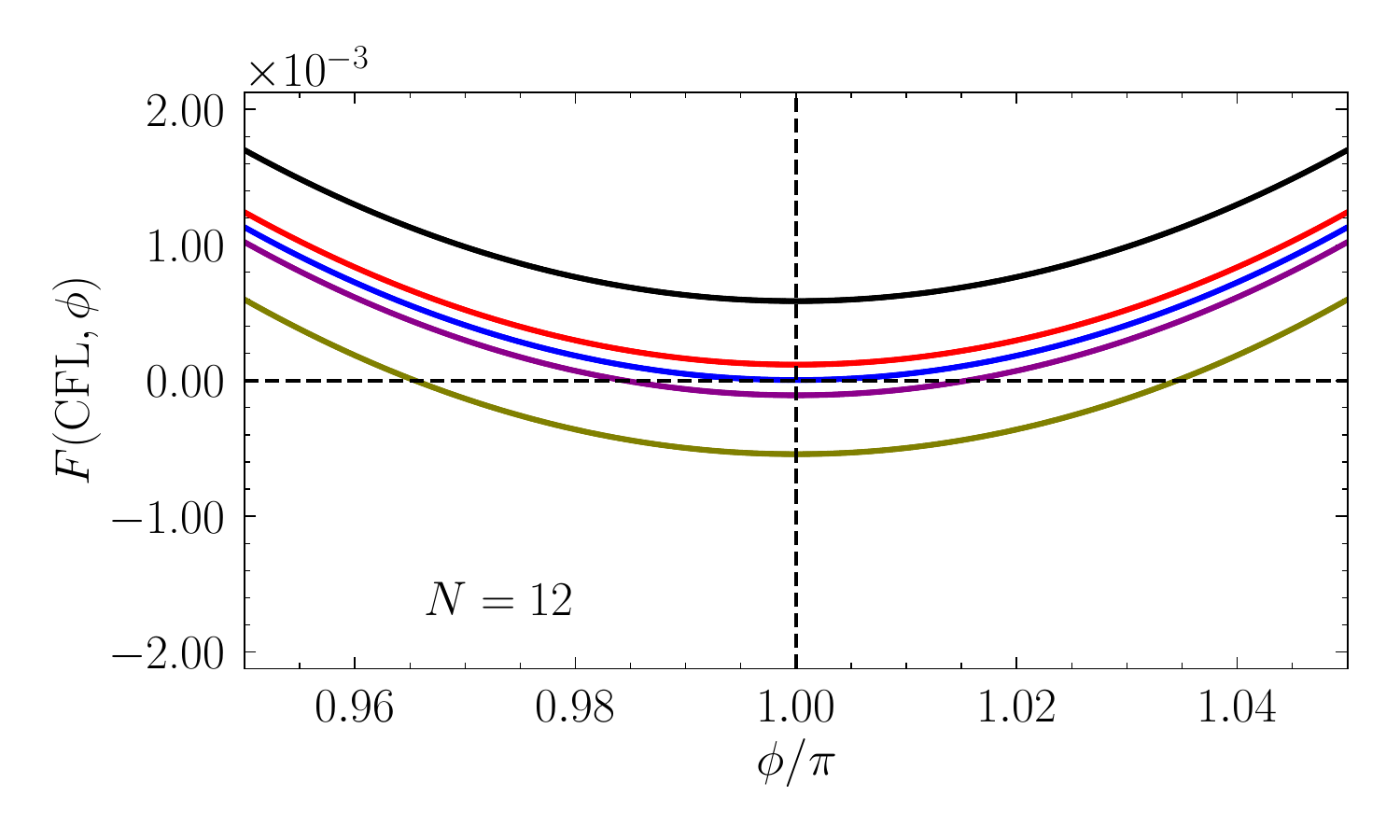}
\includegraphics[width=0.24\textwidth]{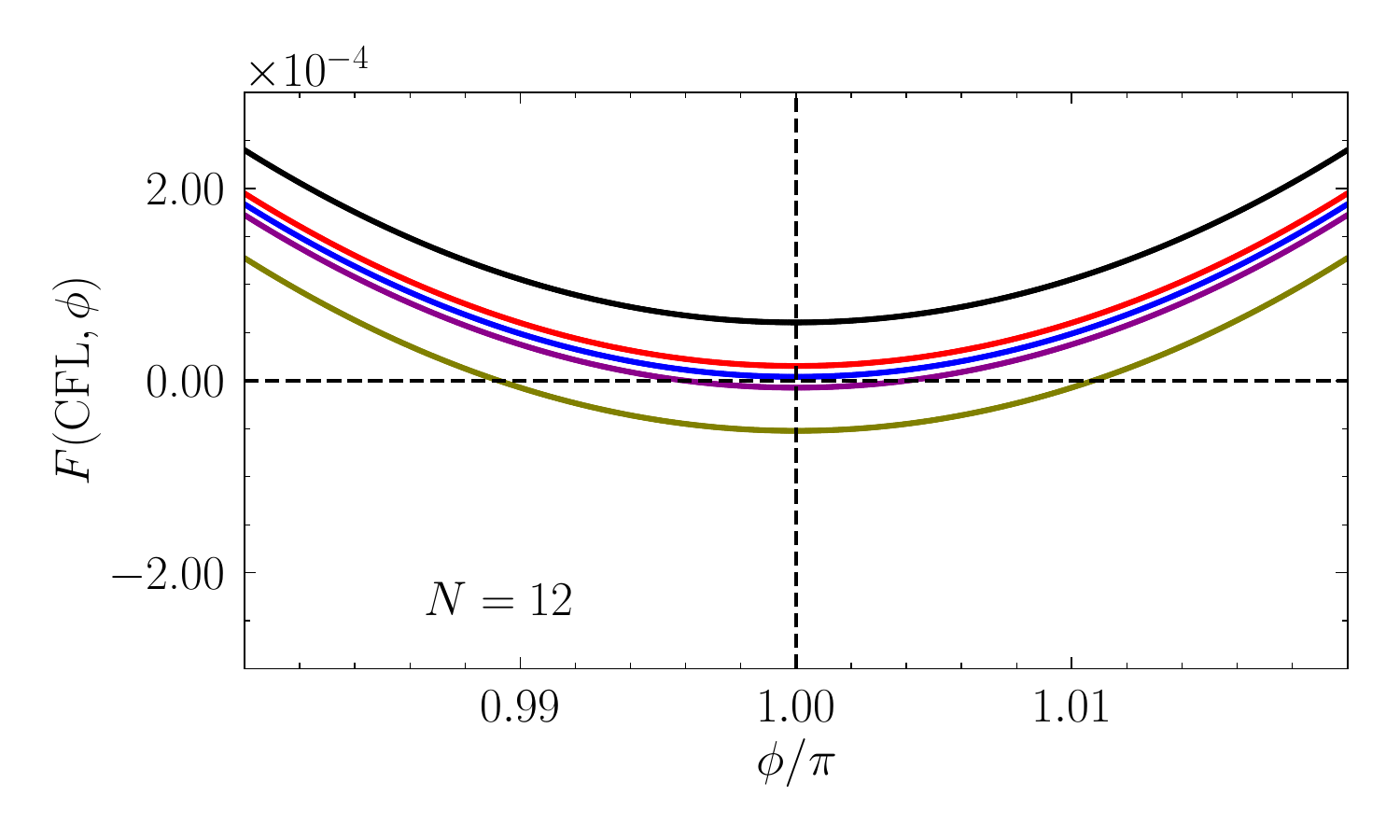}
\caption{%
The dependence of function $F(\mathrm{CFL}, \phi)$ (\ref{eq:f_def_plus}) on phase $\phi\in[0, 2\pi]$ for several values of the Courant number $\mathrm{CFL}$, selected in the vicinity of the stability boundary $\mathrm{CFL}_{\rm max}$, defined further in Table~\ref{tab:cfls_max_data} and in Figure~\ref{fig:cfls_max_data}, for polynomial degrees $N = 7, \ldots, 12$. The graph legends are distributed among columns and located at the top of the columns. From left to right, the graphs represent increasingly narrower neighborhoods of the stability boundary $\mathrm{CFL}_{\rm max}$, from $\mathrm{CFL} \in [0.90\, \mathrm{CFL}_{\rm max},\, 1.10\, \mathrm{CFL}_{\rm max}]$ in the left column to $\mathrm{CFL} \in [0.9995\, \mathrm{CFL}_{\rm max},\, 1.0005\, \mathrm{CFL}_{\rm max}]$ in the right column; the range of phase $\phi$ variation also narrows when moving from the left to the right columns of the graphs to ensure a correct and clear presentation of the root $\phi = \pi$ of the equation $F(\mathrm{CFL}, \pi) = 0$. Additional intersections of the dependencies of functions $F(\mathrm{CFL}, \phi)$ (\ref{eq:f_def_plus}) on phase $\phi$ are observed in the vicinity of points $0$ and $2\pi$, which corresponds to the ``false'' instability violations described in the main text. The horizontal dotted line indicates $F = 0$, the vertical dotted line indicates $\phi = \pi$ ($\lambda = \exp(\pm i\phi) = -1$). \textit{Note}: the phase $\phi$ of eigenvalue $\lambda_{k} = |\lambda_{k}|\exp(i\phi)$ is not the phase $\theta$.
}
\label{fig:dep_f_on_phi_diff_cfls_degrees_7_12}
\end{figure}

Before analyzing the resulting equations (\ref{eq:h_abs_equation_plus_def}) in case $a > 0$ or (\ref{eq:h_abs_equation_minus_def}) in case $a < 0$ in the vicinity of the point $\lambda = -1$ where instability is expected, it is necessary to study the behavior of the equations (\ref{eq:h_abs_equation_plus_def}) and (\ref{eq:h_abs_equation_minus_def}) for arbitrary $\lambda$ on the unit circle $|\lambda| = 1$. For this purpose, dependencies of the following function
\begin{equation}\label{eq:f_def_plus}
F(\mathrm{CFL}, \theta) = |H^{+}(0;\, \mathrm{CFL}, \lambda) - 1|^{2} - |H^{+}(1;\, \mathrm{CFL}, \lambda)|^{2},
\end{equation}
as a function of the phase $\theta$, which determines $\lambda = \exp(i\theta)$, are studied. The squares of the absolute values in the function $F(\mathrm{CFL}_{\rm max}, \theta)$ definition (\ref{eq:f_def_plus}) are chosen to reduce computational costs --- to avoid the need to calculate the square roots. From the structure of the polynomials $H^{+}(\xi;\, \mathrm{CFL}, \lambda)$ (\ref{eq:h_plus_def}), $H^{-}(\xi;\, \mathrm{CFL}, \lambda)$ (\ref{eq:h_minus_def}) and one of the properties of the nodes $\tau_{p} = \tau_{N-p}$ it is clear that a different definition
\begin{equation}\label{eq:f_def_minus}
F(\mathrm{CFL}, \theta) = |H^{-}(1;\, \mathrm{CFL}, \lambda) + 1|^{2} - |H^{-}(0;\, \mathrm{CFL}, \lambda)|^{2},
\end{equation}
can also be written for the function $F(\mathrm{CFL}, \theta)$, therefore the results of the study of the behavior of the function $F(\mathrm{CFL}, \theta)$ (\ref{eq:f_def_plus}) for polynomials $H^{+}(\xi;\, \mathrm{CFL}, \lambda)$ (\ref{eq:h_plus_def}) and the equation (\ref{eq:h_abs_equation_plus_def}) are also directly applicable to the analysis of the behavior of the polynomials $H^{-}(\xi;\, \mathrm{CFL}, \lambda)$ (\ref{eq:h_minus_def}) and the equation (\ref{eq:h_abs_equation_minus_def}) (differences arose only before taking the absolute value from the left and right sides of the equations (\ref{eq:h_equation_plus_def}) and (\ref{eq:h_equation_minus_def}), where it was necessary to make a substitution $\exp(+i\theta) \leftrightarrow \exp(-i\theta)$, which is also unimportant for the domain of definition of the phase $\theta\in[0, 2\pi]$).

\begin{wrapfigure}[28]{l}{0.5\textwidth}
\centering
\includegraphics[width=\linewidth]{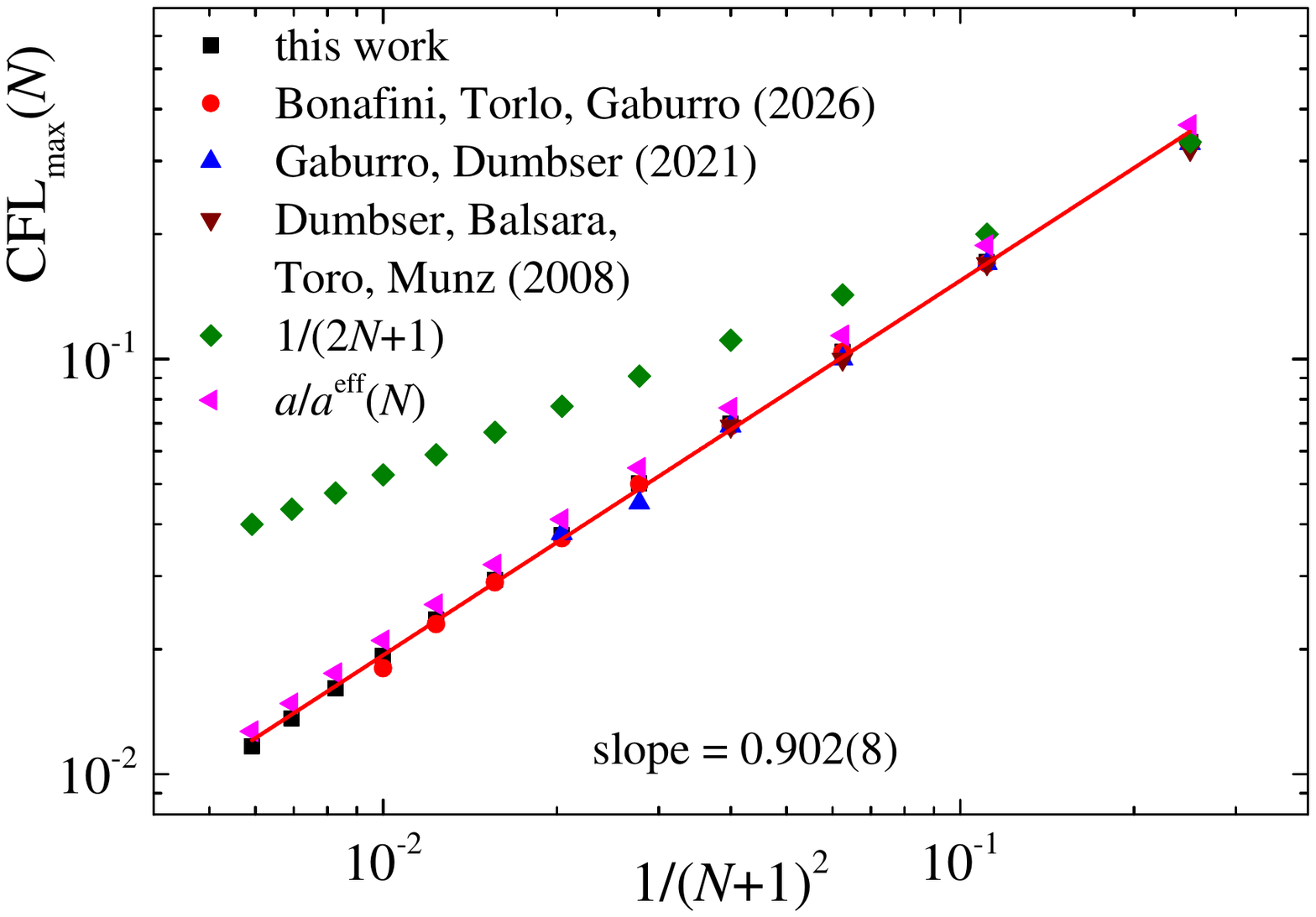}
\caption{\label{fig:cfls_max_data}
Boundary values of the Courant number $\mathrm{CFL}_{\rm max}$ determining the stability boundary of the numerical method ADER-DG with the LST-DG predictor for polynomial degrees $N = 1, \ldots, 12$.
}
\end{wrapfigure}

Figures~\ref{fig:dep_f_on_phi_diff_cfls_degrees_1_6} and~\ref{fig:dep_f_on_phi_diff_cfls_degrees_7_12} show the dependence of function $F(\mathrm{CFL}, \phi)$ (\ref{eq:f_def_plus}) on phase $\phi\in[0, 2\pi]$ for several values of the Courant number $\mathrm{CFL}$, selected in the vicinity of the stability boundary $\mathrm{CFL}_{\rm max}$ (defined further in Table~\ref{tab:cfls_max_data} and in Figure~\ref{fig:cfls_max_data}): Figure~\ref{fig:dep_f_on_phi_diff_cfls_degrees_1_6} shows the results for polynomial degrees $N = 1, \ldots, 6$, and Figure~\ref{fig:dep_f_on_phi_diff_cfls_degrees_7_12} shows the results for polynomial degrees $N = 7, \ldots, 12$. All calculations to obtain these results are performed using a developed program in the \texttt{python} programming language. Floating-point calculations are performed using the \texttt{mpmath} module (with \texttt{gmp2} module), which enables calculations with arbitrarily high precision representation of real numbers as floating-point numbers. The presented results are calculated using \texttt{mpmath.mp.dps = 1000}. The phase $\phi$, which determines the eigenvalue $\lambda_{k} = |\lambda_{k}|\exp(i\phi)$, is discretized using a uniform grid with $1000$ nodes on each graph (taking into account the narrowing of the phase range). It is important to note the following: the phase $\phi$ is not the phase $\theta$. It is also important to note one important feature of the dependence of function $F(\mathrm{CFL}, \phi)$ (\ref{eq:f_def_plus}) on phase $\phi$: at points $\phi = 0$ and $\phi = 2\pi$, the function diverges --- $F(\mathrm{CFL}, 0)\rightarrow\pm\infty$ and $F(\mathrm{CFL}, 2\pi)\rightarrow\pm\infty$. However, this is consistent with the constant-preserving property (\ref{eq:cont_presiv_prop}). Direct construction of the function $F(\mathrm{CFL}, \phi)$ for polynomial degrees $N = 1$, $2$ and $3$ shows that the divergence is related to the denominator $[1 - \cos(\phi)]$, while for higher degrees $N \geqslant 4$, the divergence structure is more complex. However, sequential calculation of all divergent contributions of types $\phi^{-s}$, $s\in\mathcal{Z}_{+}$, to the final expression shows that all divergences in $F(\mathrm{CFL}, \phi)$ cancel out, but this cannot be calculated numerically due to division by zero. The obtained results clearly demonstrate that instability occurs precisely in the vicinity of point $\lambda = -1$, which corresponds to phase $\phi = \pi$ in Figures~\ref{fig:dep_f_on_phi_diff_cfls_degrees_1_6} and~\ref{fig:dep_f_on_phi_diff_cfls_degrees_7_12}. For polynomial degrees $N \geqslant 4$, additional intersections of the dependencies of functions $F(\mathrm{CFL}, \phi)$ (\ref{eq:f_def_plus}) on phase $\phi$ are observed in the vicinity of points $0$ and $2\pi$, which corresponds to the above-described phenomenon of ``false'' violation of the stability condition.

These results clearly confirm the empirical result stated above: it is necessary to search a violation of the stability criterion precisely in the region of $\mathrm{CFL} \in [0, 1]$ where one of the eigenvalues $\lambda_{k}$ of the matrix $\mathrm{R}(\mathrm{CFL}, \theta)$ (\ref{eq:r_matrix_in_matrix_form}) of the evolution operator $R$ reaches point $\lambda = -1$. The equations (\ref{eq:h_abs_equation_plus_def}) in the case $a > 0$ and (\ref{eq:h_abs_equation_minus_def}) in the case $a < 0$ will be further considered in the following form:
\begin{equation}\label{eq:h_abs_equations_in_m1}
|H^{+}(1;\, \mathrm{CFL}, -1)| = |H^{+}(0;\, \mathrm{CFL}, -1) - 1|,\quad
|H^{-}(0;\, \mathrm{CFL}, -1)| = |H^{-}(1;\, \mathrm{CFL}, -1) + 1|.
\end{equation}
These equations could already be used to calculate the maximum value of the Courant number $\mathrm{CFL}_{\rm max}$; however, equations (\ref{eq:h_abs_equations_in_m1}) are written in a rather complex and essentially native (relative to the potential for generalization) form. 
By substituting value $\lambda = -1$ and expression $\lambda I - \exp(-\mathrm{CFL}\cdot\mathrm{D}) = -[I + \exp(-\mathrm{CFL}\cdot\mathrm{D})]$ into the expression (\ref{eq:h_plus_def}) for $H^{+}(\xi;\, \mathrm{CFL}, \lambda)$ and the expression (\ref{eq:h_plus_coeffs_def}) for $\mathbf{h}^{+}(\, \mathrm{CFL}, \lambda)$, it is obtained the following expressions.
\begin{equation}\label{eq:h_plus_def_final}
\begin{split}
H^{+}(0;\, \mathrm{CFL}, -1) = -\big[\mathrm{B}(\mathrm{CFL})(I + \exp(-\mathrm{CFL}\cdot\mathrm{D}))^{-1}m^{-1}\uvecphi\big]^{T}\uvecphi,\\
H^{+}(1;\, \mathrm{CFL}, -1) = -\big[\mathrm{B}(\mathrm{CFL})(I + \exp(-\mathrm{CFL}\cdot\mathrm{D}))^{-1}m^{-1}\uvecphi\big]^{T}\ovecphi,\\
\end{split}
\end{equation}
for the expressions $H^{+}(0;\, \mathrm{CFL}, -1)$ and $H^{+}(1;\, \mathrm{CFL}, -1)$ included in the equations (\ref{eq:h_abs_equations_in_m1}). Direct calculation of the sum:
\begin{equation}
\mathrm{D}\mathrm{B}(\mathrm{CFL}) = -\sum\limits_{s = 0}^{N}(-1)^{s+1}\frac{\mathrm{CFL}^{s+1}}{(s+1)!}\, \mathrm{D}^{s+1} =
I - \sum\limits_{s = 0}^{N}\frac{(-\mathrm{CFL})^{s}}{s!}\, \mathrm{D}^{s} =
I - \exp\left(-\mathrm{CFL}\cdot\mathrm{D}\right),
\end{equation}
allowed us to conveniently express the matrix $\mathrm{B}(\mathrm{CFL})$ in terms of the differentiation matrix $\mathrm{D}$:
\begin{equation}\label{eq:b_by_exp_of_d}
\mathrm{B}(\mathrm{CFL}) = \mathrm{D}^{-1}\Big[I - \exp\left(-\mathrm{CFL}\cdot\mathrm{D}\right)\Big] = \Big[I - \exp\left(-\mathrm{CFL}\cdot\mathrm{D}\right)\Big]\mathrm{D}^{-1},
\end{equation}
where the inverse matrix $\mathrm{D}^{-1}$ cannot be calculated directly --- the differentiation matrix $\mathrm{D}$ is nilpotent. However, it can be used for intermediate calculations in which $\mathrm{D}^{-1}$ must be reduced. Technically, this term $\mathrm{D}^{N+1}$ is not present in the sum at all, since it is $\mathrm{D}^{N+1} = 0$. However, this representation of $\mathrm{B}(\mathrm{CFL})$ is necessary for the formal derivation, and it is clarified below.

To clarify the meaning of the resulting expression with the inverse matrix $\mathrm{D}^{-1}$, it is necessary to consider the effect of the matrix $\mathrm{B}(\mathrm{CFL})$ on the vector of expansion coefficients $\mathbf{f} = \{f_{s}\}_{s}$, where $f_{s} = f(\xi_{s})$, of any function $f\in\mathcal{P}_{N}(\omega)$:
\begin{equation}\label{eq:invert_b_semantic}
\begin{split}
[\mathrm{B}(\mathrm{CFL}) \mathbf{f}](\xi) & = \sum\limits_{s = 0}^{N} \frac{(-1)^{s}\mathrm{CFL}^{s+1}}{(s+1)!} [\mathrm{D}^{s}\mathbf{f}](\xi) =
\sum\limits_{s = 0}^{N} \frac{(-1)^{s}\mathrm{CFL}^{s+1}}{(s+1)!} f^{(s)}(\xi) =
\sum\limits_{s = 0}^{N} \frac{f^{(s)}(\xi)}{s!} \frac{(-1)^{s}\mathrm{CFL}^{s+1}}{s+1}\\ & =
\sum\limits_{s = 0}^{N} \frac{f^{(s)}(\xi)}{s!} (-1)^{s} \int\limits_{0}^{\mathrm{CFL}}d\varsigma\, \varsigma^{s} =
\int\limits_{0}^{\mathrm{CFL}}d\varsigma\, \left[\sum\limits_{s = 0}^{N} \frac{f^{(s)}(\xi)}{s!} (-\varsigma)^{s}\right] =
\int\limits_{0}^{\mathrm{CFL}}d\varsigma\, f(\xi - \varsigma) = \int\limits_{\xi-\mathrm{CFL}}^{\xi}d\varsigma\, f(\varsigma),
\end{split}
\end{equation}
where the notation $[\mathbf{f}](\xi) \equiv f(\xi)$ is introduced. The semantics of the operator represented by the matrix $\exp(-\mathrm{CFL}\cdot\mathrm{D})$ is similarly clarified:
\begin{equation}
\begin{split}
[\exp(-\mathrm{CFL}\cdot\mathrm{D}) \mathbf{f}](\xi) & =
\sum\limits_{p = 0}^{N}\left[\sum\limits_{s = 0}^{N}\frac{(-\mathrm{CFL})^{s}}{s!}\, ([\mathrm{D}^{s}]\mathbf{f})\right]\varphi_{p}(\xi) =
\sum\limits_{p = 0}^{N}\left[\sum\limits_{s = 0}^{N}\frac{(-\mathrm{CFL})^{s}}{s!}\, \sum\limits_{q = 0}^{N}[\mathrm{D}^{s}]_{pq}f(\xi_{q})\right]\varphi_{p}(\xi)\\ & =
\sum\limits_{p = 0}^{N}\left[\sum\limits_{s = 0}^{N}\frac{(-\mathrm{CFL})^{s}}{s!}\, f^{(s)}(\xi_{p})\right]\varphi_{p}(\xi)= 
\sum\limits_{s = 0}^{N}\frac{(-\mathrm{CFL})^{s}}{s!}\, \left[\sum\limits_{p = 0}^{N}f^{(s)}(\xi_{p})\varphi_{p}(\xi)\right]\\ & =
\sum\limits_{s = 0}^{N}\frac{(-\mathrm{CFL})^{s}}{s!}\, f^{(s)}(\xi) = f(\xi-\mathrm{CFL}) \in \mathcal{P}_{N}([-1, 1] \supset \omega),
\end{split}
\end{equation}
where it is clear that this operator is a shift operator $T_{-\mathrm{CFL}}$: $T_{-\mathrm{CFL}} f(\xi) = f(\xi-\mathrm{CFL}) \in \mathcal{P}_{N}([-1, 1]\supset\omega)$. This expression clarifies the meaning of the operator represented by the matrix $(I - \exp(-\mathrm{CFL}\cdot\mathrm{D}))$, which can be expressed as follows:
\begin{equation}
\big[(I - \exp(-\mathrm{CFL}\cdot\mathrm{D}))\mathbf{f}\big](\xi) = f(\xi) - f(\xi-\mathrm{CFL}),
\end{equation}
which immediately implies the relation:
\begin{equation}
\big[\mathrm{D}^{-1}(I - \exp(-\mathrm{CFL}\cdot\mathrm{D}))\mathbf{f}\big](\xi) = \int\limits_{\xi-\mathrm{CFL}}^{\xi}d\varsigma\, f(\varsigma),
\end{equation}
when compared with the formula (\ref{eq:invert_b_semantic}). It is concluded that the matrix $\mathrm{D}^{-1}$ formally if the representation matrix of the integration operator. The difficulty in defining the matrix $\mathrm{D}^{-1}$ of a simple integration operator for any functions $f\in\mathcal{P}_{N}(\omega)$ arose because the integrals of functions in $\mathcal{P}_{N}(\omega)$ are in the space $\mathcal{P}_{N+1}(\omega)$, and therefore the functional basis $\{\varphi_{p}\}_{p}$ (\ref{eq:phi_def}) chosen in $\mathcal{P}_{N}(\omega)$ is insufficient to accurately represent the integrals of functions in $\mathcal{P}_{N}(\omega)$ (see~\cite{ader_dg_imp_ode} for a detailed discussion of these integration features for ADER-DG ODE solvers, and~\cite{ader_improving_2024, ader_proofs_2025, ader_dg_ivp_ode_sinum}).

Using the expression (\ref{eq:b_by_exp_of_d}), the matrix in the expression (\ref{eq:h_abs_equations_in_m1}) can be represented in the following convenient form:
\begin{equation}
\begin{split}
&\mathrm{B}(\mathrm{CFL})\Big[I + \exp\left(-\mathrm{CFL}\cdot\mathrm{D}\right)\Big]^{-1} =
\Big[I - \exp\left(-\mathrm{CFL}\cdot\mathrm{D}\right)\Big]\mathrm{D}^{-1}\Big[I + \exp(-\mathrm{CFL}\cdot\mathrm{D})\Big]^{-1}\\ & =
\mathrm{D}^{-1}\Big[I - \exp\left(-\mathrm{CFL}\cdot\mathrm{D}\right)\Big]\Big[I + \exp(-\mathrm{CFL}\cdot\mathrm{D})\Big]^{-1}\\ & =
\mathrm{D}^{-1}\left[\exp\left(\frac{\mathrm{CFL}}{2}\,\mathrm{D}\right) - \exp\left(-\frac{\mathrm{CFL}}{2}\,\mathrm{D}\right)\right]
\left[\exp\left(\frac{\mathrm{CFL}}{2}\,\mathrm{D}\right) + \exp\left(-\frac{\mathrm{CFL}}{2}\,\mathrm{D}\right)\right]^{-1} =
\mathrm{D}^{-1}\mathrm{tanh}\left(\frac{\mathrm{CFL}}{2}\,\mathrm{D}\right),
\end{split}
\end{equation}
which completely eliminates native dependencies on the structure of the basis functions $\{\varphi_{p}\}_{p}$ (\ref{eq:phi_def}), except through the differentiation matrix $\mathrm{D}$. Further simplification of the expression (\ref{eq:h_abs_equations_in_m1}) requires identifying all universal functional constructs, which include the delta function $\delta_{\mathcal{P}_{N}}(\xi)$ (\ref{eq:delta_func_varphi_action}) represented by vector $\boldsymbol{\delta}^{(0)} = m^{-1}\uvecphi = \{\uphi_{p}/w_{p}\}_{p}$ (\ref{eq:delta_func_varphi_coeffs}), leading to the following expression: 
\begin{equation}\label{eq:h_plus_by_tanh_and_delta}
\begin{split}
H^{+}(\xi;\, \mathrm{CFL}, -1) & =
-\sum\limits_{p = 0}^{N} \left[\mathrm{D}^{-1} \mathrm{tanh}\left(\frac{\mathrm{CFL}}{2}\,\mathrm{D}\right)\boldsymbol{\delta}^{(0)}\right]_{p} \varphi_{p}(\xi) =
-\sum\limits_{p = 0}^{N} \left[\mathrm{tanh}\left(\frac{\mathrm{CFL}}{2}\,\mathrm{D}\right)\mathrm{D}^{-1}\boldsymbol{\delta}^{(0)}\right]_{p} \varphi_{p}(\xi),
\end{split}
\end{equation}
which explicitly identifies the following interesting function:
\begin{equation}\label{eq:gamma_plus_def}
\Gamma^{+}(\xi) = \Big[\mathrm{D}^{-1}\boldsymbol{\delta}^{(0)}\Big](\xi) =
\pi^{\mathrm{span}(\{\varphi_{p}\}_{p})}_{\mathcal{P}_{N}}\left[\,\int\limits_{\tilde{\xi}^{(0)}}^{\xi} d\xi'\, \delta_{\mathcal{P}_{N}}(\xi')\right],
\end{equation}
where $\pi^{\mathrm{span}(\{\varphi_{p}\}_{p})}_{\mathcal{P}_{N}}$ is the operator of projection of the function from $\mathcal{P}_{N+1}$ onto the space $\mathcal{P}_{N}$, the elements of which are defined by the functional basis $\{\varphi_{p}\}_{p}$ (\ref{eq:phi_def}), and $\tilde{\xi}^{(0)}$ is an arbitrarily chosen coordinate, the value of which does not in any way affect further transformations (due to at least one differentiation of the expression). The resulting expressions (\ref{eq:h_plus_by_tanh_and_delta}) and (\ref{eq:gamma_plus_def}) clearly demonstrate all the above-mentioned features of the application of the inverse differentiation matrix $\mathrm{D}^{-1}$, which cannot be calculated explicitly: it allows for a formally correct description of the universal form of the operator (\ref{eq:h_plus_by_tanh_and_delta}) representing the expression $H^{+}(\xi;\, \mathrm{CFL}, -1)$, and it itself is not present in the final calculations due to cancellation. Substituting expression (\ref{eq:gamma_plus_def}) into (\ref{eq:h_plus_by_tanh_and_delta}) yields the following final expression
\begin{equation}\label{eq:h_plus_by_tanh_and_gamma}
H^{+}(\xi;\, \mathrm{CFL}, -1) = -\mathrm{tanh}\left(\frac{\mathrm{CFL}}{2}\,\frac{d}{d\xi}\right)\Gamma^{+}(\xi),
\end{equation}
where $\mathrm{tanh}(\mathrm{const}\,d/d\xi)$ is a pseudo-differential operator acting on the elements of the space $\mathcal{P}_{N}(\omega)$, which must be calculated in the form of a Taylor series expansion:
\begin{equation}\label{eq:tanh_taylor_exp}
\begin{split}
&\mathrm{tanh}(x) = \sum\limits_{m = 0}^{\infty} t_{2m+1}x^{2m+1},\quad t_{2m+1} = 2^{2m+2}\frac{2^{2m+2} - 1}{(2m+2)!}B_{2m+2},\\
&\mathrm{tanh}(x) = x - \frac{1}{3} x^{3} + \frac{2}{15} x^{5} - \frac{17}{315} x^{7} + \frac{62}{2835} x^{9} - \frac{1382}{155925} x^{11} + \ldots,
\end{split}
\end{equation}
where $B_{2m+2}$ is the even Bernoulli number, and which in this case represents a finite sum (or when pseudo-differential operator in (\ref{eq:h_plus_by_tanh_and_gamma}) acting on eigenfunctions of the operator $d/d\xi$ that are not exists in the space $\mathcal{P}_{N}(\omega)$).

Completely similar calculations, using the delta function $\delta_{\mathcal{P}_{N}}(\xi-1)$ (\ref{eq:delta_func_varphi_coeffs}) represented by vector $\boldsymbol{\delta}^{(1)} = m^{-1}\ovecphi = \{\ophi_{p}/w_{p}\}_{p}$ (\ref{eq:delta_func_varphi_coeffs}), lead to the following expression (\ref{eq:h_abs_equations_in_m1}) for $H^{-}(\xi;\, \mathrm{CFL}, -1)$: 
\begin{equation}
\begin{split}
H^{-}(\xi;\, \mathrm{CFL}, -1) & =
-\sum\limits_{p = 0}^{N} \left[\mathrm{D}^{-1} \mathrm{tanh}\left(\frac{\mathrm{CFL}}{2}\,\mathrm{D}\right)\boldsymbol{\delta}^{(1)}\right]_{p} \varphi_{p}(\xi) =
-\sum\limits_{p = 0}^{N} \left[\mathrm{tanh}\left(\frac{\mathrm{CFL}}{2}\,\mathrm{D}\right)\mathrm{D}^{-1}\boldsymbol{\delta}^{(1)}\right]_{p} \varphi_{p}(\xi)
\end{split}
\end{equation}
which explicitly identifies the following function analogous to $\Gamma^{+}(\xi)$ (\ref{eq:gamma_plus_def}):
\begin{equation}
\Gamma^{-}(\xi) = \Big[\mathrm{D}^{-1}\boldsymbol{\delta}^{(1)}\Big](\xi) =
\pi^{\mathrm{span}(\{\varphi_{p}\}_{p})}_{\mathcal{P}_{N}}\left[\,\int\limits_{\tilde{\xi}^{(1)}}^{\xi} d\xi'\, \delta_{\mathcal{P}_{N}}(\xi'-1)\right],
\end{equation}
which leads to the following expression analogous to (\ref{eq:h_plus_by_tanh_and_gamma}):
\begin{equation}\label{eq:h_minus_by_tanh_and_gamma}
H^{-}(\xi;\, \mathrm{CFL}, -1) = -\mathrm{tanh}\left(\frac{\mathrm{CFL}}{2}\,\frac{d}{d\xi}\right)\Gamma^{-}(\xi).
\end{equation}
The main difference between the expressions (\ref{eq:h_plus_by_tanh_and_gamma}) for $H^{+}(\xi;\, \mathrm{CFL}, -1)$ and (\ref{eq:h_minus_by_tanh_and_gamma}) for $H^{-}(\xi;\, \mathrm{CFL}, -1)$ is the inversion of the coordinate $\xi$: $\xi \leftrightarrow 1-\xi$.

Taking into account the obtained expressions (\ref{eq:h_plus_by_tanh_and_gamma}) and (\ref{eq:h_minus_by_tanh_and_gamma}), and taking into account the relations (\ref{eq:h_abs_equations_in_m1}), the following expressions for functions $S_{0}(\mathrm{CFL})$ and $S_{1}(\mathrm{CFL})$ are introduced:
\begin{equation}\label{eq:s0_and_s1_def}
\begin{split}
H^{+}(0;\, \mathrm{CFL}, -1) = H^{-}(1;\, \mathrm{CFL}, -1) \equiv S_{0}(\mathrm{CFL}),\quad
H^{+}(1;\, \mathrm{CFL}, -1) = H^{-}(0;\, \mathrm{CFL}, -1) \equiv S_{1}(\mathrm{CFL}),
\end{split}
\end{equation}
using which the condition (\ref{eq:cfl_by_h_src}) is rewritten in the following form:
\begin{equation}\label{eq:cfl_by_s_src}
\mathrm{CFL}_{\rm max} = \inf\left\{\mathrm{CFL}\, \big|\, \mathrm{CFL}\in[0, 1]\subset\mathcal{R}:\, |S_{0}(\mathrm{CFL}) \pm 1| = |S_{1}(\mathrm{CFL})|\right\}.
\end{equation}
From the construction, it is clear that the functions $S_{0}(\mathrm{CFL})$ and $S_{1}(\mathrm{CFL})$ (\ref{eq:s0_and_s1_def}) are polynomials. Numerically, $+1$ can be used in the expressions instead of $\pm 1$, allowing for expansion of the expressions for absolute values: $|S_{0}(\mathrm{CFL}) \pm 1| = |S_{1}(\mathrm{CFL})|\,\rightarrow\, S_{0}(\mathrm{CFL}) \pm S_{1}(\mathrm{CFL}) = 1$. Consequently, it becomes convenient to introduce new functions $F_{1}(\mathrm{CFL})$ and $F_{2}(\mathrm{CFL})$:
\begin{equation}\label{eq:F1_and_F2_def}
F_{1}(\mathrm{CFL}) = S_{0}(\mathrm{CFL}) + S_{1}(\mathrm{CFL}) - 1,\quad
F_{2}(\mathrm{CFL}) = S_{0}(\mathrm{CFL}) - S_{1}(\mathrm{CFL}) - 1,
\end{equation}
using which the condition (\ref{eq:s0_and_s1_def}) is rewritten in the following form:
\begin{equation}\label{eq:cfl_by_f_src}
\mathrm{CFL}_{\rm max} = \inf\left\{\mathrm{CFL}\, \big|\, \mathrm{CFL}\in[0, 1]\subset\mathcal{R}:\, F_{1}(\mathrm{CFL}) = 0\ \vee\,F_{2}(\mathrm{CFL}) = 0\right\}.
\end{equation}
This form is convenient for calculations of the boundary value of the Courant number $\mathrm{CFL}_{\rm max}$, and subsequent results are based on it.

The only non-universal component of the obtained relations (\ref{eq:s0_and_s1_def}) is delta function $\delta_{\mathcal{P}_{N}}(\xi)$ (\ref{eq:delta_func_varphi_action}), the definition of which is essentially related to the choice of a functional basis $\{\varphi_{p}\}_{p}$ (\ref{eq:phi_def}) in space $\mathcal{P}_{N}(\omega)$. However, it is obvious that the delta function $\delta_{\mathcal{P}_{N}}(\xi)$ acting in space $\mathcal{P}_{N}(\omega)$ should not depend on the choice of a functional basis $\{\varphi_{p}\}_{p}$ (\ref{eq:phi_def}) --- only the specific values of the expansion coefficients $\{\ophi_{p}/w_{p}\}_{p}$ (\ref{eq:delta_func_varphi_coeffs}) in terms of the functional basis $\{\varphi_{p}\}_{p}$ (\ref{eq:phi_def}) chosen for representing the function depend on its choice. Therefore, below it is using a representation of the delta function $\delta_{\mathcal{P}_{N}}(\xi)$ in space $\mathcal{P}_{N}(\omega)$ convenient for analytical calculations, which is based on a functional basis $\{\tilde{L}_{k}\}_{k = 0}^{N}$, consisting of shifted Legendre polynomials $\tilde{L}_{k}$ (\ref{eq:phi_def}):
\begin{equation}\label{eq:delta_func_legendre_coeffs}
\delta_{\mathcal{P}_{N}}(\xi) = \sum\limits_{k = 0}^{N} (-1)^{k}(2k+1)\tilde{L}_{k}(\xi).
\end{equation}
It is easy to rigorously prove that this representation (\ref{eq:delta_func_legendre_coeffs}) of the delta function $\delta_{\mathcal{P}_{N}}(\xi)$ is identical to representation (\ref{eq:delta_func_varphi_coeffs}), with the differences arising only from the choice of functional basis. To do this, the expansion coefficients for the delta function $\delta_{\mathcal{P}_{N}}(\xi)$ defined by representation (\ref{eq:delta_func_varphi_coeffs}) is calculated in functional basis $\{\tilde{L}_{k}\}_{k}$:
\begin{equation}
\delta_{\mathcal{P}_{N}}(\xi) = \sum\limits_{k = 0}^{N} c_{k}\tilde{L}_{k}(\xi),\qquad
c_{k} = \frac{\displaystyle\intrefdom{\xi}\,f(\xi)\tilde{L}_{k}(\xi)}{\displaystyle\intrefdom{\xi}\,\tilde{L}_{k}^{2}(\xi)} =
(2k+1)\intrefdom{\xi}\,f(\xi)\tilde{L}_{k}(\xi),
\end{equation}
using the well-known $\mathcal{L}_{2}(\omega)$-norm $\|\tilde{L}_{k}\|_{\mathcal{L}_{2}(\omega)}^{-2} = 2k+1$ for shifted Legendre polynomials. Direct calculation, using property (\ref{eq:delta_func_varphi_action}), yields the following expression for the expansion coefficients:
\begin{equation}
\begin{split}
&\delta^{\tilde{L}}_{k} = (2k+1)\intrefdom{\xi}\,\delta_{\mathcal{P}_{N}}(\xi)\tilde{L}_{k}(\xi) = 
\Big\{\tilde{L}_{k}\in\mathcal{P}_{k}(\omega)\subseteq\mathcal{P}_{N}(\omega)\Big\} = (2k+1)\tilde{L}_{k}(0) = (-1)^{k}(2k+1),
\end{split}
\end{equation}
which concludes the proof. Using well-known expressions for the values of the derivatives of Legendre polynomials $\tilde{L}_{k}^{(m)}$ (recalculated taking into account the shift in the domain of the Legendre polynomials $[-1, +1] \rightarrow [0, 1]$):
\begin{equation}
\tilde{L}_{k}(0) = (-1)^{k-m}\frac{(k+m)!}{m!(k-m)!},\qquad
\tilde{L}_{k}^{(m)}(1) = \frac{(k+m)!}{m!(k-m)!},
\end{equation}
the following expressions for the derivatives of the delta functions at points $\xi = 0$ and $1$ is obtained:
\begin{equation}
\begin{split}
&\delta_{\mathcal{P}_{N}}^{(2m)}(0) = \sum\limits_{k = 0}^{N} (-1)^{k}(2k+1)\tilde{L}_{k}^{(2m)}(0) =
\sum\limits_{k = 2m}^{N} (-1)^{k}(2k+1) (-1)^{k-2m}\frac{(k+m)!}{(2m)!(k-2m)!} = \sum\limits_{k = 2m}^{N} \frac{(2k+1)(k+m)!}{(2m)!(k-2m)!},\\
&\delta_{\mathcal{P}_{N}}^{(2m)}(1) = \sum\limits_{k = 0}^{N} (-1)^{k}(2k+1)\tilde{L}_{k}^{(2m)}(1) =
\sum\limits_{k = 2m}^{N} (-1)^{k}(2k+1) \frac{(k+m)!}{(2m)!(k-2m)!} = \sum\limits_{k = 2m}^{N} (-1)^{k}\frac{(2k+1)(k+m)!}{(2m)!(k-2m)!}.
\end{split}
\end{equation}
Substitution these into expressions (\ref{eq:s0_and_s1_def}), (\ref{eq:h_plus_by_tanh_and_gamma}), (\ref{eq:h_minus_by_tanh_and_gamma}), (\ref{eq:tanh_taylor_exp}) and using the nilpotency property $\mathrm{D}^{N+1} = 0$ of the differentiation matrix $\mathrm{D}$ allowed to calculate the representations of functions $S_{0}(\mathrm{CFL})$ and $S_{1}(\mathrm{CFL})$ in polynomial form:
\begin{equation}\label{eq:s0_and_s1_calc}
\begin{split}
S_{0}(\mathrm{CFL}) &= \hspace{-2mm}\sum\limits_{m = 0}^{\left\lfloor N/2 \right\rfloor}
t_{2m+1} \left(\frac{\mathrm{CFL}}{2}\right)^{2m+1} \hspace{-1mm}\delta_{\mathcal{P}_{N}}^{(2m)}(0) =
\hspace{-2mm}\sum\limits_{m = 0}^{\left\lfloor N/2 \right\rfloor}\left[B_{2m+2}\sum\limits_{k = 2m}^{N}
\frac{(2k+1)(2^{2m+2} - 1)(k+m)!}{(2m)!(2m+2)!(k-2m)!}\right] \mathrm{CFL}^{2m+1},\\
S_{1}(\mathrm{CFL}) &= \hspace{-2mm}\sum\limits_{m = 0}^{\left\lfloor N/2 \right\rfloor}
t_{2m+1} \left(\frac{\mathrm{CFL}}{2}\right)^{2m+1} \hspace{-1mm}\delta_{\mathcal{P}_{N}}^{(2m)}(1) =
\hspace{-2mm}\sum\limits_{m = 0}^{\left\lfloor N/2 \right\rfloor}\left[B_{2m+2}\sum\limits_{k = 2m}^{N}
(-1)^{k}\frac{(2k+1)(2^{2m+2} - 1)(k+m)!}{(2m)!(2m+2)!(k-2m)!}\right] \mathrm{CFL}^{2m+1},
\end{split}
\end{equation}
where $\lfloor\ldots\rfloor$ denotes the operation of taking the integer part of a number. Table~\ref{tab:s0_and_s1} presents the calculated polynomials $S_{0}(\mathrm{CFL})$ and $S_{1}(\mathrm{CFL})$ for degrees $N = 1, \ldots, 12$ of basis polynomials.

\begin{table}[h!]
\centering
\normalsize
\caption{%
Expressions for polynomials $S_{0}(\mathrm{CFL})$, $S_{1}(\mathrm{CFL})$ (\ref{eq:s0_and_s1_def}) for degrees $N = 1, \ldots, 12$ of basis polynomials $\{\varphi_{p}\}_{p}$.
\label{tab:s0_and_s1}
}
\setlength{\tabcolsep}{3.5pt}
\begin{tabular}{@{}|l|l|@{}}
\hline
$N$ & $S_{0}(\mathrm{CFL})$, $S_{1}(\mathrm{CFL})$ \\
\hline
1 & $S_{0}(\mathrm{CFL}) = 2\,\mathrm{CFL}$,$\quad$
    $S_{1}(\mathrm{CFL}) = -\mathrm{CFL}$ \\
\hline
2 & $S_{0}(\mathrm{CFL}) = - \frac{5}{2}\,\mathrm{CFL}^{3} + \frac{9}{2}\,\mathrm{CFL}$,\\
  & $S_{1}(\mathrm{CFL}) = - \frac{5}{2}\,\mathrm{CFL}^{3} + \frac{3}{2}\,\mathrm{CFL}$ \\
\hline
3 & $S_{0}(\mathrm{CFL}) = - 20\,\mathrm{CFL}^{3} + 8\,\mathrm{CFL}$,\\
  & $S_{1}(\mathrm{CFL}) = 15\,\mathrm{CFL}^{3} - 2\,\mathrm{CFL}$ \\
\hline
4 & $S_{0}(\mathrm{CFL}) = 63\,\mathrm{CFL}^{5} - \frac{175}{2}\,\mathrm{CFL}^{3} + \frac{25}{2}\,\mathrm{CFL}$, \\
  & $S_{1}(\mathrm{CFL}) = 63\,\mathrm{CFL}^{5} - \frac{105}{2}\,\mathrm{CFL}^{3} + \frac{5}{2}\,\mathrm{CFL}$ \\
\hline
5 & $S_{0}(\mathrm{CFL}) = 756\,\mathrm{CFL}^{5} - 280\,\mathrm{CFL}^{3} + 18\,\mathrm{CFL}$, \\
  & $S_{1}(\mathrm{CFL}) = - 630\,\mathrm{CFL}^{5} + 140\,\mathrm{CFL}^{3} - 3\,\mathrm{CFL}$ \\
\hline
6 & $S_{0}(\mathrm{CFL}) = - \frac{7293}{2}\,\mathrm{CFL}^{7} + 4851\,\mathrm{CFL}^{5} - 735\,\mathrm{CFL}^{3} + \frac{49}{2}\,\mathrm{CFL}$, \\
  & $S_{1}(\mathrm{CFL}) = - \frac{7293}{2}\,\mathrm{CFL}^{7} + 3465\,\mathrm{CFL}^{5} - 315\,\mathrm{CFL}^{3} + \frac{7}{2}\,\mathrm{CFL}$ \\
\hline
7 & $S_{0}(\mathrm{CFL}) = - 58344\,\mathrm{CFL}^{7} + 22176\,\mathrm{CFL}^{5} - 1680\,\mathrm{CFL}^{3} + 32\,\mathrm{CFL}$, \\
  & $S_{1}(\mathrm{CFL}) = 51051\,\mathrm{CFL}^{7} - 13860\,\mathrm{CFL}^{5} + 630\,\mathrm{CFL}^{3} - 4\,\mathrm{CFL}$ \\
\hline
8 & $S_{0}(\mathrm{CFL}) = 376805\,\mathrm{CFL}^{9} - \frac{984555}{2}\,\mathrm{CFL}^{7} + 81081\,\mathrm{CFL}^{5} - 3465\,\mathrm{CFL}^{3} + \frac{81}{2}\,\mathrm{CFL}$, \\
  & $S_{1}(\mathrm{CFL}) = 376805\,\mathrm{CFL}^{9} - \frac{765765}{2}\,\mathrm{CFL}^{7} + 45045\,\mathrm{CFL}^{5} - 1155\,\mathrm{CFL}^{3} + \frac{9}{2}\,\mathrm{CFL}$ \\
\hline
9 & $S_{0}(\mathrm{CFL}) = 7536100\,\mathrm{CFL}^{9} - 2917200\,\mathrm{CFL}^{7}$ \\
  & $\hphantom{S_{0}(\mathrm{CFL})} + 252252\,\mathrm{CFL}^{5} - 6600\,\mathrm{CFL}^{3} + 50\,\mathrm{CFL}$, \\
  & $S_{1}(\mathrm{CFL}) = - 6782490\,\mathrm{CFL}^{9} + 2042040\,\mathrm{CFL}^{7}$ \\
  & $\hphantom{S_{1}(\mathrm{CFL})} - 126126\,\mathrm{CFL}^{5} + 1980\,\mathrm{CFL}^{3} - 5\,\mathrm{CFL}$ \\
\hline
10 & $S_{0}(\mathrm{CFL}) = - 60931689\,\mathrm{CFL}^{11} + 78752245\,\mathrm{CFL}^{9} - 13637910\,\mathrm{CFL}^{7}$ \\
   & $\hphantom{S_{0}(\mathrm{CFL})} + 693693\,\mathrm{CFL}^{5} - \frac{23595}{2}\,\mathrm{CFL}^{3} + \frac{121}{2}\,\mathrm{CFL}$, \\
   & $S_{1}(\mathrm{CFL}) = - 60931689\,\mathrm{CFL}^{11} + 64433655\,\mathrm{CFL}^{9} - 8678670\,\mathrm{CFL}^{7}$ \\
   & $\hphantom{S_{1}(\mathrm{CFL})} + 315315\,\mathrm{CFL}^{5} - \frac{6435}{2}\,\mathrm{CFL}^{3} + \frac{11}{2}\,\mathrm{CFL}$ \\
\hline
11 & $S_{0}(\mathrm{CFL}) = - 1462360536\,\mathrm{CFL}^{11} + 572743600\,\mathrm{CFL}^{9}$ \\
   & $\hphantom{S_{0}(\mathrm{CFL})} - 53559792\,\mathrm{CFL}^{7} + 1729728\,\mathrm{CFL}^{5} - 20020\,\mathrm{CFL}^{3} + 72\,\mathrm{CFL}$, \\
   & $S_{1}(\mathrm{CFL}) = 1340497158\,\mathrm{CFL}^{11} - 429557700\,\mathrm{CFL}^{9}$ \\
   & $\hphantom{S_{1}(\mathrm{CFL})} + 31243212\,\mathrm{CFL}^{7} - 720720\,\mathrm{CFL}^{5} + 5005\,\mathrm{CFL}^{3} - 6\,\mathrm{CFL}$ \\
\hline
12 & $S_{0}(\mathrm{CFL}) = 14199419150\,\mathrm{CFL}^{13} - 18218575011\,\mathrm{CFL}^{11} + 3257479225\,\mathrm{CFL}^{9}$ \\
   & $\hphantom{S_{0}(\mathrm{CFL})} - 183739842\,\mathrm{CFL}^{7} + 3981978\,\mathrm{CFL}^{5} - \frac{65065}{2}\,\mathrm{CFL}^{3} + \frac{169}{2}\,\mathrm{CFL}$, \\
   & $S_{1}(\mathrm{CFL}) = 14199419150\,\mathrm{CFL}^{13} - 15415717317\,\mathrm{CFL}^{11} + 2255177925\,\mathrm{CFL}^{9}$ \\
   & $\hphantom{S_{1}(\mathrm{CFL})} - 98936838\,\mathrm{CFL}^{7} + 1531530\,\mathrm{CFL}^{5} - \frac{15015}{2}\,\mathrm{CFL}^{3} + \frac{13}{2}\,\mathrm{CFL}$ \\
\hline
\end{tabular}
\end{table}

\begin{table}[h!]
\centering
\normalsize
\caption{%
Expressions for equations $F_{1}(\mathrm{CFL}) = 0$, $F_{2}(\mathrm{CFL}) = 0$ (\ref{eq:F1_and_F2_def}), the smallest values $\mathrm{CFL}_{\rm max}^{(1, 2)}$ of their solutions $F_{1, 2}(\mathrm{CFL}_{\rm max}^{(1, 2)}) = 0$, and the resulting values $\mathrm{CFL}_{\rm max} = \min(\mathrm{CFL}_{\rm max}^{(1)}, \mathrm{CFL}_{\rm max}^{(2)})$, for degrees $N = 1, \ldots, 12$ of basis polynomials $\{\varphi_{p}\}_{p}$.
\label{tab:f1_and_f2}
}
\setlength{\tabcolsep}{3.5pt}
\begin{tabular}{@{}|l|l|l|l|@{}}
\hline
$N$ & $F_{1}(\mathrm{CFL}) = 0$, $F_{2}(\mathrm{CFL}) = 0$ & $\mathrm{CFL}_{\rm max}^{(1, 2)}$ & $\mathrm{CFL}_{\rm max}$ \\ 
\hline
1 & $3\,\mathrm{CFL} - 1 = 0$, & $0.333333$ & $0.333333$\\
  & $\mathrm{CFL} - 1 = 0$ & $1.000000$ & \\
\hline
2 & $3\,\mathrm{CFL} - 1 = 0$, & $0.333333$ & $0.170820$\\
  & $- 5\,\mathrm{CFL}^{3} + 6\,\mathrm{CFL} - 1 = 0$ & $0.170820$ & \\
\hline
3 & $- 35\,\mathrm{CFL}^{3} + 10\,\mathrm{CFL} - 1 = 0$, & $0.103929$ & $0.103929$\\
  & $- 5\,\mathrm{CFL}^{3} + 6\,\mathrm{CFL} - 1 = 0$ & $0.170820$ & \\
\hline
4 & $- 35\,\mathrm{CFL}^{3} + 10\,\mathrm{CFL} - 1 = 0$, & $0.103929$ & $0.0698309$\\
  & $126\,\mathrm{CFL}^{5} - 140\,\mathrm{CFL}^{3} + 15\,\mathrm{CFL} - 1 = 0$ & $0.0698309$ & \\
\hline
5 & $1386\,\mathrm{CFL}^{5} - 420\,\mathrm{CFL}^{3} + 21\,\mathrm{CFL} - 1 = 0$, & $0.0501156$ & $0.0501156$\\
  & $126\,\mathrm{CFL}^{5} - 140\,\mathrm{CFL}^{3} + 15\,\mathrm{CFL} - 1 = 0$ & $0.0698309$ & \\
\hline
6 & $1386\,\mathrm{CFL}^{5} - 420\,\mathrm{CFL}^{3} + 21\,\mathrm{CFL} - 1 = 0$, & $0.0501156$ & $0.0377012$\\
  & $- 7293\,\mathrm{CFL}^{7} + 8316\,\mathrm{CFL}^{5} - 1050\,\mathrm{CFL}^{3} + 28\,\mathrm{CFL} - 1 = 0$ & $0.0377012$ & \\
\hline
7 & $- 109395\,\mathrm{CFL}^{7} + 36036\,\mathrm{CFL}^{5} - 2310\,\mathrm{CFL}^{3} + 36\,\mathrm{CFL} - 1 = 0$, & $0.0293838$ & $0.0293838$\\
  & $- 7293\,\mathrm{CFL}^{7} + 8316\,\mathrm{CFL}^{5} - 1050\,\mathrm{CFL}^{3} + 28\,\mathrm{CFL} - 1 = 0$ & $0.0377012$ & \\
\hline
8 & $- 109395\,\mathrm{CFL}^{7} + 36036\,\mathrm{CFL}^{5} - 2310\,\mathrm{CFL}^{3} + 36\,\mathrm{CFL} - 1 = 0$, & $0.0293838$ & $0.0235415$\\
  & $753610\,\mathrm{CFL}^{9} - 875160\,\mathrm{CFL}^{7} + 126126\,\mathrm{CFL}^{5}$ & $0.0235415$ & \\
  & $- 4620\,\mathrm{CFL}^{3} + 45\,\mathrm{CFL} - 1 = 0$ & & \\
\hline
9 & $14318590\,\mathrm{CFL}^{9} - 4959240\,\mathrm{CFL}^{7} + 378378\,\mathrm{CFL}^{5}$ & $0.0192819$ & $0.0192819$ \\
  & $- 8580\,\mathrm{CFL}^{3} + 55\,\mathrm{CFL} - 1 = 0$, & & \\
  & $753610\,\mathrm{CFL}^{9} - 875160\,\mathrm{CFL}^{7} + 126126\,\mathrm{CFL}^{5}$ & $0.0235415$ & \\
  & $- 4620\,\mathrm{CFL}^{3} + 45\,\mathrm{CFL} - 1 = 0$ & & \\
\hline
10 & $14318590\,\mathrm{CFL}^{9} - 4959240\,\mathrm{CFL}^{7} + 378378\,\mathrm{CFL}^{5}$ & $0.0192819$ & $0.0160813$ \\
   & $- 8580\,\mathrm{CFL}^{3} + 55\,\mathrm{CFL} - 1 = 0$, & & \\
   & $- 121863378\,\mathrm{CFL}^{11} + 143185900\,\mathrm{CFL}^{9} - 22316580\,\mathrm{CFL}^{7}$ & $0.0160813$ & \\
   & $+ 1009008\,\mathrm{CFL}^{5} - 15015\,\mathrm{CFL}^{3} + 66\,\mathrm{CFL} - 1 = 0$ & & \\
\hline
11 & $- 2802857694\,\mathrm{CFL}^{11} + 1002301300\,\mathrm{CFL}^{9} - 84803004\,\mathrm{CFL}^{7}$ & $0.0136158$ & $0.0136158$ \\
   & $+ 2450448\,\mathrm{CFL}^{5} - 25025\,\mathrm{CFL}^{3} + 78\,\mathrm{CFL} - 1 = 0$, & & \\
   & $- 121863378\,\mathrm{CFL}^{11} + 143185900\,\mathrm{CFL}^{9} - 22316580\,\mathrm{CFL}^{7}$ & $0.0160813$ & \\
   & $+ 1009008\,\mathrm{CFL}^{5} - 15015\,\mathrm{CFL}^{3} + 66\,\mathrm{CFL} - 1 = 0$ & & \\
\hline
12 & $- 2802857694\,\mathrm{CFL}^{11} + 1002301300\,\mathrm{CFL}^{9} - 84803004\,\mathrm{CFL}^{7}$ & $0.0136158$ & $0.0116764$ \\
   & $+ 2450448\,\mathrm{CFL}^{5} - 25025\,\mathrm{CFL}^{3} + 78\,\mathrm{CFL} - 1 = 0$, & & \\
   & $28398838300\,\mathrm{CFL}^{13} - 33634292328\,\mathrm{CFL}^{11} + 5512657150\,\mathrm{CFL}^{9}$ & $0.0116764$ & \\
   & $- 282676680\,\mathrm{CFL}^{7} + 5513508\,\mathrm{CFL}^{5} - 40040\,\mathrm{CFL}^{3} + 91\,\mathrm{CFL} - 1 = 0$ & & \\
\hline
\end{tabular}
\end{table}

It should be noted that, despite the fact that all expressions (\ref{eq:s0_and_s1_calc}), (\ref{eq:F1_and_F2_def}) and (\ref{eq:cfl_by_f_src}) are derived under the initial assumption of a specific polynomial basis $\{\varphi_{p}\}_{p}$ (\ref{eq:phi_def}), the resulting expressions turned out to be completely free of a specific polynomial basis, and the correctness of the obtained results for the ADER-DG methods with a local DG predictor is expected in the case of an arbitrary choice of functional basis.

Table~\ref{tab:f1_and_f2} presents the expressions for equations $F_{1}(\mathrm{CFL}) = 0$, $F_{2}(\mathrm{CFL}) = 0$ (\ref{eq:F1_and_F2_def}), the smallest values $\mathrm{CFL}_{\rm max}^{(1, 2)}$ of their solutions $F_{1, 2}(\mathrm{CFL}_{\rm max}^{(1, 2)}) = 0$, and the resulting values $\mathrm{CFL}_{\rm max} = \min(\mathrm{CFL}_{\rm max}^{(1)}, \mathrm{CFL}_{\rm max}^{(2)})$ that correspond to the boundary values of the Courant number $\mathrm{CFL}_{\rm max}$ by the criterion (\ref{eq:cfl_by_f_src}), for degrees $N = 1, \ldots, 12$ of basis polynomials. The resulting boundary values of the Courant number $\mathrm{CFL}_{\rm max}$, their comparisons with the known values~\cite{ader_dg_stab, ader_dg_PNPM, PNPM_DG_2008} and other estimates are presented in Table~\ref{tab:cfls_max_data} and Figure~\ref{fig:cfls_max_data}. An interesting feature is that functions $F_{1} = F_{1}(\mathrm{CFL})$, $F_{2} = F_{2}(\mathrm{CFL})$ (\ref{eq:F1_and_F2_def}) exhibit a grouping for different successive degrees $N$ of basis polynomials: $F_{1}(\mathrm{CFL})$ for an even degree $N$ of basis polynomials completely coincides with $F_{1}(\mathrm{CFL})$ for the previous (and, accordingly, odd) degree $N$ of basis polynomials; $F_{2}(\mathrm{CFL})$ for an even degree $N$ of basis polynomials completely coincides with $F_{2}(\mathrm{CFL})$ for the next (and, accordingly, odd) degree $N$ of basis polynomials (except for case $N = 1$, for which there is no previous degree $N$ of basis polynomials considered in this work; however, in case $N = 0$, a trivial calculation of $F_{2}(\mathrm{CFL})$ yields $\mathrm{CFL} - 1$, and $\mathrm{CFL}_{\rm max} = 1$ holds in the case $N = 0$, therefore, this property is also satisfied in case $N = 1$). A similar behavior is consequently demonstrated by the values $\mathrm{CFL}_{\rm max}^{(1, 2)}$ of Courant number, but due to different combinations of functions $F_{1} = F_{1}(\mathrm{CFL})$, $F_{2} = F_{2}(\mathrm{CFL})$ for different degrees $N$ of basis polynomials, the resulting boundary values of the Courant number $\mathrm{CFL}_{\rm max}$ are not repeated for different degrees $N$ of basis polynomials.

\begin{table}[h!]
\centering
\normalsize
\caption{\label{tab:cfls_max_data}
Boundary values of the Courant number $\mathrm{CFL}_{\rm max}$ determining the stability boundary of the numerical method ADER-DG with the LST-DG predictor for polynomial degrees $N = 1, \ldots, 12$.
}
\begin{tabular}{@{}|c|c|c|c|c|c|c|@{}}
\hline
$N$ & \texttt{this work} & \cite{ader_dg_stab} & \cite{ader_dg_PNPM} & \cite{PNPM_DG_2008} & $\frac{1}{2N+1}$ (\ref{eq:old_eff_cfl}) & $\frac{a}{a^{\rm eff}(N)}$ (\ref{eq:a_eff}) \\ 
\hline
$1$  & $0.333333$ & $0.333$ & $0.33$  & $0.32$  & $0.333$ & $0.3660$ \\
$2$  & $0.170820$ & $0.170$ & $0.17$  & $0.17$  & $0.200$ & $0.1878$ \\
$3$  & $0.103929$ & $0.104$ & $0.10$  & $0.10$  & $0.143$ & $0.1139$ \\
$4$  & $0.069831$ & $0.069$ & $0.069$ & $0.069$ & $0.111$ & $0.0764$ \\
$5$  & $0.050116$ & $0.050$ & $0.045$ & --      & $0.091$ & $0.0547$ \\
$6$  & $0.037701$ & $0.037$ & $0.038$ & --      & $0.077$ & $0.0411$ \\
$7$  & $0.029384$ & $0.029$ & --      & --      & $0.067$ & $0.0320$ \\
$8$  & $0.023542$ & $0.023$ & --      & --      & $0.059$ & $0.0256$ \\
$9$  & $0.019282$ & $0.018$ & --      & --      & $0.053$ & $0.0210$ \\
$10$ & $0.016081$ & --      & --      & --      & $0.048$ & $0.0175$ \\
$11$ & $0.013616$ & --      & --      & --      & $0.043$ & $0.0148$ \\
$12$ & $0.011676$ & --      & --      & --      & $0.040$ & $0.0127$ \\
\hline
\end{tabular}
\end{table}

The results of the works~\cite{ader_dg_stab, ader_dg_PNPM, PNPM_DG_2008} presented for comparison in Table~\ref{tab:cfls_max_data} and Figure~\ref{fig:cfls_max_data} are in good agreement with each other in the intersection of the set of polynomial degrees $N \leqslant 4$, however, for polynomial degrees $N \leqslant 5$, the results of work~\cite{ader_dg_stab} slightly diverge from the results of work~\cite{ader_dg_PNPM}. A detailed description of the similarities and differences in the results of the works~\cite{ader_dg_stab, ader_dg_PNPM, PNPM_DG_2008} is presented in work~\cite{ader_dg_stab}. It is important to note the following --- in all works~\cite{ader_dg_stab, ader_dg_PNPM, PNPM_DG_2008}, different functional bases were used: in work~\cite{ader_dg_stab}, the classical power basis was used, in work~\cite{ader_dg_PNPM}, the nodal basis was used $\{\varphi_{p}\}_{p}$ (\ref{eq:phi_def}), which is also used in the present work, and in work~\cite{PNPM_DG_2008}, a modal basis was used, but all the results of boundary values of the Courant number $\mathrm{CFL}_{\rm max}$ are relatively comparable. It should be noted that the results $\mathrm{CFL}_{\rm max}$ in the works~\cite{ader_dg_stab, ader_dg_PNPM, PNPM_DG_2008} were obtained based on an analysis of the spectral radius of the matrix $\mathrm{R}(\mathrm{CFL}, \theta)$ (\ref{eq:r_matrix_def}) for a specific partition of the phase determination domain $\theta\in[0, 2\pi]$, and it is very difficult to obtain sufficiently accurate results $\mathrm{CFL}_{\rm max}$ within this approach, especially for high degrees $N$ of the basis polynomials. The approach proposed in this work avoids this complexity, which could lead to incorrect results at certain decimal places.

The values $\mathrm{CFL}_{\rm max}$ of the Courant number obtained in this study agree well with the results of the work~\cite{ader_dg_stab} for polynomial degrees $N \leqslant 3$, $N = 5$ and $7$ within the limits of numerical rounding. However, already in case $N \geqslant 4$, a contradiction arises in the fourth decimal place after the decimal point (which in this case corresponds to the third significant digit of the number). Up to case $N = 8$, the contradiction between $\mathrm{CFL}_{\rm max}$ obtained in this work and in the work~\cite{ader_dg_stab} is approximately of the same order of magnitude, but for case $N = 9$, the difference reaches the second significant digit of the number. This difference for degree $N = 9$ is already quite significant, and this degree is the upper limit of the intersection of the results from the work~\cite{ader_dg_stab} and this work, therefore, to clarify the correctness of the obtained results, it is necessary to turn to the analysis of the direct results of the numerical experiment, which is done in the following Section~\ref{sec:app} ``Applications'' in Subsection~\ref{sec:app:lin_adv} ``Linear advection equation''.

Comparison with the estimate of the boundary values of the Courant number $\mathrm{CFL}_{\rm max} = 1/(2N+1)$ (\ref{eq:old_eff_cfl}), which corresponds to an ``effective Courant number'' $\mathrm{C} = 1$ in the expression (\ref{eq:old_eff_cfl}), is significantly higher than more precise values $\mathrm{CFL}_{\rm max}$ of the Courant number. Therefore, using this estimate (\ref{eq:old_eff_cfl}) requires values of the ``effective Courant number'' $\mathrm{C}$ significantly less than one. Comparison with the estimate of the boundary values of the Courant number $\mathrm{CFL}_{\rm max} = a/a^{\rm eff}(N)$ obtained above in this work in expression (\ref{eq:a_eff}) is also presented. Despite the extremely strong assumptions underlying the derivation of $\mathrm{CFL}_{\rm max}$ by relation (\ref{eq:a_eff}), the comparison reveals an unexpectedly high accuracy in agreement with high-precision values. This is particularly evident in Figure~\ref{fig:cfls_max_data}, which plots a power-law approximation (\ref{eq:cfls_asymp_est}) $\mathrm{CFL}_{\rm max} \propto [(N+1)^{-2}]^{\alpha}$. It is evident that even considering only polynomial degrees $N \leqslant 12$, the slope coefficient is $\alpha \approx 0.902(8)$ and is very close to expected $\alpha = 1$, although the derivation of the asymptotics (\ref{eq:cfls_asymp_est}) assumed very high polynomial degrees $N \gg 1$. Therefore, it can be concluded that the qualitative analysis of the stability of the numerical method is at least qualitatively correct.

\section{Applications}
\label{sec:app}

In the previous Section~\ref{sec:approx_anal} ``Approximation analysis'', a rigorous direct analysis of the approximation properties of the ADER-DG numerical method with the LST-DG predictor was performed, and in the preceding Section~\ref{sec:stab_anal} ``Stability analysis'', a stability analysis of the numerical method was performed based on a set of empirical considerations. However, in computational physics, it is clear that the computational experiment is everything, so this final Section~\ref{sec:app} ``Applications'' presents the results of applying the ADER-DG numerical method with the LST-DG predictor to solving both the original linear advection equation (\ref{eq:adv_eq_src}) and the nonlinear problem for the Euler system as a characteristic example of a quasilinear hyperbolic system of equations. Direct checks of the obtained approximation orders $p = N+1$ and the obtained boundary values of the Courant numbers $\mathrm{CFL}_{\rm max}(N)$ defining the stability boundary are performed. Although empirical convergence orders have been calculated numerous times in previous studies of the ADER-DG numerical method with the LST-DG predictor (see~\cite{ader_dg_ideal_flows, ader_dg_ale, ader_dg_grmhd, ader_dg_gr_prd, ader_dg_simple_mod_2016, fron_phys, ader_dg_axioms, exahype, ader_stiff_3, ader_stiff_4, ader_eff_blas, PNPM_DG_2008, PNPM_DG_2009, ader_dg_PNPM, ader_dg_stab}), in this work, the calculations are performed for a high accurately defined stability boundary $\mathrm{CFL}_{\rm max}(N)$. All calculations are performed using a computational framework developed for these purposes, a detailed description of which is presented in~\cite{ader_stiff_3, ader_stiff_4, ader_eff_blas}. For the calculations, only the ADER-DG numerical method with the LST-DG predictor is used, and the possibility of a posteriori limitation using a subcell finite-volume limiter is disabled.

Subsection~\ref{sec:app:lin_adv} ``Linear advection equation'' presents a set of numerical experiments for a linear one-dimensional advection equation (\ref{eq:adv_eq_src}). Analysis of the obtained results showed that the boundary values of the Courant number calculated $\mathrm{CFL}_{\rm max}(N)$ in the previous Section~\ref{sec:stab_anal} ``Stability analysis'' are determined with sufficient accuracy and correctness, and that going beyond the stability boundary $\mathrm{CFL} > \mathrm{CFL}_{\rm max}(N)$ determined by them immediately leads to instability of the ADER-DG numerical method with the LST-DG predictor. Moreover, in the stability region up to the stability boundary, the empirical convergence orders correspond well to the approximation orders $p = N+1$ of the ADER-DG numerical method with the LST-DG predictor calculated in Section~\ref{sec:approx_anal} ``Approximation analysis''. Subsection~\ref{sec:app:euler} ``Nonlinear Euler equations'' presents a set of numerical experiments for a system of quasilinear Euler equations. Analysis of the obtained results showed that the results obtained for a linear one-dimensional advection equation can, to a certain extent, also be applied to a quasilinear system of Euler equations. However, the use of dissipative Riemann solvers leads to a slight expansion of the stability region of the numerical method.

\subsection{Linear advection equation}
\label{sec:app:lin_adv}

\begin{figure}[h!]
\includegraphics[width=0.245\textwidth]{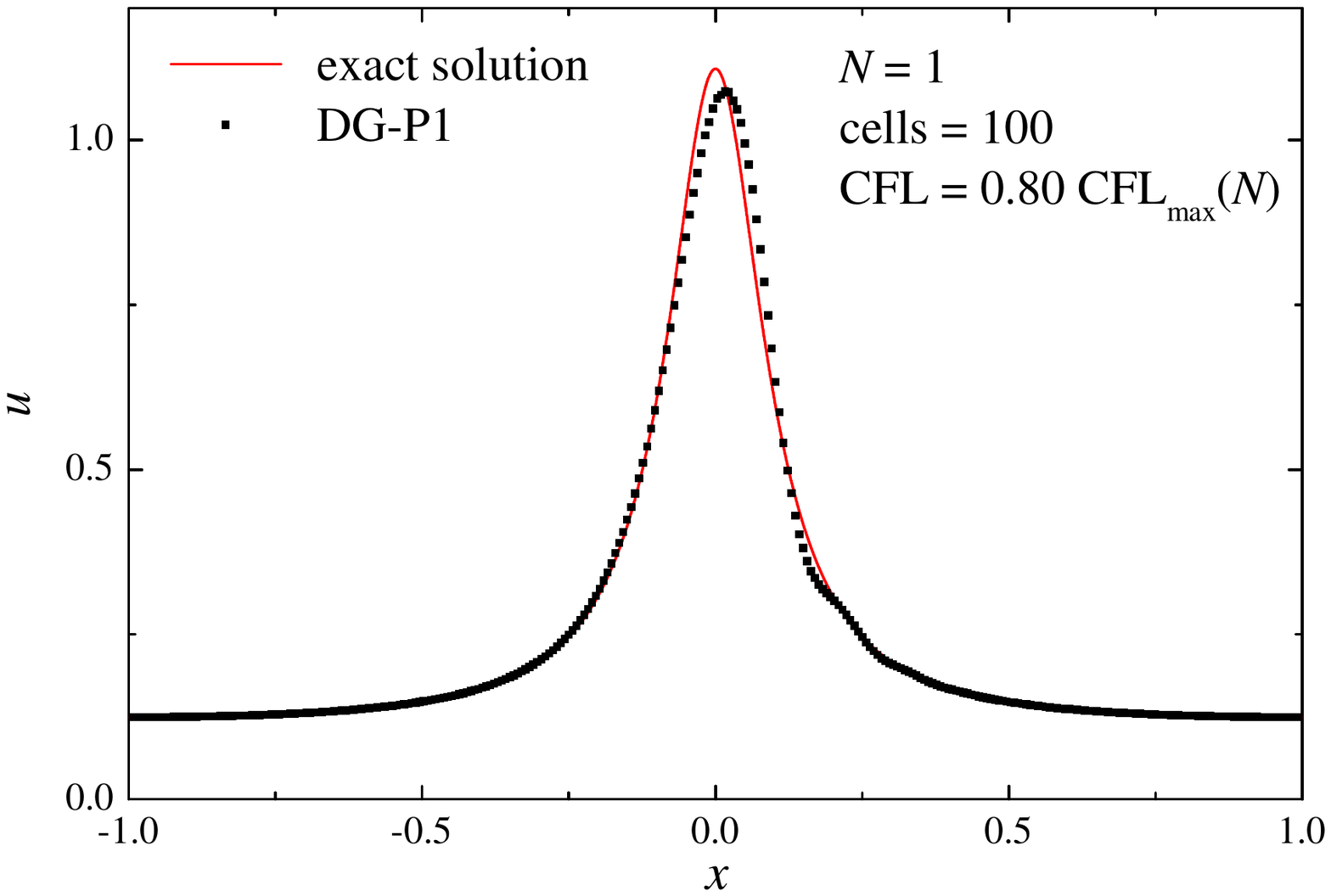}
\includegraphics[width=0.245\textwidth]{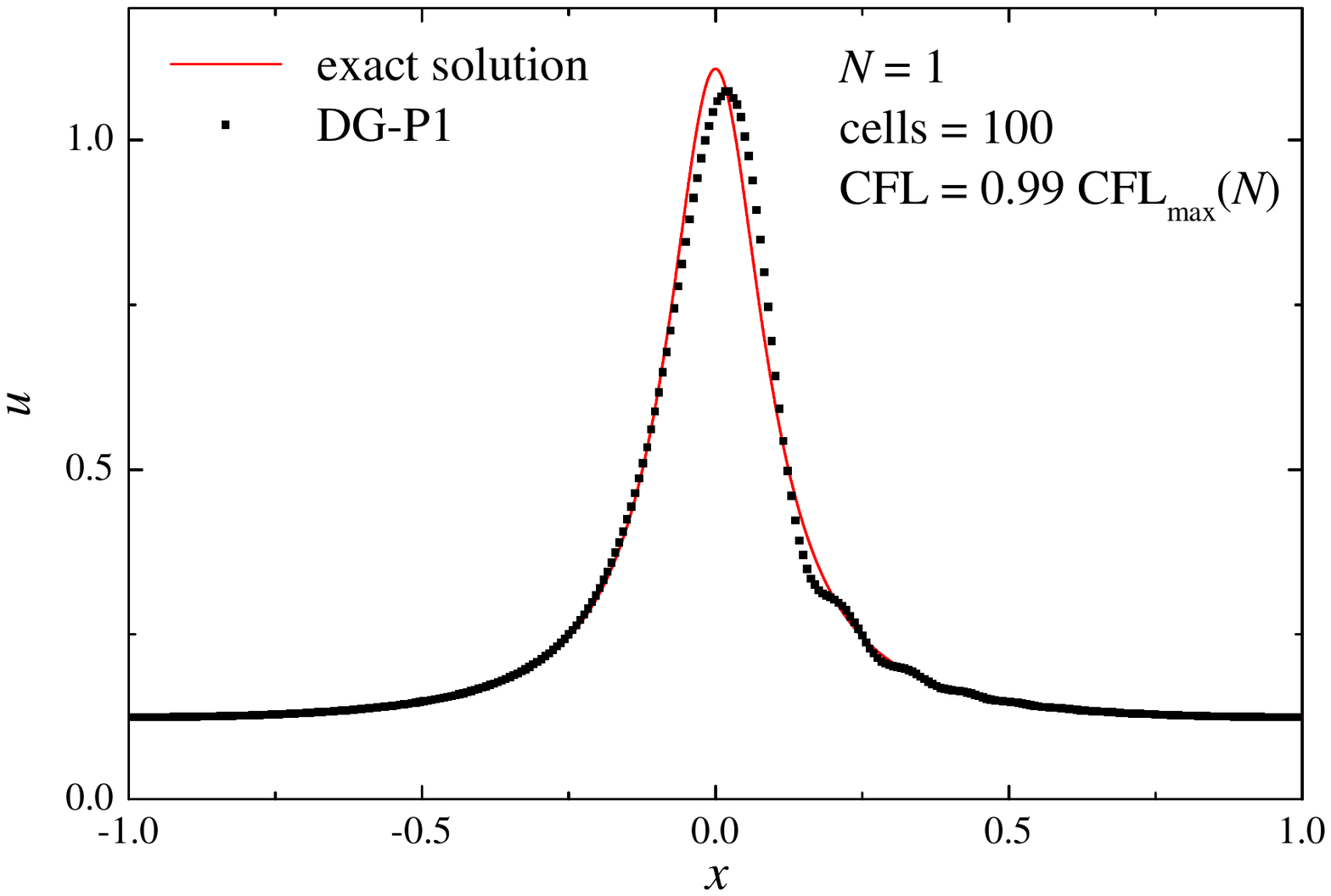}
\includegraphics[width=0.245\textwidth]{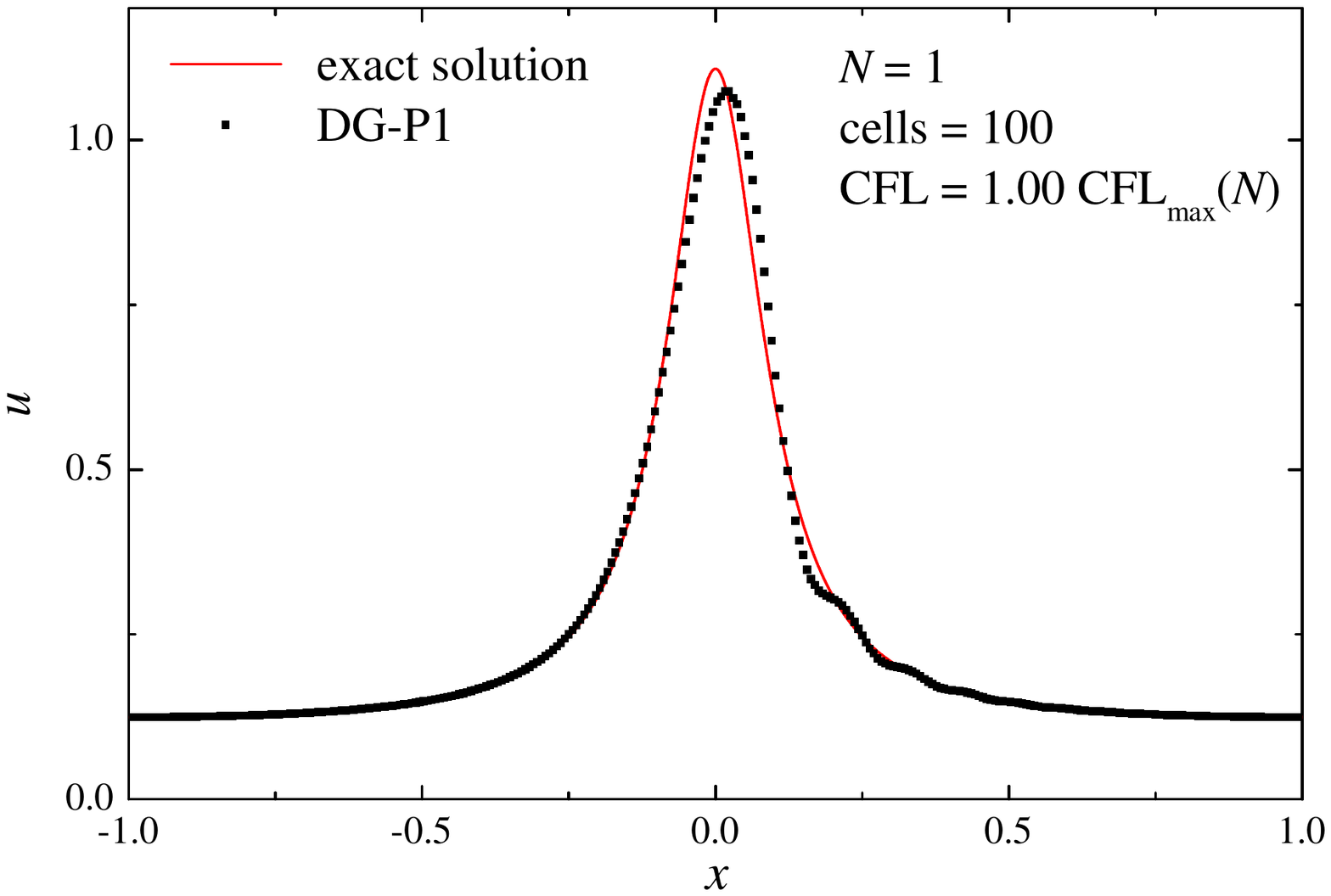}
\includegraphics[width=0.245\textwidth]{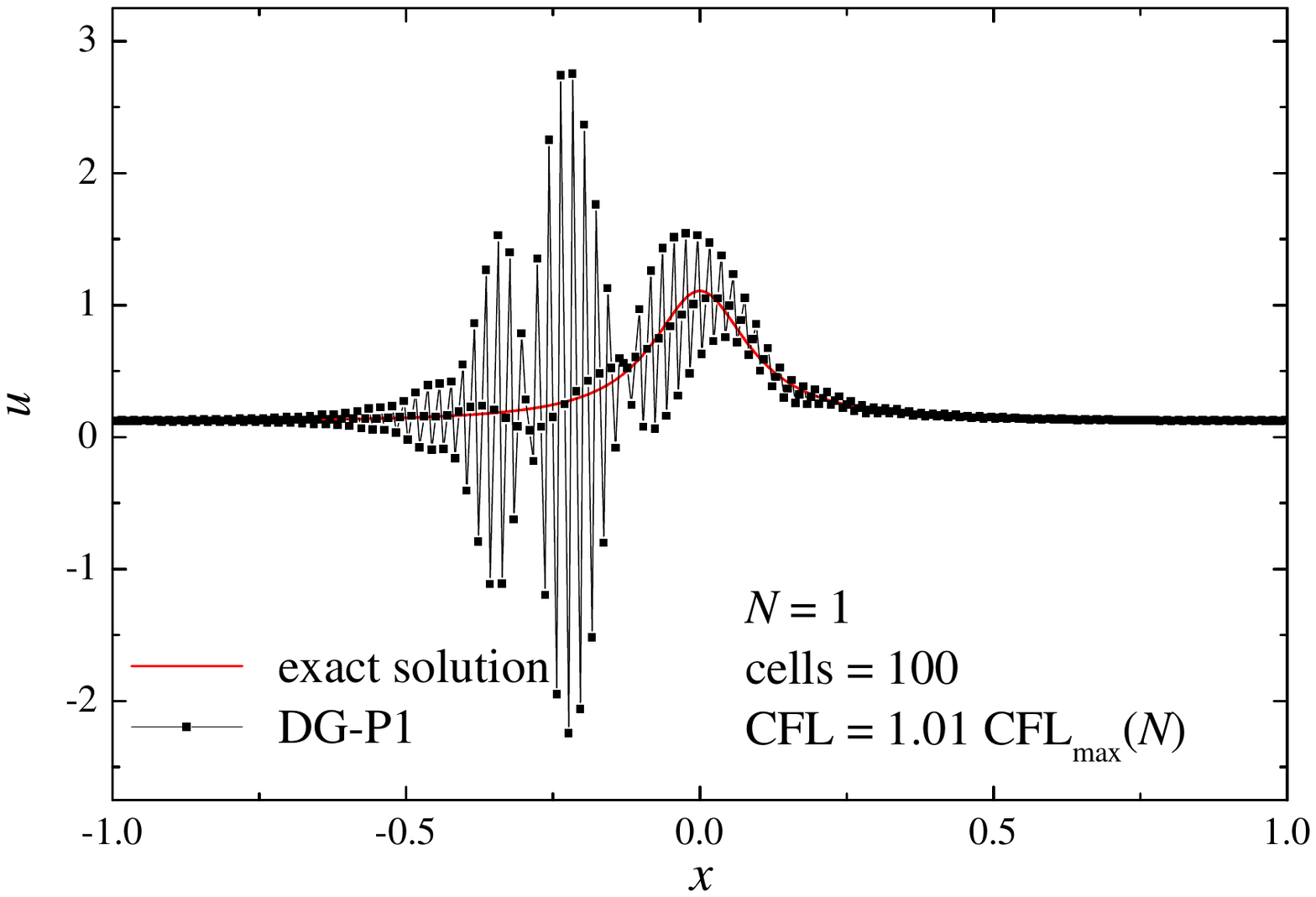}\\
\includegraphics[width=0.245\textwidth]{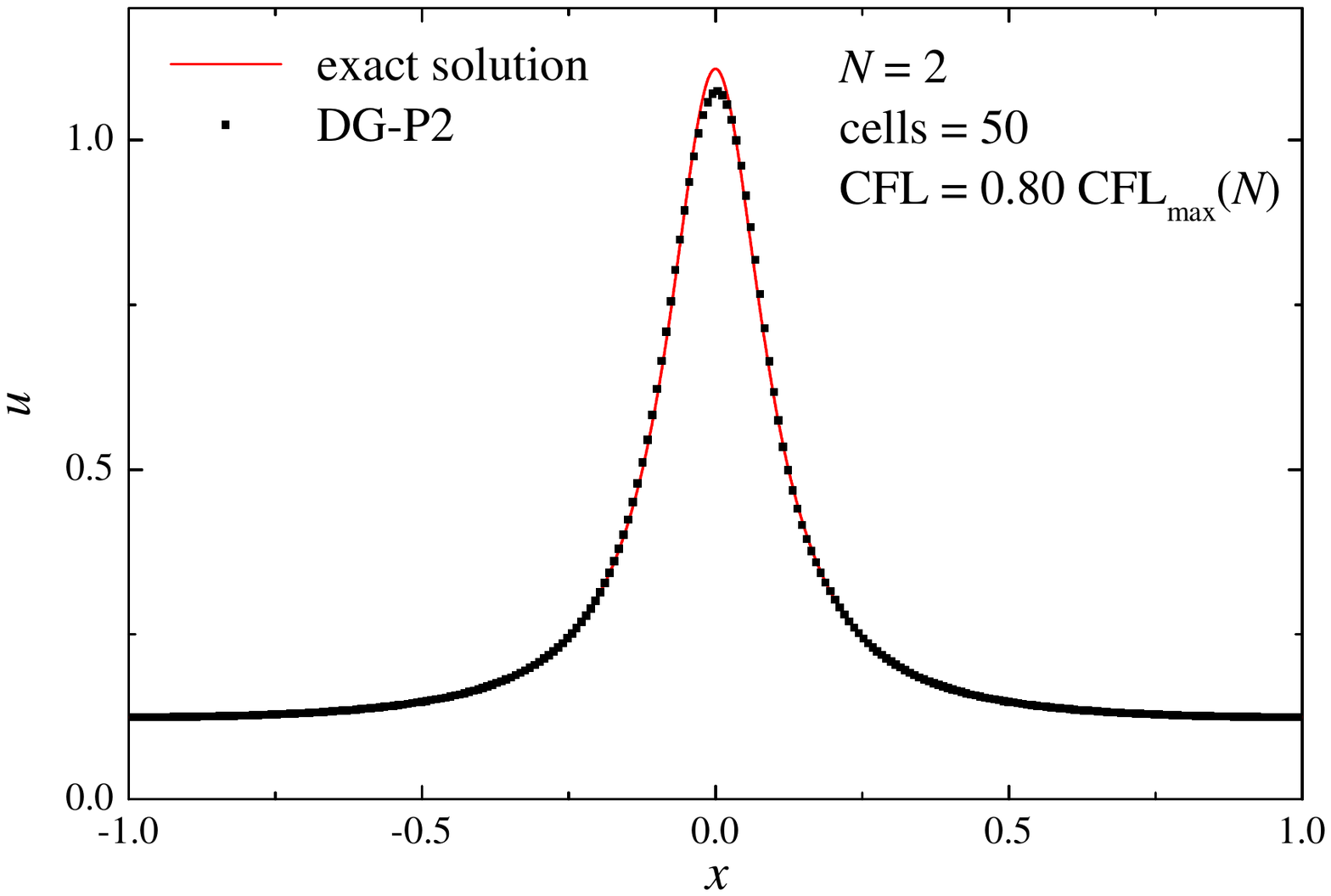}
\includegraphics[width=0.245\textwidth]{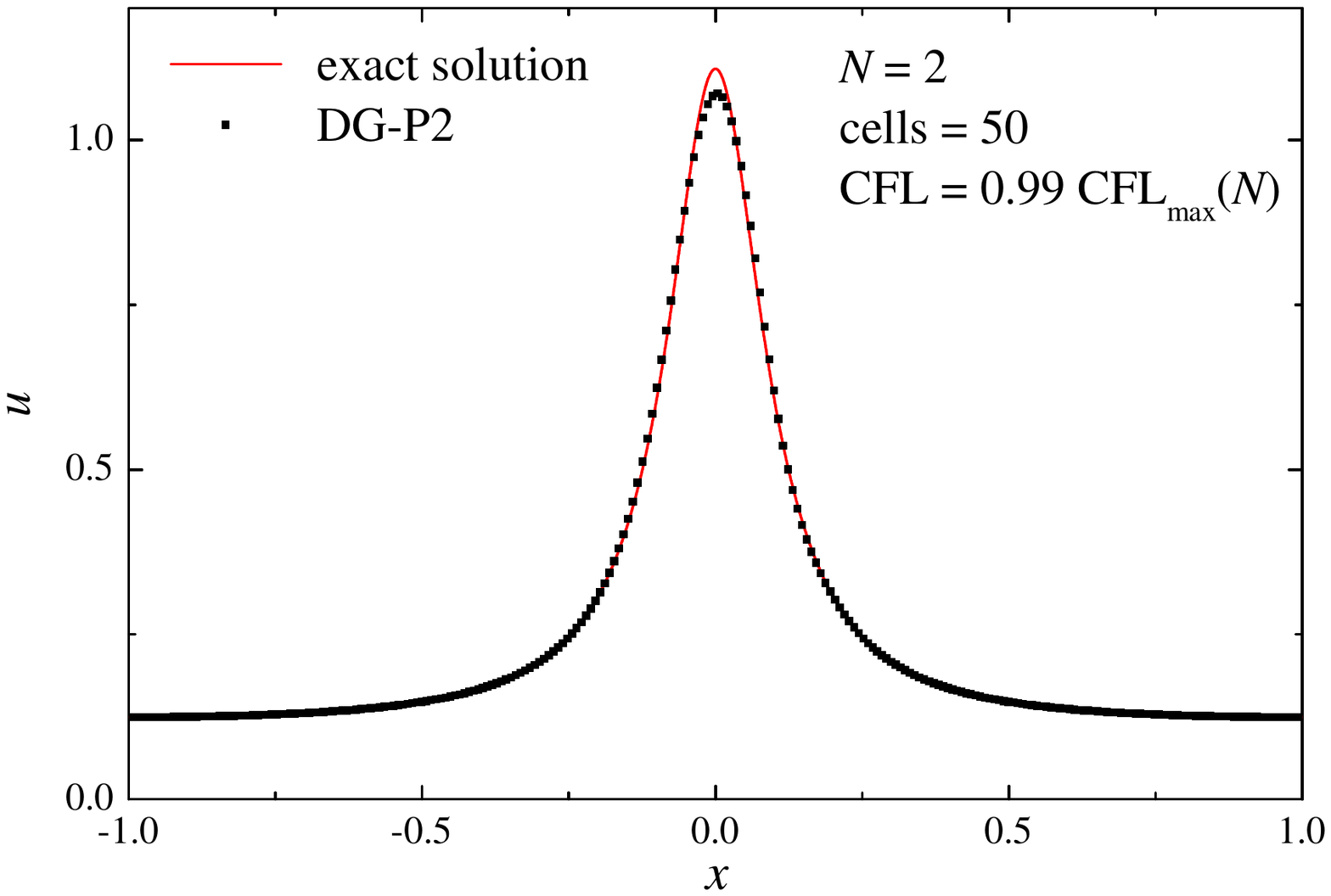}
\includegraphics[width=0.245\textwidth]{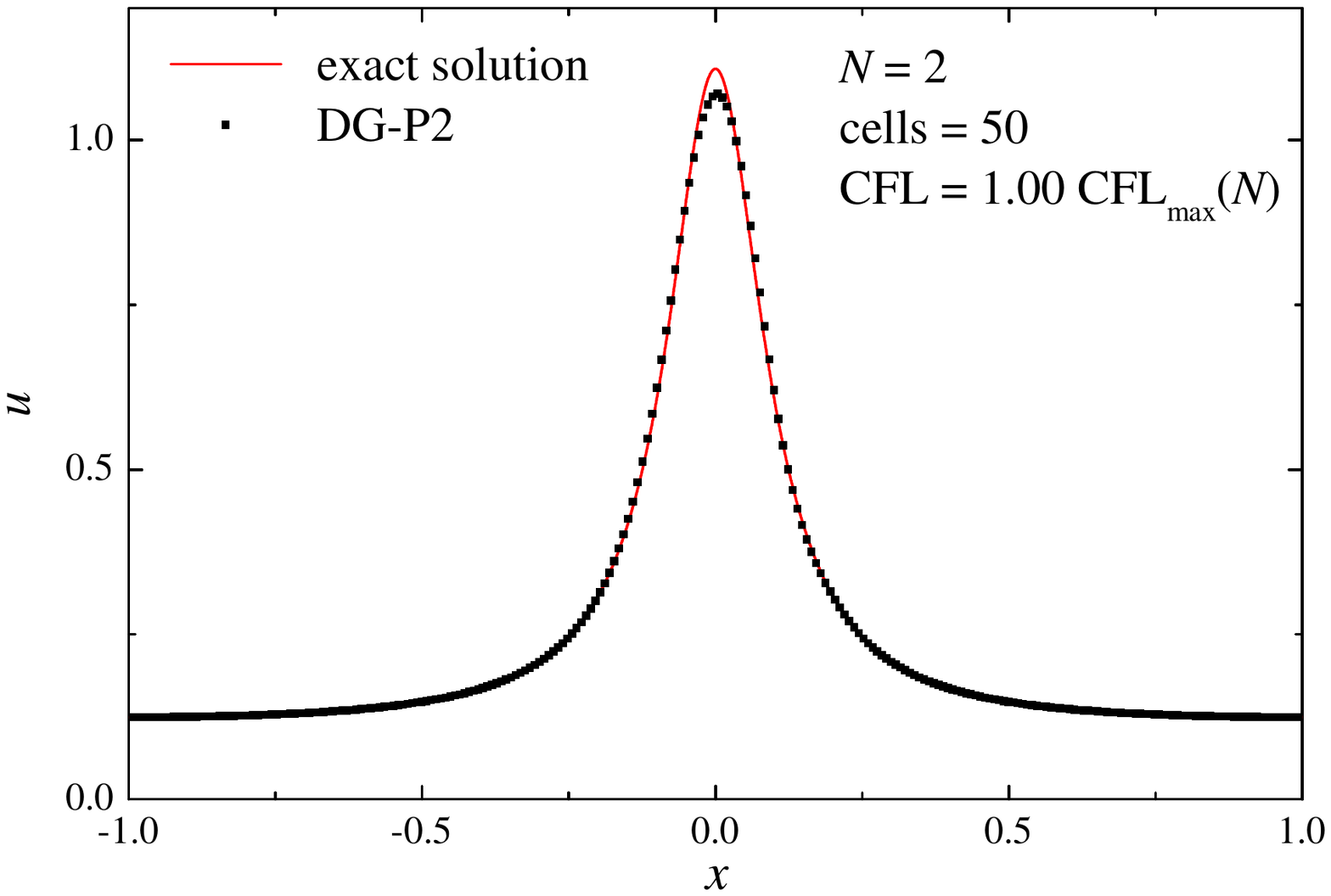}
\includegraphics[width=0.245\textwidth]{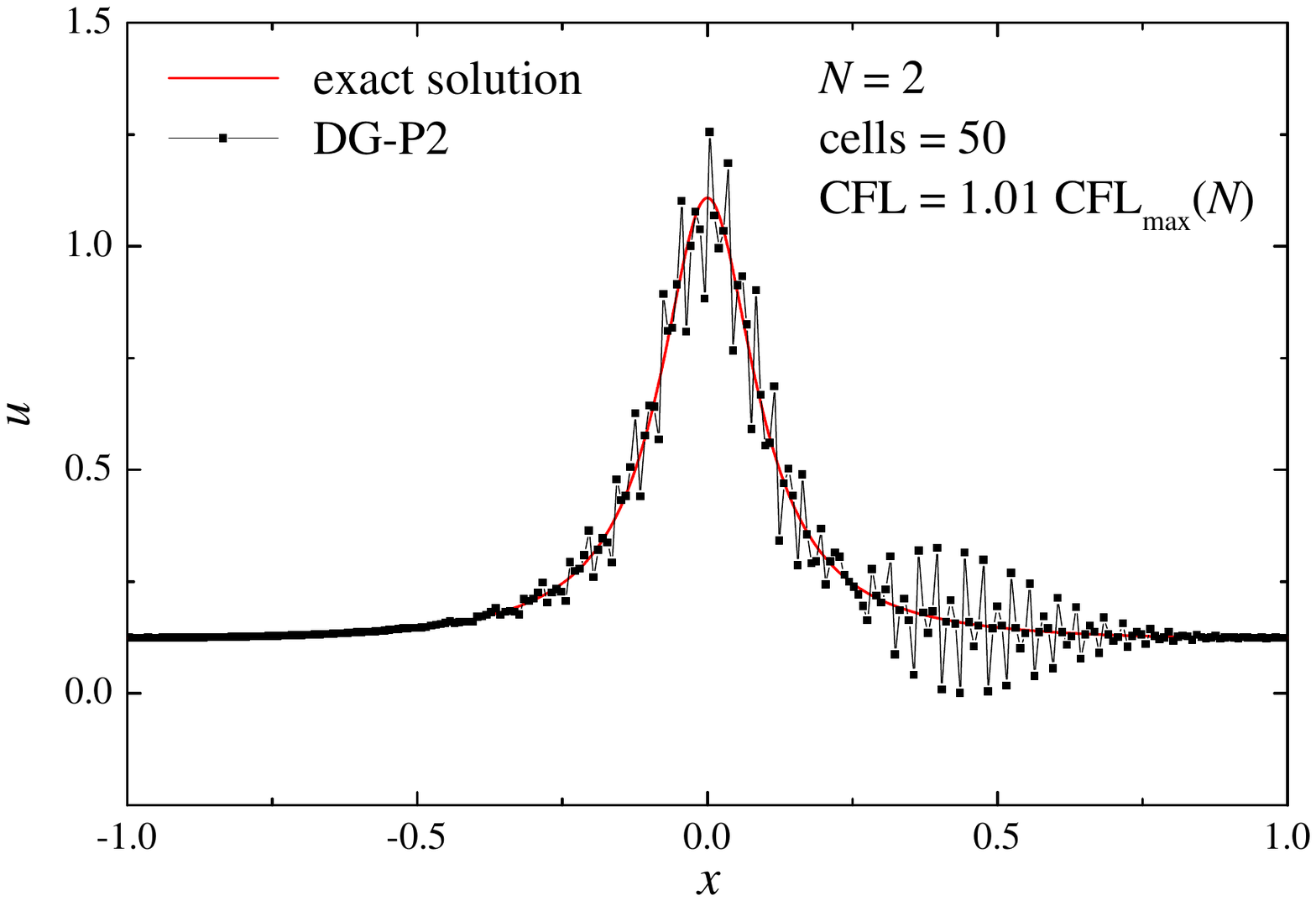}\\
\includegraphics[width=0.245\textwidth]{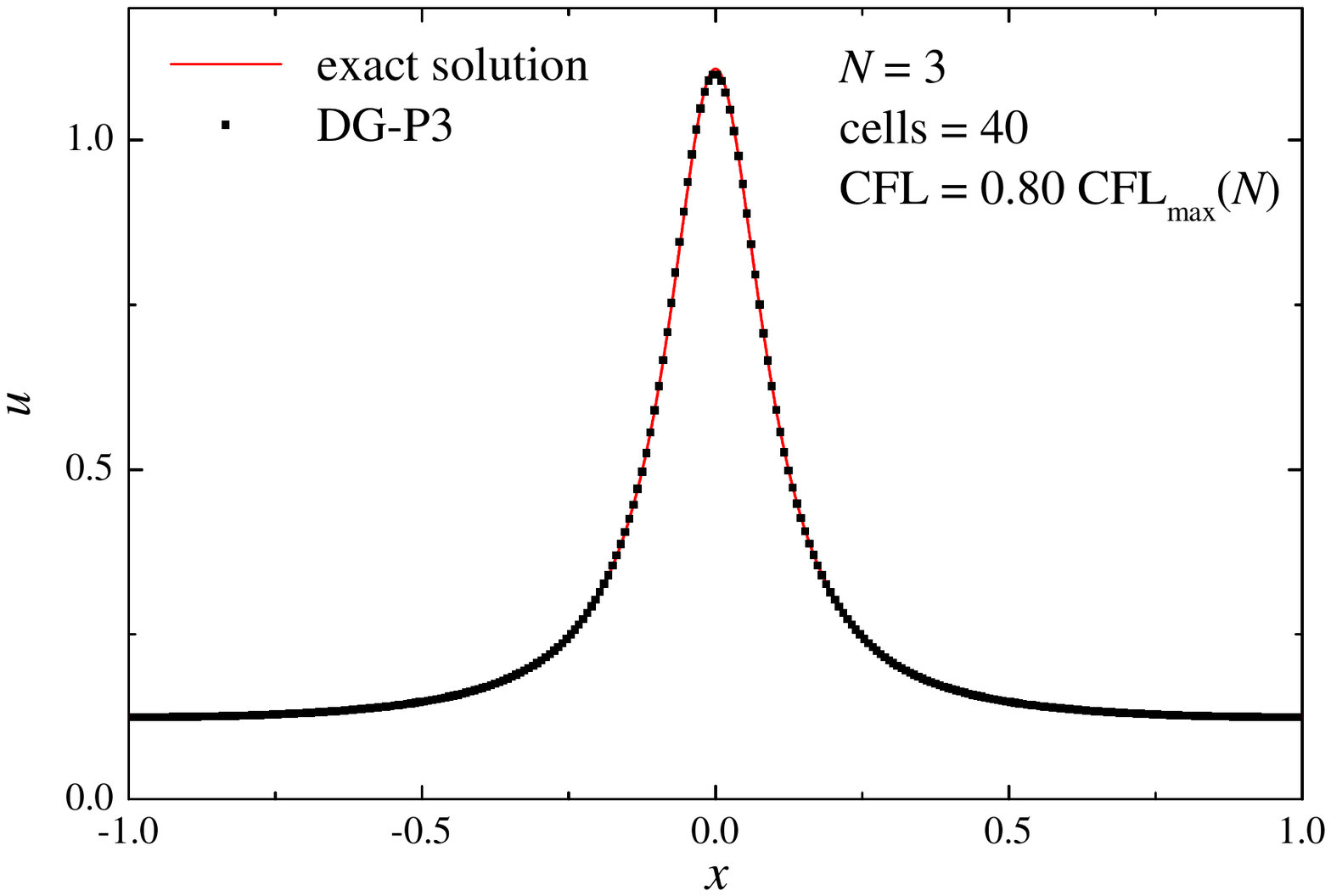}
\includegraphics[width=0.245\textwidth]{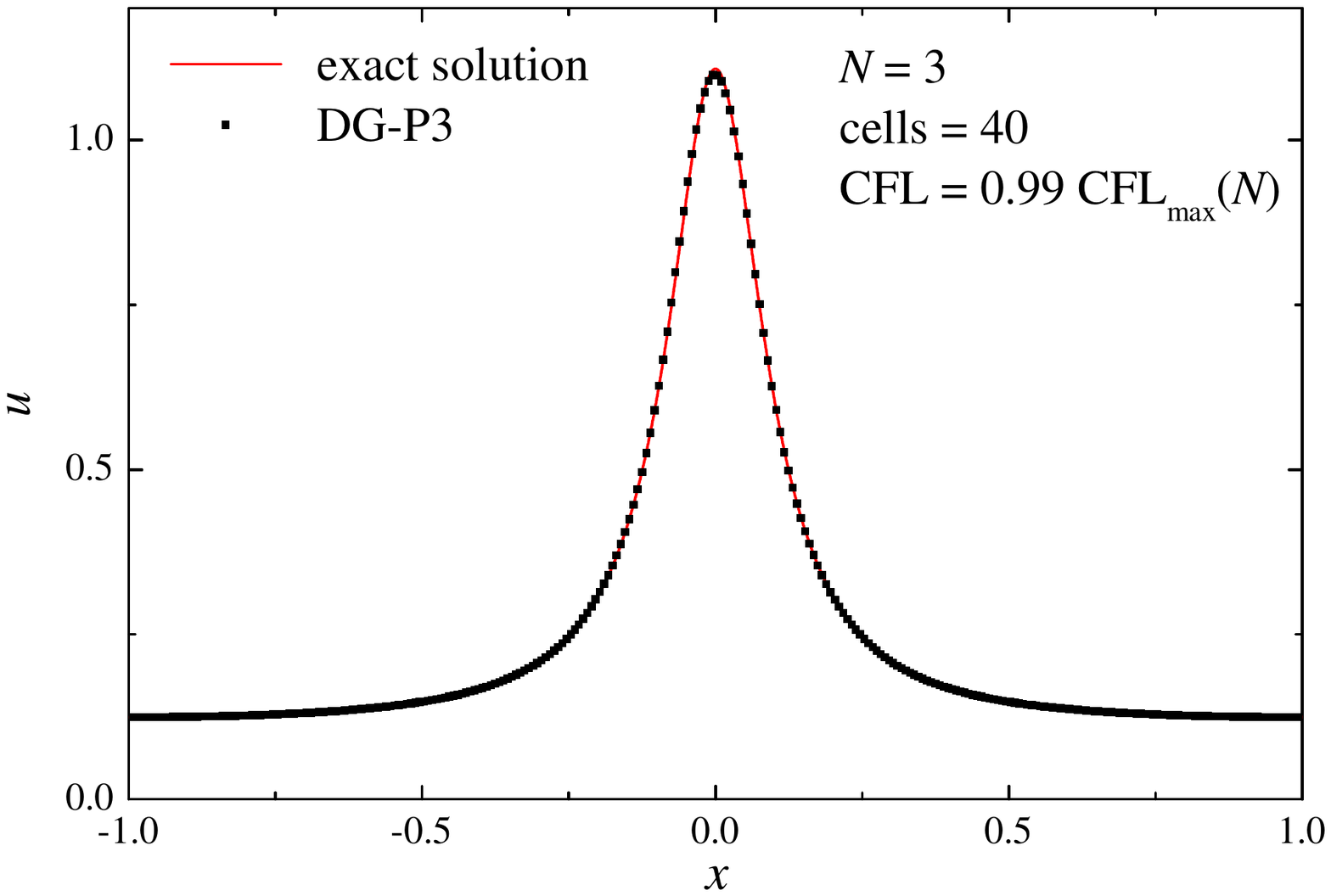}
\includegraphics[width=0.245\textwidth]{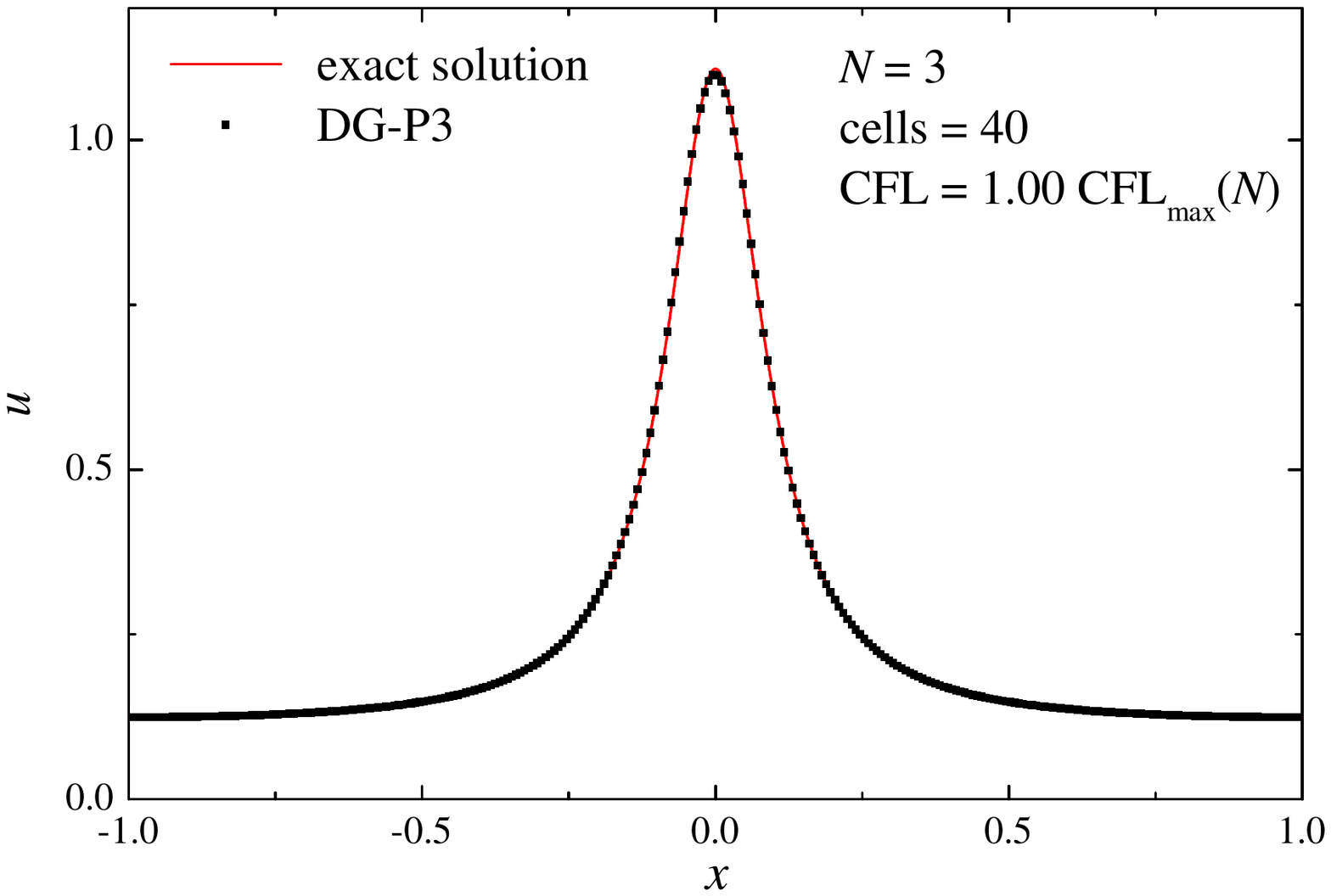}
\includegraphics[width=0.245\textwidth]{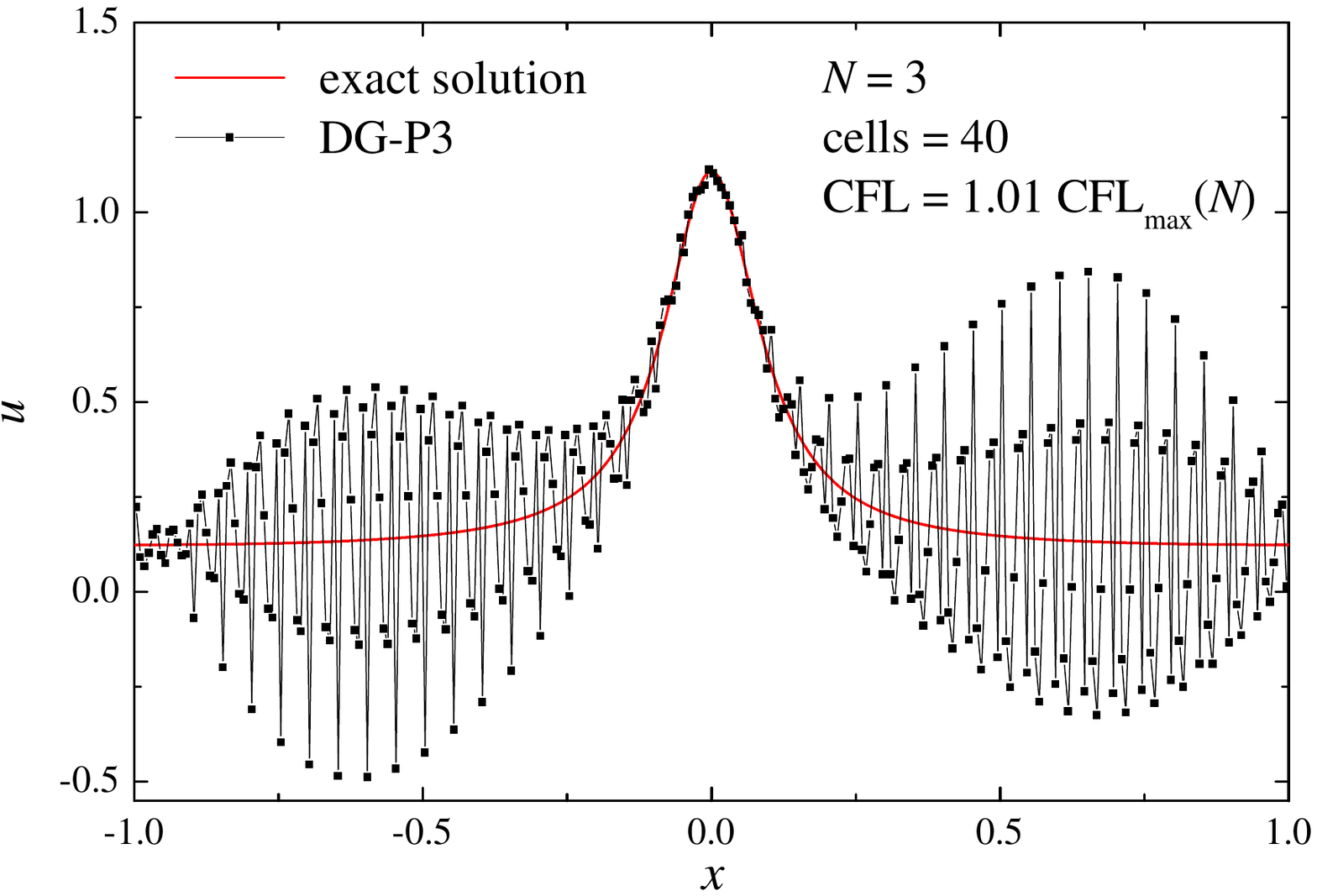}\\
\includegraphics[width=0.245\textwidth]{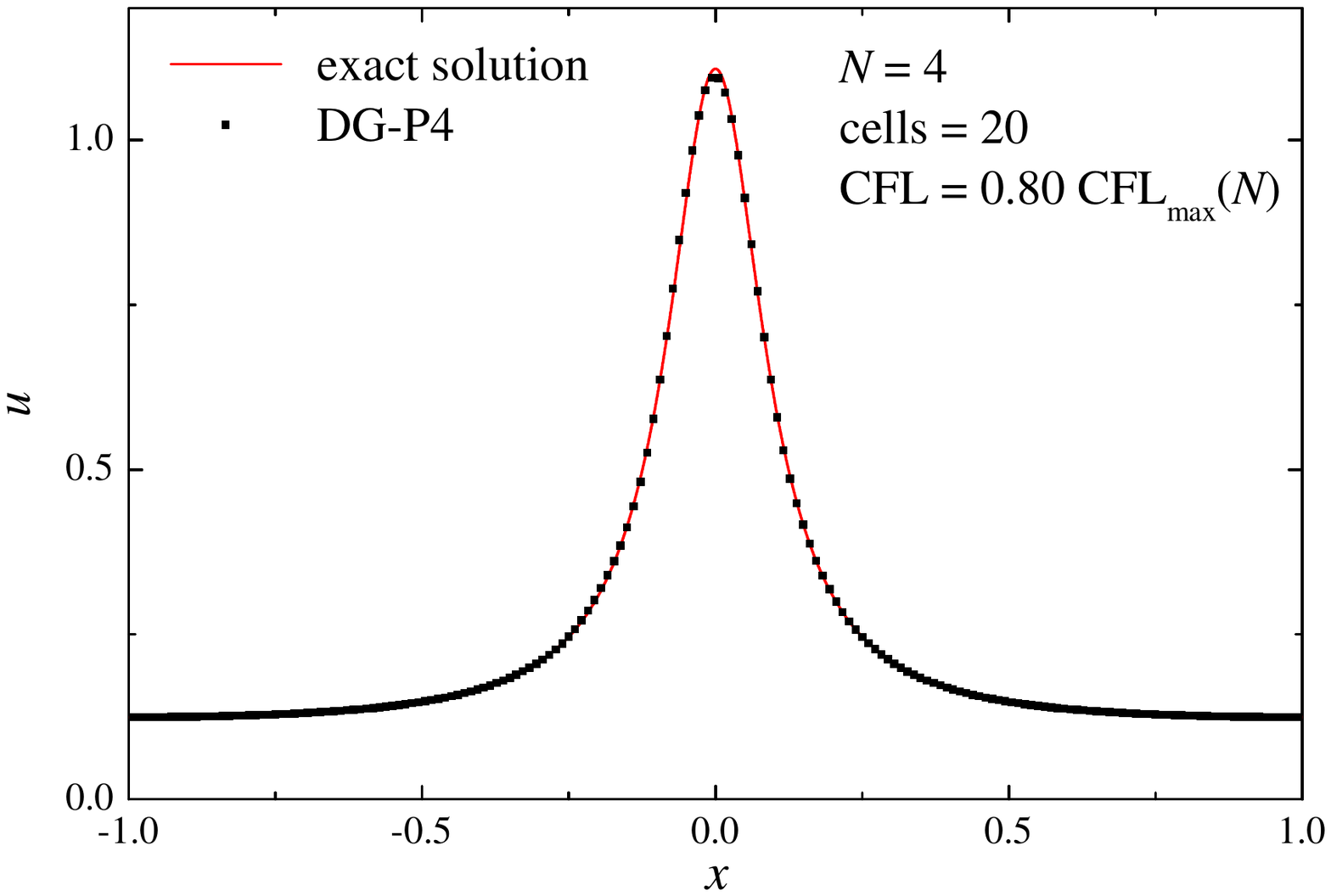}
\includegraphics[width=0.245\textwidth]{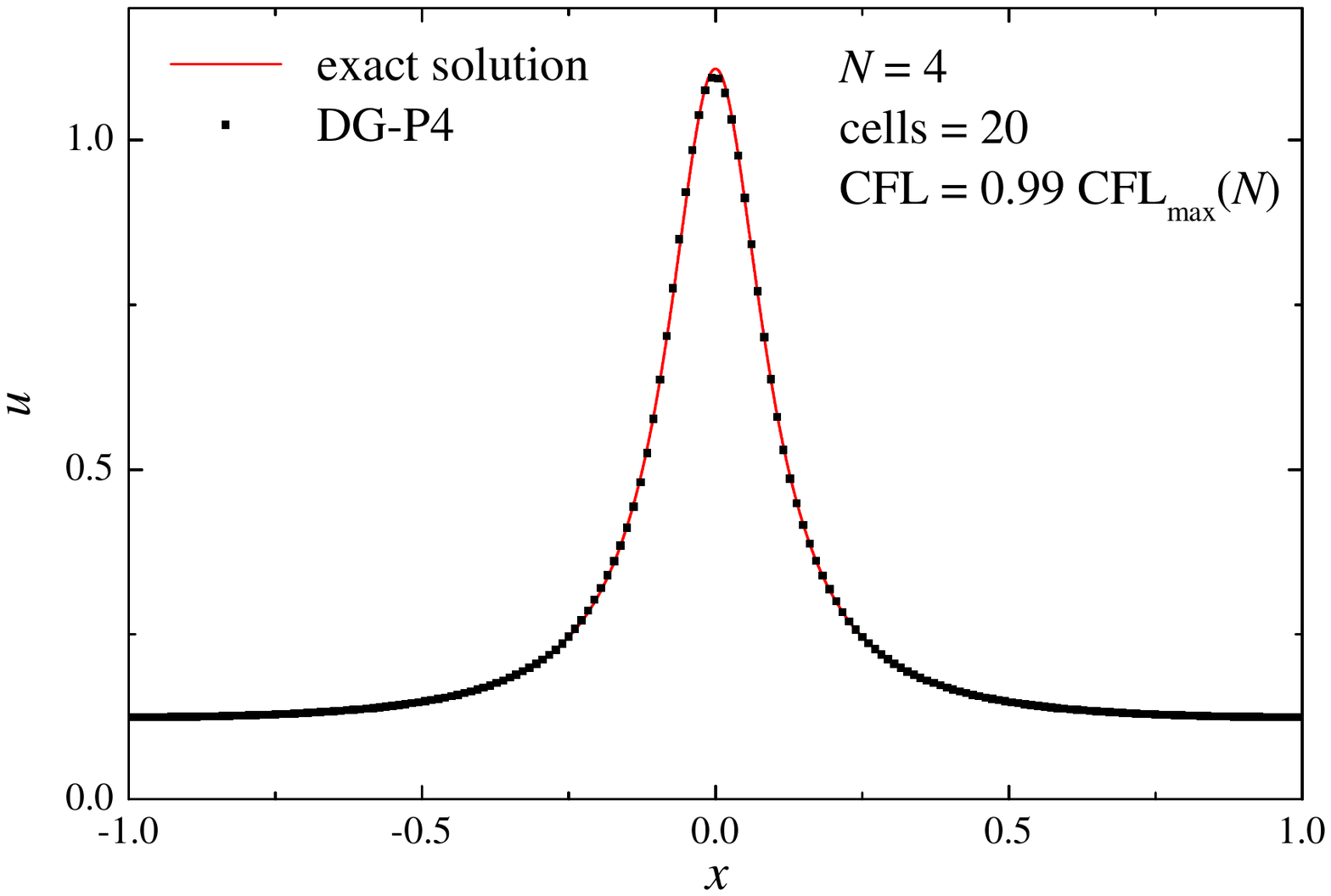}
\includegraphics[width=0.245\textwidth]{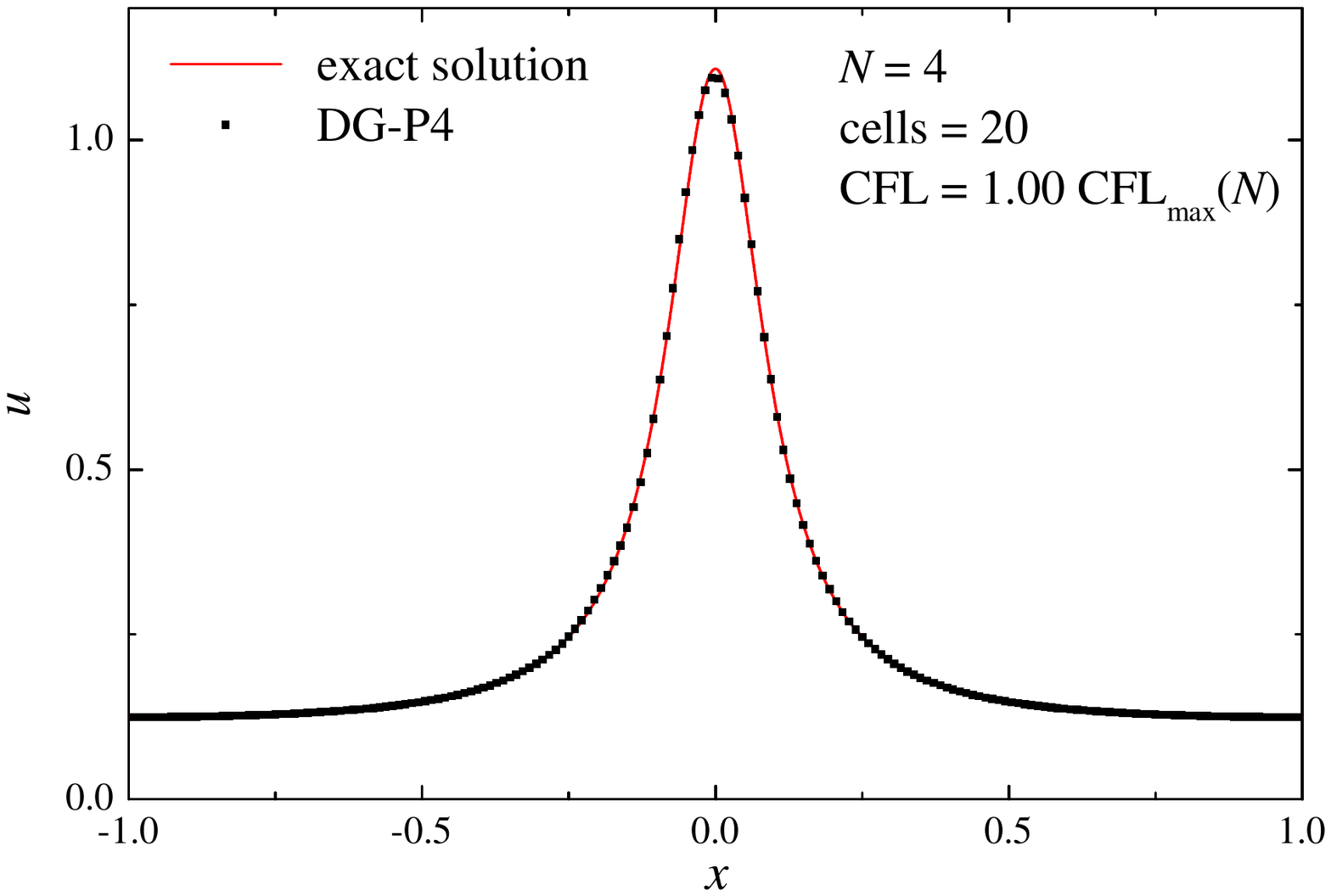}
\includegraphics[width=0.245\textwidth]{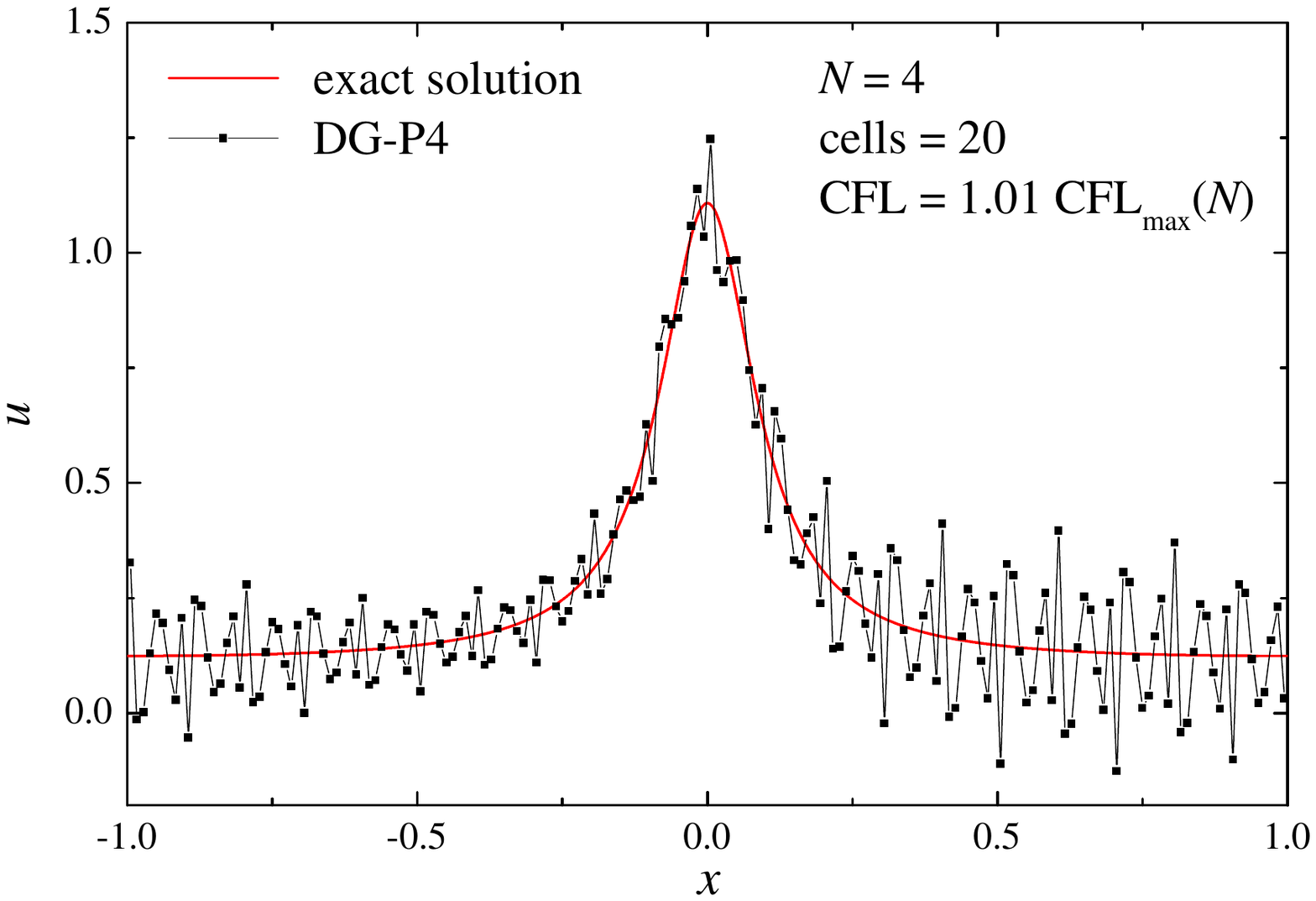}\\
\includegraphics[width=0.245\textwidth]{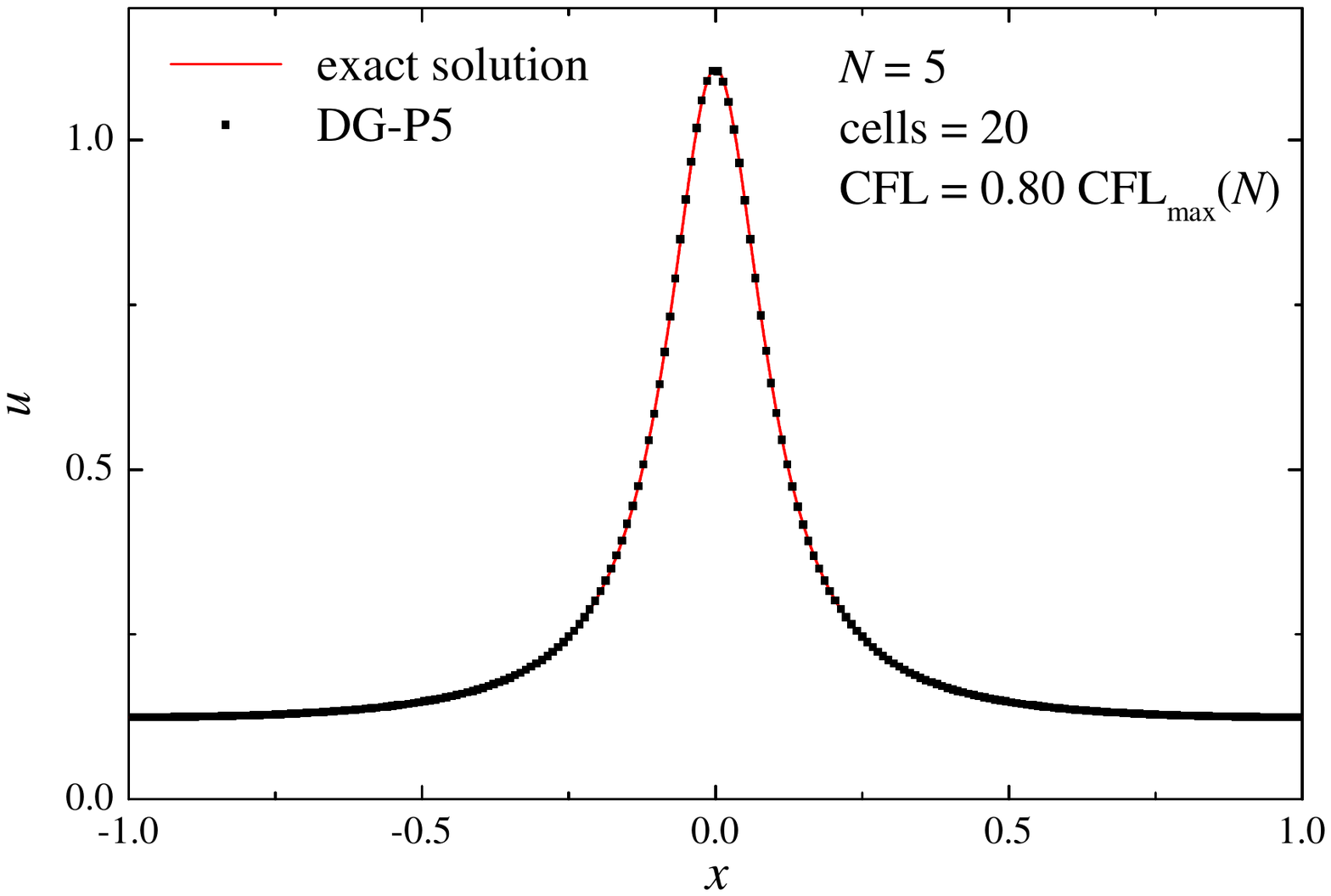}
\includegraphics[width=0.245\textwidth]{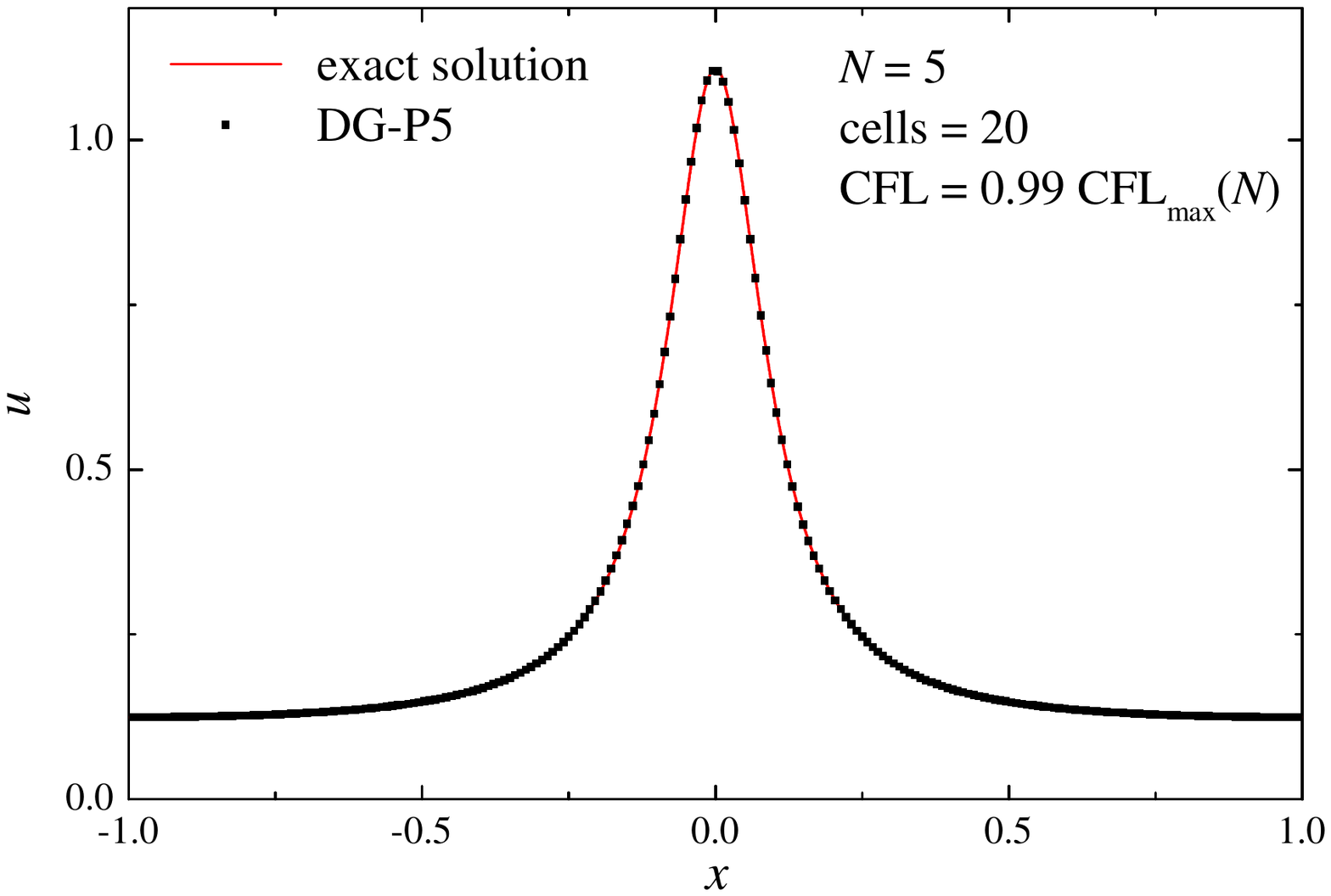}
\includegraphics[width=0.245\textwidth]{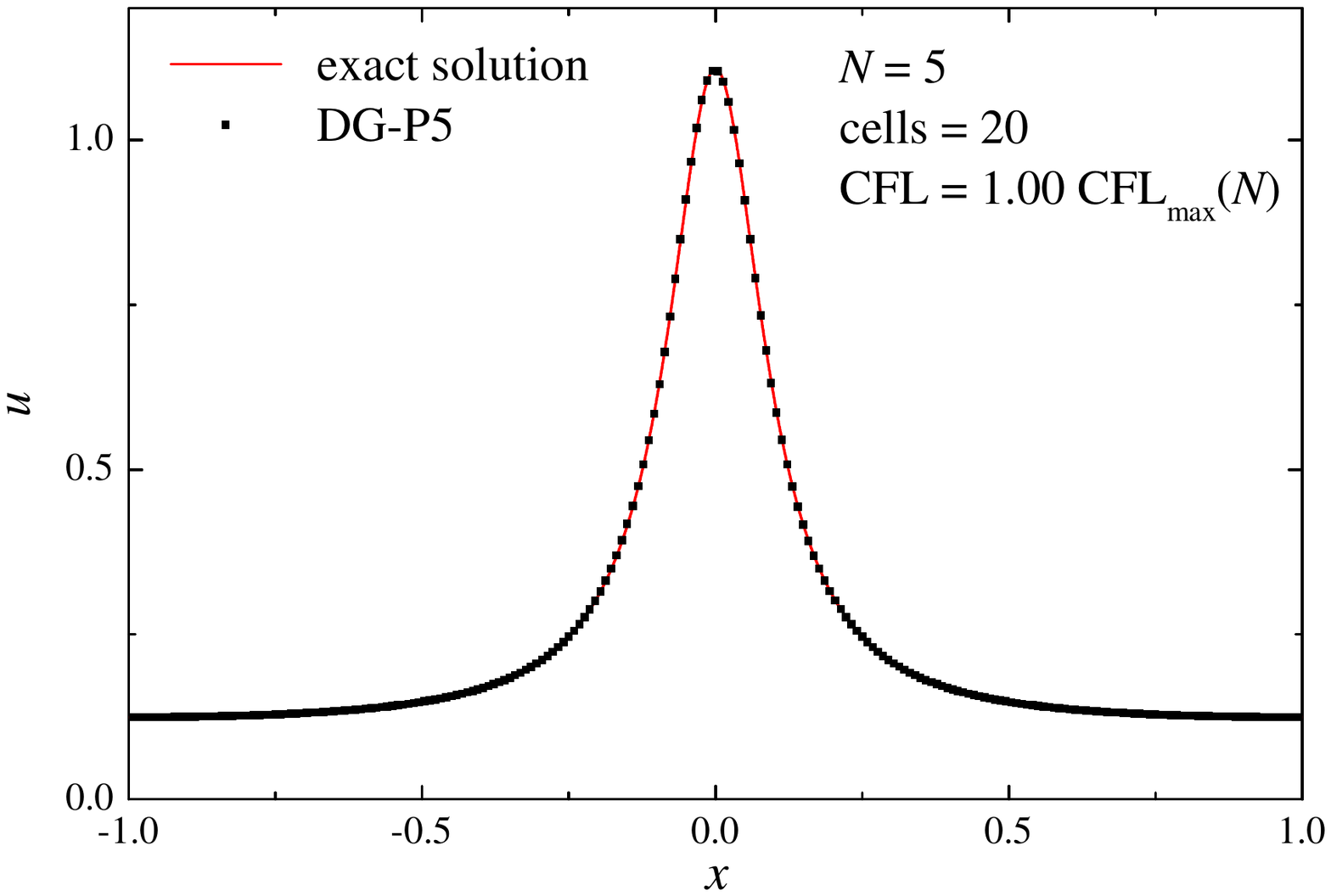}
\includegraphics[width=0.245\textwidth]{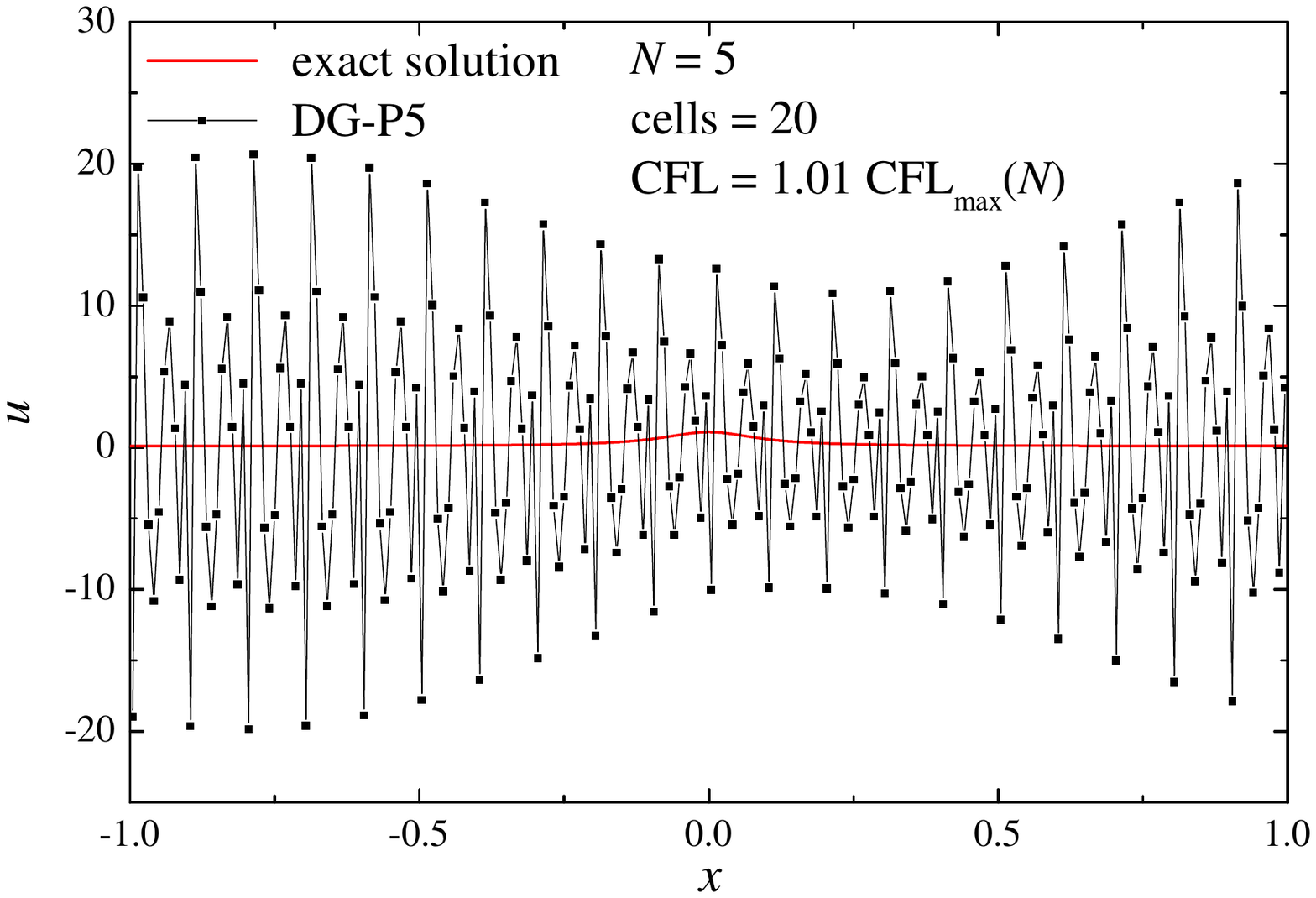}\\
\includegraphics[width=0.245\textwidth]{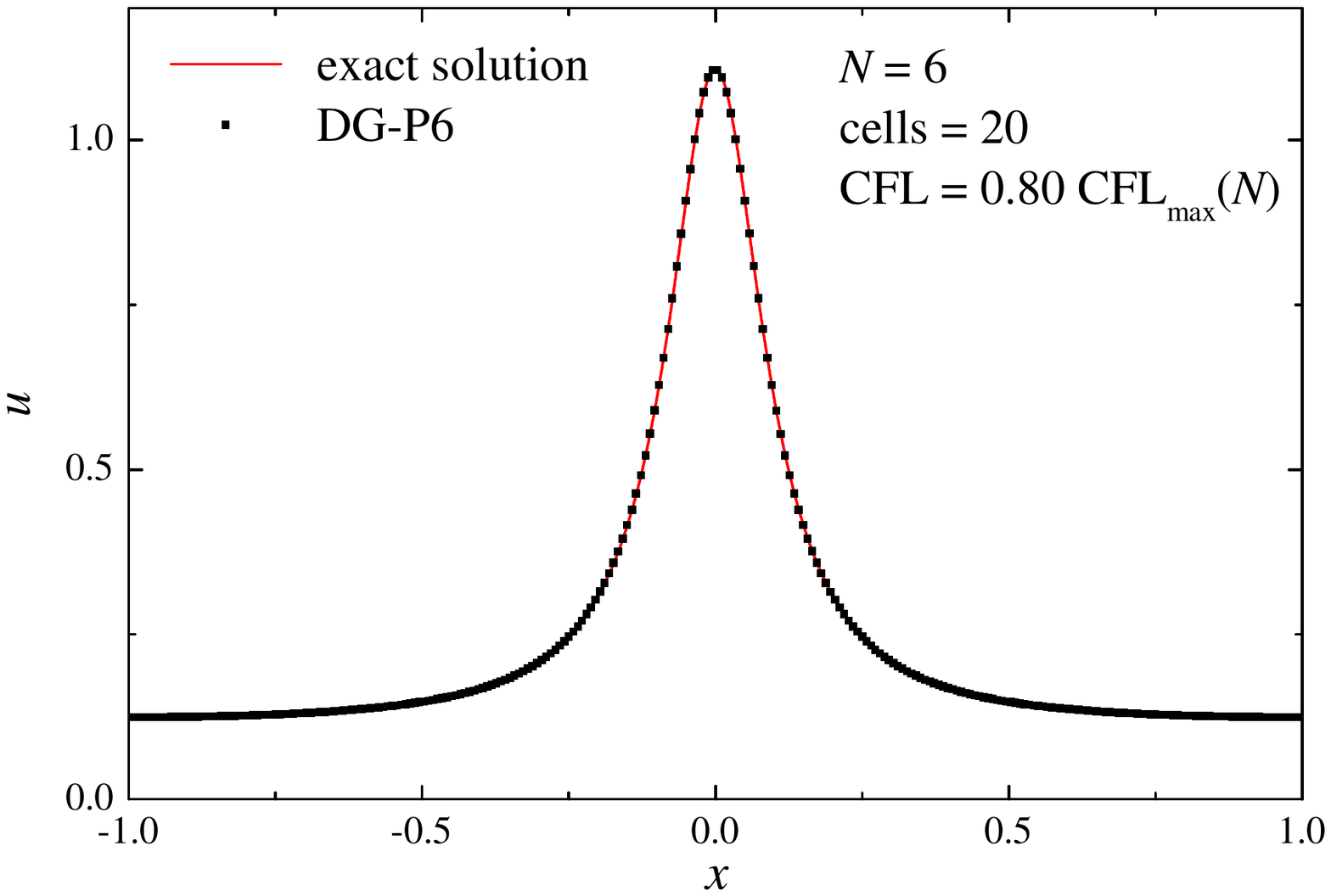}
\includegraphics[width=0.245\textwidth]{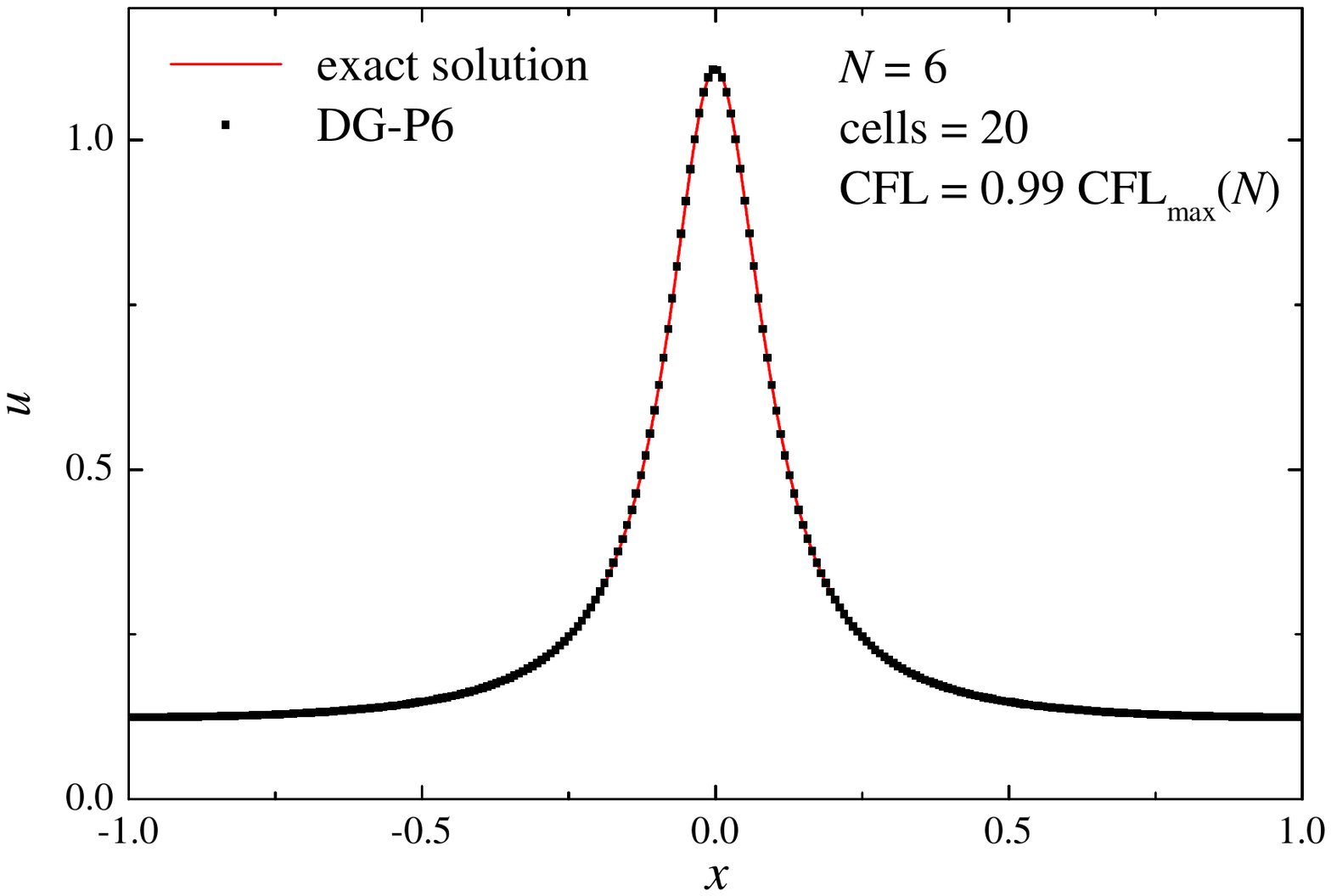}
\includegraphics[width=0.245\textwidth]{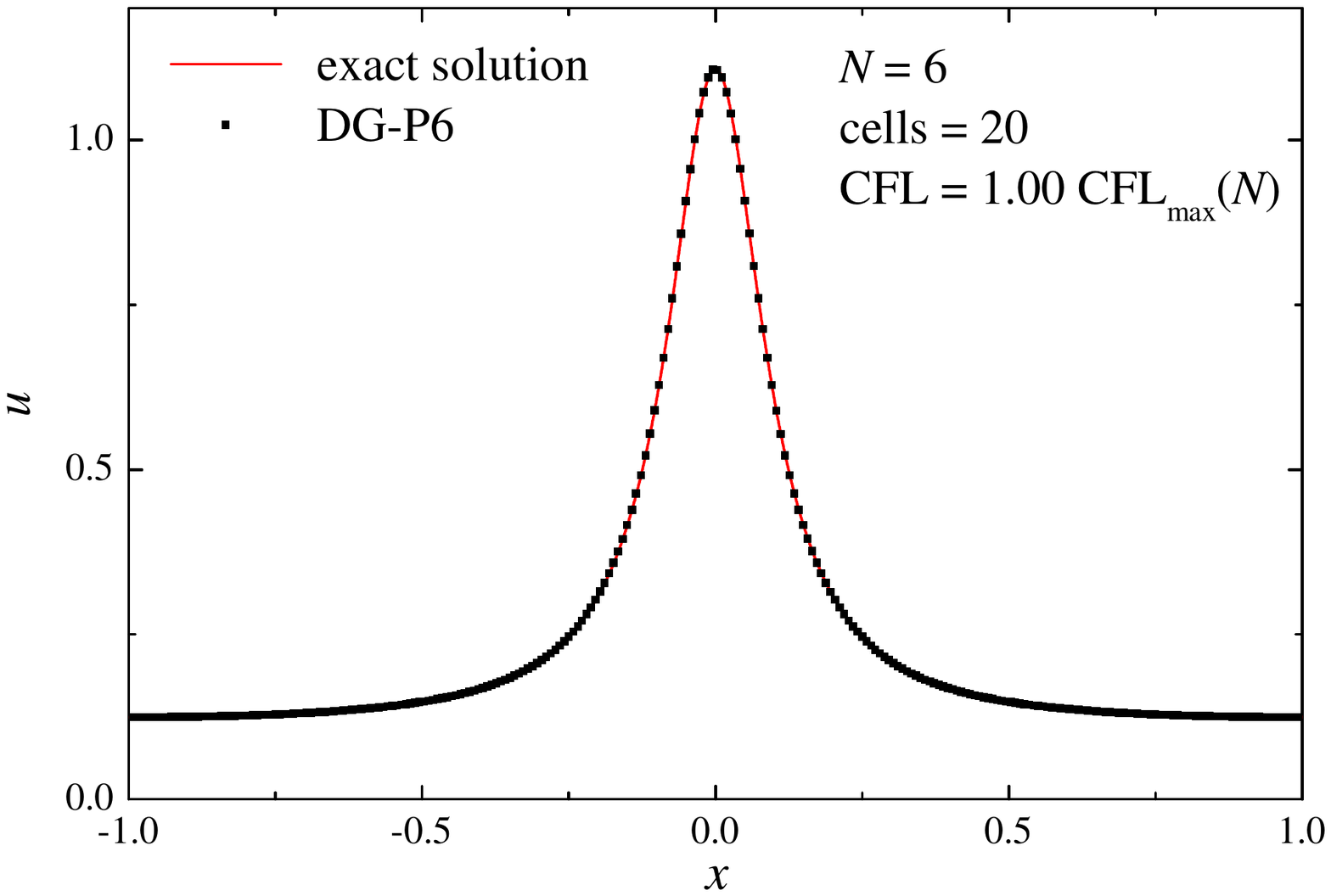}
\includegraphics[width=0.245\textwidth]{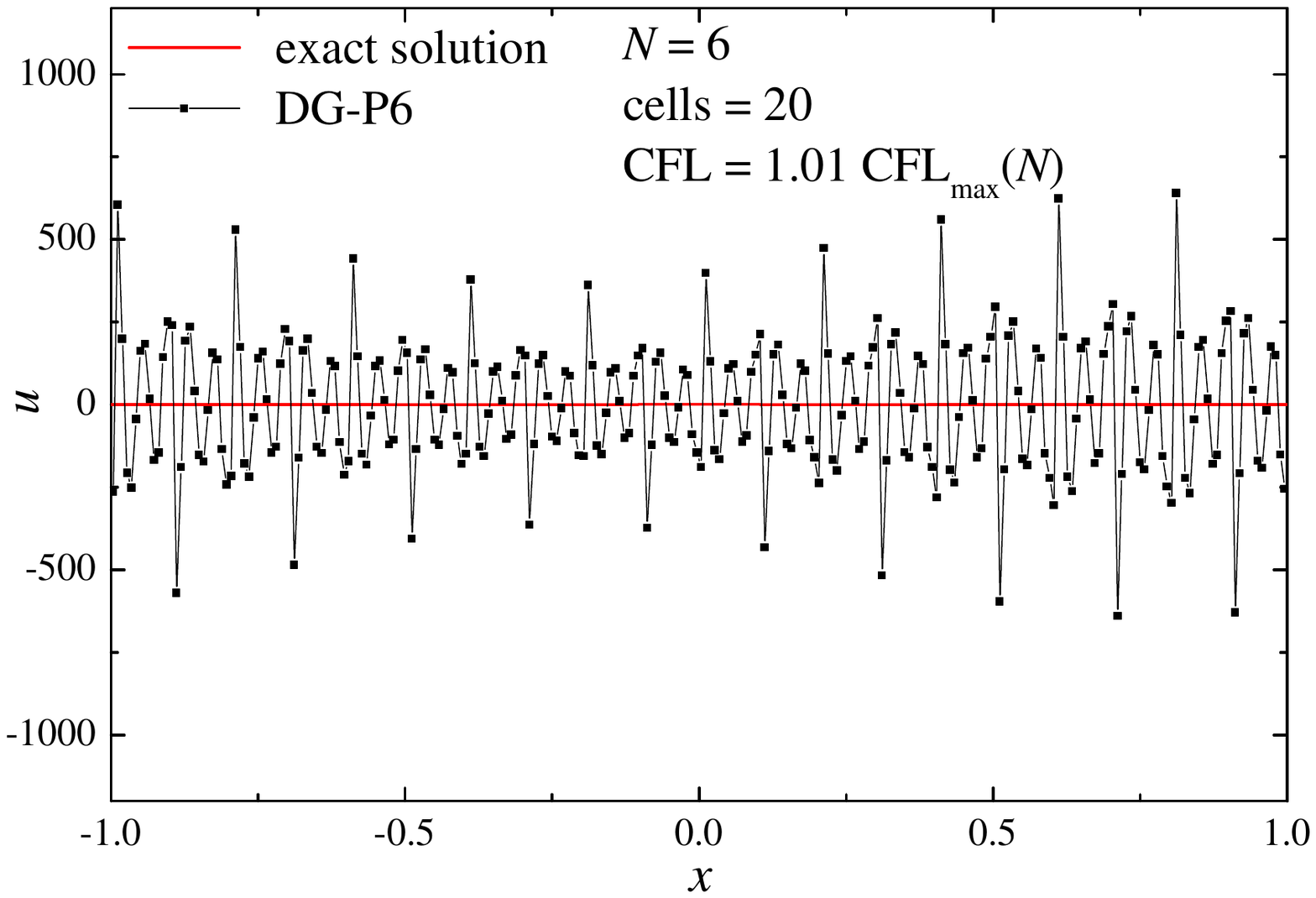}
\caption{%
Coordinate dependencies of the solution $u(x, t_{f})$ to the linear advection equation (\ref{eq:adv_eq_src}) at the final time $t_{f}$, obtained by the ADER-DG numerical method with the LST-DG predictor, for the degrees $N = 1, \ldots, 6$ of the basis polynomials (top to bottom) for a set of Courant number values $\mathrm{CFL} = 0.80$, $0.99$, $1.00$, $1.01\, \mathrm{CFL}_{\rm max}(N)$ (left to right). The first two columns correspond to the interior $\mathrm{CFL} < \mathrm{CFL}_{\rm max}(N)$ of the stability region, the third column correspond to the boundary $\mathrm{CFL} = \mathrm{CFL}_{\rm max}(N)$ of the stability region, and the right column correspond to the outside $\mathrm{CFL} > \mathrm{CFL}_{\rm max}(N)$ of the boundary of the stability region (the Courant number $\mathrm{CFL}$ is 1\% higher than the boundary value $\mathrm{CFL}_{\rm max}(N)$). The red line represents the exact analytical solution.
}
\label{fig:test_adveq_lorentz_1d_cfls_degrees_1_6}
\end{figure}

\begin{figure}[h!]
\includegraphics[width=0.245\textwidth]{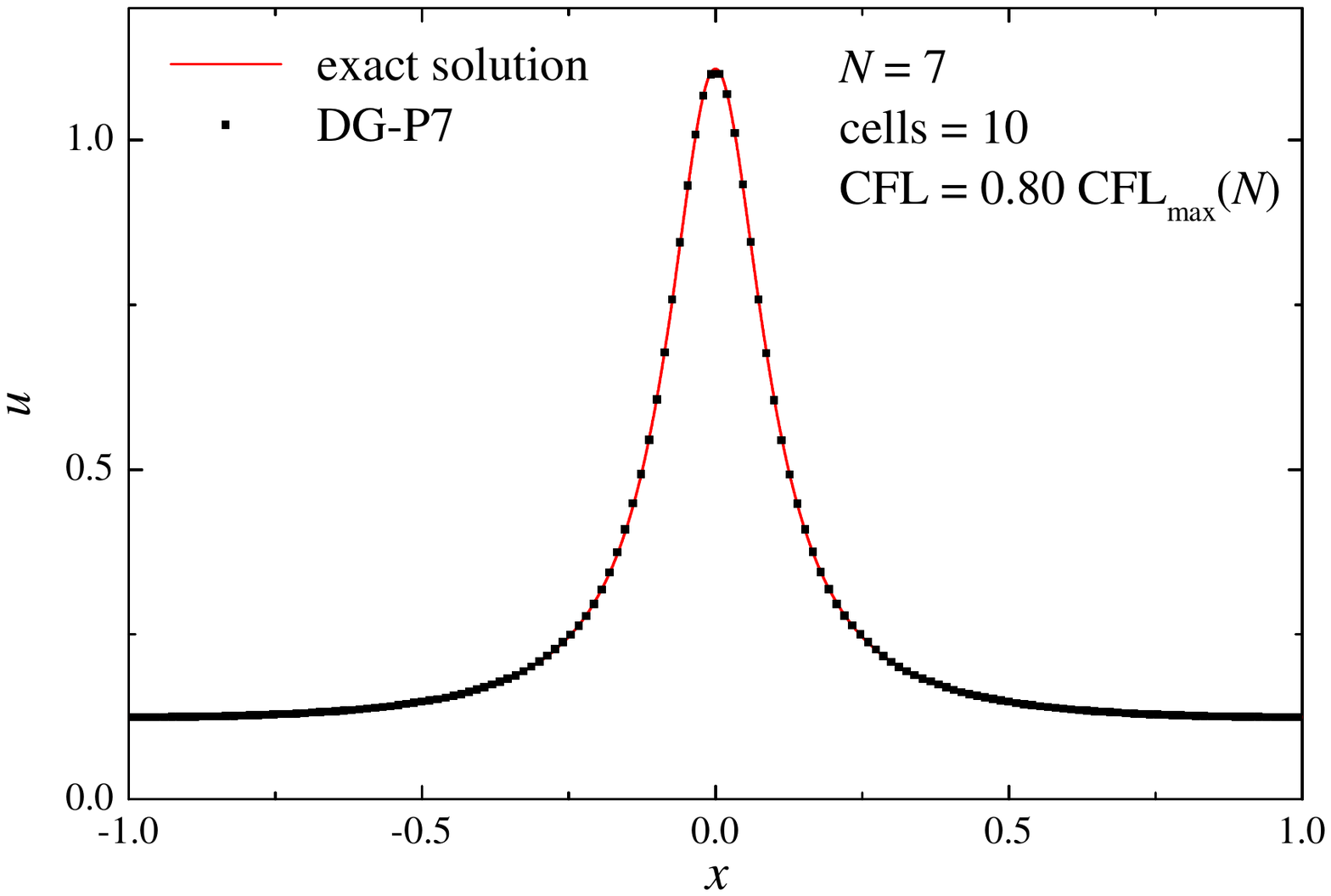}
\includegraphics[width=0.245\textwidth]{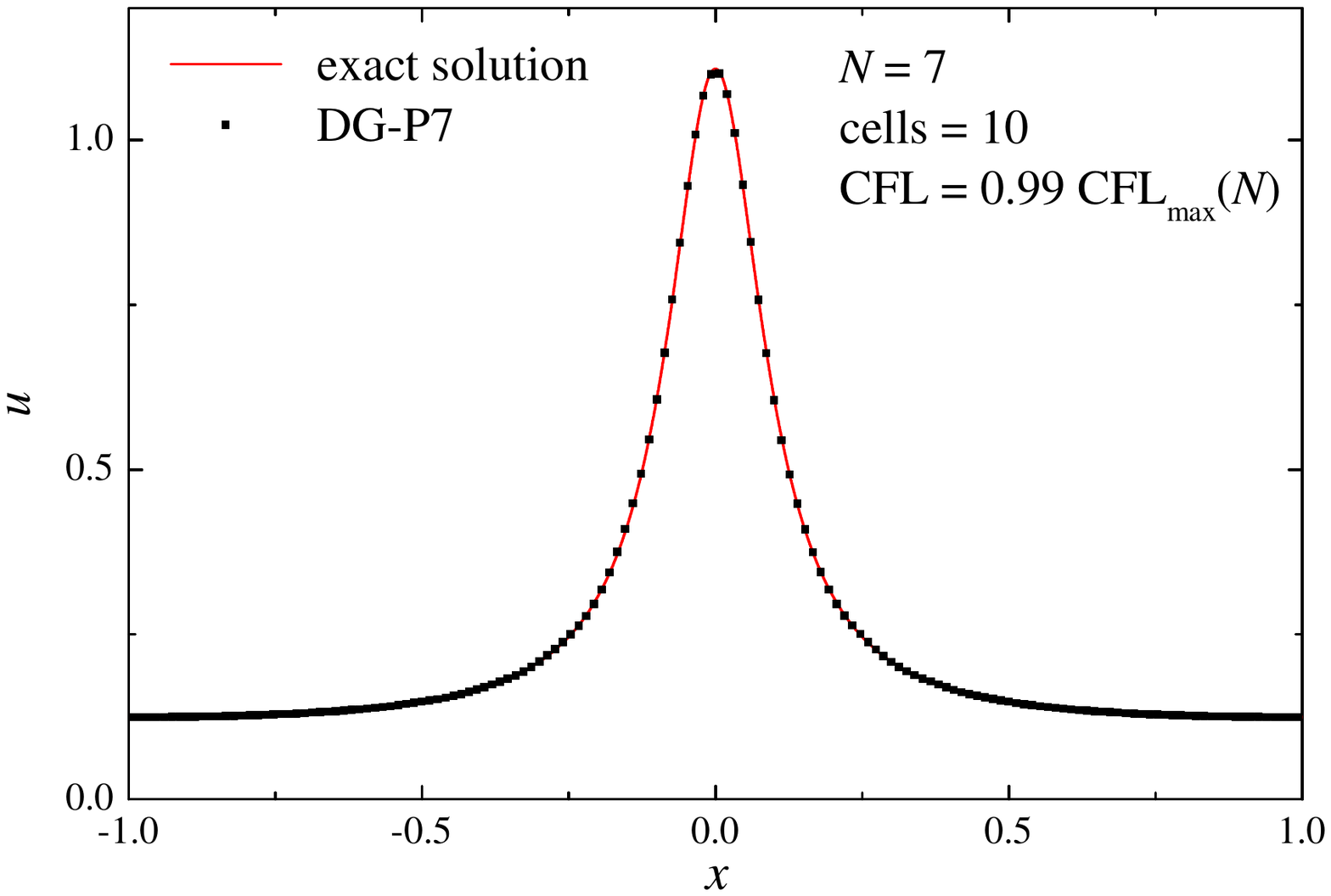}
\includegraphics[width=0.245\textwidth]{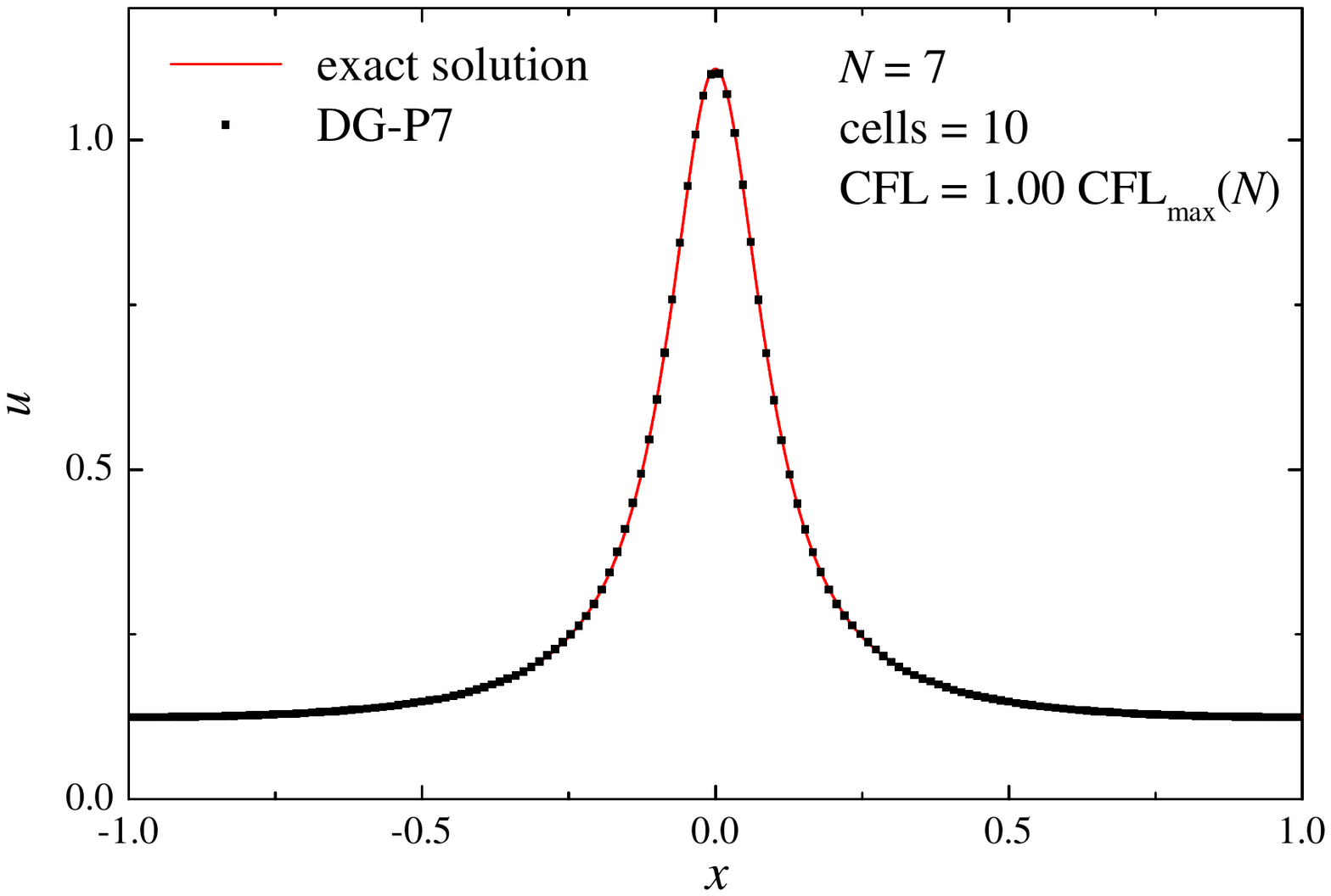}
\includegraphics[width=0.245\textwidth]{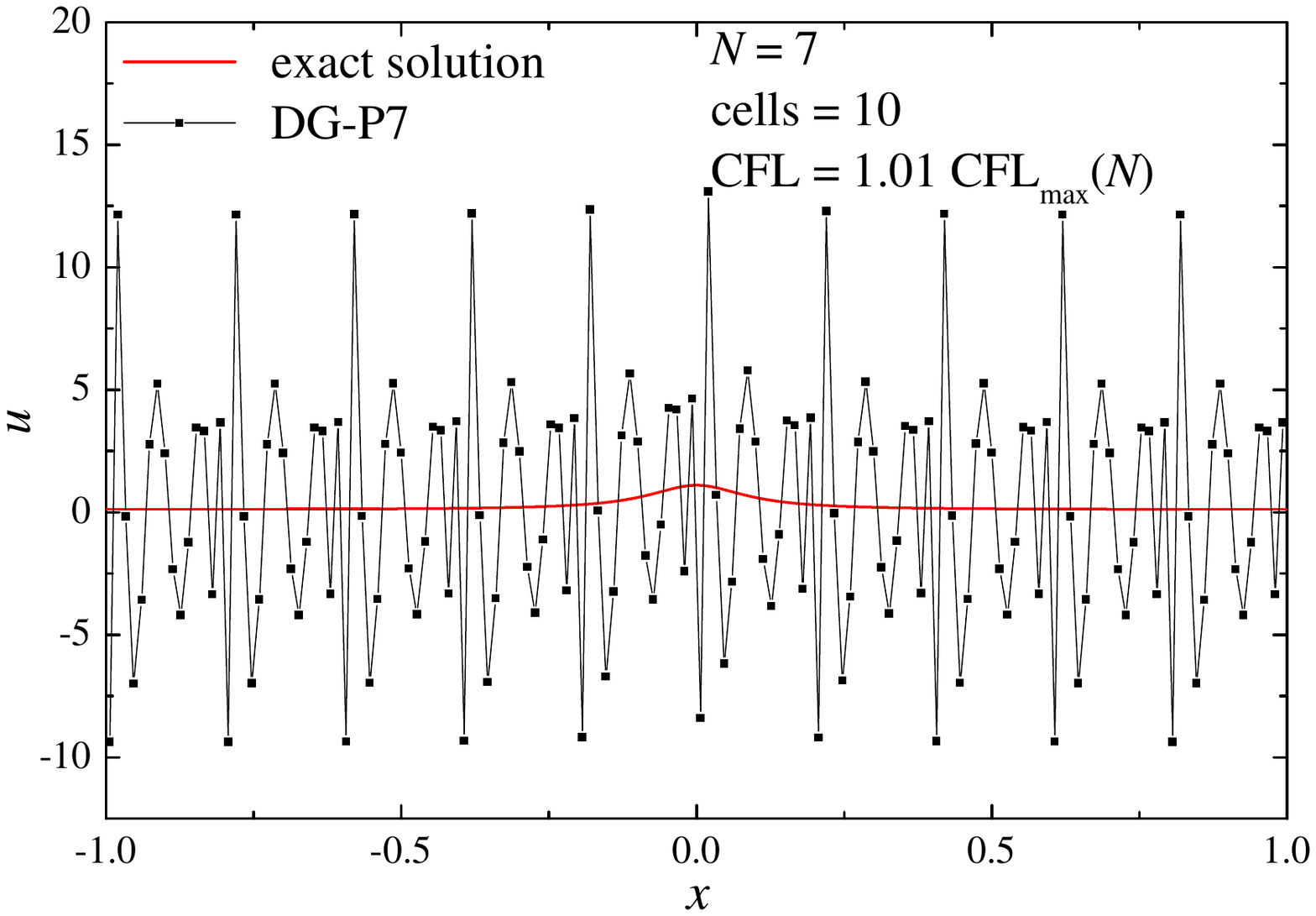}\\
\includegraphics[width=0.245\textwidth]{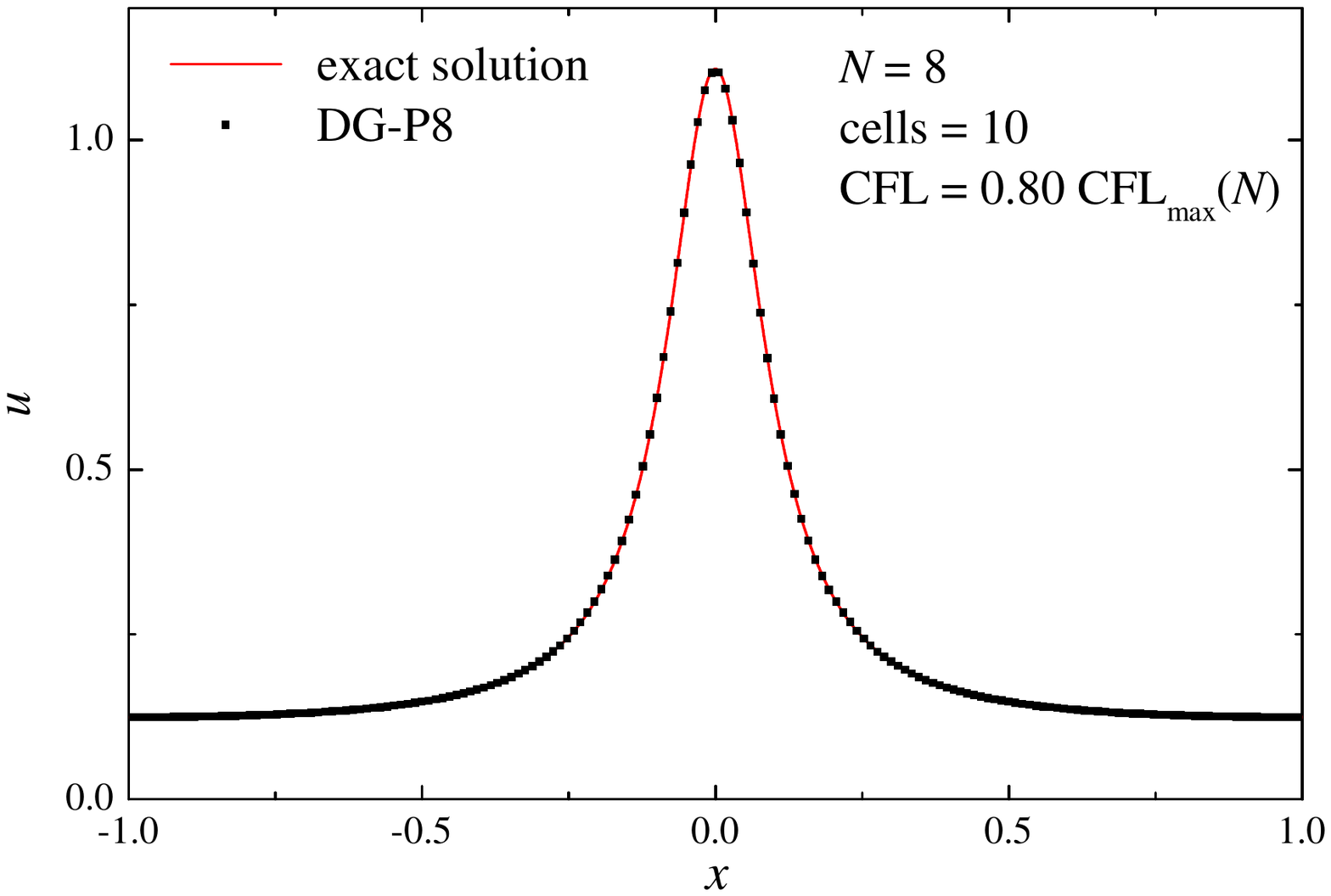}
\includegraphics[width=0.245\textwidth]{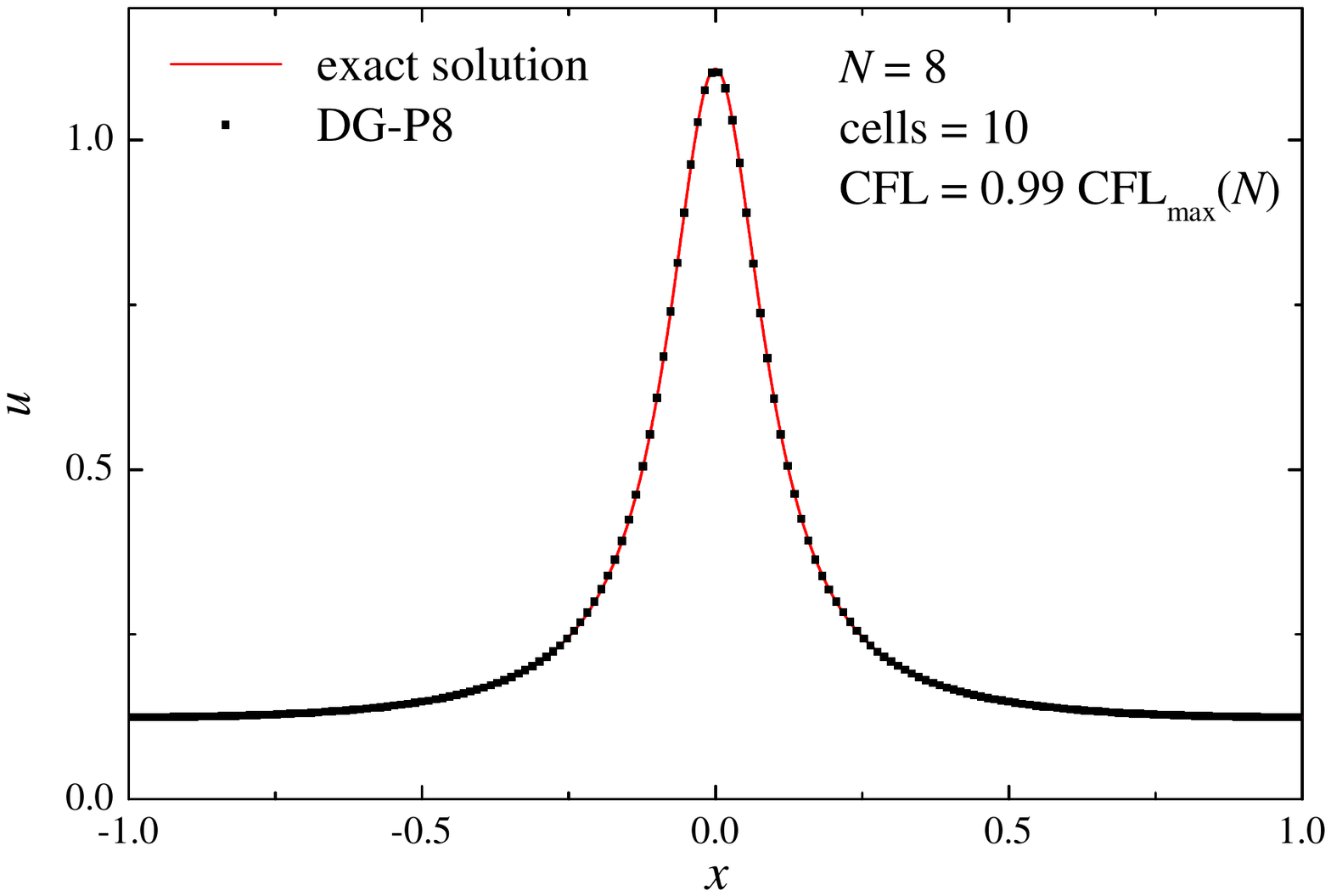}
\includegraphics[width=0.245\textwidth]{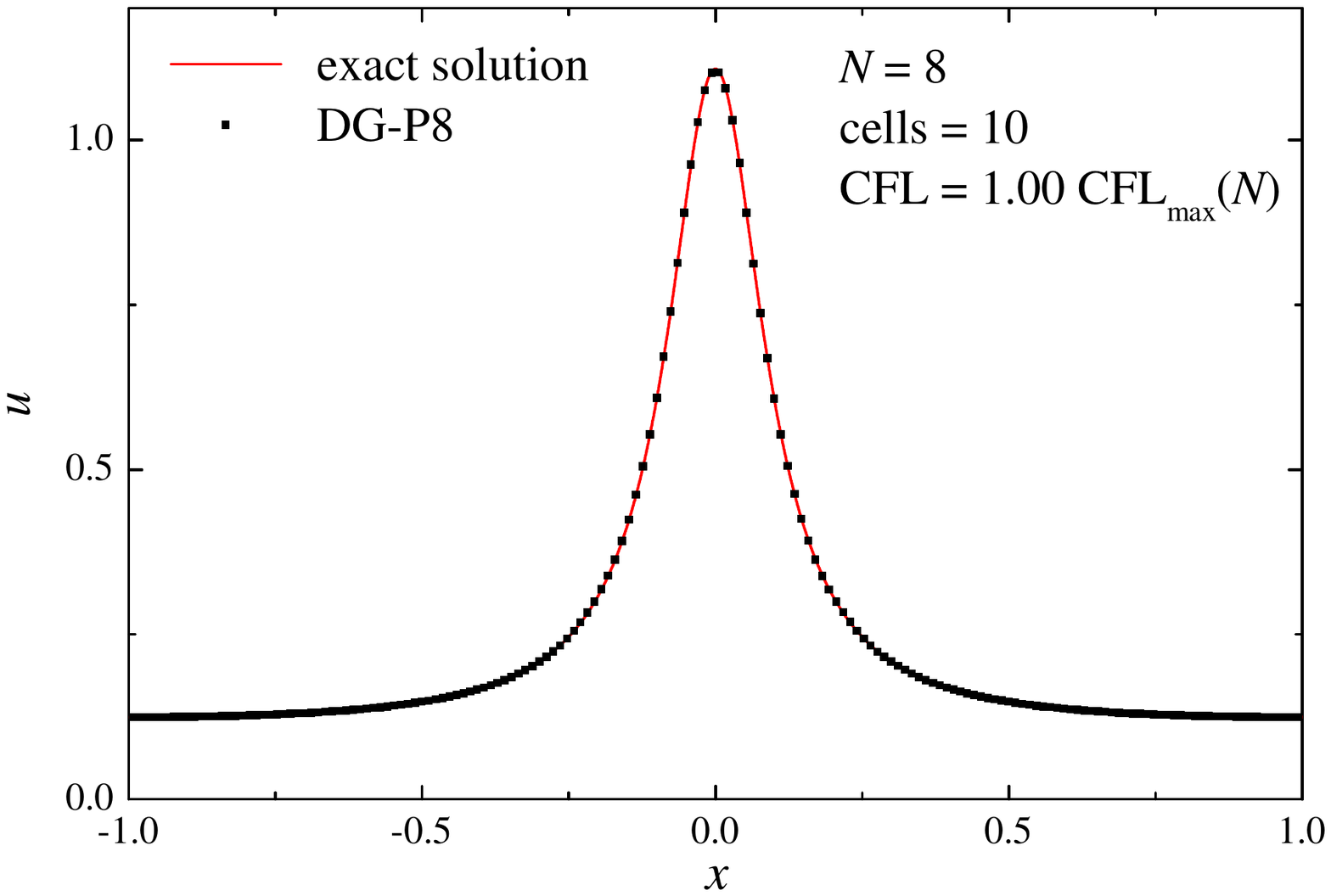}
\includegraphics[width=0.245\textwidth]{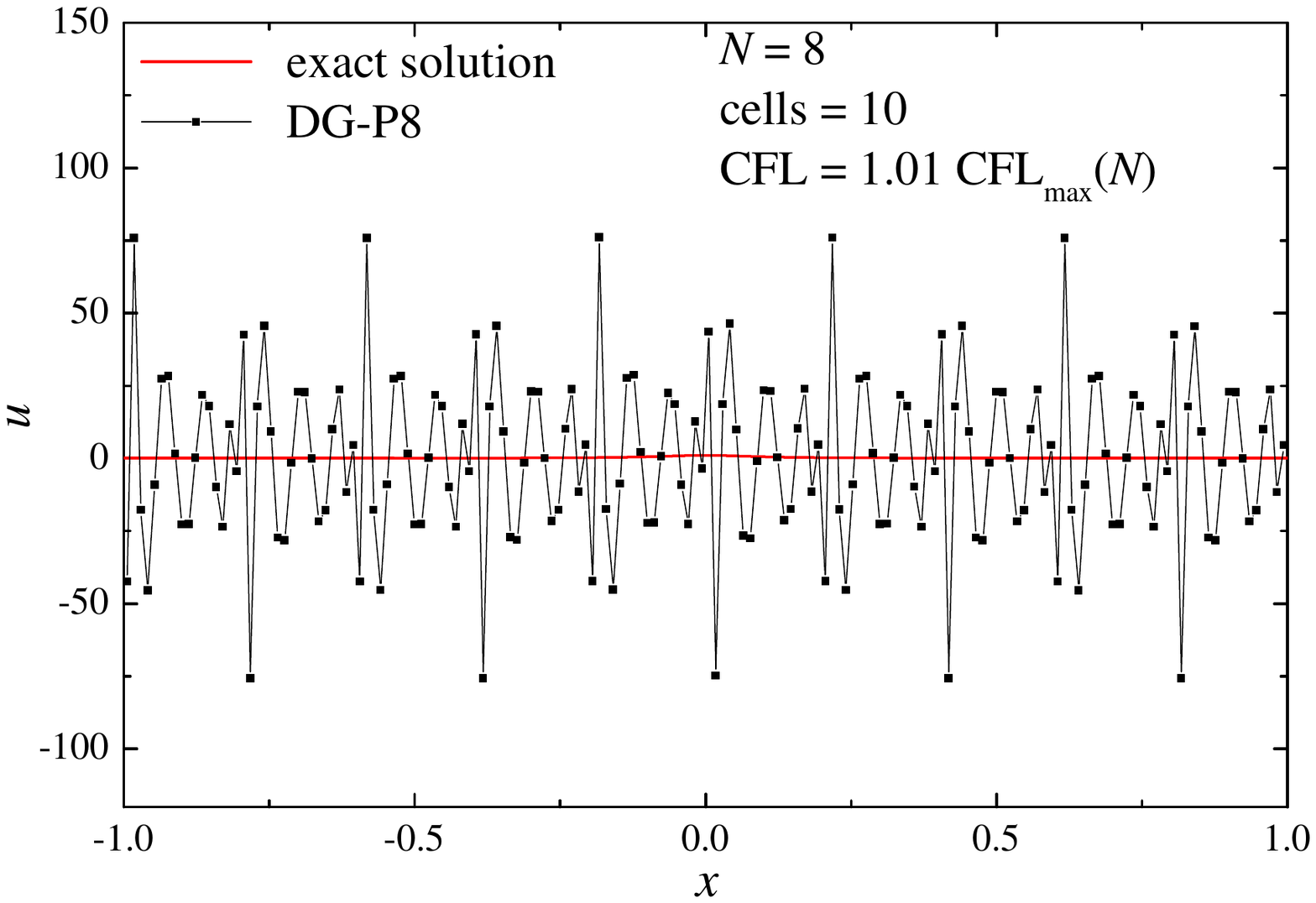}\\
\includegraphics[width=0.245\textwidth]{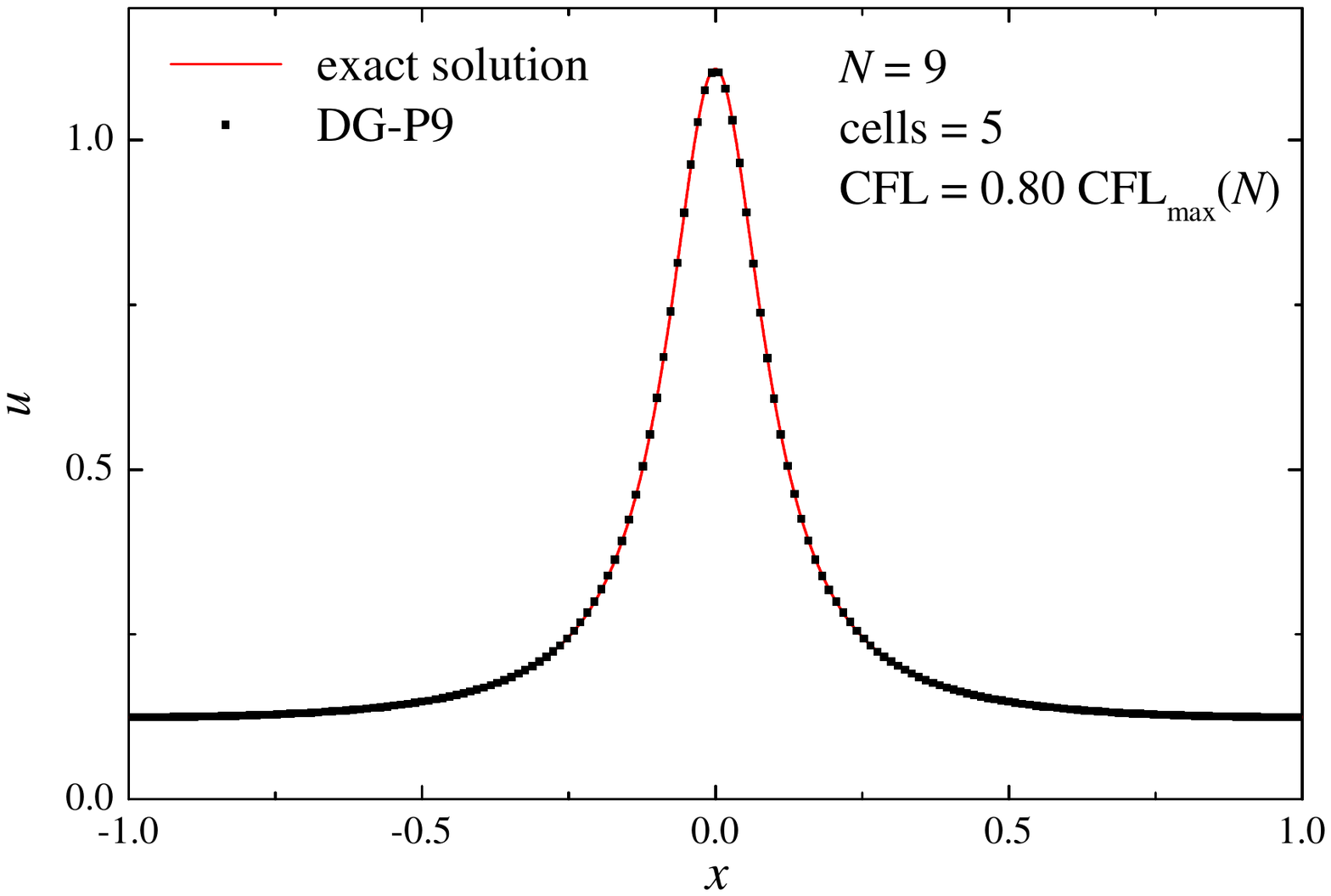}
\includegraphics[width=0.245\textwidth]{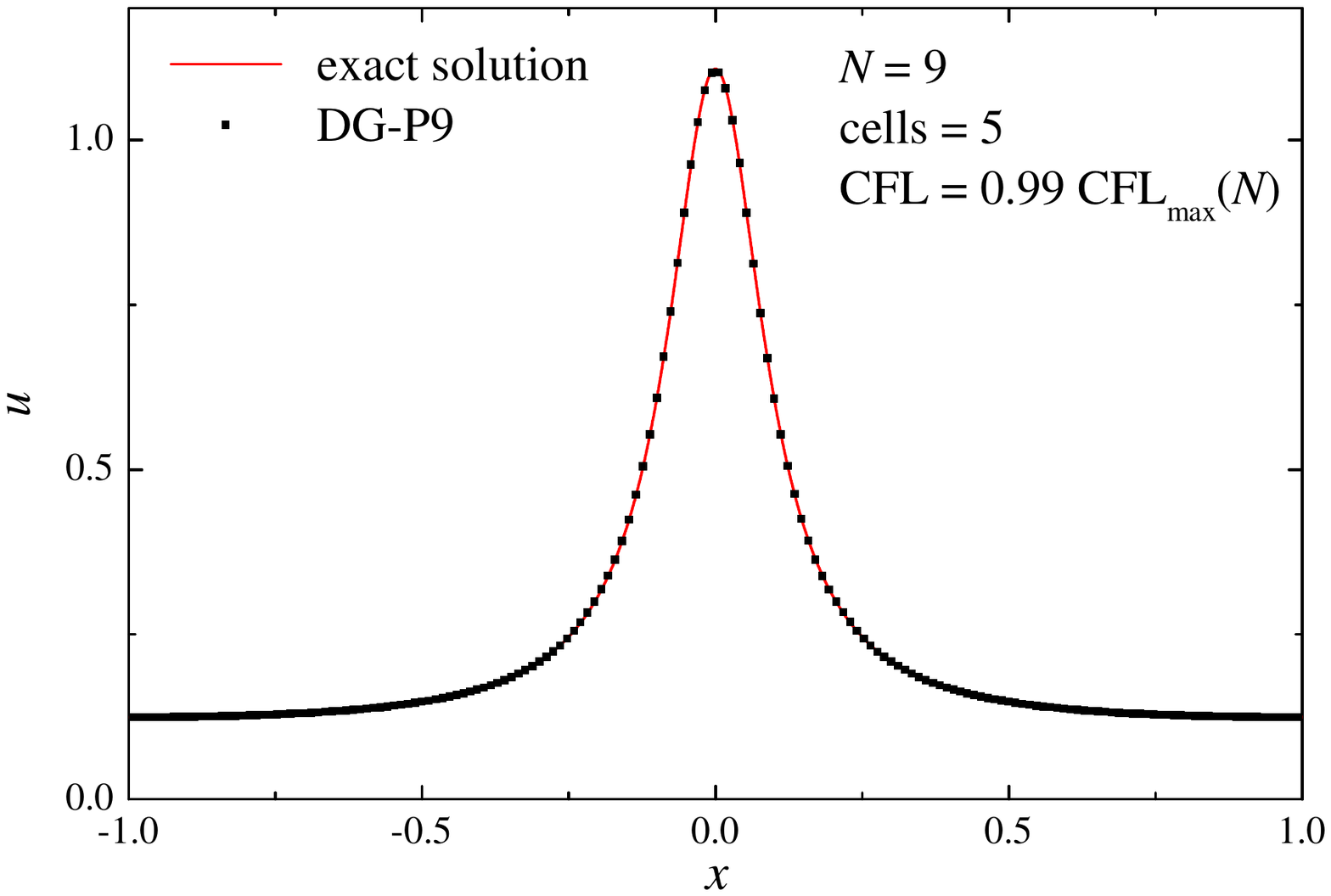}
\includegraphics[width=0.245\textwidth]{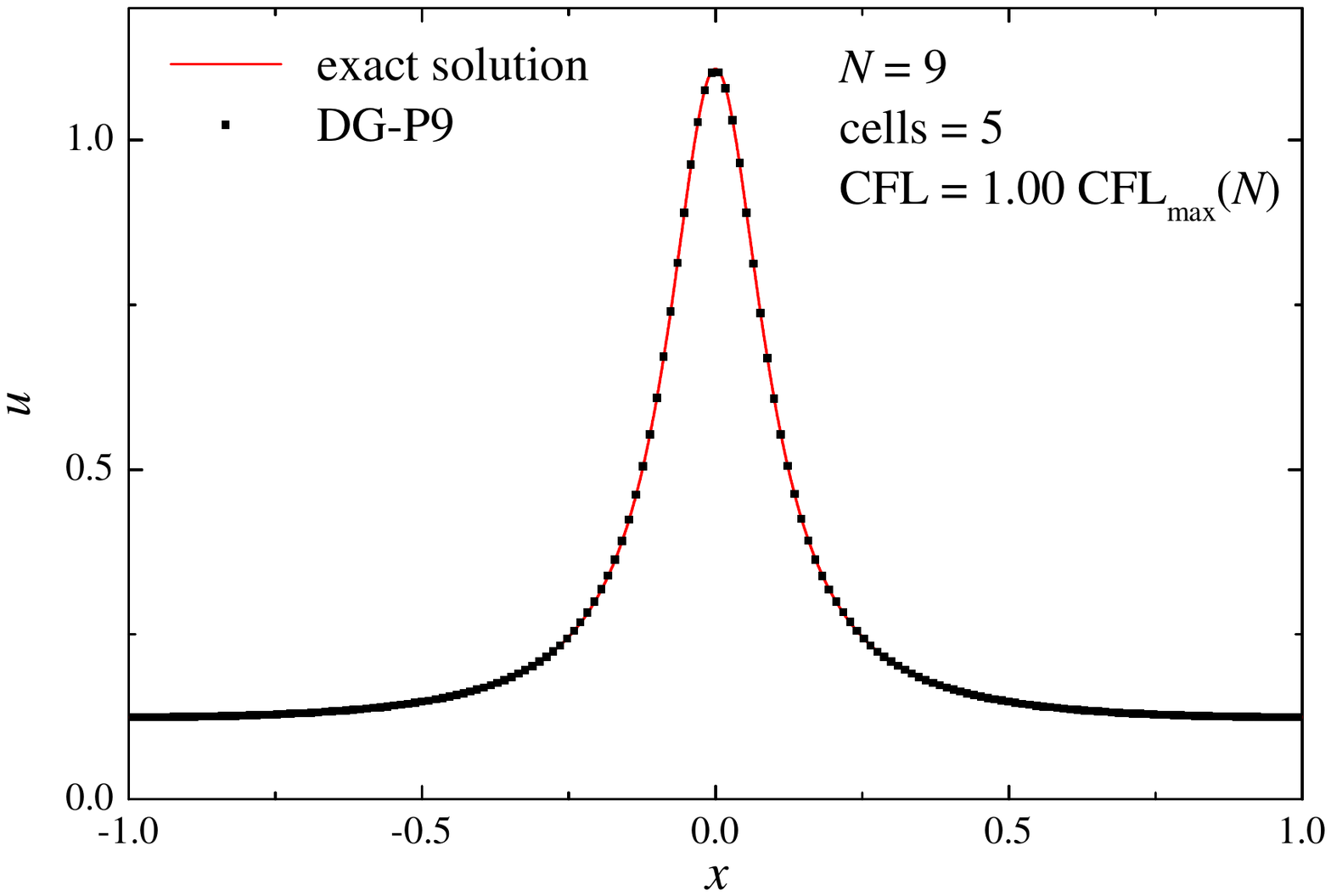}
\includegraphics[width=0.245\textwidth]{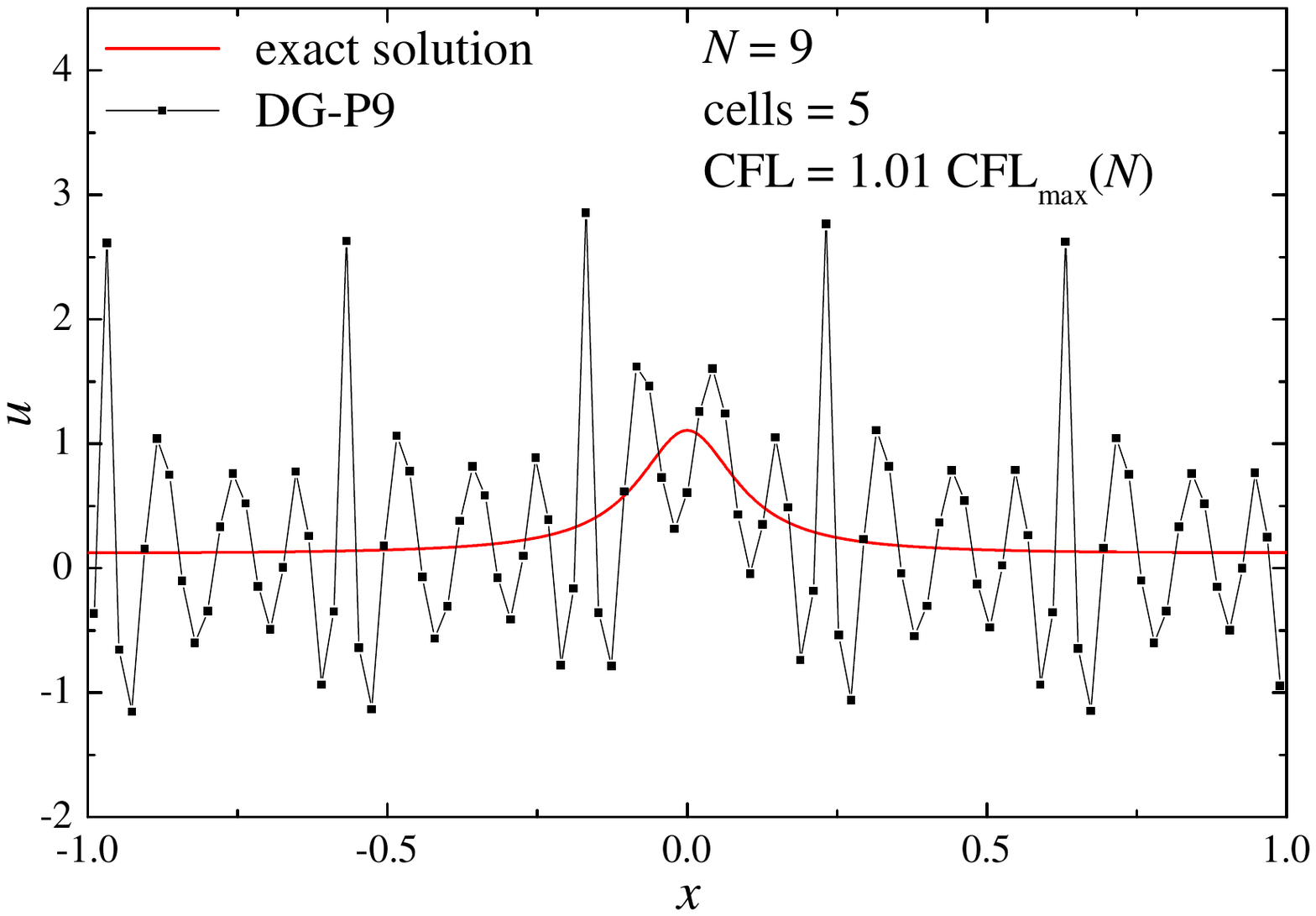}\\
\includegraphics[width=0.245\textwidth]{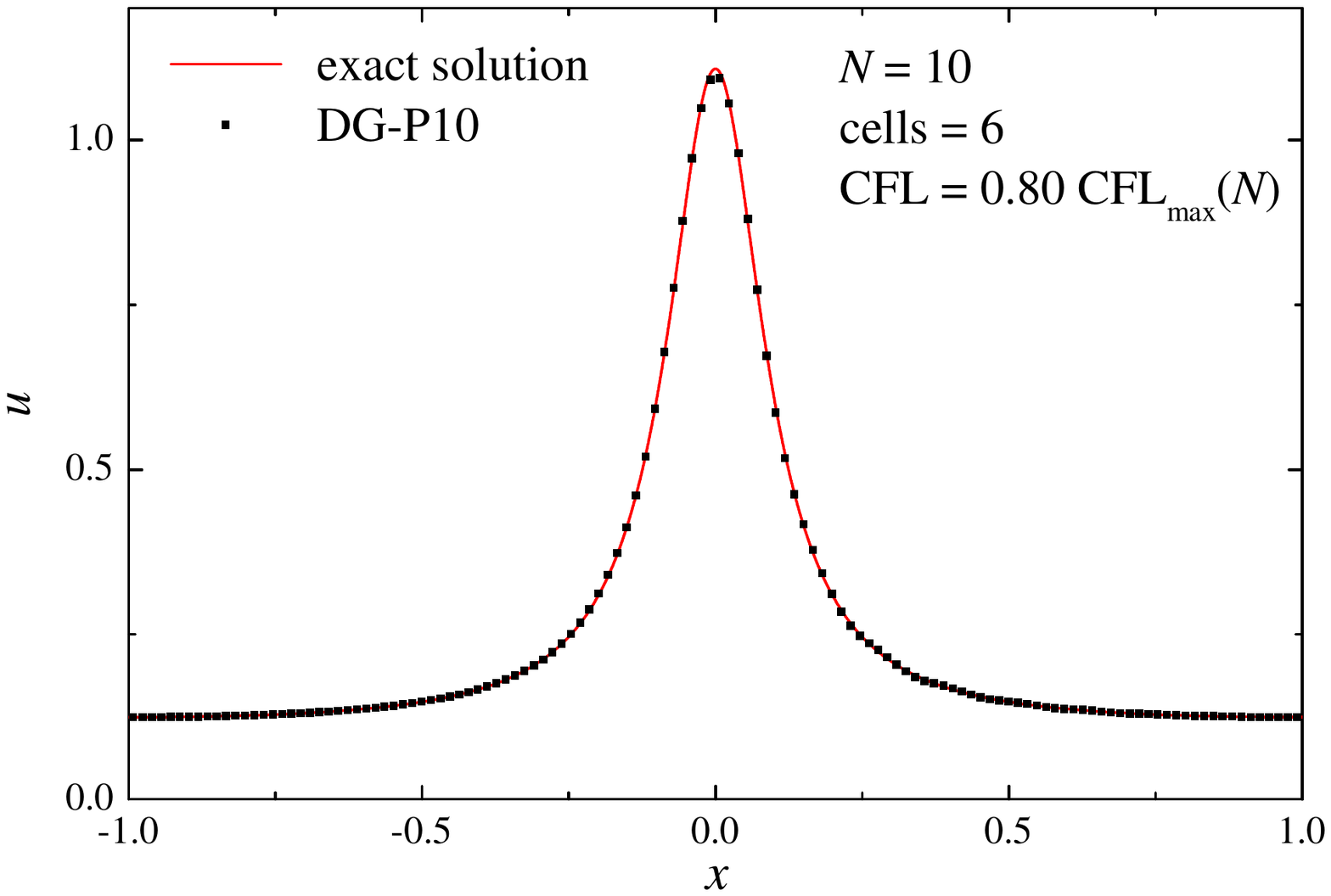}
\includegraphics[width=0.245\textwidth]{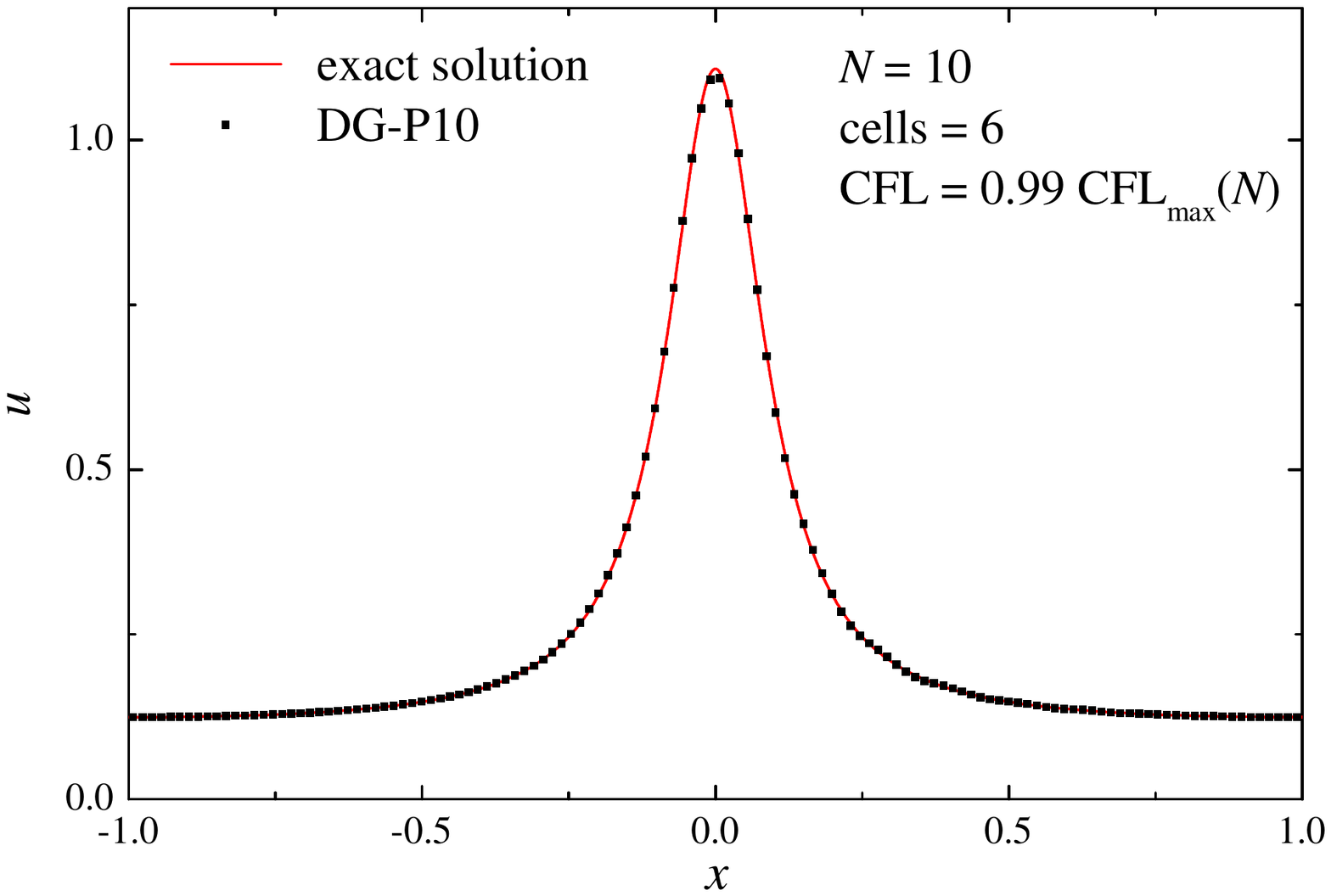}
\includegraphics[width=0.245\textwidth]{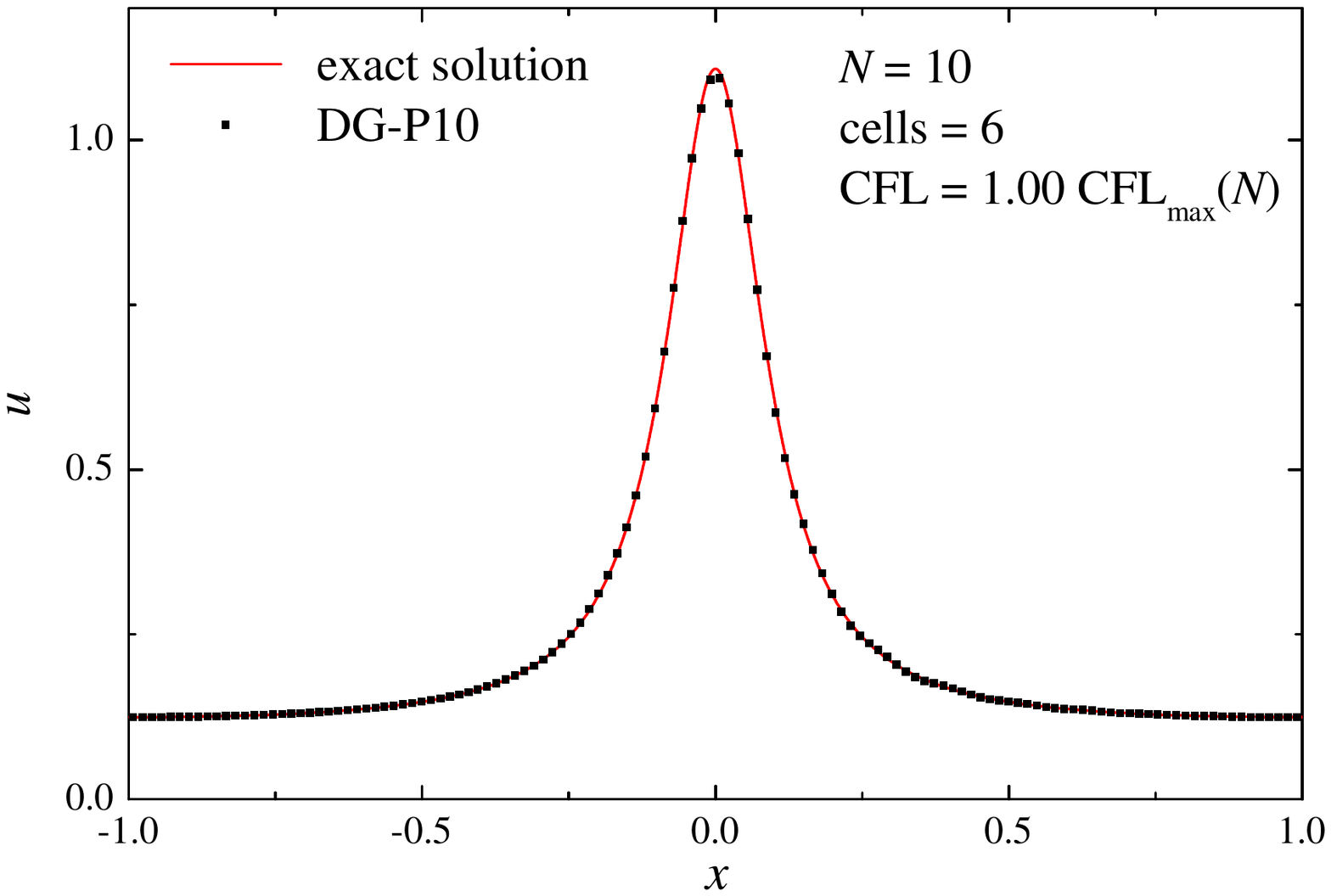}
\includegraphics[width=0.245\textwidth]{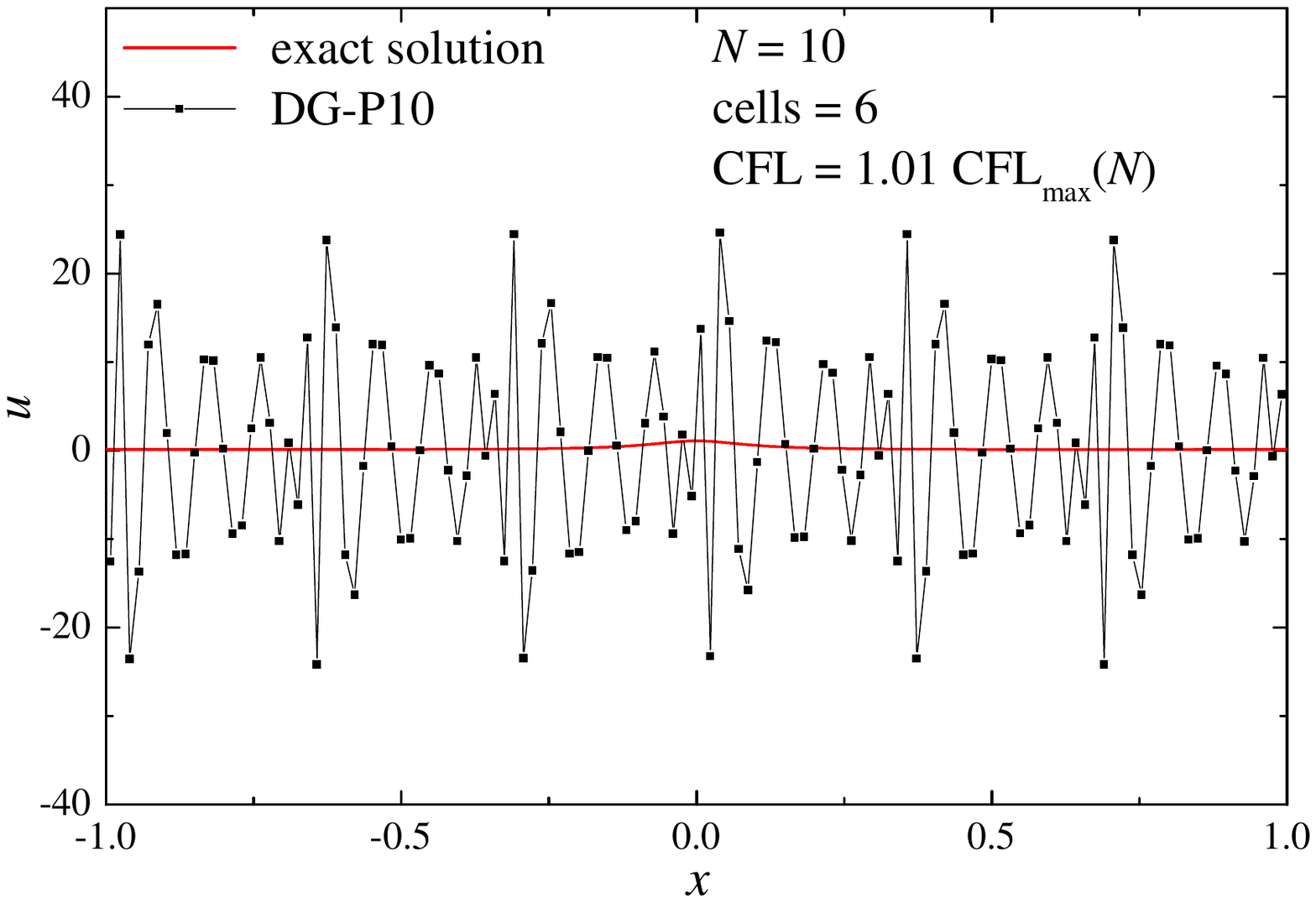}\\
\includegraphics[width=0.245\textwidth]{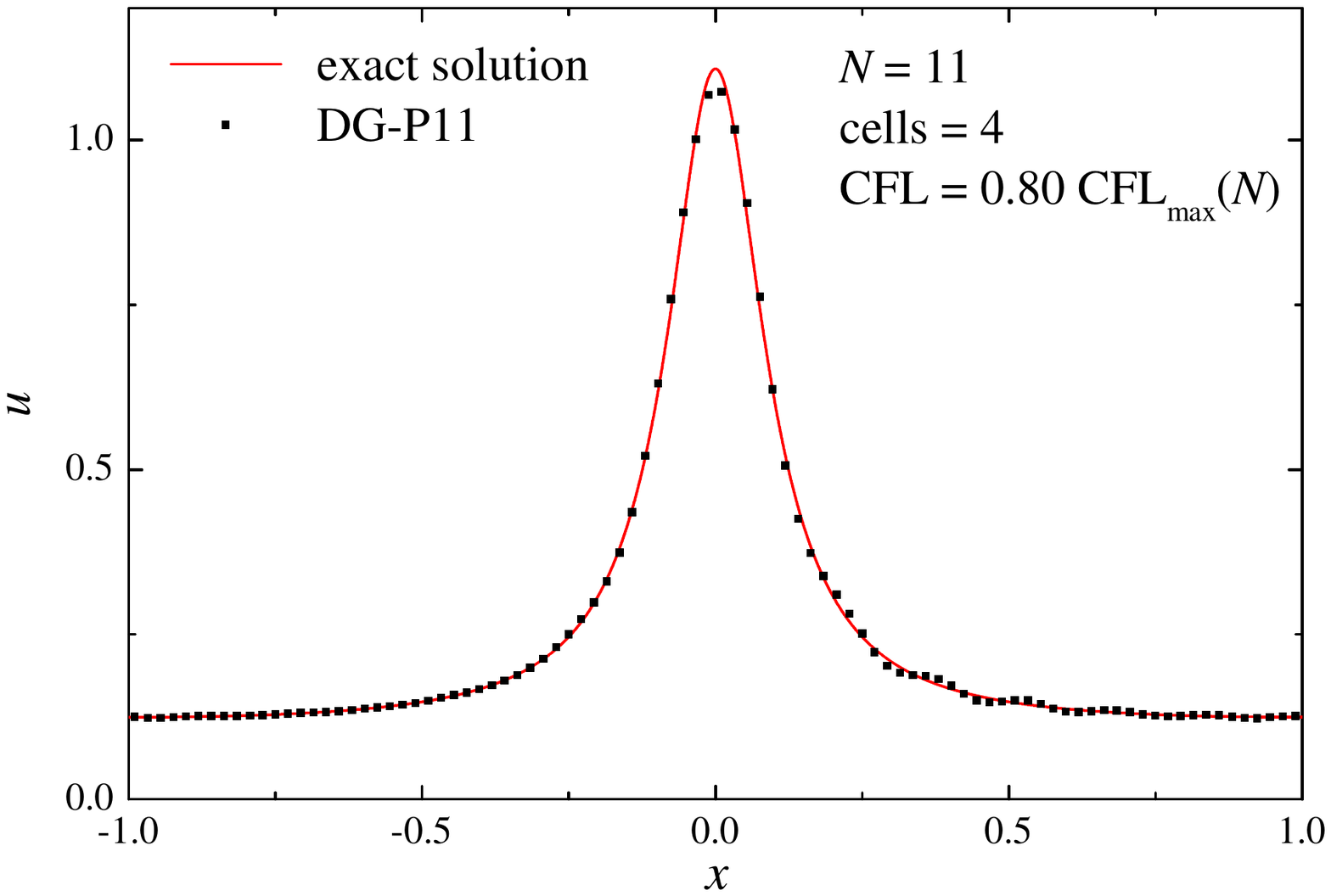}
\includegraphics[width=0.245\textwidth]{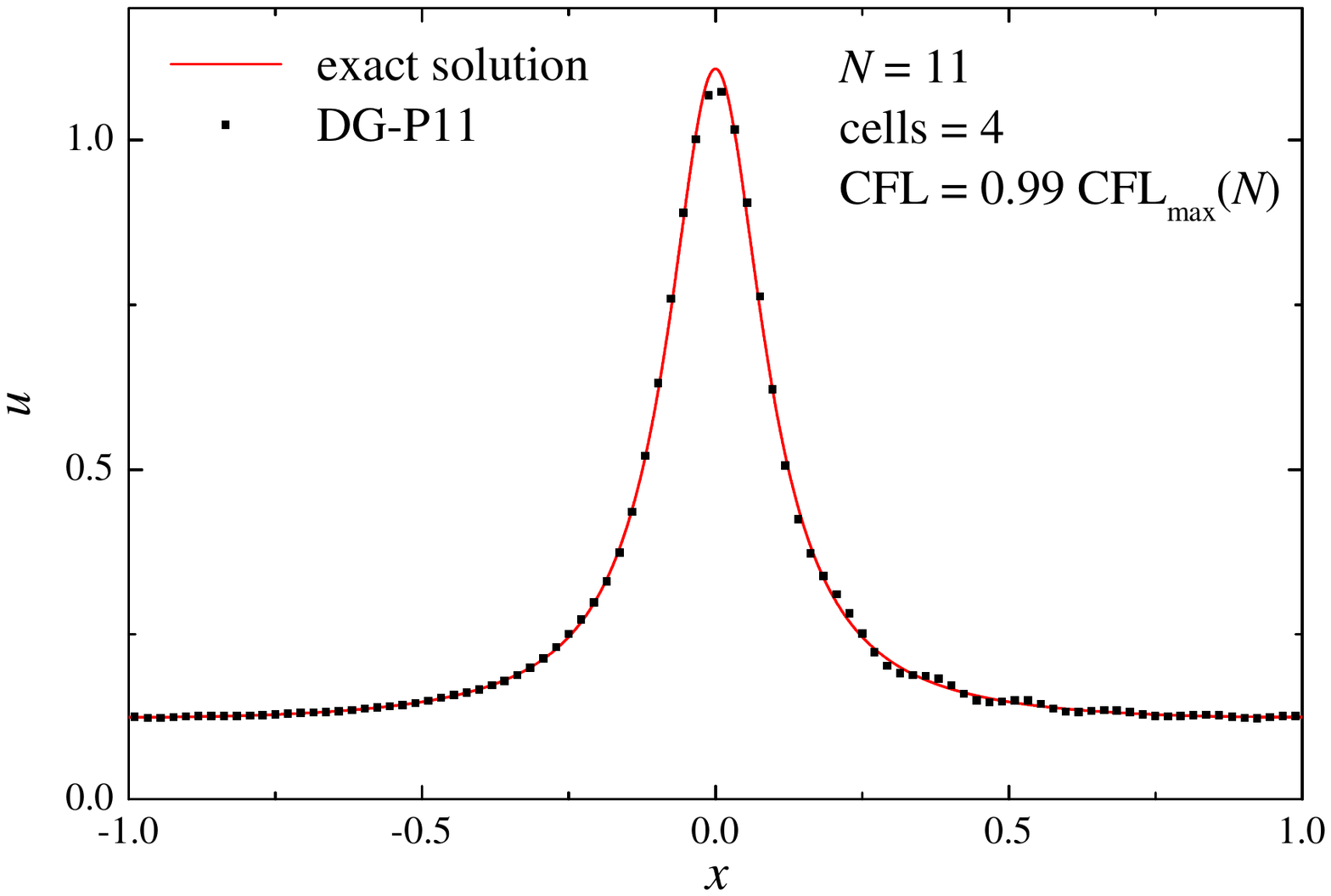}
\includegraphics[width=0.245\textwidth]{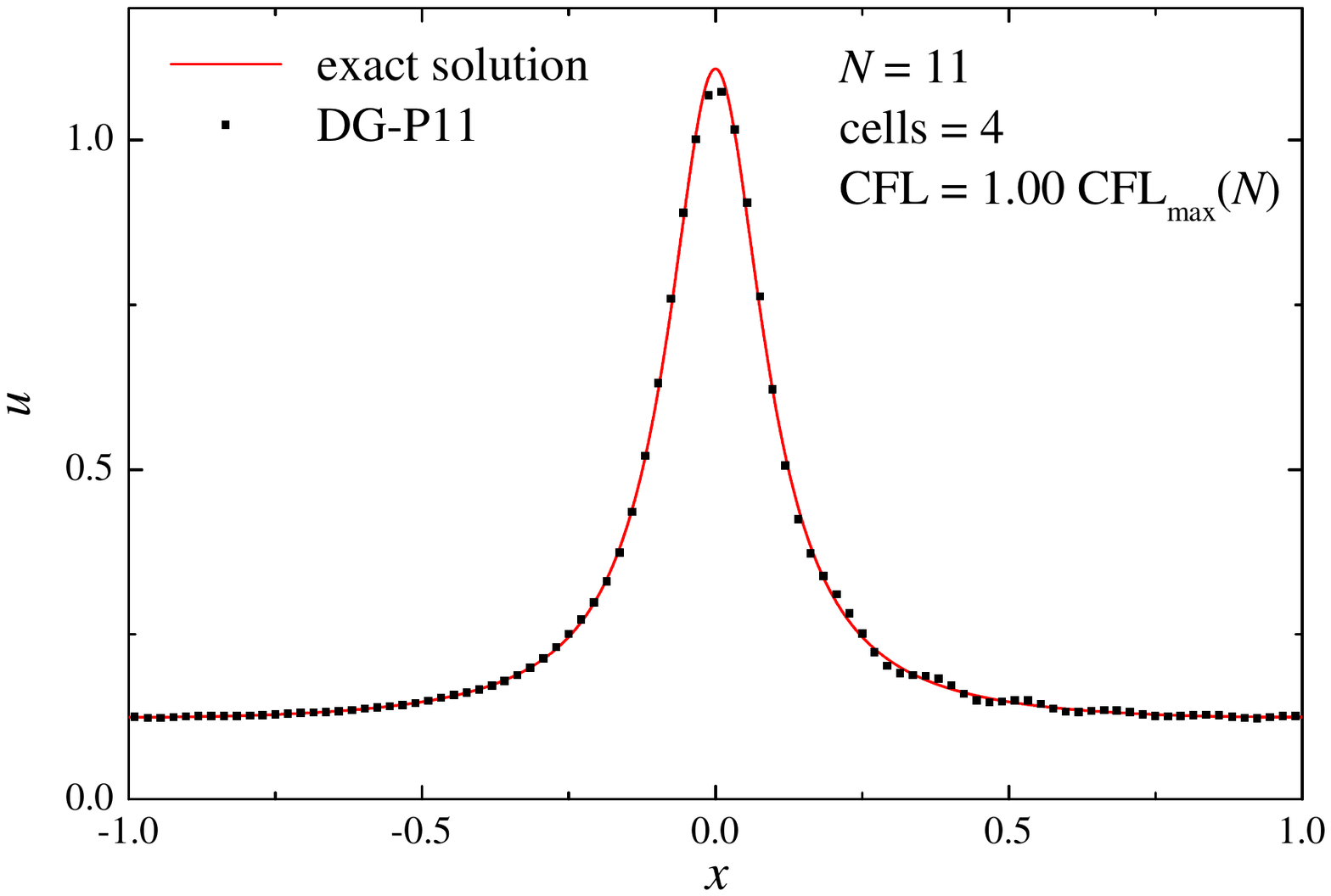}
\includegraphics[width=0.245\textwidth]{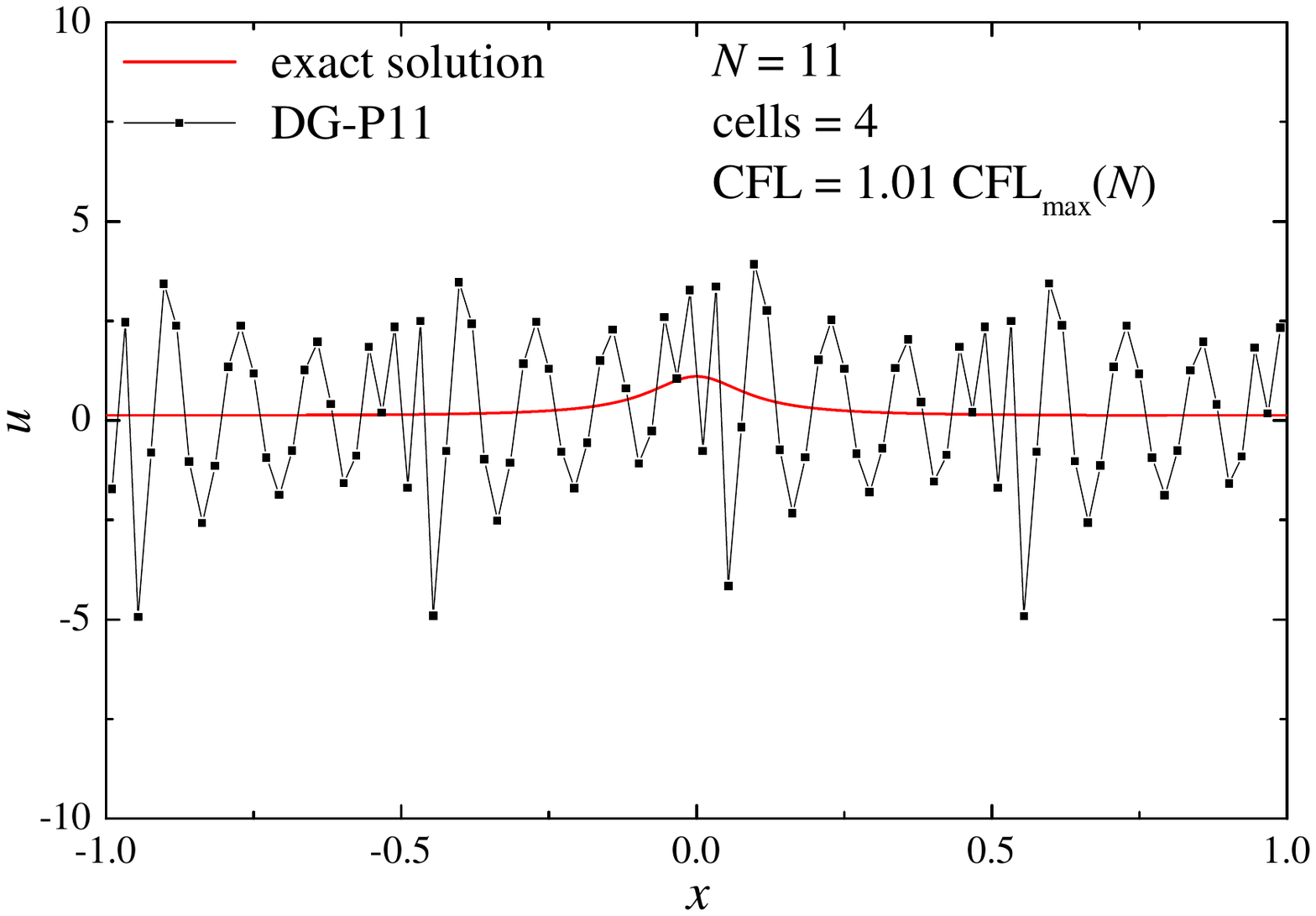}\\
\includegraphics[width=0.245\textwidth]{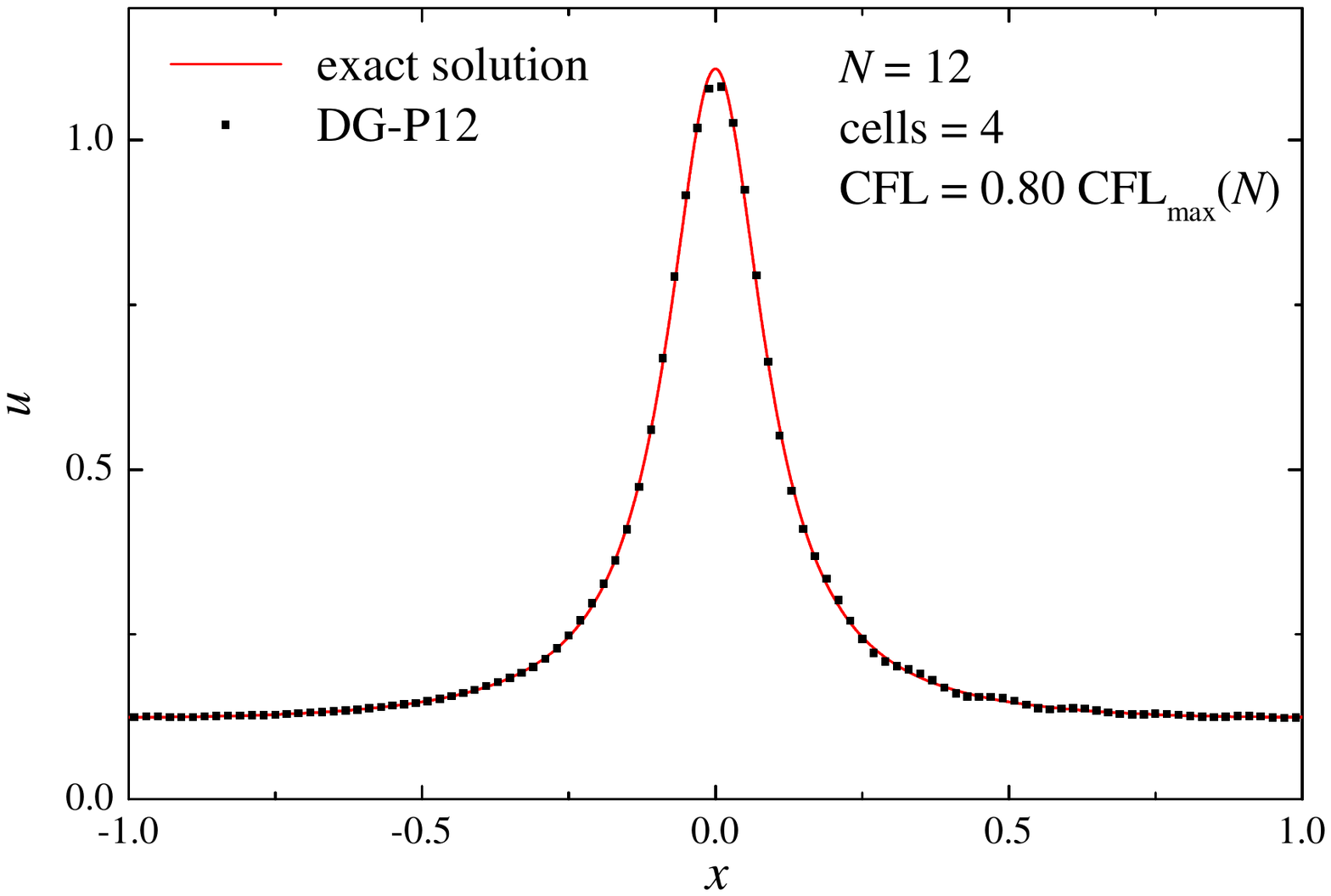}
\includegraphics[width=0.245\textwidth]{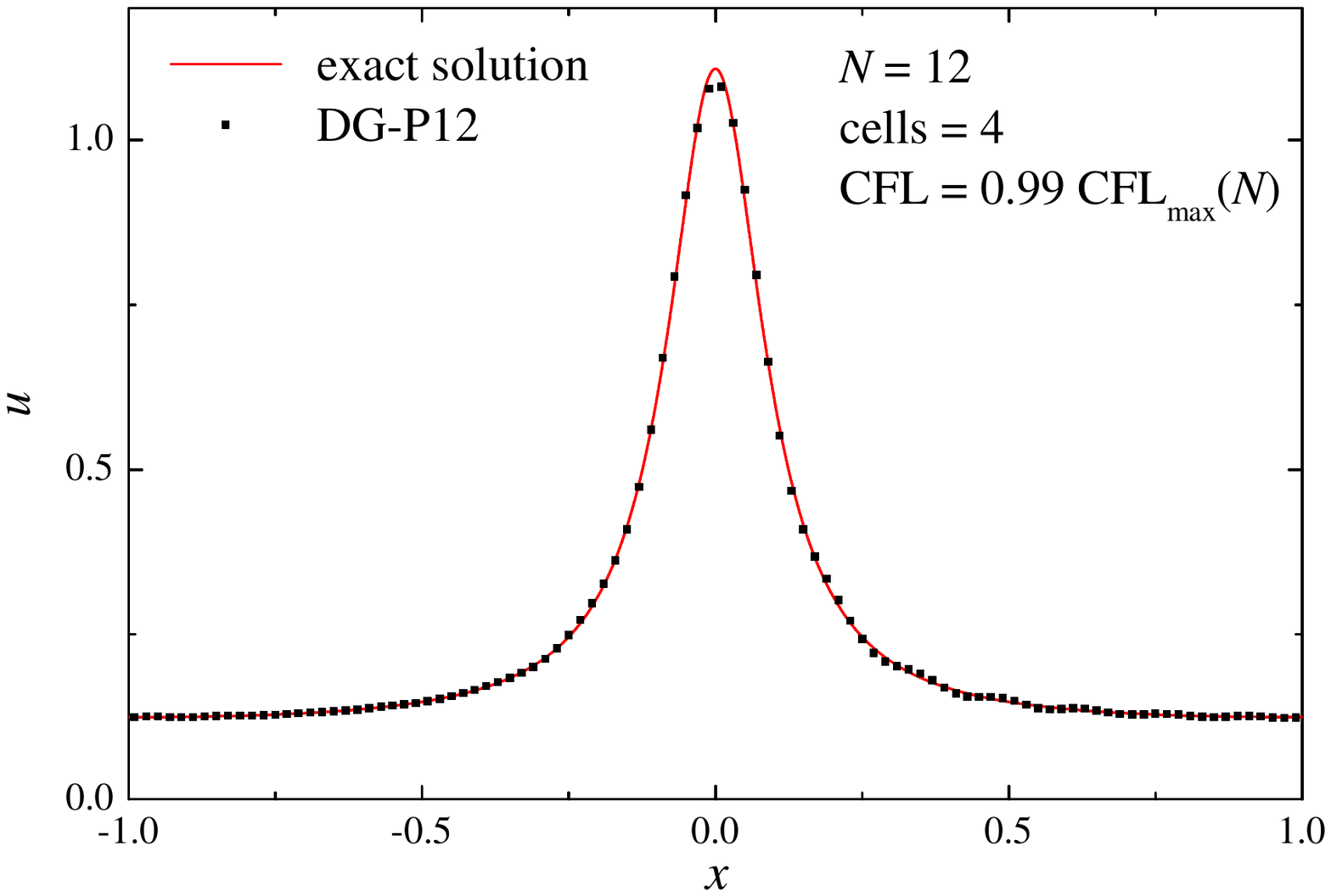}
\includegraphics[width=0.245\textwidth]{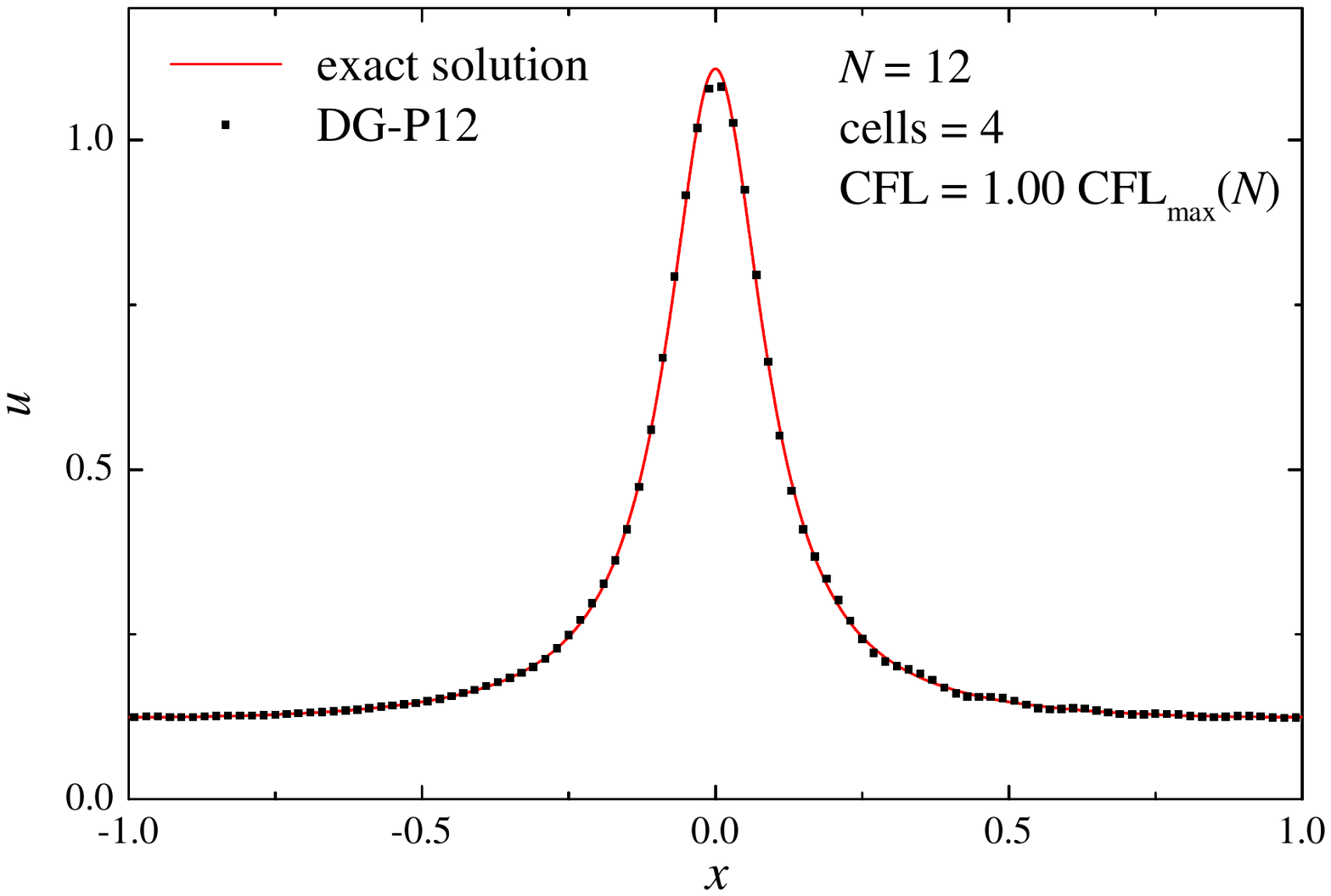}
\includegraphics[width=0.245\textwidth]{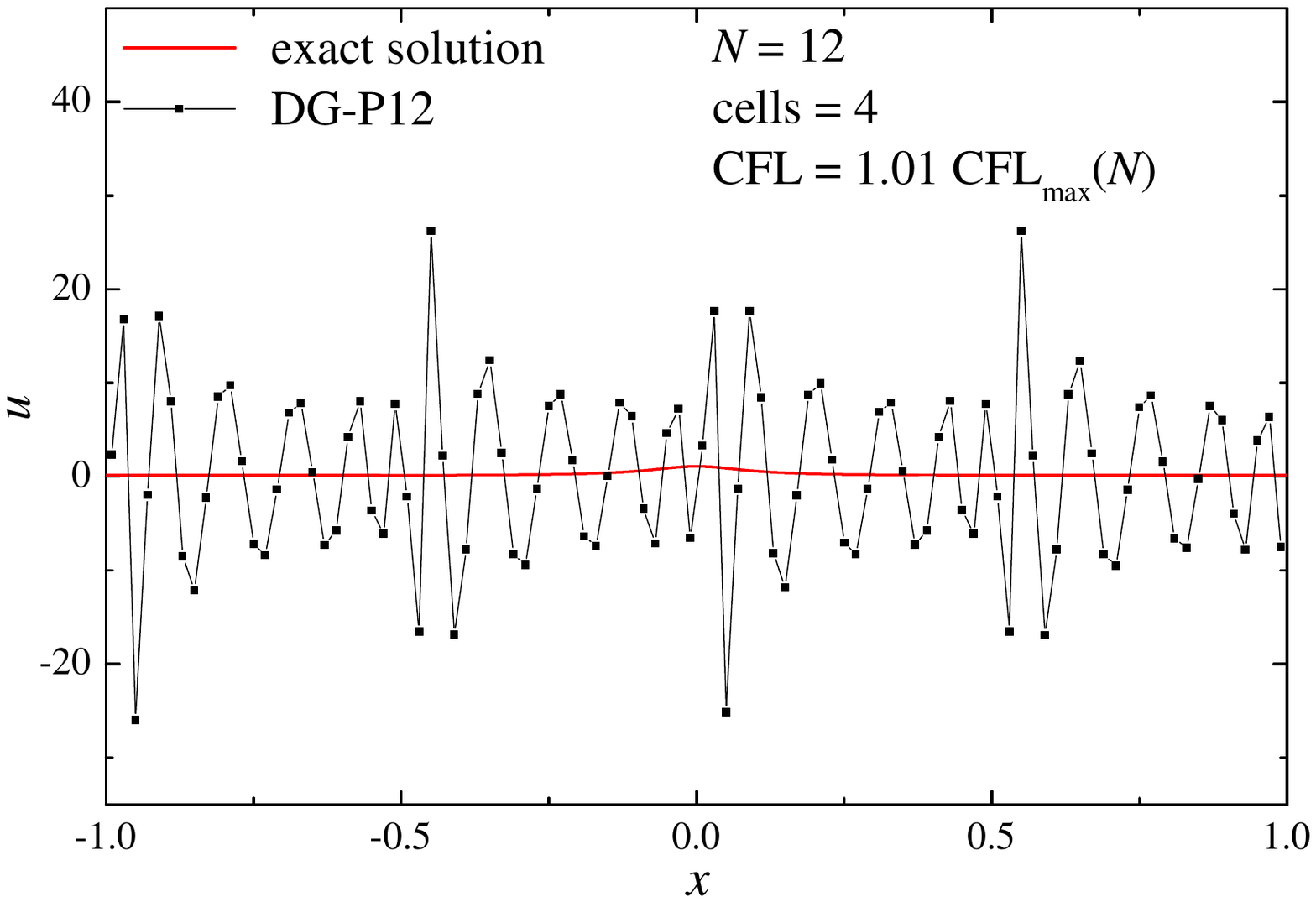}
\caption{%
Coordinate dependencies of the solution $u(x, t_{f})$ to the linear advection equation (\ref{eq:adv_eq_src}) at the final time $t_{f}$, obtained by the ADER-DG numerical method with the LST-DG predictor, for the degrees $N = 7, \ldots, 12$ of the basis polynomials (top to bottom) for a set of Courant number values $\mathrm{CFL} = 0.80$, $0.99$, $1.00$, $1.01\, \mathrm{CFL}_{\rm max}(N)$ (left to right). The first two columns correspond to the interior $\mathrm{CFL} < \mathrm{CFL}_{\rm max}(N)$ of the stability region, the third column correspond to the boundary $\mathrm{CFL} = \mathrm{CFL}_{\rm max}(N)$ of the stability region, and the right column correspond to the outside $\mathrm{CFL} > \mathrm{CFL}_{\rm max}(N)$ of the boundary of the stability region (the Courant number $\mathrm{CFL}$ is 1\% higher than the boundary value $\mathrm{CFL}_{\rm max}(N)$). The red line represents the exact analytical solution.
}
\label{fig:test_adveq_lorentz_1d_cfls_degrees_7_12}
\end{figure}

\begin{table}[h!]
\centering
\normalsize
\caption{%
Empirical orders of convergence $p$ of the ADER-DG numerical method with the LST-DG predictor, obtained based on the solution to the profile (\ref{eq:lorentz_bell}) advection problem for equation (\ref{eq:adv_eq_src}), calculated in functional norms $\mathcal{L}_{1}$, $\mathcal{L}_{2}$, $\mathcal{L}_{\infty}$ of errors $\epsilon$ (\ref{eq:norms_of_errors}), for degrees $N = 1, \ldots, 6$ of basis polynomials for $\mathrm{CFL} = 0.80\, \mathrm{CFL}_{\rm max}(N)$ and $1.00\, \mathrm{CFL}_{\rm max}(N)$. Column Th. corresponds to $p = N+1$.
\label{tab:adveq_conv_orders_degrees_1_6}
}
\setlength{\tabcolsep}{2.5pt}
\begin{tabular}{@{}|r|r|ccc|lll|ccc|lll|l|@{}}
\hline
& & \multicolumn{6}{c|}{$\mathrm{CFL} = 0.80\, \mathrm{CFL}_{\rm max}(N)$} & \multicolumn{6}{c}{$\mathrm{CFL} = 1.00\, \mathrm{CFL}_{\rm max}(N)$} & \\
\hline
$N$ & cells 
& $\epsilon_{\mathcal{L}_{1}}$ & $\epsilon_{\mathcal{L}_{2}}$ & $\epsilon_{\mathcal{L}_{\infty}}$ 
& $p_{\mathcal{L}_{1}}$ & $p_{\mathcal{L}_{2}}$ & $p_{\mathcal{L}_{\infty}}$
& $\epsilon_{\mathcal{L}_{1}}$ & $\epsilon_{\mathcal{L}_{2}}$ & $\epsilon_{\mathcal{L}_{\infty}}$ 
& $p_{\mathcal{L}_{1}}$ & $p_{\mathcal{L}_{2}}$ & $p_{\mathcal{L}_{\infty}}$ & Th. \\
\hline
1		&	$50$	&	4.99E--02	&	7.86E--02	&	2.17E--01	&	--		&	--		&	--		&	6.66E--02	&	9.24E--02	&	2.45E--01	&	--		&	--		&	--		&	2\\
						&	$100$	&	1.51E--02	&	2.96E--02	&	1.04E--01	&	1.72	&	1.41	&	1.06	&	1.82E--02	&	3.41E--02	&	1.19E--01	&	1.87	&	1.44	&	1.05	&	\\
						&	$150$	&	6.92E--03	&	1.46E--02	&	5.54E--02	&	1.93	&	1.75	&	1.55	&	7.94E--03	&	1.67E--02	&	6.41E--02	&	2.05	&	1.77	&	1.52	&	\\
						&	$200$	&	3.90E--03	&	8.47E--03	&	3.23E--02	&	2.00	&	1.89	&	1.88	&	4.45E--03	&	9.66E--03	&	3.72E--02	&	2.01	&	1.90	&	1.89	&	\\
\hline
2		&	$50$	&	4.58E--03	&	9.56E--03	&	3.68E--02	&	--		&	--		&	--		&	5.11E--03	&	1.06E--02	&	4.05E--02	&	--		&	--		&	--		&	3\\
						&	$100$	&	6.33E--04	&	1.51E--03	&	6.82E--03	&	2.86	&	2.66	&	2.43	&	7.41E--04	&	1.75E--03	&	7.84E--03	&	2.79	&	2.59	&	2.37	&	\\
						&	$150$	&	1.90E--04	&	4.62E--04	&	2.18E--03	&	2.96	&	2.92	&	2.81	&	2.26E--04	&	5.46E--04	&	2.56E--03	&	2.93	&	2.88	&	2.76	&	\\
						&	$200$	&	8.05E--05	&	1.96E--04	&	9.38E--04	&	2.99	&	2.98	&	2.93	&	9.60E--05	&	2.33E--04	&	1.11E--03	&	2.97	&	2.95	&	2.90	&	\\
\hline
3		&	$20$	&	8.71E--03	&	1.53E--02	&	5.37E--02	&	--		&	--		&	--		&	9.25E--03	&	1.63E--02	&	5.68E--02	&	--		&	--		&	--		&	4\\
						&	$40$	&	8.23E--04	&	1.91E--03	&	7.98E--03	&	3.40	&	3.00	&	2.75	&	9.24E--04	&	2.14E--03	&	8.78E--03	&	3.32	&	2.93	&	2.70	&	\\
						&	$60$	&	1.66E--04	&	4.24E--04	&	1.82E--03	&	3.94	&	3.72	&	3.65	&	1.91E--04	&	4.83E--04	&	2.04E--03	&	3.89	&	3.67	&	3.60	&	\\
						&	$80$	&	5.21E--05	&	1.36E--04	&	5.82E--04	&	4.04	&	3.96	&	3.96	&	5.98E--05	&	1.56E--04	&	6.66E--04	&	4.03	&	3.94	&	3.89	&	\\
\hline
4		&	$20$	&	1.80E--03	&	3.43E--03	&	1.27E--02	&	--		&	--		&	--		&	1.92E--03	&	3.66E--03	&	1.36E--02	&	--		&	--		&	--		&	5\\
						&	$40$	&	6.96E--05	&	1.86E--04	&	9.44E--04	&	4.69	&	4.20	&	3.74	&	7.91E--05	&	2.10E--04	&	1.05E--03	&	4.60	&	4.13	&	3.69	&	\\
						&	$60$	&	1.05E--05	&	2.82E--05	&	1.50E--04	&	4.67	&	4.66	&	4.54	&	1.21E--05	&	3.24E--05	&	1.71E--04	&	4.64	&	4.61	&	4.48	&	\\
						&	$80$	&	2.62E--06	&	7.16E--06	&	3.83E--05	&	4.80	&	4.76	&	4.74	&	3.05E--06	&	8.30E--06	&	4.41E--05	&	4.78	&	4.73	&	4.71	&	\\
\hline
5		&	$20$	&	3.33E--04	&	6.75E--04	&	2.57E--03	&	--		&	--		&	--		&	3.63E--04	&	7.48E--04	&	2.91E--03	&	--		&	--		&	--		&	6\\
						&	$40$	&	7.80E--06	&	2.14E--05	&	1.00E--04	&	5.41	&	4.98	&	4.68	&	9.21E--06	&	2.52E--05	&	1.17E--04	&	5.30	&	4.89	&	4.63	&	\\
						&	$60$	&	7.58E--07	&	2.17E--06	&	1.06E--05	&	5.75	&	5.65	&	5.55	&	8.85E--07	&	2.53E--06	&	1.25E--05	&	5.78	&	5.67	&	5.53	&	\\
						&	$80$	&	1.36E--07	&	3.95E--07	&	1.97E--06	&	5.96	&	5.92	&	5.84	&	1.58E--07	&	4.58E--07	&	2.28E--06	&	5.99	&	5.95	&	5.91	&	\\
\hline
6		&	$20$	&	6.47E--05	&	1.33E--04	&	5.42E--04	&	--		&	--		&	--		&	7.56E--05	&	1.57E--04	&	6.46E--04	&	--		&	--		&	--		&	7\\
						&	$40$	&	7.92E--07	&	2.32E--06	&	1.17E--05	&	6.35	&	5.84	&	5.54	&	9.02E--07	&	2.66E--06	&	1.37E--05	&	6.39	&	5.88	&	5.56	&	\\
						&	$60$	&	5.05E--08	&	1.51E--07	&	8.23E--07	&	6.79	&	6.73	&	6.54	&	5.74E--08	&	1.72E--07	&	9.41E--07	&	6.79	&	6.75	&	6.60	&	\\
						&	$80$	&	7.14E--09	&	2.13E--08	&	1.19E--07	&	6.79	&	6.82	&	6.71	&	8.17E--09	&	2.43E--08	&	1.37E--07	&	6.77	&	6.80	&	6.71	&	\\
\hline
\end{tabular}
\end{table}

\begin{table}[h!]
\centering
\normalsize
\caption{%
Empirical orders of convergence $p$ of the ADER-DG numerical method with the LST-DG predictor, obtained based on the solution to the profile (\ref{eq:lorentz_bell}) advection problem for equation (\ref{eq:adv_eq_src}), calculated in functional norms $\mathcal{L}_{1}$, $\mathcal{L}_{2}$, $\mathcal{L}_{\infty}$ of errors $\epsilon$ (\ref{eq:norms_of_errors}), for degrees $N = 7, \ldots, 12$ of basis polynomials for $\mathrm{CFL} = 0.80\, \mathrm{CFL}_{\rm max}(N)$ and $1.00\, \mathrm{CFL}_{\rm max}(N)$. Column Th. corresponds to $p = N+1$.
\label{tab:adveq_conv_orders_degrees_7_12}
}
\setlength{\tabcolsep}{2.5pt}
\begin{tabular}{@{}|r|r|ccc|lll|ccc|lll|l|@{}}
\hline
& & \multicolumn{6}{c|}{$\mathrm{CFL} = 0.80\, \mathrm{CFL}_{\rm max}(N)$} & \multicolumn{6}{c}{$\mathrm{CFL} = 1.00\, \mathrm{CFL}_{\rm max}(N)$} & \\
\hline
$N$ & cells 
& $\epsilon_{\mathcal{L}_{1}}$ & $\epsilon_{\mathcal{L}_{2}}$ & $\epsilon_{\mathcal{L}_{\infty}}$ 
& $p_{\mathcal{L}_{1}}$ & $p_{\mathcal{L}_{2}}$ & $p_{\mathcal{L}_{\infty}}$
& $\epsilon_{\mathcal{L}_{1}}$ & $\epsilon_{\mathcal{L}_{2}}$ & $\epsilon_{\mathcal{L}_{\infty}}$ 
& $p_{\mathcal{L}_{1}}$ & $p_{\mathcal{L}_{2}}$ & $p_{\mathcal{L}_{\infty}}$ & Th. \\
\hline
7		&	$20$	&	1.46E--05	&	3.18E--05	&	1.27E--04	&	--		&	--		&	--		&	1.66E--05	&	3.68E--05	&	1.49E--04	&	--		&	--		&	--		&	8\\
						&	$40$	&	9.03E--08	&	2.72E--07	&	1.41E--06	&	7.34	&	6.87	&	6.49	&	1.04E--07	&	3.13E--07	&	1.62E--06	&	7.31	&	6.88	&	6.52	&	\\
						&	$60$	&	3.95E--09	&	1.21E--08	&	6.32E--08	&	7.72	&	7.67	&	7.66	&	4.59E--09	&	1.40E--08	&	7.35E--08	&	7.70	&	7.65	&	7.63	&	\\
						&	$80$	&	4.04E--10	&	1.25E--09	&	6.87E--09	&	7.92	&	7.89	&	7.72	&	4.68E--10	&	1.45E--09	&	7.80E--09	&	7.94	&	7.89	&	7.80	&	\\
\hline
8		&	$20$	&	3.71E--06	&	8.82E--06	&	3.73E--05	&	--		&	--		&	--		&	4.12E--06	&	9.94E--06	&	4.17E--05	&	--		&	--		&	--		&	9\\
						&	$40$	&	9.59E--09	&	3.06E--08	&	1.83E--07	&	8.59	&	8.17	&	7.67	&	1.10E--08	&	3.51E--08	&	2.08E--07	&	8.55	&	8.15	&	7.65	&	\\
						&	$60$	&	2.72E--10	&	8.66E--10	&	5.38E--09	&	8.79	&	8.79	&	8.69	&	3.08E--10	&	9.82E--10	&	6.08E--09	&	8.82	&	8.82	&	8.71	&	\\
						&	$80$	&	2.18E--11	&	6.79E--11	&	4.26E--10	&	8.77	&	8.85	&	8.82	&	2.46E--11	&	7.70E--11	&	4.76E--10	&	8.79	&	8.85	&	8.85	&	\\
\hline
9		&	$20$	&	8.52E--07	&	2.16E--06	&	1.03E--05	&	--		&	--		&	--		&	9.43E--07	&	2.42E--06	&	1.16E--05	&	--		&	--		&	--		&	10\\
						&	$40$	&	1.07E--09	&	3.41E--09	&	1.82E--08	&	9.63	&	9.31	&	9.15	&	1.23E--09	&	3.89E--09	&	2.07E--08	&	9.59	&	9.28	&	9.13	&	\\
						&	$60$	&	2.20E--11	&	6.93E--11	&	3.89E--10	&	9.59	&	9.61	&	9.48	&	2.49E--11	&	7.99E--11	&	4.44E--10	&	9.61	&	9.58	&	9.48	&	\\
						&	$80$	&	2.51E--12	&	4.73E--12	&	2.33E--11	&	7.54	&	9.33	&	9.79	&	1.53E--12	&	4.75E--12	&	2.66E--11	&	9.69	&	9.81	&	9.79	&	\\
\hline
10		&	$10$	&	1.09E--04	&	1.83E--04	&	6.08E--04	&	--		&	--		&	--		&	1.15E--04	&	1.92E--04	&	6.40E--04	&	--		&	--		&	--		&	11\\
						&	$20$	&	1.76E--07	&	4.65E--07	&	2.43E--06	&	9.28	&	8.62	&	7.97	&	1.96E--07	&	5.18E--07	&	2.71E--06	&	9.20	&	8.53	&	7.88	&	\\
						&	$30$	&	2.43E--09	&	7.80E--09	&	4.62E--08	&	10.57	&	10.08	&	9.77	&	2.80E--09	&	8.93E--09	&	5.27E--08	&	10.47	&	10.02	&	9.72	&	\\
						&	$40$	&	1.22E--10	&	4.00E--10	&	2.33E--09	&	10.39	&	10.33	&	10.38	&	1.41E--10	&	4.60E--10	&	2.69E--09	&	10.39	&	10.31	&	10.35	&	\\
\hline
11		&	$10$	&	4.59E--05	&	7.66E--05	&	2.52E--04	&	--		&	--		&	--		&	4.81E--05	&	8.00E--05	&	2.62E--04	&	--		&	--		&	--		&	12\\
						&	$20$	&	3.39E--08	&	9.57E--08	&	5.03E--07	&	10.40	&	9.65	&	8.97	&	3.78E--08	&	1.08E--07	&	5.66E--07	&	10.31	&	9.54	&	8.86	&	\\
						&	$30$	&	3.56E--10	&	1.18E--09	&	6.97E--09	&	11.23	&	10.84	&	10.56	&	4.08E--10	&	1.35E--09	&	7.89E--09	&	11.17	&	10.80	&	10.54	&	\\
						&	$40$	&	1.38E--11	&	4.51E--11	&	2.80E--10	&	11.30	&	11.35	&	11.17	&	1.50E--11	&	5.12E--11	&	3.13E--10	&	11.48	&	11.38	&	11.21	&	\\
\hline
12		&	$10$	&	1.79E--05	&	2.94E--05	&	9.08E--05	&	--		&	--		&	--		&	1.85E--05	&	3.05E--05	&	9.45E--05	&	--		&	--		&	--		&	13\\
						&	$20$	&	6.74E--09	&	2.01E--08	&	1.07E--07	&	11.37	&	10.52	&	9.73	&	7.67E--09	&	2.29E--08	&	1.23E--07	&	11.24	&	10.38	&	9.59	&	\\
						&	$30$	&	5.36E--11	&	1.80E--10	&	1.06E--09	&	11.92	&	11.63	&	11.38	&	6.12E--11	&	2.05E--10	&	1.22E--09	&	11.91	&	11.63	&	11.37	&	\\
						&	$40$	&	2.60E--12	&	5.73E--12	&	3.68E--11	&	10.52	&	11.98	&	11.68	&	2.60E--12	&	6.34E--12	&	3.57E--11	&	10.98	&	12.08	&	12.28	&	\\
\hline
\end{tabular}
\end{table}

In this Subsection, a linear one-dimensional advection equation (\ref{eq:adv_eq_src}) is numerically solved using the ADER-DG numerical method with the LST-DG predictor. The primary aim of this computational experiment is to determine the stability boundary $\mathrm{CFL}_{\rm max}(N)$ accurately and to calculate empirical convergence orders $p$ both within the stability region $\mathrm{CFL} < \mathrm{CFL}_{\rm max}(N)$ and at the boundary itself $\mathrm{CFL} = \mathrm{CFL}_{\rm max}(N)$. Clearly, instability depends significantly on the spectral composition $u_{0}(k)$ of the solution $u(x, 0)$. Therefore, trivial tests for sine wave advection may ``miss'' the instability boundary if the wave vector $k$ for the selected sine wave in the initial conditions $u(x, 0) \sim \sin(kx)$ and coordinate discretization steps $\Dx$ are chosen such that the corresponding phases $\theta = k\Dx$ (\ref{eq:r_matrix_def}) for the boundary values of the Courant number $\mathrm{CFL}_{\rm max}(N)$ do not characterized eigenvalues $\lambda_{k}$ on or outside the circle of stability $|\lambda_{k}| \leqslant 1$. The choice of the Gaussian profile $u(x, 0) \sim \exp(-x^{2})$ (or Gaussian bell), as was done in the work~\cite{ader_dg_stab}, also revealed certain limitations: despite the very narrow profile (and, accordingly, a very broad solution spectrum at the initial instant of time), the Gaussian profile exhibits a very rapid decay (also having the form of a Gaussian profile) of the spectrum $k$ with increasing distance from zero $k = 0$. Moreover, mathematically rigorously, the standard Gaussian profile is not a periodic function, although it can, of course, attain very small solution values at the solution domain $\Omega$ boundary at the initial and reference instants of time.

In this work, a compromise variant is chosen --- the well-known Lorentzian profile $u(x, 0) \sim 1/(\alpha^{2}+x^{2})$ (or Lorentzian bell), which is characterized by a significantly slower decay with increasing coordinate $x$, compared to a Gaussian profile, but is also characterized only by an exponential decay of the spectrum with increasing wave vector $k$: $u_{0}(k) \propto \exp(-\alpha|k|)$. To achieve periodicity, the initial condition is chosen in the form of a superposition of an infinite number of Lorentzian profiles:
\begin{equation}\label{eq:lorentz_bell}
u(x, 0) = u_{0} + \sum\limits_{k = -\infty}^{+\infty} \cfrac{\alpha A}{\alpha^{2} + [x - L k]^{2}}
        = u_{0} + \frac{\pi\alpha A}{L}\cfrac{\sinh\left(\cfrac{2\pi\alpha}{L}\right)}{\cosh\left(\cfrac{2\pi\alpha}{L}\right) - \cos\left(\cfrac{2\pi x}{L}\right)},\quad
u_{0}(k) \propto \pi A \exp(-\alpha|k|),
\end{equation}
where the summation is carried out based on trivial considerations of the representation of the cotangent in the complex plane $\mathcal{C}$ and regarding the residue theory, $\alpha\in\mathcal{R}_{+}$ is the width of the Lorentz profile at half the level, $L\in\mathcal{R}_{+}$ is the coordinate period, $A\in\mathcal{R}_{+}$ is the intensity of the Lorentz profile (reduced height), and the wave vector $k$ has only a discrete (countable) set of values due to the periodicity of the function $u(x, 0)$ (therefore, of course, the exact expression for the spectrum $u_{0}(k)$ contains an additional factor in the form of a sum of delta functions $\delta(k - k_{l})$).

A set of numerical experiments on this Subsection is based on the numerical solution of a linear one-dimensional advection equation (\ref{eq:adv_eq_src}) with an initial condition (\ref{eq:lorentz_bell}) for a wide range of coordinate discretization steps $\Dx$ and a wide range of Courant numbers $\mathrm{CFL} \in [0.80\,\mathrm{CFL}_{\rm max}(N),\, 1.20\,\mathrm{CFL}_{\rm max}(N)]$, including those within the stability region $\mathrm{CFL} \leqslant \mathrm{CFL}_{\rm max}(N)$ and those extending beyond the stability region $\mathrm{CFL} > \mathrm{CFL}_{\rm max}(N)$. The following simulation parameters are chosen:
\begin{equation}
\Omega = [-1.0, +1.0],\quad L = 2.0,\quad A = 1.0,\quad \alpha = 0.1,\quad
u_{0} = 0.1,\quad t_{0} = 0.0,\quad t_{f} = 4.0,\quad a = 1.0.
\end{equation}
Calculations are performed for degrees $N = 1, \ldots, 12$ of basis polynomials $\{\varphi_{p}\}_{p}$ (\ref{eq:phi_def}). The boundary conditions is periodic: $u(-1.0, t) = u(+1.0, t)$. The time $t_{f} - t_{0} \equiv t_{f} = 4.0$ is chosen so that the periodic coordinate dependence of the solution $u(x, t)$ transferred the solution domain $\Omega$ exactly twice.

Figures~\ref{fig:test_adveq_lorentz_1d_cfls_degrees_1_6} and~\ref{fig:test_adveq_lorentz_1d_cfls_degrees_7_12} show examples of the obtained coordinate dependencies of the solution $u(x, t_{f})$ to the linear advection equation (\ref{eq:adv_eq_src}) at the final time $t_{f}$, obtained by the ADER-DG numerical method with the LST-DG predictor, for degrees $N = 1, \ldots, 12$ of the basis polynomials $\{\varphi_{p}\}_{p}$: in Figure~\ref{fig:test_adveq_lorentz_1d_cfls_degrees_1_6} for degrees $N = 1, \ldots, 6$ of the basis polynomials, in Figure~\ref{fig:test_adveq_lorentz_1d_cfls_degrees_7_12} for degrees $N = 7, \ldots, 12$ of the basis polynomials. The Courant numbers $\mathrm{CFL}$ are chosen equal to $0.80\, \mathrm{CFL}_{\rm max}(N)$, $0.99\, \mathrm{CFL}_{\rm max}(N)$, $1.00\, \mathrm{CFL}_{\rm max}(N)$ and $1.01\, \mathrm{CFL}_{\rm max}(N)$. The presented results clearly demonstrate the following: even when the Courant number $\mathrm{CFL}$ exceeds its boundary value $\mathrm{CFL}_{\rm max}(N)$ by 1\%, a strong instability of the numerical solution occurs. Moreover, the development of the instability has the character described in Section~\ref{sec:stab_anal} ``Stability analysis'': the real part of the eigenvalue $\lambda_{k}$ (or set $\{\lambda_{k}\}$ of eigenvalues) is negative, as a result of which the dependence of the numerical solution has an ``accordion'' character. It should be noted that the calculations performed show that even in the case of a boundary value of the Courant number $\mathrm{CFL} = \mathrm{CFL}_{\rm max}(N)$, the numerical solution remains stable for $250$ periods of profile advection through the solution domain $\Omega$. However, after $2$ periods, the solution obtained with a Courant number of $\mathrm{CFL} = 1.01\, \mathrm{CFL}_{\rm max}(N)$ clearly exhibits instability. This also confirms the assumptions made in Section~\ref{sec:stab_anal} ``Stability analysis'' regarding the insignificance of small deviations from the stability boundary of the eigenvalues $\lambda_{k}$ with a phase $|\arg\lambda_{k}| < \pi/4$. The dependence of the ``exponent'' $\ln|\lambda_{k}|$ of instability growth on the Courant number is quite strong $\mathrm{CFL}$, and all solutions (even for $1$ period of profile advection through the solution's domain) obtained with a Courant number of $\mathrm{CFL} = 1.02\, \mathrm{CFL}_{\rm max}(N)$ already consisted of \texttt{nan} (``Not a number''~\cite{ieee754}). From the presented results, it can be concluded that the boundary values of the Courant numbers $\mathrm{CFL}_{\rm max}(N)$ obtained in Section~\ref{sec:stab_anal} ``Stability analysis'' (in Table~\ref{tab:cfls_max_data} and Figure~\ref{fig:cfls_max_data}) are correctly and sufficiently accurately determined, and the considerations underlying their calculation are correct and justified.

It is necessary to note an interesting feature in the presented coordinate dependencies of the solution $u(x, t_{f})$ to the linear advection equation (\ref{eq:adv_eq_src}) at the final time $t_{f}$ in the case of instability development at the Courant number $\mathrm{CFL} = 1.01\, \mathrm{CFL}_{\rm max}(N)$: with an increase of the degree $N$ of the basis polynomials, the amplitude of the emerging oscillations oscillates. This effect has an obvious nature and is associated with the behavior of the absolute values $|\lambda_{k}|$ of the eigenvalues $\lambda_{k}$ of the matrix $\mathrm{R}(\mathrm{CFL}, \theta)$ (\ref{eq:r_matrix_in_matrix_form}) of the evolution operator $R$ with a change in the degree $N$ of the basis polynomials, which is presented above in Figures~\ref{fig:rhos_on_theta_degrees_1_6} and~\ref{fig:rhos_on_theta_degrees_7_12}. Despite the fact that the spectrum $u_{0}(k)$ of the initial condition function $u(x, 0)$ (\ref{eq:lorentz_bell}) is quite wide, it is still the same for different degrees $N$ of the polynomials, but the eigenvalues $\lambda_{k}$ of the matrix $\mathrm{R}(\mathrm{CFL}, \theta)$ (\ref{eq:r_matrix_in_matrix_form}) of the evolution operator $R$ that cross the stability boundary $|\lambda_{k}| > 1$, for different degrees $N$ of the basis polynomials, correspond to different phases $\theta$. And despite the fact that the eigenvalue $\lambda_{k}$ responsible for instability is in the vicinity $\lambda = -1$ (and thus generates oscillations with exponential amplitude growth, rather than simply constant-sign exponential growth), this eigenvalue corresponds to different phases $\theta = k\Dx$ for different degrees $N$ of basis polynomial, and therefore to different harmonics $k$ of the initial condition function $u(x, 0)$ spectrum $u_{0}(k)$. The results presented in Figures~\ref{fig:rhos_on_theta_degrees_1_6} and~\ref{fig:rhos_on_theta_degrees_7_12} show that as the degree $N$ of the basis polynomials increases by one, the phase region $\Delta\theta$, where the absolute value $|\lambda_{k}|$ of the eigenvalues $\lambda_{k}$ intersects stability boundary $|\lambda| = 1$, changes by almost $\sim\pi$, which determines the significantly different harmonics $k$ of the initial condition function $u(x, 0)$ spectrum $u_{0}(k)$ for which the amplitude increases.

Empirical convergence orders $p$ are calculated based on the dependence $\epsilon \propto \Dx^{p}$ of the numerical solution error $\epsilon$ on the coordinate grid step $\Dx$. Three classical norms $\mathcal{L}_{1}$, $\mathcal{L}_{2}$, $\mathcal{L}_{\infty}$ of the numerical solution errors $\epsilon$ are used to determine the empirical orders $p$ of convergence:
\begin{equation}\label{eq:norms_of_errors}
\begin{split}
\epsilon_{\mathcal{L}_{1}} = \int\limits_{\Omega}dx\, |u(x, t_{f}) - u_{\rm exact}(x)|,\qquad
\epsilon_{\mathcal{L}_{2}}^{2} = \int\limits_{\Omega}dx\, |u(x, t_{f}) - u_{\rm exact}(x)|^{2},\qquad
\epsilon_{\mathcal{L}_{\infty}} = \operatorname{ess}\sup\limits_{\hspace{-5mm}x\in\Omega} |u(x, t_{f}) - u_{\rm exact}(x)|,\vphantom{\int\limits_{\Omega}}
\end{split}
\end{equation}
where $u(x, t_{f})$ is the numerical solution, $u_{\rm exact}(x)$ is the exact analytical solution to the problem. The integrals in (\ref{eq:norms_of_errors}) are calculated using the GL quadrature formula (\ref{eq:gl_rule}) with very high degree $N = 60$ of polynomial $\tilde{L}_{N}$, and the $\operatorname{ess}\sup$ is calculated as $\max$ on a subgrid containing $1000$ nodes in each discretization domain $\Omega_{i}$ of the coordinate mesh.

The empirical convergence orders $p$ are calculated for studied wide range of Courant numbers $\mathrm{CFL} \in [0.80\,\mathrm{CFL}_{\rm max}(N),\, 1.20\,\mathrm{CFL}_{\rm max}(N)]$, including those within the stability region $\mathrm{CFL} \leqslant \mathrm{CFL}_{\rm max}(N)$ and those extending beyond the stability region $\mathrm{CFL} > \mathrm{CFL}_{\rm max}(N)$. For all studied Courant numbers beyond the stability region $\mathrm{CFL} > \mathrm{CFL}_{\rm max}(N)$, the orders $p$ are negative or \texttt{nan}. Tables~\ref{tab:adveq_conv_orders_degrees_1_6} and~\ref{tab:adveq_conv_orders_degrees_7_12} present examples of the calculated empirical convergence orders $p$ for Courant numbers $\mathrm{CFL} = 0.80\, \mathrm{CFL}_{\rm max}(N)$ and $1.00\, \mathrm{CFL}_{\rm max}(N)$ (boundary of stability region) for degrees $N = 1, \ldots, 12$ of the basis polynomials: in Table~\ref{tab:adveq_conv_orders_degrees_1_6} for degrees $N = 1, \ldots, 6$ of the basis polynomials, in Table~\ref{tab:adveq_conv_orders_degrees_7_12} for degrees $N = 7, \ldots, 12$ of the basis polynomials. It is evident that as the Courant number $\mathrm{CFL}$ changes in the stability region $\mathrm{CFL} \leqslant \mathrm{CFL}_{\rm max}(N)$, the empirical convergence orders $p$ are agree well with the approximation orders $p = N+1$ calculated in Section~\ref{sec:approx_anal} ``Approximation analysis'', which are presented in the Th. column of Tables~\ref{tab:adveq_conv_orders_degrees_1_6} and~\ref{tab:adveq_conv_orders_degrees_7_12}. For degrees $N = 11$ and $12$ of basis polynomials, the empirical convergence orders $p$ are slightly lower than expected $p = N+1$. This is due to the accuracy $\epsilon_{\rm LST-DG} = 10^{-12}$ of obtaining a local discrete space-times solution $q(\tau, \xi)$ (\ref{eq:local_sol_def}) using the LST-DG predictor (\ref{eq:lstdg_pred_eq_src}) and the accuracy of round-off errors in representing double-precision floating-point numbers within the intermediate calculations. The LST-DG predictor is implemented not by an exact solver for a system of linear algebraic equations, but by Picard iterations~\cite{ader_stiff_3, ader_stiff_4, ader_eff_blas}, which is used in the general case of nonlinear fluxes and sources (see~\cite{ader_dg_ideal_flows, ader_dg_gr_prd, fron_phys, ader_dg_axioms, exahype, ader_dg_PNPM, Zanotti_lects_2016}). The obtained results well confirm the theoretical results obtained in Section~\ref{sec:approx_anal} ``Approximation analysis'' and Section~\ref{sec:stab_anal} ``Stability analysis''.

\begin{figure}[h!]
\includegraphics[width=0.32\textwidth]{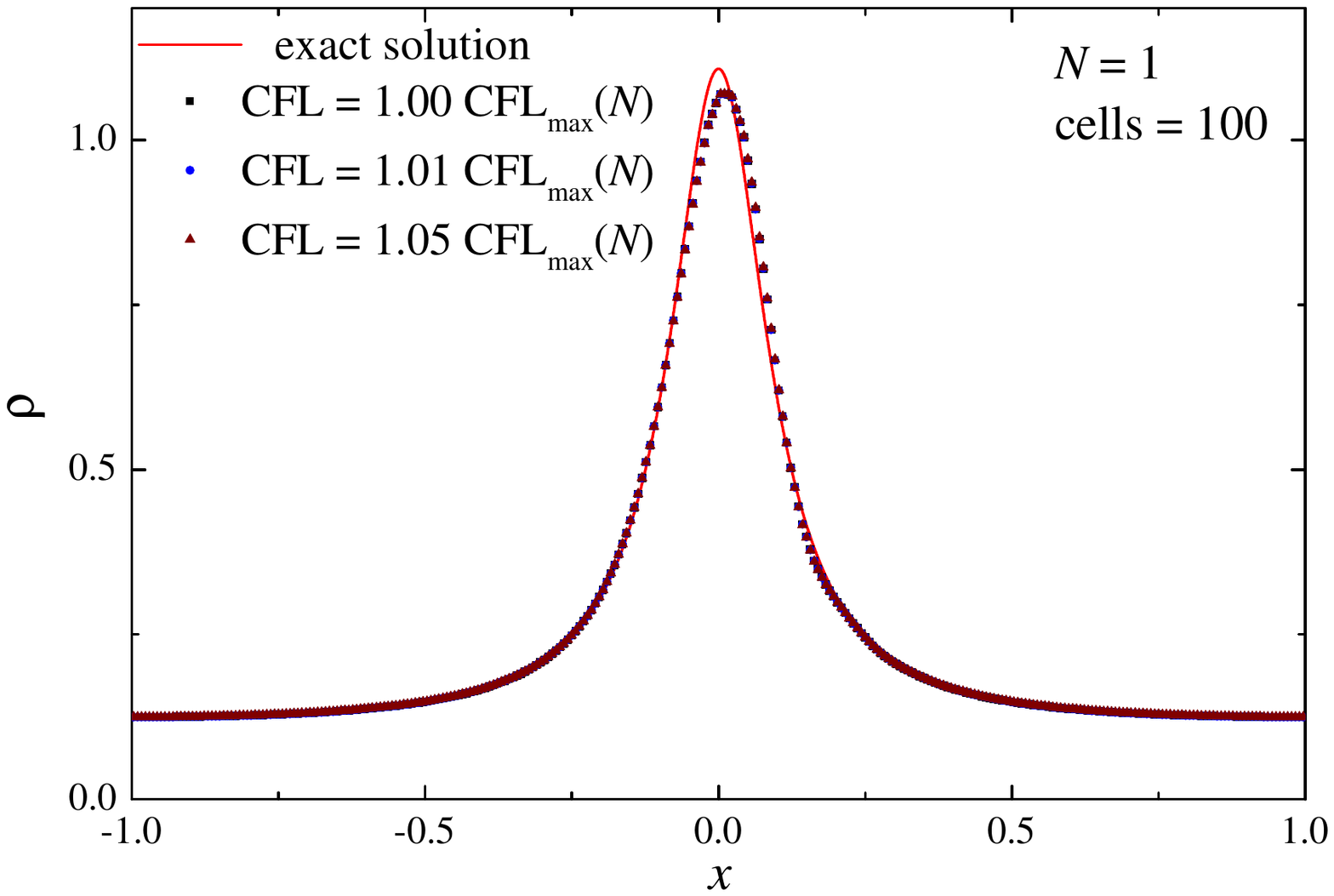}
\includegraphics[width=0.32\textwidth]{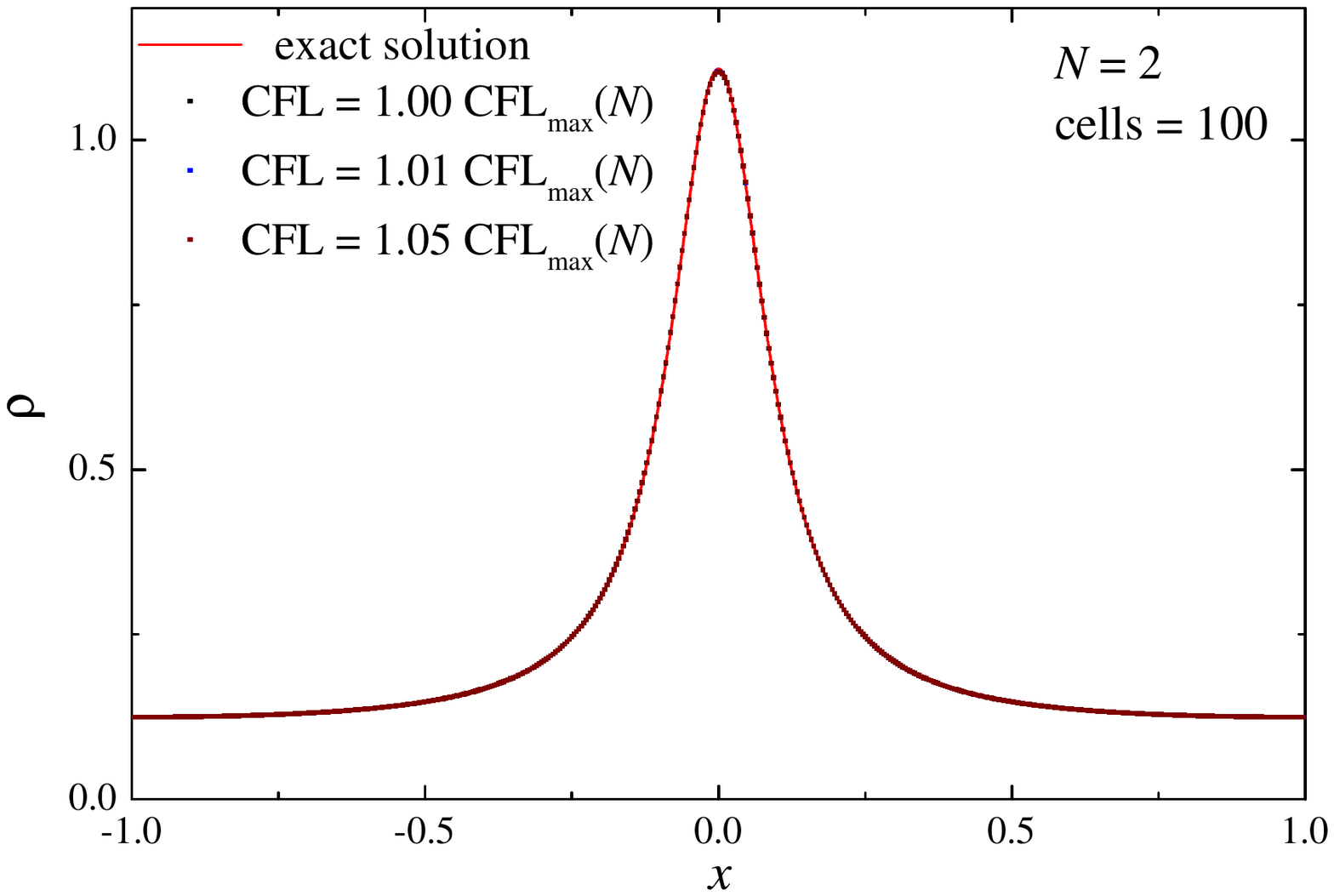}
\includegraphics[width=0.32\textwidth]{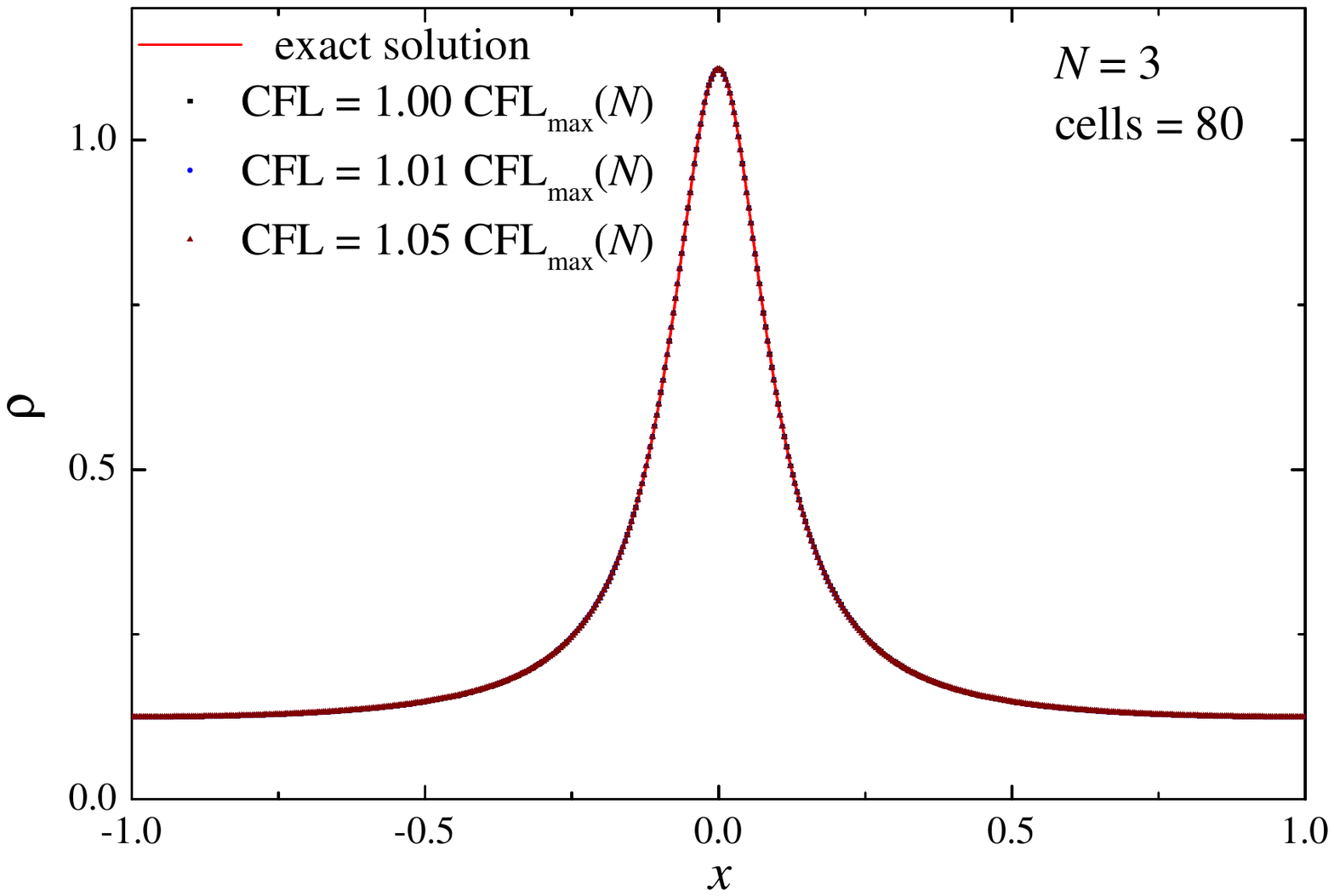}\\
\includegraphics[width=0.32\textwidth]{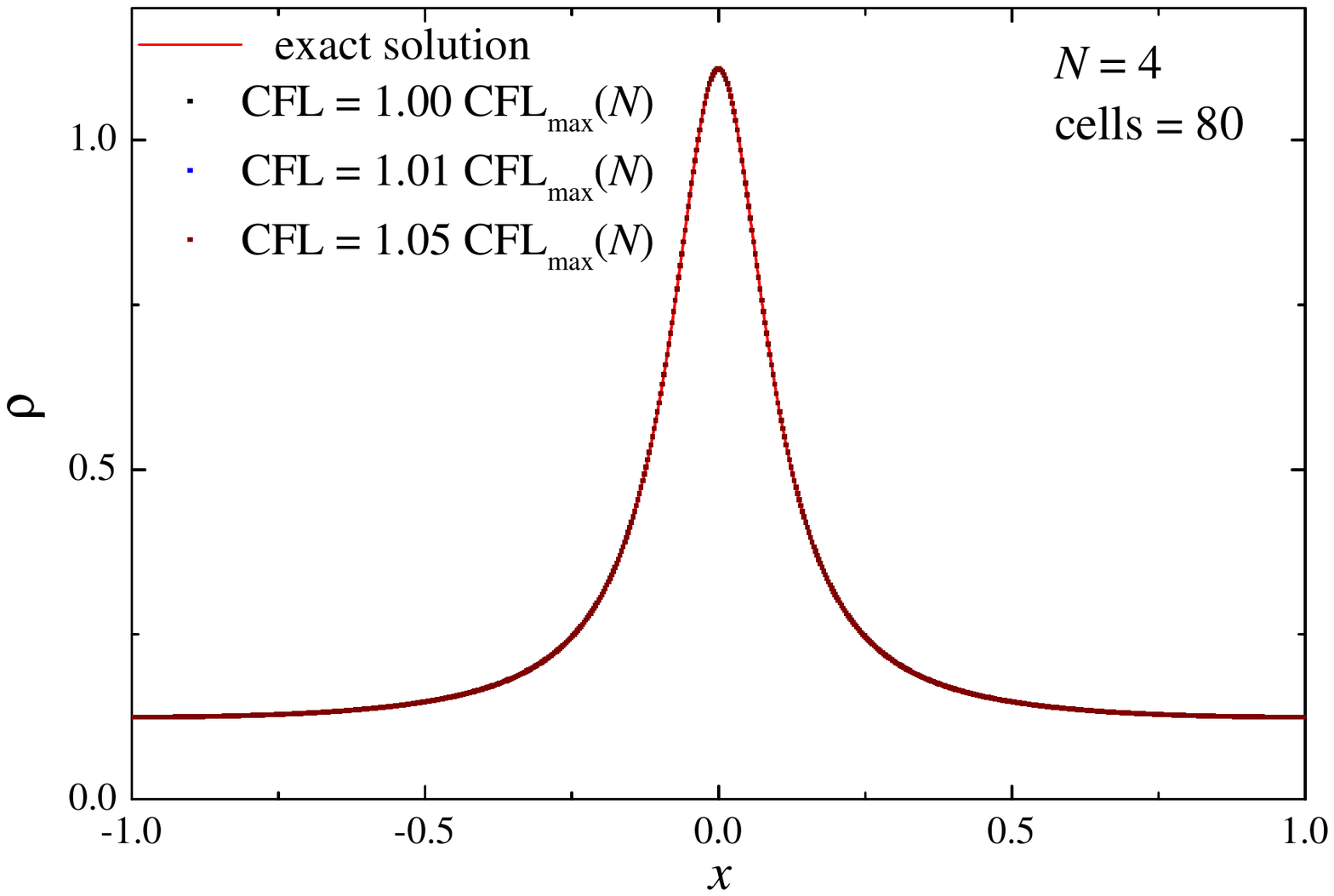}
\includegraphics[width=0.32\textwidth]{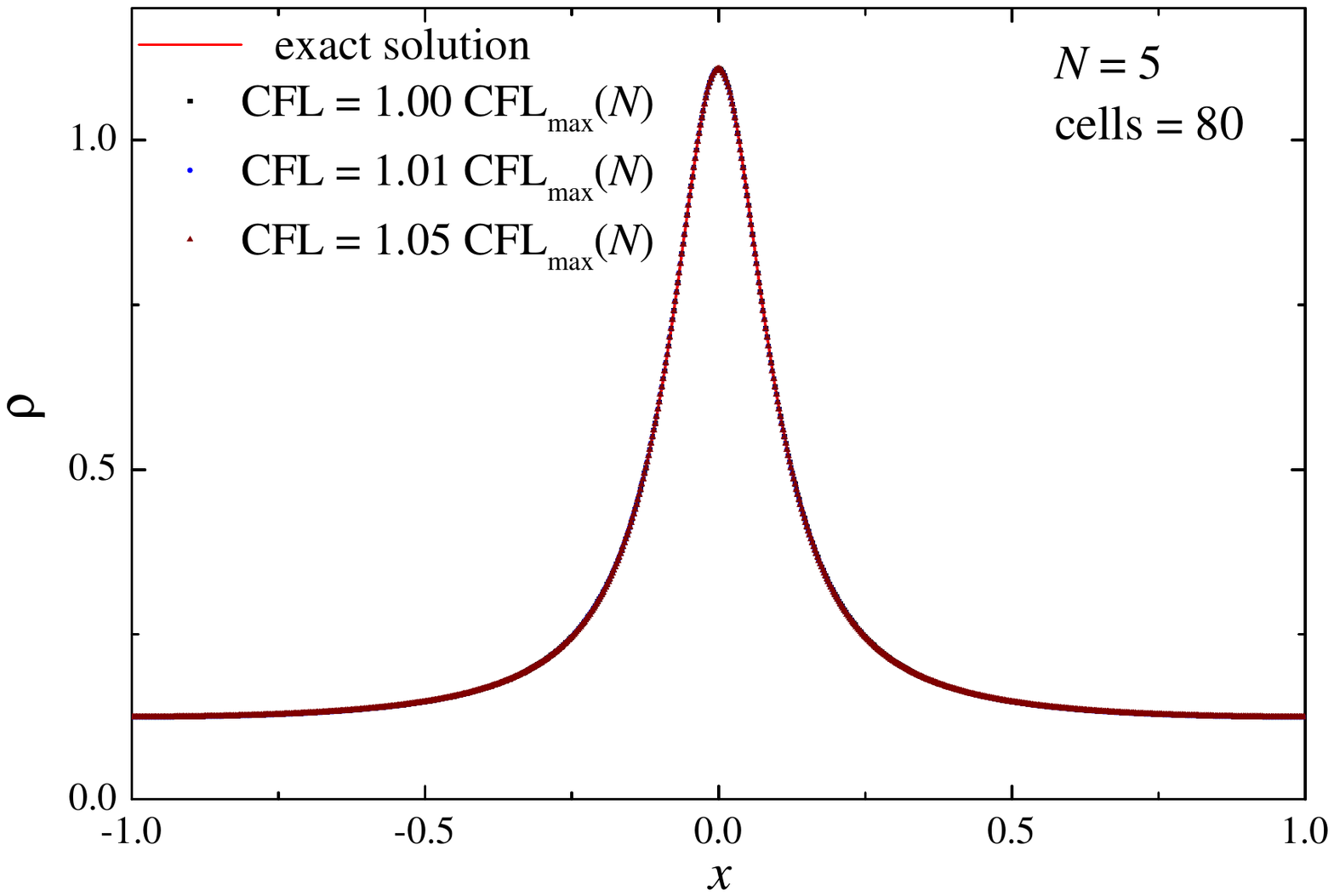}
\includegraphics[width=0.32\textwidth]{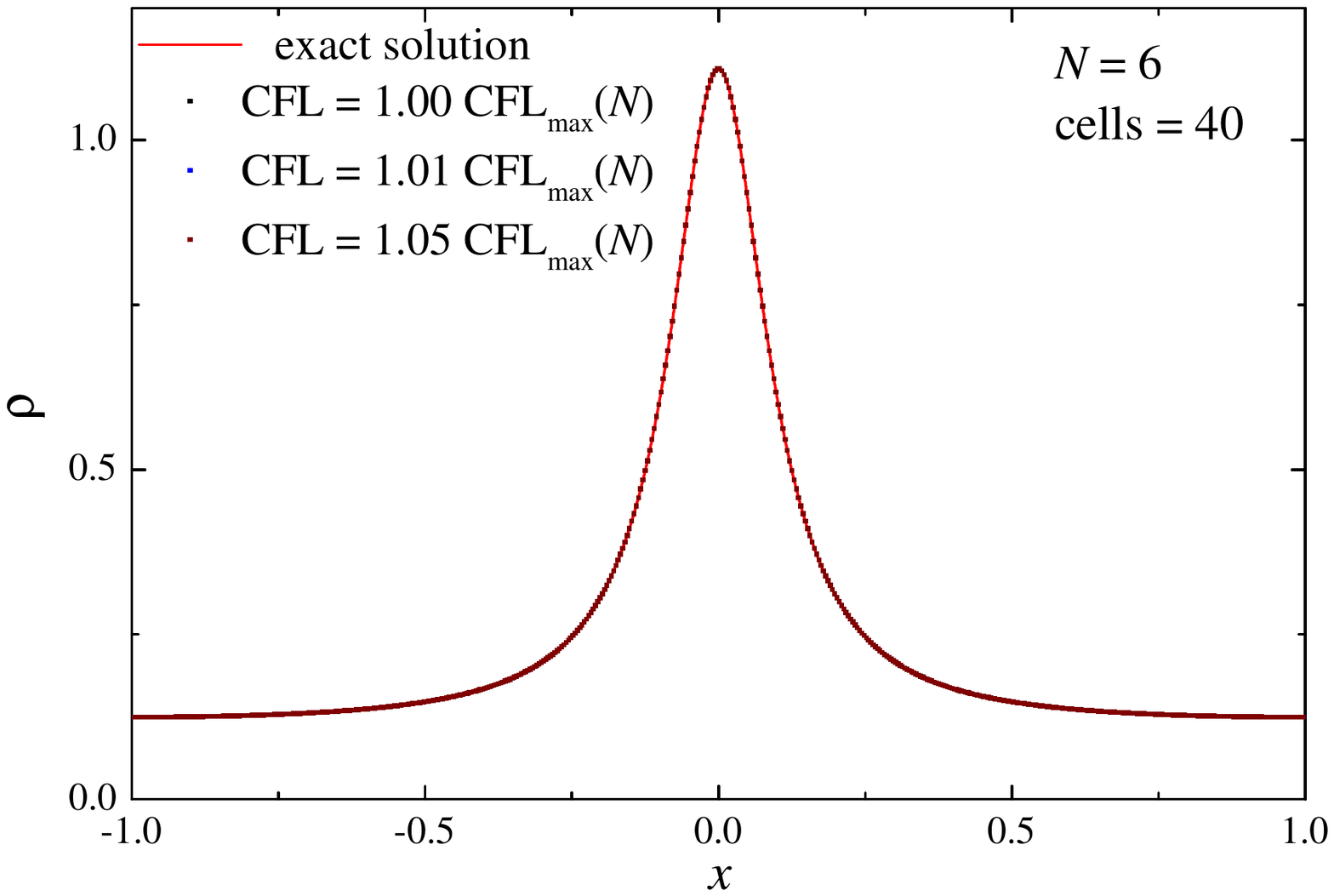}\\
\includegraphics[width=0.32\textwidth]{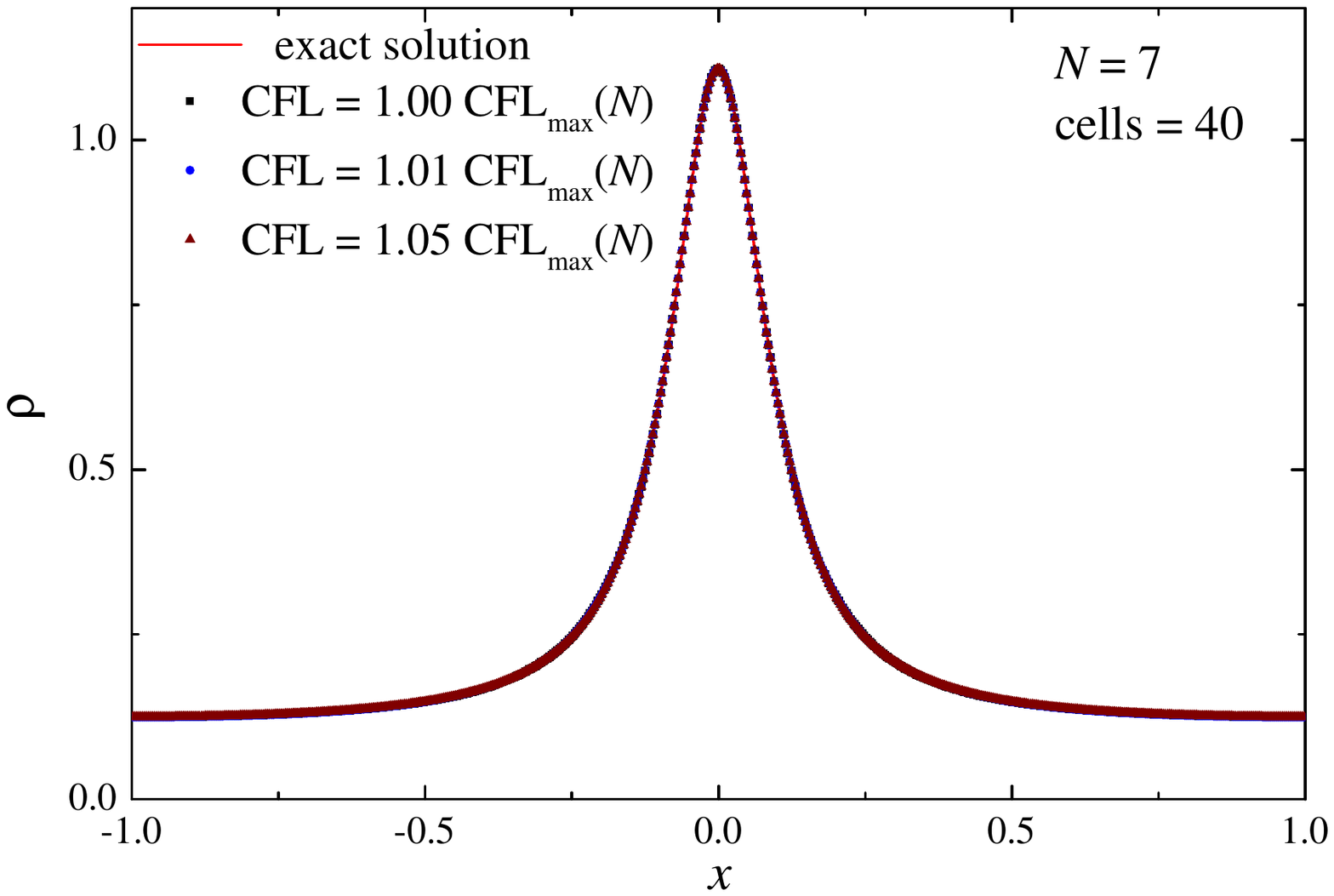}
\includegraphics[width=0.32\textwidth]{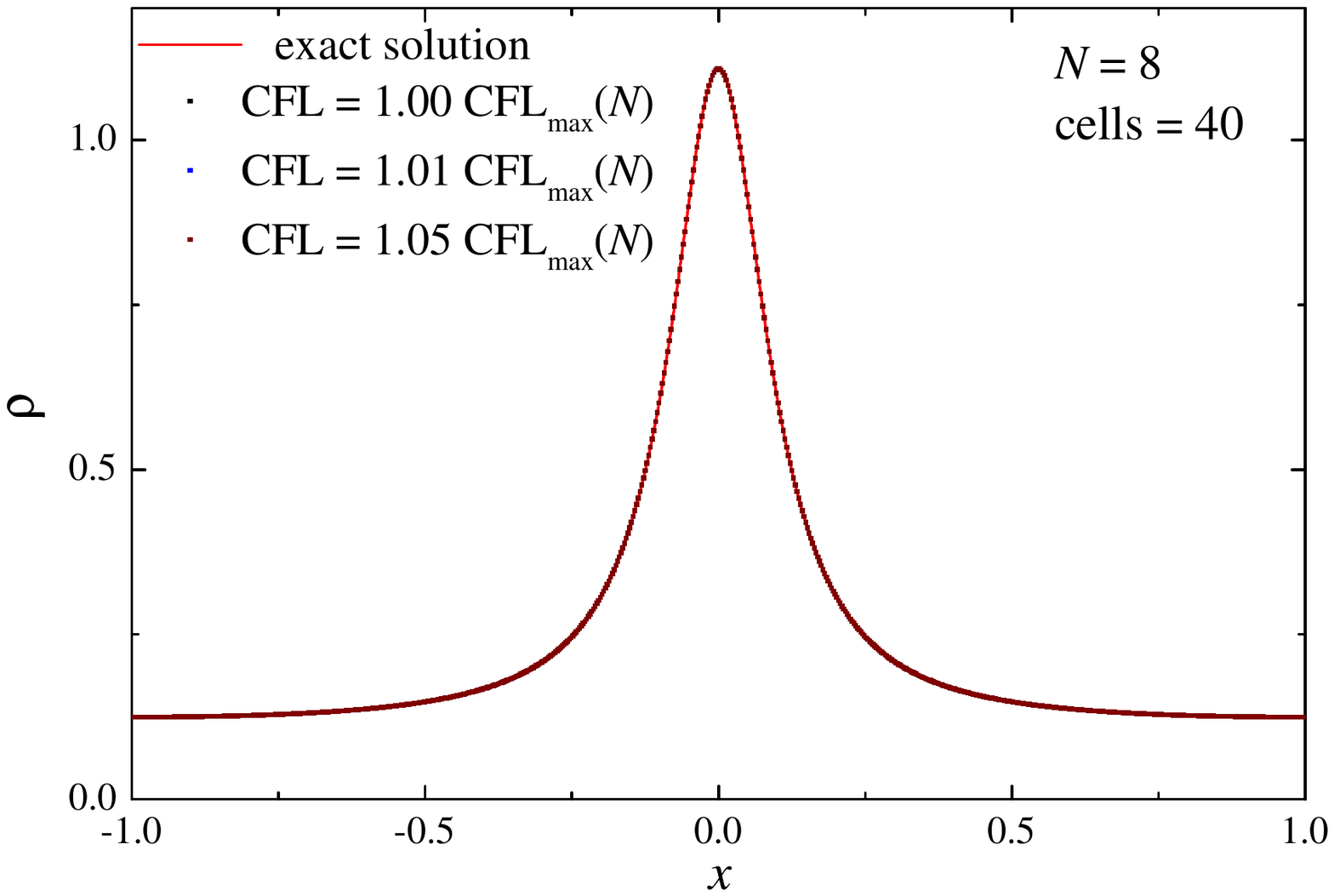}
\includegraphics[width=0.32\textwidth]{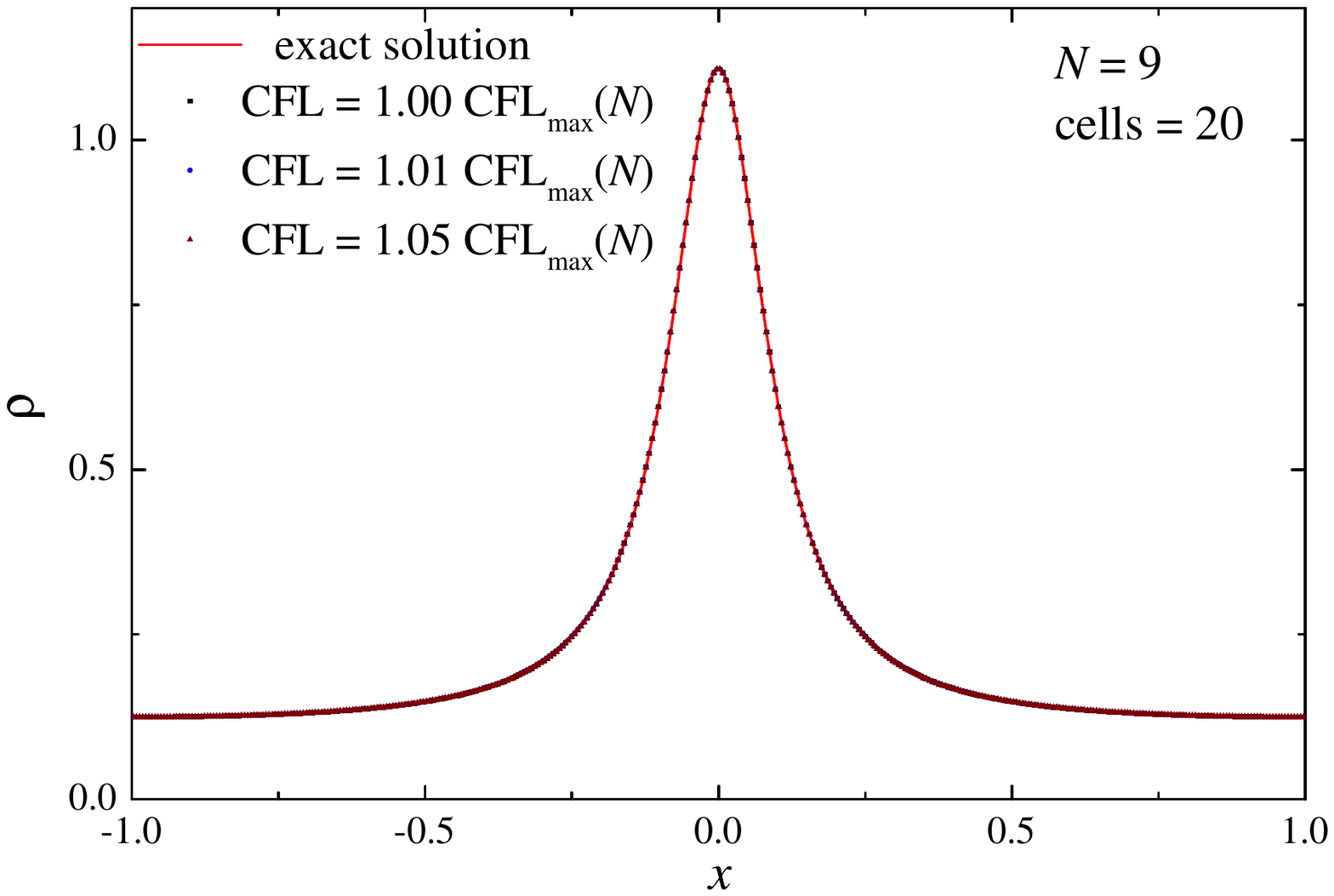}\\
\includegraphics[width=0.32\textwidth]{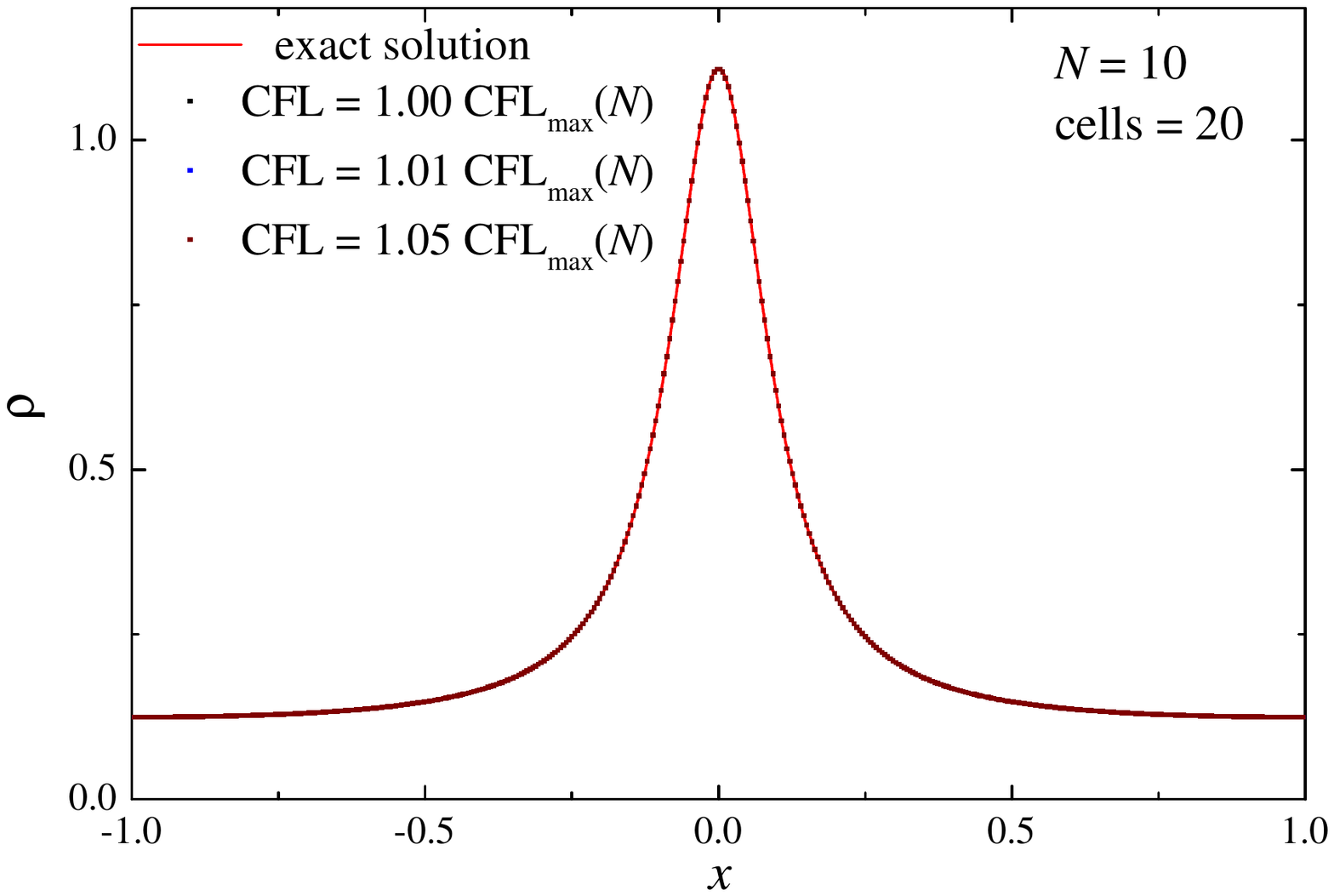}
\includegraphics[width=0.32\textwidth]{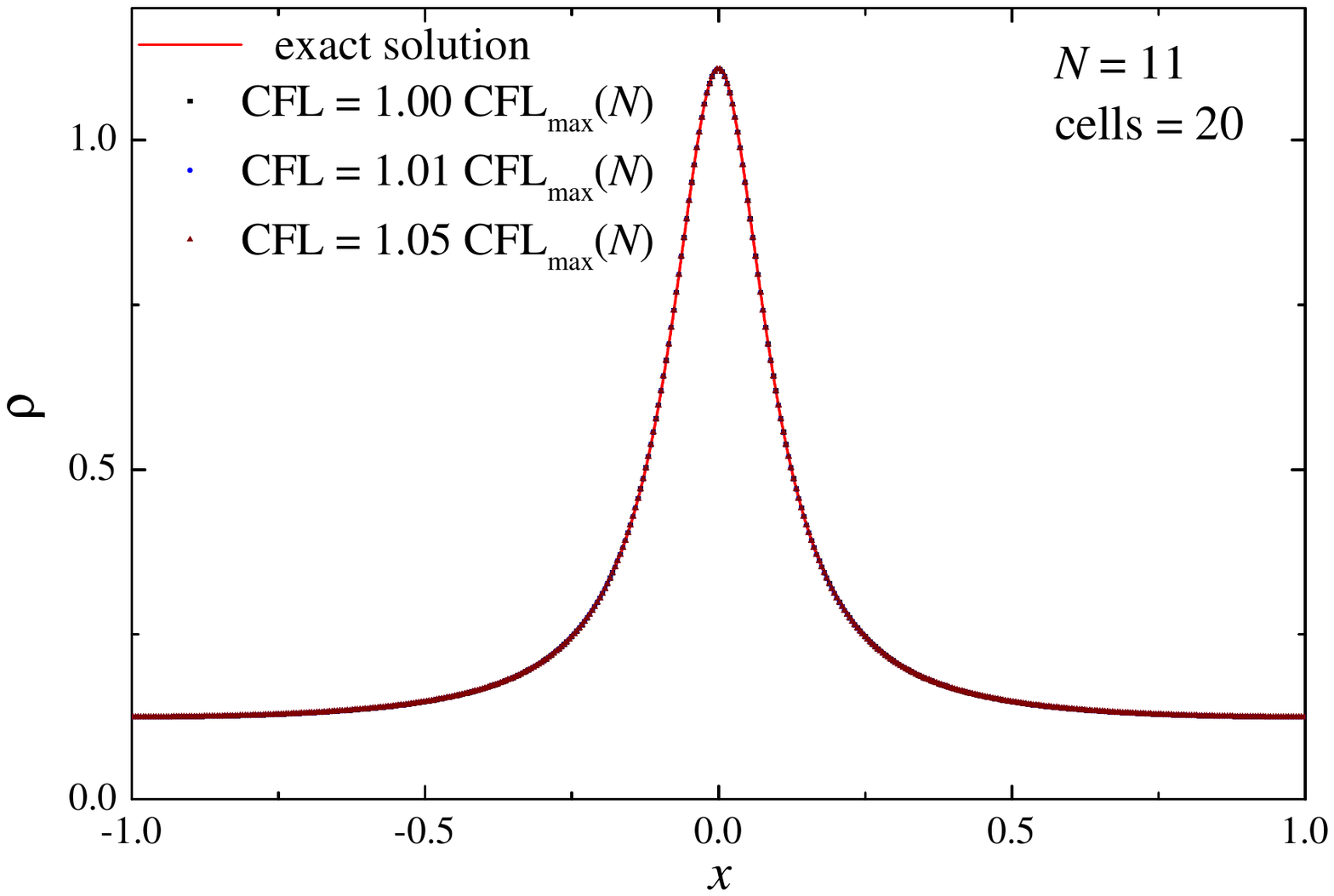}
\includegraphics[width=0.32\textwidth]{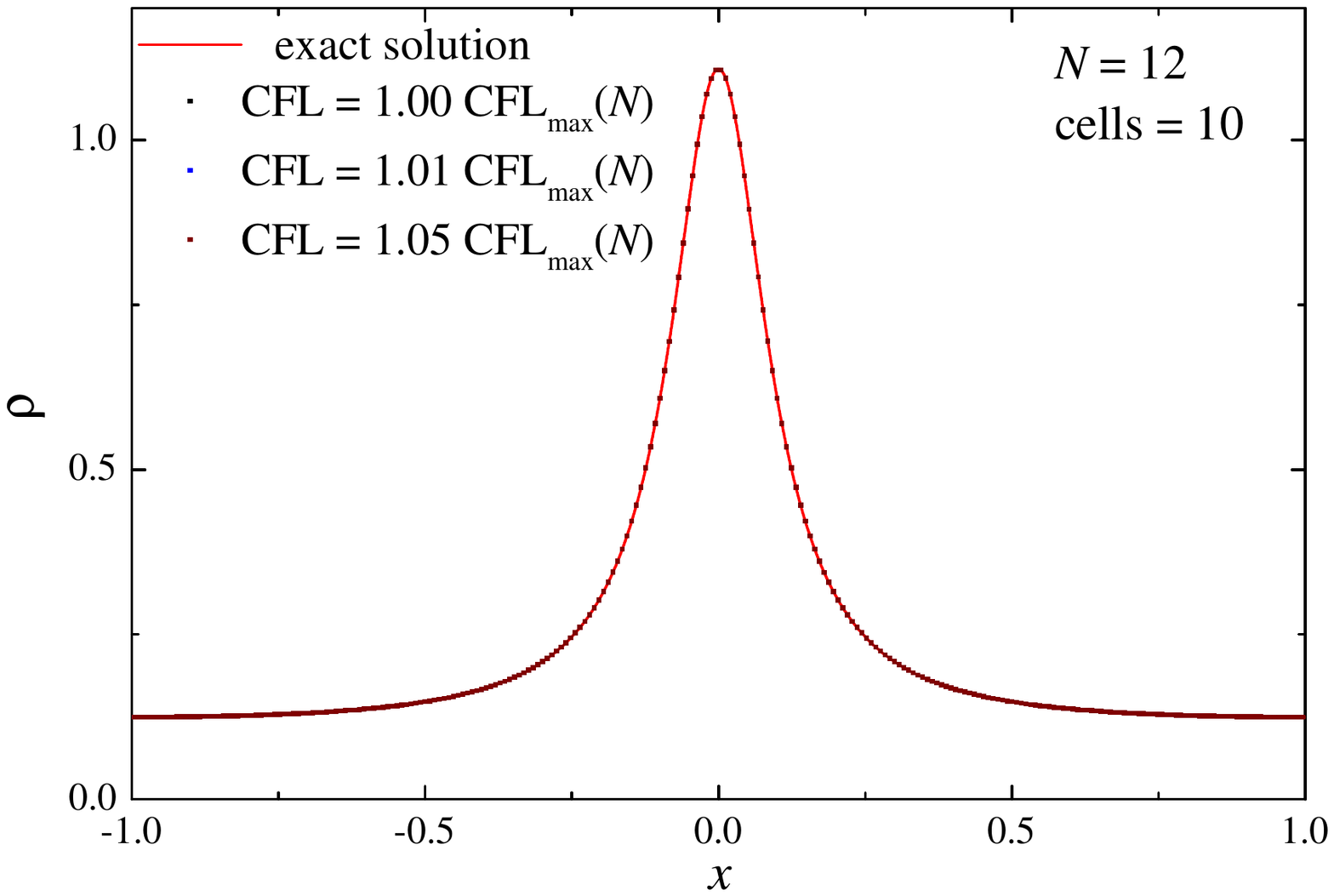}
\caption{%
Coordinate dependencies of the density $\rho(x, t_{f})$ to the Euler system of equation (\ref{eq:euler_eqs_def}) at the final time $t_{f}$, obtained by the ADER-DG numerical method with the LST-DG predictor, for the degrees $N = 1, \ldots, 12$ of the basis polynomials for a set of Courant number values $\mathrm{CFL}$ (left to right).
}
\label{fig:test_euler_lorentz_1d_cfls}
\end{figure}

\subsection{Nonlinear Euler equations}
\label{sec:app:euler}

\begin{table}[h!]
\centering
\normalsize
\caption{%
Empirical orders of convergence $p$ of the ADER-DG numerical method with the LST-DG predictor, obtained based on the solution to the profile (\ref{eq:lorentz_bell}) advection problem for Euler system of equations (\ref{eq:euler_eqs_def}), calculated in functional norms $\mathcal{L}_{1}$, $\mathcal{L}_{2}$, $\mathcal{L}_{\infty}$ of errors $\epsilon$ (\ref{eq:norms_of_errors}), for degrees $N = 1, \ldots, 6$ of basis polynomials for $\mathrm{CFL} = 1.00\, \mathrm{CFL}_{\rm max}(N)$ and $1.05\, \mathrm{CFL}_{\rm max}(N)$. Column Th. corresponds to $p = N+1$.
\label{tab:euler_conv_orders_degrees_1_6}
}
\setlength{\tabcolsep}{2.5pt}
\begin{tabular}{@{}|r|r|ccc|lll|ccc|lll|l|@{}}
\hline
& & \multicolumn{6}{c|}{$\mathrm{CFL} = 1.00\, \mathrm{CFL}_{\rm max}(N)$} & \multicolumn{6}{c}{$\mathrm{CFL} = 1.05\, \mathrm{CFL}_{\rm max}(N)$} & \\
\hline
$N$ & cells 
& $\epsilon_{\mathcal{L}_{1}}$ & $\epsilon_{\mathcal{L}_{2}}$ & $\epsilon_{\mathcal{L}_{\infty}}$ 
& $p_{\mathcal{L}_{1}}$ & $p_{\mathcal{L}_{2}}$ & $p_{\mathcal{L}_{\infty}}$
& $\epsilon_{\mathcal{L}_{1}}$ & $\epsilon_{\mathcal{L}_{2}}$ & $\epsilon_{\mathcal{L}_{\infty}}$ 
& $p_{\mathcal{L}_{1}}$ & $p_{\mathcal{L}_{2}}$ & $p_{\mathcal{L}_{\infty}}$ & Th. \\
\hline
1		&	$50$	&	3.42E-02	&	5.81E-02	&	1.61E-01	&	--		&	--		&	--		&	3.52E-02	&	5.96E-02	&	1.66E-01	&	--		&	--		&	--		&	2\\
						&	$100$	&	1.05E-02	&	2.12E-02	&	7.37E-02	&	1.70	&	1.46	&	1.13	&	1.09E-02	&	2.18E-02	&	7.64E-02	&	1.69	&	1.45	&	1.12	&	\\
						&	$150$	&	4.84E-03	&	1.03E-02	&	3.85E-02	&	1.91	&	1.78	&	1.60	&	5.02E-03	&	1.07E-02	&	3.99E-02	&	1.91	&	1.77	&	1.60	&	\\
						&	$200$	&	2.72E-03	&	5.95E-03	&	2.24E-02	&	2.00	&	1.91	&	1.88	&	2.82E-03	&	6.17E-03	&	2.33E-02	&	2.00	&	1.90	&	1.87	&	\\
\hline
2		&	$50$	&	3.53E-03	&	7.50E-03	&	2.94E-02	&	--		&	--		&	--		&	3.62E-03	&	7.68E-03	&	3.00E-02	&	--		&	--		&	--		&	3\\
						&	$100$	&	4.35E-04	&	1.06E-03	&	4.88E-03	&	3.02	&	2.82	&	2.59	&	4.51E-04	&	1.10E-03	&	5.04E-03	&	3.00	&	2.81	&	2.57	&	\\
						&	$150$	&	1.27E-04	&	3.11E-04	&	1.48E-03	&	3.04	&	3.03	&	2.94	&	1.32E-04	&	3.23E-04	&	1.54E-03	&	3.03	&	3.02	&	2.92	&	\\
						&	$200$	&	5.28E-05	&	1.29E-04	&	6.24E-04	&	3.04	&	3.04	&	3.01	&	5.51E-05	&	1.35E-04	&	6.50E-04	&	3.03	&	3.03	&	3.00	&	\\
\hline
3		&	$20$	&	7.61E-03	&	1.34E-02	&	4.71E-02	&	--		&	--		&	--		&	7.71E-03	&	1.35E-02	&	4.78E-02	&	--		&	--		&	--		&	4\\
						&	$40$	&	6.26E-04	&	1.47E-03	&	6.38E-03	&	3.60	&	3.19	&	2.88	&	6.43E-04	&	1.51E-03	&	6.54E-03	&	3.58	&	3.17	&	2.87	&	\\
						&	$60$	&	1.17E-04	&	3.03E-04	&	1.38E-03	&	4.13	&	3.89	&	3.78	&	1.22E-04	&	3.14E-04	&	1.42E-03	&	4.11	&	3.87	&	3.77	&	\\
						&	$80$	&	3.61E-05	&	9.46E-05	&	4.08E-04	&	4.10	&	4.05	&	4.24	&	3.75E-05	&	9.82E-05	&	4.23E-04	&	4.09	&	4.04	&	4.20	&	\\
\hline
4		&	$20$	&	1.55E-03	&	2.94E-03	&	1.09E-02	&	--		&	--		&	--		&	1.57E-03	&	2.98E-03	&	1.09E-02	&	--		&	--		&	--		&	5\\
						&	$40$	&	4.89E-05	&	1.34E-04	&	7.04E-04	&	4.99	&	4.46	&	3.96	&	5.07E-05	&	1.38E-04	&	7.26E-04	&	4.95	&	4.43	&	3.91	&	\\
						&	$60$	&	7.08E-06	&	1.92E-05	&	1.05E-04	&	4.77	&	4.78	&	4.69	&	7.38E-06	&	2.00E-05	&	1.09E-04	&	4.76	&	4.77	&	4.68	&	\\
						&	$80$	&	1.77E-06	&	4.83E-06	&	2.64E-05	&	4.83	&	4.80	&	4.81	&	1.84E-06	&	5.03E-06	&	2.74E-05	&	4.82	&	4.80	&	4.80	&	\\
\hline
5		&	$20$	&	2.76E-04	&	5.38E-04	&	2.03E-03	&	--		&	--		&	--		&	2.81E-04	&	5.49E-04	&	2.07E-03	&	--		&	--		&	--		&	6\\
						&	$40$	&	5.11E-06	&	1.39E-05	&	6.62E-05	&	5.76	&	5.27	&	4.94	&	5.34E-06	&	1.46E-05	&	6.91E-05	&	5.72	&	5.23	&	4.90	&	\\
						&	$60$	&	5.07E-07	&	1.45E-06	&	7.20E-06	&	5.70	&	5.59	&	5.47	&	5.29E-07	&	1.51E-06	&	7.48E-06	&	5.70	&	5.60	&	5.49	&	\\
						&	$80$	&	9.25E-08	&	2.68E-07	&	1.37E-06	&	5.92	&	5.87	&	5.77	&	9.63E-08	&	2.79E-07	&	1.41E-06	&	5.92	&	5.87	&	5.80	&	\\
\hline
6		&	$20$	&	4.62E-05	&	9.22E-05	&	3.79E-04	&	--		&	--		&	--		&	4.75E-05	&	9.51E-05	&	3.93E-04	&	--		&	--		&	--		&	7\\
						&	$40$	&	5.49E-07	&	1.58E-06	&	7.58E-06	&	6.39	&	5.86	&	5.64	&	5.71E-07	&	1.65E-06	&	7.95E-06	&	6.38	&	5.85	&	5.63	&	\\
						&	$60$	&	3.51E-08	&	1.05E-07	&	5.70E-07	&	6.78	&	6.68	&	6.38	&	3.65E-08	&	1.09E-07	&	5.93E-07	&	6.78	&	6.69	&	6.40	&	\\
						&	$80$	&	4.93E-09	&	1.47E-08	&	8.23E-08	&	6.83	&	6.84	&	6.73	&	5.13E-09	&	1.53E-08	&	8.56E-08	&	6.82	&	6.84	&	6.73	&	\\
\hline
\end{tabular}
\end{table}

\begin{table}[h!]
\centering
\normalsize
\caption{%
Empirical orders of convergence $p$ of the ADER-DG numerical method with the LST-DG predictor, obtained based on the solution to the profile (\ref{eq:lorentz_bell}) advection problem for Euler system of equations (\ref{eq:euler_eqs_def}), calculated in functional norms $\mathcal{L}_{1}$, $\mathcal{L}_{2}$, $\mathcal{L}_{\infty}$ of errors $\epsilon$ (\ref{eq:norms_of_errors}), for degrees $N = 7, \ldots, 12$ of basis polynomials for $\mathrm{CFL} = 1.00\, \mathrm{CFL}_{\rm max}(N)$ and $1.05\, \mathrm{CFL}_{\rm max}(N)$. Column Th. corresponds to $p = N+1$. ``\texttt{neg.}'' denotes a negative value.
\label{tab:euler_conv_orders_degrees_7_12}
}
\setlength{\tabcolsep}{2.5pt}
\begin{tabular}{@{}|r|r|ccc|lll|ccc|lll|l|@{}}
\hline
& & \multicolumn{6}{c|}{$\mathrm{CFL} = 1.00\, \mathrm{CFL}_{\rm max}(N)$} & \multicolumn{6}{c}{$\mathrm{CFL} = 1.05\, \mathrm{CFL}_{\rm max}(N)$} & \\
\hline
$N$ & cells 
& $\epsilon_{\mathcal{L}_{1}}$ & $\epsilon_{\mathcal{L}_{2}}$ & $\epsilon_{\mathcal{L}_{\infty}}$ 
& $p_{\mathcal{L}_{1}}$ & $p_{\mathcal{L}_{2}}$ & $p_{\mathcal{L}_{\infty}}$
& $\epsilon_{\mathcal{L}_{1}}$ & $\epsilon_{\mathcal{L}_{2}}$ & $\epsilon_{\mathcal{L}_{\infty}}$ 
& $p_{\mathcal{L}_{1}}$ & $p_{\mathcal{L}_{2}}$ & $p_{\mathcal{L}_{\infty}}$ & Th. \\
\hline
7		&	$20$	&	1.08E-05	&	2.26E-05	&	9.21E-05	&	--		&	--		&	--		&	1.11E-05	&	2.33E-05	&	9.38E-05	&	--		&	--		&	--		&	8\\
						&	$40$	&	6.16E-08	&	1.86E-07	&	9.76E-07	&	7.45	&	6.92	&	6.56	&	6.41E-08	&	1.94E-07	&	1.01E-06	&	7.44	&	6.91	&	6.53	&	\\
						&	$60$	&	2.67E-09	&	8.18E-09	&	4.31E-08	&	7.74	&	7.71	&	7.70	&	2.78E-09	&	8.52E-09	&	4.48E-08	&	7.74	&	7.70	&	7.69	&	\\
						&	$80$	&	2.75E-10	&	8.50E-10	&	4.78E-09	&	7.90	&	7.87	&	7.64	&	2.87E-10	&	8.85E-10	&	4.99E-09	&	7.89	&	7.87	&	7.63	&	\\
\hline
8		&	$20$	&	2.83E-06	&	6.57E-06	&	2.81E-05	&	--		&	--		&	--		&	2.90E-06	&	6.76E-06	&	2.89E-05	&	--		&	--		&	--		&	9\\
						&	$40$	&	6.60E-09	&	2.11E-08	&	1.28E-07	&	8.74	&	8.28	&	7.78	&	6.86E-09	&	2.19E-08	&	1.33E-07	&	8.72	&	8.27	&	7.77	&	\\
						&	$60$	&	1.90E-10	&	6.04E-10	&	3.83E-09	&	8.75	&	8.76	&	8.66	&	1.98E-10	&	6.28E-10	&	3.97E-09	&	8.74	&	8.76	&	8.66	&	\\
						&	$80$	&	1.48E-11	&	4.73E-11	&	3.02E-10	&	8.88	&	8.86	&	8.83	&	6.81E-11	&	1.38E-10	&	1.91E-09	&	3.71	&	5.26	&	2.54	&	\\
\hline
9		&	$20$	&	6.58E-07	&	1.63E-06	&	7.53E-06	&	--		&	--		&	--		&	6.76E-07	&	1.67E-06	&	7.78E-06	&	--		&	--		&	--		&	10\\
						&	$40$	&	7.46E-10	&	2.37E-09	&	1.26E-08	&	9.79	&	9.42	&	9.22	&	7.75E-10	&	2.46E-09	&	1.31E-08	&	9.77	&	9.41	&	9.21	&	\\
						&	$60$	&	1.46E-11	&	4.72E-11	&	2.66E-10	&	9.71	&	9.66	&	9.53	&	1.95E-11	&	5.04E-11	&	2.76E-10	&	9.09	&	9.59	&	9.53	&	\\
						&	$80$	&	8.59E-13	&	2.80E-12	&	1.62E-11	&	9.84	&	9.82	&	9.72	&	2.50E-10	&	6.85E-10	&	1.23E-08	&	\texttt{neg.}	&	\texttt{neg.}	&	\texttt{neg.}	&	\\
\hline
10		&	$10$	&	9.71E-05	&	1.64E-04	&	5.47E-04	&	--		&	--		&	--		&	9.82E-05	&	1.65E-04	&	5.52E-04	&	--		&	--		&	--		&	11\\
						&	$20$	&	1.35E-07	&	3.49E-07	&	1.81E-06	&	9.50	&	8.87	&	8.24	&	1.38E-07	&	3.59E-07	&	1.86E-06	&	9.47	&	8.85	&	8.21	&	\\
						&	$30$	&	1.66E-09	&	5.38E-09	&	3.21E-08	&	10.84	&	10.29	&	9.94	&	1.73E-09	&	5.60E-09	&	3.33E-08	&	10.81	&	10.26	&	9.92	&	\\
						&	$40$	&	8.35E-11	&	2.74E-10	&	1.58E-09	&	10.39	&	10.35	&	10.46	&	8.70E-11	&	2.85E-10	&	1.65E-09	&	10.39	&	10.35	&	10.45	&	\\
\hline
11		&	$10$	&	4.10E-05	&	6.92E-05	&	2.30E-04	&	--		&	--		&	--		&	4.15E-05	&	6.99E-05	&	2.32E-04	&	--		&	--		&	--		&	12\\
						&	$20$	&	2.54E-08	&	6.96E-08	&	3.65E-07	&	10.66	&	9.96	&	9.30	&	2.61E-08	&	7.19E-08	&	3.78E-07	&	10.63	&	9.92	&	9.26	&	\\
						&	$30$	&	2.45E-10	&	8.15E-10	&	4.94E-09	&	11.45	&	10.97	&	10.61	&	2.54E-10	&	8.48E-10	&	5.13E-09	&	11.42	&	10.95	&	10.61	&	\\
						&	$40$	&	9.14E-12	&	3.14E-11	&	1.97E-10	&	11.43	&	11.32	&	11.20	&	9.90E-12	&	3.27E-11	&	2.04E-10	&	11.28	&	11.32	&	11.21	&	\\
\hline
12		&	$10$	&	1.64E-05	&	2.71E-05	&	8.67E-05	&	--		&	--		&	--		&	1.65E-05	&	2.73E-05	&	8.71E-05	&	--		&	--		&	--		&	13\\
						&	$20$	&	4.79E-09	&	1.40E-08	&	7.33E-08	&	11.74	&	10.91	&	10.21	&	4.95E-09	&	1.46E-08	&	7.63E-08	&	11.70	&	10.87	&	10.16	&	\\
						&	$30$	&	3.76E-11	&	1.25E-10	&	7.18E-10	&	11.95	&	11.65	&	11.41	&	3.90E-11	&	1.30E-10	&	7.49E-10	&	11.95	&	11.64	&	11.40	&	\\
						&	$40$	&	1.08E-12	&	3.66E-12	&	2.31E-11	&	12.35	&	12.27	&	11.94	&	1.67E-12	&	4.04E-12	&	2.43E-11	&	10.97	&	12.06	&	11.92	&	\\
\hline
\end{tabular}
\end{table}

The results presented in the previous Subsection~\ref{sec:app:lin_adv} ``Linear advection equation'' pertain only to the accuracy and stability of ADER-DG numerical method with the LST-DG predictor for linear problems (\ref{eq:adv_eq_src}). Clearly, the results of linear analysis based on the study of the linear advection equation (\ref{eq:adv_eq_src}) can be extended to the more general nonlinear case of systems of quasilinear hyperbolic equations, since they can be represented as a system of advection equations for Riemann invariants (though, of course, a quasilinear system of equations). However, certain differences in the results from the linear case can be expected in nonlinear case. This is primarily due to the fact that for quasilinear systems, it is common to use approximate Riemann solvers~\cite{Toro_solvers_2009} in calculations, which can exhibit significant numerical dissipation, which can alter the boundaries of the stability region $\mathrm{CFL}_{\rm max}(N)$ of the ADER-DG numerical method with the LST-DG predictor. This Subsection considers a nonlinear problem, for which the Euler system of equations, a common system of quasilinear hyperbolic equations, is chosen. The Euler system of equations is chosen in the following form:
\begin{equation}\label{eq:euler_eqs_def}
\begin{split}
\pderfoper{t}\left[
\begin{array}{c}
\rho\\ \rho u\\ \varepsilon
\end{array}
\right] + \pderfoper{x}\left[
\begin{array}{c}
\rho u\\ \rho u^{2} + p\\ (\varepsilon + p)u
\end{array}
\right] = 0,\qquad
\begin{split}
&(x, t)\in\Omega\times\mathcal{T},\quad
\Omega\subset\mathcal{R},\quad
\mathcal{T} = \left[t_{0}, t_{f}\right]\subset\mathcal{R},\\
&\varepsilon = e + \frac12 \rho u^{2},\quad
p = (\gamma - 1) e,\quad
\gamma = 1.4,
\end{split}
\end{split}
\end{equation}
where $\rho$ is the density, $u$ is the flow velocity, $p$ is the pressure, $\varepsilon$ is the volumetric energy density, $e$ is the volumetric internal energy density, and $\gamma$ is the adiabatic exponent. Similarly to the previous Subsection~\ref{sec:app:lin_adv} ``Linear advection equation'', the problem of advection of a density profile, defined similar to (\ref{eq:lorentz_bell}), is studied:
\begin{equation}\label{eq:euler_init_cond}
\rho(x, 0) = \rho_{0} + \frac{\pi\alpha A}{L}\cfrac{\sinh\left(\cfrac{2\pi\alpha}{L}\right)}{\cosh\left(\cfrac{2\pi\alpha}{L}\right) - \cos\left(\cfrac{2\pi x}{L}\right)},\quad
p(x, 0) = p_{0},\quad u(x, 0) = u_{0},
\end{equation}
where now the constant flow velocity $u_{0}$ is the advection velocity. The following simulation parameters are chosen:
\begin{equation}
\begin{split}
&\Omega = [-1.0, +1.0],\quad L = 2.0,\quad A = 1.0,\quad \alpha = 0.1,\\
&u_{0} = 1.0,\quad \rho_{0} = 0.1,\quad p_{0} = 0.1,\quad t_{0} = 0.0,\quad t_{f} = 4.0.
\end{split}
\end{equation}
Calculations are performed for degrees $N = 1, \ldots, 12$ of basis polynomials $\{\varphi_{p}\}_{p}$ (\ref{eq:phi_def}). The boundary conditions is periodic. The time $t_{f} - t_{0} \equiv t_{f} = 4.0$ is chosen so that the periodic coordinate dependence of the solution $u(x, t)$ transferred the solution domain $\Omega$ exactly twice. The most important difference between this case and the one studied in the previous Subsection~\ref{sec:app:lin_adv} ``Linear advection equation'', except for the nonlinearity of the system of equations, is the use of the HLLE solver~\cite{HLLE_solver_1, HLLE_solver_2, Toro_solvers_2009} to calculate the fluxes, which is known for its high dissipativity and stability. It should be noted that similar results are obtained using Rusanov solver~\cite{Rusanov_solver, Toro_solvers_2009}, sometimes referred to as the local Lax-Friedrichs flux~\cite{ader_dg_ideal_flows}.

Examples of the obtained coordinate dependencies of the density $\rho(x, t_{f})$ to the Euler system of equations (\ref{eq:euler_eqs_def}) at the final time $t_{f}$, obtained by the ADER-DG numerical method with the LST-DG predictor, for degrees $N = 1, \ldots, 12$ of the basis polynomials $\{\varphi_{p}\}_{p}$ are shown in Figure~\ref{fig:test_euler_lorentz_1d_cfls}. The Courant numbers $\mathrm{CFL}$ are chosen equal to $1.00\, \mathrm{CFL}_{\rm max}(N)$, $1.01\, \mathrm{CFL}_{\rm max}(N)$ and $1.05\, \mathrm{CFL}_{\rm max}(N)$. For all Courant numbers $\mathrm{CFL}$ in the stability region $\mathrm{CFL} \leqslant \mathrm{CFL}_{\rm max}(N)$, the numerical solution is correct, and unlike the case of the linear advection equation (\ref{eq:adv_eq_src}), in this case the stability boundary is wider --- a correct numerical solution is obtained up to a Courant number $1.05\,\mathrm{CFL}_{\rm max}(N)$. This feature is not likely due to the nonlinearity of the Euler system of equations(\ref{eq:euler_eqs_def})  compared to the linearity of the advection equation (\ref{eq:adv_eq_src}), but to the use of a dissipative HLLE flux, which suppresses (``damps'') eigenvalues that slightly exceed the stability region $|\lambda| > 1$ characteristic of the linear case. For Courant numbers $\mathrm{CFL} > 1.05\,\mathrm{CFL}_{\rm max}(N)$, coordinate discretization steps $\Dx$ are always existed for which the solution is obtained in the form \texttt{nan}. Unlike the results presented in Figures~\ref{fig:test_adveq_lorentz_1d_cfls_degrees_1_6} and~\ref{fig:test_adveq_lorentz_1d_cfls_degrees_7_12} for the numerical solution of the linear advection equation, in the case of the Euler system of equations, it is not possible to obtain coordinate dependencies of the solution with an ``accordion'' character --- when oscillations of the numerical solution arose, a non-physical negative density $\rho < 0$ immediately arose, as a result of which \texttt{nan} arose from conservative terms in the internal calculations of non-linear flux terms.

Empirical convergence orders $p$ are calculated based on the dependence $\epsilon \propto \Dx^{p}$ of the numerical solution error $\epsilon$ on the coordinate grid step $\Dx$. Error calculations are performed based on density $\rho(x, t_{f})$ using three classical functional error norms $\mathcal{L}_{1}$, $\mathcal{L}_{2}$, $\mathcal{L}_{\infty}$ (\ref{eq:norms_of_errors}). Tables~\ref{tab:euler_conv_orders_degrees_1_6} and~\ref{tab:euler_conv_orders_degrees_7_12} present examples of the calculated empirical convergence orders $p$ for Courant numbers $\mathrm{CFL} = 1.00\, \mathrm{CFL}_{\rm max}(N)$ (boundary of stability region for linear case) and $1.05\, \mathrm{CFL}_{\rm max}(N)$ for degrees $N = 1, \ldots, 12$ of the basis polynomials: in Table~\ref{tab:euler_conv_orders_degrees_1_6} for degrees $N = 1, \ldots, 6$ of the basis polynomials, in Table~\ref{tab:euler_conv_orders_degrees_7_12} for degrees $N = 7, \ldots, 12$ of the basis polynomials. The presented results for empirical convergence orders $p$ are in good agreement with the theoretical estimates $p = N+1$ obtained in Section~\ref{sec:approx_anal} ``Approximation analysis'', do not systematically differ quantitatively from the empirical convergence orders for the linear case in Tables~\ref{tab:adveq_conv_orders_degrees_1_6} and~\ref{tab:adveq_conv_orders_degrees_7_12}.

The obtained results demonstrate that the results of the approximation analysis of the ADER-DG numerical method with the LST-DG predictor and the approximation estimates for the linear case are also adequately applied to a nonlinear system of equations, such as the Euler system of equations. The stability analysis results of the ADER-DG numerical method with the LST-DG predictor obtained in the linear case are slightly underestimated, but by no more than 5\%, and this is due to the dissipativity of approximate Riemann solvers used for nonlinear systems.

\section{Conclusion}
\label{concl}

In conclusion, it should be noted that this study yielded very accurate values for the stability boundary of the Courant numbers $\mathrm{CFL}_{\rm max}$ of the ADER-DG numerical method with the LST-DG predictor within the framework of the Von Neumann stability criterion (spectral stability criterion) for arbitrary degrees $N$ of basis polynomials $\{\varphi\}$ (\ref{eq:phi_def}). It was found that, in the linear case, instability occurs precisely when one of the matrix $\mathrm{R}(\mathrm{CFL}, \theta)$ (\ref{eq:r_matrix_in_matrix_form}) of the evolution operator $R$ eigenvalues $\lambda_{k}$ reaches $\lambda = -1$, regardless of the phase $\theta$ (\ref{eq:r_matrix_def}) of the numerical solution. It is important to note that the overall instability of the numerical solution is associated with the eigenvalue $\lambda = -1$, which begins to significantly exceed unity, while the phase determines $\theta = k\Dx$ the harmonic wave number $k$ for which instability occurs.

Based on the obtained and substantiated empirical reasoning, it was constructed a rigorous mathematical theory (generally, with some limitations, free from the choice of specific basis polynomials $\{\varphi\}$ (\ref{eq:phi_def})) and significantly simplify the stability condition $\rho[\mathrm{R}(\mathrm{CFL}, \theta)] \leqslant 1$, reducing it from the complex problem of finding the maximum spectral radius $\rho[\mathrm{R}(\mathrm{CFL}, \theta)]$ of the matrix $\mathrm{R}(\mathrm{CFL}, \theta)$, dependent on the continuous phase $\theta\in[0, 2\pi]$, to the problem of calculating the roots of $(N+1)$-degree polynomials $F_{1}(\mathrm{CFL}) = 0$ and $F_{2}(\mathrm{CFL}) = 0$ (\ref{eq:F1_and_F2_def}) in the Courant number $\mathrm{CFL}$. This is technically very simple and can be performed with arbitrarily high accuracy, even without significant computational resources. It was found that, despite instability within the expected boundaries of the Courant number when one of the matrix $\mathrm{R}(\mathrm{CFL}, \theta)$ (\ref{eq:r_matrix_in_matrix_form}) eigenvalues $\lambda_{k}$ reaches $\lambda = -1$, small unstable eigenvalues with phase $\arg\lambda_{k} \in [-\pi/4, +\pi/4]$ still arise. However, their absolute values $|\lambda_{k}|$ exceeding one are very small, and the corresponding eigenvalues $\lambda_{k}$ have only a positive real part $\Re{\lambda_{k}} > 0$, which does not lead to the development of unstable oscillations.

The boundary values of the Courant numbers $\mathrm{CFL}_{\rm max}(N)$ for the ADER-DG numerical method with the LST-DG predictor were calculated for degrees $N = 1, \ldots, 12$ of the basis polynomials. A comparison of the obtained results with existing literature data~\cite{ader_dg_stab, ader_dg_PNPM, PNPM_DG_2008} revealed differences in $\mathrm{CFL}_{\rm max}$ for certain degrees $N$ of the basis polynomials, especially for large degrees $N \sim 9$, that could be significant for the selection of calculation parameters using the ADER-DG numerical method with the LST-DG predictor. It was shown that widely used existing estimates of the boundary value of the Courant number $\mathrm{CFL}_{\rm max}(N)$ are somewhat overestimated, necessitating the introduction of an ``effective Courant number'' $\mathrm{C}$ (\ref{eq:old_eff_cfl}), which will be significantly less than unity. The choice of this effective Courant number $\mathrm{C}$ is related precisely to the discrepancy between this common estimate and the correct one $\mathrm{CFL}_{\rm max}(N)$. An interesting qualitative asymptotic $\mathrm{CFL}_{\rm max}(N) \propto 1/(N+1)^{2}$ (\ref{eq:cfls_asymp_est}) dependence of the boundary value of the Courant number $\mathrm{CFL}_{\rm max}(N)$ for the ADER-DG numerical method with the LST-DG predictor is obtained. This allows for computational cost estimates when using very high-degree basis polynomials and is in relatively good agreement with the quantitative results $\mathrm{CFL}_{\rm max}(N)$ for $N = 1, \ldots, 12$ obtained in this work.

A rigorous direct proof of the approximation of the ADER-DG numerical method with the LST-DG predictor is presented. It is shown that, despite the implicit nature of the LST-DG predictor in the linear case, the local discrete space-time solution is in fact calculated using an explicit scheme with a numerical evolution operator representing the first $N+1$ terms of the Taylor series for the solution. Approximation orders $p = N+1$ for arbitrary degrees $N$ of basis polynomials are rigorously derived.

A set of numerical experiments was carried out to apply the ADER-DG numerical method with the LST-DG predictor to solving a linear advection equation (\ref{eq:adv_eq_src}) to verify the obtained theoretical results. Results were obtained for degrees $N = 1, \ldots, 12$ of the basis polynomials. The results obtained in these calculations clearly confirm the theoretical results. In particular, even a 1\% excess of the Courant number $\mathrm{CFL}$ above the boundary value $\mathrm{CFL}_{\rm max}(N)$ immediately leads to significant instability of the numerical solution. Empirical convergence orders $p$ were obtained, which are in good agreement with the theoretical results $p = N+1$ for the approximation orders.

To determine the applicability of the obtained results to ``real-world problems'', a set of numerical experiments was carried out for the Euler system of equations (\ref{eq:euler_eqs_def}) as a good special case of a nonlinear problem. It was shown that the obtained quantitative estimates of the boundary value of the Courant number $\mathrm{CFL}_{\rm max}(N)$ are also applicable in this nonlinear case; however, they are somewhat underestimated --- by no more than 5\%. This is due to the use of the HLLE approximate Riemann solver, which is characterized by strong diffusivity and high stability, so it suppresses unstable harmonics until their increment becomes too large. In general, these features of the transfer of stability analysis results from the case of linear problems (\ref{eq:adv_eq_src}) to nonlinear problem (\ref{eq:euler_eqs_def}) are expected. In the nonlinear case (\ref{eq:euler_eqs_def}), empirical convergence orders $p$ were also obtained, which are in good agreement with the theoretical results $p = N+1$ for the orders of approximation.

\section*{Acknowledgments}
The author would like to thank Popova A.P. for help in correcting the English text.


\end{document}